\documentclass[11pt,twoside]{cernrep}
\usepackage{graphicx,epsfig}
\usepackage{here,cite}
\usepackage{rotating} 

\begin{document}

\thispagestyle{empty}

\vspace*{-2.3cm}
\begin{tabbing}
\hspace*{12.3cm} \= CERN-2003-002-corr\\
\> 10 October 2003\\
\> Experimental\\
\> Physics Division
\end{tabbing}

\vspace*{2.7cm}

\hspace*{-7mm}
{\large\bf ORGANISATION EUROP\'EENNE POUR LA RECHERCHE NUCL\'EAIRE}

\vspace*{2mm}

\hspace*{-7mm}
{\LARGE\bf CERN} {\large\bf EUROPEAN ORGANIZATION FOR NUCLEAR RESEARCH}

\vspace*{4.5cm}

\begin{center}
\hspace*{-15mm}
{\Large\bf THE CKM MATRIX AND THE UNITARITY TRIANGLE}
\end{center}

\vspace*{1.5cm}

\begin{center}
\hspace*{-15mm}
{\large Based on the workshop held at CERN, 13--16 February 2002}
\end{center}

\vspace*{1.5cm}

{\large 
\begin{tabbing}
\hspace*{8cm} \= {\Large\bf PROCEEDINGS} \\[1cm]
%
%
\> Editors: M.~Battaglia,~ A.J.~Buras,\\
\> ~~~~~~~~~~~~~~P.~Gambino and  A.~Stocchi
\end{tabbing}
}

\vspace*{4cm}

\begin{center}
\hspace*{-25mm} {GENEVA} \\
\hspace*{-25mm} {2003}
\end{center}

\renewcommand{\thepage}{\roman{page}}

\vglue4cm

\begin{center}
{\Large\bf Abstract}
\end{center}

\vspace{7mm}

\noindent \noindent 
This report contains the results of the Workshop on the CKM Unitarity
Triangle, held at CERN on 13--16 February 2002, to study the determination
of the Cabibbo--Kobayashi--Maskawa (CKM) matrix from the available data of
K, D, and B physics. This is a coherent document with chapters covering
the determination of CKM elements from tree-level decays and K- and
B-meson mixing and the global fits of the unitarity triangle parameters.
The impact of future measurements is also discussed.


\newpage

\newcommand{\dmd}{\Delta M_d}
\newcommand{\dms}{\Delta M_s}
\newcommand{\dmq}{\Delta M_q}     

\renewcommand{\baselinestretch}{1.3}
\renewcommand{\thefootnote}{\fnsymbol{footnote}}
 \def\OEE{\Omega_{\rm IB}}

\newcommand {\bb} {\ifmmode {b\bar{b}}\else {$b\bar{b}$}\fi}
\newcommand {\cc} {\ifmmode {c\bar{c}}\else {$c\bar{c}$}\fi}
\newcommand {\uds}{\ifmmode {uds}\else {$uds$}\fi}
\newcommand{\ra} {\rightarrow}

\newcommand{\errparen}[2]{%
{{\renewcommand{\arraystretch}{0.4}%
\ensuremath{(\raisebox{0.1\height}{\scriptsize
$\begin{array}{@{}r@{}}#1\\#2\end{array}$})}}}%
}
\let\errp\errparen

\newcommand{\fbdsqbd}{F_{B_d} \sqrt{\hat B_{B_d}}}
\newcommand{\fbssqbs}{F_{B_s} \sqrt{\hat B_{B_s}}}
\newcommand{\RE}{{\rm Re}}
\newcommand{\IM}{{\rm Im}}
\newcommand{\vcb}{|V_{cb}|}
\newcommand{\vtd}{|V_{td}|}
\newcommand{\vubvcb}{|V_{ub}/V_{cb}|}
\newcommand{\vub}{|V_{ub}|}
\newcommand{\vts}{|V_{ts}|}
\newcommand{\vus}{|V_{us}|}
\newcommand{\vud}{|V_{ud}|}
\newcommand{\vucb}{|V_{ub}/V_{cb}|}
\newcommand{\Bs}{\mbox{\rm B}^0_s}
\newcommand{\Bd}{\mbox{\rm B}^0_d}
\newcommand{\Bp}{\mbox{\rm B}^{+}}
\newcommand{\Bm}{\mbox{\rm B}^{-}}
\newcommand{\Bdb}{\overline{\rm B}^0_d}  
%
\def\Ks    {\ensuremath{K_s^0}}
\def\Dbar  {\kern 0.2em\overline{\kern -0.2em \rm D}{}}
\def\Db    {\ensuremath{\Dbar}}
\def\Dzb   {\ensuremath{\Dbar^0}}
\def\DzDzb {\ensuremath{\rm D^0 {\kern -0.16em \Dzb}}}
\def\Dstar   {\ensuremath{\rm D^*}}
\def\Dstarb   {\ensuremath{\Dbar^*}}
\def\Dstarz  {\ensuremath{\rm D^{*0}}}
\def\Dstarzb  {\ensuremath{\Dbar^{*0}}}
\def\Ds    {\ensuremath{{\rm D}^+_s}}
\def\Dsb   {\ensuremath{\Dbar^+_s}}
\def\Dss   {\ensuremath{{\rm D}^*_s}}
\def\Bz    {\ensuremath{\rm B^0}}
\def\B     {\ensuremath{\rm B}}
\def\Bbar  {\kern 0.18em\overline{\kern -0.18em {\rm B}}{}}
\def\Bb    {\ensuremath{\Bbar}}
\def\Bzb   {\ensuremath{\Bbar^0}}
\def\Bu    {\ensuremath{{\rm B}^+}}
\def\Bub   {\ensuremath{{\rm B}^-}}
\def\Bpm   {\ensuremath{{\rm B}^\pm}}
\def\Bs    {\ensuremath{{\rm B}^0_s}}
\def\Bsb   {\ensuremath{\overline{\rm B}^0_s}}
\def\Bd    {\ensuremath{{\rm B}^0_d}}
\def\Bdb   {\ensuremath{\Bbar^0_d}}
\def\Bq    {\ensuremath{{\rm B}^0_q}}
\def\Bqb   {\ensuremath{\Bbar^0_q}}
\def\BB    {\ensuremath{\rm B\Bbar}} 
\def\BzBzb {\ensuremath{B^0 {\kern -0.16em \Bzb}}}
\def\fours {\ensuremath{\Upsilon{\rm( 4S)}}} 
\def\bbbar {\ensuremath{b\overline b}}
\def\ccbar {\ensuremath{c\overline c}}
\def\btol       {\ensuremath{b\to\ell}}
\def\btocl      {\ensuremath{b\to c\to\ell}}
\def\ctol       {\ensuremath{c\to\ell}}
\def\BDDX       {\ensuremath{B\to D\overline{D}X}}
\def\Ztohad     {\ensuremath{Z^0\to\mathrm{hadrons}}}
\def\bmix       {\ensuremath{B^0 \mbox{--} {\Bbar^0}}}
\def\bdmix      {\ensuremath{B_d^0 \mbox{--} {\Bbar_d^0}}}
\def\bqmix      {\ensuremath{B_q^0 \mbox{--} {\Bbar_q^0}}}
\def\bsmix      {\ensuremath{B_s^0 \mbox{--} {\Bbar_s^0}}}
\def\fB#1{F_{B_{#1}}}
\def\bbhat#1{\hat B_{B_{#1}}}
\newcommand{\uu}{\langle\bar uu\rangle}

\newcommand{\tvs}{\vbox{\vskip 3mm}}
\newcommand{\svs}{\vbox{\vskip 5mm}}
\newcommand{\mvs}{\vbox{\vskip 8mm}}
\newcommand{\msvs}{\vbox{\vskip 7mm}}

\def\ps{\rm ps}
\def\R1{\varepsilon_1}
\def\E8{\varepsilon_8}
\def\gat{\tilde{\gamma}}
\def\gh{\hat{g}}
\def\gt{\tilde{g}}
\def\gah{\hat{\gamma}}
\def\ga{\gamma}
\def\gaf{\gamma_{5}}
\def \branch{{\cal B}}
\def\r#1{(\ref{#1})}
\def\eps{\varepsilon}
\def\epe{\varepsilon'/\varepsilon}
\def\as{\alpha_s}
\newcommand{\eqn}{\ref}
\def\Heff{{\cal H}_{\rm eff}}
\def \lu{\lambda_u}
\def \cl#1{{#1\%\ \mathrm{C.L.}}}
\def \fig#1{Fig.~\ref{#1}}
\newcommand{\nn}{\nonumber}
\newcommand{\mt}{m_{\rm t}}
\newcommand{\mtb}{\overline{m}_{\rm t}}
\newcommand{\mcb}{\overline{m}_{\rm c}}
\newcommand{\mc}{m_{\rm c}}
\newcommand{\ms}{m_{\rm s}}
\newcommand{\md}{m_{\rm d}}
\newcommand{\mb}{m_{\rm b}}
\newcommand{\mw}{M_{\rm W}}
\newcommand{\mz}{M_{\rm Z}}
\newcommand{\gev}{{\rm GeV}}
\newcommand{\mev}{{\rm MeV}}
\newcommand{\tev}{{\rm TeV}}
\def\gevc{\ensuremath{\gev\!/c}}
\def\mevc{\ensuremath{\mev\!/c}}
\def\gevcc{\ensuremath{\gev\!/c^2}}
\def\mevcc{\ensuremath{\nev\!/c^2}}
\newcommand{\bsi}{B_6^{(1/2)}}
\newcommand{\bei}{B_8^{(3/2)}}
\newcommand{\Lms}{\Lambda_{\overline{\rm MS}}}
\newcommand{\bsg}{$b \to s \gamma$ }
\newcommand{\Bsg}{$B \to X_s \gamma$ }
\newcommand{\newsection}[1]{\section{#1}\setcounter{equation}{0}}
\newcommand{\bea}{\begin{eqnarray}}
\newcommand{\eea}{\end{eqnarray}}
\newcommand{\bd}{\begin{displaymath}}
\newcommand{\ed}{\end{displaymath}}
\newcommand{\aem}{\alpha}
\newcommand{\Bsee}{$B \to X_s e^+ e^-$ }
\newcommand{\bsee}{$b \to s e^+ e^-$ }
\newcommand{\bcenu}{$b \to c e \bar\nu $ }
\newcommand{\beq}{\begin{equation}}
\newcommand{\eeq}{\end{equation}}
\newcommand{\be}{\begin{equation}}
\newcommand{\ee}{\end{equation}}
\newcommand{\bi}{\begin{itemize}}
\newcommand{\ei}{\end{itemize}}
\newcommand{\ord}{{\cal O}}
\newcommand{\order}{{\cal O}}
\newcommand{\f}{\frac}
\newcommand{\Ctilde}{\tilde{C}}
\def\kpn{\rm K^+\rightarrow\pi^+\nu\bar\nu}
\def\klpn{\rm K_{\rm L}\rightarrow\pi^0\nu\bar\nu}
\newcommand{\kmm}{\rm K_{\rm L} \to \mu^+ \mu^-}
\newcommand{\kpe}{{\rm K}_{\rm L} \to \pi^0 e^+ e^-}

\def \beqa{\begin{eqnarray}}
\def \eeqa{\end{eqnarray}}
\def\aspi{\frac{\as}{4\pi}}
\def\gf{\gamma_5}
\def \appros{A_{\pi \pi}}
\def \ocb{\overline{{\cal B}}}
\def \bo{{\rm B}^0}
\def \cpp{C_{\pi \pi}}
\def \lpp{\lambda_{\pi \pi}}
\def \ob{\overline{\rm B}^0}
\def \rpp{R_{\pi \pi}}
\def \spp{S_{\pi \pi}}

\newcommand{\imlt}{\Im\lambda_t}
\newcommand{\relt}{\Re\lambda_t}
\newcommand{\relc}{\Re\lambda_c}
\renewcommand{\theequation}{\arabic{equation}}
\renewcommand{\baselinestretch}{1.3}
\renewcommand{\thefootnote}{\fnsymbol{footnote}}

\def\lsim{\mathrel{\mathpalette\Zoversim<}}
\def\gsim{\mathrel{\mathpalette\Zoversim>}}
\def\Zoversim#1#2{\vcenter{\offinterlineskip
     \ialign{$\m@th#1\hfil##\hfil$\crcr#2\crcr\noalign{\kern.2ex}\sim\crcr}}}
\let\lesssim=\lsim

\def\point#1 #2 #3{\put(#1,#2){\makebox(0,0){$#3$}}}

%

\newcommand{\eD} {\ifmmode \varepsilon{\cal D}^2 \else 
                          $\varepsilon{\cal D}^2$\fi}

\setlength{\unitlength}{0.0022\textwidth}

\def\P{{\rm{P}}}
\def\Pdot{\dot{\P}}
\def\Pdotdot{{\ddot{\P}}}
\def\xs{{x_\mathrm{s}}}
\def\xsmes{{x_\mathrm{s}^\mathrm{mes}}}
\def\xshyp{{x_\mathrm{s}^\mathrm{hyp}}}
\def\fs{f_\mathrm{s}}
\def\Gt{{\rm{G}_t}}
\def\dtmes{\mathrm{d}t_\mathrm{mes}}
\def\Lik{{\cal L}}
\def\CL{\mathrm{CL}}
\def\CLlo{\mathrm{CL}_\mathrm{lo}}
\def\Prob{\mathrm{Prob}}
\def\A{\mathrm{A}}
\def\C{\mathrm{C}}
\def\ao{\mathrm{a_0}}
\def\at{\mathrm{a_3}}
\def\nlo{\mathrm{nlo}}
\def\Philo{\Phi_\mathrm{lo}}
\def\Phinlo{\Phi_{\nlo}}
\def\d{\mathrm{d}}

\newcommand{\rhobar}{\bar {\rho}}
\newcommand{\etabar}{\bar{\eta}}
\def\rfit{{\em R}fit}
\def\CKMfitter{{\em CKMfitter}}\def\CP{$ C \! P$ } 
\def\babar{BaBar}
\newcommand{\epsilonk}{\left|\varepsilon_K \right|}


\addcontentsline{toc}{section}{Contributors}
\vspace{2mm}

\begin{center}
{\Large\bf Contributors}
\vspace{12mm}
\end{center}

\noindent
{\it 
D.~Abbaneo$^{1}$,
A.~Ali$^{2}$,  
P.~Amaral$^{3}$,  
V.~Andreev$^{4}$,
M.~Artuso$^{5}$, 
E.~Barberio$^{6}$,
M.~Battaglia$^{1}$,
C.~Bauer$^{6}$,
D.~Becirevic$^{8}$,
M.~Beneke$^{9}$,
I.~Bigi$^{10}$,
C.~Bozzi$^{11}$,
T.~Brandt$^{12}$,
G.~Buchalla$^{13}$,
A.J.~Buras$^{14,a}$,
M.~Calvi$^{15}$, 
D.~Cassel$^{16}$,
V.~Cirigliano$^{17,b}$,
M.~Ciuchini$^{18}$,
G.~Colangelo$^{19,b,c}$,
A.~Dighe$^{20}$,
G.~Dubois-Felsmann$^{21,d}$,
G.~Eigen$^{22,d}$,
K.~Ecklund$^{16}$,
P.~Faccioli$^{23}$,
R.~Fleischer$^{1}$,
J.~Flynn$^{24}$,
R.~Forty$^{1}$,
E.~Franco$^{8}$,
P.~Gagnon$^{25}$,
P.~Gambino$^{1,e}$,
R.~Gupta$^{26,f}$,
S.~Hashimoto$^{27}$,
R.~Hawkings$^{1}$,
D.~Hitlin$^{21,d}$,
A.~Hoang$^{20}$,
A.~Hocker$^{28}$,
T.~Hurth$^{1}$,
G.~Isidori$^{1,29,b}$,
T.~Iijima$^{30}$,
D.~Jaffe$^{31}$,
M.~Jamin$^{32}$,
Y.Y.~Keum$^{30,g}$,
A.~Khodjamirian$^{33}$,
C.S.\ Kim$^{34,h}$,
P.~Kluit$^{35}$,
A.~Kronfeld$^{36,i}$,
H.~Lacker$^{28}$, 
S.~Laplace$^{28}$,
F.~Le~Diberder$^{28}$,
L.~Lellouch$^{37,l}$,
A.~Lenz$^{38}$,
C.~Leonidopoulos$^{36,i}$, 
A.~Le~Yaouanc$^{39,b}$,
Z.~Ligeti$^{40}$,
D.~Lin$^{24}$,
G.~Lopez-Castro$^{41,k}$,
V.~Lubicz$^{18}$,
D.~Luc\-che\-si$^{42}$,
T.~Mannel$^{33}$,
M.~Margoni$^{42}$, 
G.~Martinelli$^{8}$,
D.~Melikhov$^{32}$, 
M.~Misiak$^{43,l}$,
V.~Mor\'enas$^{44}$,
H.G.~Mo\-ser$^{20}$,
U.~Nierste$^{36,i}$,
J.~Ocariz$^{51}$, 
L.~Oliver$^{39,b}$,
F.~Parodi$^{45}$,
M.~Paulini$^{46,m}$,
C.~Paus$^{47}$,  
O.~P\`ene$^{39,b}$,
M.~Pierini$^{8}$,
F.~Porter$^{21,d}$,
D.~Po\v{c}ani\'c$^{48}$,
J.-C.~Raynal$^{39,b}$, 
J.L.~Rosner$^{3,n}$,
P.~Roudeau$^{28}$,
Y.~Sakai$^{27}$, 
K.R.~Schubert$^{12}$,
C.~Schwanda$^{27,o}$,
B.~Sciascia$^{29}$,
O.~Schneider$^{49}$,
B.~Serfass$^{12}$,
L.~Silvestrini$^{8}$,
M.\ ~Smi\-zans\-ka$^{50}$,
J.~Stark$^{51}$,
B.~Stech$^{32}$,
A.~Stocchi$^{28}$,
N.~Uraltsev$^{15}$,
M.~Villa$^{23}$,
A.~Warburton$^{16,52}$,
C.~Weiser$^{33}$,
L.H.~Wil\-den$^{12}$,
G.~Wilkinson$^{53}$,
S.~Willocq$^{54}$,
N.~Yamada$^{27}$}

\vspace{1.5cm}
\noindent
{\it
~$^{1}$CERN, CH-1211 Geneva, Switzerland
~$^{2}$DESY, D-22603 Hamburg, Germany
~$^{3}$Chicago Univ., Chicago, IL 60637, USA
~$^{4}$INP, 188350 St. Petersburg, Russia
~$^{5}$Syracuse Univ., Syracuse, NY 13244, USA
~$^{6}$Southern Methodist Univ., Dallas, TX 75275, USA
~$^{7}$Univ. of California, San Diego, La Jolla, CA 92093 USA
~$^{8}$Univ. Roma La Sapienza, Roma, Italy
~$^{9}$RWTH, D-52056 Aachen, Germany
~$^{10}$Univ. of Notre Dame, Notre Dame, IN 46556, USA
~$^{11}$INFN, I-44100 Ferrara, Italy
~$^{12}$Technische Univ.\ Dresden, D-01069 Dresden, Germany
~$^{13}$Ludwig-Maximilians-Univ. M\"unchen, D-80799 Munich, Germany
~$^{14}$Technische Univ.\ M\"unchen, D-85748 Garching, Germany
~$^{15}$Univ. Milano Bicocca and INFN Milano, Milano, Italy
~$^{16}$Cornell Univ., Ithaca, NY 14853, USA
~$^{17}$IFIC, Univ. de Valencia, E-46071 Valencia, Spain
~$^{18}$Univ. Roma Tre, Roma, I-00146 Rome, Italy
~$^{19}$University of Bern,  CH-3012 Bern, Switzerland
~$^{20}$MPI, D-80805 Munich, Germany
~$^{21}$California Inst. of Technology, Pasadena, CA 91125, USA
~$^{22}$Univ. of Bergen, 5020 Bergen, Norway
~$^{23}$INFN, I-40126 Bologna, Italy
~$^{24}$Univ. of Southampton, Southampton S017 1BJ, England, UK
~$^{25}$Indiana Univ., Bloomington, IN 47405 USA
~$^{26}$LANL, Los Alamos, NM 87545, USA
~$^{27}$KEK, Tsukuba, 305-0801, Japan
~$^{28}$LAL, F-91898 Orsay, France
~$^{29}$LNF, I-00044 Frascati, Italy
~$^{30}$Nagoya Univ., Nagoya, 464-6801, Japan
~$^{31}$BNL, Upton, NY 11973, USA
~$^{32}$Univ. Heidelberg, D-69120 Heidelberg, Germany
~$^{33}$Univ. Karlsruhe, D-76128 Karlsruhe, Germany
~$^{34}$Yonsei Univ., Seoul 120-749, South Korea
~$^{35}$NIKHEF, 1009 DB Amsterdam. Netherlands
~$^{36}$Fermi National Accelerator Laboratory, 
  P.O.\ Box 500, Batavia, IL 60510, USA
~$^{37}$CPT, F-13288 Marseille, France                   
~$^{38}$Univ. Regensburg, D-93053 Regensburg, Germany
~$^{39}$LPTHE, F-91405 Orsay, France
~$^{40}$LBL, Berkeley, CA 94720, USA
~$^{41}$CINVESTAV, 07000 Mexico, Mexico
~$^{42}$INFN, I-35100 Padova, Italy
~$^{43}$Univ. of Warsaw, PL-00 681 Warsaw, Poland
~$^{44}$LPC, F-63177 Clermont Ferrand, France
~$^{45}$INFN, I-16146 Genova, Italy
~$^{46}$Carnegie Mellon Univ., Pittsburgh, PA 15213, USA
~$^{47}$MIT, Cambridge, MA 02139, USA
~$^{48}$Univ. of Virgina, Charlottesville, VA 22901, USA
~$^{49}$Univ. of Lausanne, CH-1015 Lausanne, Switzerland
~$^{50}$Univ. of Lancaster, Lancaster LA1 4YB, England, UK
~$^{51}$LPNHE, 75252 Paris, France
~$^{52}$McGill Univ., Montr\'eal, Qu\'ebec  H3A 2T8, Canada
~$^{53}$Univ. of Oxford, Oxford OX13RH, England, UK
~$^{54}$Univ. of Massachusetts, Amherst, MA 01003, USA }
\vspace{1cm}

\newpage

\noindent {\bf Grant Acknowledgements} \ \
$^a$   Contract 05HT1WOA3 of  German Bundesministerium 
f\"ur Bildung und Forschung  
         and DFG Project  Bu.~706/1-1.
~$^b$  EU contract HPRN-CT-2002-00311 (EURIDICE).
~$^c$  Partially supported by Schweizerische Nationalfonds. 
~$^d$  Grant DE-FG03-92-ER40701 of the  U.S.\ Department of Energy.
~$^e$  Marie Curie Fellowship No.~HPMF-CT-2000-01048.
~$^f$  Grant KA-04-01010-E161 of the U.S.\ Department of Energy.
~$^g$  Grant No.~NSC-90-2811-M-002 of Science Council of R.O.C.\ and  
Grant in Aid for 
       Scientific Research from Ministry of Education, 
Science and Culture of Japan.
~$^h$  Grant No. 2001-042-D00022 of the KRF.
~$^i$  Fermilab is operated by Universities Research Association, 
Inc., under contract 
         with the U.S.\ Department of Energy.
~$^j$ Contract HPRN-CT-2000-00145, Hadrons/Lattice QCD, 
of the EU Human Potential
  Program.
~$^k$ Conacyt, Mexico.
~$^l$ Grant $2~P03B~121~20$ of the Polish Committee for
Scientific Research.
~$^m$ DOE grant DE-FG02-91ER40682.
~$^n$ Grant No.\ DE FG02 90ER40560 of the U.S.\ Department of Energy.   
~$^o$ Japan Society for the Promotion of Science.

\vskip 1.2cm  
\noindent

\noindent
{\bf Local Organizing Committee:} 
{\it
E~. Barberio,~
M.~ Battaglia,~
R.~ Forty,~
P.~ Gambino,~
P.~ Kluit,~\\
M.~ Mangano,~
G.~ Martinelli,~
O.~ Schneider,~
A.~ Stocchi,~
G.~ Wilkinson.
}
\noindent
\vskip 1.2cm
\noindent
{\bf International Advisory Committee:} 
{\it
H.~ Aihara
% ...
W.~ Yao.}

\addcontentsline{toc}{section}{Participants}

\newpage
\addcontentsline{toc}{section}{Foreword}
\noindent
{\Huge\bf  Foreword}\\
\vspace{3mm}

\noindent
This report contains the results of the Workshop on the CKM Unitarity Triangle
that was held at CERN on 13-16 February 2002. There had been several
Workshops on B physics 
that concentrated on
studies at $e^+e^-$ machines, at the Tevatron, or at LHC separately.
Here we brought together experts of different fields, 
both theorists and experimentalists, 
to study the determination of the CKM matrix from 
all the  available data of K, D, and B physics.
The analysis of LEP data for B physics is reaching its end,
and one of the goals of the Workshop was to underline the results that
have been achieved at LEP, SLC,  and CESR.
Another goal was to prepare for
the transfer of responsibility for averaging B physics properties,
that has developed within the LEP community, to the present main 
actors of these 
studies, from the B factory and the Tevatron experiments. The optimal way to
combine the various experimental and theoretical inputs and to fit for
the apex of the Unitarity Triangle has been a contentious issue. A
further goal of the Workshop was to bring together the proponents of
different fitting strategies, and to compare their approaches when
applied to the same inputs. 

Since lattice QCD plays a very important role in the determination 
of the non-perturbative parameters needed to constrain the CKM
unitarity triangle, the first Workshop was seen as an excellent
opportunity to bring together lattice theorists with the aim of
establishing a working group to compile averages for
phenomenologically relevant quantities. Representatives from lattice
collaborations around the world were invited to attend a meeting
during the Workshop. A consensus was reached to set up three test 
working groups, collectively known as the \emph{CKM Lattice Working Group},
 to review a number of well-studied quantities: quark masses, the kaon
$B$-parameter, and the matrix elements relevant for neutral
B-meson mixing. 

This report is
organized as a coherent document 
with chapters covering the domains of activity of the working groups.
It deals mainly with the
present determination of the CKM matrix in the Standard Model
with a brief outlook on the near future.
The impact of future measurements and of physics beyond the Standard Model
will be developed further in  forthcoming Workshops with the same title.
Indeed, the Workshop was conceived as the first of a series.
The second one will take place on  5-9~April~2003 in Durham and will focus 
on the results from the B-factories.

\begin{flushright} 
Geneva, March~2003 \ \ \ \ \ 
\end{flushright}

\newpage

\thispagestyle{empty}
~

\newpage





\newpage
\tableofcontents

\newpage

\renewcommand{\thepage}{\arabic{page}} \setcounter{page}{1}

\chapter{INTRODUCTION}
\label{chap:I}
{\it M.~Battaglia, A.J.~Buras, J.~Flynn, R.~Forty, P.~Gambino,
     P.~Kluit, P.~Roudeau, O.~Schneider, A.~Stocchi}

\section{Setting the scene}
\newenvironment{comment}[1]{}{}

The understanding of flavour dynamics, and of the related origin of
quark and lepton masses and mixings, is among the most important goals
in elementary particle physics. In this context, weak decays of
hadrons, and in particular the CP violating and rare decay processes,
play an important role as they are sensitive to short distance
phenomena. Therefore the determination of the
Cabibbo-Kobayashi-Maskawa (CKM) 
matrix~[\ref{1CAB},\ref{1KM}] that parametrizes
the weak charged current interactions of quarks is currently a central
theme in particle physics. Indeed, the four parameters of this matrix
govern all flavour changing transitions involving quarks in the
Standard Model (SM). These include tree level decays mediated by
W bosons, which are essentially unaffected by new physics
contributions, as well as a vast number of one-loop induced flavour
changing neutral current (FCNC) transitions responsible for rare and
CP violating decays in the SM, which involve gluons, photons, W$^\pm$,
Z$^0$ and H$^0$, and are sensitive probes of new physics. This
role of the CKM matrix is preserved in most extensions of
the SM, even if they contain new sources of flavour and CP
violation.

An important goal is then to find out whether the SM is able to
describe the flavour and CP violation observed in nature. All the
existing data on weak decays of hadrons, including rare and
CP violating decays, can at present be described by the SM within the
theoretical and experimental uncertainties. On the other hand,
the SM is an incomplete theory: some kind of new physics is
required in order to understand the patterns of quark and lepton
masses and mixings, and generally to understand flavour dynamics.
There are also strong theoretical arguments suggesting that 
new physics cannot be far from the electroweak scale, and  
new sources of flavour and CP violation appear in most
extensions of the SM, such as supersymmetry. Consequently, for several reasons,
it is likely that the CKM picture of 
flavour physics is modified at accessible energy scales. In addition, 
the studies of  dynamical generation of the baryon asymmetry in
the universe show that the size of CP violation in the SM is too small 
to generate a 
matter-antimatter asymmetry as large as that observed in the 
universe today. Whether the physics responsible for the baryon asymmetry 
involves only very short distance scales
like the GUT or the Planck scales, or  it is related to CP violation
observed in experiments performed by humans, is an important question that
still has to be answered.

To shed light on these questions the CKM matrix has to be determined
with high accuracy and well understood errors. Tests of its structure,
conveniently represented by the unitarity triangle, have to be
performed; they  will allow a precision determination of the SM
contributions to various observables and possibly reveal the onset of 
new physics contributions.

The major theoretical problem in this program is the presence of
strong interactions. Although the gluonic contributions at scales
${\cal O} (\mw, \mz, \mt)$ can be calculated within the perturbative
framework, owing to the smallness of the effective QCD coupling at
short distances, the fact that hadrons are bound states of quarks and
antiquarks forces us to consider QCD at long distances as well. Here
we have to rely on the existing non-perturbative methods, which are not
yet fully satisfactory. On the experimental side, the basic problem in
extracting CKM parameters from the relevant rare and CP violating
transitions is the smallness of the branching ratios, which are 
often very difficult to measure. 
As always in the presence of large theoretical and systematic
uncertainties, their treatment in the context of global fits is a  
problematic and divisive issue. 

In the last decade considerable progress in the determination of the
unitarity triangle and the CKM matrix has been achieved through more
accurate experiments, short distance higher order QCD calculations,
novel theoretical methods like Heavy Quark Effective Theory (HQET) and
 Heavy Quark Expansion (HQE), and progress in non-perturbative
methods such as lattice gauge simulation and QCD sum rules. It is the
purpose of these proceedings to summarize the present status of these
efforts, to identify the most important remaining challenges, and to
offer  an outlook for the future.

While many decays used in the determination of the CKM matrix are
subject to significant hadronic uncertainties, there are a handful of
quantities that allow the determination of the CKM parameters
with reduced or no
hadronic uncertainty. The prime examples are the
CP asymmetry $a_{\psi K_S}$, certain strategies in B decays relevant
for the angle $\gamma$ in the unitarity triangle, the branching ratios
for $\kpn$ and $\klpn$, and suitable ratios of the branching ratios for
rare decays ${\rm B}_{d,s}\to\mu^+\mu^-$ and 
${\rm B}\to X_{d,s}\nu\overline\nu$
relevant for the determination of $\vtd/\vts$.
Also the ratio $\Delta M_d/\Delta M_s$ is important in this respect.

The year 2001 opened a new era of theoretically clean measurements of
the CKM matrix through the discovery of CP violation in the B system
$(a_{\psi K_S}\not=0)$ and further evidence for the decay $\kpn$. In
2002 the measurement of the angle $\beta$ in the unitarity triangle by
means of $a_{\psi K_S}$ has been considerably improved. It is an
exciting prospect that new data on CP violation and rare decays as
well as ${\rm B}_s^0-\overline {\rm B}_s^0$ mixing coming 
from a number of laboratories
in Europe, USA and Japan will further improve the determination of the
CKM matrix, possibly modifying the SM description of flavour physics.

Recently, there have  been several workshops on B 
physics~[\ref{1BABAR}--\ref{1FERMILAB}] 
that concentrated on studies at $e^+e^-$ machines, 
at the Tevatron or at LHC, separately.
 Here we focus instead 
 on the discussion of the CKM matrix and its determination from all available
data at different machines.

\section{CKM matrix and the Unitarity Triangle}
\label{sec:ckmgen}
\setcounter{equation}{0}

\subsection{General remarks}
The unitary CKM matrix [\ref{1CAB},\ref{1KM}] connects the {\it weak
eigenstates} $(d^\prime,s^\prime,b^\prime)$ and the corresponding {\it
mass eigenstates} $d,s,b$:
\begin{equation}\label{2.67}
\left(\begin{array}{c}
d^\prime \\ s^\prime \\ b^\prime
\end{array}\right)=
\left(\begin{array}{ccc}
V_{ud}&V_{us}&V_{ub}\\
V_{cd}&V_{cs}&V_{cb}\\
V_{td}&V_{ts}&V_{tb}
\end{array}\right)
\left(\begin{array}{c}
d \\ s \\ b
\end{array}\right)\equiv\hat V_{\rm CKM}\left(\begin{array}{c}
d \\ s \\ b
\end{array}\right).
\end{equation}
Several parametrizations of the CKM matrix have been proposed in the
literature; they are classified in~[\ref{1FX1}]. We will use two
in these proceedings: the standard parametrization~[\ref{1CHAU}] 
recommended by the Particle Data Group~[\ref{1PDG}]  
and a generalization of the Wolfenstein parametrization~[\ref{1WO}] as
presented in~[\ref{1BLO}].

\subsection{Standard parametrization}
            \label{sec:sewm:stdparam}

With $c_{ij}=\cos\theta_{ij}$ and $s_{ij}=\sin\theta_{ij}$
($i,j=1,2,3$), the standard parametrization is given by:
\begin{equation}\label{2.72}
\hat V_{\rm CKM}=
\left(\begin{array}{ccc}
c_{12}c_{13}&s_{12}c_{13}&s_{13}e^{-i\delta}\\ -s_{12}c_{23}
-c_{12}s_{23}s_{13}e^{i\delta}&c_{12}c_{23}-s_{12}s_{23}s_{13}e^{i\delta}&
s_{23}c_{13}\\ s_{12}s_{23}-c_{12}c_{23}s_{13}e^{i\delta}&-s_{23}c_{12}
-s_{12}c_{23}s_{13}e^{i\delta}&c_{23}c_{13}
\end{array}\right)\,,
\end{equation}
where $\delta$ is the phase necessary for {\rm CP} violation. $c_{ij}$
and $s_{ij}$ can all be chosen to be positive and $\delta$ may vary in
the range $0\le\delta\le 2\pi$. However, measurements of CP violation
in $K$ decays force $\delta$ to be in the range $0<\delta<\pi$.

From phenomenological studies we know that $s_{13}$ and $s_{23}$ are
small numbers: $\ord(10^{-3})$ and ${\cal O}(10^{-2})$, respectively.
Consequently to an excellent accuracy the four independent parameters
can be chosen as
\begin{equation}\label{2.73}
s_{12}=| V_{us}|, \quad s_{13}=| V_{ub}|, \quad s_{23}=|
V_{cb}| \quad {\rm and}  ~~\delta.
\end{equation}
As discussed in detail in Chapters~\ref{chap:II} and~\ref{chap:III}, 
the first three
parameters can be extracted from tree level decays mediated by the
transitions $s \to u$, $b \to u$ and $b \to c$, respectively. The
phase $\delta$ can be extracted from CP violating transitions or loop
processes sensitive to $| V_{td}|$. We will analyse this in detail in
Chapters 4--6.


\subsection{Wolfenstein parametrization and its generalization}
\label{Wolf-Par}

The absolute values of the elements of the CKM matrix show a
hierarchical pattern with the diagonal elements being close to unity,
the elements $\vus$ and $|V_{cd}|$ being of order $0.2$, the elements
$\vcb$ and $\vts$ of order $4\cdot 10^{-2}$ whereas $|V_{ub}|$ and
$\vtd$ are of order $5\cdot 10^{-3}$. The Wolfenstein 
parametrization~[\ref{1WO}] exhibits this hierarchy in a transparent 
manner. It is an
approximate parametrization of the CKM matrix in which each element is
expanded as a power series in the small parameter $\lambda=|
V_{us}|\approx 0.22$,
\begin{equation}\label{2.75} 
\hat V=
\left(\begin{array}{ccc}
1-{\lambda^2\over 2}&\lambda&A\lambda^3(\varrho-i\eta)\\ -\lambda&
1-{\lambda^2\over 2}&A\lambda^2\\ A\lambda^3(1-\varrho-i\eta)&-A\lambda^2&
1\end{array}\right)
+\ord(\lambda^4)\,,
\end{equation}
and the set (\ref{2.73}) is replaced by
\begin{equation}\label{2.76}
\lambda, \qquad A, \qquad \varrho, \qquad {\rm and}~~ \eta \, .
\end{equation}
Because of the
smallness of $\lambda$ and the fact that for each element 
the expansion parameter is actually
$\lambda^2$, it is sufficient to keep only the first few terms
in this expansion. 

The Wolfenstein parametrization is certainly more transparent than
the standard parametrization. However, if one requires sufficient 
level of accuracy, the terms of $\ord(\lambda^4)$ and $\ord(\lambda^5)$ 
have to be included in phenomenological applications.
This can be done in many ways~[\ref{1BLO}]. The
point is that since (\ref{2.75}) is only an approximation the {\em exact}
definition of the parameters in (\ref{2.76}) is not unique in terms of the 
neglected order ${\cal O}(\lambda^4)$. 
This situation is familiar from any perturbative expansion, where
different definitions of expansion parameters (coupling constants) 
are possible.
This is also the reason why in different papers in the
literature different ${\cal O}(\lambda^4)$ terms in (\ref{2.75})
 can be found. They simply
correspond to different definitions of the parameters in (\ref{2.76}).
Since the physics does not depend on a particular definition, it
is useful to make a choice for which the transparency of the original
Wolfenstein parametrization is not lost.

In this respect
a useful definition adopted by most authors in the literature 
is to go back to the standard parametrization (\ref{2.72}) and to
 {\it define} the parameters $(\lambda,A,\varrho,\eta)$ 
through ~[\ref{1BLO},\ref{1schubert}]
\begin{equation}\label{2.77} 
s_{12}=\lambda\,,
\qquad
s_{23}=A \lambda^2\,,
\qquad
s_{13} e^{-i\delta}=A \lambda^3 (\varrho-i \eta)
\end{equation}
to {\it  all orders} in $\lambda$. 
It follows  that
\begin{equation}\label{2.84} 
\varrho=\frac{s_{13}}{s_{12}s_{23}}\cos\delta,
\qquad
\eta=\frac{s_{13}}{s_{12}s_{23}}\sin\delta.
\end{equation}
The expressions (\ref{2.77}) and (\ref{2.84}) represent simply
the change of variables from (\ref{2.73}) to (\ref{2.76}).
Making this change of variables in the standard parametrization 
(\ref{2.72}) we find the CKM matrix as a function of 
$(\lambda,A,\varrho,\eta)$ which satisfies unitarity exactly.
Expanding next each element in powers of $\lambda$ we recover the
matrix in (\ref{2.75}) and in addition find explicit corrections of
$\ord(\lambda^4)$ and higher order terms. Including $\ord(\lambda^4)$ 
and $\ord(\lambda^5)$ terms we find
\begin{equation}\label{2.775} 
\hat V=
\left(\begin{array}{ccc}
1-\frac{1}{2}\lambda^2-\frac{1}{8}\lambda^4               &
\lambda+\ord(\lambda^7)                                   & 
A \lambda^3 (\varrho-i \eta)                              \\
-\lambda+\frac{1}{2} A^2\lambda^5 [1-2 (\varrho+i \eta)]  &
1-\frac{1}{2}\lambda^2-\frac{1}{8}\lambda^4(1+4 A^2)     &
A\lambda^2+\ord(\lambda^8)                                \\
A\lambda^3(1-\overline\varrho-i\overline\eta)                       &  
-A\lambda^2+\frac{1}{2}A\lambda^4[1-2 (\varrho+i\eta)]   &
1-\frac{1}{2} A^2\lambda^4                           
\end{array}\right)
\end{equation}
where~[\ref{1BLO}]
\begin{equation}\label{2.88d}
\overline\varrho=\varrho (1-\frac{\lambda^2}{2}),
\qquad
\overline\eta=\eta (1-\frac{\lambda^2}{2}).
\end{equation}
We emphasize that by definition the expression for
$V_{ub}$ remains unchanged relative to the original Wolfenstein 
parametrization and the
corrections to $V_{us}$ and $V_{cb}$ appear only at $\ord(\lambda^7)$ and
$\ord(\lambda^8)$, respectively.
The advantage of this generalization of the Wolfenstein parametrization
over other generalizations found in the literature 
is the absence of relevant corrections to $V_{us}$, $V_{cd}$, $V_{ub}$ and 
$V_{cb}$ and an  elegant
change in $V_{td}$ which allows a simple generalization of the 
so-called unitarity triangle to higher orders in $\lambda$~[\ref{1BLO}] 
as discussed below. 

\subsection{Unitarity Triangle}
The unitarity of the CKM-matrix implies various relations between its
elements. In particular, we have
\begin{equation}\label{2.87h}
V_{ud}^{}V_{ub}^* + V_{cd}^{}V_{cb}^* + V_{td}^{}V_{tb}^* =0.
\end{equation}
Phenomenologically this relation is very interesting as it involves
simultaneously the elements $V_{ub}$, $V_{cb}$ and $V_{td}$ which are
under extensive discussion at present. Other relevant unitarity 
relations will be presented as we proceed. 

The relation (\ref{2.87h})  can be
represented as a {\it unitarity triangle} in the complex 
$(\overline\varrho,\overline\eta)$ plane. 
The invariance of (\ref{2.87h})  under any phase-transformations
implies that the  corresponding triangle
is rotated in the $(\overline\varrho,\overline\eta)$  
plane under such transformations. 
Since the angles and the sides
(given by the moduli of the elements of the
mixing matrix)  in this triangle remain unchanged, they
 are phase convention independent and are  physical observables.
Consequently they can be measured directly in suitable experiments.  
One can construct  five additional unitarity triangles~[\ref{1Kayser}] 
corresponding to other orthogonality relations, like the one in (\ref{2.87h}).
Some of them should be useful
when the data on rare and CP violating decays improve.
The areas ($A_{\Delta}$) of all unitarity triangles are equal and 
related to the measure of CP violation 
$J_{\rm CP}$~[\ref{1CJ}]:
$\mid J_{\rm CP} \mid = 2\cdot A_{\Delta}$.

\begin{figure}[hbt]
\begin{center}
\includegraphics[width=9cm]{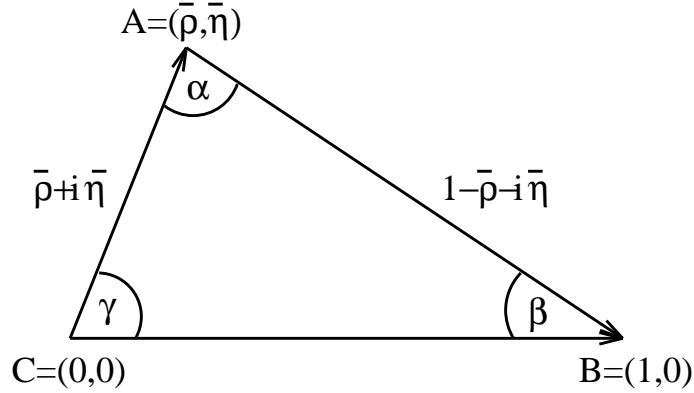}
\end{center}
\vspace{-1cm}
\caption{\it Unitarity Triangle.}
\label{fig:1utriangle}
\end{figure}

Noting that to an excellent accuracy $V_{cd}^{}V_{cb}^*$ in the 
parametrization (\ref{2.72}) is real with
$| V_{cd}^{}V_{cb}^*|=A\lambda^3+\ord(\lambda^7)$ and
rescaling all terms in (\ref{2.87h}) by $A \lambda^3$ 
we indeed find that the relation (\ref{2.87h}) can be represented 
as the triangle 
in the complex $(\overline\varrho,\overline\eta)$ plane 
as shown in Fig.~\ref{fig:1utriangle}.

Let us collect useful formulae related to this triangle:
\bi
\item
We can express $\sin(2\alpha_i$), $\alpha_i=
\alpha, \beta, \gamma$, in terms of $(\overline\varrho,\overline\eta)$ as follows:
\begin{equation}\label{2.89}
\sin(2\alpha)=\frac{2\overline\eta(\overline\eta^2+\overline\varrho^2-\overline\varrho)}
  {(\overline\varrho^2+\overline\eta^2)((1-\overline\varrho)^2
  +\overline\eta^2)},  
\end{equation}
\begin{equation}\label{2.90}
\sin(2\beta)=\frac{2\overline\eta(1-\overline\varrho)}{(1-\overline\varrho)^2 + \overline\eta^2},
\end{equation}
 \begin{equation}\label{2.91}
\sin(2\gamma)=\frac{2\overline\varrho\overline\eta}{\overline\varrho^2+\overline\eta^2}=
\frac{2\varrho\eta}{\varrho^2+\eta^2}.
\end{equation}
\item
The lengths $CA$ and $BA$ to be denoted by $R_b$ and $R_t$,
respectively, are given by
\begin{equation}\label{2.94}
R_b \equiv \frac{| V_{ud}^{}V^*_{ub}|}{| V_{cd}^{}V^*_{cb}|}
= \sqrt{\overline\varrho^2 +\overline\eta^2}
= (1-\frac{\lambda^2}{2})\frac{1}{\lambda}
\left| \frac{V_{ub}}{V_{cb}} \right|,
\end{equation}
\begin{equation}\label{2.95}
R_t \equiv \frac{| V_{td}^{}V^*_{tb}|}{| V_{cd}^{}V^*_{cb}|} =
 \sqrt{(1-\overline\varrho)^2 +\overline\eta^2}
=\frac{1}{\lambda} \left| \frac{V_{td}}{V_{cb}} \right|.
\end{equation}
\item
The angles $\beta$ and $\gamma=\delta$ of the unitarity triangle are related
directly to the complex phases of the CKM-elements $V_{td}$ and
$V_{ub}$, respectively, through
\beq\label{e417}
V_{td}=|V_{td}|e^{-i\beta},\quad V_{ub}=|V_{ub}|e^{-i\gamma}.
\eeq
\item
The unitarity relation (\ref{2.87h}) can be rewritten as
\be\label{RbRt}
R_b e^{i\gamma} +R_t e^{-i\beta}=1~.
\ee
\item
The angle $\alpha$ can be obtained through the relation
\beq\label{e419}
\alpha+\beta+\gamma=180^\circ
\eeq
expressing the unitarity of the CKM-matrix.
\ei
Formula (\ref{RbRt}) shows transparently that the knowledge of
$(R_t,\beta)$ allows to determine $(R_b,\gamma)$ through~[\ref{1BCRS1}]
\be\label{VUBG}
R_b=\sqrt{1+R_t^2-2 R_t\cos\beta},\qquad
\cot\gamma=\frac{1-R_t\cos\beta}{R_t\sin\beta}.
\ee
Similarly, $(R_t,\beta)$ can be expressed through $(R_b,\gamma)$:
\be\label{VTDG}
R_t=\sqrt{1+R_b^2-2 R_b\cos\gamma},\qquad
\cot\beta=\frac{1-R_b\cos\gamma}{R_b\sin\gamma}.
\ee
These formulae relate strategies $(R_t,\beta)$ and $(R_b,\gamma)$ for the 
determination of the  unitarity triangle that we will discuss in Chapter 6. 

The triangle depicted in Fig. \ref{fig:1utriangle}, together with
$|V_{us}|$ and $\vcb$, gives the full description of the CKM matrix.
Looking at the expressions for $R_b$ and $R_t$, we observe that within
the SM the measurements of four CP {\it conserving } decays sensitive
to $|V_{us}|$, $|V_{ub}|$, $|V_{cb}|$ and $|V_{td}|$ 
can tell us whether CP violation ($\overline\eta \not= 0$) is
predicted in the SM. This fact is often used to determine the angles
of the unitarity triangle without the study of CP violating
quantities.

Indeed, 
$R_b$ and $R_t$ determined in tree-level B decays and through 
${\rm B}^0_d-\overline {\rm B}^0_d$ mixing respectively, 
satisfy (see Chapters 3 and 4)
\be\label{con}
          1-R_b < R_t <1+R_b,
\ee
and $\overline\eta$ is predicted to be non-zero on the basis of CP conserving
transitions in the B-system alone without any reference  
 to CP violation discovered in $\rm K_L\to\pi^+\pi^-$
in 1964~[\ref{1CRONIN}]. Moreover one finds
\be\label{eta}
\overline\eta=\pm\sqrt{R^2_b-\overline\varrho^2}~, \qquad
\overline\varrho=\frac{1+R^2_b-R^2_t}{2}.
\ee  
Several expressions for $\overline\varrho$ and $\overline\eta$ in terms of 
$R_b$, $R_t$, $\alpha$, $\beta$ and $\gamma$ are collected in Chapter 6.

\boldmath
\subsection{The special role of {$|V_{us}|$}, {$|V_{ub}|$}
and {$|V_{cb}|$} elements}
\unboldmath
What do we know about the CKM matrix and the unitarity triangle on the
basis of {\it tree level} decays? 
Here the semi-leptonic K and B decays play the decisive role. As discussed 
in detail in Chapters 2 and 3 the 
present situation can be summarized by 
\begin{equation}\label{vcb}
|V_{us}| = \lambda =  0.2240 \pm 0.0036\,
\quad\quad
\vcb=(41.5\pm 0.8)\cdot 10^{-3},
\end{equation}
\begin{equation}\label{v13}
\frac{|V_{ub}|}{\vcb}=0.086\pm0.008, \quad\quad
|V_{ub}|=(35.7\pm 3.1)\cdot 10^{-4} 
\end{equation}
implying
\be
 A=0.83\pm0.02,\qquad R_b=0.37\pm 0.04~.
\ee
\begin{figure}[htb]
\begin{center}
\includegraphics[angle=-90,width=9cm]{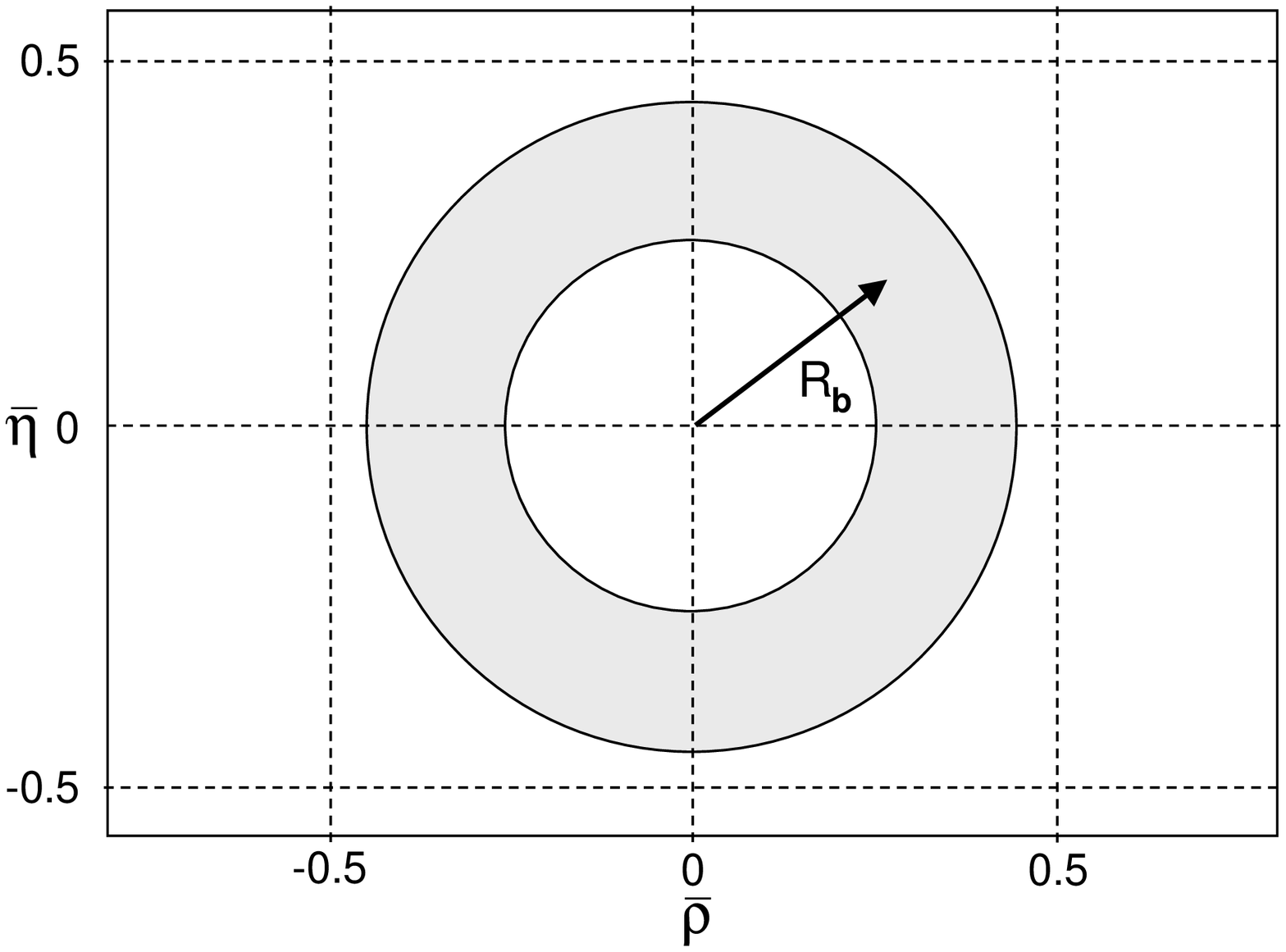}
\end{center}\vspace{-.8cm}
\caption{\it ``Unitarity Clock''}
\label{L:2}
\end{figure}
This tells us only that the apex $A$ of the unitarity triangle lies
in the band shown in Fig.~\ref{L:2}. 
While this information appears at first sight to be rather limited, 
it is very important for the following reason. As $|V_{us}|$, $\vcb$, 
 $|V_{ub}|$ and $R_b$ are determined here from tree level decays, their
values given above are to an excellent accuracy independent of any 
new physics contributions. That is, they are universal fundamental 
constants valid in any extension of the SM. Therefore their precise 
determination is of utmost importance. 
To  find where the apex $A$ lies on the {\it unitarity clock} in
Fig.~\ref{L:2} we have to look at other decays. Most promising in this
respect are the so-called {\it loop induced} decays and transitions and
CP violating B decays which will be discussed in Chapters~4--6. They
should allow us to answer the important question of whether the
Cabibbo-Kobayashi-Maskawa picture of CP violation is correct and more
generally whether the Standard Model offers a correct description of
weak decays of hadrons. In the language of the unitarity triangle the
question is whether the various curves in the 
$(\overline\varrho,\overline\eta)$
plane extracted from different decays and transitions using the SM
formulae cross each other at a single point, as shown in
Fig.~\ref{fig:2011}, and whether the angles $(\alpha,\beta,\gamma)$ in
the resulting triangle agree with those extracted from CP asymmetries
in B decays and from CP conserving B decays.
\vspace*{-0.5cm}
\begin{figure}[hbt]
\begin{center}
\includegraphics[width=17.5cm]{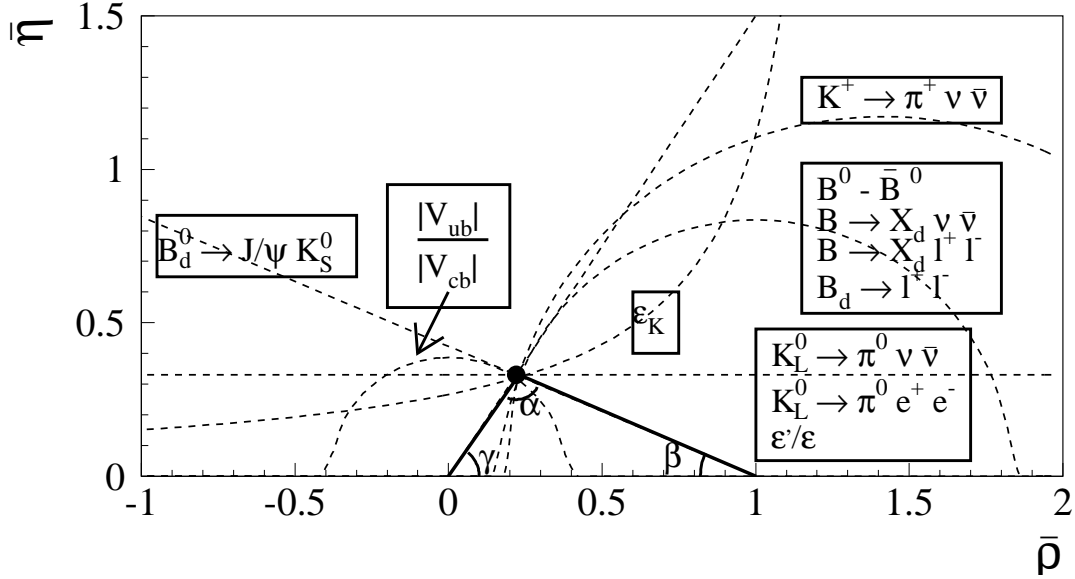}
\end{center}
\vspace*{-1.0cm}
\caption{\it The ideal Unitarity Triangle}
\label{fig:2011}
\end{figure}

\vspace*{3mm}

Any inconsistencies in the $(\overline\varrho,\overline\eta)$ plane will then
give us some hints about the physics beyond the SM. One obvious
inconsistency would be the violation of the constraint (\ref{con}).
Another signal of new physics would be the inconsistency between the
unitarity triangle constructed with the help of rare K decays alone
and the corresponding one obtained by means of B decays. Also
$(\overline\varrho,\overline\eta)$ extracted from loop induced processes and CP
asymmetries lying outside the unitarity clock in Fig.~\ref{L:2} would
be a clear signal of new physics.

In this context the importance of precise measurements of $|V_{ub}|$
and $\vcb$ should be again emphasised. Assuming that the SM with three
generations and a unitary CKM matrix is a part of a bigger theory, the
apex of the unitarity triangle has to lie on the unitarity clock
obtained from tree level decays. That is, even if SM expressions for
loop induced processes put $(\overline\varrho,\overline\eta)$ outside the
unitarity clock, the corresponding expressions of the grander theory
must include appropriate new contributions so that
$(\overline\varrho,\overline\eta)$ is shifted back to the band in
Fig.~\ref{L:2}. In the case of CP asymmetries, this could be 
achieved by realizing that in the presence of new physics contributions
 the measured angles
$\alpha$, $\beta$ and $\gamma$ are not the true angles of the
unitarity triangle but sums of the true angles and new complex phases
present in extensions of the SM. Various possibilities will be
discussed in the forthcoming CKM workshops. The better $|V_{ub}|$ and
$\vcb$ are known, the thinner the band in Fig.~\ref{L:2} becomes,
improving the selection of the correct theory. Because the branching
ratios for rare and CP violating decays depend sensitively on the
parameter $A$, precise knowledge of $\vcb$ is very important.

In order for us to draw such thin curves as in Fig.~\ref{fig:2011}, we
require both experiments and theory to be under control. Let us then
briefly discuss the theoretical framework for weak decays.

\section{Theoretical framework}
\subsection{Operator Product Expansion}
The present framework describing weak decays is based on the operator
product expansion (OPE) that allows short $(\mu_{SD}$) and long
distance $(\mu_{LD}$) contributions to weak amplitudes to be
separated, and on renormalization group (RG) methods that allow us to
sum large logarithms $\log \mu_{SD}/\mu_{LD}$ to all orders in
perturbation theory. A full exposition of these methods can be found
in~[\ref{1AJBLH},\ref{1BBL}].

The OPE allows us to write the effective weak Hamiltonian for $\Delta
F=1$ transitions as an expansion in inverse powers of $M_W$. The
leading term is simply
\be\label{intro:b1}
{\cal H}_\mathrm{eff}=\frac{G_F}{\sqrt{2}}\sum_i V^i_{\rm CKM}C_i(\mu)Q_i
\ee
with an analogous expression for $\Delta F=2$ transitions. Here $G_F$
is the Fermi constant and $Q_i$ are the relevant local operators,
built out of quark, gluon, photon and lepton fields, which govern the 
decays in question. The Cabibbo-Kobayashi-Maskawa factors $V^i_{\rm CKM}$
[\ref{1CAB},\ref{1KM}] and the Wilson coefficients $C_i(\mu)$ describe the
strength with which a given operator enters the Hamiltonian. The
latter coefficients can be considered as scale dependent {\it couplings}
related to {\it vertices} $Q_i$ and as discussed below can be calculated
using perturbative methods, as long as $\mu$ is not too small. A well
known example of $Q_i$ is the $(V-A)\otimes(V-A)$ operator relevant
for ${\rm K}^0-\overline {\rm K}^0$ mixing
\be\label{QKK}
Q(\Delta S=2)=
\overline s \gamma_\mu(1-\gamma_5) d \otimes \overline s \gamma^\mu(1-\gamma_5) d.
\ee
 We will encounter other examples later on.

An amplitude for a decay of a given meson 
$M= \rm K, B,..$ into a final state $F=\pi\nu\overline\nu,~\pi\pi,~ \rm D\,K$ 
is then simply given by
\be\label{amp5}
A(M\to F)=\langle F|{\cal H}_\mathrm{eff}|M\rangle
=\frac{G_F}{\sqrt{2}}\sum_i V^i_\mathrm{CKM}C_i(\mu)\langle 
  F|Q_i(\mu)|M\rangle,
\ee
where $\langle F|Q_i(\mu)|M\rangle$ are the matrix elements of $Q_i$
between $M$ and $F$, evaluated at the renormalization scale $\mu$. An
analogous formula exists for particle-antiparticle mixing.

The essential virtue of the OPE is that it allows the problem of 
calculating the amplitude
$A(M\to F)$ to be separated into two distinct parts: the {\it short distance}
(perturbative) calculation of the coefficients $C_i(\mu)$ and 
the {\it long-distance} (generally non-perturbative) calculation of 
the matrix elements $\langle Q_i(\mu)\rangle$. The scale $\mu$
separates, roughly speaking, the physics contributions into short
distance contributions contained in $C_i(\mu)$ and the long distance 
contributions contained in $\langle Q_i(\mu)\rangle$. 
Thus $C_i$ include the top quark contributions and
those from other heavy particles such as W-, Z-bosons and charged
Higgs or supersymmetric particles in the supersymmetric extensions
of the SM. Consequently $C_i(\mu)$ depend generally 
on $m_t$ and also on the masses of new particles if extensions of the 
SM are considered. This dependence can be found by evaluating 
so-called {\it box} and {\it penguin} diagrams with full W-, Z-, top- and 
new particles exchanges and properly including short distance QCD 
effects. The latter govern the $\mu$-dependence of~$C_i(\mu)$.

The value of $\mu$ can be chosen arbitrarily but the final result must
be $\mu$-independent. Therefore the $\mu$-dependence of $C_i(\mu)$ has
to cancel the $\mu$-dependence of $\langle Q_i(\mu)\rangle$. In other
words it is a matter of choice what exactly belongs to $C_i(\mu)$ and
what to $\langle Q_i(\mu)\rangle$. This cancellation of the
$\mu$-dependence generally involves several terms in the expansion in
(\ref{amp5}). The coefficients $C_i(\mu)$ depend also on the
renormalization scheme. This scheme dependence must also be cancelled
by that of $\langle Q_i(\mu)\rangle$, so that physical amplitudes are
renormalization scheme independent. Again, as in the case of the
$\mu$-dependence, cancellation of the renormalization scheme
dependence generally involves several terms in the 
expansion~(\ref{amp5}).

Although $\mu$ is in principle arbitrary, it is customary to choose
$\mu$ to be of the order of the mass of the decaying hadron. This is
$\ord (\mb)$ and $\ord(\mc)$ for B decays and D decays respectively.
For K decays the typical choice is $\mu=\ord$(1-2~\gev) rather than
$\ord(m_K)$ that would be much too low for any perturbative calculation of
the couplings $C_i$. Now since $\mu\ll M_{W,Z},~ m_t$, large
logarithms $\ln\mw/\mu$ compensate in the evaluation of $C_i(\mu)$ the
smallness of the QCD coupling constant $\alpha_s$, and terms
$\alpha^n_s (\ln\mw/\mu)^n$, $\alpha^n_s (\ln\mw/\mu)^{n-1}$ etc. have
to be resummed to all orders in $\alpha_s$ before a reliable result
for $C_i$ can be obtained. This can be done very efficiently by
renormalization group methods. The resulting {\it renormalization
group improved} perturbative expansion for $C_i(\mu)$ in terms of the
effective coupling constant $\alpha_s(\mu)$ does not involve large
logarithms. The related technical issues are discussed in detail
in~[\ref{1AJBLH}] and~[\ref{1BBL}].

Clearly, in order to calculate the amplitude $A(M\to F)$ the matrix 
elements $\langle Q_i(\mu)\rangle$ have to be evaluated. 
Since they involve long distance contributions one is forced in
this case to use non-perturbative methods such as lattice calculations, the
1/N expansion (where N is the number of colours), QCD sum rules, 
hadronic sum rules,
chiral perturbation theory and so on. In the case of certain B-meson decays,
the {\it Heavy Quark Effective Theory} (HQET) and {\it Heavy Quark
Expansion} (HQE) also turn out to be useful tools.
These approaches will be described in Chapter 3.
Needless to say, all these non-perturbative methods have some limitations.
Consequently the dominant theoretical uncertainties in the decay amplitudes
reside in the matrix elements $\langle Q_i(\mu)\rangle$ and non-perturbative 
parameters present in HQET and HQE.

The fact that in many cases the matrix elements $\langle Q_i(\mu)\rangle$
 cannot be reliably
calculated at present is very unfortunate. The main goal of the
experimental studies of weak decays is the determination of the CKM factors 
($V_{\rm CKM}$)
and the search for the physics beyond the SM. Without a reliable
estimate of $\langle Q_i(\mu)\rangle$ these goals cannot be achieved unless 
these matrix elements can be determined experimentally or removed from the 
final measurable quantities
by taking suitable ratios and combinations of decay amplitudes or branching
ratios. Classic examples are the extraction of the angle $\beta$ from the
CP asymmetry in $\rm B \to\psi K_S$ and the determination of the unitarity 
triangle by means of ${\rm K}\to\pi \nu\overline\nu$ decays. 
Flavour symmetries like $SU(2)_{\rm F}$ and 
$SU(3)_{\rm F}$ relating various
matrix elements can also be useful in this respect, provided flavour
breaking effects can be reliably calculated. 
However, the elimination of hadronic uncertainties from measured quantities 
can be achieved rarely  and often  one has to face directly the calculation of 
$\langle Q_i(\mu)\rangle$. 

One of the outstanding issues in the calculation of $\langle
Q_i(\mu)\rangle$ is the compatibility ({\it matching}) of $\langle
Q_i(\mu)\rangle$ with $C_i(\mu)$. $\langle Q_i(\mu)\rangle$ must have
the correct $\mu$ and renormalization scheme dependence to ensure that
physical results are $\mu$- and scheme-independent.
Non-perturbative methods often struggle with this problem, but lattice
calculations using non-perturbative matching techniques can meet this
requirement.

 Finally,  we would  like to emphasize that in addition to the hadronic 
uncertainties, any analysis of weak decays, and in particular of rare 
decays, is sensitive to possible contributions from physics beyond the SM.
Even if the latter are not discussed 
in this document and will be the subject of future workshops, it is 
instructive to describe how new physics would enter into the formula 
(\ref{amp5}). 
This can be done efficiently by using the master formula for weak decay 
amplitudes given in~[\ref{1Pisa}]. It
follows from the OPE and RG, in particular from (\ref{amp5}), but is
more useful for phenomenological applications than the formal
expressions given above. This formula incorporates SM contributions
but also applies to any extension of the SM: 
\be\label{master}
{\rm A(Decay)}= \sum_i B_i \eta^i_{\rm QCD}V^i_{\rm CKM} 
\lbrack F^i_{\rm SM}+F^i_{\rm New}\rbrack +
\sum_k B_k^{\rm New} \lbrack\eta^k_{\rm QCD}\rbrack^{\rm New} V^k_{\rm New} 
\lbrack G^k_{\rm New}\rbrack\, .
\ee
The non-perturbative parameters $B_i$ represent the matrix elements 
$\langle Q_i(\mu)\rangle$ of local 
operators present in the SM. For instance in the case of 
${\rm K}^0-\overline {\rm K}^0$ mixing, the matrix element of the operator 
$Q(\Delta S=2)$ 
in (\ref{QKK}) is represented by the parameter $\hat B_K$. An explicit 
expression is given in Chapter~4. 
There are other non-perturbative parameters in the SM that represent 
matrix elements of operators $Q_i$ with different colour and Dirac 
structures. Explicit expressions for these operators and their matrix 
elements will be given in later chapters.

The objects $\eta^i_{\rm QCD}$ are the QCD factors resulting 
from RG-analysis of the corresponding operators. They summarise the 
contributions from scales $m_b\le\mu\le m_t$ and 1-2~GeV$\le\mu\le m_t$ 
in the case of B and K  decays, respectively.
Finally, $F^i_{\rm SM}$ stand for 
the so-called Inami-Lim functions~[\ref{1IL}] that result from the 
calculations of various box and penguin diagrams. They depend on the 
top-quark mass. $V^i_{\rm CKM}$ are 
the CKM-factors we want to determine. 
 
New physics can contribute to our master formula in two ways. First,
it can modify the importance of a given operator, already relevant in
the SM, through a new short distance function $F^i_{\rm New}$ that
depends on new parameters in extensions of the SM, such as the masses
of charginos, squarks, and charged Higgs particles, or the value of
$\tan\beta=v_2/v_1$, in the Minimal Supersymmetric Standard Model (MSSM). 
These new particles enter the new
box and penguin diagrams. Second, in more complicated extensions of
the SM new operators (Dirac structures) that are either absent or very
strongly suppressed in the SM, can become important. Their
contributions are described by the second sum in (\ref{master}) with
$B_k^{\rm New}, \lbrack\eta^k_{\rm QCD}\rbrack^{\rm New}, V^k_{\rm
New}, G^k_{\rm New}$ the analogues of the corresponding objects in the
first sum of the master formula. The $V^k_{\rm New}$ show explicitly
that the second sum describes generally new sources of flavour and CP
violation beyond the CKM matrix. This sum may, however, also include
contributions governed by the CKM matrix that are strongly suppressed
in the SM but become important in some extensions of the SM. A typical
example is the enhancement of the operators with Dirac structures
$(V-A)\otimes(V+A)$, $(S-P)\otimes (S\pm P)$ and $\sigma_{\mu\nu}
(S-P) \otimes \sigma^{\mu\nu} (S-P)$ contributing to $\rm K^0$-$\rm 
\overline {K}^0$
and ${\rm B}_{d,s}^0$-$\overline {\rm B}_{d,s}^0$ mixings in the 
MSSM with large
$\tan\beta$ and in supersymmetric extensions with new flavour
violation. The latter may arise from the misalignment of quark and
squark mass matrices.
The most recent compilation of references to existing 
next-to-leading (NLO) 
calculations of $\eta^i_{\rm QCD}$ and 
$\lbrack\eta^k_{\rm QCD}\rbrack^{\rm New}$ can be found in~[\ref{1Erice}].
 
The new functions $F^i_{\rm New}$ and $G^k_{\rm New}$ as well as the
factors $V^k_{\rm New}$ may depend on new CP violating phases, making
the phenomenological analysis considerably more complicated. 
 On the other hand, in the simplest class of the extensions of the SM where
 the flavour mixing is still entirely given by the CKM matrix and  
only the SM low energy operators are relevant~[\ref{1UUT}] 
the formula (\ref{master}) simplifies to
\be
{\rm A(Decay)}= \sum_i B_i \eta^i_{\rm QCD}V^i_{\rm CKM} 
\lbrack F^i_{\rm SM}+F^i_{\rm New}\rbrack 
\label{mmaster}
\ee 
with $F^i_{\rm SM}$ and $F^i_{\rm New}$ real.
This scenario is often called 
{\it Minimal Flavour Violation} (MFV)~[\ref{1UUT}],
although one should be mindful that for
some authors MFV means a more general framework in which also new
operators can give significant contributions~[\ref{1BOEWKRUR}].

The simplicity of (\ref{mmaster}) allows to eliminate the new physics 
contributions by 
taking suitable ratios of various quantities, so that the CKM matrix can be 
determined in this class of models without any new physics uncertainties. 
This implies a universal unitarity triangle~[\ref{1UUT}] and a number of 
relations  between various quantities that are universal in this class of 
models~[\ref{1REL}]. Violation of these relations would indicate the relevance 
of new low energy operators and/or the presence of new sources of flavour 
violation.
In order to see possible violations of these 
relations and consequently the signals of new sources of flavour violation it
is essential to have a very precise determination of the CKM 
parameters. We hope that the material presented in this document is a
relevant step towards this goal.

\subsection{Importance of lattice QCD}
\def\gev{\,\mathrm{Ge\kern-0.1em V}}

Lattice calculations of the matrix elements $\langle Q_i(\mu)\rangle$
are based on a first-principles evaluation of the path integral for
QCD on a discrete space-time lattice. They have the advantage of being
systematically improvable to approach continuum QCD results with no
additional parameters beyond those of QCD itself. Indeed, lattice QCD
can be applied to determine these QCD parameters --- the quark masses
and the coupling constant. The most notable application of lattice QCD
for CKM-fitting is to the mixing parameters for neutral kaons ($B_K$)
and neutral B-mesons ($F_B$ and $B_B$). Uncertainties in these
quantities are now dominant in CKM fits. Lattice calculations are also
important for determining form factors used to extract $|V_{ub}|$ from
exclusive semileptonic B decays to light pseudoscalars or vectors, and
for providing the endpoint form factor normalization needed to extract
$|V_{cb}|$ from semileptonic B to $\mathrm{D}^{(*)}$ decays. With the
advent of CLEO-c, the current round of lattice calculations for charm
physics will be tested at the few-percent level. The charm
calculations share several features with their analogues in the $b$
sector, so a favourable outcome would bolster confidence in lattice
techniques.

In recent years much effort has been devoted to non-perturbative
techniques for improvement, to reduce discretization errors, and for
renormalization and matching, to relate lattice results either
directly to physical quantities or to quantities defined in some
continuum renormalization scheme. With non-perturbative matching, the
$\mu$- and scheme-dependence of the matrix elements $\langle
Q_i(\mu)\rangle$ is correctly matched with that of the $C_i(\mu)$.

The outstanding issue for the lattice is the inclusion of dynamical
quark effects or {\it unquenching}. Many phenomenologically important
quantities have been or are being calculated with dynamical quarks.
However the dynamical quarks cannot be simulated with light enough
masses to describe physical up and down quarks (the state-of-the-art
is a mass of about $m_s/5$). Likewise, the {\it valence} quarks, whose
propagators are used to evaluate matrix elements, also cannot be
simulated with physical up and down masses. The combined
extrapolations (chiral extrapolations) of both kinds of masses to
realistic values are a major current focus of activity.

For heavy quarks the issue is to avoid discretization errors
proportional to positive powers of $m_Q\, a$ where $m_Q$ is the mass of
the heavy quark and $a$ the lattice spacing. Since  present-day inverse
lattice spacings are in the range $2\gev < a^{-1} < 4\gev$ or so, $m_b\, a$
is intolerably large for the $b$-quark. One approach is to restrict
calculations to masses around that of charm and extrapolate to the
$b$-quark regime guided by HQET, but the extrapolation can be
significant and may amplify the $m_Q\, a$ errors, unless a continuum
limit is taken first. In the last few years much has been learned
about how to disentangle heavy quark mass-dependence and
discretization effects using an effective theory approach where QCD is
expanded in powers of $\mu/m_Q$, where $\mu$ denotes other
dimensionful scales in the problem, and discretization errors are
proportional to powers of $\mu a$ (so that $\mu$ should be smaller
than $m_Q$ and $a^{-1}$). This has been pioneered by the Fermilab
group and implemented by them and others in numerical simulations for
B-meson decay constants and semileptonic decay form factors. Lattice
discretizations of HQET and NRQCD are also effective theory approaches
which are used in simulations. In the effective theories one has to
ensure that corrections in powers of $1/m_Q$ are calculated
accurately, which involves issues of renormalization and the
proliferation of terms as the power of $1/m_Q$ increases. By combining
lattice HQET with direct simulations around the charm mass, the
$b$-quark can be reached by interpolation, but this makes sense only
if the continuum limit is taken for both calculations first.
Currently, results obtained with different approaches to treating
heavy quarks agree fairly well for $b$-physics.

An important theoretical advance in 1998 was the realization that full
chiral symmetry could be achieved at finite lattice spacing, allowing
the continuum and chiral limits to be separated. Lattice actions
incorporating chiral symmetry are being used notably in calculations
for kaon physics, including $B_K$, the $\Delta I=1/2$ rule and
$\epsilon'/\epsilon$, where the symmetry can be used to simplify the
structure of the calculation. However, these calculations are
currently quenched and have not yet had much impact on phenomenology.

\section{Experimental aspects of B physics and the CKM matrix elements} 

In this report we will review B decay properties relevant for the
measurement of the $\vub$ and $\vcb$ CKM matrix elements, and
${\rm B}^0-\overline{\rm B}^0$ oscillations which constrain  $\vtd$ and
$\vts$, allowing to test the Standard Model through the CKM
Unitarity Triangle. However, many additional measurements of B mesons
properties (masses, branching fractions, lifetimes etc.) are necessary
to constrain Heavy Quark theories to enable a precise extraction of
the CKM parameters. These measurements are also important because they
propagate to the CKM-related measurements as systematic errors.

\subsection{B physics at colliders} 

In the last 15 years --- before the start of asymmetric B-factories
--- the main contributors to B hadron studies have been symmetric
$e^+e^-$ colliders operating at the $\Upsilon(4S)$ and at the $Z^0$
resonance, and also the Tevatron $p \overline{p}$ collider (see
Table~\ref{tab:stat}).
\begin{table}[htbt] 
\begin{center}
\begin{tabular}{|c|ccc|} 
\hline
 Experiments & Number of $b\overline{b}$ events &       Environment            &      Characteristics      \\ 
  & ($\times$ 10$^6$) &             &           \\ 
\hline
 ALEPH, DELPHI  & $\sim 1 $ per expt. &       Z$^0$ decays           &  back-to-back 45 GeV b-jets   \\
 OPAL, L3        &                   & ($\sigma_{bb} \sim~$ 6nb)   &    all B hadrons produced     \\ 
\hline
    SLD      & $\sim 0.1 $           &       Z$^0$ decays           &  back-to-back 45 GeV b-jets   \\
             &                        & ($\sigma_{bb} \sim~$ 6nb)   &    all B hadrons produced     \\
             &                        &                              &        beam polarized         \\  
\hline
   ARGUS     & $\sim 0.2 $           &  $\Upsilon(4S)$ decays       &    mesons produced at rest    \\
             &                        & ($\sigma_{bb} \sim~$ 1.2nb) &       ${\rm B}^0_d$ and ${\rm B}^+$       \\
\hline
   CLEO      & $\sim  9  $           &  $\Upsilon(4S)$ decays       &    mesons produced at rest    \\
             &                        & ($\sigma_{bb} \sim~$ 1.2nb) &       ${\rm B}^0_d$ and ${\rm B}^+$       \\
\hline
    CDF      & $\sim~ \rm{several} $ & $p \overline{p}$ collisions    & events triggered with leptons \\
             &                        &    $\sqrt s$ = 1.8 TeV       &    all B hadrons produced     \\  
\hline
\end{tabular}
\caption{\it { Summary of the recorded statistics for experiments 
               at different facilities and their main characteristics.}}
\label{tab:stat}
\end{center}
\end{table}

At the $\Upsilon(4S)$ peak, $\rm B^+ B^-$ and 
${\rm B}_d^0$ $\overline{\rm B}_d^0$ 
meson pairs
are produced on top of a continuum background, with a cross section of
about 1.2~nb. At the energy used, only $\rm B^{\pm}$ and ${\rm B}_d^0$ 
mesons are
produced, almost at rest, with no additional hadrons. The constraint
that the energy taken by each B meson is equal to the beam energy is
useful for several measurements which rely on kinematic
reconstruction.

At the $Z^0$ resonance the production cross section is $\sim$ 6~nb,
about five times larger than at the $\Upsilon(4S)$, and the fraction
of $b \overline{b}$ in hadronic events, is $\sim 22\%$, very similar
to that obtained at the $\Upsilon(4S)$. Further, at the $Z^0$ peak
${\rm B}_s^0$ mesons and B baryons are produced in addition to ${\rm B}^{\pm}$
and ${\rm B}_d$ mesons. B hadrons carry, on average, about 70$\%$ of the
available beam energy, resulting in a significant boost which confines
their decay products within well-separated jets. The resulting flight
distance of a B hadron, $L = \gamma \beta c \tau$, is on average about
3~mm at these energies. Since the mean charged multiplicity in B
decays is about five, it is possible to tag B hadrons using a lifetime
tag based on the track topology. Additional hadrons are created in the
fragmentation process which can be distinguished from the heavy hadron
decay products using similar procedures.

Finally, at $p \overline{p}$ colliders $b$ quarks are produced
predominantly through gluon-gluon fusion. At the Tevatron energy of
$\sqrt s$ = 1.8 TeV the $b$-production cross section is observed to be
around 100~$\mu b$, which is huge. As the B decay products are
contained in events with a much greater multiplicity than at the $Z^0$
pole and as backgrounds are important, only specific channels, such as
fully reconstructed final states, can be studied with a favourable
signal-to-background ratio.

\begin{figure}[htbt] 
\begin{center}
\includegraphics[angle=-90,width=15cm,clip]{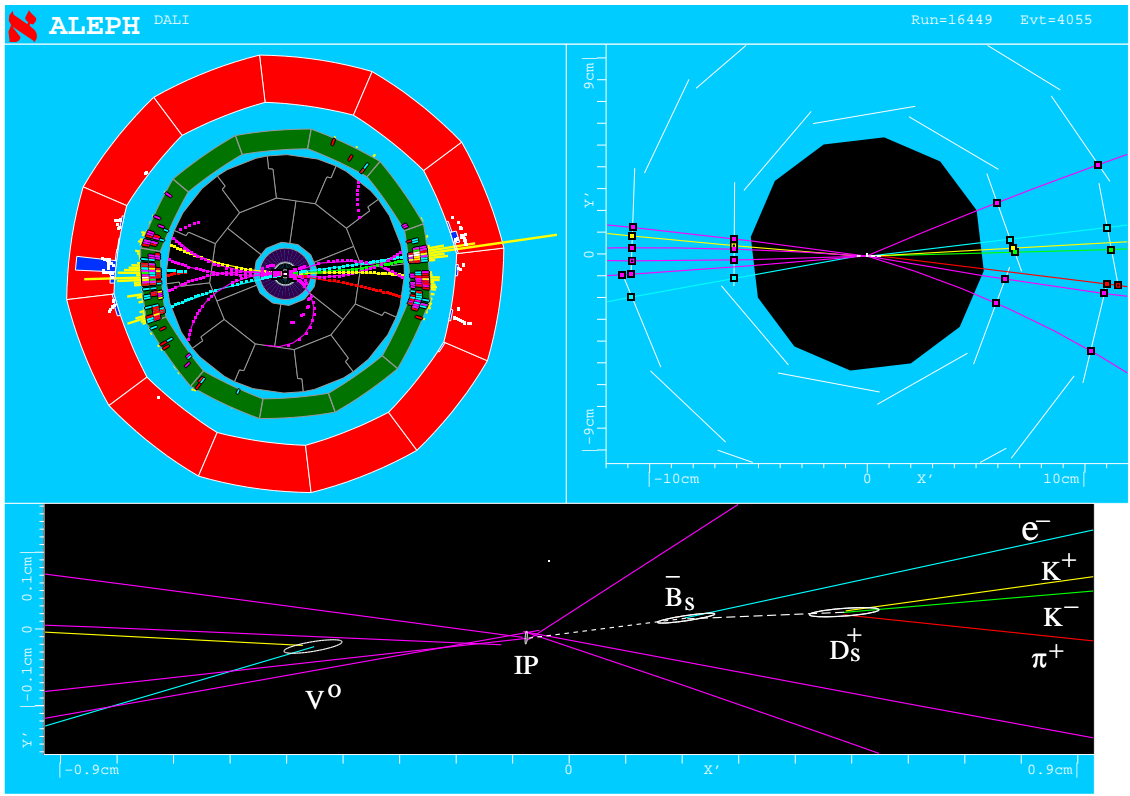}
\caption[]{\it{A $b \overline{b}$ event at LEP recorded by the ALEPH detector. 
The event consists of two jets containing the decay products of the 
two B hadrons and other particles. 
In one hemisphere a $\overline{{\rm B}}_s^0$ decays semileptonically : 
$\overline{{\rm B}}^0_s \rightarrow D_s^+ e^- \overline{\nu}_e X$, 
$ D_s^+ \rightarrow K^+ K^- \pi^+$ (tertiary vertex).}}
\label{fig:chap1_aleph_event}
\end{center}
\end{figure}

Most of the precision measurements in B~physics performed since
SLC/LEP startup have been made possible by the development of high
resolution Vertex Detectors, based on Silicon sensors. As the average
flight distance of the $b$ quark is of the order of 3~mm at $Z^0$
energies and as the typical displacement of secondary charged
particles from the event primary vertex is of the order of 200~$\mu$m,
secondary particles can be identified and the decay topology of
short-lived B~hadrons can be measured. The typical resolution of
silicon detectors varies between a few and a few tens of microns
depending on particle momentum and on detector geometry. A typical LEP
$b\overline{b}$ event is shown in Fig.~\ref{fig:chap1_aleph_event}. In
spite of a smaller $Z^0$ data set, the SLD experiment has proven to be
highly competitive, due to a superior CCD-based vertex detector,
located very close to the interaction point.

The physics output from the data taken on the  $Z^0$ resonance at LEP and SLC
has continued to improve, with a better understanding of the detector
response and new analysis techniques. Better-performing statistical
treatments of the information have been developed. As a result, the
accuracy of several measurements and the reach of other analyses have
been considerably enhanced.

In 1984, six years after the discovery of the $b \overline{b}$ bound state
 $\Upsilon$, the first experimental evidence for the existence of $\rm
 B_d$ and $\rm B^+$ mesons was obtained by ARGUS at DORIS and CLEO at
 CESR and the B mesons joined the other known hadrons in the Review of
 Particle Physics listings. By the time LEP and SLC produced their
 first collisions in 1989, the inclusive $b$ lifetime was known with 
 about 20$\%$ accuracy from measurements at PEP and PETRA. The
 relatively long $b$ lifetime provided a first indication for the
 smallness of the $\vcb$ matrix element. Branching fractions of $\rm
 B_d$ and $\rm B^+$ meson decays with values larger than about few
 $10^{-3}$ had been measured.

\begin{figure}[ht]
\vspace{-.6cm}
\begin{center}
\includegraphics[width=10.05cm]{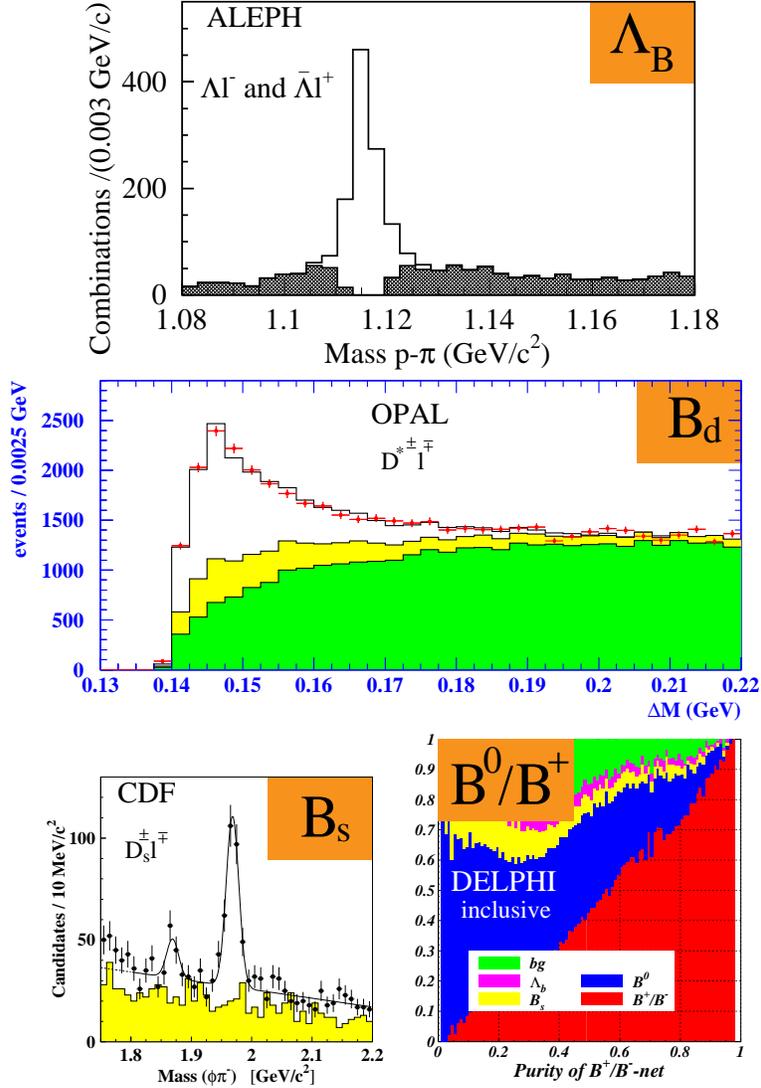}
\vspace{-.5cm}
\caption[]{\it{Signals from B hadrons. From top left to bottom left are the 
invariant mass spectra of $\rm{\Lambda}$, $\rm{((D^0 \pi)-D^0)}$, $\rm{D_s}$ which 
are obtained in correlation with 
an opposite sign lepton. These events are attributed mainly to the semileptonic decays 
of $\rm{\Lambda_b}$, $\rm{B_d^0}$ and $\rm{B_s^0}$ hadrons, respectively. 
The bottom right figure shows the possibility of distinguishing charged from
neutral B mesons based on inclusive techniques.}}
\label{fig:chap1_4figures}
\end{center}
\end{figure}
\noindent

In the early 90's the B sector landscape was enriched by the
observation of new states at LEP. Evidence of the $\Lambda_b$ baryon
was obtained in the $\Lambda_b \rightarrow \Lambda \ell \nu X$ decay
mode~[\ref{1ref:lambdab}]. This was followed by the observations of the
${\rm B}_s^0$ meson, in the decay 
$\overline{{\rm B}}_s^0 \rightarrow {\rm D}_s^+ \ell^-
\overline{\nu}_{\ell}$, in 1992 and of the $\Xi_b$ baryon in 1994.
These analyses used semileptonic decays with a relatively large
branching ratio of the order of a few $\%$ in combination with a clean
exclusive final state (${\rm D}_s$ , $\Lambda$ or $\Xi$). Using right and
wrong sign combinations, the background could be controlled and
measured using the data. Selection of those signals is shown in Fig.
\ref{fig:chap1_4figures}.\ The orbitally excited B hadrons ($L=1$)
(${\rm B}^{**}$)~[\ref{1ref:bstst}] were also found and studied starting in
1994. These analyses were mostly based on partial reconstruction,
profiting from the characteristic decay topology, and estimated the
backgrounds relying to a large extent on the data themselves.

\begin{figure}[ht]
\begin{center}
\includegraphics[width=10cm]{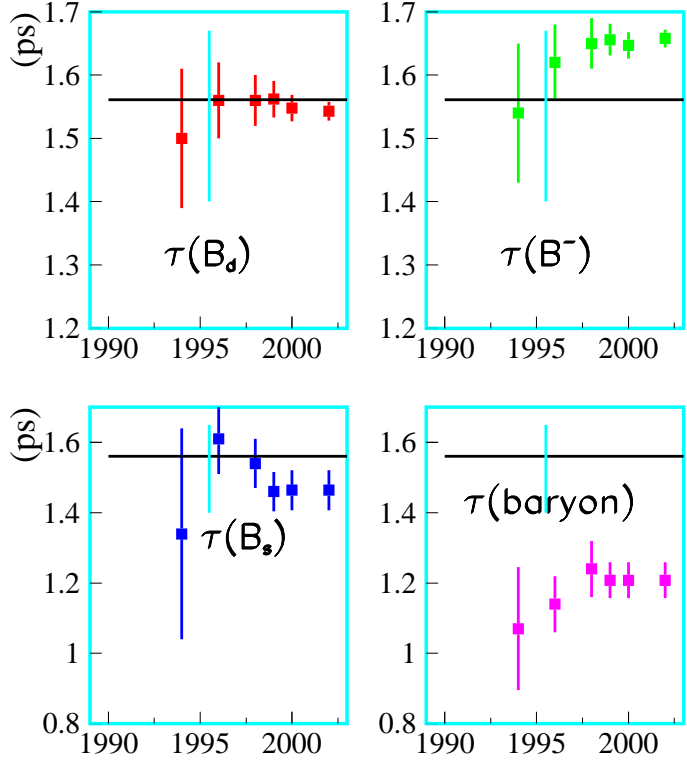} 
\caption{\it{Evolution of the combined measurement of the different B hadron 
lifetimes over the years (the last point for ${\rm B}_d^0$ and 
${\rm B}^-$ meson lifetimes includes measurement obtained at $b$-factories). 
The horizontal lines indicate the values of the inclusive $b$ lifetime, 
while the vertical lines
indicate the end of LEP data taking at the $Z^0$ resonance.}}
\label{fig:chap1_life_story}
\end{center}
\end{figure}

In parallel with studies on B spectroscopy, inclusive and individual
$\Bd$, $\Bu$, $\Bs$ and $b$-baryon exclusive lifetimes were measured
at LEP, SLD and CDF with increasing accuracies (as shown in
Fig.~\ref{fig:chap1_life_story}) down to the present final
precision, of a few percent.

\begin{figure}[ht]
\begin{center}
\includegraphics[width=7.2cm]{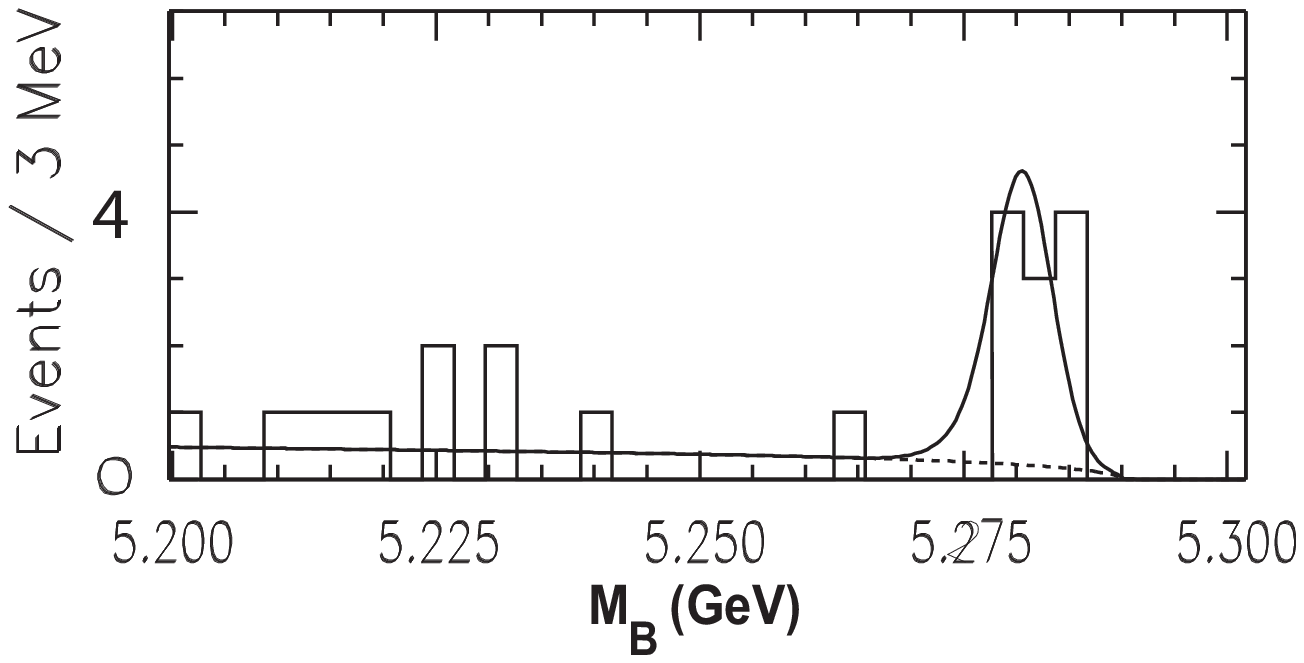}\        
\includegraphics[width=7.2cm]{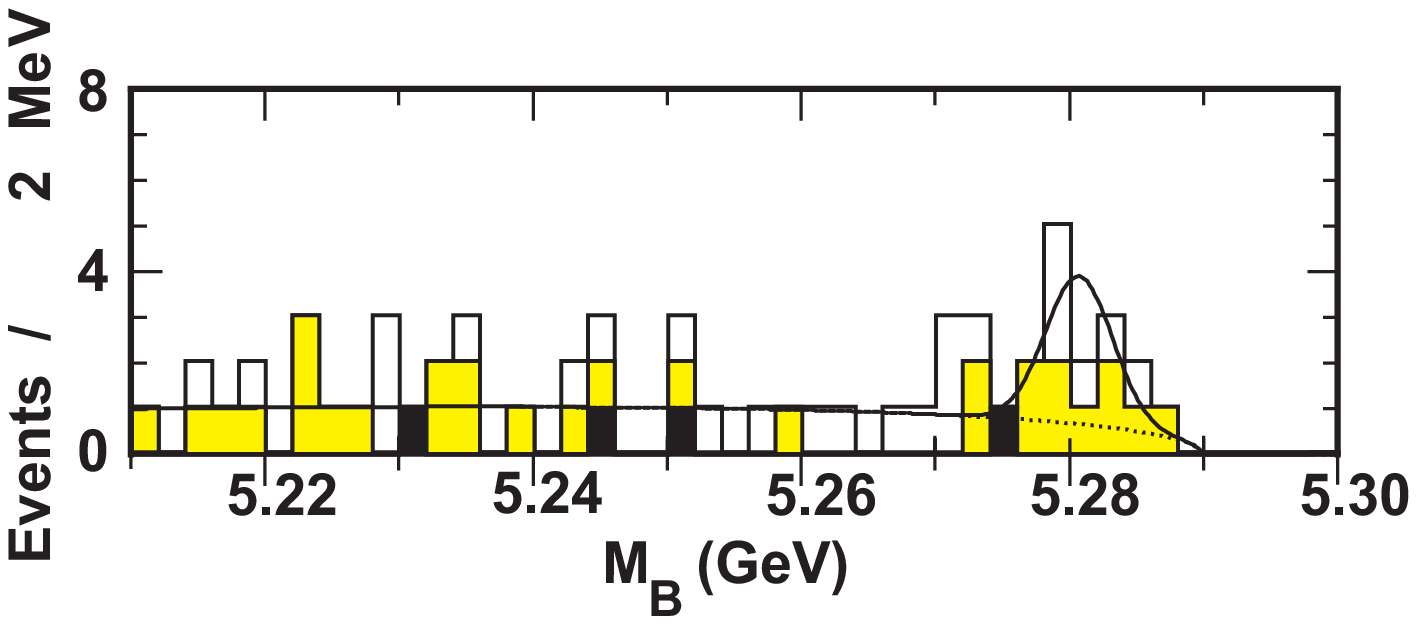}
\caption[]{\it{Left Plot: the ${\rm K}^* \gamma$ mass spectrum reconstructed
by CLEO in the ${\rm B} \rightarrow {\rm K}^* \gamma$ decay modes. Right Plot :
charmless B decays observed by CLEO : $\pi \pi$ and K$\pi$ mass
spectrum in the ${\rm B} \rightarrow \pi \pi (\rm K \pi)$ decay modes. In dark
the events with an identified pion. The plots show the updated signals
in 1995.}}
\label{fig:bsgcleo}
\end{center}
\end{figure}

Rare decays have been traditionally a hunting ground for the CLEO
experiment, which benefited from the large statistics recorded at
CESR. With about 9M ${\rm B}\overline{{\rm B}}$ meson pairs registered, 
B decay modes
with branching fractions down to 10$^{-5}$ could be observed. The
first signal for the loop-mediated ${\rm B} \rightarrow K^* \gamma$ decay
was obtained in 1993. Evidence for charmless decay of B mesons
followed [\ref{1ref:vubcleo}] (see Fig.~\ref{fig:bsgcleo}).
At LEP, where the data sets were smaller, topological decay
reconstruction methods and the efficient separation of decay products
from the two heavy mesons allowed access to some transitions having
branching fractions of order 10$^{-4}$--10$^{-5}$~[\ref{1ref:rarelep}].

A value close to 10$\%$ for the semileptonic (s.l.) $b$ branching fraction
was not expected by theorists in the early 90's. More recent
theoretical work suggests measuring both the s.l.\ 
branching fraction and the number of charmed particles in B decays. In
fact, a s.l.\  branching ratio of $10 \%$ favours a low value of
the charm mass and a value for the B branching ratio into
double charm $b \rightarrow c \overline{c} s$ of about 20\%. 
Much experimental effort has been made in recent years by
the CLEO and LEP collaborations, allowing a coherent picture to emerge.
The interplay among data analyses and phenomenology has promoted these
studies to the domain of precision physics. The s.l.\  B
branching fraction is presently known with about 2$\%$ accuracy and
much data has become available for fully inclusive, semi-inclusive and
exclusive decays. Inclusive and exclusive s.l.\  decays allow
the extraction of $|V_{cb}|$ and $|V_{ub}|$ with largely independent
sources of uncertainties and underlying assumptions. The inclusive
method is based on the measured inclusive s.l.\  widths
for $b \to X_{c,u} \ell
\overline{\nu}_{\ell}$ interpreted on the basis of the Operator Product
Expansion predictions. The exclusive method uses processes such as
$\overline{\rm B}^0_d \rightarrow \rm D^{*+} \ell^- \overline{\nu}$ and
${\rm B^-} \rightarrow \rm \rho \ell \overline{\nu}$ and relies on Heavy
Quark Effective Theory and form factor determinations. The
requirements of precision tests of the unitarity triangle are now
setting objectives for further improving our understanding of these
decays and their application in the extraction of the CKM parameters.

The second major source of information on the magnitude of the
relevant elements in the CKM matrix comes from oscillations of neutral
B mesons. A $\rm{B}^0$ meson is expected to oscillate into a
$\overline{\rm{B}}^0$ with a probability given by: $ P_{\rm{B}^0_q
\rightarrow \rm{B}^0_q(\overline{\rm{B}}^0_q)}=
\frac{1}{2}e^{-t/\tau_q} (1 \pm \cos \Delta {M}_q t)$ where
 $\Delta M_q$ is proportional to the magnitude of the $V_{tq}$ element
squared. The first signals for ${\rm B}_d$ mixing were obtained in 1987 by
the ARGUS~[\ref{1ref:argusll}] and CLEO~[\ref{1ref:cleoll}] experiments. 
The UA1 experiment at the CERN S$p\overline{p}$S collider showed evidence for
mixing due to combined contributions from both ${\rm B}_d^0$ and 
${\rm B}_{s}^0$ mesons~[\ref{1ref:ua1ll}].

At energies around the $Z^0$ peak, where both ${\rm B}_d^0$ and 
${\rm B}^0_s$ mesons
are produced with fractions $f_{B_d}$ and $f_{B_s}$, the mixing
parameter $\chi$ is given by $\chi = f_{B_d} \chi_d + f_{B_s} \chi_s$
(where $\chi_{d(s)}$ is the probability to observe a
$\overline{{\rm B}}^0_{d(s)}$ meson starting from a ${\rm B}^0_{d(s)}$ 
meson and
$f_{B_{d(s)}}$ is the ${\rm B}_{d(s)}^0$ production fraction). Owing to the
fast ${\rm B}^0_s$ oscillations, the $\chi_s$ value is close to $0.5$ and
becomes very insensitive to $\Delta M_s$. Therefore even a very
precise measurement of $\chi_s$ does not provide a determination of
$|V_{ts}|$. 

\begin{figure}[htbp!]
\begin{center}
\includegraphics[width=9cm]{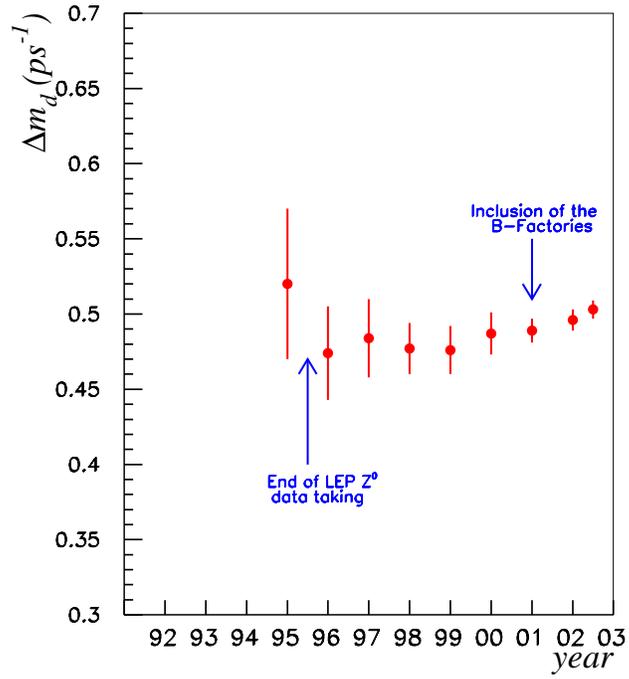}
\caption[]{\it Evolution of the average of
 $\Delta M_d$ frequency measurements over the years.}
\label{fig:chap1_dmd_story}
\end{center}
\end{figure}

It became clear that only the observation of time
evolution of ${\rm B}^0-\overline{\rm B}^0$ oscillations, for ${\rm B}_d$ and ${\rm B}_s$
mesons separately, would allow measurement of $\Delta M_d$ and $\Delta
M_s$. Time dependent ${\rm B}^0_d-\overline{{\rm B}}^0_d$ oscillation was first
observed~[\ref{1ref:dmdtime}] in 1993. The precision of the measurement
of the ${\rm B}_d$ oscillation frequency has significantly improved in
recent years. Results have been extracted from the combination of more
than thirty-five analyses which use different event samples from the
LEP/SLD/CDF experiments. At present, new results from the B-factories are
also being included. The evolution of the combined results for the
$\Delta M_d$ frequency measurement over the years is shown in
Fig.~\ref{fig:chap1_dmd_story}, reaching, before the contribution from 
the B-factories, an accuracy of $\sim$ 2.5$\%$. New, precise 
measurements performed at the B-factories
further improved this precision by a factor of~2.

\begin{figure}[htbp]
\begin{center}
\includegraphics[width=10cm]{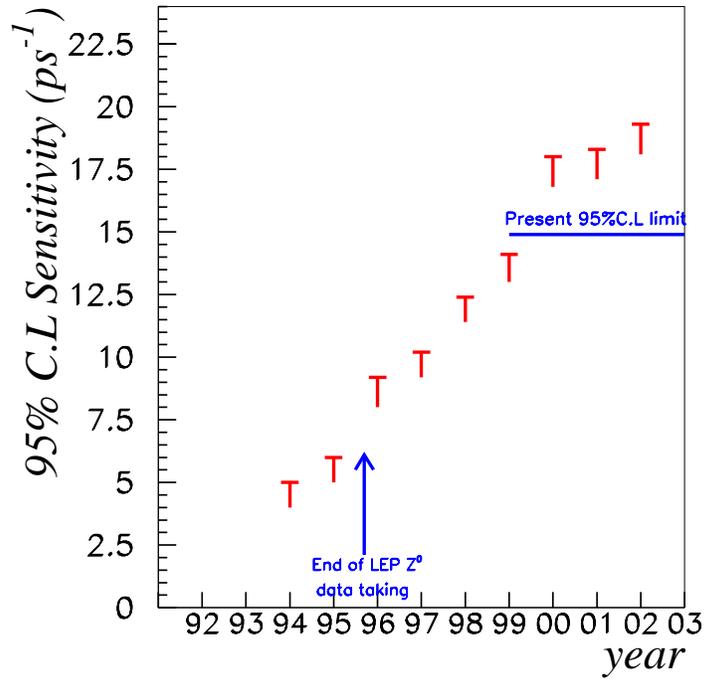}
\caption{\it Evolution of the combined $\Delta M_s$ sensitivity 
over the last decade.}
\label{fig:chap1_dms_story}
\end{center}
\end{figure}

As the ${\rm B}_s^0$ meson is expected to oscillate more than 20 times
faster than the ${\rm B}_d^0$ meson ($\sim 1/\lambda^2$) 
and as ${\rm B}_s$ mesons
are less abundantly produced, the search for 
${\rm B}^0_s-\overline{{\rm B}}^0_s$
oscillations is much more difficult. To observe these fast
oscillations, excellent resolution on the proper decay time is
mandatory. Improvements in the $\Delta M_s$ sensitivity are depicted
in Fig.~\ref{fig:chap1_dms_story}. As no signal for
${\rm B}^0_s-\overline{{\rm B}}^0_s$ oscillations has been observed 
so far, the present
limit implies that ${\rm B}_s^0$ mesons oscillate at least 30 times faster
than ${\rm B}_d^0$ mesons. The impact of such a limit on the determination
of the unitarity triangle parameters is already significant.

\section{Heavy flavour averages}
\subsection{Motivation and history}
Averaging activities have played an important role in the LEP community
 and 
several different working groups were formed to address the issue of combining 
LEP results.
The first working group to appear was the LEP Electroweak WG with members 
from ALEPH, DELPHI, L3 and OPAL, soon followed in 1994 by the $b$-hadron 
lifetime WG.
They both rather quickly felt the need to enlarge their scope, 
and provide {\it world averages} rather than just LEP averages, 
so these groups have grown to include also representatives from 
the SLD collaboration, as well as from the CDF collaboration in the case 
of the lifetime WG. 
The B oscillations WG was formed in 1996 (once the need for 
combining B mixing results arose), and was also joined by 
SLD and CDF a year later.
 
In fall 1998, the four LEP collaborations decided to create the 
Heavy Flavour Steering Group (HFS), with members from the ALEPH, 
CDF, DELPHI, L3, OPAL and SLD collaborations. 
Within the scope of heavy flavour physics --- in particular beauty 
physics --- its mandate was to identify new areas where combined results
are useful, and coordinate the averaging activities.

The HFS quickly spawned three new working groups on $\Delta \Gamma_s$,
$|V_{cb}|$ and $|V_{ub}|$, and also supported or initiated activities
in other areas like charm-counting in $b$-hadron decays, determination
of the $b$-fragmentation function, and extraction of the CKM
parameters. The coordination of all these activities resulted in
better communication between experimenters and theorists and, as a
product, a more coherent set of averages in $b$ physics updated on a
regular basis~[\ref{1LEPHFS-web}].
In order to provide world averages,
contacts have also been established with representatives of other
collaborations (CLEO, and more recently BABAR and BELLE).

The results of the $b$-lifetime WG were 
 used by the Particle Data Group from 1996 onwards; later also averages
from the B oscillation and $b$-fractions (1998),
the $|V_{cb}|$ and the $|V_{ub}|$ Working Groups (2000) were also included.  
During this Workshop an Open Forum was organised for an orderly hand-over of
the responsibility for heavy flavour physics world averages. This
forum was chaired by  HFS and PDG members. As a result, in the
future, after the HFS group disbands, these averaging activities will
be continued in the framework of a new Heavy Flavour Averaging Group
[\ref{1HFAG}], in which the Particle Data Group is also taking part.

In 2000 and 2001, the HFS group has produced
reports~[\ref{1LEPHFS_preprints}] containing combined results on
$b$-hadron production rates and decay properties from the ALEPH,
CDF-1, DELPHI, L3, OPAL and SLD experiments. A final report is
expected soon after all major results from these experiments have been
published. In the remainder of this chapter, we will give some
information on the combination procedures used for extracting averages
for the $b$-hadron lifetimes, oscillations parameters and $b$-hadron
fractions, $|V_{cb}|$ and $|V_{ub}|$. More details as well as
technical aspects can be found in~[\ref{1LEPHFS_preprints}].

\boldmath
\subsection{Averages of $b$-hadron lifetimes}
\unboldmath
Methods for combining $b$-hadron lifetime results were established in
1994, following a study~[\ref{1RForty}] triggered by a rather puzzling
fact: the world averages for the ${\rm B}_s^0$ lifetime quoted by
independent reviewers at the 1994 Winter Conferences differed 
significantly, although they were based essentially on the same data.
Different combination methods have been developed~[\ref{1Blife_method}]
in the $b$-hadron lifetime WG to take into account the underlying
exponential behaviour of the proper time distribution, as well as
handling the resulting asymmetric uncertainties and biases in low
statistics measurements.

The $b$-hadron lifetime WG provides the following averages: 
the ${\rm B}^+$ lifetime, the mean ${\rm B}^0$ lifetime, 
the ${\rm B}^+/{\rm B}^0$ lifetime ratio, the mean ${\rm B}^0_s$ lifetime, 
the $b$-baryon lifetime (averaged over all $b$-baryon species), 
the $\Lambda_b^0$ 
lifetime, the $\Xi_b$ lifetime (averaged over the two isospin states), 
and various average $b$-hadron lifetimes (e.g.\ for an unbiased mixture of 
weakly decaying $b$-hadrons). 
These averages take into account all known systematic correlations, 
which are most important for the inclusive and semi-inclusive analyses:
physics backgrounds (e.g.\ ${\rm B}\to {\rm D}^{**}\ell \nu$ 
branching ratios), bias in momentum estimates (from $b$ fragmentation, 
decay models and multiplicities, branching ratios of 
$b$- and $c$-hadrons, $b$-baryon polarization, etc.), and 
the detector resolution. For the ${\rm B}^+$ and ${\rm B}^0$ lifetimes, 
the fractions of weakly-decaying $b$-hadrons determined by 
the B oscillation WG (see Sec.~\ref{bfractions} below)
are used as an input to the averaging procedure. 
The $b$-lifetime averages are used as input by the other working groups 
for the determination of other $b$-physics averages. 

\boldmath
\subsection{Averages of  B oscillation frequencies}
\unboldmath
The main motivation for the creation of the B oscillation WG was to
combine the different lower limits obtained on $\Delta M_s$. In 1995,
the ALEPH collaboration proposed the so-called {\it amplitude
method}~[\ref{1amplitude_method}], as a way to present the $\Delta M_s$
results in a form which allowed them to be combined in a
straightforward manner. Each analysis provides the measured value of
the ${\rm B}_s$ oscillation amplitude as a function of the oscillation
frequency, normalized in such a way that a value of 1 is expected for
a frequency equal to $\Delta M_s$, and 0 for a frequency much below
$\Delta M_s$. A limit on $\Delta M_s$ can be set by excluding a value
of 1 in a certain frequency range, and the results can be combined by
averaging the measurements of this amplitude at each test frequency,
using standard techniques.

The B oscillation working group played a major role in 
promoting this method, which was eventually adopted by 
each experiment studying ${\rm B}_s$ oscillations. As a result, 
all published papers on $\Delta M_s$ since 1997 
give the {\it amplitude spectrum}, i.e.\ the ${\rm B}_s$ 
oscillation amplitude as a function of the oscillation frequency.
As the individual $\Delta M_s$ results are limited by the available 
statistics (rather than by systematics), 
the overall sensitivity to $\Delta M_s$ is greatly 
increased by performing a combination of the results of the
ALEPH, CDF, DELPHI, OPAL and SLD experiments.  

It should be noted that the sensitivity of the inclusive analyses 
depends on the assumed value for the fraction of ${\rm B}_s$ mesons in 
a sample of weakly decaying $b$-hadrons. This is taken into 
account in the combination procedure, 
which is performed assuming the latest average value for 
this fraction (see Sec.~\ref{bfractions} below).

The B oscillation working group also combines the many measurements 
of $\Delta M_d$: in February 2002, 34 measurements were available 
from 8 different experiments. Several correlated systematic 
(and statistical) uncertainties are taken into account. 
Systematic uncertainties come from two main sources: 
experimental effects (which may be correlated amongst analyses
from the same experiment), and imperfect knowledge of physics parameters
like the $b$-hadron lifetimes and $b$-hadron production fractions 
which are common to all analyses. Since different individual results are 
assuming different values for the physics parameters, all measurements are 
re-adjusted to a common (and recent) set of average values for these 
parameters before being combined. 

The average $\Delta M_d$ value is also combined with the ${\rm B}^0$ lifetime 
to get a value for $x_d$, and with the time-integrated measurements of 
$\chi_d$ performed at ARGUS and CLEO, to get world averages of 
$\Delta M_d$ and $\chi_d$. 

\boldmath
\subsection{Averages of $b$-hadron fractions in $b$-jets} 
\label{bfractions}
\unboldmath
Knowledge of the fractions of the different hadron species in an
unbiased sample of weakly-decaying $b$ hadrons produced in high-energy
$b$ jets is important for many $b$ physics measurements.
These fractions are rather poorly known from direct branching ratio
measurements: for example the fraction of ${\rm B}_s$ mesons is only known
with a $\sim 25\%$ uncertainty. However, mixing measurements allow
this uncertainty to be reduced significantly, roughly by a
factor~$2$.

Because these fractions play an important role in time-dependent
mixing analyses, the B oscillation WG was also committed to provide
$b$-hadron fractions (as well as a complete covariance matrix) that
incorporate all the available information. A procedure was developed
by this group, in which the determinations from direct measurements
are combined with the world average of $\chi_d$ and the value of
$\overline{\chi}$ (the mixing probability averaged over all $b$-hadron
species) provided by the LEP electroweak WG, under the assumption that
$\chi_s =1/2$ (as is known from the limit on $\Delta M_s$).

The $b$-hadron fractions are used as input for the $\Delta M_d$ combination 
procedure. Because the final fractions can only be known once the average 
$\Delta M_d$ is computed (and vice versa), 
the calculation of the $b$-hadron fractions and the $\Delta M_d$ averaging 
are part of the same fitting procedure, in such a way that the final 
results form a consistent set.
The fractions are also used as input for the $\Delta M_s$ combination, 
for the lifetime averages, and for the $|V_{cb}|$ average.

\boldmath
\subsection{Averages of $|V_{cb}|$ and $\vub$ elements} \label{Vcb}
\unboldmath
The $|V_{cb}|$ working group started to combine LEP results and has by now 
evolved in a  worldwide effort
including results from the collaborations BABAR, BELLE, CDF, and CLEO.
Only the case of exclusive $b\to c $ transitions presents specific problems.
To combine the different results, central values and uncertainties on 
${\cal{F}}(1) |V_{cb}|$ and $\rho^2$ have been rescaled to a common set 
of input parameters and ranges of values. The ${\cal{F}}(1) |V_{cb}|$ 
central value has then been extracted using the parametrization of 
Ref.~[\ref{Briere:2002ew}], which is based on the experimental determination 
of the $R_1$ and $R_2$ vector and axial form factors. LEP results have 
been rescaled 
accordingly. In the averaging, the correlations between the different 
measurements and that between ${\cal{F}}(1) |V_{cb}|$ and $\rho^2$ have 
been taken into account. 
The working group also provides the combination of inclusive and exclusive 
determinations. 

In order to average the inclusive charmless semileptonic branching fraction 
results from the LEP experiments, uncorrelated and correlated systematic errors
are carefully examined.  The correlated systematical errors 
come from the description of background 
$b \rightarrow c$ and from the theoretical modelling of 
signal $b \rightarrow u$ transitions. They are assumed to be fully 
correlated between the different measurements. 
The four measurements of BR($b \rightarrow X_u \ell \overline \nu$) have been 
averaged using the Best Linear Unbiased Estimate technique~[\ref{1blue}].

From this average branching 
fraction, using as an input the average $b$~lifetime value,
the probability density function for $|V_{ub}|$ has been derived. 
To obtain this function all the errors have been 
convoluted assuming that they are Gaussian in 
BR($b \rightarrow X_u \ell \overline \nu$) with the exception of 
the HQE theory error which is assumed 
to be Gaussian in $|V_{ub}|$. The negligible part of this function 
in the negative unphysical $|V_{ub}|$ region is 
discarded and the 
probability density function renormalised accordingly.
The median of this function has been chosen as the best estimate of the 
$|V_{ub}|$ value and the corresponding errors are obtained from  
the probability density function.

\section{Outline}
This document is organized as follows:

Chapters 2 and 3 are dedicated to the determination of the elements $V_{ud}$,
$V_{us}$, $V_{cb}$ and $V_{ub}$ by means of tree level decays. In Chapter~2 
we summarize the present status of the elements $V_{ud}$ and $V_{us}$. In 
Chapter 3 we discuss in detail the experimental and theoretical issues 
related to the determination of $V_{cb}$ and $V_{ub}$ from 
semileptonic inclusive and exclusive B decays and we discuss status and perspectives for  
${\rm B}^0$-$\overline{\rm B}^0$ lifetime differences and for the ratios  
of the lifetime of B hadrons.

In Chapter 4 we consider the  determination of the elements 
$\vts$ and $\vtd$, or
equivalently of  $\overline\varrho,\overline\eta$ 
by means of $\rm K^0-\overline K^0$ and 
${\rm B}^0_{d,s}-\overline {\rm B}^0_{d,s}$ mixings. 
The first part of this chapter recalls 
the formalism for  $\varepsilon_K$ and the mass differences 
$\Delta M_d$ and $\Delta M_s$. Subsequently, the present status of the 
non-perturbative calculations of $\hat B_K$, $\sqrt{\hat B_{B_d}} F_{B_d}$, 
 $\sqrt{\hat B_{B_s}} F_{B_s}$, and  $\xi$ is reviewed. The final part 
of this chapter deals with the measurements of 
${\rm B}^0_{d,s}-\overline {\rm B}^0_{d,s}$ 
oscillations, parameterized by the mass differences $\Delta M_{d,s}$.

In Chapter 5 we describe two different  statistical methods for the 
analysis of the unitarity triangle: the Bayesian approach and the frequentist
method. Subsequently,  we compare the results 
obtained in the two approaches, using 
in both cases the same  inputs from Chapters 2-4.   
We also investigate  the impact of theoretical uncertainties
on the CKM fits.

Chapter 6  deals with  topics that will be the focus of  
future CKM workshops. In 
this respect it  differs significantly 
from the previous chapters and   consists of self-contained separate 
contributions by different authors.
After a general discussion of future strategies for the determination 
of the Unitarity Triangle, a few possibilities 
 for the determination of its
angles $\alpha$, $\beta$ and $\gamma$ in B decays are reviewed. 
The potential of 
radiative and rare leptonic B decays and of $\rm K\to\pi\nu\overline \nu$ 
for the CKM determination is also considered.

Finally, Chapter 7 has a summary of the main results of 
this workshop and the conclusion.

\vspace{-2mm}

\section*{References}
\addcontentsline{toc}{section}{References}

\vspace{5mm}

\renewcommand{\labelenumi}{[\theenumi]}
\begin{enumerate}

\item \label{1CAB}
N. Cabibbo, Phys. Rev. Lett. {\bf 10} (1963) 531.

\vspace{3mm}

\item \label{1KM}
  M. Kobayashi and T. Maskawa, Prog. Theor. Phys. {\bf 49} (1973) 652.

\vspace{3mm}

\item \label{1BABAR}
 The BaBar Physics Book, eds. P. Harrison and H. Quinn, (1998),
                SLAC report 504.

\vspace{3mm}

\item \label{1LHCB}
   B Decays at the LHC, eds. P. Ball, R. Fleischer, 
  G.F. Tartarelli, P. Vikas and G. Wilkinson, \\
  hep-ph/0003238.

\vspace{3mm}

\item \label{1FERMILAB} B Physics at the Tevatron, Run II and Beyond, 
  K. Anikeev et al., hep-ph/0201071.

\vspace{3mm}

\item \label{1FX1}  H. Fritzsch and Z.Z. Xing, Prog. Part. Nucl. Phys. 
  {\bf 45} (2000) 1.

\vspace{3mm}

\item \label{1CHAU} { L.L. Chau and W.-Y. Keung}, { Phys. Rev. Lett.} 
  {\bf 53} (1984) 1802.

\vspace{3mm}

\item \label{1PDG}  { Particle Data Group,} 
 { Euro. Phys. J.} C~{\bf 15} (2000) 1.

\vspace{3mm}

\item \label{1WO} {L. Wolfenstein}, {Phys. Rev. Lett.} {\bf 51} (1983) 1945.

\vspace{3mm}

\item \label{1BLO} {A.J. Buras, M.E. Lautenbacher and G. Ostermaier,} 
               { Phys. Rev.} D~{\bf  50} (1994) 3433.

\vspace{3mm}

\item \label{1schubert} { M. Schmidtler and K.R. Schubert}, 
  { Z. Phys.} C~{\bf  53} (1992) 347.

\vspace{3mm}

\item \label{1Kayser} R. Aleksan, B. Kayser and D. London, Phys. Rev. Lett. 
  {\bf 73} (1994) 18; \\
 J.P. Silva and L. Wolfenstein, { Phys. Rev.} D~{\bf  55} (1997) 5331; \\
                 I.I. Bigi and A.I. Sanda, hep-ph/9909479.

\vspace{3mm}

\item \label{1CJ} {C. Jarlskog,} {Phys. Rev. Lett.} {\bf 55}, (1985) 1039;
                 { Z. Phys.} C~{\bf 29} (1985) 491.

\vspace{3mm}

\item \label{1BCRS1}   A.J. Buras, P.H. Chankowski, J. Rosiek and {\L}. 
 S{\l}awianowska, { Nucl. Phys.} B~{\bf 619} (2001) 434.

\vspace{3mm}

\item \label{1CRONIN}  { J.H. Christenson, J.W. Cronin, 
                         V.L.~Fitch and R.~Turlay},
                  {Phys. Rev. Lett.} {\bf 13} (1964) 128.

\vspace{3mm}

\item \label{1AJBLH}   A.J. Buras, hep-ph/9806471, in {\it Probing the Standard
                  Model of Particle Interactions}, eds. R.~Gupta, A. Morel,
                  E. de Rafael and F. David  
   (Elsevier Science B.V., Amsterdam, 1998), 
                  p.~281.

\vspace{3mm}

\item \label{1BBL}     { G. Buchalla, A.J. Buras and M. Lautenbacher,} 
                  { Rev. Mod. Phys} {\bf 68} (1996) 1125.

\vspace{3mm}

\item \label{1Pisa}    A.J. Buras, in {\it Kaon 2001}, eds. F. Constantini, G.
                  Isidori  and M. Sozzi, Frascati Physics Series, p.~15, 
                  hep-ph/0109197.

\vspace{3mm}

\item \label{1IL} { T. Inami and C.S. Lim,} {Progr. Theor. Phys.} {\bf 65} 
                  (1981) 297.

\vspace{3mm}

\item \label{1Erice}  A.J. Buras, in {\it Theory and Experiment Heading for 
                  New Physics}, ed. A. Zichichi, World Scientific, 2001, 
                  page 200, hep-ph/0101336. 

\vspace{3mm}

\item \label{1UUT}     A.J. Buras, P. Gambino, M. Gorbahn, 
  S. J\"ager and L. Silvestrini, 
                  { Phys. Lett.} B~{\bf500} (2001) 161.

\vspace{3mm}

\item \label{1BOEWKRUR} C. Bobeth, T. Ewerth, F. Kr{\"u}ger and J. Urban,
         { Phys. Rev.} D:~{\bf 64} (2001) 074014; {\bf 66} (2002) 074021; \\
 G. D'Ambrosio, G.F. Giudice, G. Isidori and A.  Strumia, 
                    { Nucl. Phys.} B~{\bf 645} (2002) 155.

\vspace{3mm}

\item \label{1REL}
S. Bergmann and G. Perez, { Phys. Rev.} D~{\bf 64} (2001) 115009, 
{ JHEP} {\bf 0008} (2000) 034; \\
A.J. Buras and R. Fleischer, { Phys. Rev.} D~{\bf 64} (2001) 115010; \\
S. Laplace, Z. Ligeti, Y. Nir and G. Perez, 
{ Phys. Rev.} D~{\bf 65} (2002) 094040; \\
A.J. Buras, hep-ph/0303060.

\vspace{3mm}

\item \label{1ref:lambdab} 
                      ALEPH Coll.,  { Phys. Lett.} B~{\bf297} (1992) 449;
                           { Phys. Lett.} B~{\bf294} (1992) 145; \\
                      OPAL Coll.,   { Phys. Lett.} B~{\bf316} (1992) 435; 
                      { Phys. Lett.} B~{\bf281} (1992) 394; \\
                      DELPHI Coll., { Phys. Lett.} B~{\bf311} (1993) 379.

\vspace{3mm}

\item \label{1ref:bstst} OPAL Coll., { Zeit. Phys.} C~{\bf 66} (1995) 19; \\
 DELPHI Coll., { Phys. Lett.} B~{\bf345} (1995) 598; \\
 ALEPH Coll.,  { Zeit. Phys.} C~{\bf 69} (1996) 393.

\vspace{3mm}

\item \label{1ref:vubcleo} CLEO Coll., {Phys. Rev. Lett.} {\bf 71} (1993) 3922.

\vspace{3mm}

\item \label{1ref:rarelep} P. Kluit, Nucl Instr. Meth. A~{\bf 462} (2001) 108.

\vspace{3mm}

\item \label{1ref:argusll} ARGUS Coll., { Phys. Lett.} B~{\bf192} (1987) 245.

\vspace{3mm}

\item \label{1ref:cleoll} CLEO Coll., { Phys. Rev. Lett.} {\bf 58} (1987) 18.

\vspace{3mm}

\item \label{1ref:ua1ll} UA1 Coll., { Phys. Lett.} B~{\bf186} (1987) 247, 
                        erratum ibid B~{\bf197} (1987) 565.

\vspace{3mm}

\item \label{1ref:dmdtime} ALEPH Coll., { Phys. Lett.} B~{\bf313} (1993)  498; 
           DELPHI Coll., {Phys. Lett.} B~{\bf332} (1994)  488; 
           OPAL Coll., { Phys. Lett.} B~{\bf336} (1994)  585.

\vspace{3mm}

\item \label{1LEPHFS-web}  The LEP Heavy Flavour Steering group maintains 
  a web page at\\ 
  {\tt http://www.cern.ch/LEPHFS/} with links to the web sites of each 
  of the different working groups, where the latest averages can be found.

\vspace{3mm}

\item \label{1HFAG} The HFAG Steering group maintains a web page at\\ 
               {\tt http://www.slac.stanford.edu/xorg/hfag/} 
 with links to the web sites of each of the different working groups,
 where the latest averages can be found.

\vspace{3mm}

\item \label{1LEPHFS_preprints} D.~Abbaneo {\it et al}, LEP Heavy Flavour 
Steering group and the different Heavy Flavour working groups 
(for the ALEPH, CDF, DELPHI, L3 OPAL, and SLD collaborations),
CERN-EP/2000-096 and update in {\sl idem}, CERN-EP/2001-050.

\vspace{3mm}

\item \label{1RForty}  R.\ Forty, CERN-PPE/94-144.

\vspace{3mm}

\item \label{1Blife_method}  L.\ Di Ciaccio {\it et al.},
  OUNP 96-05, ROM2F/96/09.

\vspace{3mm}

\item \label{1amplitude_method} ALEPH Coll., 
 contributed paper EPS 0410 to 
 Int.\ Europhysics Conf. on High Energy Physics, Brussels, July 1995; 
 the amplitude method is described in 
 H.-G.~Moser and A.~Roussarie, Nucl.\ Instrum.\ Meth. A~{\bf 384} (1997) 491.  

\vspace{3mm}

\item \label{Briere:2002ew}
R.~A.~Briere {\it et al.}  [CLEO Coll.],
Phys.\ Rev.\ Lett.\  {\bf 89} (2002) 081803, 
[hep-ex/0203032].

\vspace{3mm}


\item \label{1blue} L. Lyons, D. Gibaut and G. Burdman,  
Nucl.\ Instrum.\ Meth. A~{\bf 270} (1988) 110. 
The fit program code blue.f (author: P.~Checchia) has
been employed by the  COMBOS program (authors: O.~Schneider and H.~Seywerd) 
developed by the LEP B Oscillation Working Group \\
(see {\tt http://www.cern.ch/LEPOSC/combos/}).  


\end{enumerate}

\boldmath
\chapter{DETERMINATION OF THE~CABIBBO~ANGLE}
\unboldmath
\label{chap:II}
{\it Convener : G. Isidori \\
Contributors : V.~Cirigliano, G.~Colangelo, G.~Lopez-Castro,  
               D.~Po\v{c}ani\'c, and B.~Sciascia \\
}

\section{Introduction}

The determinations of $|V_{us}|$ and $|V_{ud}|$ provide, 
at present, the most precise constraints on the size of CKM 
matrix elements. This high-precision information is extracted 
from the semileptonic transitions $s\rightarrow u$ and $d\rightarrow u$
which, although occurring in low-energy hadronic environments,
in a few cases can be described with excellent theoretical accuracy. 
In particular, the best determination of $|V_{us}|$ is obtained 
from $\rm K \to \pi \ell \nu$ decays ($K_{\ell3}$), 
whereas the two most stringent constraints on $|V_{ud}|$ are
obtained from superallowed Fermi transitions (SFT), i.e.
beta transitions among members of a $J^P=0^+$ 
isotriplet of nuclei, and from the neutron beta decay. 
From a theoretical point of view, the beta decay of
charged pions could offer a third clean alternative 
to determine $|V_{ud}|$; however, at present 
this is not competitive with the first two because 
of the experimental difficulty in measuring 
the tiny  $\pi_{e 3}$ branching fraction 
($\sim 10^{-8}$) at the desired level of precision.

In all cases, the key observation which allows a
precise extraction of the CKM factors is the non-renormalization 
of the vector current at zero momentum transfer
in the $SU(N)$ limit (or the conservation of the vector current) 
and the Ademollo Gatto theorem [\ref{ag}]. The latter 
implies that the relevant hadronic form factors
are completely determined up to tiny isospin-breaking
corrections (in the $d\rightarrow u$ case) or $SU(3)$-breaking 
corrections (in the $s\rightarrow u$ case) of second order. 
As a result of this fortunate situation, the accuracy on
$|V_{us}|$ has reached the 1\% level and the one on 
$|V_{ud}|$ can be pushed below 0.1\%. 

Interestingly enough, if we make use of the unitarity relation 
\begin{equation}
U_{uu}=|V_{ud}|^2+|V_{us}|^2+|V_{ub}|^2 =1 \ ,
\label{eq:unitarity}
\end{equation}
the present level of accuracy on $|V_{ud}|$ and $|V_{us}|$ is such 
that the contribution of $|V_{ub}|$ to Eq.~(\ref{eq:unitarity})
can safely be neglected, and 
the uncertainty of the first two terms is comparable. 
In other words, $|V_{ud}|$ and $|V_{us}|$ lead 
to two independent determinations of the Cabibbo angle 
both at the 1\% level. 

In the following four sections we review the 
determinations of $|V_{us}|$ and $|V_{ud}|$ from 
the four main observables mentioned above. These 
results are then summarized and combined in the last section,
where we shall discuss the accuracy to which  Eq.~(\ref{eq:unitarity})
is satisfied and we shall provide a final global 
estimate of the Cabibbo angle.

\section{Determination of  $|V_{us}|$}

The amplitudes of  ${\rm K}(k) \to \pi(p) \ell \nu$  decays can be expressed in
terms of the two form factors ($f_\pm$) that determine the matrix element 
of the vector current between a pion and a kaon:
\begin{eqnarray} 
{\cal M}(K_{\ell 3}) &=& {G_\mu \over \sqrt{2}} V_{us}^* C_K
\left[f_+(t)(k+p)_\mu +f_-(t)(k-p)_\mu \right]  L^\mu~,  \qquad 
 t=(k-p)^2~,  
\label{matrix element}
\end{eqnarray}
Here $C_{K}$ is a Clebsh-Gordan coefficient, equal 
to 1 ($2^{-1/2}$) for neutral (charged) kaon decays, and 
$L^\mu$ is the usual leptonic part of the matrix
element. The corresponding decay rate reads
\be
\Gamma(K_{\ell 3}) = { G_\mu^2 \over 192 \pi^3} M_K^5 |V_{us}|^2 C_K^2
|f_+(0)|^2 I(f_+,f_-) \; \; ,
\label{decay rate}
\ee
where $I(f_+,f_-)$ is the result of the phase space integration
after factoring out $f_+(0)$. 
We recall that the dependence of $I(f_+,f_-)$ on $f_-$ 
is proportional to $(m_\ell/M_K)^2$, thus $f_-$ is completely irrelevant 
for the electron modes ($K_{e3}$). Moreover, it is customary to trade 
$f_- (t)$ for the so-called scalar form factor $f_0 (t)$, defined 
as: 
\be
f_0 (t) = f_+ (t) \ + \ \frac{t}{M_K^2 - M_\pi^2} \, f_- (t) \ . 
\ee
The momentum dependence of the form factors, which is relevant for the
integral over the phase space is often described in terms of a single
parameter, the slope at $t=0$
\be
f_{+,0} (t) = f_+ (0)\left(1+\lambda_{+,0} {t \over M_\pi^2} \right) \; \; .
\ee
In this approximation the phase space integral depends explicitly 
only on the slope parameters, and we use the notation 
$I(f_+,f_-) \rightarrow  I(\lambda_+,\lambda_0)$.  

The steps necessary to extract $|V_{us}|$ from the 
experimental determination of $K_{\ell 3}$ decay rates 
can be summarized as follows:
\begin{itemize}

\item[1.] theoretical evaluation of $f_+(0)$, including strong isospin
  violations; 
\item[2.]  measurement (or, if not available, theoretical evaluation)
  of the momentum dependence of $f_\pm(t)$;
\item[3.] theoretical treatment of photonic radiative corrections 
[note that Eq.~(\ref{decay rate})
  is not yet general enough to account for these effects, see below].
\end{itemize}
The first analysis that included all these ingredients was performed by
Leutwyler and Roos [\ref{LeuRo}]. In summary, they
\begin{itemize}
\item[1.] relied on Chiral Perturbation Theory (CHPT) 
  to $O(p^4)$ for the evaluation of $f_+(0)$, and on a
  quark model for the estimate of higher-order corrections (see below for
  more details), obtaining
\be
f_+^{K^0\pi^-}(0)=0.961\pm 0.008 ~~~ \mbox{and}  ~~~
{f_+^{K^+\pi^0}(0) / f_+^{K^0 \pi^-}(0) }=1.022~;
\ee
\item[2.] relied on CHPT at $O(p^4)$ for the evaluation of $\lambda_{+,0}$ 
      (obtaining, in particular,  $\lambda_+ = 0.031$); 
\item[3.] relied on previous work on the photonic radiative corrections, both
  for the short- (Sirlin [\ref{sirlin}]) and the long-distance (Ginsberg
  [\ref{ginsberg}]) part of this contribution;
  for the latter, they estimated an effect on the rate of the form 
\be
\Gamma(K_{e3}) \rightarrow  \Gamma(K_{e3}) (1 +  \delta)~, \quad  \delta
\simeq \pm 1 \%~.
\ee
\end{itemize}
Using all these ingredients, they finally obtained
\be
 |V_{us}|=0.2196 \pm 0.0023~. \label{eq:LR_fin}
\ee
An update of the same analysis, with substantially unchanged 
final numerical outcome, has recently been performed by Calderon
and Lopez-Castro [\ref{CLC}]. On the other hand,
new analytical ingredients were brought to this kind of analysis by
Cirigliano {\em et al.}~[\ref{CKNRT}], who did the first complete 
$O(p^4, \epsilon
p^2)$ analysis of isospin breaking corrections in the framework of CHPT
($\epsilon$ stands for both $e^2$ and $m_u-m_d$).

\subsection{Electromagnetic corrections}
The first observation to be made in order to account for 
electromagnetic corrections is the fact that photon loops
modify the very  structure of the amplitude: 
\begin{itemize}
\item[1.] The form factors now depend on another kinematical variable 
$$f_{\pm} (t) \rightarrow  f_{\pm}(t,v) \simeq \left[ 1 + 
\frac{\alpha}{4 \pi} \Gamma_c (v, \lambda_{IR}) \right] 
\, f_{\pm} (t) \ , $$
where $v=(p_K - p_l)^2$ in ${\rm K}^+$ decays and 
$v=(p_\pi + p_l)^2$ in ${\rm K}^0$ decays.
The function $\Gamma_c (v, \lambda_{IR})$ encodes universal long
distance corrections, depending only on the charges of the external
particles.  This contribution is infrared divergent, hence it depends
upon the regulator $\lambda_{IR}$.  Since the dependence on the second
kinematical variable can be factored out (to a very good approximation),
the notion of effective form factor $f_{\pm} (t) $ survives and proves
useful in the subsequent analysis.
\item[2.]  New local contributions appear in the effective form factors 
$f_{\pm} (0)$. These, together with the chiral logarithms, are truly
structure dependent corrections, which can be described in a model
independent way within the CHPT approach. 
Let us note here that the universal short-distance electroweak 
corrections to semileptonic charged-current 
amplitudes~[\ref{sirlin},\ref{Sirlin2},\ref{MS93}] 
belong in principle to this class of corrections. In fact, they can be 
related to one of the local couplings of CHPT ($X_6$)~[\ref{CKNRT}]. However, 
for consistency with previous literature, we keep the short distance 
correction explicit, and denote it by $S_{\rm ew}$. Its 
numerical value is fixed to 1.0232, corresponding to renormalization 
group evolution between $M_Z$ and $M_\rho$. 
\end{itemize}
The second observation is that one has to consider how radiation 
of real photons affects the various observables (e.g. Dalitz Plot 
density, spectra, branching ratios). For the purpose of extracting 
$|V_{us}|$, we need to assess the effect of real photon emission 
to the partial widths. As is well known, a given experiment measures 
an inclusive sum of the parent mode and radiative modes:
$$ d \Gamma_{\rm obs} = d \Gamma (K_{\ell 3}) \ + \ 
 d \Gamma (K_{\ell 3 \gamma}) \ + \ \cdots $$ 
From the theoretical point of view, only such an inclusive sum 
is free of infrared singularities. 
At the precision we aim to work at, a meaningful comparison of theory
and experiment can be done only once a clear definition of the inclusive
observable is given. In practice this means that the phase space
integrals are calculated using the same cuts on real photons employed
in the experimental analysis.

In summary, all long distance QED effects, due to both virtual photons
[$\Gamma_c (v,\lambda_{IR})$] and real photons [$d \Gamma (K_{\ell 3
\gamma})$], can be combined to produce a correction to the phase space
factor in the expression for the decay width.  This term comes, in
principle, with no theoretical uncertainty.  The structure dependent
electromagnetic corrections, as well as the chiral corrections 
to the $SU(3)$ results, are in the form factor $f_{+} (0)$, 
where all of the theoretical uncertainty concentrates.  

Based on the above considerations, we can write the partial widths 
for the $K_{\ell 3}$ modes as:
\begin{equation}
 \Gamma_{i} = {\cal N}_{i} \,  |V_{us}|^2 \, S_{\rm ew} \, 
  | f_{+}^{i} (0) |^2 \, I_{i} (\lambda_+, \lambda_0, \alpha)~, 
\end{equation} 
where the index $i$ runs over the four $K_{\ell 3}$ modes 
($K^{\pm,0}_{e 3}, K^{\pm,0}_{\mu 3}$) and we defined 
\begin{eqnarray}
{\cal N}_{i} & = & \frac{G_{\mu}^2 M_{K_i}^5}{192 \pi^3} \, C_i^2 \\
I_{i} (\lambda_+, \lambda_0, \alpha) & = & 
I_{i} (\lambda_+, \lambda_0, 0) \, \bigg[ 
1 + \Delta I_{i} (\lambda_+, \lambda_0) \bigg] \ . 
\end{eqnarray}
In the above relation $G_\mu$ indicates the Fermi constant 
as extracted from the muon decay rate after inclusion of 
radiative corrections at order $\alpha$. 

\subsection{Estimates of $f_{+}^{i} (0)$} 
The estimate of the four $f_{+}^{i} (0)$ is the key (and most delicate)
theoretical ingredient in the extraction of $|V_{us}|$. We choose to
``normalize'' them to $f_{+}^{K^0 \pi^-} (0)$, evaluated in absence of
electromagnetic corrections. Differences between the various form factors
are due to isospin breaking effects, both of strong ($\delta^{i}_{SU(2)}$)
and electromagnetic ($\delta^{i}_{e^2 p^2}$) origin, which have been
evaluated at $ {\cal O} (\epsilon p^2)$ in the chiral expansion (see
Ref.~[\ref{CKNRT}]):
\begin{eqnarray}
f_{+}^{i} (0) & = & f_{+}^{K^0 \pi^-} (0) \  (1 + \delta^{i}_{SU(2)})  \ 
(1 + \delta^{i}_{e^2 p^2})~.
\end{eqnarray}
The expansion of $f_{+}^{K^0 \pi^-} (0)$ in the quark
masses has been analysed up to the next-to-next-to-leading order
[\ref{LeuRo}]. At this level of accuracy we write
\begin{eqnarray}
f_{+}^{K^0 \pi^-} (0) & = & 1 + f_2 + f_4 + {\cal O}(p^6) \ ,   
\end{eqnarray}
where the identity $f_0=1$ follows from current conservation in the 
chiral limit.  
Because of the Ademollo-Gatto theorem [\ref{ag}], which states that 
corrections to $f_+(0)=1$ have
to be quadratic in the $SU(3)$ breaking, local terms are not allowed to
contribute to $f_2$. An explicit calculation gives
\begin{equation}
f_2=H_{K^0 \pi}+{1\over 2} H_{K^+ \pi} + {3 \over 2} H_{K^+ \eta} +
  \varepsilon \sqrt{3} \left( H_{K\pi} - H_{K \eta}  \right)\; \; ,
\end{equation}
where $H_{PQ}$ is a loop function
\begin{equation}
H_{PQ}= -{1 \over 128 \pi^2 F_\pi^2} \left[ M_P^2+M_Q^2+ {2 M_P^2 M_Q^2
    \over M_P^2 - M_Q^2} \ln {M_Q^2 \over M_P^2} \right] \; \; ,
\end{equation}
and $\varepsilon$ is the $\pi^0-\eta$ mixing angle, $\tan 2
\varepsilon=\sqrt{3}/2(m_d-m_u)/(m_s-\hat m)$. The absence of low-energy
constants in the expression for $f_2$ allows a numerical evaluation which
is practically free of uncertainties:
\begin{equation}
f_2 = -0.023 \; \; .
\end{equation}

As for $f_4$ the situation is much less clear, because low energy constants
(LECs) of the $p^6$ Lagrangian can now contribute. Before entering the
discussion of various evaluations of $f_4$ which can be found in the
literature, it is useful to recall a model-independent bound on
$f_+(0)$. A sum rule discovered in the sixties [\ref{FLRS}] implies:
\begin{equation}
\left| f_{+}^{K^0 \pi^-} (0) \right|^2=1-\sum_{n\neq \pi^-} \left| \langle K^0 |
Q^{us} | n \rangle \right|^2 \; \; ,
\label{eq:sum_rule}
\end{equation}
where $Q^{us}$ is the vector charge, thus $f_+(0)$ has to be smaller than one.

Various estimates of the size of $f_4$ have been given:
\paragraph{1}
Leutwyler and Roos [\ref{LeuRo}] relate this form factor at zero momentum
transfer to the matrix element of the vector charge between a kaon and a
pion in the infinite-momentum limit. The latter matrix element is then
given as a superposition of the wave functions of the constituents of the
kaon and of the pion. If one defines the asymmetry between the two
wave functions with a function $\delta$:
\begin{equation}
\varphi_K=(1+\delta) \varphi_\pi \; \; ,
\end{equation}
the average of the square of the asymmetry (calculated with the pion wave
function as a weight) gives the deviation of $f_+(0)$ from one:
\begin{equation}
f_+(0)=1-{1 \over 2} \langle \delta^2 \rangle_\pi \; \; .
\label{eq:delta_LR}
\end{equation}
Instead of performing an explicit model calculation of the asymmetry $\delta$,
they only made a simple para\-me\-tri\-zation of $\delta$, and estimated the
parameters on the basis of the $SU(3)$ rule of thumb: $SU(3)$ breaking
effects are expected to be at the 30\% level. Despite this large 
amount of symmetry breaking (which yields a consistent description of the 
ratio $F_K/F_\pi$), the numerical estimate of $f_4$ turns out to be very
suppressed because of the quadratic dependence from $\delta$ in Eq.~(\ref{eq:delta_LR}).
Assigning a conservative 50\% error to the effect thus found, Leutwyler and Roos finally 
quoted
\be
f_4=-0.016 \pm 0.008 \; \; .
\label{LRf4}
\ee

\paragraph{2}
Evaluations of $f_+(0)$ in a constituent quark model description of the
kaon and the pion have been made by several authors (see 
Refs.~[\ref{choiji},\ref{Jaus:zv}] and references therein). 
In particular, the result 
\be 
f_+(0) = 0.96 \pm 0.013
\ee 
can be found in Ref.~[\ref{choiji}],
which also provides a good summary of earlier literature on the subject
(as an estimate of the uncertainty we took the sensitivity
of the result to the constituent strange quark mass).\footnote{~A very 
similar result, with a larger uncertainty, has been reported more 
recently in~Ref.~[\ref{Vito}].} 
It should be stressed that in this framework no chiral logs are 
generated -- the quark masses which
appear as parameters of the model are constituent quark masses and do not
vanish in the chiral limit. In this sense, these models can only give an
estimate of the local part of $f_4$, whereas a complete estimate of $f_+(0)$
seems to require also the contribution from $f_2$.  The approach followed in constituent 
quark models is internally consistent, since the parameters of the model are also fixed 
ignoring the presence of chiral logs. However, the potential difference of chiral logs 
in $f_+(0)$ and the physical observables used to constrain the model 
could induce sizable uncertainties. As far as we
could see, nobody has addressed this point. 
Note that this problem is absent in the estimate of Leutwyler and Roos,
which consistently took into account the chiral logs.

\paragraph{3}
From the point of view of the pure chiral expansion, the only
parameter-free prediction which one can make for $f_4$ concerns the chiral
logs. A first step in this direction was made by Bijnens, Ecker and
Colangelo [\ref{doublelogs}], who calculated the double chiral logs
contribution to this quantity. The size of this term, however, depends on
the renormalization scale $\mu$. By varying the latter within a reasonable
range, the numerical estimate $|f_4|_{\rm chiral\ logs} \le 0.5 \%$ was obtained. 

Post and Schilcher~[\ref{post}] recently completed a full two-loop
evaluation of $f_4$, which besides the double chiral logs contains single
ones and polynomial contributions. The latter contain LECs of the $p^6$
Lagrangian, whose value is basically unknown, and make a numerical estimate
difficult. The authors simply set to zero the LECs of order $p^6$ (at $\mu
=M_\rho$) and obtained 
\be 
f_4 \Big|_{[O(p^6)~{\rm LECs}~=~0]} = 0.018 \; \; .
\ee
Notice the sign difference with respect to Leutwyler and Roos [and most
model calculations, which according to the sum rule (\ref{eq:sum_rule}) 
obtain negative results]. There is, however, no
contradiction between the two calculations: the neglected $O(p^6)$
constants may well give a negative contribution roughly twice as large 
as the part evaluated by Post and Schilcher.

\medskip

This brief summary clearly indicates the need for more theoretical work on
this issue. In particular it appears within reach an analysis that combines
model calculations with the contribution of the chiral logs in a consistent
way at next-to-next-to-leading order in the chiral expansion. Since the
parameters of the constituent quark models are usually fixed with simple
observables (like the decay constants), whose chiral expansion is already
known at the required level of precision, it appears to us that all the
ingredients for such an analysis are already available in the literature.

In absence of such a complete analysis, for the time being we believe that the best choice 
is to stick to Leutwyler and Roos' estimate -- keeping in mind that, e.g., a value of
$f_4$ two sigmas away from the central value in (\ref{LRf4}) is not
strictly forbidden, but rather unlikely.

Let us close this section with a comment on lattice calculations:
we are not aware of any attempts to calculate $f_+(0)$ on the lattice. 
We believe, however, that in the long run this is the only method that
offers any hope to improve the model independent estimate of Leutwyler and
Roos, and eagerly wait for the first calculations. It goes without saying
that the precision required to have an impact on the determination of
$V_{us}$ is extremely challenging.

\subsection{Numerical evaluation of $|V_{us}|$}

\begin{table}[t]
\centering
\begin{tabular}{|c|c|c|c|}
\hline
Mode            &     BR ($\%$)    &     $\lambda_+$  &   $\lambda_0$  \\
\hline
$K^{+}_{e3}$    &  4.87  $\pm$ 0.06  &  0.0278 $\pm$ 0.0019 &    \\
$K^{0}_{e3}$    &  38.79 $\pm$ 0.27  &  0.0291 $\pm$ 0.0018 &    \\
$K^{+}_{\mu 3}$ &  3.27  $\pm$ 0.06  &  0.033  $\pm$ 0.010  & 0.004 $\pm$
0.009
\\
$K^{0}_{\mu 3}$ &  27.18 $\pm$ 0.25  &  0.033  $\pm$ 0.005  & 0.027 $\pm$
0.006
\\ \hline
\end{tabular}
\caption{\it  $K_{\ell 3}$ branching ratios (BR) and slopes 
from Ref.~[\ref{pdg2002}].
The lifetimes used as input are:
$\tau_{K^\pm} = (1.2384 \pm 0.0024) \times  10^{-8} \, s$ and
$\tau_{K_L} = (5.17 \pm 0.04) \times  10^{-8} \, s$. }
\label{tab:PDGinput}
\end{table}

The $K_{\ell 3}$ widths reported in Table~\ref{tab:PDGinput}
allow us to obtain four determinations of 
$|V_{us}| \cdot f_{+}^{K^0 \pi^-} (0)$ which are independent,
up to the small correlations of theoretical uncertainties 
from isospin-breaking corrections $\delta_{e^2 p^2}$ and $\delta_{SU(2)}$
(almost negligible at present, see Table~\ref{tab:iso-brk}).
The master formula for a combined analysis of these modes is:
\begin{equation}
|V_{us}| \cdot f_{+}^{K^0 \pi^-} (0) = \left[ \frac{\Gamma_i}{ {\cal N}_i \,
S_{\rm ew} \,  I_{i} (\lambda_+, \lambda_0, 0) } \right]^{1/2} \, \frac{1}{1
+ \delta^{i}_{SU(2)} +   \delta^{i}_{e^2 p^2} + \frac{1}{2}
\Delta I_{i} (\lambda_+, \lambda_0) }
\end{equation}

\begin{table}[t]
\centering
\begin{tabular}{|c|c|c|c|c|}
\hline
  & $\delta_{SU(2)} (\%)$ & $\delta_{e^2 p^2} (\%)$ &
$\Delta I (\lambda_+, \lambda_0)  (\%)$
&  $\delta_K (\%) $ \\
\hline
$K^{+}_{e3}$ & 2.4 $\pm$ 0.2  & 0.32 $\pm$ 0.16 & -1.27 & -0.63 $\pm$ 0.32
\\
$K^{0}_{e 3}$ &    0    & 0.46 $\pm$ 0.08 & +0.37 & + 1.30 $\pm$ 0.16 \\
$K^{+}_{\mu 3}$ & 2.4 $\pm$ 0.2  & 0.006 $\pm$ 0.16 &  &   \\
$K^{0}_{\mu 3}$ &  0  & 0.15 $\pm$ 0.08 &  &   \\ \hline
\end{tabular}
\caption{\it Summary of isospin-breaking factors from Ref.~[\ref{CKNRT}].}
\label{tab:iso-brk}
\end{table}

The radiative correction factor $\delta_{K} (i)$, often quoted in the
literature, is recovered in our framework by combining the phase space
correction and the structure dependent electromagnetic corrections:
\begin{equation}
\delta_{K} (i) = 2 \, \delta^{i}_{e^2 p^2} \ + \
\Delta I_{i} (\lambda_+, \lambda_0)   \ .
\end{equation}
In the Table~\ref{tab:iso-brk} we report estimates of these
isospin-breaking parameters based on Ref.~[\ref{CKNRT}],
and subsequent work on the $K^0_{e 3}$ mode.\footnote{~Very recently 
a new calculation of electromagnetic corrections to the $K^+_{e 3}$  mode   
has been presented [\ref{Bytev}]. 
Although the approach followed in Ref.~[\ref{Bytev}] is not coherent with 
the model-independent CHPT approach of Ref.~[\ref{CKNRT}], 
the results obtained  are numerically consistent with 
those reported in  Table~\ref{tab:iso-brk}.}

The phase space corrections refer to the definition of
photon-inclusive width given by Ginsberg 
(see Refs.~[\ref{ginsberg},\ref{CKNRT}]).
The analysis of muonic modes is incomplete, as the phase space corrections
have not yet been evaluated (hence the blank spaces
in Table~\ref{tab:iso-brk}).
In order to include these modes in the phenomenological analysis,
we can use the estimates  $\delta_K (K^{+}_{\mu 3}) = - 0.06 \%$ and
$\delta_K (K^{0}_{\mu 3}) = + 2.02 \%$ obtained by Ginsberg~[\ref{ginsberg}].
However, there are systematic uncertainties in Ginsberg's approach
which cannot be easily estimated. They arise
because these results depend on the UV cutoff
($\Lambda = m_p$ was used in obtaining the above numbers),
and on the ratio $f_- (0) / f_+ (0)$ (set to zero to obtain the
above numbers). Therefore, in the following analysis,
we assign an uncertainty of $\pm 1 \% $ to  $\delta_K (K^{+,0}_{\mu 3})$.
\footnote{~As for the electronic modes, let us note that Ginsberg's
results $\delta_K (K^{+}_{e 3}) = - 0.45 \% $ and
$\delta_K (K^{0}_{e 3}) = + 1.5 \%$  are consistent with the
ranges reported in  Table~\ref{tab:iso-brk}.
}

Using the above input for the isospin-breaking factors and the 
Particle Data Group (PDG)
averages for branching ratios and slopes (see Table
\ref{tab:PDGinput}), we obtain the following results for $|V_{us}|
\cdot f_{+}^{K^0 \pi^-} (0)$:
\begin{eqnarray}
|V_{us}| \cdot f_{+}^{K^0 \pi^-} (0) & = & 0.2133 \pm 0.0016
  \qquad (K^{+}_{e 3}) \\
|V_{us}| \cdot f_{+}^{K^0 \pi^-} (0) & = & 0.2095 \pm 0.0013
\qquad (K^{0}_{e 3}) \\
|V_{us}| \cdot f_{+}^{K^0 \pi^-} (0) & = & 0.2142 \pm 0.0040
         \qquad (K^{+}_{\mu 3}) \\
|V_{us}| \cdot f_{+}^{K^0 \pi^-} (0) & = & 0.2109 \pm 0.0026
         \qquad (K^{0}_{\mu 3})
\end{eqnarray}

\begin{figure}[t]
\begin{center}
\leavevmode
\begin{picture}(100,180)
\put(10,65){\makebox(50,50){\includegraphics[height= 2.1 in]{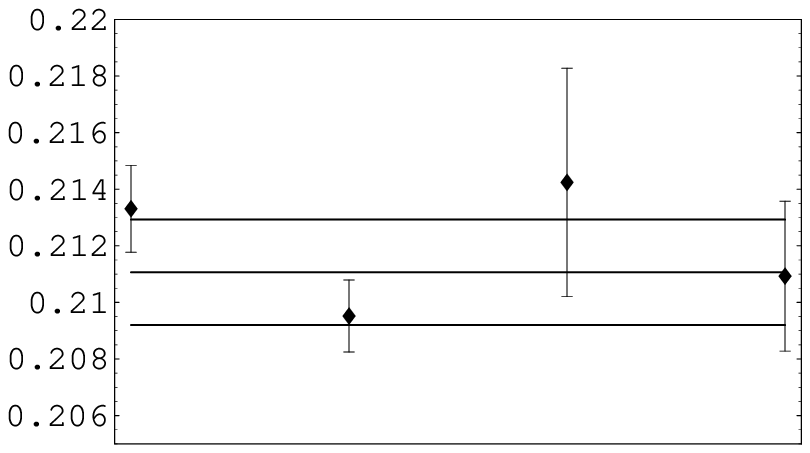}}}
\put(-170,140){\scriptsize{  $|V_{us}| \cdot f_{+}^{K^0 \pi^-} (0)$  }}
\put(-45,10){\scriptsize{$K^+_{e3}$}}
\put(12,10){\scriptsize{$K^0_{e3}$}}
\put(75,10){\scriptsize{$K^+_{\mu 3}$}}
\put(140,10){\scriptsize{$K^0_{\mu 3}$}}
\end{picture}
\caption{\it  $|V_{us}| \cdot f_{+}^{K^0 \pi^-} (0)$ from the
four $K_{\ell 3}$ modes (and average over the electronic modes).  }
\label{fig:average}
\end{center}
\end{figure}
In the above determination we have added in quadrature all the
uncertainties involved. The final error is completely dominated
by experimental uncertainties (rates and slopes).
We note that the muonic decays imply larger uncertainties
compared to the electronic modes (see also Fig.~\ref{fig:average}), 
as a consequence of larger uncertainties
in branching ratios and slopes.  In a weighted average the muonic
modes are seen to be irrelevant.  Therefore, also in view of the
incomplete knowledge of radiative corrections, we conclude that at
present the inclusion of muonic decays does not allow us to improve
the extraction of $|V_{us}|$. An average over the electronic modes brings to:
\begin{equation}
|V_{us}| \cdot f_{+}^{K^0 \pi^-} (0)  =  0.2110 \pm 0.0018 \ , 
\end{equation}
where the error has been multiplied by a scale factor $S=1.85$, 
as defined by the PDG~[\ref{pdg2002}]. 
Finally, if we use the Leutwyler-Roos estimate of $f_4$, or
$f_{+}^{K^0 \pi^-} (0)=0.961 \pm 0.008$, we obtain 
\begin{eqnarray}
|V_{us}| &=&  0.2196 \pm 0.0019_{\rm exp}  \pm 0.0018_{{\rm th}(f_4)} \nonumber \\
  &=&  0.2196 \pm 0.0026  \ .
\label{eq:Vus_fin}
\end{eqnarray}

As can be noted, the above result has an error larger 
with respect to the estimate made almost 20 years ago 
by Leutwyler and Roos. The reasons is the decreased consistency of 
the two $|V_{us}| \cdot f_{+}^{K^0 \pi^-}(0)$ determinations 
from $K_{e 3}$ modes (and the corresponding large scale 
factor) after a revised average of (old) measurements 
of $K_{e 3}$ branching ratios [\ref{pdg2002}]. 
This seems to suggest that some experimental errors have 
been underestimated, or unduly reduced by the PDG averaging procedure,
as different experimental input might have common systematic errors
(especially the ones related to radiative corrections).
We stress, in particular, that the latest rate measurements 
were performed in the 70's and it is not clear 
if these correspond to photon-inclusive widths. 

This issue should soon be clarified by the 
high-statistics measurements of $K_{\ell 3}$ widths 
expected from recently-completed or ongoing experiments,
such as BNL-E865 [\ref{E865}] and KLOE   
(which is presently analysing the four $K_{\ell 3}$ modes 
in the same experimental setup, with systematic uncertainties 
rather different from those of the existing measurements) [\ref{KLOE}].
In our opinion, a reliable extraction of
$|V_{us}|$ might come in the near future by using experimental input
coming solely from such high-statistics measurements.  Indeed, apart from 
the great improvement in the statistical signal, we can expect a substantial
improvement also in the treatment of radiative corrections.
As an illustration of this scenario, we use the preliminary result
from E865 [\ref{E865}] 
\begin{eqnarray}
{\rm BR}(K^+_{e 3}) &=&  ( 5.13 \pm 0.02_{\rm stat.} \pm 0.08_{\rm syst.}  
\pm 0.04_{\rm norm.}) \%  \\
  &=&  (5.13 \pm 0.09) \%  \ , 
\label{eq:BR_BNL}
\end{eqnarray}
to obtain 
$|V_{us}| \cdot f_{+}^{K^0 \pi^-} (0)  =  0.2189 \pm 0.0021 $, 
and hence:
\begin{eqnarray}
|V_{us}| &=&  0.2278 \pm 0.0022_{\rm exp}  \pm 0.0019_{{\rm th}(f_4)} \nonumber \\
  &=&  0.2278 \pm 0.0029  \ .
\label{eq:Vus_BNL}
\end{eqnarray} 
We show this number only for illustrative purpose 
(we do not include it in the average) since it is based 
on a preliminary result, and it is not clear if it corresponds
to the photon-inclusive branching ratio. On the other hand, it should be 
noted how a single present-day measurement allows us 
to extract $|V_{us}|$ at the $\sim 1 \% $ level 
[same as in Eq.(\ref{eq:Vus_fin})], and could offer 
the advantage of a clear understanding 
of all sources of uncertainty. 

\subsection{Other determinations of $|V_{us}|$}
To conclude the discussion about $|V_{us}|$, we briefly 
comment about other possibilities to determine this quantity,
alternative to $K_{\ell 3}$ decays:

\noindent
$\underline{\mbox{Tau decays}}$.
A novel strategy  to determine $V_{us}$
has been proposed very recently in Ref.~[\ref{pich02}]. 
The method relies on the following facts:

\begin{itemize}
\item The possibility to express theoretically, via the OPE,\footnote{~In addition, 
a phenomenological parameterisation 
for the longitudinal contribution is adopted.} the hadronic width
of the $\tau$ lepton ($R_\tau$) and the appropriate moments
($R_\tau^{kl}$)~[\ref{pich92}], for both Cabibbo-allowed ($\bar{u} d$)
and Cabibbo-suppressed ($\bar{u} s$) sectors.  The relevant moments
are denoted respectively by $R_{\tau, V+A}^{kl}$ and $R_{\tau,S}^{kl}$. 

\item The measurements of hadronic branching fractions (and moments) 
in $\tau$ decays~[\ref{davier02}].

\item The strong sensitivity of the flavour-breaking difference 
\begin{equation}
\delta R_{\tau}^{kl} = \frac{R_{\tau, V+A}^{kl}}{|V_{ud}|^2} -
\frac{R_{\tau,S}^{kl}}{|V_{us}|^2} 
\label{eq:taudiff}
\end{equation} 
to the strange-quark mass ($m_s$) and the CKM matrix elements. 
\end{itemize}
Originally these features have been exploited to determine
$m_s$, using $|V_{us}|$ as input.  The authors of
Ref.~[\ref{pich02}] have inverted this line of reasoning: they have
determined $|V_{us}|$ from (\ref{eq:taudiff}), with $k=l=0$, 
using the theoretical input $\delta R_{\tau}^{00} = 0.229 \pm 0.030$
and the range $m_s (2 \, {\rm GeV}) = 105 \pm 20$ MeV,   
derived from other observables.  
Assuming CKM unitarity to relate $|V_{ud}|$
to $|V_{us}|$ in (\ref{eq:taudiff}) one finds~[\ref{pich02}]:
\begin{equation}
|V_{us}| = 0.2179 \pm 0.0044_{\rm exp} \pm 0.0009_{\rm th} = 
0.2179 \pm 0.0045 \ .  
\label{eq:tauvus1}
\end{equation}  
The theoretical error reflects the uncertainty in $m_s$, while the
experimental one (by far the dominant error), reflects the inputs
$R_{\tau,S} = 0.1625 \pm 0.0066$ and $R_{\tau,V+A} = 3.480 \pm
0.014$~[\ref{davier02}].  
Relaxing the  assumption of CKM unitarity 
and employing the extremely safe range 
$|V_{ud}|= 0.9739 \pm 0.0025$, leads~to:
\begin{equation}
|V_{us}| = 0.2173 \pm 0.0044_{\rm exp} \pm 0.0009_{\rm th} \pm 
0.0006_{V_{ud}} = 
0.2173 \pm 0.0045 \ .  
\label{eq:tauvus2}
\end{equation} 
A reduction in the uncertainty of $R_{\tau,S}$ by a factor of two
would make this extraction of $|V_{us}|$ competitive with the one
based on $K_{e 3}$ decays. To this purpose, it would be highly
desirable also to estimate the systematic uncertainty of the method 
(e.g. extracting $V_{us}$ from higher $R_{\tau,S}^{kl}$ moments, 
and obtaining additional constraints on the $m_s$ range). 
Future precise measurements of
$R_{\tau,S}^{kl}$ and of the $SU(3)$-breaking differences $\delta
R_{\tau}^{kl}$ at $B$ factories could allow to reach this goal.
However, at the moment we believe it is safer not to include 
this (weak) constraint in the average value of $|V_{us}|$.

\medskip

\noindent
$\underline{\mbox{Hyperon semileptonic decays}}$. 
As in the case of the neutron beta decay, both
vector and axial vector currents contribute to 
Hyperon semileptonic decays (HSD). However, data 
on decay rates and asymmetries of three different HSD can not be
adequately fitted by existing models of form factors 
(see e.g. Ref~[\ref{HSD})]. On the basis of this conclusion, 
and given the large discrepancies
among different calculations of the leading vector form factors 
(of up to 13\% in
the case of $\Sigma^-$ [\ref{HSD}]) at zero momentum transfer, we decided 
not to include the HSD constraints on $|V_{us}|$ in this review.

\section{$|V_{ud}|$ from superallowed Fermi transitions}

Currently, SFT provide the most precise determination of $|V_{ud}|$.
Several features combine for this purpose. First, the calculation of SFT
are simplified by the fact that  only the nuclear matrix elements of the vector
current contribute to the decay amplitude. Second, since the
SFT occur within members of a given isotriplet, the conservation of the 
vector current helps to fix the normalization of the nuclear form factors. 
A third important theoretical ingredient is
that the calculation of isospin breaking and radiative corrections 
have achieved a level suited to match the accuracy of
experimental data. 

From the experimental side, the input information has reached 
an accuracy that challenges present theory calculations. Accurate
measurements of the half-lives $t$,  branching fractions
and  $Q$ values for nine different $J^P=0^+ \rightarrow 0^+$ beta
nuclear transitions have been reported so far [\ref{ft}]. These
experimental input allows to  compute the $ft$ values, where 
$f$ is essentially the nuclear-dependent phase space factor,
with high accuracy  for each transition (see the first column 
in Table~\ref{tab:SFT2}). 
At present, the final  uncertainty in the determination of $|V_{ud}|$ is
further reduced by taking the average
over the nine different measurements of SFT. Thus, the high
accuracy attained in $|V_{ud}|$ can largely be attributed to a statistical
origin despite the fact that a nuclear environment is being used for this
purpose.

Due to the spin and parity quantum numbers of the initial and final
nuclei, only the vector current is involved in SFT at the tree-level.
In the limit where isospin is an exact symmetry, the nuclear matrix
elements are fixed by the conservation of the vector current  
and are given by:
\begin{equation}
\langle p_f; 0^+ | \bar{u} \gamma_{\mu} d| p_i; 0^+ \rangle =\sqrt{2}
(p_i+p_f)_{\mu}\ .
\label{eq:nme}
\end{equation}
An important check in the determination of $|V_{ud}|$ from SFT is
to verify that after removing nuclear-dependent corrections from
the observables, the common nuclear matrix elements share the 
universality predicted by Eq.~(\ref{eq:nme}).

Isospin breaking corrections introduce a (nucleus-dependent) 
correction factor  $(1-\delta_C/2)$ on the right-hand 
side of Eq.~(\ref{eq:nme}). This correction arises from the
incomplete overlap of initial and final nuclear wave functions. 
This mismatch effect has  its origin in
the Coulomb distortions due to the different number of protons in the
decaying and daughter nuclei [\ref{deltac1},\ref{deltac2}]. The two different
calculations of $\delta_C$ used by the authors of Ref.~[\ref{wein98}] for
the nine measured decaying nuclei  are shown in Table~\ref{tab:SFT1}. 
The error assigned to $\delta_C$ is chosen to cover the typical 
spread between the central values obtained in both calculations. 
As we shall show in the following, at present this error is not 
the dominant source of uncertainty on $|V_{ud}|$: 
doubling it, for a more conservative approach, would induce 
an increase of the total uncertainty on $|V_{ud}|$ of about 25\%.

\begin{table}[t]
\begin{center}
\begin{tabular}{|c| c c c|}
\hline
Decaying & $\delta_C$ (in \%) & $\delta_C$ (in \%) & Averages \\
nucleus & (Ref.~[\ref{deltac1}])& (Ref.~[\ref{deltac2}]) & adopted in
[\ref{wein98}]
\\ 
 \hline
$^{10}$C & 0.18 & 0.15 & 0.16(3) \\
 $^{14}$O & 0.28 & 0.15 & 0.22(3) \\
 $^{26m}$Al & 0.33 & 0.30 & 0.31(3) \\
 $^{34}$Cl & 0.64 & 0.57 & 0.61(3) \\
 $^{38m}$K & 0.64 & 0.59 & 0.62(3) \\
 $^{42}$Sc & 0.40 & 0.42 & 0.41(3) \\
 $^{46}$V & 0.45 & 0.38 & 0.41(3) \\
 $^{50}$Mn & 0.47 & 0.35 & 0.41(3) \\
 $^{54}$Co & 0.61 & 0.44 & 0.52(3) \\ \hline
\end{tabular}
\caption{\it Isospin breaking corrections ($\delta_C$, in \% units) to the
decay rates of SFT. \label{tab:SFT1} }
\end{center}
\end{table}

After including radiative and isospin breaking 
corrections, it becomes convenient to define the ${\cal F}t$ values for
each transition, namely:  
\be
{\cal F}t  \equiv 
ft(1+\delta_R)(1-\delta_C)\nonumber  = \frac{\pi^3\ln 2}{G_\mu^2m_e^5(1+\Delta_R) 
|V_{ud}|^2}~ . \label{eq:ft}
\ee
By this way all the nucleus-dependent corrections are included 
in the definition of the  ${\cal F}$ constants and the 
right-hand side contains only the nucleus-independent piece 
of radiative corrections $\Delta_R$ (see below) and fundamental
constants. Thus, the equality of the ${\cal
F}t$ values of the nine SFT provides a good consistency 
check of the conservation of the vector current.

The radiative corrections are split into a nucleus-dependent piece
$\delta_R$ (also called {\it outer} corrections) and a nucleus-independent
piece $\Delta_R$ (also called {\it inner} corrections). The outer
corrections have been computed by two different groups [\ref{outer}]
finding good agreement with each other. These corrections include
basically model-independent virtual and real QED corrections and the
calculations include terms up to $O(Z^2\alpha^3)$ as required by
experiments (see Table~\ref{tab:SFT2}).  The inner corrections on the other hand,
include the short distance electroweak corrections and other pieces of
model-dependent (but nucleus-independent) radiative corrections
[\ref{Sirlin2},\ref{inner}]. Since a large $\ln(m_Z/m_p)$ term appears at 
leading order,
a resummation of higher order logarithms is required. This
improved calculation lead to the following updated 
numerical value of inner corrections [\ref{inner},\ref{nos01}]: 
\begin{equation}
\vspace{-0.3cm}
\Delta_R =(2.40 \pm 0.08)\%~.
\label{eq:DeltaR}
\end{equation}
The uncertainty in the inner corrections has its origin in the lower
value for the cutoff used for the axial-induced photonic corrections
[\ref{inner}] and it turns out to be the dominant theoretical
uncertainty in the current determination of $|V_{ud}|$ from SFT
[\ref{wein98}]. As a final comment, we should mention that the  factorization
of inner and outer radiative corrections introduces spurious terms 
of $O(\alpha^2)$  in the {\it r.h.s} of Eq.~(\ref{eq:ft}). However,
these additional terms are not relevant at the present level of accuracy,
since they affect the determination of $|V_{ud}|$ at the level of
$\delta_R \Delta_R/2 \sim 0.0001$.

Table~\ref{tab:SFT2} illustrates how quantum corrections are crucial to offer a
high-precision test of the conservation of the vector current. As it was mentioned 
above, this implies that the ${\cal F}t$ values should be the same
for the nine nuclear transitions under consideration. This test is evident
from the $\chi^2/$dof  of the fit, which substantially improves in going from the 
tree-level $ft$ values to the quantum corrected ${\cal F}t$ values.

\begin{table}[t] 
\begin{center}
\begin{tabular}{|c| c c c c|}
\hline
 Decaying & $ft$ & $\delta_R$ & $\delta_C$ & ${\cal F}t$ \\ nucleus &
(sec.)
& (\%) & (\%) & (sec.) \\
 \hline 
 $^{10}$C & 3038.7(45) & 1.30(4)&0.16(3)& 3072.9(48) \\
 $^{14}$O & 3038.1(18) & 1.26(5)&0.22(3)& 3069.7(26) \\
  $^{26m}$Al & 3035.8(17) & 1.45(2)&0.31(3)& 3070.0(21) \\
 $^{34}$Cl & 3048.4(19) &1.33(3)&0.61(3)& 3070.1(24) \\
 $^{38m}$K & 3049.5(21) &1.33(4)&0.62(3)& 3071.1(27) \\ 
 $^{42}$Sc & 3045.1(14) &1.47(5)&0.41(3)& 3077.3(23) \\
 $^{46}$V & 3044.6(18) &1.40(6)&0.41(3)& 3074.4(27) \\
  $^{50}$Mn & 3043.7(16) &1.40(7)&0.41(3)& 3073.8(27) \\
 $^{54}$Co & 3045.8(11) &1.40(7)&0.52(3)& 3072.2(27) \\
 \hline
 Average & 3043.9(6) & & & 3072.3(9) \\ $\chi^2/$dof & 6.38 & & & 1.10\\ \hline
 \end{tabular}
\caption{\it \label{tab:SFT2} 
Tree-level ($ft$) and corrected (${\cal F}t$) values for SFT
[\ref{wein98}].}
\end{center}
\end{table}
\vspace{3mm}

Inserting the weighted average of ${\cal F}t$, reported in 
Table~\ref{tab:SFT2}, into Eq.~(\ref{eq:ft}) we obtain:
\begin{eqnarray}
 |V_{ud}| &=& 0.9740 \pm
\sqrt{(0.0001)_{\rm exp}^2+(0.0004)^2_{\Delta_R}+(0.0002)^2_{\delta_c}}
\nonumber \\
&=& 0.9740 \pm 0.0005\ .
 \label{eq:vudsft}
\end{eqnarray}
In the above result we have explicitly separated the different contributions 
to the total uncertainty in $|V_{ud}|$. As can be noted, at present 
the dominant error is induced by ${\Delta_R}$, or by the 
choice of a low-energy cutoff for the axial-induced photonic 
corrections [\ref{inner}].

Based on the work of Ref.~[\ref{saito}], the Particle Data Group
[\ref{pdg2002}] adopt the conservative approach of doubling the error
in Eq.~(\ref{eq:vudsft}). According to Ref.~[\ref{saito}], isospin
breaking at the quark level would increase the size of the 
corrections $\delta_C$. However, we stress that this proposed manifestation 
of quarks degrees of freedom at the level of nuclear structure may lead to a
double counting of isospin breaking effects [\ref{wein98}]. Moreover, 
we note that  $\delta_C$ is not the dominant uncertainty at present. 
For these reasons, we prefer not to modify the  uncertainty
in  Eq.~(\ref{eq:vudsft}) and to quote this result as 
the best information on  $|V_{ud}|$ available 
at present from SFT.

\section{$|V_{ud}|$ from neutron beta decay}

The neutron beta decay ($n \rightarrow p e^- \overline{\nu}_e$) is another
place where measurements and theory are getting accurate enough to provide
a determination of $|V_{ud}|$ at the level of 0.1\%. This happens
despite the fact that both axial and vector weak currents
contribute to the hadronic matrix element. Indeed the 
axial form factor normalized to the vector one ($g_A/g_V$) 
can be determined from data and, once this is known, theory can
provide a calculation of the decay rate 
as a function of  $|V_{ud}|$ at the level of a few
parts in $10^{-4}$ [\ref{nos01}]. Similarly to SFT, the main theoretical
uncertainty arises from the low-energy cut-off of 
the axial-induced photonic corrections [see Eq.~(\ref{eq:DeltaR})].
However, the present major source of uncertainty in this 
determination of $|V_{ud}|$ comes from the measurements of $g_A/g_V$.

In the last twenty years several authors contributed to derive 
a master formula for the beta decay rate of the neutron precise at the
level of a few parts in $10^{-4}$ 
[\ref{inner},\ref{nos01},\ref{wilkinson},\ref{paver}]. 
At this level of accuracy, only two momentum-independent 
form factors, $g_{A,V}$, are required to describe the hadronic amplitude:
\begin{equation}
\langle p | \overline{u}\gamma_{\mu}(1-\gamma_5) d| n \rangle =
\overline{u}_p \gamma_{\mu} (g_V+g_A\gamma_5) u_n~.
 \label{eq:neutronme}
\end{equation}
By including the radiative corrections up to order $O(\alpha^2)$ in an
additive form [\ref{nos01}], and assuming that the beta decay (inclusive
of photon corrections) is the only relevant neutron decay mode, we can write 
\begin{equation}
|V_{ud}|^2= \frac{2\pi^3}{G_\mu^2 m_e^5 (1+\Delta_R)g_V^2 [1+3(g_A/g_V)^2] f_R
\tau_n}\ , 
\label{eq:neu-rate}
\end{equation}
where $\tau_n$ denotes the lifetime of the neutron.  The
factor $f_R =1.71312 \pm 0.00002$ is essentially the integrated electron
spectrum folded with the energy-dependent radiative corrections which are
required  up to order $\alpha^2$ [\ref{nos01},\ref{wilkinson}]. Here, an
important numerical remark is in order. Our expression for $f_R$ contains
the radiative
corrections in an additive form, contrary to some other expressions
currently used in the literature 
(see for example Ref.~[\ref{wein98},\ref{abele02}]).
In fact, other analyses use a factorization prescription of the dominant
Coulomb term (or Fermi-function) and the remaining $O(\alpha, \alpha^2)$
corrections (see the discussion in Ref.~[\ref{nos01}]). This factorization
introduces spurious terms at the  order $\alpha^2$ which affect 
the decay rate at the level of $10^{-4}$. For instance, using the 
factorized formula of Ref.~[\ref{abele02}] and 
Ref.~[\ref{wein98},\ref{wilkinson}] 
would lead to a decrease of $|V_{ud}|$ of about $5\times 10^{-4}$ 
and $10\times 10^{-4}$, respectively.

Let us now focus on the value of $g_V$. In the limit where isospin
is an exact symmetry, the CVC hypothesis is
useful to fix the value of the vector form factor, namely $g_V^{CVC}=1$.
Owing to the Ademollo-Gatto theorem [\ref{ag}], 
the correction $\delta g_V$  to this value is expected to be very small.
The evaluations of $\delta g_V$ present in the 
literature indicate that $\delta g_V$ is of order $10^{-5}$ [\ref{paver}]. 
Thus, at the accuracy level of $10^{-4}$, we can
safely set $g_V=1$ in Eq.~(\ref{eq:neu-rate}).

As already mentioned, the determination of
$|V_{ud}|$ requires two experimental inputs: $\tau_n$ and $g_A/g_V$. 
At present, the set of data used to extract the average of the neutron lifetime is
consistent and very accurate, leading to the world average [\ref{pdg2002}]
\begin{equation}
\tau_n^{\rm exp} = (885.7 \pm 0.8)\ {\rm sec}\ ,
\label{eq:lifetime}
\end{equation}
with an associated scale factor of $S=1.07$. 
The situation concerning $g_A/g_V$ does not share the same consistency.
In Table~\ref{tab:nbeta} we show a collection of the most precise measurements of
this ratio, together with the corresponding determinations of $|V_{ud}|$.  
Although other measurements have been reported so far 
(see Ref.~[\ref{pdg2002}]), 
those shown in Table~\ref{tab:nbeta} are the most relevant ones 
for the determination of $|V_{ud}|$ in neutron beta decay [\ref{nos01}].

\begin{table}[t]
\begin{center}
\begin{tabular}{|l|c|c|}
\hline
Reference & $g_A/g_V$ & $|V_{ud}|$  \\ 
\hline
$\left [\ref{bopp86}\right ]$  & $-1.262 \pm 0.005$ & $0.9794 \pm 0.0032$ \\ 
$\left [\ref{ero97}\right ]$   & $-1.2594 \pm 0.0038$ & $0.9811 \pm 0.0025$ \\ 
$\left [\ref{liaud97}\right ]$~[*] & 
$-1.266 \pm 0.004$ & $0.9768 \pm 0.0026$ \\ 
$\left [\ref{mostovoi01}\right ]$~[*] & 
$-1.2686 \pm 0.0046$ & $0.9752 \pm 0.0030$ \\
$\left [\ref{abele02}\right ]$~[*]    & 
$-1.2739 \pm 0.0019$ & $0.9718 \pm 0.0013$ \\ 
\hline 
Total average & $-1.2694 \pm 0.0014\ (0.0028) $ & 
$0.9746 \pm 0.0010\ (0.0019)$ \\
 Average [*]  & $-1.2720 \pm 0.0016\ (0.0022) $ & 
$0.9731 \pm 0.0011\ (0.0015)$ \\
\hline
\end{tabular}
\caption{\it \label{tab:nbeta} Experimental values of   $g_A/g_V$
and the determination of $|V_{ud}|$ from neutron beta decay; the 
[*] denotes recent experiments with a polarization larger than 90\%; the 
errors between brackets in the last two rows are multiplied by 
the corresponding scale factor.}
\end{center} 
\end{table}


The quoted uncertainty for $|V_{ud}|$ in the third column of Table~\ref{tab:nbeta} 
has been computed by adding in quadrature theoretical and experimental
uncertainties, following the formula
\begin{equation}
\Delta |V_{ud}| = \pm
\sqrt{(0.00044)^2_{\tau_n}+(0.00036)^2_{\rm r.c.}+[0.64\Delta(g_A/g_V)]^2}\ ,
\end{equation}
where the subscript `r.c.' denote the irreducible theoretical 
error of axial-induced photonic corrections. Similarly to the SFT case, 
the latter contribute at the level of 0.04\%. 

Clearly, the present uncertainty in $|V_{ud}|$ from neutron beta decay is
largely dominated by the error in the measurements of $g_A/g_V$.
It should also be noted that there is some inconsistency 
among the measurements of this ratio (in fact, this is at the origin 
of the large scale factor quoted in the total average). 
On the other hand, if only the recent measurements of $g_A/g_V$ performed 
with a high degree of polarization are considered (entries denoted 
by [*] in Table~\ref{tab:nbeta}), as recommended by the PDG [\ref{pdg2002}],
then a good consistency is recovered. For this reason, we consider 
the result in the last row (scale factor included), 
\begin{equation}
  |V_{ud}| = 0.9731 \pm 0.0015~,
\label{eq:nbfin}
\end{equation}
as the best information available at present from neutron beta decay.

\section{$|V_{ud}|$ from   $\pi_{e 3}$ decay}

Another interesting possibility to extract $|V_{ud}|$, 
which shares the advantages of both Fermi transitions (pure vector transition, 
no axial-vector contribution) and neutron $\beta$-decay (no nuclear structure
dependent radiative corrections) is provided by the $\beta$-decay of the 
charged pion. The difficulty here lies in the extremely small branching 
ratio, of order $10^{-8}$. Nevertheless, such a measurement is presently
being performed at PSI by the PIBETA collaboration, with the aim of 
measuring the branching ratio with 0.5\% accuracy. At this level of 
precision, radiative corrections have to be taken into account, and 
Ref.~[\ref{cknp02}] addresses this problem, within the effective 
theory formalism for processes involving light pseudoscalar mesons, 
photons and leptons~[\ref{lept}]. 

The decay amplitude for $ \pi^+ (p_+) \to \pi^0 (p_0) 
\, e^+ (p_e) \, \nu_e (p_\nu) $ is determined by the vector pion 
form factor  $f_{+}^{\pi^\pm \pi^0} (t)$, entering in the 
current matrix element 
\begin{equation}
\langle  \pi^0(p_0) | \bar{u} \gamma_\mu d | \pi^+(p_+) \rangle = \sqrt{2} \, 
\bigg[ f_{+}^{\pi^\pm \pi^0} (t) \, (p_+ + p_0)_{\mu} 
+  f_{-}^{\pi^\pm \pi^0} (t) \, (p_+ - p_0)_{\mu} \bigg] ,
\end{equation}
where $t = (p_+ - p_0)^2$. As usual, in the isospin limit
$f_{+}^{\pi^\pm \pi^0} (0) = 1$  and the kinematical 
dependence from $t$ is parameterised by a linear slope $\lambda$. 
The effect of $f_{-}$, suppressed by $(m_e/M_\pi)^2$ 
and by isospin breaking, can safely be neglected.

Accounting for isospin-breaking and radiative corrections (see
analogous discussion for $K_{\ell 3}$ decays), the decay rate can be
written as
\begin{equation}
 \Gamma_{\pi_{e 3 [\gamma]}} = {\cal N}_{\pi} \,  |V_{ud}|^2 \, S_{\rm ew} \, 
  | f_{+}^{\pi^\pm \pi^0} (0) |^2   \, I_{\pi} (\lambda, \alpha) 
\end{equation} 
with $~{\cal N}_{\pi} = G_{\mu}^2 M_{\pi^{\pm}}^5 /(64 \pi^3)$ and 
\be
I_{\pi} (\lambda, \alpha) = 
I_{\pi} (\lambda, 0) \, \bigg[ 
1 + \Delta I_{\pi} (\lambda) \bigg] \; \; ,  ~~~~~
f_{+}^{\pi^\pm \pi^0} (0) =  (1 + \delta^{\pi}_{SU(2)})  \ 
(1 + \delta^{\pi}_{e^2 p^2}) \; \; .
\ee
Thus we obtain: 
$$
|V_{ud}| = \left[ \frac{\Gamma_{\pi_{e 3 [\gamma]}}}{ {\cal N}_\pi \,
S_{\rm ew} \,  I_{\pi} (\lambda, 0) } \right]^{1/2} \, \frac{1}{1 + 
\delta^{\pi}_{SU(2)} +   \delta^{\pi}_{e^2 p^2} + \frac{1}{2} 
\Delta I_{\pi} (\lambda) } \ . 
$$
The recent analysis of Ref.~[\ref{cknp02}] shows that 
\begin{equation}
\delta^{\pi}_{SU(2)} \sim 10^{-5} \ , \qquad  
\delta^{\pi}_{e^2 p^2} = (0.46 \pm 0.05) \% \ , \qquad 
 \Delta I_{\pi} (\lambda) = 0.1 \%  \ ,
\end{equation}
with a total effect of radiative corrections consistent 
with previous estimates [\ref{Sirlin2},\ref{JAUS}].

The present experimental precision for the branching ratio of the pionic 
beta decay cannot compete yet with the very small theoretical 
uncertainties of SFT and neutron beta decay: 
using the latest PDG value 
${\rm BR} = (1.025 \pm 0.034) \times 10^{-8}$ [\ref{pdg2002}], we find
\begin{eqnarray}
|V_{ud}| &=& 0.9675 \pm 0.0160_{\rm exp} \pm 0.0005_{\rm th}
\nonumber \\
         &=& 0.9675 \pm 0.0161 .
\end{eqnarray}
However, a substantial improvement of the experimental accuracy is to be 
expected in the near future. Inserting the present preliminary result
obtained by the PIBETA Collaboration [\ref{PSI}],
$ {\rm BR} = (1.044 \pm 0.007 ({\rm stat.}) \pm 0.009 ({\rm syst.})) \times 
10^{-8}$, we find
\begin{eqnarray}
|V_{ud}| &=& 0.9765 \pm 0.0056_{\rm exp} \pm 0.0005_{\rm th}
\nonumber \\
         &=& 0.9765 \pm 0.0056~,
\end{eqnarray}
where the error should be reduced by about a factor of 3 at the 
end of the experiment.

\section{CKM unitarity and the determination of the Cabibbo angle}
The two measurements of $|V_{ud}|$ from SFT and nuclear beta decay, 
reported in  Eqs.~(\ref{eq:vudsft}) and (\ref{eq:nbfin}) respectively,
are perfectly compatible. Combining them in quadrature we obtain 
\be
|V_{ud}| = 0.9739 \pm 0.0005~,
\label{eq:Vud_fin}
\ee
a result which is not modified 
by the inclusion in the average of the present $\pi_{e 3}$ data.
Due to the small differences in the treatment of radiative 
corrections and theory errors discussed in the previous sections, 
this value is slightly different, but perfectly compatible,
with the one quoted by the PDG: $|V_{ud}| = 0.9735 \pm 0.0008$ [\ref{pdg2002}].

The compatibility of SFT and nuclear beta decay results 
is clearly an important consistency check of Eq.~(\ref{eq:Vud_fin}).
However, it should also be stressed that the theoretical uncertainty 
of inner radiative corrections (which contribute at the level of 
$\pm 0.04\% $) can be considered to a good extent 
a common systematic error for both determinations. 
Thus the uncertainty quoted in Eq.~(\ref{eq:Vud_fin}) is mainly 
of theoretical nature and should be taken with some care.

Using the unitarity relation (\ref{eq:unitarity}) we can translate 
Eq.~(\ref{eq:Vud_fin}) into a prediction for $|V_{us}|$:
\be
|V_{us}|_{\rm unit.} = 0.2269  \pm 0.0021~,
\label{eq:Vus_unit}
\ee
to be compared with the direct determination in Eq.~(\ref{eq:Vus_fin}). 
The $2.2\sigma$ discrepancy between these two determinations could
be attributed to: i) an underestimate of theoretical and, more in general,
systematic errors; ii) 
an unlikely statistical fluctuation; iii) the existence of new degrees 
of freedoms which spoil the unitarity of the CKM matrix. 
Since theoretical errors provide a large fraction of the total
uncertainty in both cases, the solution i), or at least a combination of i) and 
ii), appears to be the most likely scenario. Barring the possibility iii),
and in absence of a clear indication of which of the errors 
is underestimated, a conservative approach is obtained by 
treating the two determinations on the same footing and introducing 
the PDG scale factor. Following this procedure, 
our final estimate of $|V_{us}|$ imposing the unitarity constraint is

\begin{equation}
\begin{array}{|c|}\hline\left.
|V_{us}|_{{\rm unit.}+K_{\ell 3}} = 0.2240 \pm 0.0036~  \ .\right.
\\ \hline\end{array}
\label{eq:Vus_global}
\end{equation}

\section*{References}
\addcontentsline{toc}{section}{References}

\vspace{7mm}

\renewcommand{\labelenumi}{[\theenumi]}
\begin{enumerate}

\item \label{ag} 
M.~Ademollo and R.~Gatto,
Phys.\ Rev.\ Lett.\  {\bf 13} (1964) 264.

\vspace{3mm}

\item \label{LeuRo}
H.~Leutwyler and M.~Roos,
The Kobayashi-Maskawa Matrix,''
Z.\ Phys.\ C {\bf 25} (1984) 91.

\vspace{3mm}

\item \label{sirlin}
A.~Sirlin,
Nucl.\ Phys.\ B {\bf 196} (1982) 83, and references therein.

\vspace{3mm}

\item \label{ginsberg}
E.~S.~Ginsberg,
Phys.\ Rev.\  {\bf 162} (1967) 1570
[Erratum, ibid.\  {\bf 187} (1969) 2280]; \\
Phys.\ Rev.\  {\bf 171} (1968) 1675
[Erratum, ibid.\  {\bf 174} (1968) 2169];  
Phys.\ Rev.\ D {\bf 1} (1970) 229.

\vspace{3mm}

\item \label{CLC}
G.~Calderon and G.~Lopez Castro,
Phys.\ Rev.\ D {\bf 65} (2002) 073032
[hep-ph/0111272].

\vspace{3mm}

\item \label{CKNRT} 
V.~Cirigliano, M.~Knecht, H.~Neufeld, H.~Rupertsberger and P.~Talavera,
Eur.\ Phys.\ J.\ C {\bf 23} (2002) 121
[hep-ph/0110153]; 
V. Cirigliano, M. Knecht, H. Neufeld, H. Pichl, 
in preparation.

\vspace{3mm}

\item \label{Sirlin2}
A. Sirlin, Rev. Mod. Phys. {\bf 50} (1978) 579. 

\vspace{3mm}

\item \label{MS93}
W.J. Marciano, A. Sirlin, Phys. Rev. Lett. {\bf 71} (1993) 3629.

\vspace{3mm}

\item \label{FLRS}
G.~Furlan, F.G.~Lannoy, C.~Rossetti, and G.~Segr\'e, Nuovo Cim. {\bf 38}
(1965) 1747.

\vspace{3mm}

\item \label{choiji}
H.~M.~Choi and C.~R.~Ji,
Phys.\ Rev.\ D {\bf 59} (1999) 034001
[hep-ph/9807500].

\vspace{3mm}

\item \label{Jaus:zv}
W.~Jaus,
Phys.\ Rev.\ D {\bf 60} (1999) 054026.

\vspace{3mm}

\item \label{Vito} V. Antonelli,  R. Ferrari, M. Picariello, 
E. Torrente-Lujan, A. Vicini,  
IFUM-726-FT.

\vspace{3mm}

\item \label{doublelogs}
J.~Bijnens, G.~Colangelo and G.~Ecker,
Phys.\ Lett.\ B {\bf 441} (1998) 437
[hep-ph/9808421].

\vspace{3mm}

\item \label{post}
P.~Post and K.~Schilcher,
hep-ph/0112352.

\vspace{3mm}

\item \label{E865}
A.~Sher (E865 Collaboration), talk given at the DPF 2002 Meeting
(College of William \& Mary, Williamsburg, May 2002),
see  \texttt{http://www.dpf2002.org}

\vspace{3mm}

\item \label{KLOE}
E. De Lucia,  talk given at the Workshop On Quark Mixing and CKM Unitarity 
(Heidelberg, Germany, September 2002), to appear in the proceedings.

\vspace{3mm}

\item \label{pdg2002}
K. Hagiwara {\em et al.},  Particle Data Group
Phys. Rev. D {\bf 66} (2002) 010001.

\vspace{3mm}

\item \label{Bytev}
V.~Bytev, E.~Kuraev, A.~Baratt and J.~Thompson,
hep-ph/0210049.

\vspace{3mm}

\item \label{pich02}
E.~Gamiz, M.~Jamin, A.~Pich, J.~Prades and F.~Schwab,
hep-ph/0212230.

\vspace{3mm}

\item \label{pich92}
F.~Le Diberder and A.~Pich,
Phys.\ Lett.\ B {\bf 289} (1992) 165.

\vspace{3mm}

\item \label{davier02}
M.~Davier and C.~Yuan,
hep-ex/0211057.

\vspace{3mm}

\item \label{HSD}
R.~Flores-Mendieta, A.~Garcia and G.~Sanchez-Colon,
Phys.\ Rev.\ D {\bf 54} (1996) 6855
[hep-ph/9603256].


\vspace{3mm}

\item \label{ft}
J.~C.~Hardy {\em et al.},
Nucl.\ Phys.\ A {\bf 509} (1990) 429; \\
B.~K.~Fujikawa {\it et al.},
life of C-10 with  GAMMASPHERE,''
nucl-ex/9806001; \\
I.~S.~Towner {\it et al.},
nucl-th/9507004.

\vspace{3mm}

\item \label{deltac1} I.~S.~Towner and J.~C.~Hardy, 
Nucl.\ Phys.\ A {\bf 284} (1977) 269.

\vspace{3mm}

\item \label{deltac2} 
W.~E.~Ormand and B.~A.~Brown,
nucl-th/9504017.

\vspace{3mm}

\item \label{wein98} 
I.~S.~Towner and J.~C.~Hardy,
nucl-th/9809087.

\vspace{3mm}

\item \label{outer} 
A.~Sirlin and R.~Zucchini,
Phys.\ Rev.\ Lett.\  {\bf 57} (1986) 1994; \\
W.~Jaus and G.~Rasche,
Phys.\ Rev.\ D {\bf 35} (1987) 3420.

\vspace{3mm}

\item \label{inner}
W.~J.~Marciano and A.~Sirlin,
Phys.\ Rev.\ Lett.\  {\bf 56} (1986) 22; \\
I.~S.~Towner,
Nucl.\ Phys.\ A {\bf 540} (1992) 478.

\vspace{3mm}

\item \label{nos01} A. Garc\'\i a, J. L. Garc\'\i a-Luna and G. L\'opez
Castro, Phys. Lett. B {\bf 500} (2001) 66. 

\vspace{3mm}

\item \label{saito}
K. Saito and A. W. Thomas, Phys. Lett. B {\bf 363} (1995) 157.


\vspace{3mm}

\item \label{wilkinson}
D.~H.~Wilkinson,
Nucl.\ Phys.\ A {\bf 377} (1982) 474.

\vspace{3mm}

\item \label{paver}
N.~Paver and Riazuddin,
Phys.\ Lett.\ B {\bf 260} (1991) 421; \\
J.~F.~Donoghue and D.~Wyler,
Phys.\ Lett.\ B {\bf 241} (1990) 243; \\
N.~Kaiser, Phys. Rev. C {\bf 64}
(2001) 028201.

\vspace{3mm}

\item \label{abele02} H. Abele {\em et al.}, 
Phys. Rev. Lett. {\bf 88} (2002) 211801.

\vspace{3mm}

\item \label{bopp86}
P. Bopp {\em et al.}, Phys. Rev. Lett. {\bf 56} (1986) 919 
[Erratum, ibid. {\bf 57} (1986) 1192].

\vspace{3mm}

\item \label{ero97}
B. Erozolimsky, {\it et al.}, 
Phys. Lett. B~{\bf 412} (1997) 240.

\vspace{3mm}

\item \label{liaud97}
P. Liaud, {\em et al.}, Nucl. Phys. A {\bf 612} (1997) 53.

\vspace{3mm}

\item \label{mostovoi01} Yu. A. Mostovoi {\em et al.}, 
Phys. Atom. Nucl. {\bf 64}  (2001) 1955.


\vspace{3mm}

\item \label{cknp02}
V. Cirigliano, M. Knecht, H. Neufeld, H. Pichl, hep-ph/0209226.

\vspace{3mm}

\item \label{lept}
M. Knecht, H. Neufeld, H. Rupertsberger, P. Talavera, 
Eur. Phys. J. C {\bf 12} (2000) 469.

\vspace{3mm}

\item \label{JAUS}
W. Jaus, Phys. Rev. D {\bf 63} (2001) 053009.

\vspace{3mm}


\item \label{PSI} PIBETA Collaboration, M. Bychkov {\em et al.}, 
PSI Scientific Report 2001, Vol.1, p. 8, 
eds. J. Gobrecht {\em et al.}, Villigen PSI (2002).

\end{enumerate}


%
\def\subcubesection#1{\bigskip\noindent{\bf #1}\nopagebreak\\[0.1cm]
\nopagebreak}

\def\dec{\rightarrow}
\def\mus{$m_b^{\rm 1S}$}
\def\epm{$e^+e^-$}
\def\ups{$\Upsilon{(\rm 4S)}$}
\def\elep{$e^{\pm}$}
\def\mulep{$\mu^{\pm}$}
%
\newcommand{\Bstar}{\mbox{\rm B}^{\ast}}


\def\app#1#2#3{{ Act. Phys. Pol. }{\bf B #1} (#2) #3}
\def\apa#1#2#3{{ Act. Phys. Austr.}{\bf #1} (#2) #3}
\def\lhc{Proc. LHC Workshop, CERN 90-10}
\def\npb#1#2#3{{ Nucl. Phys. }{\bf B #1} (#2) #3}
\def\nP#1#2#3{{ Nucl. Phys. }{\bf #1} (#2) #3}
\def\pra#1#2#3{{ Phys. Rev. }{\bf A #1} (#2) #3}
\def\pR#1#2#3{{ Phys. Rev. }{\bf #1} (#2) #3}
\def\prc#1#2#3{{ Phys. Reports }{\bf #1} (#2) #3}
\def\cpc#1#2#3{{ Comp. Phys. Commun. }{\bf #1} (#2) #3}
\def\nim#1#2#3{{ Nucl. Inst. Meth. }{\bf #1} (#2) #3}
\def\pr#1#2#3{{ Phys. Reports }{\bf #1} (#2) #3}
\def\sovnp#1#2#3{{ Sov. J. Nucl. Phys. }{\bf #1} (#2) #3}
\def\sovpJ#1#2#3{{ Sov. Phys. LETP }{\bf #1} (#2) #3}
\def\jl#1#2#3{{ JETP Lett. }{\bf #1} (#2) #3}
\def\jet#1#2#3{{ JETP Lett. }{\bf #1} (#2) #3}
\def\zpc#1#2#3{{ Z. Phys. }{\bf C #1} (#2) #3}
\def\ptp#1#2#3{{ Prog.~Theor.~Phys.~}{\bf #1} (#2) #3}
\def\nca#1#2#3{{ Nuovo~Cim.~}{\bf #1A} (#2) #3}
\def\ap#1#2#3{{ Ann. Phys. }{\bf #1} (#2) #3}
\def\hpa#1#2#3{{ Helv. Phys. Acta }{\bf #1} (#2) #3}
\def\ijmpA#1#2#3{{ Int. J. Mod. Phys. }{\bf A #1} (#2) #3}
\def\ZETF#1#2#3{{ Zh. Eksp. Teor. Fiz. }{\bf #1} (#2) #3}
\def\jmp#1#2#3{{ J. Math. Phys. }{\bf #1} (#2) #3}
\def\yf#1#2#3{{ Yad. Fiz. }{\bf #1} (#2) #3}
\def\ufn#1#2#3{{ Usp. Fiz. Nauk }{\bf #1} (#2) #3}
\def\spu#1#2#3{{ Sov. Phys. Usp.}{\bf #1} (#2) #3}
\def\epjc#1#2#3{{ Eur. Phys. J. C }{\bf #1} (#2) #3}
\def\Journal#1#2#3#4{#1 {\bf #3} (#2) #4}
\def\NCA{ Nuovo Cimento}
\def\NIM{ Nucl. Instrum. Methods}
\def\NIMA{{ Nucl. Instrum. Methods} A}
\def\npbs{{ Nucl. Phys.} B}
\def\plb{{ Phys. Lett.}  B}
\def\prl{{ Phys. Rev. Lett.}}
\def\prd{ Phys. Rev. D}
\def\zpc{{ Z. Phys.} C}
\def\epj{{ Eur. Phys. J.} C}
\def\lsim{\mathrel{\raise.3ex\hbox{$<$\kern-.75em\lower1ex\hbox{$\sim$}}}}
\def\gsim{\mathrel{\raise.3ex\hbox{$>$\kern-.75em\lower1ex\hbox{$\sim$}}}}
\def\Li2{{\rm Li}_2}
\def\Litri{{\rm Li}_3}
\catcode`\@=11
\def\slash{\mathpalette\make@slash}
\def\make@slash#1#2{\setbox\z@\hbox{$#1#2$}%
  \hbox to 0pt{\hss$#1/$\hss\kern-\wd0}\box0}
\catcode`\@=12 
\def\ms{$\overline{\rm MS}$ }



\def\simleq{\; \raise0.3ex\hbox{$<$\kern-0.75em
      \raise-1.1ex\hbox{$\sim$}}\; }
\def\simgeq{\; \raise0.3ex\hbox{$>$\kern-0.75em
      \raise-1.1ex\hbox{$\sim$}}\; }
\def\noi{\noindent}
\def\bfrho{{\fam=9 \myrho}\fam=1}
\def\R{ {\rm R \kern -.31cm I \kern .15cm}}
\def\C{ {\rm C \kern -.15cm \vrule width.5pt
\kern .12cm}}
\def\Z{ {\rm Z \kern -.27cm \angle \kern .02cm}}
\def\N{ {\rm N \kern -.26cm \vrule width.4pt \kern .10cm}}
\def\1{{\rm 1\mskip-4.5mu l} }

\def\p{\mbox{$p$}}
\newcommand {\bc} {\begin{center}}
\newcommand {\bt} {\begin{table}}
\newcommand {\et} {\end{table}}
\newcommand {\ec} {\end{center}}
\newcommand {\Rb} {\ifmmode {\rm{R}_b} \else $\rm{R}_b$ \fi }
\newcommand {\fd} {\ifmmode {\rm{f}_d} \else $\rm{f}_d$ \fi }
\newcommand {\als} {\mbox{$\alpha_S$}}
\newcommand {\alphas} {\mbox{$\alpha_S(M_{\rm{Z}^0})$}}
\newcommand {\alphasmu} {\mbox{$\alpha_S(\mu)$}}
\newcommand {\eec}{\mbox{$\Sigma_{EEC}$}}
\newcommand {\eecchi}{\mbox{$\Sigma_{EEC}(\chi)$}}
\newcommand {\aeec}{\mbox{$\Sigma_{AEEC}$}}
\newcommand {\aeecchi}{\mbox{$\Sigma_{AEEC}(\chi)$}}
\newcommand {\scalef}{\mbox{$x_{\mu}$}}
\newcommand {\Dstars} {\ifmmode {\rm{D}^{\star +}} \else {D$^{\star +}$} \fi}
\newcommand {\Dz} {\ifmmode {\rm{D}^{0}} \else {D$^{0}$} \fi}
\newcommand {\bl}    {BR({\rm{b \rightarrow \ell}})}
\newcommand {\brcl}    {BR({\rm{c \rightarrow \ell}})}
\newcommand {\bcbl}   {BR({\rm{b \rightarrow \bar{c} \rightarrow \ell}})}
\newcommand {\bcl}   {BR({\rm{b \rightarrow c \rightarrow \ell}})}
\newcommand {\btaul}  {BR({\rm{b \rightarrow \tau \rightarrow \ell}})}
\newcommand {\bpsill} {BR({\rm{b \rightarrow J/\psi\rightarrow \ell^+\ell^-}})}
\newcommand {\glcc}   {\rm{g \rightarrow c \bar c}}
\newcommand {\glbb}   {\rm{g \rightarrow b \bar b}}
\newcommand {\bu} {\rm {b\ra u \ra \ell^+}}
\def\GeV{{\mbox{\rm GeV}}}
\def\Dstar{\ifmmode {{\rm D}^*} \else {${\rm D}^*$} \fi}
\def\Dzero{\ifmmode {{\rm D}^0} \else {${\rm D}^0$} \fi}
\def\md0c{M_{{\rm D}^0}^{\rm cand}}
\def\etal{{\it et al.}}
\def\bsbar{${\rm \overline{B_s^0}}$}
\def\bsbs{${\rm B_s^0}$}
\def\bd{${\rm B^0_d}$}
\def\bu{${\rm B^-_u}$}
\def\bdbar{$ {\overline {\rm B}}_d^0$}
\def\bbar{$ {\overline {\rm B}}^0$}
\def\dmd{$\Delta M_d$}
\def\dstara{${\rm D^{\star \pm}}$}
\def\dstarp{${\rm D^{\star +}}$}
\def\dsp{${\rm D^{\star +}}$}
\def\dstarm{${\rm D^{\star -}}$}
\def\dstar{${\rm D^{\star}}$}
\def\epem{${ e^+e^-}$}
\def\vcb{\mbox{$|V_{cb}|$}}
\def\fw{${\cal F}(w)$}
\def\fone{${\cal F}(1)$}
\def\fvcb{${\cal F}(1)|V_{\rm cb}|$}
\def\btods{${\rm \overline B_d}^0\to {\rm D}^{*+}\ell^-{\overline \nu}_\ell$}
\def\btodss{$ b \to \rm D^{\star \star}\ell^-{\overline \nu}_\ell$}
\newcommand {\Bzerod} {\ifmmode {\rm \bar{B}^0_d} \else {$\rm \bar{B}^0_d$} \fi}
\newcommand {\BtoD} {\ifmmode {\Bz \to \Ds \ell \bar{\nu}}
                \else ${\Bz \to \Ds \ell \bar{\nu}}$ \fi }
\newcommand {\Btau} {\ifmmode {\bar{\rm{B}} \to \tau \bar{\nu_\tau} \Ds}
                \else ${\bar{\rm{B}} \to \tau \bar{\nu_\tau} \Ds }$ \fi }
\newcommand {\Bxc} {\ifmmode {\bar{\rm{B}} \to \rm{X}_c \Ds }
                \else ${\bar{\rm{B}} \to \rm{X}_c \Ds }$ \fi }
\def\dec{\rightarrow}

\newcommand{\deltaE}{$\ensuremath{\Delta E}$}
\newcommand{\brhoenu}{$\ensuremath{{\rm B}^+ \rightarrow \rho^0 e^+ \nu}$}
\newcommand{\brhochgenu}{$\ensuremath{{\rm B}^0 \rightarrow \rho^- e^+ \nu}$}
\newcommand{\bpienu}{$\ensuremath{{\rm B}^+ \rightarrow \pi^0 e^+ \nu}$}
\newcommand{\bpichgenu}{$\ensuremath{{\rm B}^0 \rightarrow \pi^- e^+ \nu}$}
\newcommand{\bomegaenu}{$\ensuremath{{\rm B}^+ \rightarrow \omega e^+ \nu}$}
\newcommand{\brholnu}{$\ensuremath{{\rm B}^+ \rightarrow \rho^0 \ell^+ \nu}$}
\newcommand{\brhochglnu}{$\ensuremath{{\rm B}^0 \rightarrow \rho^- \ell^+ \nu}$}
\newcommand{\bpilnu}{$\ensuremath{{\rm B}^+ \rightarrow \pi^0 \ell^+ \nu}$}
\newcommand{\bpichglnu}{$\ensuremath{{\rm B}^0 \rightarrow \pi^- \ell^+ \nu}$}
\newcommand{\bomegalnu}{$\ensuremath{{\rm B}^+ \rightarrow \omega \ell^+ \nu}$}
\def\btodslnu{\rm B\to D^*\ell\nu}
\def\btodsnslnu{\rm B\to D^{(*)}\ell\nu}
\def\btodlnu{\rm B\to D\ell\nu}
\def\cF{\mathcal{F}}
\def\cK{\mathcal{K}}
\def\cG{\mathcal{G}}
\def\y4s{\Upsilon(4S)}
\def\rhoAone{\rho_{A_1}}
\def\rhoG{\rho_\cG}

\def\la{\langle}
\def\ra{\rangle}
\def\l{\left}
\def\r{\right}

\def\mev{\mbox{\rm MeV}}
\def\gev{\mbox{\rm GeV}}
\def\fm{\mbox{\rm fm}}
\def\eq#1{Eq.~(\ref{#1})}
\def\eqs#1#2{Eqs.~(\ref{#1}) and (\ref{#2})}
\def\eqss#1#2#3{Eqs.~(\ref{#1}), (\ref{#2}) and (\ref{#3})}
\def\tab#1{Table \ref{#1}}
\def\tabs#1#2{Tables \ref{#1} and \ref{#2}}
\def\sec#1{Sec.~\ref{#1}}
\def\secs#1#2{Secs.~\ref{#1} and \ref{#2}}
\def\fig#1{Fig.~\ref{#1}}
\def\figs#1#2{Figs.~\ref{#1} and \ref{#2}}
\def\figss#1#2#3{Figs.~\ref{#1}, \ref{#2} and \ref{#3}}

\newcommand{\DG}{\mbox{$\rm \Delta\Gamma_{s}\ $}}
\newcommand{\aBs}{\mbox{$\rm \overline B^{0}_{s}\ $}}
\newcommand{\Bh}{\mbox{$\rm B^{0}_{H}\ $}}
\newcommand{\Bl}{\mbox{$\rm B^{0}_{L}\ $}}
\newcommand{\Gh}{\mbox{$\rm \Gamma_{H}\ $}}
\newcommand{\Gl}{\mbox{$\rm \Gamma_{L}\ $}}
\newcommand{\xis}{\mbox{$\rm x_{s}\ $}}
\newcommand{\dMss}{\mbox{$\rm \small{\Delta} m_{s}  \ $}}
\newcommand{\dMt}{\mbox{$\rm \small{\Delta} m t \ $}}
\newcommand{\dMsst}{\mbox{$\rm \small{\Delta} m_{s} t \ $}}

\newcommand{\Bsto}{\mbox{$\rm \Bs \rightarrow \JpsiPhi \rightarrow
 \mu^{+} \mu^{-} K^{+}K^{-} \ $}}
\newcommand{\Bst}{\mbox{$\rm \Bs \rightarrow \JpsiPhi \   $}}
\newcommand{\aBst}{\mbox{$\rm \aBs \rightarrow \JpsiPhi \   $}}
\newcommand{\JpsiPhi}{\mbox{$\rm J/\psi \phi\ $}}
\newcommand{\BstoDs}{\mbox{$\rm \Bs \rightarrow D_{s} \pi \ $}}
\newcommand{\BstoDsa}{\mbox{$\rm \Bs \rightarrow D_{s} a_{1} \ $}}
\newcommand{\BtoKs}{\mbox{$\rm B_{d} \rightarrow J/\psi K^{*}  \ $ }}
\newcommand{\BtoK} {\mbox{$\rm B_{d} \rightarrow J/\psi K^{0}_{s}  \ $ }}
\newcommand{\Bpipi} {\mbox{$\rm B_{d} \rightarrow \pi^{+} \pi^{-}  \ $ }}
\newcommand{\bbJpsi}{\mbox{$\rm  \bab \rightarrow J/\psi X\ $}}
\newcommand{\bbJpsimm}{\mbox{$\rm  \bab \rightarrow J/\psi(\mu^{+}\mu^{-}) X\ $}}
\newcommand{\bbJpsiee}{\mbox{$\rm  \bab \rightarrow J/\psi(e^{+}e^{-}) X\ $}}

\newcommand{\Lamzero}{\mbox{$\rm \Lambda^{0}\ $}}
\newcommand{\Lamzerob}{\mbox{$\rm \Lambda^{0}_{b}\ $}}
\newcommand{\Lamzeroto}{\mbox{$\rm \Lamzero \rightarrow p \pi^{-}\ $}}
\newcommand{\Lamzerobto}{\mbox{$\rm \Lamzerob \rightarrow \LamzeroJpsi\ $}}
\newcommand{\Lamzerobtoo}{\mbox{$\rm \Lamzerob \rightarrow \LamzeroJpsi\ \rightarrow
 p\pi^{-}\mu^{+}\mu^{-}\ $}}
\newcommand{\aLamzero}{\mbox{$\rm \overline{\Lambda^{0}}\ $}}
\newcommand{\LamzeroJpsi}{\mbox{$\rm \Lamzero J/\psi\ $}}
\newcommand{\LamzeroJpsitol}{\mbox{$\rm \LamzeroJpsi \rightarrow
 p\pi^{-}\mu^{+}\mu^{-}\ $}}
\newcommand{\LamzeroJpsitomu}{\mbox{$\rm \LamzeroJpsi \rightarrow
 p\pi^{-}\mu^{+}\mu^{-}\ $}}
\newcommand{\LamzeroJpsitoe}{\mbox{$\rm \LamzeroJpsi \rightarrow p\pi^{-}
 e^{+}e^{-}\ $}}
\newcommand{\aLamzeroJpsi}{\mbox{$\rm \aLamzero J/\psi\ $}}
\newcommand{\Xizerob}{\mbox{$\rm \Xi^{0}_{b}\ $}}
\newcommand{\Ximinusb}{\mbox{$\rm \Xi^{-}_{b}\ $}}
\newcommand{\aXizerob}{\mbox{$\rm \overline{\Xi^{0}_{b}}\ $}}
\newcommand{\aXiplus}{\mbox{$\rm \overline{\Xi^{+}_{b}}\ $}}
\newcommand{\aSigmaminus}{\mbox{$\rm \overline{\Sigma^{-}_{b}}\ $}}

\newcommand{\Xizerobto}{\mbox{$\rm \Xizerob \rightarrow \LamzeroJpsi\ $}}
\newcommand{\Xinulabto}{\mbox{$\rm \Xizerob \rightarrow \Xi^{0} J/\psi\ $}}

\newcommand{\Xizeroto}{\mbox{$\rm \Xi^{0} \rightarrow \Lamzero \pi^{0}\ $}}
\newcommand{\Ximinusbto}{\mbox{$\rm \Xi^{-}_{b} \rightarrow \Xi^{-} J/\psi\ $}}
\newcommand{\Ximinusto}{\mbox{$\rm \Xi^{-} \rightarrow \Lamzero \pi^{-}\ $}}
\newcommand{\aXizerobto}{\mbox{$\rm \aXizerob \rightarrow \aLamzeroJpsi\ $}}
\newcommand{\Bto}{\mbox{$\rm B_{b} \rightarrow \LamzeroJpsi\ $}}
\newcommand{\Jpsitomm}{\mbox{$\rm J/\psi \rightarrow \mu^{+} \mu^{-}\ $}}
\newcommand{\Jpsitoee}{\mbox{$\rm J/\psi \rightarrow e^{+} e^{-}\ $}}
\newcommand{\Jpsitol}{\mbox{$\rm J/\psi \rightarrow \mu^{+} \mu^{-}\ $}}
\newcommand{\mumuppi}{\mbox{$\rm \mu^{+}\mu^{-}p\pi^{-}\ $}}
\newcommand{\mueeppi}{\mbox{$\rm \mu eep\pi^{-}\ $}}
\newcommand{\bquark}{\mbox{$\rm b\ $}}
\newcommand{\abquark}{\mbox{$\rm \overline{b}\ $}}
\newcommand{\bab}{\mbox{$\rm b\bar{b}\ $}}
\newcommand{\bbarb}{\mbox{$\rm pp \rightarrow \bquark \abquark X\ $}}
\newcommand{\Dmumumu}{\mbox{$\rm \Delta\phi(J/\psi-\mu)\ $}}
\newcommand{\Vubd}{\mbox{$V_{ub}^{\ast} V_{ud}\ $}}
\newcommand{\Vcbd}{\mbox{$V_{cb}^{\ast} V_{cd}\ $}}
\newcommand{\AsAo}{\mbox{$A_{s} A_{p 1/2}^{\ast}\ $}}
\newcommand{\AsAt}{\mbox{$A_{s} A_{p 3/2}^{\ast}\ $}}
\newcommand{\AtAt}{\mbox{$A_{p 3/2}^{2}\ $}}
\newcommand{\AoAt}{\mbox{$A_{p 1/2} A_{p 3/2}^{\ast}\ $}}
\newcommand{\AlAj}{\mbox{$A_{l} A_{j}^{\ast}\ $}}
\newcommand{\aAlAj}{\mbox{$\overline A_{l} \overline A_{j}^{\ast}\ $}}
\newcommand{\AsAoAt}{\mbox{$2 A_{s}^{2} + 2 A_{p 1/2}^{2} + A_{p 3/2}^{2}\ $}}
\newcommand{\asaoat}{\mbox{$2 a_{s}^{2} + 2 a_{p 1/2}^{2} + a_{p 3/2}^{2}\ $}}
\newcommand{\sppd}{\mbox{$s, p_{1/2}, p_{3/2}, d_{3/2}$}}
\newcommand{\jnn}{\mbox{${1 \over N}$}}
\newcommand{\jnd}{\mbox{${1 \over 2}$}}
\newcommand{\jnt}{\mbox{${1 \over 3}$}}
\newcommand{\jnp}{\mbox{${1 \over 5}$}}
\newcommand{\jnsty}{\mbox{${1 \over 4}$}}

\newcommand{\jnde}{\mbox{${1 \over 9}$}}
\newcommand{\jnpt}{\mbox{${1 \over 15}$}}
\newcommand{\jnsp}{\mbox{${1 \over 45}$}}
\newcommand{\snstp}{\mbox{${16 \over 135}$}}
\newcommand{\dnstp}{\mbox{${2 \over 135}$}}
\newcommand{\dnsp}{\mbox{${2 \over 45}$}}
\newcommand{\jns}{\mbox{${1 \over 4 \pi}$}}
\newcommand{\jnstt}{\mbox{${1 \over {(4 \pi)}^{2}}$}}
\newcommand{\jnst}{\mbox{${1 \over {(4 \pi)}^{3}}$}}
\newcommand{\tnd}{\mbox{${3 \over 2}$}}
\newcommand{\jnsd}{\mbox{${1 \over\sqrt{2}}$}}
\newcommand{\tnsd}{\mbox{${3 \over\sqrt{2}}$}}
\newcommand{\tnsN}{\mbox{${3 \over\sqrt{N}}$}}
\newcommand{\onde}{\mbox{${1 \over 9}$}}
\newcommand{\dendo}{\mbox{${9 \over 8}$}}
\newcommand{\psubt}{\mbox{$p_{\rm T}\ $}}  

\newcommand{\adi}{{\cal A}_{\rm CP}^{\rm dir}}
\newcommand{\ami}{{\cal A}_{\rm CP}^{\rm mix}}
\newcommand{\adg}{{\cal A}_{\rm \Delta\Gamma}}
\newcommand{\eff}{\text{eff}}
\newcommand{\Order}{{\mathcal O}}
\newcommand{\bra}{\langle}
\newcommand{\ket}{\rangle}
\newcommand{\s}{\hat{s}}
\newcommand{\bsll}{b\to s\ell^{+}\ell^{-}}
\newcommand{\Bsll}{B\to X_s\ell^{+}\ell^{-}}
\newcommand{\pole}{\text{pole}}

\def\NPB{{ Nucl. Phys.} B}
\def\PLB{{ Phys. Lett.}  B}
\def\PRL{ Phys. Rev. Lett.}
\def\PRD{{ Phys. Rev.} D}
\def\dg{\Delta \Gamma_d}
\def\bea{\begin{eqnarray}}
\def\eea{\end{eqnarray}}
\def\nn{\nonumber}
\def\barr{\begin{eqnarray}}
\def\earr{\end{eqnarray}}

\def\lsim{\raise0.3ex\hbox{$\;<$\kern-0.75em\raise-1.1ex\hbox{$\sim\;$}}}
\def\gsim{\raise0.3ex\hbox{$\;>$\kern-0.75em\raise-1.1ex\hbox{$\sim\;$}}}
\def\mf{{\cal M}_F}
\def\md{{\cal M}_D}
\def\irad{{\cal I}_\tau}
\def\dm{\Delta m}
\def\dmsq{\Delta m^2}
\def\gh{\Gamma_H}
\def\gl{\Gamma_L}
\def\dg{\Delta \Gamma_d}
\def\aa{{\cal A}}
\def\lt{\left}
\def\rt{\right}


\newcommand{\mtms}{m_t^{\overline{\rm MS}}(m_t^{\overline{\rm MS}})}
\newcommand{\msms}{m_s^{\overline{\rm MS}}(\mu=2\, \mbox{GeV})}
\newcommand{\DB}{\Delta B}
\newcommand{\ep}{\varepsilon}
\newcommand{\msbar}{\overline{\rm MS}}
\newcommand{\msb}{\mbox{NDR-}\overline{\rm{MS}}}
\newcommand{\dgamma}{(\Delta \Gamma/\Gamma)_{B_s}}
\newcommand{\rp}{\tau({\rm B}_u)/\tau({\rm B}_d)}
\newcommand{\rs}{\tau({\rm B}_s)/\tau({\rm B}_d)}
\newcommand{\rl}{\tau(\Lambda_b)/\tau({\rm B}_d)}

\def\dfrac#1#2{{\displaystyle {#1 \over #2}}}
\def\dsum{\mathop{\displaystyle \sum }}
\def\dint{\displaystyle \int }
\def\simge{\mathrel{\rlap{\raise 0.511ex \hbox{$>$}}{\lower 0.511ex 
  \hbox{$\sim$}}}}
\def\simle{\mathrel{\rlap{\raise 0.511ex \hbox{$<$}}{\lower 0.511ex 
  \hbox{$\sim$}}}} 
\def\slash#1{\setbox0=\hbox{$#1$}\dimen0=\wd0
  \setbox1=\hbox{/} \dimen1=\wd1 \ifdim\dimen0>\dimen1
  \rlap{\hbox to \dimen0{\hfil/\hfil}} #1 \else 
  \rlap{\hbox to \dimen1{\hfil$#1$\hfil}}/ \fi}

\chapter{CKM ELEMENTS FROM TREE-LEVEL B DECAYS AND LIFETIMES}
\label{chap:III}

\noindent
{\it Conveners:     E.~Barberio, L.~Lellouch, K.R.~Schubert\\ 
Contributors: M.~Artuso, M.~Battaglia, 
C.~Bauer, D.~Becirevic, M.~Beneke, I.~Bigi, T.~Brandt, D.~Cassel, 
M.~Calvi, M.~Ciuchini,
A.~Dighe~,K.~Ecklund, P.~Gagnon, P.~Gambino, S.~Hashimoto, A.~Hoang, 
T.~Hurth, A.~Khodjamirian, C.S.~Kim, A.~Kronfeld, A.~Lenz, 
A. Le Yaouanc, Z.~Ligeti, 
V.~Lubicz, D.~Lucchesi, T.~Mannel, M.~Margoni, G.~Martinelli,
D. Melikhov, V. Mor\'enas,
H.G.~Moser,  L.~Oliver, O.~P\`ene, J.-C.~Raynal, P.~Roudeau, C.~Schwanda, 
B.~Serfass, M.~Smizanska, J.~Stark, B.~Stech, A.~Stocchi, N.~Uraltsev, 
A.~Warburton, L.H.~Wilden. }

\vspace{1cm}
Tree level semileptonic (s.l.) decays of B~mesons are crucial for 
determining the $\vub$ and $\vcb$ elements of the CKM matrix.
In this Chapter we review our present understanding of inclusive 
and exclusive s.l.\ B decays and give an overview of the experimental 
situation. The second part of the Chapter is devoted to B~mesons lifetimes, 
whose measurement are important for several reasons. Indeed, 
these lifetimes are necessary 
to extract the s.l.\ widths, while the $\rm B^0$ lifetime differences and
the ratios of lifetimes of individual species provide a test of the OPE. 

After a brief introduction to the main concepts involved in 
theoretical analysis of the inclusive  decays, we discuss the 
determination of the relevant parameters --- 
$b$ quark mass and non-perturbative parameters of the Operator Product 
Expansion (OPE) ---  and underlying assumption of quark-hadron duality.
We then review the inclusive determination of $\vub$ and $\vcb$.
The extraction of these two CKM elements from exclusive s.l.\ B decays 
is discussed in 
the two following sections, after which we  review the theoretical framework 
and the measurements of the lifetimes and lifetime differences.

\section{Theoretical tools}
\subsection{The Operator Product Expansion for inclusive decays}
\label{sec:OPE}
\def\lqcd{\Lambda_{\rm QCD}}
\def\bra#1{\langle#1|}
\def\ket#1{|#1\rangle}\def\ds{\displaystyle}

Sometimes, instead of identifying all particles in a decay, it is convenient to
be ignorant about some details.  For example, we might want to specify the
energy of a charged lepton or a photon in the final state, without 
looking at the specific accompanying hadron. 
  These decays are inclusive in the sense that we
sum over final states which can be produced as a result of a given short 
distance interaction.  Typically,  we are interested in a quark-level 
transition, such as
$b\to c\ell\bar\nu$, $b\to s\gamma$, etc., and we would like to extract the
corresponding short distance parameters, $|V_{cb}|$, $C_7(m_b)$, etc., from the
data.  To do this, we need to be able to model independently relate the
quark-level operators to the experimentally accessible observables.

In the large $m_b$ limit, we have $M_W\gg m_b \gg \lqcd$ and
we can hope to use this hierarchy to organize an expansion in $\lqcd/m_b$, 
analogous to
the one in $1/M_W$ introduced in Chapter 1, already based on the OPE.
Since the energy released in the decay is large,
a simple heuristic argument shows that the inclusive rate may be 
modelled simply by
the decay of a free $b$ quark.   
The $b$ quark decay mediated by weak interactions
takes place on a time scale that is much shorter than the time it takes the
quarks in the final state to form physical hadronic states.  Once the $b$ quark
has decayed on a time scale $t \ll \lqcd^{-1}$, the probability that the final
states will hadronize somehow is unity, and we need not know the 
probability of hadronization into specific final states. 
Moreover, since the energy release in the decay is much larger than the 
hadronic scale, the decay is largely insensitive to the details of the 
initial state hadronic structure.
This intuitive picture is formalized by the OPE, 
which expresses the inclusive rate 
as an expansion in inverse powers of the heavy quark mass, 
with the leading term corresponding to the free quark decay 
[\ref{ope},\ref{opebigi}] (for a pedagogical introduction to the OPE 
and its applications, see [\ref{Manohar:dt},\ref{Uraltsev:2000qw}]). 

Let us consider, as an example, the inclusive s.l.\  $b\to c$ decay, 
mediated by the operator
$O_{\rm sl} = -{4G_F/ \sqrt2}\, V_{cb}\, 
  (J_{bc})^\alpha\, (J_{\ell\nu})_\alpha $,
where $J^\alpha_{bc} = (\overline c\, \gamma^\alpha P_L\, b)$ and
$J^\beta_{\ell\nu} = (\overline\ell\, \gamma^\beta P_L\, \nu)$.
The decay rate is given by the square of the matrix element, integrated over
phase space and summed over final states,
\be
\ds \Gamma({\rm B}\to X_c\ell\bar\nu) \sim \sum_{X_c} \int\! \d [{\rm PS}]\,
  \big| \bra{X_c\ell\bar\nu} O_{\rm sl} \ket{\rm B} \big|^2 \,.
\ee
Since the leptons have  no strong interaction, it is convenient to
factorize the phase space into ${\rm B}\to X_c W^*$ and a perturbatively
 calculable leptonic part, $W^* \to \ell\bar\nu$.  The nontrivial part is
the hadronic tensor,
\bea
W^{\alpha\beta} &\sim & \sum_{X_c} \delta^4(p_B-q-p_{X_c})\,
  \big| \bra{\rm B} J^{\alpha\dagger}_{bc} \ket{X_c}\, 
  \bra{X_c} J^\beta_{bc} \ket{\rm B} \big|^2 
\nn\\*
&\sim & {\rm Im} \int\! \d x\, e^{-iq\cdot x}\,
  \bra{\rm B}\, T \big\{ J^{\alpha\dagger}_{bc}(x)\, 
  J^\beta_{bc}(0) \big\}\, \ket{\rm B} \,,
\eea
where the second line is obtained using the optical theorem, and $T$ 
denotes the
time ordered product of the two operators.  This is convenient because 
the time ordered product can be expanded in local operators in the $m_b
\gg \lqcd$ limit.  In this limit the time ordered product is
dominated by short distances, $x \ll \lqcd^{-1}$, and one can express the
nonlocal hadronic tensor $W^{\alpha\beta}$ as a sum of local operators. 
Schematically,
\beq\label{opesketch}
\raisebox{-36pt}{\includegraphics*[width=.34\textwidth]{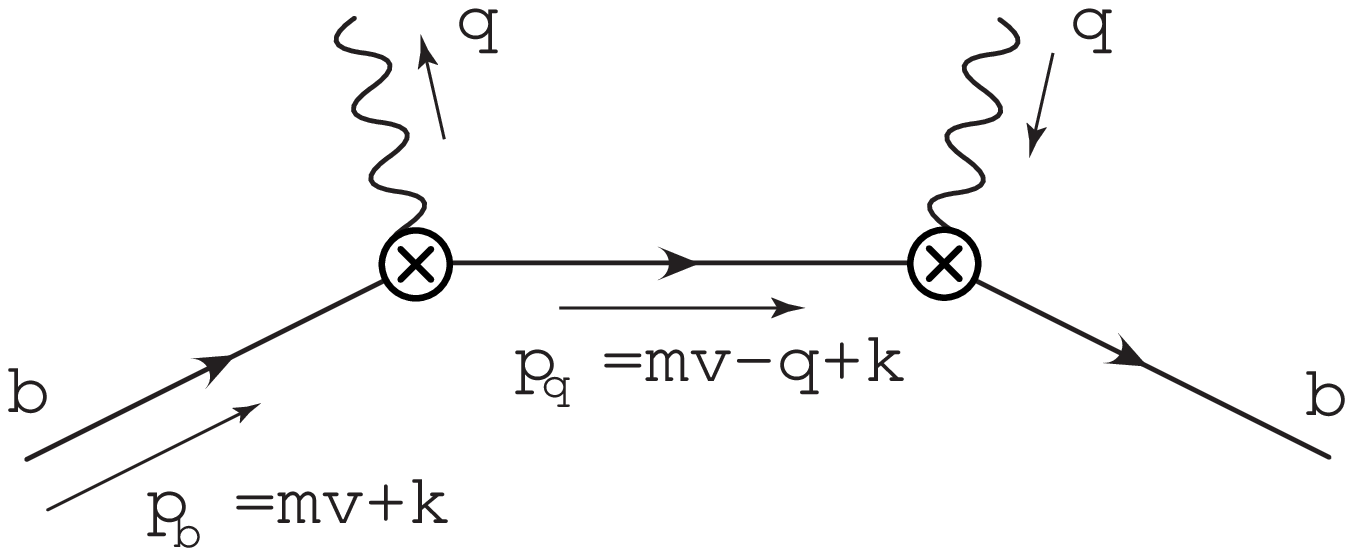}} = 
  \raisebox{-22pt}{\includegraphics*[width=.12\textwidth]{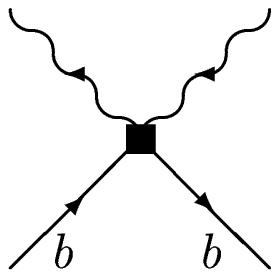}}
 + {0\over m_b} \raisebox{-22pt}{\includegraphics*[width=.12\textwidth]{ChapterIII/ope2}}
 + {1\over m_b^2} \raisebox{-22pt}{\includegraphics*[width=.12\textwidth]{ChapterIII/ope2}}
 + \ldots \,.
\eeq
At leading order the decay rate is
determined by the $b$ quark content of the initial state, while subleading
effects are parametrized by matrix elements of operators with increasing
number of derivatives that are sensitive to the structure of the B meson.
There are no ${ O}(\lqcd/m_b)$ corrections, because the B meson matrix 
element of any dimension-4 operator vanishes.
As the coefficients in front of each operator are calculable in perturbation 
theory, this leads 
to a simultaneous expansion in powers of the strong coupling constant 
$\alpha_s(m_b)$  and inverse powers of the heavy $b$ quark mass 
(more precisely, of $m_b-m_q$). The leading order of  this expansion is the 
parton model s.l.\ width
\begin{eqnarray}
\Gamma_0 = \frac{G_F^2 |V_{cb}|^2 m_b^5}{192 \pi^3} \left( 1-8 \rho + 8 \rho^3 - \rho^4 -12 \rho^2
 \ln \rho \right)\,,
\end{eqnarray}
where $\rho = m_q^2/m_b^2$. Non-perturbative corrections 
are suppressed by at least two
 powers of $m_b$ [\ref{opebigi}]. 
The resulting expression for the total rate of the 
s.l.\ ${\rm B} \to X_c \ell \bar\nu$ has the form
\begin{eqnarray}\label{Gammaincl}
\Gamma^{b \to c} = \Gamma_0 \left[ 1+ A \left[\frac{\alpha_s}{\pi}\right] +
 B \left[\left(\frac{\alpha_s}{\pi}\right)^2 \beta_0\right] + 0 \left[\frac{\Lambda}{m_b}\right] +
 C \left[\frac{\Lambda^2}{m_b^2}\right] + { O} \left(\alpha_s^2, \frac{\Lambda^3}{m_b^3}, 
\frac{\alpha_s}{m_b^2} \right) \right]\,,
\end{eqnarray}
where the  coefficients $A,\, B, \, C$ depend on the quark masses $m_{c,b}$. 
The perturbative corrections are known up to order $\alpha_s^2\beta_0$. 
Non-perturbative corrections
are parameterized by matrix elements of local operators. 
The $O(\Lambda^2/m_b^2)$ corrections are given in terms of the two matrix elements 
\begin{eqnarray}
\lambda_1  &=& \frac{1}{2 M_B} \left<{\rm B}|\bar{h}_{v}(iD)^2 h_{v}|{\rm B}\right>\,,\nn\\
\lambda_2  &=& 
\frac{1}{6 M_B}\left<{\rm B}\left|\bar{h}_{v}\frac{g}{2}\sigma_{\mu \nu} G^{\mu \nu} h_{v}
\right|{\rm B}\right>\,.
\end{eqnarray}
The dependence on these matrix elements is contained in the coefficient 
$C \equiv C(\lambda_1, \lambda_2)$. Up to higher-order corrections, the connection
to an alternative notation is $\lambda_1=-\mu_\pi^2$
and $\lambda_2= \mu^2_G/3$.
 At order $1/m_b^3$ there are two additional matrix elements.
Thus,
the total decay rate depends on a set of non-perturbative parameters, including the quark masses,
with the number 
of such parameters depending on the order in $\Lambda_{\rm QCD}/m_b$ one is working. 

Similar results can be derived for differential distributions, as long as the distributions are 
sufficiently inclusive. To quantify this last statement, it is crucial to remember that the 
OPE does not apply to fully differential distributions but  requires that
such distributions be smeared over enough final state phase space. The size of the 
smearing region $\Delta$ introduces a new scale into the expressions for differential rates and 
can lead to non-perturbative corrections being suppressed by powers of 
$\Lambda_{\rm QCD}^n/\Delta^n$ rather than $\Lambda_{\rm QCD}^n/m_b^n$. 
Thus, a necessary requirement for the OPE to converge is $\Delta \gg \Lambda_{\rm QCD}$, 
although a quantitative understanding of how experimental cuts  affect the size of 
smearing regions is  difficult. 

\def\lqcd{\Lambda_{QCD}}
\def\l{\left}
\def\r{\right}
\def\cL{\mathcal{L}}
\def\cO{\mathcal{O}}

\def\diracslash#1{\setbox0=\hbox{$#1$}           
   \dimen0=\wd0                                 
   \setbox1=\hbox{/} \dimen1=\wd1               
   \ifdim\dimen0>\dimen1                        
      \rlap{\hbox to \dimen0{\hfil/\hfil}}      
      #1                                        
   \else                                        
      \rlap{\hbox to \dimen1{\hfil$#1$\hfil}}   
      /                                         
   \fi}                                         %

\subsection{Heavy Quark Effective Theory}

\def\lqcd{\Lambda_{QCD}}
\def\l{\left}
\def\r{\right}
\def\cL{\mathcal{L}}
\def\cO{\mathcal{O}}

\def\diracslash#1{\setbox0=\hbox{$#1$}           
   \dimen0=\wd0                                 
   \setbox1=\hbox{/} \dimen1=\wd1               
   \ifdim\dimen0>\dimen1                        
      \rlap{\hbox to \dimen0{\hfil/\hfil}}      
      #1                                        
   \else                                        
      \rlap{\hbox to \dimen1{\hfil$#1$\hfil}}   
      /                                         
   \fi}                                         %

The bound state problem for exclusive decays of hadrons composed of a
heavy quark $Q$ and light degrees of freedom simplifies in the limit
$m_Q\gg\lqcd$. The size of such heavy-light hadrons is 
$\sim 1/\lqcd$ and hence, the
typical momenta exchanged between the heavy and light degrees of
freedom are of order $\lqcd$. Such momenta do not permit the light
constituents to resolve the quantum numbers of the heavy quark, whose
Compton wavelength is $\sim 1/m_Q$.  It follows that the light
constituents of hadrons which differ only by the flavour or spin of
their heavy quark have the same configuration. For
$N_Q$ heavy-quark flavours, this invariance results in an $SU(2N_Q)_v$
symmetry which acts on the spin and flavour components of the
heavy-quark multiplet and under which the strong interactions are
invariant at energies much smaller than $m_Q$ 
[\ref{Shuryak:pg},\ref{Shifman:sm},\ref{Shifman:1987rj},\ref{Isgur:vq},\ref{Isgur:ed}].  The subscript $v$ on
$SU(2N_Q)_v$ labels the velocity of the heavy quark on which the
configuration of the light constituents obviously depends.

The spin-flavour symmetry leads to many interesting relations between
the properties of hadrons containing a heavy quark. The most immediate
consequences concern the spectra of these states
[\ref{Isgur:ed}]. Indeed, since the spin of the
heavy quark decouples, states occur in mass-degenerate doublets
corresponding to the two possible orientations of the heavy-quark
spin.\footnote{An exception to this rule are the ground state baryons
$\Lambda_b$ and $\Lambda_c$: their light constituents carry no
angular momentum.} Examples are the meson doublets $(\rm B,B^*)$ and
$(\rm D,D^*)$ or the baryon doublets $(\Sigma_b,\Sigma_b^*)$ and
$(\Sigma_c,\Sigma_c^*)$. Moreover, the flavour symmetry implies that
the energy carried by the light constituents in a heavy-light hadron
must be the same whether the heavy quark is a beauty or a charm. Thus,
in the symmetry limit, we have relations such as $M_{\Lambda_b}-
M_B=M_{\Lambda_c}-M_D$ and $M_{B_s}-M_B=M_{D_s}-M_D$. All of these
relations are satisfied experimentally to the expected accuracy, that
is up to terms of order $\lqcd/m_b$ or $\lqcd/m_c$, depending on
whether the charm quark is present.

Another set of consequences of heavy quark symmetry concerns current
matrix elements and, in particular, $\rm B\to D^{(*)}$ transitions
[\ref{Shifman:1987rj},\ref{Isgur:vq},\ref{Isgur:ed}]. Consider the matrix
element of the $b$-number current between B meson states of given
velocities:
\be
\langle {\rm B}(v')|\bar b\gamma^\mu b|{\rm B}(v)\rangle = 
M_B(v+v')^\mu F_B(t_{BB})
\ee
with $t_{BB}=M_B^2(v-v')^2=2M_B^2(1-w)$, where $w=v\cdot
v'$. $F_B(t_{BB})$ simply measures the overlap of the wave-function of
light constituents around a $b$ quark of velocity $v$ with that of
light constituents around a $b$ quark of velocity $v'$. In the
heavy-quark limit, flavour symmetry implies that this same form factor
describes the matrix element obtained by replacing one or both of the
beauty quarks by a charm quark of same velocity. The spin symmetry
implies that this form factor parametrizes matrix elements in which
the initial and/or final pseudo-scalar meson is replaced by the
corresponding vector meson. It further requires the same form factor
to parametrize matrix elements in which a vector current such as $\bar
c\gamma^\mu b$ is replaced by any other $b\to c$ current.  This means
that in the heavy-quark limit, the s.l.\  decays $\rm B\to D\ell\nu$
and $\rm B\to D^*\ell\nu$, which are governed by the hadronic matrix
elements, $\langle {\rm D}^{(*)}|\bar c\gamma^\mu(\gamma^5)b|{\rm B}\rangle$, are
described by a single form factor, $\xi(w)=F_B(t_{BB}(w))+O(1/m_b)$,
instead of the six form factors allowed by Lorentz
invariance. Moreover, this form factor, known as the Isgur-Wise
function, is normalized to one at zero-recoil, i.e. $\xi(1)=1$ for
$v=v'$ or $w=1$, because the $b$-number current is conserved. 

The normalization imposed by heavy quark symmetry is the basis for the
 measurement of $|V_{cb}|$ from exclusive s.l.\  B decays
 described in Sec.~\ref{sec:vcbexc}. Symmetry is used to the same
 effect elsewhere in
 the determination of CKM matrix elements: isospin symmetry normalizes
 the form factor in $\beta$ decays, yielding $|V_{ud}|$, and $SU(3)$ flavour
 symmetry of light quarks approximatively normalizes the form factor
 in $K_{l3}$ decays, yielding $|V_{us}|$.


In order to explore the consequences of heavy quark symmetry more
systematically and compute corrections to the symmetry limit, which
are essential for reaching the accuracies required for precise
determinations of CKM parameters, it is convenient to construct an
effective field theory which displays this symmetry explicitly and
gives a simplified description of QCD at low energies
[\ref{Eichten:1980mw}]. The idea behind effective theories is a 
separation of scales such that the effective theory correctly 
reproduces the long-distance physics of the 
underlying theory. For the case at hand, we are after a 
theory which duplicates QCD 
on scales below a cutoff $\mu$ such that:
\be
\lqcd\ll\mu\ll m_Q
\label{eq:sepscale}\ .\ee
The construction of heavy quark effective 
theory (HQET)\footnote{There exist many reviews of
heavy quark effective theory. See for instance
[\ref{Flynn:1992fm}--\ref{Neubert:2000hd}].}
 begins with
the observation that the heavy quark bound inside a heavy-light hadron
is nearly on-shell and that its four-velocity is approximately the
hadron's velocity, $v$. Its momentum can thus be written
\be
p^\mu=m_Qv^\mu+k^\mu
\ ,
\ee
where the components of the residual momentum $k^\mu$ are much smaller
than $m_Q$ and where $v^2=1$. The heavy-quark field is then decomposed
into its ``particle'' and ``anti-particle'' components, $h_v$ and $H_v$, as
\be
Q(x)=e^{-im_Qv\cdot x}\l[\frac{1+\diracslash{v}}2h_v(x)+\frac{1-\diracslash{v}}2H_v(x)\r]
\label{eq:hvdef}\ .\ee
This decomposition shifts the zero of four-momentum in such a way that
the heavy-quark degrees of freedom become massless while the
anti-quark degrees of freedom acquire a mass $2m_Q$.\footnote{A
description of heavy anti-quarks is obtained by performing a shift in
four-momentum of opposite sign.}  The latter are the heavy degrees of
freedom which are integrated out in the construction of the effective
theory. Performing this operation in the path integral and expanding
the result in powers of terms of order $1/2m_Q$, one finds  the following
leading order effective Lagrangian:
\be
\cL_{eff}=\cL_0+O\l(\frac{1}{2m_Q}\r)=\bar h_viv\cdot D h_v+O\l(\frac{1}{2m_Q}\r)
\ .\ee
At subleading order it becomes:
\be
\cL_{eff}=\cL_0+\cL_1+O\l(\frac{1}{4m_Q^2}\r)=\cL_0+
\frac{1}{2m_Q}\bar h_v(iD_\perp)^2h_v+\frac{g}{2m_Q}\bar h_v\sigma_{\mu\nu}G^{\mu\nu}
h_v+O\l(\frac{1}{4m_Q^2}\r)
\ ,
\label{eq:leff1}
\ee
with $D_\perp^\mu=D-v^\mu v\cdot D$.

The absence of Dirac structure and of masses in $\cL_0$ signals the
existence of the heavy quark spin-flavour symmetry. This symmetry is
broken at order $1/m_Q$. In Eq.~(\ref{eq:leff1}), the first correction
corresponds to the gauge-invariant extension of the kinetic energy
arising from the residual motion of the heavy quark and breaks only
the flavour component of the symmetry. The second term describes the
colour-magnetic coupling of the heavy-quark spin to the gluons and
breaks both the spin and the flavour components of the symmetry.

\medskip

In order to incorporate the weak interactions of heavy quarks, one
must also consider the expansion of weak operators in powers of
$1/2m_Q$.  Introducing a source in the path integral for the quark
field, $Q(x)$, one finds that this source couples to
\bea
Q(x)&=&e^{-im_Qv\cdot x}\l[1+\frac1{iv\cdot D+2m_Q}i\diracslash{D}_\perp\r]h_v(x)\nn\\
&=& e^{-im_Qv\cdot x}\l[1+\frac{i\diracslash{D}_\perp}{2m_Q}+O(1/4m_Q^2)\r]h_v(x)\ ,
\label{eq:qexp}\eea
once the substitution of Eq.~(\ref{eq:hvdef}) and 
the integral over the ``anti-quark'' mode $H_v$ are performed. Thus,
the expansion of weak currents involving heavy quarks in powers of
$1/2m_Q$ is obtained by replacing occurrences of $Q(x)$ by the
expansion of Eq.~(\ref{eq:qexp}).

\medskip

The construction described up until now correctly reproduces the
long-distance physics of QCD, (below $\mu$ of
Eq.~(\ref{eq:sepscale})). However, this procedure does not take into
account the effects of hard gluons whose virtual momenta can be of the
order of the heavy-quark mass, or even larger~[\ref{Shifman:sm}].  Such
gluons can resolve the flavour and the spin of the heavy quark and thus
induce symmetry breaking corrections. Schematically, the relation between matrix
elements of an operator $\cO$ in the full and in the effective theory is
\be
\langle \cO(\mu)\rangle_{QCD}=C_0(\mu,\bar\mu)\langle \bar\cO_0(\bar\mu)\rangle_{HQET}
+\frac{C_1(\mu,\bar\mu)}{2m_Q}\langle \bar\cO_1(\bar\mu)\rangle_{HQET}
\ ,\label{eq:hqexp}
\ee
where $\bar\mu\sim\mu$ and where we have assumed, for simplicity, that
only one HQET operator appears at leading and at sub-leading order in
the $1/2m_Q$ expansion. The short-distance coefficients
$C_i(\mu,\bar\mu)$ are defined by this equation, and should be
accurately calculable order by order in perturbation theory because
$\alpha_s$ is small in the region between $\mu$ and $m_Q$.
 One typically obtains $C_i=1+O(\alpha_s)$. The
way in which these virtual processes break the heavy quark symmetry is
by inducing a logarithmic dependence of the $C_i$ on $m_Q$ and by
causing mixing with operators which have a different spin structure
(not shown here).

Since the effective theory is constructed to reproduce the low-energy
behaviour of QCD, the matching procedure must be independent of
long-distance effects such as infrared singularities or the nature of
the external states used. It is therefore possible and convenient to
perform the matching using external on-shell quark states. Furthermore,
if the logarithms of $m_Q/\mu$ which appear in the short-distance
coefficients are uncomfortably large, it is possible to resum them
using renormalization group techniques.

It is important to note that the matrix elements in the effective theory, such as 
$\bar\cO_0(\bar\mu)\rangle_{HQET}$ and $\langle
\bar\cO_1(\bar\mu)\rangle_{HQET}$ in Eq.~(\ref{eq:hqexp}), 
involve long-distance strong-interaction effects and therefore require
non-perturbative treatment. It is also important to note that the
separation between short-distance perturbative and long-distance
non-perturbative contributions is ambiguous, though these ambiguities
must cancel in the calculation of physical observables. These
ambiguities require one to be careful in combining results for
short-distant coefficients and for the non-perturbative HQET matrix
elements. In particular, one has to make sure that these coefficients
are combined with matrix elements which are defined at the same order
and, of course, in the same renormalization scheme.

\boldmath
\section{Inclusive semileptonic $b$ decays}
\unboldmath

\subsection{Bottom  and charm quark mass determinations}
\label{sec:mb}

In the framework of B physics the bottom quark
mass parameter is particularly important because theoretical
predictions of many quantities strongly depend on $m_b$. Thus,
uncertainties on $m_b$ can affect the determination of other
parameters. However, due to confinement and the non-perturbative aspect of the 
strong interaction
the concept of quark masses cannot be tied to an intuitive picture of
the weight or the rest mass of a particle, such as for leptons, which
are to very good approximation insensitive to the strong
interactions. Rather, quark masses have to be considered as couplings
of the Standard Model Lagrangian that have to be determined from
processes that depend on them. As such, the bottom quark mass is a
scheme-dependent, renormalized quantity. 
For recent reviews on the determination of the $b$ quark mass, 
see~[\ref{El-Khadra:2002wp}].

\subsubsection{Quark mass definitions in perturbation theory}
\label{subsectionmassdefs}

In principle, any renormalization scheme, or definition for quark
masses is possible. In the framework of QCD perturbation theory the
difference between two mass schemes can be determined as a series in
powers of $\alpha_s$. Therefore, higher-order terms in the
perturbative expansion of a quantity that depends on quark masses are
affected by which scheme is employed. There are schemes that
are more appropriate and more convenient for some purposes than
others. In this section 
we review  the prevalent perturbative quark  mass
definitions, focusing on the case of the bottom quark.

\subsubsection*{Pole mass}
\label{subsubsectionpolemass}
The bottom quark pole mass $m_b$ is defined as the solution to
\begin{equation}
\slash{p}-m_b-\Sigma(p,m_b)\Big|_{p^2=m_b^2} 
\, = \, 0
\,,
\end{equation}
where $\Sigma(p,m_b)$ is the bottom quark self energy. The pole mass definition
is gauge-invariant and infrared-safe~[\ref{Kronfeld1}] to all
orders in perturbation theory and has been used as the standard mass
definition of many perturbative computations in the past. By
construction, the pole mass is directly related to the concept of the
mass of a free quark, which is, however, problematic because of
confinement. In practical applications the pole mass has the
disadvantage that the perturbative series relating it to physical
quantities are in general quite badly behaved, 
due to a strong sensitivity of the pole mass definition itself to infrared
gluons [\ref{Bigi1}]. 

There is nothing wrong to use the pole mass as an intermediate
quantity, as long as it is used in a consistent way. In particular,
the presence of a renormalon ambiguity~[\ref{Bigi1}] 
requires considering the numerical value of the pole mass as an
order-dependent quantity. Because this makes estimates of
uncertainties difficult, the pole mass definition should be avoided
for analyses where quark mass uncertainties smaller than $\Lambda_{\rm
QCD}$ are necessary. 
The problems of the pole mass definition can be avoided if one 
uses quark mass definitions that are less sensitive to small momenta
and do not have an ambiguity of order $\Lambda_{\rm QCD}$. Such quark
mass definitions are generically called ``short-distance'' masses. They 
have a parametric ambiguity or order $\Lambda_{\rm QCD}^2/m_b$ or
smaller.

\subsubsection*{\ms mass}
\label{subsubsectionmsbarmass}
The most common short-distance mass parameter is the \ms mass
$\overline m_b(\mu)$, which is defined by regulating QCD with
dimensional regularization and subtracting the divergences in the \ms 
scheme. Since the subtractions do not contain any infrared
sensitive terms, the \ms mass is only sensitive to scales of order or
larger than $m_b$. The relation
between the pole mass and the \ms mass is known to 
${\cal O}(\alpha_s^3)$~[\ref{Broadhurst1},\ref{Melnikov1}] and
reads ($\bar\alpha_s\equiv\alpha_s^{(n_l=4)}(\overline m_b(\overline m_b))$)
\be
\frac{m_{b,{\rm pole}}}{\overline m_b(\overline m_b)}  = 
1
+\frac{4\bar\alpha_s}{3\pi} +
\left(\frac{\bar\alpha_s}{\pi}\right)^2\,\Big( 13.44 - 1.04\, n_f\Big)
+ \Big(\frac{\bar\alpha_s}{\pi}\Big)^3\,
\Big( 190.8 - 26.7\, n_f+0.65\,
n_f^2\Big)
+\ldots
\,.
\nonumber
\label{polemsbar}
\ee
The bottom quark \ms mass arises naturally in processes where the
bottom quark is far off-shell. The scale $\mu$ in the \ms mass is
typically chosen of the order of the characteristic energy scale of 
the process under consideration since perturbation theory contains
logarithmic terms  $\sim\alpha_s(\mu)^n\ln(Q^2/\mu^2)$ that would
be large otherwise. Using 
the renormalization group equation for $\overline m_b(\mu)$ the value
of the \ms mass for different $\mu$ can be related to each other. 
The \ms mass definition is less useful for processes where the bottom
quark is close to its mass-shell, i.e. when the bottom quark has
non-relativistic energies.

\subsubsection*{Threshold masses}
\label{subsubsectionthresholdmass}
The shortcomings of the pole and the \ms masses in describing
non-relativistic bottom quarks can be resolved by so-called threshold 
masses~[\ref{Hoang1}]. The threshold masses are free of an ambiguity of
order $\Lambda_{\rm QCD}$ and, at the same time, are defined through
subtractions that contain contributions that are
universal for the dynamics of non-relativistic quarks. Since the
subtractions are not unique, an arbitrary number of threshold masses
can be constructed. In the following the threshold mass
definitions that appear in the literature are briefly reviewed.

\vspace{0.2cm}\noindent
{\em Kinetic mass}\\
The kinetic mass is defined as~[\ref{Bigi2},\ref{Bigi3}] 
\begin{eqnarray}
m_{b,{\rm kin}}^{}(\mu_{\rm kin}^{}) 
& = &
m_{b,{\rm pole}} - \left[\bar\Lambda(\mu_{\rm kin}^{})\right]_{\rm pert} 
- \left[\frac{\mu^2_\pi(\mu_{\rm kin}^{})}
{2 m_{b,{\rm kin}}(\mu_{\rm kin}^{})}\right]_{\rm pert}
+\ldots
\,,
\label{kindef}
\end{eqnarray}
where $\left[\bar\Lambda(\mu^{}_{\rm kin})\right]_{\rm pert}$ and
$\left[\mu_\pi^2(\mu^{}_{\rm kin})\right]_{\rm pert}$ are perturbative
evaluations of HQET matrix elements that describe the difference
between the pole and the B meson mass. 

The relation between the kinetic mass and the $\overline{\rm MS}$ mass
is known to ${\cal O}(\alpha_s^2)$ and ${\cal O}(\alpha_s^3 \beta_0)$
[\ref{Czarnecki5},\ref{Melnikov2}]. 
The formulae for $[\bar\Lambda(\mu_{\rm kin})]_{\rm pert}$ and
$[\mu^2_\pi(\mu_{\rm kin})]_{\rm pert}$
at ${\cal O}(\alpha_s^2)$ read [\ref{Melnikov2}]
\begin{eqnarray}
\left [\bar\Lambda (\mu) \right ]_{\rm pert} &=&
\frac {4}{3}C_F\mu_{\rm kin} \frac {\alpha_s(\bar m)}{\pi} \left \{
1 + \frac {\alpha_s}{\pi} \left [
\left (\frac {4}{3} - \frac {1}{2}\ln\frac{2\mu_{\rm kin}}{\bar m} \right
)\beta_0 - C_A \left (\frac {\pi^2}{6} - \frac {13}{12} \right )
\right ] \right \},\label{eq:Lambda_pert}
\\
\left [\mu_{\pi}^2 (\bar m) \right ]_{\rm pert} &=&
C_F\mu^2 \frac {\alpha_s(\bar m)}{\pi} \left \{
1 + \frac {\alpha_s}{\pi} \left [
\left (\frac {13}{12} - \frac {1}{2} \ln \frac{2\mu_{\rm kin}}{\bar m}
\right
)\beta_0 - C_A \left (\frac {\pi^2}{6} - \frac {13}{12} \right )
\right ] \right \}.
\label{eq:mu2pert}
\end{eqnarray}
 where $\bar m=\overline m_b(\overline m_b)$, $C_F=4/3$, and $\beta_0=11-\frac{2}{3}\, n_f$.
For $\mu_{\rm kin}\to 0$ the kinetic mass reduces to the pole mass. 

\newpage
\noindent {\em Potential-subtracted mass}\\
The potential-subtracted (PS) mass is similar to the kinetic mass, but
arises considering the static energy of a bottom-antibottom quark pair
in NRQCD~[\ref{Beneke1}].The PS mass is known to ${\cal O}(\alpha_s^3)$  and
its relation to the pole mass reads 
\begin{eqnarray}
\lefteqn{
m_{b,{\rm PS}}(\mu^{}_{\rm PS})
\, = \,
m_{b,{\rm pole}} - 
\frac{C_F\alpha_s(\mu)}{\pi}\,\mu_{\rm PS}^{}\Bigg[
1+\frac{\alpha_s(\mu)}{4\pi}\left(a_1-\beta_0\left(\ln\frac{\mu_{\rm PS}^2}{\mu^2} 
-2\right)\right)
}
\\
&&\hspace*{-1.5cm}+\left(\frac{\alpha_s(\mu)}{4\pi}\right)^2 
\Bigg(a_2-\left(2 a_1 \beta_0+\beta_1\right)\left(\ln\frac{\mu_{\rm PS}^2}{\mu^2} 
-2\right)
+\beta_0^2\left(\ln^2\frac{\mu_{\rm PS}^2}{\mu^2}-4 \ln\frac{\mu_{\rm PS}^2}{\mu^2}+8
\right)\Bigg)\Bigg],
\nonumber 
\end{eqnarray}
where $\beta_0=11-\frac{2}{3}\, n_f$ and
$\beta_1=102-\frac{38}{3}\, n_f$ are the one- and two-loop beta
functions, and $a_1=\frac{31}{3}-\frac{10}{9}\, n_f$, 
$a_2=456.749 - 66.354 \,n_f + 1.235\, n_f^2$
(see Refs.~[\ref{Schroeder1}]). 
For $\mu_{\rm PS}\to 0$
the PS mass reduces to the pole mass.

\vspace{0.2cm}\noindent
{\em 1S mass}\\
The kinetic and the potential-subtracted mass depend on an explicit
subtraction scale to remove the universal infrared sensitive
contributions associated with the non-relativistic heavy quark
dynamics. The 1S mass~[\ref{Hoang2},\ref{Hoang3}] achieves the same task
without a factorization scale, since it is directly related to a
physical quantity. The bottom 1S mass is 
defined as one half of the perturbative contribution to 
the mass of the $n=1$,
${}^{2s+1}L_j={}^3S_1$ quarkonium bound state in the limit 
$m_b\gg m_b v\gg m_b v^2\gg\Lambda_{\rm QCD}$. To three loop order
the 1S mass is defined as
\begin{eqnarray}
\frac{m_{b,{\rm 1S}}^{}}{m_{b,{\rm pole}}} 
& = & 
1  - \frac{(C_F \alpha_s(\mu))^2}{8}\,
\bigg\{\,1+
\Big(\frac{\alpha_s(\mu)}{\pi}\Big)\,\bigg[\,
\beta_0\,\bigg( L + 1 \,\bigg) + \frac{a_1}{2} 
\,\bigg]
\nonumber\\ & & \qquad
+\,\Big(\frac{\alpha_s(\mu)}{\pi}\Big)^2\,
\bigg[\,
\beta_0^2\,\bigg(\, \frac{3}{4} L^2 +  L + 
                             \frac{\zeta_3}{2} + \frac{\pi^2}{24} +
                             \frac{1}{4} 
\,\bigg) + 
\beta_0\,\frac{a_1}{2}\,\bigg(\, \frac{3}{2}\,L + 1
\,\bigg)
\nonumber\\ & & \qquad\qquad
 + \frac{\beta_1}{4}\,\bigg(\, L + 1
\,\bigg) +
\frac{a_1^2}{16} + \frac{a_2}{8} + 
\bigg(\, C_A - \frac{C_F}{48} \,\bigg)\, C_F \pi^2 
\,\bigg]
\,\bigg\}
\,,
\label{M1Sdef}
\end{eqnarray}
where
$ L \equiv \ln(\mu/(C_F\alpha_s(\mu)\,m_{b,{\rm pole}}))$ and $\zeta_3=1.20206$.
The expression for the 1S mass is derived in the framework of the
non-relativistic expansion, where powers of the bottom quark velocity  
arise as powers of $\alpha_s$ in the 1S mass definition.
Thus, to achieve the renormalon cancellation for B decays in the 1S
mass scheme it is
mandatory to treat terms of order $\alpha_s^{n+1}$  
in Eq.\,(\ref{M1Sdef}) as being of order $\alpha_s^n$. This
prescription is called ``upsilon expansion''~[\ref{Hoang2}] and arises
because 
of the difference between the non-relativistic power counting and the
usual counting in numbers of loops of powers of $\alpha_s$.

\vspace{0.2cm}\noindent
{\em Renormalon-subtracted mass}\\
The renormalon-subtracted mass~[\ref{Pineda6}] is defined as the 
perturbative series that results from subtracting all non-analytic pole 
terms from the Borel
transform of the pole-\ms mass relation at $u=1/2$ with a fixed choice
for the renormalization scale $\mu=\mu_{\rm RS}^{}$. The scale
$\mu_{\rm RS}^{}$ is then kept independent from the renormalization scale 
used for the computation of the quantities of
interest. To order $\alpha_s$ the relation between RS mass and pole
mass reads, \begin{eqnarray}
M_{\rm RS}^{}(\mu_{\rm RS}^{}) 
& = &
m_{\rm pole} - c\,\alpha_s\,\mu_{\rm RS}^{}
\, + \, \ldots
\,,
\label{RSdef}
\end{eqnarray}
where the constant $c$ depends on the number of light quark species
and has an uncertainty because the residue at $u=1/2$ in  the Borel
transform of the pole-\ms mass relation is known only approximately.

In Table~\ref{tabmasses} the various $b$ quark mass parameters are
compared numerically taking the \ms mass $\overline m_b(\overline
m_b)$ as a reference value  for different values for
the strong coupling.
Each entry corresponds to the mass using the respective
1-loop/2-loop/3-loop relations.  
\begin{table}[t] 
\vskip 0mm
\begin{center}
\begin{tabular}{|c|c|c|c|c|} \hline
$\overline m_b(\overline m_b)$ & 
$m_{b,\rm pole}$ &
$m_{b,\rm kin}(1\,\mbox{GeV})$ &
$m_{b,\rm PS}(2\,\mbox{GeV})$ & 
$m_{b,{\rm 1S}}$
\\[0.05cm]
\hline
\multicolumn{5}{|c|}{{\small $\alpha_s^{(5)}(m_Z)=0.116$ 
}}\\
\hline
{\small 4.10} & 
{\small 4.48/4.66/4.80} &
{\small 4.36/4.42/4.45$^*$} & 
{\small 4.29/4.37/4.40} & 
{\small 4.44/4.56/4.60}\\
\hline
{\small 4.15} & 
{\small 4.53/4.72/4.85} &
{\small 4.41/4.48/4.50$^*$} & 
{\small 4.35/4.42/4.45} & 
{\small 4.49/4.61/4.65}\\
\hline
{\small 4.20} & 
{\small 4.59/4.77/4.90} &
{\small 4.46/4.53/4.56$^*$} & 
{\small 4.40/4.48/4.51} & 
{\small 4.54/4.66/4.71}\\
\hline
{\small 4.25} & 
{\small 4.64/4.83/4.96} & 
{\small 4.52/4.59/4.61$^*$} & 
{\small 4.46/4.53/4.56} & 
{\small 4.60/4.72/4.76}\\
\hline
{\small 4.30} & 
{\small 4.69/4.88/5.01} &
{\small 4.57/4.64/4.67$^*$} & 
{\small 4.51/4.59/4.62} & 
{\small 4.65/4.77/4.81}\\
\hline
\multicolumn{5}{|c|}{{\small $\alpha_s^{(5)}(m_Z)=0.118$ 
}}\\
\hline
{\small 4.10} & 
{\small 4.49/4.69/4.84} & 
{\small 4.37/4.44/4.46$^*$} & 
{\small 4.30/4.38/4.41} & 
{\small 4.45/4.57/4.62}\\
\hline
{\small 4.15} & 
{\small 4.55/4.74/4.89} & 
{\small 4.42/4.49/4.52$^*$} & 
{\small 4.36/4.43/4.47} & 
{\small 4.50/4.63/4.67}\\
\hline
{\small 4.20} & 
{\small 4.60/4.80/4.94} & 
{\small 4.47/4.55/4.57$^*$} & 
{\small 4.41/4.49/4.52} & 
{\small 4.55/4.68/4.73}\\
\hline
{\small 4.25} & 
{\small 4.65/4.85/5.00} & 
{\small 4.52/4.60/4.63$^*$} & 
{\small 4.46/4.54/4.58} & 
{\small 4.61/4.73/4.78}\\
\hline
{\small 4.30} & 
{\small 4.71/4.91/5.05} & 
{\small 4.58/4.66/4.69$^*$} & 
{\small 4.52/4.60/4.63} & 
{\small 4.66/4.79/4.84}\\
\hline
\multicolumn{5}{|c|}{{\small $\alpha_s^{(5)}(m_Z)=0.120$ 
}}\\
\hline
{\small 4.10} & 
{\small 4.51/4.72/4.88} & 
{\small 4.37/4.45/4.48$^*$} & 
{\small 4.31/4.39/4.43} & 
{\small 4.46/4.59/4.64}\\
\hline
{\small 4.15} & 
{\small 4.56/4.77/4.93} & 
{\small 4.43/4.51/4.54$^*$} & 
{\small 4.36/4.45/4.48} & 
{\small 4.51/4.64/4.70}\\
\hline
{\small 4.20} & 
{\small 4.61/4.83/4.99} & 
{\small 4.48/4.56/4.59$^*$} & 
{\small 4.42/4.50/4.54} & 
{\small 4.56/4.70/4.75}\\
\hline
{\small 4.25} & 
{\small 4.67/4.88/5.04} & 
{\small 4.54/4.62/4.65$^*$} & 
{\small 4.47/4.56/4.59} & 
{\small 4.62/4.75/4.80}\\
\hline
{\small 4.30} & 
{\small 4.72/4.94/5.10} & 
{\small 4.59/4.67/4.71$^*$} & 
{\small 4.53/4.61/4.65} & 
{\small 4.67/4.81/4.86}\\
\hline
\end{tabular}
\caption{\it \label{tabmasses}    
Numerical values of b quark masses in units of GeV for a given \ms 
mass for $\overline m_b(\overline m_b)$ for  $\mu=m_b(m_b)$, 
$n_l=4$ and three values of 
$\alpha_s^{(5)}(m_Z)$. Flavor matching was carried out at 
$\mu=\overline m_b(\overline m_b)$. 
Numbers with a star are given in the large-$\beta_0$
approximation. 
}
\end{center}
\vskip 3mm
\end{table}

\subsubsection{Bottom quark mass from spectral sum rules}
\label{subsectionsumrules}

The spectral sum rules for $\sigma(e^+e^-\to b\bar b)$ start from the
correlator of two electromagnetic bottom quark currents
\begin{eqnarray}
(g_{\mu\nu}\,q^2-q_\mu\,q_\nu)\,\Pi(q^2) & = &
-\,i \int dx\,e^{i\,qx}\,
   \langle\, 0\,|\,T\,j^b_\mu(x)\,j^b_\nu(0)\,|\,0\, \rangle
\,,
\label{vacpoldef}
\end{eqnarray}
where $j^b_\mu(x)\equiv\bar b(x)\gamma_\mu b(x)$.
Using analyticity and the optical theorem one can relate 
theoretically calculable derivatives of $\Pi$ at $q^2=0$ to moments of
the total cross section $\sigma(e^+e^-\to b\bar b)$,
\begin{equation}
{\cal M}_n 
\, = \,
\frac{12\,\pi^2\,Q_b^2}{n!}\,
\bigg(\frac{d}{d q^2}\bigg)^n\,\Pi(q^2)\bigg|_{q^2=0}
\, = \,
\int \frac{d s}{s^{n+1}}\,R(s)
\,,
\label{Mdef}
\end{equation}
where $R=\sigma(e^+e^-\to b\bar b)/\sigma(e^+e^-\to\mu^+\mu^-)$.  From
Eq.\,(\ref{Mdef}) it is possible to determine the bottom quark
mass~[\ref{Novikov1}]. From the theoretical point of view $n$ cannot be
too large because the effective energy range contributing to the
moment becomes of order or smaller than $\Lambda_{\rm QCD}$ and
non-perturbative effects become uncontrollable. Since the effective
range of $\sqrt{s}$ contributing to the spectral integral is of order
$m_b/n$ one finds the range
\begin{equation}
n\lsim 10
\,,
\end{equation} 
where a reliable extraction of the bottom quark mass is feasible.
In this range one can distinguish two regions. In the large-$n$ region,
$4\lsim n\lsim 10$, the $b\bar b$-dynamics is predominantly
non-relativistic and threshold masses are the suitable mass parameters
that can be determined. In the small-$n$ region, $1\le n\lsim 4$,
the $b\bar b$ dynamics is predominantly relativistic and the \ms mass
is the appropriate mass parameter. In the following the advantages and
disadvantages of the two types of sum rules are reviewed. 
Results for bottom quark masses obtained in recent sum rule analyses
have been collected in Table~\ref{tabcollection}.

\begin{table}[t!]  
\vskip 0mm
\begin{center}
\begin{small}
\begin{tabular}{|l|l|c|l|} \hline
 author & $\overline m_b(\overline m_b)$ & other mass  & comments, Ref.
\\ \hline\hline
\multicolumn{4}{|c|}{ spectral sum rules }\\
\hline\hline
  Voloshin  \hfill 95
    &  
    & $m^{}_{\rm pole}=4.83\pm 0.01$
    & {$8<n<20$, NLO; no theo.uncert.}~\protect[\ref{Voloshin2}]
\\ \hline
  K\"uhn  \hfill 98 
    & 
    & $m^{}_{\rm pole}=4.78\pm 0.04$
    & $10<n<20$, NLO~\protect[\ref{Kuhn1}]
\\ \hline
  Penin \hfill 98
    & 
    & $m^{}_{\rm pole}=4.78\pm 0.04$
    & $10<n<20$, NNLO~\protect[\ref{Penin1}]
\\ \hline
  Hoang \hfill 98
    & 
    & $m^{}_{\rm pole}=4.88\pm 0.13$
    & $4<n<10$, NLO~\protect[\ref{Hoang14}] 
\\ \hline
  Hoang  \hfill 98
    & 
    & $m^{}_{\rm pole}=4.88\pm 0.09$
    & $4<n<10$, NNLO~\protect[\ref{Hoang14}] 
\\ \hline
  Melnikov \hfill  98
    & $4.20\pm 0.10$
    & $M^{1\mbox{\tiny GeV}}_{\rm kin}=4.56\pm 0.06$
    & $x<n<x$, NNLO~\protect[\ref{Melnikov2}] 
\\ \hline
  Penin  \hfill  98 
    & 
    & $m^{}_{\rm pole}=4.80\pm 0.06$
    & $8<n<12$, NNLO~\protect[\ref{Penin1}] 
\\ \hline
  Jamin  \hfill 98
    & $4.19\pm 0.06$
    & 
    & $7<n<15$~\protect[\ref{Jamin1}] 
\\ \hline
  Hoang  \hfill 99
    & $4.20\pm 0.06$ 
    & $M^{}_{\rm 1S}=4.71\pm 0.03$
    & $4<n<10$, NNLO~\protect[\ref{Hoang15}] 
\\ \hline
  Beneke  \hfill 99
    & $4.26\pm 0.09$ 
    & $M^{2\mbox{\tiny GeV}}_{\rm PS}=4.60\pm 0.11$
    & $6<n<10$, NNLO~\protect[\ref{Beneke6}]
\\ \hline
  Hoang \hfill  00
    & $4.17\pm 0.05$ 
    & $M^{}_{\rm 1S}=4.69\pm 0.03$
    & {$4<n<10$, NNLO, $m_c\neq 0$}~\protect[\ref{Hoang8}]
\\ \hline
  K\"uhn  \hfill 01  
    & $4.21\pm 0.05$ 
    & 
    & $1<n<4$, ${\cal O}(\alpha_s^2)$~\protect[\ref{Kuhn2}] 
\\ \hline
  Erler  \hfill 02  
    & $4.21\pm 0.03$ 
    & 
    & ${\cal O}(\alpha_s^2)$~\protect[\ref{Erler1}] 
\\ \hline
  Eidem\"uller  \hfill 02  
    & $4.24\pm 0.10$ 
    & $M^{2\mbox{\tiny GeV}}_{\rm PS}=4.56\pm 0.11$
    & $3<n<12$~\protect[\ref{Eidemuller1}] 
\\ \hline
  Bordes  \hfill 02  
    & $4.19\pm 0.05$ 
    & 
    & ${\cal O}(\alpha_s^2)$~\protect[\ref{Bordes1}] 
\\ \hline
  Corcella  \hfill 02  
    & $4.20\pm 0.09$ 
    & 
    & $1<n<3$, ${\cal O}(\alpha_s^2)$~\protect[\ref{Corcellamb}] 
\\ \hline\hline
\multicolumn{4}{|c|}{ $\Upsilon({\rm 1S})$ mass }\\
\hline\hline
  Pineda \hfill  97
    &  
    & $m^{}_{\rm pole}=5.00^{+ 0.10}_{- 0.07}$
    & NNLO~\protect[\ref{Pineda7}]
\\ \hline
  Beneke \hfill  99
    & $4.24\pm 0.09$ 
    & $M^{2\mbox{\tiny GeV}}_{\rm PS}=4.58\pm 0.08$
    & {NNLO}~\protect[\ref{Beneke6}] 
\\ \hline
  Hoang  \hfill 99
    & $4.21\pm 0.07$ 
    & \mbox{}\hspace{3.5mm}$M^{}_{\rm 1S}=4.73\pm 0.05$
    & {NNLO}~\protect[\ref{Hoang10}] 
\\ \hline
  Pineda \hfill 01
    & $4.21\pm 0.09$ 
    & $M_{\rm RS}^{2\mbox{\tiny GeV}} = 4.39\pm 0.11$
    & {NNLO}~\protect[\ref{Pineda6}]
\\ \hline
  Brambilla  01
  & $4.19\pm 0.03$ & 
  & NNLO, pert. th. only~\protect[\ref{Brambilla5}]
\\ \hline
\end{tabular}
\end{small}
\caption{\it \label{tabcollection}
Collection in historical order in units of GeV of recent bottom
quark mass determinations from spectral sum rules and 
the $\Upsilon(\mbox{1S})$ mass.
Only results where $\alpha_s$ was taken as an input are shown. The
uncertainties quoted in the respective references have been added
quadratically. All numbers have been taken from the respective
publications. 
}
\end{center}
\vskip 3mm
\end{table}

\subsubsection*{Non-relativistic sum rules}
\label{subsubsectionnonrelsumrules}
The large-$n$ sum rules have the advantage that the experimentally
unknown parts of 
the $b\bar b$ continuum cross section above the $\Upsilon$ resonance
region are suppressed. A crude  model for the continuum cross
section is sufficient and causes an uncertainty in the b quark mass
below the $10$~MeV level. Depending on which moment is used the
overall experimental uncertainties in the b quark mass are between
$15$ and $20$~MeV. Over the past years there has been a revived
interest in non-relativistic sum rules because new theoretical
developments allowed for the systematic determination of 
${\cal O}(v^2)$ (NNLO) corrections to the spectral
moments~[\ref{Penin1}--\ref{Melnikov2},\ref{Hoang15}--\ref{Hoang8}]. 
%
All analyses found that the NNLO  
corrections were as large or even larger than the NLO
corrections and various different methods were devised to
extract numerical values for the bottom quark mass. 
In Refs.~[\ref{Melnikov2},\ref{Hoang15}--\ref{Hoang8}] 
threshold masses were implemented accounting for the   
renormalon problem. This removed one source of the bad perturbative
behaviour, but it was found that a considerable theoretical uncertainty
remained, coming from the theoretical description of the production and
annihilation probability of the $b\bar b$ pair.
In Refs.~[\ref{Melnikov2}] and [\ref{Beneke6}] the kinetic and the PS
mass were determined from fits of individual moments. 
It was found that the NLO and NNLO results for the bottom mass differ by about
$200$~MeV. In Ref.~[\ref{Melnikov2}] it was argued that the results
form an alternating series and a value of $m_{b,{\rm kin}}(1\,{\rm GeV})=4.56\pm
0.06(\mbox{ex,th})$~GeV was determined.
In Ref.~[\ref{Beneke6}] only the NNLO results were accounted based on
consistency arguments with computations of the $\Upsilon({\rm 1S})$
mass and the result $m_{b,{\rm PS}}(2~\mbox{GeV})=4.60\pm 0.02(\mbox{ex})\pm
0.10(\mbox{th})$~GeV was obtained.
In Ref.~[\ref{Hoang15}] the 1S mass was employed and a $\chi^2$-fit based on
four different moments was carried out. It was found that the large
normalization uncertainties  drop out at NLO and NNLO and that the
results for the mass at NLO and NNLO showed good convergence. The
result was $m_{b,{\rm 1S}}=4.71\pm 0.02(\mbox{ex})
\pm 0.02(\mbox{th})$~GeV. A subsequent 
analysis~[\ref{Hoang8}] which included the effects of the nonzero charm
mass yielded $m_{b,{\rm 1S}}=4.69\pm 0.02(\mbox{ex})
\pm 0.02(\mbox{th})$~GeV.

\subsubsection*{Relativistic sum rules}
\label{subsubsectionrelsumrules}
The small-$n$ sum rules have the disadvantage that the unknown parts of
the $b\bar b$ continuum cross section above the $\Upsilon$ resonance
region constitute a substantial contribution to the spectral moments.
The advantage is that the computation of the theoretical moments is
less involved since usual perturbation theory in powers of
$\alpha_s$ can be employed.
In Ref.~[\ref{Kuhn2}] the theoretical moments were determined at order
${\cal O}(\alpha_s^2)$ and it was found that the perturbative behaviour
of the theoretical moments is quite good.
For the bottom quark mass determination it was assumed that the
unknown experimental continuum cross section agrees with the
perturbation theory prediction and subsequently the result
$\overline m_b(\overline m_b)=4.21\pm 0.05$~GeV was determined. 
A more conservative analysis 
in Ref.~[\ref{Corcellamb}] obtained the result 
$\overline m_b(\overline m_b)=4.20\pm 0.09$~GeV.

\subsubsection{Bottom quark mass from the mass of the $\Upsilon({\rm 1S})$}
\label{subsectionupsilonmass}

Among the earliest values of the b quark mass were determinations
that were based on analysis of the observed spectrum of the $\Upsilon$
mesons. However, since these determinations used potential
models to describe the $b\bar b$ dynamics they have little value for
present analyses in B physics. The same conceptual advances that led
to the progress in the determination of the ${\cal O}(v^2)$
corrections to the spectral moments also allowed to systematically
determine ${\cal O}(v^2)$ corrections to the spectrum of
quark-antiquark bound states, which provides another method to
determine a bottom quark threshold mass. The disadvantage of
this method is that the theoretical tools only apply to the case
in which the  binding energy $\sim m_b \,v^2$ is larger than 
$\Lambda_{\rm QCD}$, which is unlikely for higher radial excitations
and questionable for the ground state. As such, also the
theoretical methods to determine the effects of non-perturbative
corrections, which are based on Shifman et al.~[\ref{Shifman1}], could be
unreliable.  
In recent analyses (see Tab.\,\ref{tabcollection}) only the
$\Upsilon({\rm 1S})$ mass has been used for a bottom mass extraction. The
uncertainty is completely dominated by the estimate of the
non-perturbative effects.  

\subsubsection{Summary of $m_b$ determinations from sum rules}
\label{subsectionconclusion}

Comparing the results from the recent bottom quark mass determinations
(see Tab.\,\ref{tabcollection}) one finds a remarkable consistency
among the various analyses. However, the impression 
could be misleading because all methods have
problematic issues. Therefore, 
it is prudent to adopt a more conservative view in averaging and
interpreting the results. 
For the workshop is was agreed that the $\overline m_b( \overline m_b)$
shall be used as reference mass and that the respective threshold
masses shall be determined from it. This leads to an enhancement of the
theoretical error in the threshold masses,
due to their dependence on $\alpha_s$.
 An averaging prescription
for the results in Tab.\,\ref{tabcollection} 
has not been given, and it was agreed on the value
\be\label{mbworkshop}
\begin{array}{|c|}\hline
{\overline m_b(\overline m_b) \, = \, 4.21 \pm 0.08~\mbox{GeV} .}
\\ \hline
\end{array}
\ee
Future work should aim to reduce the uncertainty to a level of
50~MeV.

\subsubsection{Charm quark mass from sum rules}
The charm mass plays a less important role than $m_b$ 
in applications related to the CKM determination, although it certainly is 
a fundamental  parameter. Perhaps because of that, 
the determination of $m_c$ from $e^+ e^-\to $ hadrons 
has so far received less attention than 
that of $m_b$ and has not reached  the same level of maturity; 
we will  not discuss the subject here.
The most recent analyses can be found in 
[\ref{Eidemuller1},\ref{Eidemuller:2000rc},\ref{Kuhn2}]. Typical results 
for the $\msbar$ mass $\overline m_c(\overline m_c)$ range 
between 1.19 and 1.37 GeV, with uncertainties varying 
between 30 and  110 MeV.

\def\dfrac#1#2{{\displaystyle {#1 \over #2}}}
\def\dsum{\mathop{\displaystyle \sum }}
\def\dint{\displaystyle \int }

\newcommand{\mcc}{\overline{m}_c(\overline{m}_c)}
\newcommand{\mbb}{\overline{m}_b(\overline{m}_b)}

\subsubsection{Charm and bottom quark masses from Lattice QCD}

The determination of both heavy and light quark masses is one of the most
important field of activity of lattice QCD simulations. Two major theoretical 
advances have allowed to increase the accuracy of these determinations. The 
first one has been the development of non-perturbative renormalization 
techniques. The renormalized quark mass $m_q(\mu)$, in a given renormalization 
scheme, is related to the bare quark mass $m_q(a)$, which is a function of the 
lattice spacing $a$, through a multiplicative renormalization constant,
\beq
m_q(\mu)=Z_m(\mu\, a) \, m_q(a) \, .
\eeq
The bare quark mass $m_q(a)$ (with $q=u,\,d,\,s,\,c,\ldots$) is a free parameter
of the QCD Lagrangian. It can be computed on the lattice by requiring the mass 
of some physical hadron ($\pi,\,{\rm K},\,{\rm D},\,{\rm B},\ldots$), 
determined from the 
numerical simulation, to be equal to the corresponding experimental value. 
Therefore, one experimental input is needed to fix the value of the quark mass 
for each flavour of quark.

The quark mass renormalization constant, $Z_m(\mu\, a)$, can be computed in
principle in perturbation theory. Its perturbative expansion, however, is known 
only at one loop and the corresponding theoretical uncertainty is therefore 
rather large. The non-perturbative renormalization techniques allow to compute 
$Z_m$ in a non-perturbative way directly from a numerical simulation, with an
accuracy which is at the level of few per cent. The two most important 
non-perturbative renormalization methods developed so far are based on the so 
called RI/MOM~[\ref{rimom}] and Schr\"odinger functional~[\ref{sf}] schemes.

The other important theoretical progress, in lattice QCD calculations, has been
the introduction of improved actions and operators, which allow to reduce 
discretization errors (finite cut-off effects) from ${\cal O}(a)$ to 
${\cal O}(a^2)$. This improvement has been particularly relevant 
for the lattice
determination of the charm quark mass. Typical values of the lattice cut-off, 
in current numerical simulations, are in the range $a^{-1}\sim 3-4~\gev$. With 
these values, leading discretization effects proportional to 
$m_c\,a$ can be of 
the order of 30\% or larger, and they would represent the major source of 
systematic uncertainty in lattice determinations of the charm quark mass. The 
use of improved actions, combined with the extrapolation 
to the continuum limit 
$(a\to 0)$ of the results obtained at fixed lattice spacing, allows to reduce 
discretization errors well below the 10\% level.

Two lattice determinations of the charm quark mass, which use both
non-perturbative renormalization and a non-perturbatively improved action, have
been performed so far. The results, in the $\msbar$ scheme, 
read [\ref{mc_ape},\ref{mc_alpha}]
\bea
\label{mc}
&&\mcc = 1.26 \pm 0.04 \pm 0.12 \gev  \nonumber \\
&&\mcc = 1.301 \pm 0.034 \gev  \, .
\eea
The first of these results has been obtained at a fixed value of the lattice 
spacing, corresponding to $a^{-1}\simeq 2.7~\gev$. 
The second one also involves 
an extrapolation to the continuum limit, and therefore the prediction is more
accurate in this case. At fixed value of the lattice spacing the two
calculations are in very good agreement. The only uncertainty which is not 
quoted in Eq.~(\ref{mc}) is due to the use of the quenched approximation. For 
the $b$-quark mass the quenching effect has been found to be very small, 
of the 
order of 1--2\% [\ref{Davies:1994pz},\ref{mb_ape}], 
while determinations of this effect for light quarks are more 
uncertain, lying in the range between 10 and 25\%. In order to account for the
quenching error in the case of the charm quark mass, a (probably conservative) 
estimate consists in adding a systematic uncertainty of the order of 10\% to the
result of Eq.~(\ref{mc}). This gives, as best lattice estimate for the charm 
quark mass, the value
\beq
\mcc = 1.30 \pm 0.03 \pm 0.15 \gev \,.
\eeq

Lattice determinations of the $b$-quark mass have reached, at present, a very
high level of both statistical and systematic accuracy. Since the mass of the
$b$ quark is larger than the UV cut-off (the inverse of the lattice spacing)
used in current lattice calculations, the $b$ quark cannot be simulated
directly on the lattice. Therefore, one is led to use an effective theory, like 
HQET or NRQCD, in which the heavy degrees of freedom associated with the $b$ 
quark are integrated out. Within the effective theory, the pole mass of the $b$ 
quark is related to the B meson mass $M_B$ through the relation
\beq
\label{mbhqet}
M_B = m_b^{\rm pole} + \varepsilon - \delta m \, ,
\eeq
which is valid up to ${\cal O}(1/m_b^2)$ corrections. In Eq.~(\ref{mbhqet}), 
$\varepsilon$ is the so called binding energy and $\delta m$ is a mass 
counterterm induced by radiative corrections. Neither $\varepsilon$ nor $\delta 
m$ are real physical quantities, and indeed they are separately power divergent.
The binding energy $\varepsilon$ is the quantity which is directly measured in 
the numerical simulations of the effective theory on the lattice. At the same
time, an accurate determination of $\delta m$ is necessary in order to achieve 
a precise estimate of the $b$-quark mass.

The most accurate determination of the $b$-quark mass on the lattice has been 
obtained with the HQET~[\ref{mb_ape}]. It relies on the NNLO perturbative
calculation of the residual mass performed in Ref.~[\ref{martisach}]. 
The final unquenched ($N_f=2$) result for the $b$-quark mass 
in the $\msbar$ scheme reads
\beq
\label{mb}
\mbb = 4.26 \pm 0.06 \pm 0.07 \gev \, ,
\eeq
in which the combined statistical and systematic uncertainty is at the level of 
2\%. Other lattice determinations of the $b$-quark mass have been also obtained 
by using NRQCD~[\ref{mb_nrqcd}]. 
Since the systematic is rather different in the
latter case, it is quite reassuring to find that the lattice-NRQCD results are 
in very good agreement with the prediction of Eq.~(\ref{mb}).

The lattice determinations of the $b$-quark mass can be further improved. In 
the quenched case, the residual mass $\delta m$ has been computed at ${\cal O}
(\alpha_s^3)$ by implementing the so called numerical stochastic perturbation 
theory~[\ref{direnzo}]. The same NNNLO accuracy could be achieved also for the 
unquenched theory. More recently, a completely non-perturbative approach to the
calculation of $\delta m$ has been proposed. The corresponding (preliminary) 
quenched result for the $b$-quark mass is 
$\mbb = 4.53(5)(7) \gev$~[\ref{mb_alpha}], 
which is larger than the lattice determination of Eq.~(\ref{mb}) and than 
the non-lattice estimates reviewed in the previous subsection. Since the 
approach of Ref.~[\ref{mb_alpha}] is new, it deserves further investigations.
On the other hand, being completely non-perturbative, 
it is quite promising for 
future and even more accurate lattice determinations of the $b$-quark mass.

\subsection{Extraction of heavy-quark parameters from 
semileptonic  moments}
\label{sec:moments}

Important information on the parameters of the OPE can be extracted from 
the moments of the differential distributions in s.l.\ and radiative B decays,
which encode the shape of these spectra.
Recently, the first few moments of the hadronic, leptonic, 
and photonic spectra in s.l.\
and radiative B decays  have been measured by several experiments
[\ref{Cronin-Hennessy:2001fk},\ref{delphi_mx},\ref{babar9312}]. We 
define the moments of the leptonic energy distribution  as
\begin{equation}
M_1^\ell = \frac1{\Gamma}\,\int d E_\ell\, E_\ell
\frac{d\Gamma}{dE_\ell}; \quad\quad
M_n^\ell = \frac1{\Gamma}\,\int d E_\ell\,
\left(E_\ell-M_1^\ell
\right)^n  \frac{d\Gamma}{dE_\ell}\ \ \ (n>1),
\label{eq:01}
\end{equation}
and the moments of the distribution of $M_X$, the invariant hadronic
mass, as
\begin{equation}
M_1^X\!=\!\frac1{\Gamma}\,\int dM_X^2\, (M_X^2\!-\!\bar{M}_D^2)\frac{d\Gamma}{dM_X^2}
; \quad\quad
M_n^X\!=\!\frac1{\Gamma}\,\int dM_X^2\, (M_X^2\!-\!\langle M_X^2 \rangle)^n 
\frac{d\Gamma}{dM_X^2}  \ \ (n>1),
\label{eq:02}
\end{equation}
where $\bar{M}_D=1.973$ GeV is the spin averaged $D$ meson mass and
$\Gamma$ is the total s.l.\ width. In general, $n$ can also be
fractional.
Some experiments apply a lower cut on the lepton energy. 
In that case two truncated leptonic moments, 
originally suggested by Gremm {\it et al.} [\ref{gremmetal}] and
defined as
\begin{equation}
{R}_{0} =\frac{\int_{1.7}^{} (d \Gamma_{sl}/dE_l)
dE_l}{\int_{1.5}^{} (d \Gamma_{sl}/dE_l) dE_l} \ \ \ {\rm and} \ \ \
{R}_{1} =\frac{\int_{1.5}^{} E_l (d \Gamma_{sl}/dE_l)
dE_l}{\int_{1.5}^{} (d {\Gamma_{sl}} /dE_l) dE_l}, \label{r1}
\end{equation}
are often used in the experimental analysis.
The theoretical framework to interpret these data has long been known 
and is  based on the OPE. Different formulations exist,  
depending on the way the quark masses are treated. For instance, 
the $m_b$ and $m_c$ masses can be taken as independent 
parameters or subject to a constraint on $m_b-m_c$, imposed from
the measured ${\rm B}^{(*)}$ and $D^{(*)}$ meson masses. 
The second choice introduces a $1/m_c$ expansion.
Another option concerns the normalization scheme used for quark masses and 
non-perturbative parameters. As explained in the previous section, 
one can use short-distance masses, such as  the low-scale running
masses, or pole masses.

The moments $M^{\ell}_n$, $R_i$, and $M_n^X$ are highly sensitive to
the quark masses and to the non-perturbative parameters of the
OPE. For instance, the hadronic moments $M_n^X$ vanish at the parton
level and are generated  only by real gluon emission at  $O(\alpha_s)$  and
by non-perturbative effects suppressed by powers of the $b$
quark mass. The OPE expresses lepton moments through quark masses as a double
expansion in $\alpha_s$ and $1/m_b$:
\begin{equation}
M_n^{\ell} = \left(
\frac{m_b}{2}\right)^n  \left[ 
\varphi_n(r)  + \bar{a}_n(r) \frac{\alpha_s}{\pi} 
+\bar{b}_n(r)\,\frac{\mu_\pi^2}{m_b^2}
+\bar{c}_n(r)\,\frac{\mu_G^2}{m_b^2}
+\bar{d}_n(r)\,\frac{\rho_D^3}{m_b^3}
+\bar{s}_n(r)\,\frac{\rho_{LS}^3}{m_b^3} + ...
\right] ,
\label{eq:03}
\end{equation}
where $r=(m_c/m_b)^2$. Analogous expressions hold for the truncated
moments $R_i$. The higher coefficient functions $\bar{b}(r)$, 
$\bar{c}(r)$, ... are also perturbative series in $\alpha_s$. 
The functions $\varphi_n$ in Eq.~(\ref{eq:03}) are well-known parton 
expressions, given e.g.\ in [\ref{voloshin0}]. 
The expectation values of only two operators 
contribute to  $O(1/m_b^3)$: the Darwin term $\rho_D^3$ 
and the spin-orbital term $\rho_{LS}^3$. 
Due to the kinematic definition of the hadronic invariant mass $M_X^2$, the
general expression for the hadronic moments includes $M_B$ explicitly,
but it is otherwise similar to Eq.~(\ref{eq:03}):
\bea
\nonumber
{M}_n^X
&=&
m_b^{2n}\sum_{l=0} 
\left[\frac{M_B\!-\!m_b}{m_b}\right]^l \!
\left(E_{nl}(r) + a_{nl}(r)\frac{\alpha_s}{\pi}
+b_{nl}(r)\frac{\mu_\pi^2}{m_b^2} +
c_{nl}(r)\frac{\mu_G^2}{m_b^2} 
\right.\\ && \left. \hspace{5.3cm}
+d_{nl}(r)\frac{\rho_D^3}{m_b^3} +\!
s_{nl}(r)\frac{\rho_{LS}^3}{m_b^3} + ...\right).
\label{eq:04}
\eea It is possible to re-express the heavy quark masses, $m_Q$, in the above equations, 
in terms of the meson masses, $M_{H_Q}$, through the relation~[\ref{Bigi3}]: 
\begin{equation}
\label{eq:05}
M_{H_Q} = m_Q+ \bar{\Lambda} + 
\frac{\mu_\pi^2-a_{H_Q} \mu_G^2}{2 m_Q} + 
\frac{\rho_D^3 + a_{H_Q} \rho_{LS}^3 - \rho_{nl}^3}
{4 m_Q^2} + {\cal{O}}\left(\frac{1}{m_Q^3}\right),
\end{equation}
where $a_{H_Q}=1$ and $-1/3$ for pseudo-scalar and vector mesons, 
respectively.
The use of these expressions
introduces an explicit dependence on the non-local correlators contributing to
$\rho_{nl}^3$. In the notation of~[\ref{gremm-kap}], $\rho_{nl}^3$ 
corresponds to linear combinations of ${\cal T}_{1-4}$.

The moments of the photon spectrum in inclusive radiative B decays,
${\rm B}\to X_s \gamma$,  are also useful to  constrain the non-perturbative
parameters. The  relevant formulae can be found in 
[\ref{Bauer:2002sh}] and Refs.\ therein.  

Of all the possible formalisms we discuss here  only 
two extreme cases.\footnote{A few different
 possibilities are considered in [\ref{Bauer:2002sh}]. } 
 The first formalism  is based on 
the kinetic running masses, $m_Q(\mu)$, and non-perturbative parameters, 
introduced in~[\ref{Bigi2},\ref{kinmass}]. No charm mass expansion is assumed. 
The second formalism employs quark pole masses and the 
${\rm B}^{(*)}$ and $D^{(*)}$ 
meson mass relations.
Contributions through $O(\alpha_s^2 \beta_0)$ [\ref{alphas},\ref{alphas2b0}]
and $O(1/m_b^3)$ 
[\ref{ope},\ref{opebigi},\ref{voloshin0},\ref{gremm-kap},\ref{FLS},\ref{gremmetal},\ref{Falk:1997jq}] 
to the moments are available. 
Depending on the formulation adopted, the number of parameters involved at this 
order ranges from six to nine. Some of these parameters, like $m_b$ and 
$\lambda_2 \simeq \mu_G^2/3$, are relatively well known. Others, notably those which 
appear at $O(1/m_b^3)$, are virtually unknown.  

\subsubsection{The $m_{b,{\rm kin}}(\mu)$, $m_{c,{\rm kin}}(\mu)$ and $\mu_{\pi}^2(\mu)$ formalism}
\label{sec:mupiformalism}

The quark masses are here identified by the running kinetic quark
masses $m_{b,{\rm kin}}(\mu)$ and $m_{c,{\rm kin}}(\mu)$, and since no relation like
Eq.~(\ref{eq:05}) is used, they are two independent parameters.
Apart from $\mu_\pi^2(\mu)$ and $\mu_G^2(\mu)$, defined here as 
expectation values in the actual B meson, there are two 
$1/m_b^3$ parameters, $\rho_D^3$ and $\rho_{LS}^3$.
The effect of $\rho_{LS}^3$ turns out to be numerically small.
In Eqs.~(\ref{eq:03}) and (\ref{eq:04}) the mass ratio $r$ is given by 
$(m_{c,{\rm kin}}(\mu)/m_{b,{\rm kin}}(\mu))^2$, and the $b$ quark mass 
is understood as $m_{b,{\rm kin}}(\mu)$.
The perturbative coefficients additionally depend on $\mu/m_b$ and the mass 
normalization scale $\mu$ is set at $\mu=1$ GeV.
To illustrate the size of different contributions to $M_n^\ell$,  
we give the relevant coefficients for the first three moments in the case 
without a cut on the lepton energy in Table~\ref{tab2:1}, using
$m_{b,{\rm kin}}$(1~GeV) = 4.6~GeV and $r = 0.06$
[\ref{Battaglia:2002tm}] (the $O(\alpha_s^2 \beta_0)$ corrections are also
available [\ref{alphas2b0}]).
\begin{table}
\begin{center}
\begin{tabular}{|l|c|c|c|c|c|c|}
\hline 
 & $\varphi_n$ & $\bar{a}_n$ & $\bar{b}_n$ & $\bar{c}_n$ & $\bar{d}_n$ & 
$\bar{s}_n$ \\
\hline
$M_1^\ell$ & 0.6173 & 0.015  &0.31 &-0.73 &-3.7 &0.2 \\
$M_2^\ell\,(\times 10)$ & 0.3476 & 0.026  &1.7 &-1.0 & -10.2& -0.9\\
$M_3^\ell\,(\times 10^2) $ & -0.3410 &0.066 & 3.4 & 1.3&-23& -4.2\\
\hline
\end{tabular}
\end{center}
\caption{\it  Numerical values of the coefficients in Eq.(\ref{eq:03}) 
evaluated at $r$=0.06 and $m_{b,{\rm kin}}(1\,{\rm GeV})= 4.6\,{\rm GeV}$ and
without a lepton energy cut.}
\label{tab2:1}
\end{table}
In the case of hadronic moments, keeping terms up to $1/m_b^3$, 
we discard in Eq.~(\ref{eq:04})  coefficients 
$b_{nl}$, $c_{nl}$ with $l\!>\!1$, and $d_{nl}$, $s_{nl}$ with $l\!>\!0$. 
The only non-vanishing $E_{i0}$ coefficient is 
$E_{10} = r - \bar{M}_D^2/m_b^2$. The value of the other coefficients, at 
$r=0.06$ and again without a cut on the hadron energy, 
are listed in Table~\ref{tab2:2}. The $O(\alpha_s^2 \beta_0)$
corrections to  hadronic moments are not yet available in this scheme.
\begin{table}
\begin{center}
\begin{tabular}{|l|c|c|c|c|c|c|c|c|c|c|c|}
\hline 
 $i$ & $E_{i1}$ & $E_{i2}$ & $E_{i3}$ & $a_{i0}$&$a_{i1}$ & 
$b_{i0}$ & $b_{i1}$
 &$c_{i0}$ & $c_{i1}$& $d_{i0}$ & $s_{i0}$  \\
\hline
 $1$ &0.839 &1 &0 &  0.029  &0.013 & -0.58 &-0.58 &0.31 &0.87&3.2 
&-0.4  \\
 $2$ &0 &0.021 &0 &- 0.001  &-0.002& 0.16 &0.34 &0 &-0.05 &-0.8 &0.05 
\\
 $3$ &0 & 0&-0.0011 & 0.0018 &  0.0013  &0 &0.034 &0 &0 &0.15  &0\\
\hline
\end{tabular}
\end{center}
\caption{\it Numerical values of the coefficients in Eq.(\ref{eq:04}) evaluated 
at $r$=0.06 and $m_{b,{\rm kin}}(1~{\rm{GeV}})$ = 4.6~GeV and without a lepton
energy cut.}
\label{tab2:2}
\end{table}

\subsubsection{The $\bar \Lambda$ and $\lambda_1$ formalism}
\label{sec:lambdabarformalism}

This widely used scheme results from the combination of the OPE with the HQET. 
Following the notation of Ref.~[\ref{Falk:1997jq}], the moments are 
 expressed in the following general form:
\bea\label{eq:06}
M_n&=& M_B^k \left[a_0  +
a_1 \frac{\alpha_s(\overline{M}_B)}{\pi} 
 +a_2 \beta_0 \frac{\alpha_s^2}{\pi^2} 
+ b_1 \frac{ \bar{\Lambda}} {{\overline{M}_B}} 
+ b_2 \frac{\alpha_s}{\pi}\frac{ \bar{\Lambda}} {{\overline{M}_B}} 
+ \frac{c_1\,\lambda_1 + 
     c_2\, \lambda_2 +      c_3\,{ \bar{\Lambda}}^2}
     {{ {\overline{M}_B}}^2}  
\right. \nonumber\\
&& \left.+ \frac1{\overline{M}_B^3} \left(d_1\,\lambda_1\bar{\Lambda}
      +      d_2\, \lambda_2\bar{\Lambda} +  d_3\,{ \bar{\Lambda}}^3  + 
     d_4\, \rho_1 +d_5\, \rho_2 + \sum_{i=1,4} d_{5+i}{{\cal T}_i}
\right)+ O\left( \frac{\Lambda^4_{QCD}}{{m}^4_Q}\right)\right],
\eea
where $k=n$ and $k=2n$ for leptonic and hadronic moments,
respectively, while $a_0=0$ for hadronic moments. Analogous
expressions hold for the truncated moments. $\overline{M}_B=5.3135$~GeV is the
spin-averaged B meson mass,
and $\beta_0=11-2/3 n_f$, 
with $n_f=3$. The terms $O(\alpha_s^2 \beta_0)$ and
$O(\alpha_s \bar{\Lambda})$ are not known in the case
of the third hadronic moment. 
The coefficients $a_i,b_i,c_i,d_i$ 
for the first three
leptonic, $M_{1,2,3}^\ell$, and hadronic moments,
$M_{1,2,3}^X$, without a cut on the lepton energy are given 
in [\ref{Battaglia:2002tm}]. The coefficients for $i=1,2$ 
with a cut on the lepton energy and for $R_{0,1}$ 
can be found in [\ref{Bauer:2002sh}].
The non-perturbative parameters in Eq.(\ref{eq:06}) are related 
to those in \sec{sec:mupiformalism}
by the following relations, valid up to ${\cal{O}}(\alpha_s)$:
\begin{equation}
\mu_{\pi}^2 =-\lambda_1 - \frac{{\cal T}_1 + 3 {\cal T}_2}{m_b};
\quad
\mu_G^2 =3\lambda_2 + \frac{{\cal T}_3 + 3 {\cal T}_4}{m_b};
\quad 
\rho_D^3=\rho_1\, ; 
\quad
 \rho_{LS}^3=3 \rho_2\,.
\label{eq:relations}
\end{equation} 
Perturbative corrections introduce a significant numerical difference 
between the parameters in the two schemes. At $\mu=1$ GeV:
\begin{equation} 
\bar{\Lambda} \simeq M_B - m_{b,{\rm kin}}{\rm{(1~GeV)}} - \frac{\mu_{\pi}^2 - \mu_G^2}{2 m_b} - 
0.26~{\rm{GeV}}~;
\quad \quad
-\lambda_1 \simeq \mu_{\pi}^2{\rm{(1~GeV)}} -
0.17~{\rm{GeV}}^2\,.
\end{equation}
As anticipated in the previous Section, the use of the ill-defined
pole quark mass induces in this formalism large perturbative
corrections, which are however expected to cancel in the relation 
between physical observables, as long as all observables involved 
in the analysis are computed at the same order in $\alpha_s$.
We also note that, 
as a consequence of the HQET mass relations for the mesons, the intrinsic 
expansion parameter 
in Eq.(\ref{eq:06}) is $1/M_D$, rather than $1/M_B$. The convergence
of this expansion has been questioned, in view of 
indications [\ref{Uraltsev:2001ih},\ref{lattice}] that the matrix elements 
${\cal T}_i$  of some non-local operators could be larger than that 
expected from dimensional estimates. 

Higher moments are generally  more sensitive to the $1/m_b^3$ 
corrections, but the uncertainty due to unknown perturbative and
non-perturbative higher orders prevents a precision determination
of the related  parameters. Higher moments contain nonetheless
useful information: as we will see below, they have been employed in 
the first analyses based on multi-parameter fits  
[\ref{Bauer:2002sh},\ref{Battaglia:2002tm}].

\subsection*{Measurements of the moments and non-perturbative parameters} 

The study of moments in B meson s.l.\ decays and ${\rm B}\dec X_s 
\gamma$ allows to perform several independent determinations of
the non-perturbative parameters and is 
now pursued by different experiments. Here we
summarize the measurements performed by the CLEO collaboration,
taking data at the CESR \epm\ collider, and by the DELPHI Collaboration at LEP.
Measurements of the first hadronic moment in  
${\rm B} \rightarrow X_c \ell \bar \nu$ with different minimum charged lepton 
momentum  have also been reported 
by the BaBar Collaboration~[\ref{babarmom}]. 

CLEO and BaBar 
measurements have been performed at the $\Upsilon(4S)$ resonance. 
While there is an obvious advantage in measuring the 
spectra in events where the decaying B rest frame almost coincides with the 
laboratory frame, low energy particles cannot be identified there. It is thus 
necessary to rely on models for extrapolating the lepton energy spectrum to 
zero energy or to resort to computations for a truncated spectrum.  
On the other hand, performing the analysis at energies around the $Z^0$ peak, 
the large momentum of the $b$-hadrons ensures sensitivity
to almost the full lepton spectrum, thus reducing modelling assumptions. 
The main challenge put by the higher energy is the accurate determination of 
the B rest frame. 

\subsubsection{Moments of hadronic mass and $b\rightarrow s \gamma$ 
photon energy spectra at CLEO}
The first experimental determination of the HQE parameters based on
the shape variables was performed by the CLEO 
collaboration [\ref{Cronin-Hennessy:2001fk}].  The analysis 
was based on the measurement of  the photon spectrum above 2.0 GeV in $
b\rightarrow s \gamma$ inclusive decays [\ref{bsgammacleo}] and on s.l.\
inclusive decays. CLEO measured
the first two moments of the photon spectrum in radiative decays, 
$$\langle E_\gamma \rangle=2.346\pm 0.032\pm 0.011\ {\rm GeV} \ \ {\rm
and}
\ \ \langle E^2_\gamma \rangle-\langle E_\gamma \rangle^2=0.0226\pm 
0.0066\pm 0.0020\ {\rm GeV}^2
$$
the first of which 
is related to half the value of the $b$ quark pole mass, 
and thus to $\overline{\Lambda}$, of course up to $1/M_B^3$ corrections. 
The parameter $\lambda _1$ was then extracted from a measurement of the
first moment, $M_1^X$,  of the mass of the hadronic system recoiling
against the $\ell$-$\bar{\nu}$ pair in s.l.\  decays. This
measurement takes advantage of the ability of the CLEO experiment to
reconstruct the $\nu$ 4-momentum with high efficiency and resolution,
by virtue of the hermeticity of the detector and the simplicity of the
initial state in $\Upsilon (4\rm S)\rightarrow \rm B\bar{B}$. 
CLEO applied a 1.5~GeV/$c$ lower cut on
the charged lepton momentum. The
explicit relation between  $M_1^X$ and the HQE  parameters
$\overline{\Lambda}$, $\lambda _1$, etc.\  is given in that case  in
[\ref{Cronin-Hennessy:2001fk}]. 
CLEO found  
$$M_1^X 
= 0.251 \pm 0.023 \pm 0.062\ {\rm
GeV}^2 \ \ \  \ {\rm and} \ \ \ \
  M^X_2= 0.576 \pm 0.048\pm 0.163\ {\rm GeV}^4,
$$
where the first error is statistical and the second is
systematic.
\begin{figure}[t]
\centerline{\epsfig{figure=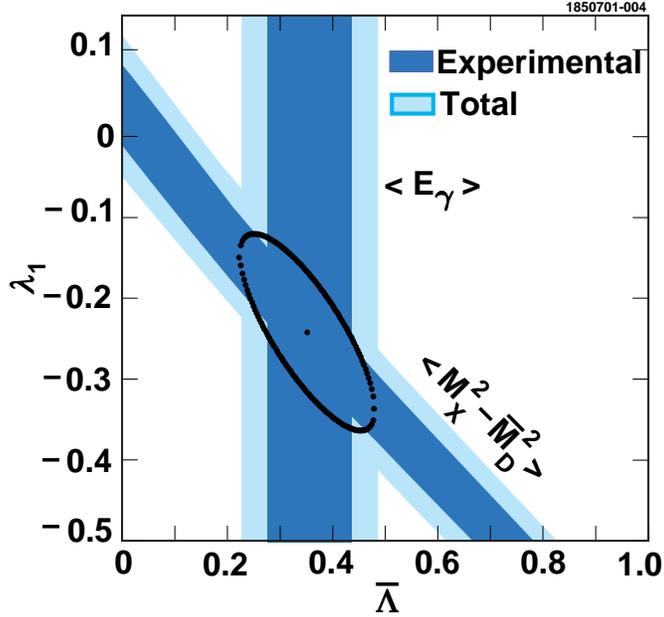,width=3.5in}}
\caption{\it Constraints on $\bar \Lambda$ (GeV), $\lambda_1$ (GeV$^2$) 
from 
the first hadronic moment and the first moment of the photon energy 
spectrum in $b
\rightarrow s \gamma$ \protect 
[\ref{Cronin-Hennessy:2001fk},\ref{bsgammacleo}]. The inner bands
show the experimental error bands.  The light gray extensions show
the theoretical errors. \label{fig:mhad}}
\end{figure}
From $M_1^X$ and $\langle E_\gamma \rangle$, 
CLEO extracted $\overline \Lambda$ and  $ \lambda_1$, obtaining
$\overline \Lambda = 0.35 \pm 0.07 \pm 0.10\ {\rm GeV}\ ,$ and 
$ \lambda_1 = -0.236 \pm 0.071 \pm 0.078\ {\rm GeV}^2\ .$
Here, the first error is governed by the experimental
measurements of the moments, and the second error reflects
theoretical uncertainties, and in particular those related to
$O(1/m_b^3)$ contributions. Figure~\ref{fig:mhad} shows the bands
corresponding to these two constraints as well as the $\Delta
\chi^2 = 1$ ellipse in the $\overline \Lambda,\lambda_1$ plane.

\subsubsection{Moments of the leptonic spectrum at CLEO} 
A recent CLEO analysis [\ref{marina-lmom}] reports the measurement of
the truncated moments of the lepton spectrum, with a momentum cut
of $p_{\ell} \ge 1.5$ GeV/$c$ in the B meson rest frame
[\ref{marina-lmom}]. This choice for the lepton momentum cut
decreases the sensitivity of the
measurement to the secondary leptons from the cascade decays ($b
\rightarrow c \rightarrow s/d \ell \bar\nu$).
The small contribution coming from charmless
s.l.\  decays $b\rightarrow u \ell \bar{\nu}$ is included
 by adding the contribution from $d\Gamma
_{u}/dE_{\ell}$, scaled by $|V_{ub}/V_{cb}|^2$ [\ref{gremmetal},\ref{chris}].
CLEO results for $R_{0,1}$ are given in Table \ref{leptmomcleo1}.
%
\begin{table}[b]
\begin{center}
\begin{tabular}{|l|c|c|}
\hline

         &   R$^{exp}_0$                 &   R$^{exp}_1$                 \\ \hline
\elep\   &$0.6184 \pm 0.0016 \pm 0.0017$ &$1.7817 \pm 0.0008 \pm 0.0010$ \\ \hline
\mulep\  &$0.6189 \pm 0.0023 \pm 0.0020$ &$1.7802 \pm 0.0011 \pm 0.0011$ \\ \hline \hline
Combined &$0.6187 \pm 0.0014 \pm 0.0016$ &$1.7810 \pm 0.0007 \pm 0.0009$ \\ \hline
\end{tabular}
\end{center}
\caption{\it \label{rmom} Measured truncated lepton moments for \elep\ and
\mulep, and for the sum.}\label{leptmomcleo1} 
\end{table}
The values of the HQE parameters and their experimental
uncertainties are obtained by calculating the $\chi^2$ from the
measured moments R$^{exp}_0$ and R$^{exp}_1$ and the covariance
matrix $\mbox{E}_{R_0 R_1}$.
\begin{figure}[t]
\center{\mbox{\epsfig{figure=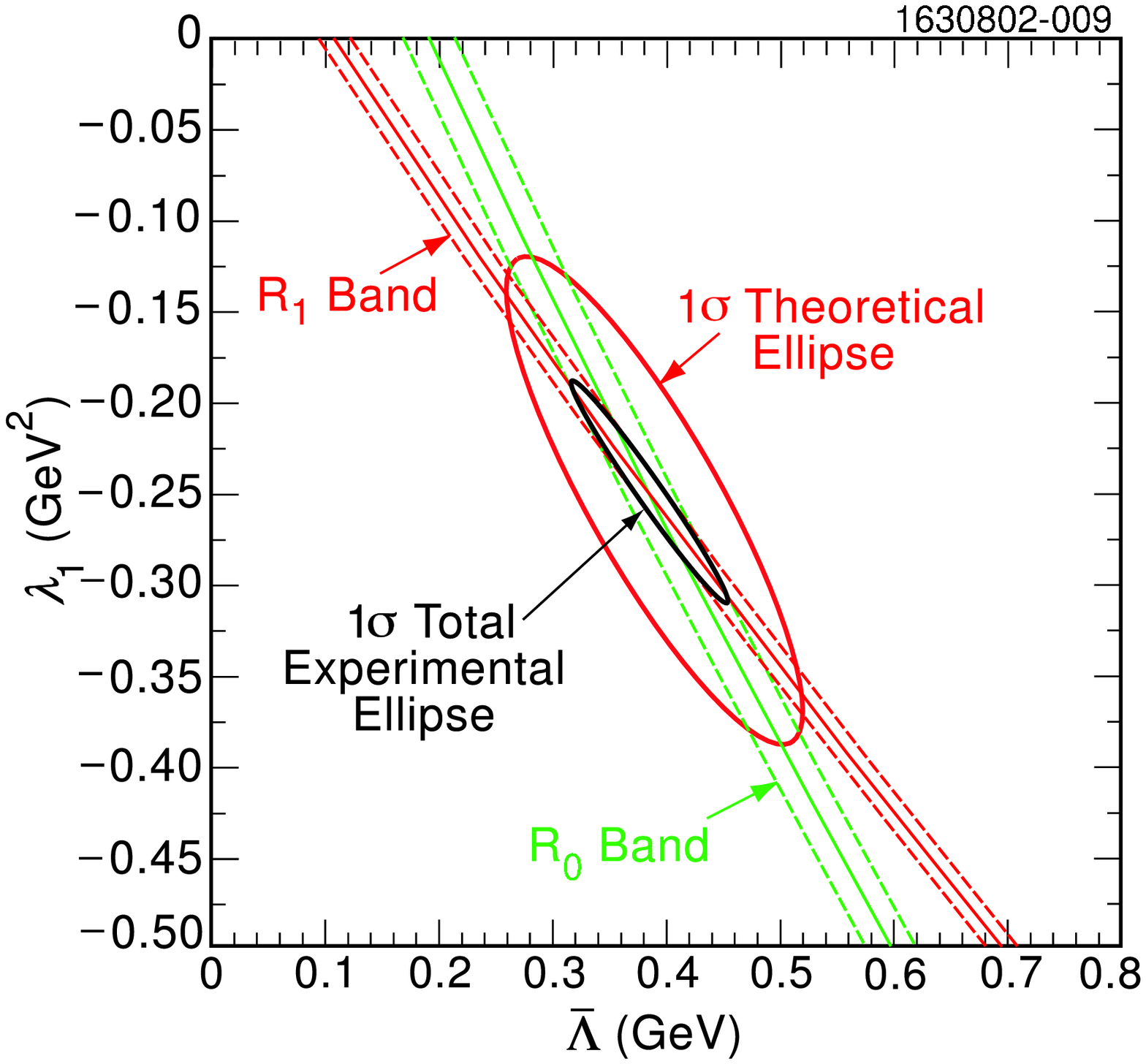,width=3.0in,height=2.9in}}
\mbox{\epsfig{figure=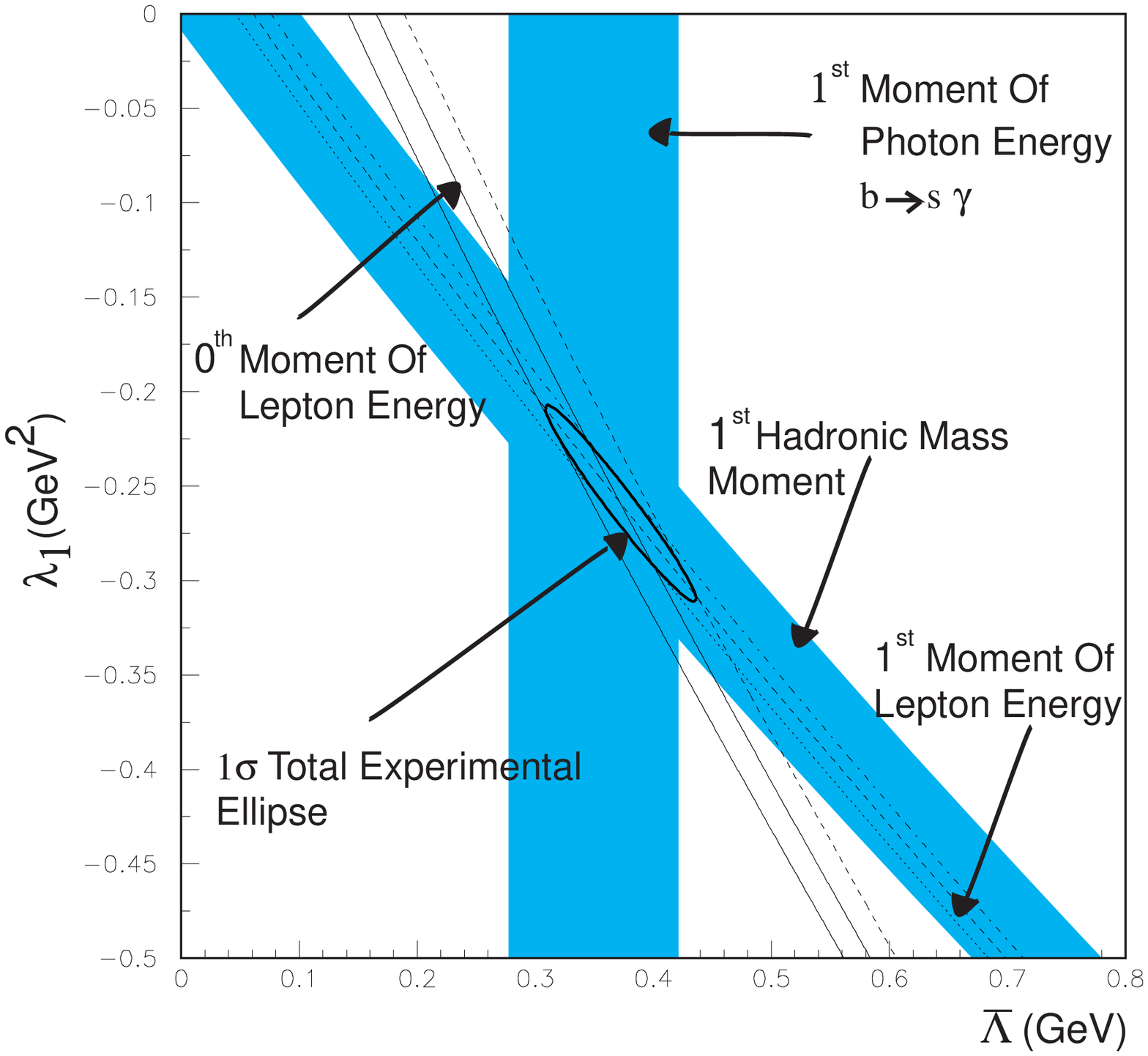,width=3.0in,height=2.9in}}}
\caption{\it Left: constraints from  combined electron and muon R$_{0,1}$
 moments, with $\Delta\chi^2=1$ contours for total
experimental and theoretical uncertainties [\ref{marina-lmom}]. 
Right: comparison
of the same constraints with those in Fig.~\ref{fig:mhad}.
$\lambda_1$ and $\bar \Lambda$ are computed in the $\overline{MS}$
scheme to order $1/M^3_B$ and $\beta_0 \alpha_s^2$. 
}
\label{therrepse}
\end{figure}
The theoretical uncertainties on the HQE parameters are determined
by varying, with flat distributions, the input parameters within
their respective errors: $|\frac{V_{ub}}{V_{cb}}|= 0.09 \pm 0.02$,
$\alpha_s = 0.22 \pm 0.027$, $\lambda_{2} = 0.128 \pm 0.010$
GeV$^2$, $\rho_1 = \frac{1}{2}(0.5)^3 \pm \frac{1}{2}(0.5)^3$
GeV$^3$, $\rho_2 = 0 \pm (0.5)^3$ GeV$^3$, and  ${\cal T}_i = 0.0 \pm
(0.5)^3$ GeV$^3$.  The contour that contains 68\% of the
probability is shown in Fig. \ref{therrepse}. This procedure for
evaluating the theoretical uncertainty from the unknown expansion
parameters that enter at order $1/M_B^3$ is similar to that used
by Gremm and Kapustin [\ref{gremm-kap}]  and Bauer and 
Trott~[\ref{chris}], but different from the procedure used in the 
CLEO analysis discussed above~[\ref{Cronin-Hennessy:2001fk}]. The
dominant theoretical uncertainty is related to the $1/M_B^3$ terms
in the non-perturbative expansion discussed
before. Ref.~[\ref{Bauer:2002sh}] has explored the convergence of the
perturbative and non-perturbative series appearing in the
expressions for the moments described in the previous Section. The most
conservative estimate gives a truncation error of at most 20\%.
The theoretical uncertainties presented in this CLEO analysis  do not
include this truncation error. The extracted ${{\lambda}_{1}}$ and
$\bar{\Lambda}$ are given in Table \ref{statsyst}.
The rhs in Fig.~\ref{therrepse} shows a comparison of these CLEO results with
the ones in Ref.~[\ref{Cronin-Hennessy:2001fk}]. 
The errors shown correspond to  the
experimental errors only: the agreement is
good, although the theoretical uncertainties do not warrant a very
precise comparison.
%
\begin{table}[ht]
\begin{center}
\begin{tabular}{|l|c|c|} \hline
                & ${\lambda}_{1}$(GeV$^2$)  & $\bar{\Lambda}$(GeV)  \\ \hline
\elep &$-0.28\pm0.03|_{stat}\pm0.06|_{syst}\pm0.14|_{th}$
&$0.41\pm 0.04|_{stat}\pm 0.06|_{syst}\pm0.12|_{th}$\\ \hline
\mulep    &$-0.22\pm0.04|_{stat}\pm0.07|_{syst}\pm0.14|_{th}$
&$0.36\pm 0.06|_{stat}\pm 0.08|_{syst}\pm0.12|_{th}$\\ \hline

$\ell ^{\pm}$ &$-0.25\pm0.02|_{stat}\pm0.05|_{syst}\pm0.14|_{th}$
&$0.39\pm 0.03|_{stat}\pm 0.06|_{syst}\pm0.12|_{th}$\\ \hline
\end{tabular}
\caption{\it Values ${\lambda}_{1}$ and $\bar{\Lambda}$ extracted from
CLEO measurement of $R_{0,1}$,
including statistical, systematic, and theoretical errors. The
last row shows the results obtained combining \elep\ and \mulep\
samples.} \label{statsyst}
\end{center}
\end{table}

CLEO also performed an analysis of the truncated leptonic moments 
in terms of the short distance \mus\ mass instead of the pole 
mass scheme implicit in
the  $\lambda_1, \overline\Lambda$ formalism.
The results in Ref. [\ref{chris}] are  used to extract 
\mus, or rather ${\bar{\Lambda}}^{1 \rm
S}\equiv {\bar M}_B-$\,\mus. Table \ref{lamb1s}
summarizes the values of $\bar{\Lambda}^{1\rm S}$ and
$m_b^{1\rm S}$ extracted from $R_{0,1}$ 
for electrons and muons samples separately, and for their
sum. The final result $m_b^{1\rm S}=(4.82\pm 0.07|_{exp}\pm
0.11|_{th}) {\rm GeV}/c^2$  is in good agreement with the estimates
of $m_b^{1\rm S}$ [\ref{Hoang15},\ref{h1}] discussed in Sec.~\ref{sec:mb}
\begin{table}[ht]
\begin{center}
\begin{tabular}{|l|c|c|} \hline
           &${\bar{\Lambda}}^{1 S}$(GeV)
                       &$m_b^{1\rm S}$(GeV/$c^2$) \\ \hline
\elep\ &$0.52\pm0.04|_{stat}\pm0.06|_{syst}\pm0.11|_{th}$
                       & $ 4.79 \pm 0.07|_{exp}\pm 0.11|_{th}$\\ \hline
\mulep\    &$0.46\pm0.05|_{stat}\pm0.08|_{syst}\pm0.11|_{th}$
                       & $ 4.85 \pm 0.09|_{exp}\pm 0.11|_{th}$\\
                       \hline
Combined &$0.49\pm0.03|_{stat}\pm0.06|_{syst}\pm0.11|_{th}$
                       & $ 4.82 \pm 0.07|_{exp}\pm 0.11|_{th}$\\ \hline
\end{tabular}
\caption{\it Values of ${\bar{\Lambda}}^{1\rm S}$ and $m_b^{1\rm
S}$ extracted from $R_{0,1}$. 
The quoted errors reflect statistical, systematic, and
theoretical uncertainties, respectively.} \label{lamb1s}
\end{center}
\end{table}

\begin{figure}[t]
\begin{center}
\center{\epsfig{figure=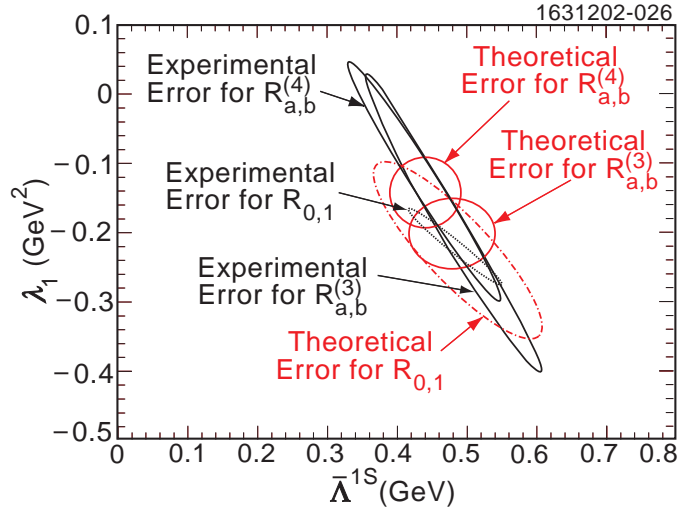,width=3.6in,height=2.7in}}\caption{\it  Constraints on the HQE parameters $\lambda_1$ and
${\bar{\Lambda}}^{1\rm S}$ from different CLEO measured spectral
moments.} \label{all1s}
\end{center}
\end{figure}

We have mentioned in the previous Section that one can also consider
fractional moments.
Bauer and Trott [\ref{chris}] have 
explored different lepton energy moments, by varying the exponent
of the energy in the integrands and the lower limits of
integration. In particular, they identify several moments that
provide constraints for
 \mus\ and $\lambda_1$ that are less sensitive to higher
order terms in the non-perturbative expansion. The
shape of the truncated lepton spectrum recently measured by CLEO 
[\ref{cleonew}]
allows to measure the following ones
\begin{equation}
{\mbox{R}}^{(3)}_{a} =\frac{\int_{1.7}^{} E_l^{0.7}(d
\Gamma_{sl}/dE_l) dE_l}{\int_{1.5}^{} E_l^2 (d \Gamma_{sl}/dE_l)
dE_l},\ \ \ \label{r3a}
{\mbox{R}}^{(3)}_{b} =\frac{\int_{1.6}^{} E_l^{0.9}(d
\Gamma_{sl}/dE_l) dE_l}{\int_{1.7}^{} (d \Gamma_{sl}/dE_l) dE_l}
,
\end{equation}
\begin{equation}
{\mbox{R}}^{(4)}_{a} =\frac{\int_{1.6}^{}E_l^{0.8} (d
\Gamma_{sl}/dE_l) dE_l}{\int_{1.7}^{} (d \Gamma_{sl}/dE_l) dE_l},\ \ \ 
 \label{r4a}
{\mbox{R}}^{(4)}_{b} =\frac{\int_{1.6}^{} E_l^{2.5} (d
\Gamma_{sl}/dE_l) dE_l}{\int_{1.5}^{}E_l^{2.9} (d {\Gamma_{sl}}
/dE_l) dE_l}. 
\end{equation}
 Tables \ref{r3ab} and \ref{r4ab} summarize the measured
values, as well as the statistical and systematic errors.
Fig.~\ref{all1s} shows the values of $\bar{\Lambda}^{1\rm S}$ and
$\lambda _1$ extracted from these two sets of observables, as well as the
constraints derived from the moments $R_0$ and $R_1$. Although
these results confirm that the  $1/M_B^3$ terms induce  much smaller
uncertainties using $R^{(3,4)}_{a,b}$, the experimental errors are
larger in this case because of the similar slopes for the two
constraints. However, the different relative importance of
experimental and theoretical errors makes these results
complementary to the previous ones reported.
\begin{table}[htpb]
\begin{center}
\begin{tabular}{|l|c|c|}
\hline
         & R$^{(3)}_a$(GeV$^{-1.3}$)  & R$^{(3)}_b$(GeV$^{0.9}$)   \\ \hline
\elep   &$0.3013 \pm 0.0006|_{stat} \pm 0.0005|_{syst}$ &$2.2632
\pm 0.0029|_{stat} \pm 0.0026|_{syst}$ \\ \hline

\mulep\  &$0.3019 \pm 0.0009|_{stat} \pm 0.0007|_{syst}$ &$2.2611
\pm 0.0042|_{stat} \pm 0.0020|_{syst}$ \\ \hline

$\ell ^{\pm}$ &$0.3016 \pm 0.0005|_{stat} \pm 0.0005|_{syst}$
&$2.2621 \pm 0.0025|_{stat} \pm 0.0019|_{syst}$ \\ \hline
\end{tabular}
\end{center}
\caption{\it \label{r3ab} Measured truncated lepton moments
$R^{(3)}_{a,b}$ for \elep , \mulep, and their weighted average.}
\end{table}
\begin{table}[htpb]
\begin{center}
\begin{tabular}{|l|c|c|}
\hline
         & R$^{(4)}_{a}$(GeV$^{0.8}$)  & R$^{(4)}_{b}$(GeV$^{-0.4}$)   \\ \hline
\elep\   &$2.1294 \pm 0.0028|_{stat} \pm 0.0027|_{syst}$ &$0.6831
\pm 0.0005|_{stat} \pm 0.0007|_{syst}$ \\ \hline

\mulep\  &$2.1276 \pm 0.0040|_{stat} \pm 0.0015|_{syst}$ &$0.6836
\pm 0.0008|_{stat} \pm 0.0014|_{syst}$ \\ \hline

$\ell ^{\pm}$ &$2.1285 \pm 0.0024|_{stat} \pm 0.0018|_{syst}$
&$0.6833 \pm 0.0005|_{stat} \pm 0.0006|_{syst}$ \\ \hline
\end{tabular}
\end{center}
\caption{\it \label{r4ab} Measured truncated $R_{4a,b}$ moments for
\elep , \mulep, and their weighted average.}
\end{table}

 Bauer and Trott [\ref{chris}] also identify moments that are
insensitive to \mus\ and $\lambda_1$. They suggest that a
comparison between a theoretical evaluations of these ``duality
moments'' and their experimental values may provide useful
constraints on possible quark-hadron duality violations in
s.l.\  processes. CLEO measures two such ``duality moments",
defined as
\begin{equation}
{\mbox{D}}_{3} =\frac{\int_{1.6}^{} E_l^{0.7} (d \Gamma_{sl}/dE_l)
dE_l}{\int_{1.5}^{} E_l^{1.5}(d {\Gamma_{sl}} /dE_l) dE_l}
\label{d3}, \ \ \ \
{\mbox{D}}_{4} =\frac{\int_{1.6}^{} E_l^{2.3} (d \Gamma_{sl}/dE_l)
dE_l}{\int_{1.5}^{}E_l^{2.9} (d {\Gamma_{sl}} /dE_l) dE_l}.
\end{equation}
The theoretical predictions from Ref.~[\ref{chris}]
are compared with the measured $D_{3,4}$ from the combined lepton
sample in Table~\ref{dualth}. The agreement is excellent and thus
no internal inconsistency of the theory is uncovered in this
analysis.
\begin{table}[htpb]
\begin{center}
\begin{tabular}{|l|c|c|} \hline
       & Experimental           &Theoretical \\ \hline
D$_{3}$&$0.5193\pm0.0008|_{exp}$& $0.5195\pm
0.0006|_{\lambda_1,\bar{\Lambda}^{1\rm S}}\pm 0.0003|_{th}$\\
\hline D$_{4}$&$0.6036\pm0.0006|_{exp}$& $0.6040\pm
0.0006|_{\lambda_1,\bar{\Lambda}^{1\rm S}}\pm 0.0005|_{th}$\\
\hline
\end{tabular}
\caption{\it Measured {\it duality} moments and theoretical predictions
using the values $\lambda_1$ and $\bar{\Lambda}^{1\rm S}$
[\ref{cleonew}]. 
The errors reflect the experimental uncertainties
in these parameters and the theoretical errors, respectively.}
\label{dualth}
\end{center}
\end{table}

\subsubsection{Moments of leptonic and hadronic mass spectra at DELPHI}
Results obtained by the DELPHI collaboration for the first three
moments of the lepton energy and the hadronic mass spectra have been
presented at ICHEP02~[\ref{delphi_mx}].  The analyses were based
on $b$-hadron s.l. decays into electrons and muons, selected from a
sample of about $3\times 10^6$ $e^+e^- \rightarrow Z^0 \rightarrow q \bar q$
events recorded with the DELPHI detector at LEP. Electrons and muons
were required to have a momentum greater than 2-3~GeV/$c$ in the
laboratory frame.
For the lepton energy spectrum measurement an inclusive reconstruction of the 
secondary vertex of the charm hadron decay was performed.
The energy of the B hadron was estimated as the energy sum of the 
identified lepton, the secondary hadronic system and 
the neutrino energy, evaluated from the event missing energy. 
The identified lepton was then boosted back to the reconstructed 
B rest frame and its energy $E_{\ell}$ re-computed in this frame. 
Results for the first three moments are summarized in Table~\ref{DELPHI_mom}.
In order to study the hadronic mass distribution the exclusive 
reconstruction of
${ \bar{{\rm B}^0_d}}\rightarrow {\rm D}^{**}\ell\bar{\nu}$ states
was performed and the total ${\rm D}^{**}$ production in $b$-hadron 
s.l. decays was determined.
Moments of the hadronic mass distribution were measured for ${\rm D}^{**}$
candidates and moments of the hadronic mass distribution 
in inclusive $b$-hadron s.l. decays, $M_X$, were derived including 
$b \rightarrow {\rm D} ~~{\rm  and}~~{\rm D}^{*} \ell^- \overline{\nu}_{\ell}$ 
channels.
Results for the first three moments are summarized in Table~\ref{DELPHI_mom}.
As we will discuss in the next subsection, 
the DELPHI results have been used in ~[\ref{Battaglia:2002tm}] 
as inputs of a multi-parameter fit 
to determine the heavy quark masses and non-perturbative parameters of
the HQE. The use of higher moments guarantees a sensitivity to the $1/m_b^3$ 
parameters and 
the simultaneous use of the hadronic and leptonic spectra ensures that 
a larger number of parameters can be kept free in the fit. 
\begin{table}[htbp] 
\begin{center}
\begin{tabular}{|l|c c c c|}
\hline 
Moment & Result & (stat) & (syst) & \\
\hline 
$M_1(E_{\ell})$    & (1.383 & $\pm$ 0.012 & $\pm$0.009)  & GeV~ \\
$M_2(E_{\ell})$    & (0.192 & $\pm$ 0.005 & $\pm$0.008)  & GeV$^2$\\
$M_3(E_{\ell})$    &(-0.029 & $\pm$ 0.005 & $\pm$0.006)  & GeV$^3$\\
\hline
$M_1(M_X)$         & (0.534 & $\pm$ 0.041 & $\pm$ 0.074) & GeV$^2$\\
$M_2(M_X)$         & (1.226 & $\pm$ 0.158 & $\pm$ 0.152) & GeV$^4$ \\
$M_3(M_X)$         & (2.970 & $\pm$ 0.673 & $\pm$ 0.478) & GeV$^6$ \\ 
\hline
\end{tabular}
\caption[]{\it \label{DELPHI_mom}
 DELPHI results for the first three leptonic and hadronic moments.}
\end{center}
\end{table}

\begin{figure}[t]
\center{\mbox{\epsfig{figure=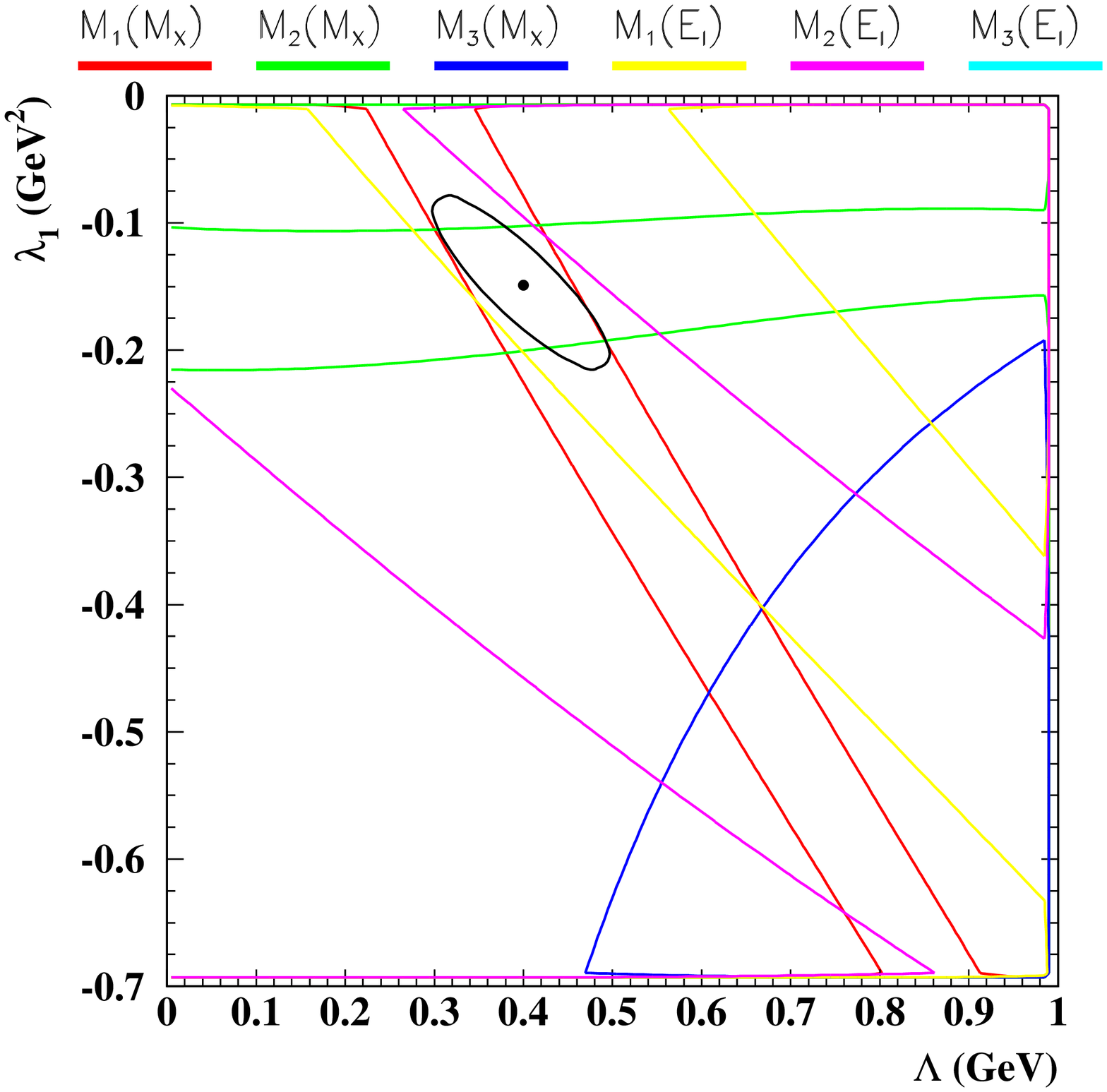,width=3.0in,height=2.25in}}
\mbox{\epsfig{figure=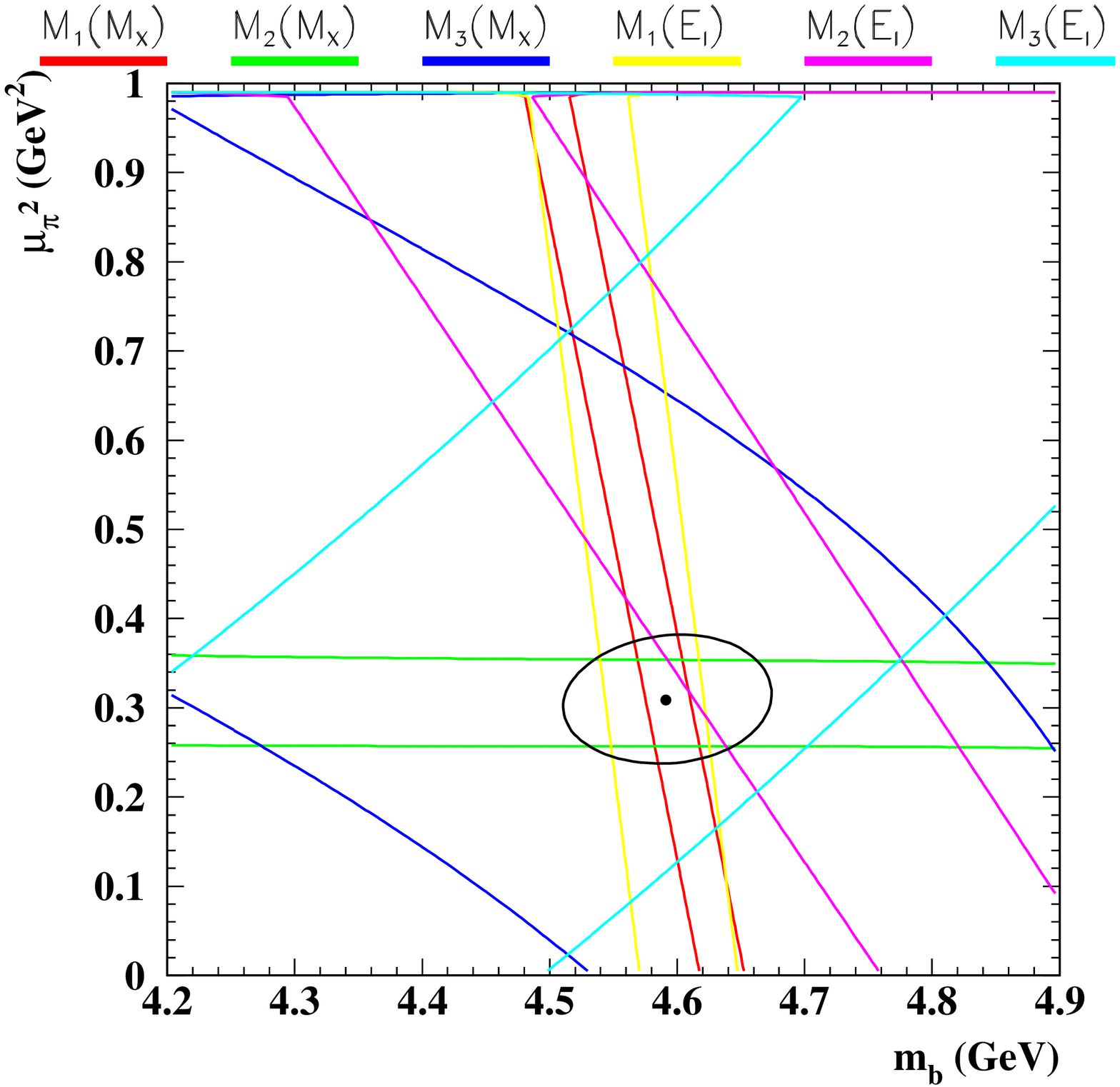,width=3.0in,height=2.25in}}}
\caption{\it Projection of the constraints from the six DELPHI moments 
on the ($m_b(\rm
1~GeV)$, $\mu_\pi^2$) and ($\overline\Lambda$, $\lambda_1$) planes
[\ref{Battaglia:2002tm}]. The bands correspond to the total
experimental error and given keeping all the other parameters at their
central values. The ellipses represent the 1$\sigma$ contour.
\label{dd1}}
\end{figure}

\vspace{3mm}

\subsubsection{Multi-parameter fits of heavy-quark parameters and outlook}
A recent and promising development, in view of the greater precision
expected at the B-factories, consists in combining  leptonic and
hadronic moments in a multi-parameter fit to determine not just $m_b$
and  $\lambda_1\sim -\mu_\pi^2$ but also the dominant $O(1/m_b^3)$ parameters.
The first comprehensive analyses that employ this approach 
[\ref{Bauer:2002sh},\ref{Battaglia:2002tm}] have shown that present data are 
consistent with each other (with the possible exception of the preliminary 
BaBar data [\ref{babarmom}]) and with our theoretical understanding,
most notably with the underlying assumption  of quark-hadron duality. 
%
%

The analysis of [\ref{Battaglia:2002tm}] is based solely on the DELPHI data in
Table~\ref{DELPHI_mom}, and performed in the two theoretical framework 
described above in 
Secs.~\ref{sec:mupiformalism} and \ref{sec:lambdabarformalism} 
The projection of the various constraints on the  ($m_{b,{\rm kin}}(\rm
1~GeV)$, $\mu_\pi^2$) and ($\overline\Lambda$, $\lambda_1$) planes are
given in Fig.~\ref{dd1}, which shows very good consistency. The
results of the fits are shown in Tables~\ref{fitdelphi1} and \ref{fitdelphi2}.
In the framework of Sec.~\ref{sec:mupiformalism} the charm mass is
a free parameter of the fit, though strongly correlated to the bottom
mass. Given  a precise determination  $\delta
m_{b,{\rm kin}}(\rm 1~GeV)\sim 50 $~MeV, the
charm mass could  therefore be extracted with $\delta m_c\sim 90$~MeV,
a competitive determination [\ref{Battaglia:2002tm}] (Cfr.\ Sec.~\ref{sec:mb}).

\begin{table}[htbp]
\begin{center}
\begin{tabular}{|l|c c c c|}
\hline
Fit           & Fit    & Fit         & Syst.      & \\
Parameter     & Values & Uncertainty & Uncertainty& \\ \hline
$m_{b,{\rm kin}}$ (1~GeV) & 4.59   & $\pm$ 0.08  & $\pm$ 0.01 & GeV~\\
$m_{c,{\rm kin}}$ (1~GeV) & 1.13   & $\pm$ 0.13  & $\pm$ 0.03 & GeV~\\
$\mu_{\pi}^2$ (1 GeV) 
              & 0.31   & $\pm$ 0.07  & $\pm$ 0.02 & GeV$^2$\\
$\rho_D^3$    & 0.05   & $\pm$ 0.04  & $\pm$ 0.01 & GeV$^3$ \\ 
\hline
\end{tabular}
\caption{\it Results of fits to the moments of Table~\ref{DELPHI_mom} 
for the $m_b(\mu)$, $m_c(\mu)$ and 
$\mu_{\pi}^2(\mu)$ formalism~[\ref{Battaglia:2002tm}].}
\label{fitdelphi1}
\end{center}
\end{table}

\vspace{-3mm}
\begin{table}[htbp] 
\begin{center}
\begin{tabular}{|l|c c c c|}
\hline
Fit             & Fit    & Fit         & Syst.      & \\
Parameter       & Values & Uncertainty & Uncertainty& \\ \hline
$\overline{\Lambda}$ & ~0.40  & $\pm$ 0.10  & $\pm$ 0.02 & GeV~\\
$\lambda_1$     & -0.15  & $\pm$ 0.07  & $\pm$ 0.03 & GeV$^2$\\
$\lambda_2$     & ~0.12  & $\pm$ 0.01  & $\pm$ 0.01 & GeV$^2$\\
$\rho_1   $     & -0.01  & $\pm$ 0.03  & $\pm$ 0.03 & GeV$^3$ \\ 
\hline
$\rho_2   $     & ~0.03  & $\pm$ 0.03  & $\pm$ 0.01 & GeV$^3$ \\ 
\hline
\end{tabular}
\caption{\it Results of fit to the moments of Table~\ref{DELPHI_mom} 
for the $\bar{\Lambda}-\lambda_1$ 
formalism~[\ref{Battaglia:2002tm}].\label{fitdelphi2} }
\end{center}
\end{table}

The analysis of Ref.~[\ref{Bauer:2002sh}] includes the first two
hadronic moments measured by CLEO and DELPHI, $R_{0,1}$ measured 
by CLEO, the first two leptonic DELPHI moments, and the first two
moments of the photon spectrum in ${\rm B}\to X_s \gamma$. The results in 
one of the formalisms adopted are shown in Table~\ref{tab2:5}. They are
in good agreement with both  CLEO and DELPHI analyses mentioned above.
The preferred ranges for the heavy quark masses
and  for the non-perturbative parameters in Tables 
\ref{fitdelphi1}, \ref{fitdelphi2}, and \ref{tab2:5} are in agreement with
theoretical expectations and with each other, although  the  
analyses [\ref{Bauer:2002sh},\ref{Battaglia:2002tm}] differ in several
respects (data employed, additional constraints, scheme adopted, 
treatment of theoretical errors).

%
\begin{table}[htbp] 
\begin{center}
\begin{tabular}{|l|c c c|}
\hline
Fit & Fit & Fit &  \\
Parameter & Values & Uncertainty &  \\ \hline
$m_b^{1S}$ & ~4.74 & $\pm$ 0.10  & GeV~\\
$\lambda_1+\frac{{\cal T}_1+3{\cal T}_3}{m_b} $   & -0.31 & $\pm$ 0.17 & GeV$^2$\\
$\rho_1$      & 0.15 & $\pm$ 0.12 & GeV$^3$ \\ 
$\rho_2$      & -0.01 & $\pm$ 0.11 & GeV$^3$ \\  \hline
\end{tabular}
\end{center}
\caption{\it  Results of fit for the $m_b^{1S}$-$\lambda_1$ 
formalism [\ref{Bauer:2002sh}].}
\label{tab2:5}
\end{table}


In summary, the experimental information appears
so far consistent with the theoretical framework,
 with the possible exception of the preliminary 
BaBar result. The emerging experimental
information  from the B factories will eventually lead to a more
complete assessment of our present understanding of inclusive s.l.\  decays. 

\subsection{Parton--hadron duality in B decays}
\label{dualitythomas}

Parton-hadron duality 
\footnote{This name might be more appropriate than the more frequently 
used {\em quark}-hadron duality since gluonic effects have to be 
included as well into the theoretical expressions.} 
-- or duality for short --
is invoked to connect quantities evaluated on the 
quark-gluon level to the (observable) world of hadrons. 
It is used all the time,  often  without explicit reference to
it. A striking example of the confidence high-energy physicists have in 
the asymptotic validity of duality was provided by the discussion 
of the width $\Gamma ( Z^0 \to H_b H_b^{\prime}X)$. 
There was about a 2\% difference between  the predicted and measured 
decay width, which lead to lively debates on its significance 
vis-a-vis the {\em experimental} error, before disappearing when the 
analysis was improved. No concern was expressed about 
the fact that the $Z^0$ width was calculated on the quark-gluon 
level, yet measured for hadrons. Likewise the strong coupling 
$\rm \alpha_s(M_Z)$ is routinely extracted from the 
perturbatively computed hadronic $Z^0$ width with a stated 
theoretical uncertainty of 0.003 which translates into a 
theoretical error in $ \Gamma _{had}(Z^0)$ of about 0.1\%.

There  are, however, several different versions and implementations of
the concept of duality. The problem with invoking duality implicitly is
that it is  very often unclear which version is used. In B physics --
in  particular when determining $\vcb$\ and $\vub$\ -- the 
measurements have become so precise that theory can no longer 
hide behind experimental errors. To estimate theoretical 
uncertainties in a meaningful way one has to give clear meaning 
to the concept of duality; only then can one analyse its 
limitations.
In response to the demands of B physics a considerable literature 
has been created on duality over the last few years, which we 
summarize here. Technical details can be found in the references.

Duality for processes involving time-like momenta was first addressed
theoretically in the late  '70's in references [\ref{PQW76}] and
[\ref{Greco:1978gy}]. 
Using the optical theorem, the cross section for $ e^+e^- \to\rm
hadrons$ at leading order in $\alpha_{em}$ can be expressed as
\begin{equation}
\sigma(s) = \frac{16 \pi^2 \alpha_{em}}{s} \mathrm{Im}\;\Pi(s)
\end{equation}
where $\Pi(s)$ is defined through the correlator of electromagnetic currents:
\begin{equation} \label{corr}
T_{\mu \nu}(q^2) = i\int d^4 x \, e^{iqx} \, \langle 0 |
         T \left( J_\mu (x) J_\nu (0) \right) | 0 \rangle
= (g_{\mu \nu} q^2- q_\mu q_\nu  ) \Pi(q^2) 
\ .\end{equation}
One might be tempted to think that by invoking QCD's asymptotic
freedom one can compute $\rm \sigma (e^+e^- \to {\rm hadrons})$ for
large c.m. energies $\sqrt{s} \gg \Lambda _{QCD}$ in terms of quarks (and
gluons) since it is shaped by short distance dynamics. However
production thresholds like those for charm induce singularities that
vitiate such a straightforward computation. Under such circumstances,
duality between the QCD-inferred cross section and the observed one
looks problematic. It was suggested in [\ref{PQW76}] 
that the equality between the two
would be restored after averaging or ``smearing'' over an energy
interval:
\begin{equation}
\langle T^{hadronic}_{\mu \nu} \rangle _w \simeq
          \langle T^{partonic}_{\mu \nu} \rangle _w
\label{ANSATZ} 
\end{equation}
where $\langle ... \rangle _w$ denotes the smearing which is
an average using a smooth weight function $w(s)$:
\begin{equation}
\langle ... \rangle _w = \int ds \, ... \, w(s)
\end{equation}

The degree to which 
$\langle T^{partonic}_{\mu \nu} \rangle _w$ can be trusted as a
theoretical description of the observable 
$\langle T^{hadronic}_{\mu \nu} \rangle _w $ depends on the weight 
function, in particular its width. It can be 
broad compared to the structures that may appear
in the hadronic spectral function, or it could be quite narrow, 
as an extreme case even $w(s) \sim \delta(s-s_0)$. It has become 
customary to refer to the first and second scenarios as 
{\em global} and {\em local} duality, respectively. Other authors 
use different names, and one can argue that this nomenclature 
is actually misleading. Below these items are described in 
more detail without attempting to impose a uniform nomenclature. 

Irrespective of names, a fundamental distinction concerning  
duality is often drawn between s.l.\ and non-leptonic widths. 
Since the former necessarily involves smearing 
with a smooth weight function due to the 
integration over neutrino momenta, it is often argued that predictions 
for the former are fundamentally more trustworthy than for the latter. 
However, such a categorical distinction is 
overstated and artificial. Of much 
more relevance is the differentiation between distributions and 
fully integrated rates.

No real progress beyond the more qualitative arguments of
Refs. [\ref{PQW76}] and [\ref{Greco:1978gy}] occurred for many
years. For as long as one has very limited control over
non-perturbative effects, there is little meaningful that can be said
about duality violation. Yet this has changed for heavy flavour
physics with the development of heavy quark expansions, since within
this OPE framework we can assess non-perturbative effects as well as
duality violation.

\subsubsection{What is parton--hadron duality?}
\label{DEF}

In order to discuss possible violations of duality one has to give
first a more precise definition of this notion, which requires 
the introduction of some theoretical tools. Here
the arguments given in the extensive reviews 
of Ref.~[\ref{shifman}] and [\ref{BU2001}]\footnote{It can be
noted that even the authors of Ref.~[\ref{shifman}] and [\ref{BU2001}]
-- although very close in the substance as well as the spirit
of their discussion -- do not use exactly the same terminology
concerning different aspects of duality.} are followed closely. 

The central ingredient in the definition of duality that will be used
here is the method of the Wilsonian OPE frequently used in field
theory to perform a separation of scales. In practical terms this
means that we can write
\begin{equation} \label{OPE}
i\,\int d^4 x \, e^{iqx} \,
\langle A | T \left( J^\mu (x) J^\nu (0) \right) | A \rangle  
\simeq \sum_n \left(\frac{1}{Q^2} \right)^n   
c_n^{\mu \nu}(Q^2;\lambda)
  \langle A |{\cal O}_n | A \rangle _{\lambda} 
\end{equation}
for $ Q^2 = -q^2 \to \infty$. The following notation has been used:
$|A\rangle$ denotes a state that could be the vacuum -- as for $
e^+e^-\to hadrons$ considered above -- or a B meson when describing
s.l.\  beauty decays. $J^\mu$ denote electro-magnetic and weak
current operators ($b\to c$ or $u$) for the former and the latter
processes, respectively; for other decays like non-leptonic or
radiative ones one employs different $\Delta B =1$ operators; the
${\cal O}_n$ are local operators of increasing dimension. The operator
of lowest dimension yields the leading contribution. In $ e^+e^-$
annihilation it is the unit operator ${\cal O}_0 = 1$, for B decays
${\cal O}_0=\bar bb$. As we have seen in Sec.~\ref{sec:OPE}, 
they lead (among other things) to the naive
partonic results. Yet the OPE allows us to systematically improve the
naive partonic result. The coefficients $c_n^{\mu \nu}$ contain the
contributions from short distance dynamics calculated perturbatively
based on QCD's asymptotic freedom. Following Wilson's prescription a
mass scale $\lambda$ has been introduced to separate long and short
distance dynamics; both the coefficients and the matrix elements
depend on it, but their product does not.

The perturbative expansion takes the form
\begin{equation} \label{pert}
c_n^{\mu \nu}  = \sum_i \left(\frac{\alpha_s (Q^2)}{\pi} \right)^i
                 a_{n,i}^{\mu \nu}
\end{equation}
and is performed in terms of quarks and gluons.
The expectation values for the local operators provide the gateways 
through which non-perturbative dynamics enters.

The crucial point is that the OPE result is obtained in the Euclidean
domain, far from any singularities induced by hadronic thresholds, and
has to be continued analytically into the Minkowskian regime relating
the OPE result to observable hadronic quantities. As long as QCD is
the theory of the strong interactions, it does not exhibit unphysical
singularities in the complex $Q^2$ plane, and the analytical
continuation will not induce additional contributions. To conclude:
{\em duality between $\langle T^{hadronic}_{\mu \nu} \rangle _w$ and
$\langle T^{partonic}_{\mu \nu} \rangle _w$ arises due to the
existence of an OPE that is continued analytically}. 
It is thus misleading to refer to duality as an additional assumption.

Up to this point the discussion was quite generic. To specify it for 
s.l.\ B decays one chooses the current $J_{\mu}$ to be the 
weak charged current related  to $b\to c$ or $b \to u$. As already noted in Sec.~\ref{sec:OPE},
the expansion 
parameter for inclusive s.l.\  decays is given by the 
energy release $\sim 1/(m_b - m_c) \; [1/m_b]$ for 
$b\to c \; [b\to u]$. 

\subsubsection{Duality violation and analytic continuation} 
\label{sec:violation}
%
%
One of the main applications of the heavy quark expansion
is the reliable extraction 
of $\vcb$ and $\vub$. One wants to be able to arrive at a 
meaningful estimate of the theoretical uncertainty in the values 
obtained. There are three obvious sources of theoretical 
errors: 

\renewcommand{\theenumi}{\arabic{enumi}}
\renewcommand{\labelenumi}{\theenumi.}
\begin{enumerate}
\item
unknown terms of higher order in $\alpha_s$; 
\item
unknown terms of higher order in $1/m_Q$;
\item
uncertainties in the input parameters $\alpha_s$, $m_Q$ 
and the expectation values of local operators which appear in the OPE.  
\end{enumerate}
Duality violations constitute additional uncertainties.
They arise from the fact that at finite order in $1/m_Q$, the
Euclidean OPE is insensitive to contributions of the type $e^{-m_Q
/\mu}$, with $\mu$ denoting some hadronic scale. While such a term is probably 
innocuous for beauty, it needs not be for charm quarks. Furthermore,
under analytic continuation these terms turn into potentially more
dangerous oscillating terms of the form $\sin$$(m_Q /\mu)$.

Though there is not (yet) a full theory for duality and its 
violations, progress has 
come about in the last few years for the following reasons:  
\begin{itemize}
\item
the understanding of the physical origins of 
duality violations has been refined as due to 

\vspace{2mm}

\begin{itemize}
\item 
hadronic thresholds; 
\item 
so-called `distant cuts'; 
\item 
the suspect validity of $1/m_c$ expansions.
\end{itemize}

\vspace{2mm}

\item
The issues surrounding the exponentially small terms discussed above
and their analytic continuation have been understood.
\item
There is an increasing array of field-theoretical 
toy models, chief among them the 't Hooft model, which is 
QCD in 1+1 dimensions in the limit of $N_c \to \infty$. It is 
solvable and thus allows an unequivocal comparison of the OPE 
result with the exact solution. 
\item 
For the analysis of $ b\to c$ transitions the small-velocity (SV)
expansion is a powerful tool [\ref{Shifman:1987rj}].
\end{itemize}

Based on general expectations as well as on analysing the models 
one finds that indeed duality violations are described by highly 
power suppressed `oscillating' terms of the form 
\begin{equation}\label{DV}
T(m_Q) \sim \left( \frac{1}{m_Q}\right)^k {\rm sin}(m_Q/\mu) 
\end{equation}
for some integer power $k$. More generally one can state: 
\begin{itemize}
\item 
Duality will not be exact at finite masses. It represents an
approximation the accuracy of which will increase with the 
energy scales in a way that depends on the process in question. 

\item 
Limitations to duality can enter only in the form of an
oscillating  function of energy or $m_Q$ (or have to be exponentially
suppressed), i.e.\  duality violation cannot
 modify all decay rates in the same way.

\item 
The OPE equally applies to s.l.\  as well as to non-leptonic 
decay rates. Likewise both widths are subject to duality violations. 
The difference here is quantitative rather than qualitative; at finite 
heavy quark masses corrections are generally expected to be
larger in the non-leptonic 
widths. In particular, duality violations there can be boosted by the 
accidental nearby presence of a narrow hadronic resonance. Similar 
effects could arise in s.l.\  rates, but are expected to be 
highly suppressed there. 
\item 
 It is not necessary to have a proliferation of decay channels 
to reach the onset of duality, either approximate or asymptotic. 
Instructive examples are provided by the SV
kinematics in s.l.\  decays and by non-leptonic rates in the 
't Hooft model.
For example, in the SV limit, the ground-state doublet of
D mesons alone saturates the inclusive s.l.\  decay
rate and is dual to the partonic rate [\ref{Boyd:1995ht}]. 
The point here is that
the large energy release would allow a large number of states to
 contribute kinematically, but only two channels are actually
 allowed by the dynamics.

\end{itemize}

\noindent
Putting everything together it has been estimated 
by the authors of Ref.~[\ref{BU2001}] -- 
that {\em duality violations in the integrated s.l.\  width of 
B mesons cannot exceed the fraction of a percent. }
As such we do not envision it to ever become the limiting factor in
extracting $\vcb$\ and $\vub$\ since the uncertainties in the expression
for the s.l.\  width due to fixed higher order contributions
will remain larger than this level.  The oscillatory nature of duality
violating contributions is a main ingredient in this conclusion. It
also shows that duality violations could become quite sizeable if an
only partially integrated width -- let alone a distribution -- is
considered. Generally, for distributions the expansion parameter is
not the heavy mass, rather it is a quantity such as $1/[m_Q (1-x)]$
where $x$ is e.g. the rescaled charged lepton energy of a s.l.\ 
decay. From Eq.~(\ref{DV}) one would expect that contributions
the form $\sin(m_Q [1-x]/\mu)/[m_Q (1-x)]^k$ would appear in
differential distributions.


\subsubsection{How can we check the validity of parton--hadron duality?}
\label{sec:validity}

If in the future a discrepancy between the measured and 
predicted values for, say, a CP asymmetry in B decays is found, 
one has  to check 
very diligently all ingredients upon which the prediction was
 based, in particular the values for $\vcb$\ and $\vub$, before 
one could make a credible claim to have uncovered New Physics. This 
means one needs a measure for potential duality violations that is 
not based purely on theoretical arguments.

Most theoretical uncertainties do not have a statistical nature.
As in the case of experimental systematics, the most convincing 
way to control  them is to determine the same quantity 
in independent ways and analyse their consistency.
The heavy quark expansions lend themselves 
naturally to such an approach since 
they allow the description of numerous decay rates in terms 
of a handful of basic parameters, namely quark masses and hadronic 
expectation values. 
Of course, such independent determinations of the same quantity 
only probe the overall theoretical control: 
by themselves they cannot tell whether a failure 
 is due to 
unusually large higher order contributions or to a breakdown of duality.

The fact that both the inclusive and exclusive methods for extracting 
$\vcb$ and $\vub$ yield consistent values 
(see Secs.~\ref{sec:vcbincl},\ref{sec:vubincl},
\ref{sec:vcbexc}, and \ref{sec:vubexcl}) is such a
test. Theoretical corrections are nontrivial and
essential for the agreement. As discussed in Sec.~\ref{sec:moments}, 
the study of moments offers another important consistency check. In particular, we
emphasize that the $b$ quark mass extracted from the shape 
variables is consistent, within errors, 
with the one extracted from sum rules and lattice calculations
(see Sec.~\ref{sec:mb}), and that 
the analyses of CLEO and DELPHI data, and those of the leptonic and hadronic moments 
point to very similar values for the kinetic energy parameter $\lambda_1\sim -\mu_\pi^2$. 
This suggests that no anomalously large higher
order corrections or unexpectedly sizeable duality violating
contributions are present in the HQE used to described inclusive
s.l.\  $b\to c$ decays. However, once again, we stress  that 
these comparisons do not represent direct tests of duality.

\subsubsection{Model based investigations of duality}

It is desirable to study in more explicit detail how duality comes about,
how it is approached and what its limitations are. This
can be done in the context of exactly soluble field theories, in
particular the 't~Hooft model, which is QCD in 1+1 dimensions with
the number of colours going to infinity [\ref{QCD2}]. There one finds
duality to be achieved very quickly, i.e. after a mere handful of
channels open up.

For detailed studies in 1+3 dimensions one is at present limited to the
use of quark models employing certain types of potential. However, 
one has to handle these models with care, as they have sometimes led
to confusion.
In particular, it has been argued in Ref.[\ref{isgur}] that within quark
models one could have an ${\cal O}(1/m_Q)$ contribution to the 
ratio of inclusive 
to free quark total decay rate. Such terms  are absent in the OPE,
and therefore violate duality. The arguments
presented in [\ref{isgur}] and similar papers have been discussed in
[\ref{BU2001}], where their internal flaws have been
pointed out. 
One of the important lessons is that such models exhibit automatically 
the proper behaviour in the Shifman-Voloshin (SV)
limit [\ref{Shifman:1987rj}], where $\Lambda_{QCD}\ll\delta m =m_b-m_c\ll m_b$.
In particular, they have to satisfy a set of sum rules. 
Once one realizes that such models are
automatically in compliance with what we know to be true in QCD, 
it becomes clear that  no $1/m_Q$ terms
can appear [\ref{articleNR},\ref{lettreNR},\ref{OH}]. Of particular importance
in this context are the Bjorken sum rule for ${\cal O }({(\delta m)^2
\over m_b^2})$ terms and the Voloshin sum rule for ${\cal O }({\delta m \over
m_b})$ terms\footnote{It also has been demonstrated explicitly that, 
contrary to what suggested in note 3 of Ref.~
[\ref{isgur}], no term of ${\cal O }({\delta m \over m_b^2})$ 
exist in QCD 
[\ref{QCD}].}. Other terms are suppressed by higher powers of
$1/m_b$ or powers of $\Delta/\delta m$, $\Delta$ being the level spacing
of $O(\Lambda_{QCD})$, {\it i.e.} the difference between the ground state and the 
first excited level.
Once such models have been brought into compliance with what we know
to be true in QCD -- like the validity of the SV sum rules -- then they
can play a significant heuristic role in educating our intuition about the
onset of duality.

In Ref.~[\ref{OH}] a detailed study of the cancellations required for duality to hold
have been performed using a  harmonic oscillator (HO) potential.
The interest of this model is that the truncation
of states to the first band of orbital excitations (lowest $\rm D^{**}$)
becomes exact to the relevant order $1/m_b^2$, which allows us to perform a
complete and explicit numerical or analytical calculation.
Furthermore, this model is close to the ones used 
in [\ref{isgur}], so that one can check precisely the various statements made there.
Using a constant for the leptonic interaction one finds in the
harmonic oscillator model
\bea
R_{sl}={\Gamma_{inclusive} \over \Gamma_{free~quark}}
=1+{3 \over R^2 m_b^2}({1 \over 4}-{\Delta \over \delta m})+
{\rm smaller~terms} \label{ROH}
\eea
where $\Delta={1 \over m_d R^2}$ is a model parameter containing the
square of the harmonic oscillator radius $R$ and the light-quark mass
$m_d$. Note that the first term inside the parentheses
originates from the kinetic energy operator. In fact,
it can be proven  [\ref{lettreNR}]
that for {\it regular} potentials the whole series,
directly calculated in the model, is {\it exactly} the one given by OPE.

What is then the explanation of the apparent disagreement with [\ref{isgur}]? 
First, there is a misunderstanding induced by the expression 
"$1/m_Q$ duality violation", used  sometimes in a misleading way. 
Ref.~[\ref{isgur}] does  not dispute that the OPE 
is basically right and that the equality with free quark decay 
is satisfied within the expected accuracy in the region of phase space 
where the energy release is large $(t_{max}^{1/2}-t^{1/2})/\Delta \gg 1$, 
i.e.\ where many states are kinematically allowed ($t=q^2$). 
This is certainly true when  $t$ is small. 
What may cause problems, according to [\ref{isgur}], 
is only the region near $t_{max}$ where this condition is not satisfied and
large effects can be generated. According to the authors of  [\ref{lettreNR}], 
one can certainly produce effects which violate the equality with free quark 
over the region of phase space {\it where only the ground state is opened},
of relative order $1/m_Q$ if this "relative order " means that one compares 
to the corresponding free quark decay {\it over the same region} of phase 
space.  But they object that such effects be related 
to the {\it total} free quark decay which is much larger.
Indeed, such effects are not of order $1/m_Q$  
with respect to the total free 
quark decay rate, but much smaller, suppressed by powers of 
$2 \Delta / \delta m$~[\ref{lettreNR}]. 
This suppression factor 
amounts, in the standard $1/m_Q$ expansion at fixed ratio of heavy masses,
to further powers of the heavy mass, because then $\delta m \propto m_b$.
Also, numerically, they are small since  $2 \Delta / \delta m$ is small. 
 
The first example given by Isgur is that the decrease of the ground state 
contribution with decreasing $t$ (or increasing $|\vec{q}{\,}|$) due to the 
form factor must be compensated by the increase of the excited states to 
maintain duality with free quarks. This is exactly guaranteed by the Bjorken 
sum rule in the heavy quark limit, but it is no longer exact at finite mass,
because there is a region below the $\rm D^{**}$ threshold where only the ground 
state $\rm D+D^{*}$ contribute.
Quantitatively, the term pointed out in [\ref{isgur}] with a constant 
leptonic interaction reads (the choice of this interaction is not crucial):
\bea
{\delta \Gamma \over \Gamma_{free}} \simeq-\rho^2 
{\int_{(\delta m - \Delta)^2}^{(\delta m)^2} dt |\vec{q}{\,}| 
{|\vec{q}{\,}|^2 \over m_b^2} \over \int_{0}^{(\delta m)^2}dt |\vec{q}{\,}|}
\eea
where $ |\vec{q}{\,}|^2 \simeq (\delta m)^2 -t $,
$-\rho^2 {|\vec{q}{\,}|^2 \over m_b^2}$ describes the falloff of the ground 
state ($\rho^2$ is the slope of the Isgur-Wise function), 
and the integration limits are approximated to the desired accuracy. 
At the lower limit of the numerator integral $t=(\delta m - \Delta)^2$ 
this falloff attains its maximum,
$- \rho^2 {2 \Delta \delta m \over m_b^2}$. This term is 
by itself the expression of a $1/m_Q$ term in the SV limit [\ref{isgur}]. 
However,   the real magnitude is
much smaller because one must integrate over a limited phase space,
while the integral of the free quark decay in the denominator
 extends over a much larger region [\ref{OH}]:
\bea
{\delta \Gamma \over \Gamma_{free}} \simeq -{3 \over 5}\rho^2 
{2 \Delta \delta m  \over m_b^2}
\left({2 \Delta  \over \delta m }\right)^{3/2}=-{3 \over 5} 
{m_d \delta m  \over m_b^2}
\left({2 \Delta  \over \delta m }\right)^{3/2}
\eea
where $\rho^2 \Delta={m_d \over 2}$ in the HO model. 
Parametrically, this is suppressed with respect to $1/m_Q$
because of the factor $({2 \Delta \over \delta m })^{3/2}$. 

In another  example relying on a model of two-body decay, Isgur  
[\ref{isgur}] tries to take into account also the larger effect due to the 
$m_d \delta m \over m_b^2$ terms present in {\it partial rates}. 
Such terms, which corresponds to $1/m_Q$, are present {\it separately}
in the various exclusive channels. For instance one has for the ratio 
of the ground state to the free quark decay rates:
\begin{equation}
R_{sl}^{(ground~state)}=1+{3 \over 2} {m_d \delta m \over m_b^2} +... ,
\end{equation}
but they cancel in the total decay rate. Then, if
the kinematical situation is such that only the ground state is produced, 
the total ratio $R_{sl}$ would depart from $1$ by  
${3 \over 2} m_d \delta m \over m_b^2$. However, 
this effect is for $t$ above the $\rm D^{**}$ threshold
$(M_B-M_{D^{**}})^2$, i.e. in a limited region of phase space. Hence,
taking the ratio of this effect to the total rate, one gets:
\bea
{\delta \Gamma \over \Gamma_{free}} \simeq {3 \over 2} 
{m_d \delta m \over m_b^2  }
{\int_{(\delta m - \Delta)^2}^{(\delta m)^2} dt 
|\vec{q}{\,}| \over \int_{0}^{(\delta m)^2}dt 
|\vec{q}{\,}|} \simeq {3 \over 2} {m_d \delta m \over m_b^2}
 \left({2 \Delta  \over \delta m }\right)^{3/2},
\eea
which is once more parametrically smaller by the factor 
$({2 \Delta \over \delta m })^{3/2}$.

In conclusion, both effects are not ${\cal O} (1/m_Q)$ but much
smaller.  Thus the model dependent investigations of possible duality violations
do not hint at any effect beyond the OPE of full QCD. In particular,
taking into account the sum rules valid in full QCD allows us to
show explicitly the absence of contributions at order $1/m_Q$, which
would be a gross violation of OPE or, likewise, of duality.

\subsubsection{Conclusion}
All  we currently know  from purely theoretical
considerations indicates that  duality violations should be safely below one percent
in the s.l.\  branching ratio. This is likely to remain in the noise
of theoretical uncertainties due 
   to  higher order perturbative and non-perturbative ($O(1/m_b^3) $ and higher) 
corrections.
Hence we will not 
assign  additional uncertainty to the extraction of $\vcb$ from
possible duality violation in inclusive decays.
As discussed above, this picture will be  tested
through an intense program of high precision measurements in the 
near future, and most notably by  the study of different moments of 
the s.l.\ distributions -- even separately in
   the decays of B$_d$, B$^-$ and B$_s$ mesons.

\setcounter{footnote}{0}
\boldmath
\subsection {Review and future prospects for the inclusive determination of $\vcb$}
\label{sec:vcbincl}
\unboldmath

The value of the CKM matrix element $\vcb$ can be obtained by comparing
the measured value of the $b$-quark s.l.\  
decay partial width with its prediction in the context of the OPE.
 Experimentally, this partial width is obtained by
measuring the inclusive s.l.\  decay rate of B-hadrons and their 
lifetime(s). Present measurements are rather accurate and experimental
uncertainties lead to a relative error of about 1$\%$ on $\vcb$.
The main limitation for a precise determination of $\vcb$ comes
from theory, as the
expression for the s.l.\  decay width depends on several poorly known
parameters that are introduced by perturbative and non-perturbative QCD effects.
Only recently, as discussed in Sec.~\ref{sec:moments},
some of the non-perturbative  parameters describing corrections
of order ${\cal O}(1/m_b)$, ${\cal O}(1/m_b^2)$, and 
${\cal O}(1/m_b^3)$ have been constrained
experimentally.
As a result, not only has the accuracy on  $\vcb$ improved, but   
also a large fraction of the previous systematic
uncertainty has changed nature. 

In the following, we briefly summarize the main ingredients 
of the evaluation of $\vcb$ from inclusive $b$ s.l.\  decay 
measurements. 
As discussed in Sec.~\ref{dualitythomas}, a possible violation of parton-hadron duality 
can be legitimately neglected at the present level of accuracy, and we will not include 
it in our estimate of the error associated with  $\vcb$.

\subsubsection{Perturbative QCD corrections}
\label{sec:common}

Using the pole mass definition
for quark masses, the first order QCD perturbative corrections 
to the s.l.\  $b$-decay width have been given in
[\ref{alphas},\ref{ref:niri}] and dominant second order 
(BLM) corrections have been obtained in [\ref{ref:lukesav}]; 
the subdominant two-loop 
corrections have been estimated in [\ref{ref:melni3}]. 
The s.l.\ width can be written as
\begin{equation}
\Gamma(b \rightarrow c \ell \overline{\nu}_{\ell})=
\frac{G_F^2 m_b^5 \vcb^2 {\cal A}_{ew}}{192 \pi^3} 
F ( z )
\left \{ 1 -\frac{\alpha_s(m_b)}{\pi}\frac{2}{3} f(z)
- \frac{\alpha_s^2}{\pi^2} 
\left [ \beta_0 \chi^{\rm BLM}(z) + \chi_0(z)\right ]
\right \}.
\label{eq:lukesav}
\end{equation}
In this expression:

\vspace{2mm}

\begin{itemize}
\item the phase space factor 
$F(z)=1-8z+8z^3-z^4-12z^2\ln{z}$,
with $z=m_c^2/m_b^2$, accounts
for the mass of the final quark, and both $m_c$ and $m_b$ are pole masses;

\item $\beta_0 = 11- \frac{2}{3}n_f$, where $n_f$ is the number of
active flavours;

\item ${\cal A}_{ew}\simeq 1+ 2\frac{\alpha}{\pi} \ln{\frac{m_Z}{m_b}}$ and  
corresponds to the electroweak correction, cfr.~Eq.~(\ref{eq:exclusiveVcb:sirlin}) below;

\item $f(x)=h(x)/F(x) $  with
\vskip -0.6cm
{\small
\begin{eqnarray}
h(x)&=&-(1-x^2)\left(\frac{25}{4}-\frac{239}{3}x+\frac{25}{4}x^2 \right )
+x\ln{x}\left ( 20 +90 x -\frac{4}{3}x^2 +\frac{17}{3}x^3 \right ) \nonumber \\
   & &+x^2\ln^2{x}(36 + x^2)+(1-x^2)
\left (\frac{17}{3}-\frac{64}{3}x+ \frac{17}{3}x^2 \right ) \ln{(1-x)}\nonumber \\
  & & -4(1+30x^2+x^4)\ln{x}\ln{(1-x)}
-(1+16x^2+x^4)\left [ 6 Li_2(x)-\pi^2\right ] \nonumber \\
  & & -32x^{3/2}(1+x)
\left [ \pi^2 -4 Li_2(\sqrt{x})+4 Li_2(-\sqrt{x})-2\ln{x}
\ln{\frac{1-\sqrt{x}}{1+\sqrt{x}}}
\right ] 
\end{eqnarray}
}
Numerical values for $f(x)$ can be found in [\ref{ref:cabi}] 
and are reported in Table~\ref{tab:cabi}.

\vspace{2mm}

\begin{table}[ht!]
  \begin{center}
    \begin{tabular}{|c|c|c|c|c|c|c|c|c|c|c|c|}
      \hline
$\frac{m_c}{m_b}$ & 0. & 0.1 & 0.2 & 0.3 & 0.4 & 0.5 & 0.6 & 0.7 & 0.8& 0.9& 1\\
      \hline
$f(\frac{m_c^2}{m_b^2})$ & 3.62 & 3.25 & 2.84 & 2.50 & 2.23 & 2.01 & 1.83 &
1.70 & 1.59 &1.53 & 1.50  \\
\hline
    \end{tabular}
  \end{center}
    \caption{\it {Values of $f(x)$ for several
values of $m_c/m_b$. 
 }
\label{tab:cabi}}
\end{table}

\vspace{2mm}

\item $ \chi^{\rm BLM}$, corresponding to the BLM 
corrections, is equal to 1.68 for $m_c/m_b=0.3$;

\item $ \chi_0$, corresponding to the non-BLM corrections, is equal
to $-1.4\pm 0.4$ for $m_c/m_b=0.3$.
\end{itemize}

\vspace{2mm}

The convergence of the perturbative series  in
Eq.~(\ref{eq:lukesav})  appears problematic. It has been
demonstrated that this expansion can be much better controlled -- within  a few
$\%$ --  using a properly normalized short-distance mass
[\ref{ref:bigim},\ref{Bigi2}]. This is the case, for instance,
of the kinetic running mass\footnote{Other definitions for quark masses 
can be adopted,
which do not suffer from problems attached to the pole mass definition,
see Sec.~\ref{sec:mb}} defined in 
Eq.~(\ref{kindef}).
Replacing in Eq.~(\ref{eq:lukesav})  the pole quark masses by kinetic running masses
through Eq.~(\ref{kindef}) and expanding in $\alpha_s$, one obtains:
\begin{equation}
\Gamma(b \rightarrow c \ell \overline{\nu}_{\ell})=
\frac{G_F^2 m_b(\mu)^5 \vcb^2{\cal A}_{ew}}{192 \pi^3}
F ( z(\mu) )
\left [ 1 +a_1(\mu)\frac{\alpha_s(m_b)}{\pi}
+a_2(\mu)\left ( \frac{\alpha_s(m_b)}{\pi} \right )^2
\right ],
\label{eq:kin}
\end{equation}
where $z(\mu)=m_c^2(\mu)/m_b^2(\mu)$. A typical value for $\mu$
is 1 GeV. The explicit expressions for $a_{1,2}(\mu)$ can be found in
[\ref{uraltsevvcb},\ref{vcb_new}].

\subsubsection{Non-perturbative QCD corrections}
Non-perturbative corrections in the OPE start at second order in
$1/m_Q$ [\ref{opebigi}]. Including those of 
$O(1/m_b^2)$ [\ref{ope},\ref{opebigi}] and $O(1/m_b^3)$
[\ref{gremm-kap}], and changing the scale at which 
$\alpha_s$ is evaluated to an arbitrary 
value $q$, Eq.~(\ref{eq:kin}) becomes:
{\small
\begin{eqnarray}
\Gamma(b \rightarrow c \ell \overline{\nu}_{\ell})=
\frac{G_F^2 m_b(\mu)^5 \vcb^2{\cal A}_{ew}}{192 \pi^3}& & \!\!\!\!\!
\left [ 1 +b_1(\mu)\frac{\alpha_s(q)}{\pi}
+b_2(\mu,q)\left ( \frac{\alpha_s(q)}{\pi} \right )^2\right]
\left \{F ( z(\mu) ) \left(1-\frac{\mu_{\pi}^2}{2 m_b^2(\mu)}\right ) \right. \nonumber\\
& & \left.-G ( z(\mu) ) \frac{1}{2 m_b^2(\mu)} 
\left ( \mu_G^2-\frac{\rho_{LS}^3}{m_b(\mu)}  \right )
+H( z(\mu) ) \frac{\rho_D^3}{6m_b^3(\mu) }
\right \}
\label{eq:kinnon}
\end{eqnarray}
}where $b_1(\mu) = a_1(\mu)$ and $b_2(\mu,q) =  a_2(\mu) + a_1(\mu)
\frac{\beta_0}{2} \ln{\frac{q}{m_b}}$, and where
we have introduced  the functions ($z=(m_c/m_b)^2$)
\begin{itemize}
\item[] $G(z)=3 -8z+24z^2-24z^3+5z^4+12z^2\ln{z}$,
\item[] $H(z)=77 -88z+24z^2-8z^3+5z^4+12(4+3z^2)\ln{z}$. 
\end{itemize}
A very  recent analysis [\ref{vcb_new}] contains a comprehensive discussion of all the 
aspects of the $\Gamma_{sl}$ calculation and several improvements.
 In particular, it includes BLM corrections to all orders in the scheme
with  running kinetic masses and non-perturbative parameters. The effect of the resummed 
BLM corrections is small, 0.1\% of the s.l.\ width, 
if compared to the perturbative corrections
calculated in Eq.~(\ref{eq:kinnon}) at $q=m_b$.
 Ref.~[\ref{vcb_new}] also 
discusses the role played by four-quark 
operators containing a pair of charm quark fields in the higher orders of the OPE.
These operators give in principle $O(1/m_b^3)$ contributions that are not necessarily 
negligible and require further study.

In the quark pole mass approach, quark masses are usually re-expressed in terms of 
heavy hadron masses, using the HQET relation of Eq.~(\ref{eq:05}):
the corresponding expression for the s.l.\  width can be found 
in [\ref{Cronin-Hennessy:2001fk}] and is quoted below for completeness:
{\small
\begin{eqnarray}
\Gamma(b \rightarrow c \ell \overline{\nu}_{\ell})=
\frac{G_F^2 {\overline M}_B^5 \vcb^2}{192 \pi^3}0.3689 && \hspace{-5mm}
\left [1 -1.54 \frac{\alpha_s}{\pi}-1.43\beta_0 \frac{\alpha_s^2}{\pi^2}
-1.648 \frac{\overline{\Lambda}}{{\overline M}_B}
\left (1 -0.87 \frac{\alpha_s}{\pi} \right ) \right. \nonumber \\
& &-0.946 \frac{\overline{\Lambda}^2}{{\overline M}_B^2}
-3.185 \frac{\lambda_1}{{\overline M}_B^2}+0.02 \frac{\lambda_2}{{\overline M}_B^2}
-0.298 \frac{\overline{\Lambda}^3}{{\overline M}_B^3}
-3.28 \frac{\overline{\Lambda}\lambda_1}{{\overline M}_B^3} \nonumber \\
& & +10.47\frac{\overline{\Lambda}\lambda_2}{{\overline M}_B^3} 
-6.153 \frac{\rho_1}{{\overline M}_B^3}+7.482 \frac{\rho_2}{{\overline M}_B^3}
-7.4 \frac{{\cal T}_1}{{\overline M}_B^3}+1.491  \frac{{\cal T}_2}{{\overline M}_B^3}\nonumber \\
& &\left.-10.41  \frac{{\cal T}_3}{{\overline M}_B^3}-7.482 \frac{{\cal T}_4}{{\overline M}_B^3}
+{\cal O}\left ( \frac{1}{{\overline M}_B^4}\right ) \right ]
\label{eq:cleoexp}
\end{eqnarray}
}In this equation, ${\overline M}_B=\frac{M_B+3M_{B^*}}{4}=5.313$~GeV
and the corresponding value for charmed mesons is taken to be
equal to $1.975$~GeV.
The relations  between the parameters used in the two formalisms
have been recalled in Eq.~(\ref{eq:relations}).
The value of $\mu_{G}^2$ is strongly constrained by  the mass
splitting between $\Bstar$ and B mesons, 
for instance one finds 
 $\mu_{G}^2(1~{\rm GeV})=0.35^{+0.03}_{-0.02}$~GeV$^2$ [\ref{Uraltsev:2001ih}].
For the other non-perturbative parameters one has to rely on theoretical
estimates. Alternately, they can be 
constrained  by measuring other observables:
as explained in Sec.~\ref{sec:moments},
the moments of differential  distributions in $b$-hadron s.l.\  decays and the
moments of the photon energy distribution in $b\rightarrow s \gamma$
decays depend on the same parameters that enter the $\vcb$
determination. Measurements of these quantities can therefore be used to
determine the OPE parameters and to verify the overall consistency 
of the formalism.

\subsubsection{$\vcb$ determination}
The value for $\vcb$ is obtained by comparing the theoretical
and experimental determinations of the inclusive s.l.\  
decay partial width:
\begin{equation}
\Gamma_{sl}|_{th} = {\rm BR}_{sl}|_{exp} \times \tau_b|_{exp}
\end{equation}
In PDG2000 [\ref{ref:pdg00}], the uncertainty attached 
to $\vcb$ was of  $O(5\%)$
and was dominated by the theoretical uncertainty related to the heavy quark 
parameters. Using the analysis of the first hadronic moment and the 
first moment of the photon energy distribution in $b\rightarrow s \gamma$
decays mentioned in Sec.~\ref{sec:moments}, together with  Eq.~(\ref{eq:cleoexp}), 
CLEO has  obtained 
[\ref{Cronin-Hennessy:2001fk}]:
\begin{equation}
\vcb = 40.4 \times (1 \pm 0.022|_{exp} \pm 0.012|_{\overline{\Lambda},\lambda_1} 
\pm 0.020|_{th})\times 10^{-3}
\label{eq:cleovcb}
\end{equation}
The first uncertainty corresponds to the  experimental measurements
of the s.l.\  branching fraction,  of the $\Bd$ and $\Bp$
fractions as obtained by CLEO, and of the $\Bd$ and $\Bp$ lifetimes
given in PDG2000. The second uncertainty corresponds to the errors on 
 $\overline{\Lambda}$ and $\lambda_1$ in the analysis of the moments.
The last uncertainty corresponds to the remaining theoretical  error
coming from contributions of ${\cal O}(1/m_b^3)$
and higher order perturbative corrections, estimated 
from the uncertainty on the scale at which $\alpha_s$
has to be evaluated \footnote{In that analysis the range is taken
to be $[m_b/2,~2m_b]$.}. It appears that the corresponding
variation of $\alpha_s=0.22\pm0.05$ gives the largest contribution
($\pm0.017$).
Remaining contributions to the theory error have been obtained by varying
the values of parameters contributing at ${\cal O}(1/m_b^3)$
within $\pm (0.5)^3$~GeV$^3$, a  rather arbitrary range, based only on 
naive dimensional analysis.

CLEO's  result on $\vcb$ was improved, at the Workshop, mainly by using
all experimental measurements on $b$-hadron s.l.\ 
branching fraction and lifetime [\ref{ref:brandt}].
Recent experimental results, made available at the
ICHEP 2002 Conference in Amsterdam, and obtained by the LEP experiments
[\ref{ref:lepbr}], by 
BaBar [\ref{ref:babarbr}] and by BELLE [\ref{ref:bellebr}]
have been combined [\ref{ref:lepvcb}], including previous measurements
of these quantities given in [\ref{ref:pdg02}]: 
\begin{eqnarray}
\Gamma_{sl}|_{\Upsilon(4S)}(b \rightarrow X_c \ell^- \overline{\nu}_{\ell})&=&
0.431\times(1 \pm 0.019 \pm0.016)\times 10^{-10} ~{\rm MeV}\nonumber\\
\Gamma_{sl}|_{LEP}(b \rightarrow X_c \ell^- \overline{\nu}_{\ell})&=&
0.438\times(1 \pm 0.024 \pm0.015)\times 10^{-10} ~{\rm MeV}\nonumber\\
\Gamma_{sl}|_{Average}(b \rightarrow X_c \ell^- \overline{\nu}_{\ell})&=&
0.434\times(1 \pm 0.018)\times 10^{-10} ~{\rm MeV} 
\label{eq:gammasl}
\end{eqnarray}
In these expressions the second contribution to the errors corresponds
to uncertainties in the decay modelling and in the subtraction of the
$b \rightarrow u \ell^- \overline{\nu}_{\ell}$ component.
Using the above  result, the corresponding
uncertainty in Eq.~(\ref{eq:cleovcb}) can be reduced by
about a factor two. 
Keeping the same values for the two remaining 
uncertainties and correcting for the slightly different central values
of ${\rm BR}_{sl}$ and $\tau_b$, one finds
\begin{equation}
\vcb = 40.7 \times (1 \pm 0.010|_{exp} \pm 0.012|_{\overline{\Lambda},\lambda_1}
 \pm 0.020|_{th})\times 10^{-3}
\label{eq:allvcb}
\end{equation}
This approach was adopted to obtain the value of $\vcb$
quoted in the  corresponding mini-review 
[\ref{ref:marina}] of PDG2002 [\ref{ref:pdg02}]\footnote{In PDG2002, 
the value given in the corresponding mini-review
for $\vcb=(40.4 \pm 0.5 \pm 0.5 \pm 0.8)\times 10^{-3}$ is slightly 
different as it depends on the values of experimental 
results available at that time.}.
However,  the result quoted in the main CKM section of the PDG2002, 
$\vcb=(41.2\pm2.0)\times 10^{-3}$, does not take into account 
this progress, and still assigns a large  uncertainty of $2.0\times 10^{-3}$, 
which is meant to account for possible parton-hadron
duality violation.

As summarized in Sec.~\ref{sec:moments}, 
progress has been achieved  soon after the Workshop  both on theoretical
and experimental aspects of the $\vcb$ determination. On the theoretical
side, the moments of the s.l.\  distributions have been studied using schemes that 
avoid the problems related to  the pole mass 
[\ref{chris},\ref{Bauer:2002sh},\ref{Battaglia:2002tm}]. 
The inclusion of higher order moments
has been reconsidered  
in [\ref{Battaglia:2002tm},\ref{Bauer:2002sh}] and, as we have seen, 
some of the corresponding measurements have been used
in these analyses.
On the experimental side, new measurements of moments have been
presented by BaBar [\ref{babarmom}], CLEO [\ref{marina-lmom},\ref{cleonew}],
and DELPHI~[\ref{delphi_mx}].

The analysis of [\ref{Bauer:2002sh}] employs first, second, and truncated  moments 
to fit the values of the $\overline{\Lambda}$, $\lambda_1$ parameters and obtain 
constraints on ${\cal O}(1/m_b^3)$ contributions. Four different definitions
of the  heavy quark masses have also been considered. A consistent picture
for inclusive $b$-hadron s.l.\  decays is obtained when theoretical  uncertainties are 
taken into account, especially if the BaBar preliminary data [\ref{babarmom}] are excluded.  
Using the average  given in  Eq.~(\ref{eq:gammasl}), the result 
of [\ref{Bauer:2002sh}]
for $\vcb$ in the  1S scheme is
\begin{equation}
\vcb = (41.2 \pm 0.9)\times 10^{-3}.
\label{eq:vcblig}
\end{equation}

In the analysis of Ref.~[\ref{Battaglia:2002tm}], 
which is  based on DELPHI data
and includes third order moments, the
low-scale running mass approach is used to extract $\mu_{\pi}^2$ and
the two parameters contributing at order ${\cal O}(1/m_b^3)$.
Neither the moments nor $\vcb$ are actually sensitive to
$\rho_{LS}^3$. The other parameter appearing at this order, 
$\rho_D^3=(0.05\pm0.05)~\rm GeV^3$,
is found to be in good agreement with some theoretical expectation 
(about 0.1~GeV$^3$ [\ref{Uraltsev:2001ih}]). 
In the low-scale running mass scheme the uncertainty related 
to the scale  at which $\alpha_s$
is computed has also been reduced  with respect to the pole mass analysis. 
Employing the average  s.l.\  width given in  Eq.~(\ref{eq:gammasl}), the result of
Ref.~[\ref{Battaglia:2002tm}] is
\begin{equation}
\vcb = 41.7 \times (1 \pm 0.010|_{exp} \pm 0.015|_{m_b,m_c,\mu_{\pi}^2,\mu_G^2,\rho_D^3,
\rho_{LS}^3}
 \pm 0.010|_{pert~QCD} \pm 0.010|_{th})\times 10^{-3}.
\label{eq:vcbnoi}
\end{equation}
The last two uncertainties in this equation are theoretical and 
correspond to the scale ambiguity
for $\alpha_s$ and to possible contributions from 
${\cal O}(1/m_b^4)$ terms for which an upper limit corresponding to the 
contribution of the previous order term has been used.
The above estimate of the overall theoretical error agrees well 
with that of [\ref{vcb_new}].

All the results presented in this Section are  preliminary.
They indicate  a  promising future for the approach where
all non-perturbative parameters, up to order ${\cal O}(1/m_b^3)$,
are experimentally constrained. 
Only the  preliminary BaBar analysis [\ref{babarmom}] 
does not seem to fit the picture:
it seems difficult to reconcile the dependence of BaBar first hadronic moment on the 
lepton momentum cut  with the 
other measurements in the context of the OPE. 
Although a high lepton momentum cut could in principle spoil the convergence
of the power expansion, this point definitely needs to be fully understood.

\subsubsection{Prospects}
Impressive improvements have been obtained in the determination of 
$\vcb$ from inclusive $b$-hadron s.l.\  decay measurements
during and just after this Workshop.
The moments in inclusive s.l. and radiative decays have been studied in
new theoretical  frameworks. 
Preliminary  analyses of recent experimental measurements of such moments
indicate that all parameters contributing
to ${\cal O}(1/m_b^3)$ included can be constrained  by experiment.
The results for $\vcb$ in Eqs.~(\ref{eq:vcblig}) and (\ref{eq:vcbnoi})
are  very similar. We can adopt a central value given by their average
with a $2.3\%$ accuracy:
\begin{equation}
\begin{array}{|c|}\hline\left.
\vcb = 41.4\cdot \left(1 \pm 0.018|_{exp} \pm 0.014|_{th}\right)\times 10^{-3} \ .\right.
\\ \hline\end{array}
\end{equation}
in which the largest fraction of the uncertainty depends on
experimental measurements.
These analyses have to be confirmed, as most of them
correspond to preliminary results, and the possible discrepancy raised
by BaBar data has to be investigated,  especially with respect to the 
impact of the lepton energy cut, by lowering the cut as much as possible.
If the present picture remains valid, more effort has to be invested in the 
control of remaining theoretical errors, namely $i)$
the uncertainty related to the truncation of the perturbative QCD 
expansion and $ii)$ 
the importance of four-quark operators containing the charm quark and of
${\cal O}(1/m_b^n),~n\geq4$ corrections.

\subsection{Review and future prospects for the inclusive determination of $\vub$}
\label{sec:vubincl}

The charmless s.l.\  decay channel $ b \rightarrow u \ell \bar \nu$
can in principle provide a clean determination of $\vub$ along the lines of that of $\vcb$.
 The main problem is the large
background from $b\to c\ell \bar{\nu}$ decay, which has a rate about 
60 times higher than that for the charmless s.l.\ decay. The experimental cuts necessary
to distinguish  the $ b \rightarrow u$  from the 
$b\to c$ transitions enhance the sensitivity 
to the non-perturbative aspects of the decay, 
like the Fermi motion of the $b$ quark inside the B meson,
and complicate the theoretical interpretation of the measurement.

The inclusive decay rate ${\rm B} \to X_u \ell \bar \nu$ 
is calculated using the OPE. 
At leading order, the decay rate is given by the parton model decay rate. 
As we have seen, non-perturbative corrections are suppressed by at least two powers of 
$1/m_b$ and to $O(1/m_b^2)$ they are parameterized by the two universal matrix elements 
$\mu_{\pi}^2$ and $\mu_G^2$ (or $\lambda_1$ and $\lambda_2$), see  
Sec.~\ref{sec:OPE} 
At $O(1/m_b^3)$, the Darwin term $\rho_D^3$ reduces $ \Gamma ({\rm B}\to l \nu X_u)$ by
 1 - 2 \%. 
Perturbative corrections are known through order $\alpha_s^2$ 
[\ref{vanRitbergen1}]. 
All this allows to relate the total inclusive decay rate directly to $\vub$ 
[\ref{uraltsevvub}] 
\begin{equation} \label{vubapprox}
|V_{ub}| = 0.0040 \times \left( 
1 \pm 0.03|_{m_b} \pm 0.025 |_{QCD}
\right) 
\left( 
\frac{{\rm BR}({\rm B} \to l \nu X_u)}{0.0016}
\right)^{\frac{1}{2}} 
\left( 
\frac{1.55 \; {\rm ps}}{\tau_B}
\right)^{\frac{1}{2}}
\end{equation}
where in the second error both perturbative and
non-perturbative uncertainties are included. 
The uncertainty due to the $b$ mass assumes
$\delta m_b\sim 60$ MeV and can easily be rescaled to accommodate the error in 
  Eq.~(\ref{mbworkshop}). 
The errors reported in Eq.~(\ref{vubapprox}) are very similar to those 
given in [\ref{Hoang2}]. In fact, despite the use of slightly different inputs, and of different 
formalisms, the results of Refs.~[\ref{uraltsevvub}] and [\ref{Hoang2}]
are remarkably consistent, 
their central values differing by only   1.7\%.
Information from  the moments of the s.l.\ distributions is unlikely to 
decrease significantly the overall uncertainty in Eq.~(\ref{vubapprox}).

The large background from ${\rm B} \to X_c \ell \bar \nu$  makes the direct 
measurement of the inclusive rate a very challenging task. In principle, 
there are several methods to suppress this background and all of them
restrict the phase space region where the decay rate is measured. 
Hence, great care must be taken to ensure that the OPE
is valid in the relevant phase space region. 

There are three main kinematical cuts which separate the $ b \to u \ell\bar \nu$ 
signal from the $ b \to c \ell\bar \nu$ background:
\begin{enumerate}
\item A cut on the lepton energy $E_\ell > (M_B^2-M_D^2)/2M_B$ [\ref{Elcut}]
\item A cut on the hadronic invariant mass $M_X < M_D$ [\ref{mXcut}]
\item A cut on the leptonic invariant mass $q^2 > M_B^2-M_D^2$ [\ref{qsqcut}]
\end{enumerate}
These cuts correspond to about 10\%, 80\% and 20\% respectively 
of the signal selected.
\begin{figure}[t]
\includegraphics[width=16cm]{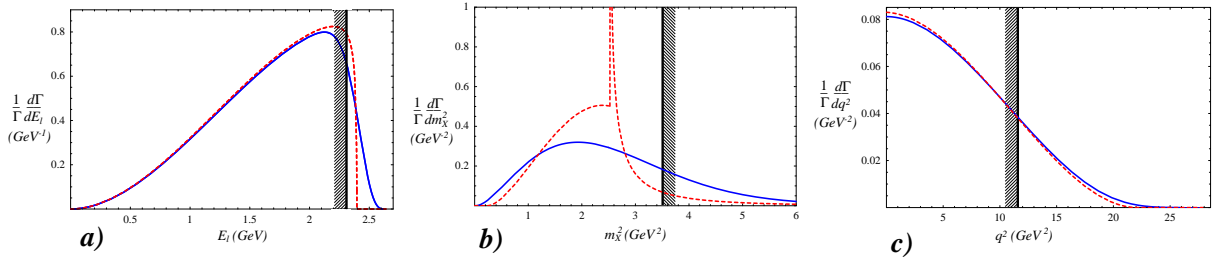}
\caption{\it The distribution of the three main discriminating variables 
in inclusive
${\rm B} \to X_u \ell \bar\nu$ analyses: lepton energy $E_{\ell}$ (left), 
hadronic invariant 
mass $M_X^2$ (center) and di-lepton invariant mass $q^2$ (right), as given by $O(\alpha_s)$
parton level decay (dashed curves), and including the Fermi motion model (solid curves) with 
typical parameters. The vertical line marks the cut necessary to eliminate the $b\to c$ 
transitions in each case.
}
\label{figv}
\end{figure}
The simplest kinematical discriminator for $ b\to u$ versus 
$b\to c $ is the endpoint in the $\ell$ inclusive spectrum,
where the first evidence for $\vub$ $\neq 0$ was seen [\ref{Elcut}]. 
However, in this case the remaining phase space is characterized by 
$\Delta E_\ell = M_D^2/2M_B = 320~{\rm MeV} \sim \Lambda_{QCD}$. 
Because of this cut on the lepton energy, the selected hadronic system has 
large energy and small invariant mass, and is placed in a kinematic 
region where the OPE is not expected to converge.
Measurements done using this method have given [\ref{Elcut}]
 $\vub / \vcb$ = $(0.08 \pm 0.02)$,
where the 25\% error is dominated by the theoretical uncertainty.

In the original analyses [\ref{Elcut}] several 
 models [\ref{accmm},\ref{isgw}] were used to estimate the 
rate at the endpoint.
In fact, the exact fraction of signal decays selected depends strongly on 
a {\sl shape function}, as illustrated in 
Fig.~\ref{figv}, where the spectrum in the parton model is 
compared to the one including the structure function. Physically,
the  {\sl shape or structure function} (sometimes also called light-cone distribution function) 
encodes the Fermi motion of the $b$ 
quark inside the B meson, which is inherently non-perturbative.
To estimate the effect of the structure function on 
the rate measured in the endpoint region, several models for the shape function have 
been constructed. They are constrained by the values of the first 
few moments of the shape function, which are related to physical 
quantities like $m_b$
and $\mu_\pi^2\sim -\lambda_1$~[\ref{shape1}--\ref{shape2}].
%
The model dependence of the measurement can be reduced by noting 
that the shape function, like the Fermi motion inside the meson, 
is a universal property of the B meson, 
independent of the decay process. 
Consequently, the shape function can in principle be extracted from a 
different 
$ \rm heavy \to light$ 
process and then employed in the inclusive ${\rm B} \to X_u \ell \bar\nu$ 
decay [\ref{shapebigi}--\ref{Bigi:2002qq}]. 
%
The best way is to use the ${\rm B} \to X_s \gamma$ decay. 
At leading order in $1/m_b$ and $\alpha_s$, the photon spectrum 
in the radiative decay
is proportional to the light cone distribution function. 
This strategy for determining $\vub$ has  three main drawbacks:
\begin{itemize}

\item The $b$ quark distribution function is the same in ${\rm B}\to l \nu X_u$ 
and ${\rm B} \to \gamma X_s$ only at leading order in $1/m_b$ and in $\as$; 
perturbative QCD corrections complicate
its extraction [\ref{aglietti},\ref{iraEl}]; 

\item There are process specific corrections of order $1/m_b$ which 
still need to be evaluated reliably. 
In Ref.~[\ref{BLM1}] it is argued that these corrections could be quite 
sizeable. Even after a precise measurement of the photon spectrum 
there are unknown and not-calculable contributions 
$\sim {\cal O}(1/m_b)$ in ${\rm B} \to l \nu X_u$ which could spoil the accurate 
extraction of $\vub$. 
It has  also been pointed out [\ref{Voloshin2001}] that there are
contributions of dimension six operators, suppressed by $1/m_b^3$, but
enhanced by a phase space factor of $16\pi^2$.  They arise from so
called {\it weak annihilation} (WA) contributions, and their total contribution
survives any cut used to reject the $ b \to c \ell \nu$
background. The size of WA contributions is hard to estimate,
as very little is known about the values of the relevant four-quark
operator matrix elements. They could in principle be constrained by a comparison of
B$^0$ and B$^\pm$ decay rates. While their impact on the integrated width
is modest ($\lsim 2 \%$), in the endpoint region WA terms could give
effects of up to 20\% [\ref{Lig_Leib_Wise}]. This conclusion, however,
is challenged in [\ref{NEUBERTVUB}], according to which the 
uncertainty induced by subleading shape functions is safely below 10\%, for lepton energy cuts
$E_\ell\le 2.2$~GeV. See also [\ref{Bigi:2002qq}].

\item Finally, the endpoint region represents such a narrow slice of the phase 
space that 
may be vulnerable to violations of {\em local} parton-hadron duality.  
\end{itemize}

The first analysis combining ${\rm B}\to l \nu X_u$ and ${\rm B} \to \gamma X_s$ 
was performed by CLEO [\ref{cleovub}]. 
To account for the distortion of the endpoint spectrum due to the motion of the 
B mesons, the initial state 
radiation and the experimental resolution, CLEO fit for the observed data 
using a theoretical momentum spectrum to model these distortions.
 They find
$$
\rm 
  |V_{ub}|= (4.12 \pm 0.34 \pm 0.44 \pm 0.23 \pm 0.24) \times 10^{-3}
$$
in the lepton momentum range 2.2--2.6 GeV$/c$.
Here the first error combines statistical and experimental 
uncertainty on the measured rate, 
the second error is the uncertainty on the fraction of leptons 
within the acceptance, derived from the uncertainty in 
the $ b\to s\gamma$ shape function, the third error is the theoretical 
uncertainty on the extraction of  $\vub$  from the total rate, 
the fourth error is an estimate of the 
uncertainty that results from the unknown power corrections 
in applying the $ b\to s\gamma$ shape function to $ b\to u\ell\nu$.  
To evaluate this last uncertainty, the parameters of the shape 
function are varied by the expected order of the corrections: 
$\Lambda_{QCD}/M_B \approx 10\%$. Clearly, this sets only the 
{\it scale} of that uncertainty.

In principle, the  hadronic recoil mass provides the single most 
efficient kinematical discriminator against the $ b\to c \ell\nu$ background. 
The $ b\to c \ell\nu$ background is separated from the signal
imposing  $M_X < M_D$. After this cut, more than 80\% of the signal survives.
However, due to the experimental resolution, 
the $b\to c \ell\nu$ transitions contaminate the $M_X < M_D$ region,
and therefore either the cut is lowered, or a different strategy has to be employed.
When the cut on the  hadronic recoil mass is used, the main theoretical issue
arises from the knowledge of the 
fraction of $ b\to u \ell\nu$ events with $M_X$ 
below a given cut-off mass, $M_{\rm cut}$: 
$$
\Phi _{SL}(M_{\rm cut}) \equiv \frac{1}{\Gamma (B\to l \nu X_u)} 
\int _0^{M_{\rm cut}} dM_X \frac{d\Gamma}{dM_X}
$$  
where $\Phi (0) =0$ and $\Phi (M_B) = 1$.  
The $M_X$ spectrum is in fact sensitive to the values of the HQE parameters 
$m_b$, $\mu _{\pi}^2$, etc. It also depends on the heavy quark 
shape function,
although the dependence is weaker than for the lepton energy in the endpoint region. 
To set the scale of the problem:  a very rough estimate for $\Phi _{SL} (1.7\; {\rm GeV})$ 
lies between 0.55 and 0.9; i.e.\ a measurement of 
$\Phi _{SL} (1.7\; {\rm GeV})$ yields 
a value for $\vub$ with {\it at most} a $\pm 12\%$ uncertainty, and possibly less.
The actual uncertainty in realistic experimental analyses  has been 
estimated by the experimental collaborations.   
Since a cut on the hadronic invariant mass allows for a much larger 
portion of the decay rate to survive, the uncertainties 
from weak annihilation contributions  are safely below
the 5\% level. The subleading shape functions contributions 
can in this case be analysed using the same method as in
[\ref{BLM1}]. A preliminary discussion can be found in  [\ref{Bigi:2002qq}].

The above observations motivated 
an intense effort to measure $\vub$ using inclusive analyses at LEP, where 
B hadrons are produced with a large and variable momentum and 
in most of the cases the B decay products are contained into narrow jets
in $Z^0 \rightarrow b \bar b$ events. These characteristics
make the LEP measurements complementary to the ones at the 
$\Upsilon(4S)$. All four LEP experiments have provided a measurement of $\vub$ 
using inclusive methods, although the actual procedures differ significantly.

DELPHI [\ref{delphivub}] perform an inclusive reconstruction of the hadronic mass of the 
system emitted together with the lepton in the B hadron decay. 
The B s.l.\  sample is 
split into $ b\to u \ell \nu$
enriched and depleted samples based on the separation 
between tertiary and secondary vertices (taking advantage of the finite charm 
lifetime) and on the presence of tagged kaons in the final state.  The mass of the
 hadronic system $M_X$ is used to subdivide further the sample into a 
$ b\to X_u\ell\nu$--favoured region 
($M_X<1.6$ GeV) and a $ b\to X_c\ell\nu$--dominated region. 
The signal is extracted from 
a simultaneous fit to the number of decays classified according to the four different 
categories and the distributions of the lepton energy in the reconstructed 
B rest frame.

The leptonic invariant mass, $q^2 = (p_\ell +p_\nu)^2$, can also suppress 
the $ b \to c$ background [\ref{qsqcut}].
This cut allows to measure $\vub$ without requiring knowledge of the
structure function of the B meson (see Fig.~\ref{figv}c). 
The acceptance of this cut on $q^2$ can be 
calculated using the usual local OPE. Depending on the value of 
the cut, the fraction of selected signal events can range between 
10 and 20\%, but  the theoretical uncertainty on $|V_{ub}|$, dominated by higher 
order power corrections,  can range from 15\% for 
$q^2_{\rm cut} = M_B^2-M_D^2 = 11.6 \,{\rm GeV}^2$ to 25\% for 
$q^2_{\rm cut} = 14 \,{\rm GeV}^2$ (see also [\ref{neubertq2}]). The $q^2$ method allows to 
measure $\vub$, albeit with  larger uncertainties than 
when one combines the lepton energy or the 
hadron invariant mass cut with data from ${\rm B} \to X_s \gamma$ decay. 

Recently, a strategy relying on the combination of $q^2$ and $M_X$ 
cuts has been proposed [\ref{doublecut}]. The $M_X$ cut is used to reject the charm background, 
while the $q^2$ cut 
is used to eliminate the high energy, low invariant mass region. 
Rejecting the region 
at small $q^2$ reduces the impact of the shape function in the 
$M_X$ analysis.  
Strong interaction effects on $M_X$ are maximal there due to the significant 
recoil [\ref{doublecut}].  Imposing, for instance, 
$q^2 \geq 0.35 m_b^2$ eliminates the impact of the primordial Fermi motion 
encoded in $M_X <1.7$ GeV events. 
Up to 50\% of all ${\rm B} \to X_u \ell \bar\nu$ events survive this cut, 
making possible to measure $\vub$ with uncertainties safely below the 10\% level.

CLEO has presented the first experimental 
attempt to implement this method [\ref{cleovubmixed}]. 
The analysis is based on a full fit to 
$q^2/(E_{\ell}+E_{\nu})^2$, $M_X$ and $\cos \theta_{W\ell}$. 
Models are needed to extract
the sample composition and to relate the regions of higher sensitivity 
and theoretically
safer to the inclusive charmless s.l. branching fraction. 
However, imposing these additional cuts has drawbacks.
The overall energy scale governing the intrinsic hardness of the 
reaction gets smaller since it is driven at large $q^2$ by 
$m_b - \sqrt{q^2}$ rather than $m_b$. This enhances the impact 
of higher-order contributions which are not calculated, like in the case of the 
direct cut on $q^2$. 
Furthermore, cutting simultaneously on $M_X$ and $q^2$ decreases the 
fraction of the full width retained in the sample, and exposes the
calculation to violations of duality.  Finally, the cut on $q^2$
removes the possibility to 
incorporate in full the constraints on the spectrum which follow from the
properties of the shape function, because 
it dissolves the connection between the $M_X$ spectrum and the shape function.

\begin{figure}[t]
\begin{center}
\includegraphics[width=11cm]{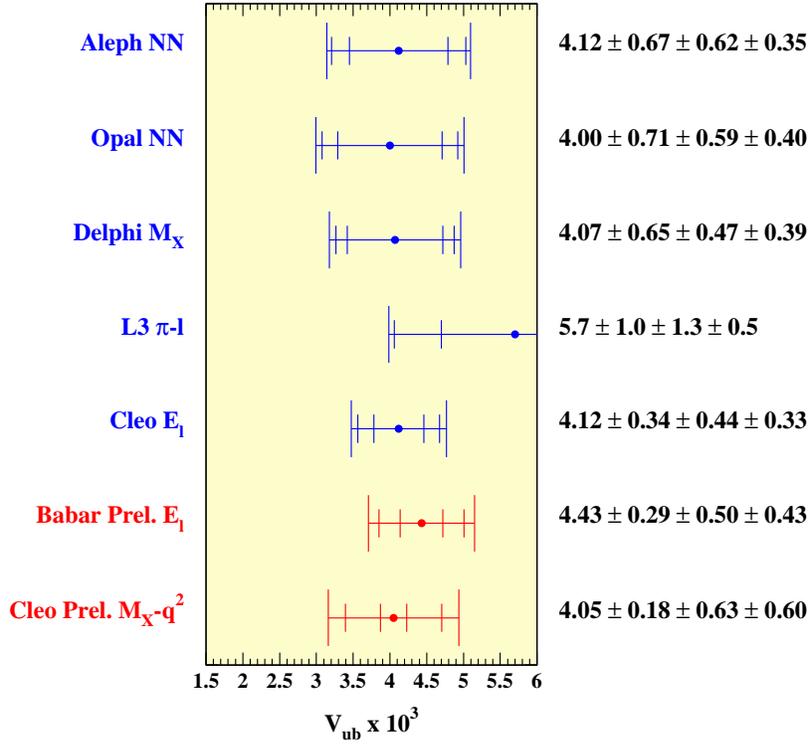}
\end{center}
\vspace{-1cm}
\caption{\it Summary of the inclusive determinations of $\vub$ .}
\label{fig:vub_summary}
\end{figure}

Yet another  approach has been followed  by the ALEPH and OPAL collaborations
in their analyses~[\ref{alephopalvub}]. 
They  use neural networks, which  input a large number
 of kinematic variables (20 in ALEPH, 7 in OPAL)
to  discriminate between 
the $ b\to c\ell\nu$ and the $ b\to u\ell\nu$ decays. 
In both experiments, the signal is extracted from 
a fit to the network output, restricted to a region enriched 
in signal events.
The observation of s.l.\  $b\to u \ell \nu$ decays 
at LEP has been very challenging. 
These analyses pioneered new approaches for extracting $\vub$ . 
Their main drawback is the S/B ratio, that requires the control of the 
background level to better than 5\% . 
Concerns,   discussed  within the 
community, include the modelling of the uncertainties on 
the non-$\rm D$ and $\rm D^*$  components of 
the background from B decays, the modelling of the $\rm B_s$ and $b$-baryon 
s.l.\  decays and the estimation of the $ b \to u \ell \nu$ 
modelling uncertainties due to the uneven sampling
of the decay phase space.

Since tight selections are needed to extract the signal, the effects of 
these experimental cuts trimming the inclusive distributions must 
be understood. In particular, it is important to make sure the
the inclusive analyses are probing the selected phase space in an even and uniform way. 
Neural network analyses bias the phase space toward the region of large
$E_{\ell}$ and low $M_X$, where the signal-to-background ratio is larger. 
The uncertainty quoted by ALEPH accounts for the range of models tested.
In this case, it would be desirable to test more unbiased methods. 
DELPHI, on the other hand,  has shown that the $M_X$ analysis has a 
reasonably uniform sensitivity in the $M_X$-$E_{\ell}$ plane
and a recent CLEO analysis, repeated 
for different  sets of $M_X$-$q^2$ selections, finds 
results compatible with LEP.

Finally, L3 applies a sequential cut analysis using the kinematics of the lepton
and of the leading hadron in the same jet for discrimination of the
signal events [\ref{l3vub}]. The uncertainty (see Fig.~\ref{fig:vub_summary})
is larger than in other analyses, mainly because 
the result depends on  a few exclusive final states only.

All the  analyses  discussed in this Section 
have an individual accuracy of about 15\% and, as can be seen in Fig.~\ref{fig:vub_summary},
 their central values agree within that uncertainty.
One can distinguish 
 two sets of inclusive determinations of $|V_{ub}|$ which rely 
on roughly the same theoretical assumptions and are extracted within the same 
OPE framework. The LEP inclusive results have been averaged accounting for 
correlated systematics. The uncertainty of the CLEO determination from the
lepton end-point and the $b \to s\gamma$ spectrum can be re-expressed in a
way corresponding to that used for the LEP averaging. The results read 

\vspace{2mm}

\bea
\begin{array}{|l|}
\hline
|V_{ub}|_{LEP}^{incl} = [4.09 ^{+0.36}_{-0.39}
\,^{+0.42}_{-0.47} 
\,^{+0.24}_{-0.26} 
\pm 0.21 
]\times 10^{-3},\\ 
|V_{ub}|_{CLEO}^{incl} =[4.08 \pm 0.44 
\pm 0.27 
\pm 0.33 
\pm 0.21 
]\times 10^{-3},\\
\hline
\end{array}
\eea

\vspace{2mm}

\noindent
where the first error corresponds to statistical and experimental systematics,
the second to the dominant $b\to c$ background, the third  to $b\to u$ modelling,
and the last one to the relation between $\vub$ and  the branching fraction, 
see Eq.~(\ref{vubapprox}). 
A first exercise aimed at understanding the relationship between 
the different sources of systematics in these determinations and
to obtain a global average was started at the workshop. A conservative
approach consists in taking the systematic uncertainties as fully 
correlated. This combined result has a total 
 uncertainty  of $\pm 14\%$ and is used in Chapter 5 (see Table \ref{tab:inputs}).
However, the uncertainties are only partly correlated and more 
precise measurements are becoming available: once the systematics and
their correlation are better understood there is room for considerable improvement.

As the B factories start focusing on the inclusive measurements of $\vub$, 
there is potential for considerable progress.  
A more precise evaluation of the $ b\to s\gamma$ photon spectrum will lead to a more 
precise effective shape function and we now have several methods to
employ it efficiently in the extraction of $\vub$.
A recent proposal [\ref{aglietti2}], for instance,  uses the  s.l.\ differential 
distribution in $M_X/E_X$ together with the  $ b\to s\gamma$ photon spectrum to build a short distance ratio 
from which $\vub/\vcb$ can be extracted, testing at the same time 
some of the underlying assumptions.  The use of event samples with one 
fully reconstructed B will 
reduce the contamination  from $ b\to c\ell \bar{\nu}$ decays
in the  reconstruction of the hadronic recoil
mass and of  $q^2$ and will allow for useful cross-checks [\ref{menke}].
Hence, experimental uncertainties should be reduced.  
If the various methods will give consistent central values 
while their precision improves, we will be confident 
that theoretical 
uncertainties are not biasing  $\vub$  beyond the level of precision which has 
been  reached in the individual measurements.

\boldmath
\section{Exclusive determination of $\vcb$}
\label{sec:vcbexc}
\unboldmath


As we have seen in the previous section, inclusive $b\to c$
semileptonic (s.l.) decay rates have a solid description via the OPE.
Exclusive s.l.\  decays have a similarly solid description in
terms of heavy-quark effective theory (HQET).  The main difference is
that the non-perturbative unknowns in the inclusive rates can be
determined from experimental measurements, while those arising in
exclusive rates must be calculated.  Thus, there is a  major theoretical challenge here
as  non-perturbative QCD calculations have to performed.  
Experimentally, the D and D$^*$ mesons have to be 
reconstructed using several decay channels, to gain in statistics, and the 
signal has to be isolated from higher excited states. 
Moreover, the theory is under best control at the kinematic endpoint,
where the rate vanishes.
Consequently, not only must the differential decay rate be measured, it 
also must be extrapolated to the endpoint.
Despite these experimental difficulties, and given the ongoing progress in
lattice QCD, these channels provide a valuable cross-check at present
and hold considerable promise for the future.

The exclusive determination of \vcb\ is obtained by studying 
$\btodslnu$ and $\btodlnu$ decays, where $\ell$ stands for either 
$e$ or~$\mu$. 
The differential rates for these decays are given by 
\begin{eqnarray}
        \frac{d\Gamma(\btodslnu)}{dw} & = &
                \frac{G_\mu^2|V_{cb}|^2}{48\pi^3} \eta_{\rm EW}^2 \,
                (M_B-M_{D^*})^2\,M_{D^*}^3 (w^2-1)^{1/2} (w+1)^2 \nonumber\\
   &&\times \bigg[ 1 + \frac{4w}{w+1}\,
    \frac{1-2w\,r_* + r_*^2}{(1 - r_*)^2}
    \bigg]\,|\cF(w)|^2 \,, 
\label{eq:exclusiveVcb:dGdw*} \\[0.4cm]
   \frac{d\Gamma(\btodlnu)}{dw}
   &=& \frac{G_\mu^2|V_{cb}|^2}{48\pi^3} \eta_{\rm EW}^2\,(M_B+M_D)^2\,M_D^3
    (w^2-1)^{3/2} |\cG(w)|^2 \,,
\label{eq:exclusiveVcb:dGdw}
\end{eqnarray}
where $w=v_B\cdot v_{D^{(*)}}$ is the product of the velocities of the 
initial and final mesons, and  $r_*=M_{D^*}/M_B$. 
The velocity transfer is related to the momentum~$q$ transferred to the
leptons by $q^2=M_B^2 - 2wM_BM_{D^{(*)}} + M_{D^{(*)}}^2$, 
and it lies in the range 
$1\leq w < (M_B^2+M_{D^{(*)}}^2)/2M_B M_{D^{(*)}}$.
Electroweak radiative corrections introduce the muon decay constant
$G_\mu=1.1664\times10^{-5}~\textrm{GeV}^{-2}$ (instead of $G_F$) and
the factor~$\eta_{\rm EW}^2$ (see \sec{sec:ewcorr}).

In the heavy-quark limit, the form factors $\cF(w)$ and $\cG(w)$ coincide
with the Isgur-Wise function $\xi(w)$, which describes the
long-distance physics associated with the light degrees of freedom in
the heavy mesons~[\ref{Isgur:vq},\ref{Isgur:ed}]. 
This function is normalized to unity at zero recoil,
corresponding to $w=1$. 
There are corrections to heavy-quark limit from short distances, which 
can be calculated in perturbation theory in $\alpha_s(\sqrt{m_cm_b})$.
There are also corrections from long distances, which are suppressed by
powers of the heavy quark masses.
The separation of the two sets of contributions can be achieved with
HQET, which is reviewed, for example, 
in [\ref{Neubert:1993mb},\ref{Neubert:2000hd}].
The calculation of the small corrections to this limit is explained 
below in Secs.~\ref{sec:exclusiveVcb:F(1)}
and~\ref{sec:exclusiveVcb:G(1)}
With a satisfactory calculation of these corrections, \vcb\ can be
determined accurately by extrapolating the differential decay rates
to $w=1$, yielding $\vcb \cF(1)$ and $\vcb \cG(1)$.  
Uncertainties associated with this extrapolation can be reduced using
model-independent constraints on the shape of the form factors,
derived with dispersive methods. 
These techniques are briefly reviewed in
\sec{sec:exclusiveVcb:extrap_in_w}

At present $\btodslnu$ transitions yield a more precise value of \vcb\ 
than $\btodlnu$.
The statistics are three times higher.
More importantly, phase space suppresses $\btodslnu$ by only $(w-1)^{1/2}$, 
but $\btodlnu$ by $(w-1)^{3/2}$.
Finally, the theoretical calculation of $\cF(1)$ is under better 
control than that of~$\cG(1)$.
Nevertheless, $\btodlnu$ provides a useful check.
For example, \vcb\ drops out of the (experimental) ratio 
$\vcb \cF(1)/\vcb\cG(1)$, which can be used to test the theoretical 
calculations.

\boldmath 
\subsection{Theory-guided extrapolation in $w$}
\label{sec:exclusiveVcb:extrap_in_w}
\unboldmath 
Dispersive methods allow the derivation of rigorous, model-independent
constraints on the form factors in exclusive s.l.\  or radiative 
decays.
The derivation is based on first principles: the analyticity
properties of two-point functions of local current operators and the
positivity of the corresponding hadronic spectral
functions. Analyticity relates integrals of these spectral functions
to the behaviour of the two-point functions in the deep Euclidean
region, where they can be calculated using the operator product
expansion. Positivity guarantees that the contributions of the states
of interest to these spectral functions are bounded from above. Constraints
on the relevant form factors are then derived, given the latter's
analyticity properties.  
The beauty of these techniques is that the bounds can be improved with 
information about the form factors, such as their value or derivatives 
at different kinematic points, or their phase along various cuts. 
These techniques also have the advantage that the
constraints they yield are optimal for any given input.

Here we focus on the application of these methods to $\btodsnslnu$
decays. The first such application was carried out in [\ref{Boyd:1995sq}],
where three-parameter descriptions of the corresponding differential
decay rates were presented.  In [\ref{Boyd:1995tg}], it was shown how a
judicious change of variables can be used to reduce the number of
parameters.  
The most recent analyses [\ref{Boyd:1997kz},\ref{Caprini:1997mu}] take two-loop
and non-perturbative corrections to the relevant two-point correlators
into account and make use of heavy-quark spin symmetry in the ground-state 
doublets $(\rm B,B^*)$  and $(\rm D,D^*)$ . 
Ref.~[\ref{Caprini:1997mu}] uses spin symmetry more extensively,
and accounts for the dominant $1/m_Q$ and radiative corrections.
The results are one-parameter descriptions of the form factors $\cG(w)$
and $A_1(w)=\cF(w)/\cK(w)$, with $\cK(w)$ defined below in \eq{eq:kw}, 
that are accurate to better than 2\% over the full kinematic range.

In the case of $\btodslnu$ transitions, it is convenient to constrain
the form factor $A_1(w)$ instead of $\cF(w)$ in order to
avoid large, kinematically enhanced corrections to the heavy-quark
limit. This yields for $\cF(w)$ [\ref{Caprini:1997mu}]:
\begin{equation}
   \frac{\cF(w)}{\cF(1)} \approx \cK(w) \l\{1 - 8\rhoAone^2 z
    + (53.\rhoAone^2 - 15.) z^2 - (231.\rhoAone^2-91.) z^3\r\} \,,
\label{A13rd}
\end{equation}
with $z$ given in \eq{eq:zdef} and where the only parameter, the slope
parameter $\rhoAone^2$ of $A_1(w)$ at zero recoil, is constrained by
the dispersive bounds to lie in the interval
$-0.14<\rhoAone^2<1.54$. This constraint on $\rhoAone^2$ is somewhat
weaker than the one derived from the inclusive heavy-quark sum rules of
Bjorken [\ref{bjorken}] and Voloshin [\ref{voloshin}] which require $0.4
\le \rhoAone^2 \le 1.3$ once $O(\alpha_s)$ corrections have been
included [\ref{alphasihqsr}]. A stronger lower bound has been derived
by Uraltsev [\ref{Uraltsev:2000ce}]. This 
is to be compared with the world experimental average 
$\rhoAone^2=1.50\pm 0.13$ given in \sec{sec:vcbfrombtodst}

In \eq{A13rd}, 
the function $\cK(w)$ is
\begin{equation}
        \cK(w)^2 = \frac{\displaystyle 2\,\frac{1-2w r_*+r_*^2}{(1-r_*)^2} 
                \left[1 + \frac{w-1}{w+1}\,R_1(w)^2 \right] + 
                \left[1 + \frac{w-1}{1-r_*}\Big( 1 - R_2(w)\Big)\right]^2}%
{\displaystyle 
                        1 + \frac{4w}{w+1}\,\frac{1-2w r_*+r_*^2}{(1-r_*)^2} },
\label{eq:kw}
\end{equation}
where $r_*$ is given after \eq{eq:exclusiveVcb:dGdw}, and $R_1(w)$ and
$R_2(w)$ describe corrections to the heavy-quark limit.
They are usually expanded in Taylor series around $w=1$.
Using  QCD sum 
rules~[\ref{Neubert:1992wq},\ref{Neubert:1992pn},\ref{Ligeti:1993hw}] 
one finds~[\ref{Caprini:1997mu}]
\begin{eqnarray}
        R_1(w) &\approx& 1.27 - 0.12(w-1) + 0.05(w-1)^2 \,, \nonumber\\
        R_2(w) &\approx& 0.80 + 0.11(w-1) - 0.06(w-1)^2 \,.
        \label{eq:exclusiveVcb:extrap:R2}
\end{eqnarray}
The sum-rule calculation is supported by measurements reported by the
CLEO Collaboration [\ref{r1r2cleo}], $R_1(1)=1.18\pm0.30\pm0.12$ and
$R_2(1)=0.71\pm0.22\pm0.07$.
These values are obtained assuming that $R_1(w)$ and $R_2(w)$ are
constant in $w$ and that $A_1(w)$ is linear in~$w$. 
CLEO also find that $R_1(1)$ and  $R_2(1)$ are not sensitive
either to the form of $A_1(w)$ or the $w$ dependence of the
form factors, consistent with the mild $w$ dependence in 
Eq.~(\ref{eq:exclusiveVcb:extrap:R2}). 
Note that the extractions of $\vcb$ by CLEO and BELLE discussed
in \sec{sec:vcbfrombtodst} use CLEO's measurements of $R_1(1)$ and~$R_2(1)$.

For $\btodlnu$ decays, the parametrization of [\ref{Caprini:1997mu}] is
\begin{equation}
   \frac{\cG(w)}{\cG(1)} \approx 1 - 8 \rhoG^2 z
    + (51.\rhoG^2 - 10.) z^2 - (252.\rhoG^2-84.) z^3 \,, 
\label{V13rd}
\end{equation}
with
\begin{equation}
z = \frac{\sqrt{w+1}-\sqrt 2}{\sqrt{w+1}+\sqrt 2} \,,
\label{eq:zdef}
\end{equation}
and where the only parameter, the slope parameter $\rhoG^2$ at zero
recoil is constrained by the dispersive bounds to lie in the interval
$-0.17<\rhoG^2<1.51$ which can be compared with the world experimental
average $\rhoG^2=1.19\pm 0.19$ given in \sec{sec:btodlnu}

\medskip

It is interesting to note that heavy quark symmetry breaking in the
difference of the slope and curvature parameters of the form factors
$\cF(w)$ and $\cG(w)$, together with measurements of the ratios $R_1$
and $R_2$ may strongly constrain the calculations which determine
$\cF(1)$ and $\cG(1)$ [\ref{Grinstein:2001yg}]. More importantly, a
better knowledge of the slope parameters will reduce the error on
$|V_{cb}|$, because of the large correlation between the two
parameters [\ref{Grinstein:2001yg}] (see Fig.~\ref{fig:vcbell}).

\boldmath 
\subsection{Theoretical calculations of the form factor $\cF(1)$
for $\btodslnu$ decays}
\unboldmath 
\label{sec:exclusiveVcb:F(1)}
The zero-recoil form factor $\cF(1)$ must be calculated
non-perturbatively in QCD.  
At zero recoil ($w=1$), all $\btodslnu$ form factors but $h_{A_1}$ 
are suppressed by phase space, and
\begin{equation}
        \cF(1) = h_{A_1}(1) =
                \langle {\rm D}^*(v) | A^\mu | \overline{\rm B}(v) \rangle,
        \label{eq:exclusiveVcb:F(1):F=hA1}
\end{equation}
where $A^\mu$ is the $b\to c$ axial vector current.
Thus, the theoretical information needed is contained in one relatively
simple hadronic matrix element, which
heavy-quark symmetry~[\ref{Shifman:1987rj},\ref{Isgur:vq},\ref{Isgur:ed}]
requires to be close to unity.
Heavy-quark \emph{spin} symmetry would imply
$\langle {\rm D}^*(v)|A^\mu|\overline{\rm B}(v)\rangle=%
\langle {\rm D}(v)|V^\mu|\overline{\rm B}(v)\rangle$, where $V^\mu$ is
the $b\to c$ vector current.  If, in addition, heavy-quark
\emph{flavor} symmetry is used, these amplitudes can be equated to
$\langle {\rm B}(v)|V^\mu|\overline{\rm B}(v)\rangle$.  The last matrix element
simply counts the number of $b$ quarks in a $\overline{\rm B}$ meson and is,
hence, exactly~1.  Deviations from the symmetry limit arise at short
distances, from the exchange of gluons with $m_c<k<m_b$, and also at
long distances.  Short-distance corrections are suppressed by powers
of $\alpha_s(\sqrt{m_cm_b})$, and long-distance corrections are
suppressed by powers of the heavy-quark masses.  The heavy-quark
symmetries also require the corrections of order $1/m_Q$ to vanish, a
result known as Luke's theorem~[\ref{Luke:1990eg}].  In summary, thanks
to heavy-quark symmetry, uncertainties from treating the
long-distance, non-perturbative QCD are suppressed by a factor of
order $(\bar{\Lambda}/2m_c)^2\sim5\%$, where $\bar\Lambda\sim 500\mev$
is the contribution of the light degrees of freedom to the mass of the
mesons.  Owing to these constraints from heavy-quark symmetry, the
exclusive technique is sometimes called
model-independent~[\ref{Neubert:td}], but in practice model
dependence could appear at order~$1/m_c^2$, through estimates of the
deviation of $h_{A_1}(1)$ from~1.

To date three methods have been used to estimate $h_{A_1}(1)-1$.
One approach starts with a rigorous inequality relating the zero-recoil
form factor to a spectral sum over excited 
states~[\ref{Shifman:1995jh},\ref{Bigi3}].
Here some contributions can be measured by moments of the inclusive
s.l.\  decay spectrum (cf.\ Sec.~\ref{sec:vcbincl}), but
others can be estimated only qualitatively.
The other two methods both start with HQET 
to separate long- and short-distance
contributions~[\ref{Falk:1993wt}].
The short-distance contributions are calculated in perturbative QCD.
The long-distance contributions are intrinsically non-perturbative.
Several years ago they were estimated in a non-relativistic quark
model~[\ref{Falk:1993wt},\ref{Neubert:1994vy}].
More recently, the HQET technique has been adapted to lattice gauge
theory~[\ref{Kronfeld:2000ck},\ref{Harada:2001fj}], and an explicit
calculation, in the so-called quenched approximation, has
appeared~[\ref{Hashimoto:2001nb}].

The three methods all quote an uncertainty on $\cF(1)$, and hence
$|V_{cb}|$, of around~4\%.  The errors arising in the sum rule and the
quark model calculations are difficult to quantify and do not appear to be
reducible.  In the lattice gauge theory calculations, there are
several ways to reduce the error, notably by removing the quenched
approximation and in improving the matching of lattice gauge theory to
HQET and continuum QCD.  It is conceivable that one could reduce the
uncertainty to the percent level over the next few years.  

\subsubsection{Sum rule method}
\label{sec:exclusiveVcb:F(1):sum_rule}

Here the main result of a sum rule that puts a rigorous bound
on~$h_{A_1}(1)$ is quoted.
For a lucid and brief derivation, the reader may consult a classic
review of the heavy-quark expansion~[\ref{Bigi:1997fj}].
Based on the optical theorem and the operator-product expansion,
one can show that
\begin{equation}
        |h_{A_1}(1)|^2 + \frac{1}{2\pi}\int_0 d\epsilon\, w(\epsilon) =
                1 - \Delta_{1/m^2} - \Delta_{1/m^3}
        \label{eq:exclusiveVcb:F(1):sum_rule}
\end{equation}
where $\epsilon=E-M_{D^*}$ is the relative excitation energy of higher
resonances and non-resonant $D\pi$ states with $J^{PC}=1^{-+}$,
and $w(\epsilon)$ is a structure function for the vector channel.
The contributions $\Delta_{1/m^n}$ describe corrections to the
axial vector current for finite-mass quarks.
The excitation integral is related to finite-mass corrections to the
bound-state wave functions---hence the ``sum'' over excited states.
The $\Delta_{1/m^n}$ and the excitation integral are positive, 
so Eq.~(\ref{eq:exclusiveVcb:F(1):sum_rule}) implies $|h_{A_1}(1)|<1$.

Let first consider the excitation integral.
For $\epsilon\gg\bar\Lambda$, the hadronic states are dual to
quark-gluon states.
Introducing a scale $\mu$ to separate this short-distance part from
the long-distance part (which must be treated non-perturbatively),
one writes
\begin{equation}
    \frac{1}{2\pi}\int_0 d\epsilon\, w(\epsilon) =
        \frac{1}{2\pi}\int^\mu_0 d\epsilon\, w(\epsilon) +
        [1-\eta_A^2(\mu)].
    \label{eq:exclusiveVcb:F(1):sum_rule_split}
\end{equation}
Here the short-distance quantity $\eta_A(\mu)$ lumps together the
short-distance ($\epsilon>\mu$) contribution.
Then, rearranging Eq.~(\ref{eq:exclusiveVcb:F(1):sum_rule}),
\begin{equation}
    h_{A_1}(1) = \eta_A(\mu) - \frac{1}{2}\Delta_{1/m^2} -
        \frac{1}{2}\Delta_{1/m^3} -
        \frac{1}{4\pi}\int^\mu_0 d\epsilon\, w(\epsilon) 
    \label{eq:exclusiveVcb:F(1):sum_rule_anatomy}
\end{equation}
and $\eta_A(\mu)$ is computed perturbatively
(to two loops~[\ref{Czarnecki:1998wy}]).
The other contributions arise from long distances and must be taken
from other considerations.
There is a good handle on the second term on the right-hand side of
Eq.~(\ref{eq:exclusiveVcb:F(1):sum_rule_anatomy}), namely,
\begin{equation}
    \Delta_{1/m^2} = \frac{\mu_G^2}{3m_c^2} +
    \frac{\mu_\pi^2(\mu)-\mu_G^2}{4}
    \left(\frac{1}{m_c^2} + \frac{2/3}{m_cm_b} + \frac{1}{m_b^2}\right),
    \label{eq:exclusiveVcb:F(1):sum_rule_1/m^2}
\end{equation}
where $\mu_G^2$ and $\mu_\pi^2(\mu)$ are matrix elements of the
chromomagnetic energy and kinetic energy (of the $b$ quark) in the
$\overline{\rm B}$ meson.
Note that the kinetic energy $\mu_\pi^2$ depends on the scale~$\mu$.
Apart from subtleties of renormalization conventions, $\mu_G^2$
and $\mu_\pi^2(\mu)$ are related to the quantities $\lambda_2$ and
$\lambda_1$, given in the discussion of inclusive s.l.\  decays.
Ignoring this subtlety for the moment,
$\mu_G^2=3\lambda_2=3(M_{B^*}^2-M_B^2)/4$ and 
$\mu_\pi^2=-\lambda_1$.
The last term in Eq.~(\ref{eq:exclusiveVcb:F(1):sum_rule_anatomy}),
from higher hadronic excitations, is unconstrained by data.

To make a numerical determination, one must choose a conventional value
for the separation scale to $\mu$, usually 1~GeV.
The choice of $\mu$ alters $\eta_A(\mu)$ and $\mu_\pi^2(\mu)$, as
well as the excitation integral, in ways that can be computed in
perturbative QCD.
A recent review~[\ref{Uraltsev:2000qw}] of the heavy quark expansion
takes
\begin{equation}
    \frac{1}{4\pi}\int^{\mathrm{1~GeV}}_0 d\epsilon\, w(\epsilon) =
        0.5 \pm 0.5,
    \label{eq:exclusiveVcb:F(1):sum_rule_guess}
\end{equation}
but emphasises that this is a heuristic estimate.
Ref.~[\ref{Uraltsev:2000qw}] found $h_{A_1}(1)=0.89\pm0.04$,
using a then-current value of $\mu_\pi^2$.
With CLEO's analysis of moments of the inclusive s.l.\  decay
spectrum in hand, one can convert that determination of $\lambda_1$ to a
determination of $\mu_\pi^2(\textrm{1~GeV})$.
The updated sum-rule becomes~[\ref{Uraltsev:2000qw}]
\begin{equation}
    \cF(1) = h_{A_1}(1) =
        0.900 \pm 0.015 \pm 0.025 \pm 0.025,
    \label{eq:exclusiveVcb:F(1):sum_rule_result}
\end{equation}
where the uncertainties are, respectively, from the two-loop calculation
of $\eta_A(\mathrm{1~GeV})$, the excitation integral [\emph{i.e.},
Eq.~(\ref{eq:exclusiveVcb:F(1):sum_rule_guess})], 
and an estimate  of~$\Delta_{1/m^3}$ based on dimensional analysis.
The uncertainty from $\eta_A(\mu)$ could be reduced, in principle, with
a three-loop calculation, but it is already smaller than the other two,
which appear to be irreducible.

\subsubsection{HQET-based methods}
\label{sec:exclusiveVcb:F(1):HQET}

The main drawback of the sum rule method is that the excitation integral is not
well constrained.
Using HQET one can characterize it in more detail.
Based on heavy-quark symmetry one can write
\begin{equation}
        h_{A_1}(1) = \eta_A \left[1 +
                \delta_{1/m^2} + \delta_{1/m^3} \right]
        \label{eq:exclusiveVcb:F(1):hqet_anatomy}
\end{equation}
where $\eta_A$ is a short-distance coefficient, which is discussed in
more detail below.
Heavy-quark symmetry implies the normalization of the first
term in brackets~[\ref{Isgur:vq},\ref{Isgur:ed}] and the absence of a
correction~$\delta_{1/m}$ of order~$1/m_Q$~[\ref{Luke:1990eg}].
The corrections $\delta_{1/m^n}$ of order $1/m_Q^n$ contain
long-distance matrix elements.
Simply from enumerating possible terms at second and third order,
they have the structure
\begin{eqnarray}
    \delta_{1/m^2} & = &
        - \frac{ \ell_V}{(2m_c)^2}
        + \frac{2\ell_A}{(2m_c)(2m_b)}
        - \frac{ \ell_P}{(2m_b)^2} ,
    \label{eq:exclusiveVcb:F(1):ell} \\
    \delta_{1/m^3} & = &
        - \frac{ \ell_V^{(3)}}{(2m_c)^3}
        + \frac{ \ell_A^{(3)}\Sigma}{(2m_c)(2m_b)}
        + \frac{ \ell_D^{(3)}\Delta}{(2m_c)(2m_b)}
        - \frac{ \ell_P^{(3)}}{(2m_b)^3} ,
    \label{eq:exclusiveVcb:F(1):ell3}
\end{eqnarray}
where $\Sigma=1/(2m_c)+1/(2m_b)$ and $\Delta=1/(2m_c)-1/(2m_b)$.

HQET is a systematic method for separating out the long- and
short-distance corrections to the symmetry limit, making efficient use
of the constraints of heavy-quark symmetry.
It provides a detailed description of the
$\ell$s~[\ref{Falk:1993wt},\ref{Mannel:1994wt}], of the form
\begin{equation}
    \ell_X = \sum_i c_i(\mu) {\cal M}_i(\mu),
    \label{eq:exclusiveVcb:F(1):ell_sum}
\end{equation}
where the $c_i(\mu)$ are short-distance coefficients and the
${\cal M}_i(\mu)$ matrix elements defined in the effective field theory.
The scale $\mu$ is now the renormalization scale of HQET.
Some contributions on the right-hand side come from the $1/m_Q$
expansion of the physical B and ${\rm D}^*$ mesons and
others from the expansion of the axial vector current.
The latter coincide with the $\lambda_1$ and $\lambda_2$ (or $\mu_\pi^2$
and $\mu_G^2$) terms in Eq.~(\ref{eq:exclusiveVcb:F(1):sum_rule_1/m^2}).
The long-distance corrections of the states are,
in Eq.~(\ref{eq:exclusiveVcb:F(1):sum_rule_anatomy}),
contained in $\int_0^\mu d\epsilon w(\epsilon)$.

It is well-known that intermediate quantities defined in effective
field theories depend on the renormalization scheme, but physical
quantities do not.
We dwell on it briefly here, for reasons that will become clear below.
At one-loop level, the short-distance coefficient is
\begin{equation}
    \eta_A(c) = 1 + \frac{4}{3} \frac{\alpha_s}{4\pi}
        \left[ 3 \frac{m_b+m_c}{m_b-m_c}\ln \frac{m_b}{m_c}
        - 8 \right]
    + \frac{4}{3} \frac{\alpha_s}{4\pi} c \mu^2
        \left(\Delta^2 + 2\Sigma^2\right)  
\end{equation}
where the constant $c$ is characteristic of the scheme for renormalizing
operators in HQET.
In minimal subtraction schemes $c=0$, whereas the energy cutoff in
Eq.~(\ref{eq:exclusiveVcb:F(1):sum_rule_split}) implies $c=4/3$ (cf. Eq.~(\ref{eq:mu2pert})).
Similarly, the scheme (and $\mu$) dependence of the $\ell$s is,
to order $\alpha_s$,
\begin{eqnarray}
        \ell_V(c) & = & \ell_V(0) + \frac{4}{3} \frac{\alpha_s}{4\pi} 3c\mu^2, \\
        \ell_A(c) & = & \ell_A(0) - \frac{4}{3} \frac{\alpha_s}{4\pi} ~c\mu^2,  \\
        \ell_P(c) & = & \ell_P(0) + \frac{4}{3} \frac{\alpha_s}{4\pi} 3c\mu^2.
\end{eqnarray}
Combining the above formulae, one can check that the
scheme dependence drops out of~$h_{A_1}(1)$.

As long as one is careful to keep track of the scheme, it does not
matter which is used.
For many purposes it is simplest to define all operator insertions in
minimal subtraction, for which $c=0$.
This is not a problem, as long as
one knows how to calculate the $\ell$s in the same scheme.
(For example, the $-\lambda_1$ and $\mu_\pi^2$ are defined by the
same HQET matrix element, renormalized such that $c=0$ and $4/3$,
respectively.)

The HQET formalism does not provide numerical estimates for the $\ell$s:
that requires a non-perturbative approach to~QCD.
The first estimates~[\ref{Falk:1993wt},\ref{Neubert:1994vy}] used the
non-relativistic quark model, which, though not QCD, can be a useful
guide and tends to yield  rather small $\delta_{1/m^2}$.
The more recent of these estimates~[\ref{Neubert:1994vy}] takes
$\delta_{1/m^2}$ to be $-0.055\pm0.025$, and 
relies on sum rule constraints.
Combining it with the two-loop calculation of
$\eta_A$~[\ref{Czarnecki:1996gu},\ref{Czarnecki:1997cf}], one obtains
\begin{equation}
    \cF(1) = h_{A_1}(1) =
        0.907 \pm 0.007 \pm 0.025 \pm 0.017,
    \label{eq:exclusiveVcb:F(1):quark_model_result}
\end{equation}
where the quoted uncertainties~[\ref{Neubert:1994vy},\ref{Czarnecki:1996gu}]
are from perturbation theory,
errors in the quark model estimate of the $1/m_Q^2$ terms,
and the omission of $1/m_Q^3$ terms.
Uncertainties from $\alpha_s$ and the quark masses are not included.
This result does not pay close attention to the scheme dependence
mentioned above, because it uses the standard ($c=0$) result for
$\eta_A$, corresponding to a minimal subtraction definition of the
matrix elements in Eq.~(\ref{eq:exclusiveVcb:F(1):ell_sum}).
The quark model, on the other hand, presumably yields the $\ell$s in
some other scheme (with unknown $c\neq0$).
In that case, Eq.~(\ref{eq:exclusiveVcb:F(1):quark_model_result})
over-~or undercounts the contribution at the interface of
long and short distances. Moreover, we note that 
estimates of the perturbative error based on BLM resummation 
[\ref{Uraltsev:1998bk},\ref{kinmass}] are larger than in
Eq.~(\ref{eq:exclusiveVcb:F(1):quark_model_result}).

Now let us turn to the recent lattice calculation of~$h_{A_1}(1)$.
A~direct calculation of the matrix element
$\langle {\rm D}^*|A^\mu|{\rm B}\rangle$
in Eq.~(\ref{eq:exclusiveVcb:F(1):F=hA1}) would be straightforward,
but not interesting:
similar matrix elements like $\langle 0|A^\mu|{\rm B}\rangle$ and 
$\langle\pi|V^\mu|{\rm B}\rangle$ have 15--20\%
errors~[\ref{Ryan:2001ej}].
One must involve heavy-quark symmetry from the outset: if one
can focus on $h_{A_1}-1$, there is a chance of success, because a 20\%
error on $h_{A_1}-1$ is interesting.
The key here is to observe that lattice gauge theory with Wilson
fermions has the same heavy-quark symmetries as continuum~QCD,
for all~$m_Qa$~[\ref{El-Khadra:1996mp}].
Consequently, one can build up a description of lattice
gauge theory using HQET, with the same logic and structure as
above~[\ref{Kronfeld:2000ck},\ref{Harada:2001fj},\ref{Harada:2001fi}].
In this description the $\ell$s in Eqs.~(\ref{eq:exclusiveVcb:F(1):ell})
and~(\ref{eq:exclusiveVcb:F(1):ell3}) are the same as for continuum QCD,
apart from lattice effects on the light quarks and gluons.
Discretization effects of the heavy quark appear at short distances,
where perturbation theory can be used.
Thus, the principal change from the usual application of HQET is in
the short-distance coefficients.

To calculate the $\ell$s in lattice gauge theory, one needs some
quantities with small statistical and normalization errors, whose
heavy-quark expansion contains the~$\ell$s.
Work on ${\rm B}\to {\rm D}$ form factors~[\ref{Hashimoto:2000yp}] showed
that certain ratios have the desired low level of uncertainty.
For the problem at hand one needs
\begin{eqnarray}
        \frac{\langle {\rm D}   |\bar{c}\gamma^4 b| {\rm B}   \rangle
                  \langle {\rm B}   |\bar{b}\gamma^4 c| {\rm D}   \rangle}
                 {\langle {\rm D}   |\bar{c}\gamma^4 c| {\rm D}   \rangle
                  \langle {\rm B}   |\bar{b}\gamma^4 b| {\rm B}   \rangle} & = &
        \left\{ \eta_V^{\rm lat} \left[ 1 - \ell_P\Delta^2 -
                \ell_P^{(3)} \Delta^2\Sigma \right] \right\}^2 ,
        \label{eq:exclusiveVcb:F(1):R+} \\
        \frac{\langle {\rm D}^* |\bar{c}\gamma^4 b| {\rm B}^* \rangle
                  \langle {\rm B}^* |\bar{b}\gamma^4 c| {\rm D}^* \rangle}
                 {\langle {\rm D}^* |\bar{c}\gamma^4 c| {\rm D}^* \rangle
                  \langle {\rm B}^* |\bar{b}\gamma^4 b| {\rm B}^* \rangle} & = &
        \left\{ \eta_V^{\rm lat} \left[ 1 - \ell_V \Delta^2 -
                \ell_V^{(3)} \Delta^2\Sigma \right] \right\}^2 ,
        \label{eq:exclusiveVcb:F(1):R1} \\
        \frac{\langle {\rm D}^* |\bar{c}\gamma^j \gamma_5 b| {\rm B}   \rangle
                  \langle {\rm B}^* |\bar{b}\gamma^j \gamma_5 c| {\rm D}   \rangle}
                 {\langle {\rm D}^* |\bar{c}\gamma^j \gamma_5 c| {\rm D}   \rangle
                  \langle {\rm B}^* |\bar{b}\gamma^j \gamma_5 b| {\rm B}   \rangle} & = &
        \left\{ \check{\eta}_A^{\rm lat} \left[ 1 - \ell_A\Delta^2 -
                \ell_A^{(3)} \Delta^2\Sigma \right] \right\}^2 .
        \label{eq:exclusiveVcb:F(1):RA1} 
\end{eqnarray}
For lattice gauge theory, the heavy-quark expansions in
Eqs.~(\ref{eq:exclusiveVcb:F(1):R+})--(\ref{eq:exclusiveVcb:F(1):RA1})
have been derived in Ref.~[\ref{Kronfeld:2000ck}], leaning heavily on
Refs.~[\ref{Falk:1993wt},\ref{Mannel:1994wt}].
One-loop perturbation theory for $\eta_V^{\rm lat}$ and 
$\check{\eta}_A^{\rm lat}$ is in Ref.~[\ref{Harada:2001fj}].
Thus, these ratios yield all three terms in $\delta_{1/m^2}$ and three
of four terms in $\delta_{1/m^3}$ (including the 
largest,~$\ell_V^{(3)}/(2m_c)^3$).

The method then proceeds as follows.
First, one computes the ratios on the left-hand sides of
Eqs.~(\ref{eq:exclusiveVcb:F(1):R+})--(\ref{eq:exclusiveVcb:F(1):RA1})
with standard techniques of lattice gauge theory,
for many combinations of the heavy quark masses.
Meanwhile one calculates the short-distance coefficients
$\eta_V^{\rm lat}$ and $\check{\eta}_A^{\rm lat}$ in perturbation
theory.
Then, one fits the numerical data to the HQET description, obtaining the
$\ell$s as fit parameters.
One can then combine these results with the perturbative calculation of
$\eta_A$ to obtain~$h_{A_1}(1)$.
The scheme mismatch that arises with the quark model calculation of
the $\ell$s is absent here, as long as one uses the same scheme to
calculate $\eta_V^{\rm lat}$ and $\check{\eta}_A^{\rm lat}$ on the
one hand, and $\eta_A$ on the other.

As expected, $\ell_V$ is the largest of the $1/m_Q^2$ matrix elements.
Because of the fit, the value of $\ell_V$ is highly correlated with 
that of $\ell_V^{(3)}$, but the physical combination is better 
determined.

Matching uncertainties arise here, as it is usually the case with HQET.
In Ref.~[\ref{Hashimoto:2001nb}] they are of order $\alpha_s^2$,
$\alpha_s\cdot(\bar{\Lambda}/m_c)^2$, and
$(\bar{\Lambda}/m_Q)^3$.
These can be improved in the future through higher-order matching
calculations.
Another uncertainty comes from the dependence of the ratios on the light
spectator quark, whose mass lies in the range $0.4\le m_q/m_s\le 1$.
There turns out to be a slight linear dependence on $m_q$, whose main
effect is to increase the statistical error.
In addition, there is a pion loop contribution~[\ref{Randall:1993qg}]
that is mistreated in the quenched approximation~[\ref{Arndt:2002ed}].
The omission of this effect is treated as a systematic error.
After reconstituting $h_{A_1}(1)$~[\ref{Hashimoto:2001nb}]
\begin{equation}
        \cF(1) = h_{A_1}(1) = 0.913 
                        {}^{+0.024}_{-0.017} 
                        \pm 0.016 
                        {}^{+0.003}_{-0.014} 
                        {}^{+0.000}_{-0.016} 
                        {}^{+0.006}_{-0.014} ,
        \label{eq:exclusiveVcb:F(1):lattice_result}
\end{equation}
where the uncertainties stem, respectively, from
statistics and fitting,
HQET matching,
lattice spacing dependence,
the chiral extrapolation,
and the effect of the quenched approximation.

\subsubsection{Comparison and summary}
\label{sec:exclusiveVcb:F(1):summary}

\begin{figure}[tp]
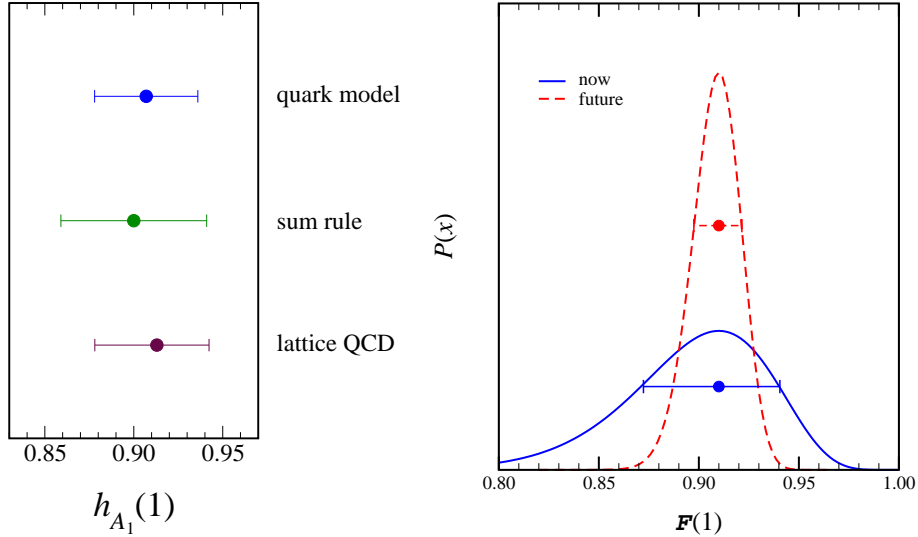

\begin{center}
\begin{tabular}{c c}
    \includegraphics[height=2.8in]{ChapterIII/hA1} &
    \includegraphics[height=2.8in]{ChapterIII/poisson} \\
\end{tabular}
    \caption[fig:2]{\it (a) Comparison of methods for~$\cF(1)$.
Note that 
the result labelled ``quark model'' actually uses sum rule constraints.
    (b) Model likelihood function for $\cF(1)$,
    now and with projected smaller errors in the future.}
    \label{fig:hA1}
\end{center}
\end{figure}

In Fig.~\ref{fig:hA1}(a) we compare the three results for
$\cF(1)$ from
Eqs.~(\ref{eq:exclusiveVcb:F(1):sum_rule_result}),
(\ref{eq:exclusiveVcb:F(1):quark_model_result}),
and~(\ref{eq:exclusiveVcb:F(1):lattice_result}).
All are compatible with
\begin{equation}
    \cF(1) = 0.91^{+0.03}_{-0.04}
    \label{eq:exclusiveVcb:F(1):workshop_result}
\end{equation}
The agreement is remarkable, even when one considers that all rely on
heavy-quark symmetry (and, so, compute the deviation from~1), and all
compute the short-distance part in perturbation theory (roughly half of
the deviation).
It is worth recalling the defects of the techniques.
The quark model omits some dynamics (more than the quenched
approximation in lattice QCD), and it is
not clear that it gives the $\ell$s in the same scheme as~$\eta_A$.
The sum rule has an incalculable contribution from excitations with
$(M-M_{D^*})^2<\mu^2$, which can only be estimated.
The present lattice result is in the quenched approximation, but
errors associated with quenching can, in this case, be estimated and are
given in the last two error bars in
Eq.~(\ref{eq:exclusiveVcb:F(1):lattice_result}).

When using Eq.~(\ref{eq:exclusiveVcb:F(1):workshop_result}) in a
global fit to the CKM matrix, one should appreciate the quality of the
theoretical information.
A~flat distribution based on
Eq.~(\ref{eq:exclusiveVcb:F(1):workshop_result}) would be incorrect:
the three methods agree well and, more significantly, part of the
uncertainty in
Eq.~(\ref{eq:exclusiveVcb:F(1):lattice_result}) is statistical, and
other uncertainties are under some control.
Also, one cannot rule out a tail for lower values,
$\cF(1)<0.87$; they are just unexpected.
Finally, we know that $\cF(1)\le1$ from the sum rule
in Eq.~(\ref{eq:exclusiveVcb:F(1):sum_rule}).
A~simple function that captures these features is the Poisson 
distribution (for $x>0$)
\begin{equation}
        P(x) = N x^7 e^{-7x}, \quad x = \frac{1-\cF(1)}{0.090},
        \label{eq:exclusiveVcb:F(1):poisson}
\end{equation}
where $N$ normalizes the distribution.
This distribution differs slightly from a synopsis of the lattice
result~[\ref{Kronfeld:2002cc}].
The most probable value has been shifted from 0.913 to 0.910,
mindful of the central value from the sum rule. 
[The average based on Eq.~(\ref{eq:exclusiveVcb:F(1):poisson}) is 0.90.]
Future work with lattice gauge theory could reduce the uncertainty by
a factor of~3, with unquenched calculations to reduce several of the
systematic errors, higher-order HQET matching to reduce the others,
and higher statistics to reduce the statistical errors.
Fig.~\ref{fig:hA1}(b) sketches how the resulting distribution would
look.
Recent developments~[\ref{Lepage:2001ym}] in the treatment of systematic
errors (except quenching) will allow lattice calculations to provide
a distribution that directly reflects statistical and systematic
uncertainties, instead of a schematic distribution 
as in~Eq.~(\ref{eq:exclusiveVcb:F(1):poisson}).

\boldmath 
\subsection{Theoretical calculations of the form factor $\cG(1)$
for $\btodlnu$ decays}
\label{sec:exclusiveVcb:G(1)}
\unboldmath 
The form factor $\cG(W)$ for $\btodlnu$ is given by
\begin{equation}
        \cG(w) = h_+(w) - \frac{M_B-M_D}{M_B+M_D} h_-(w) ,
\end{equation}
where the form factors $h_\pm(w)$ are defined by
\begin{equation}
        \langle {\rm D}(v')| V^\mu |{\rm B}(v)\rangle = \sqrt{M_BM_D}\left[
                (v'+v)^\mu h_+(w) - (v'-v)^\mu h_-(w) \right] .
\end{equation}
Even at zero-recoil both form factors remain.
With HQET one can derive expressions analogous to 
Eq.~(\ref{eq:exclusiveVcb:F(1):hqet_anatomy}).
Neglecting contributions of order $\alpha_s/m_Q^n$,
one finds~[\ref{Falk:1993wt},\ref{Mannel:1994wt},\ref{Kronfeld:2000ck}],
\begin{eqnarray}
        h_+(1) & = & \eta_V\left[1 - \left(
                \frac{1}{2m_c}-\frac{1}{2m_b}\right)^2 \ell_P \right] , \\
        h_-(1) & = & \beta_V + \left(
                \frac{1}{2m_c}-\frac{1}{2m_b}\right) \Lambda_- + \left(
                \frac{1}{(2m_c)^2}-\frac{1}{(2m_b)^2}\right) \ell_-,
\end{eqnarray}
where $\beta_V$ is of order $\alpha_s$.
Like the $\ell$s above, $\Lambda_-$ and $\ell_-$
are (combinations of) matrix elements of HQET.
They must be obtained by a non-perturbative method.
Note that the matrix element $\ell_P$ appearing in $h_+(1)$ is the 
same as in~$\cF(1)$.

Luke's theorem, applied to $\btodlnu$, explains why there is no
$1/m_Q$ term in $h_+(1)$.
The other form factor $h_-(1)$ is not protected by Luke's 
theorem, and, unfortunately, it appears in $\cG$ even at zero 
recoil~[\ref{Neubert:td}].
Moreover, although some constraint might be obtained from sum rules,
there is presently no useful bound analogous to that implied by 
Eq.~(\ref{eq:exclusiveVcb:F(1):sum_rule}).
In conclusion, there is less theoretical control over $\cG(1)-1$
than $\cF(1)-1$.

There are several calculations of $\cG(1)$.
Using the quark model, Scora and Isgur find~[\ref{ScoraIsgurPRD5295}]
\begin{equation}
        \cG(1) = 1.03 \pm 0.07.
\end{equation}
As mentioned above for $\cF(1)$, the quark model presumably has a 
problem with scheme dependence, though it may be a useful guide.
There have been a few calculations of $\ell_P$, $\Lambda_-$, 
and $\ell_-$ with QCD sum rules.
Including the full $\alpha_s^2$ correction and using the sum-rule results
of~[\ref{Ligeti:1993hw}], one finds~[\ref{Ligeti:1999yc}]:
\begin{equation}
        \cG(1) = 1.02 \pm 0.08.
\end{equation}
Although this result is based on QCD, it is unlikely that the error bar 
can be reduced further.
Finally, Hashimoto \emph{et al.}\ have used lattice QCD and a strategy 
similar to that for $\cF(1)$, which homes in on $\cG(1)-1$.
They find~[\ref{Hashimoto:2000yp}] 
\begin{equation}
        \cG(1) = 1.058^{+0.021}_{-0.017},
\end{equation}
where errors from statistics, tuning of heavy quark masses, and omitted radiative corrections have been added in quadrature.
One should also expect some uncertainty from the quenched approximation,
perhaps 15--20\% of $\cG(1)-1$.
Unlike the calculation of by $\cF(1)$ by the same
group~[\ref{Hashimoto:2001nb}], 
here the dependence on the lattice spacing was not studied.
These issues could be cleared up, by completing calculations of (lattice)
radiative corrections needed to improve the calculation of $h_-(1)$,
and then carrying out the Monte Carlo calculation of $h_-(1)$ at several 
lattice spacings.

In conclusion, the status of the theoretical calculations of $\cG(1)$ is 
less satisfactory than for $\cF(1)$. We believe that
\begin{equation}
        \cG(1) = 1.04 \pm 0.06
\end{equation}
fairly summarizes the present theoretical knowledge of $\cG(1)$.

\subsection{Electroweak corrections}
\label{sec:ewcorr}

For completeness, we close with a brief summary of electroweak 
corrections to exclusive s.l.\  decays.
Some of these effects are shared by the radiative corrections to muon 
decay, and that is why the muon decay constant~$G_\mu$ appears in
Eqs.~(\ref{eq:exclusiveVcb:dGdw*})
and~(\ref{eq:exclusiveVcb:dGdw}).
Another effect is simply radiation of photons from the outgoing 
charged lepton, which could be important in semi-electronic decays, if
the experimental acceptance is non-uniform in the electron's energy.
A~complete treatment is not available, but an adequate prescription is
given in Ref.~[\ref{Atwood:1989em}].
If the decaying B meson is electrically neutral, one must multiply
the right-hand side of Eq.~(\ref{eq:exclusiveVcb:dGdw}) with a
factor~[\ref{Ginsberg:1968pz}]
$    1 + \alpha\pi$
to account for the Coulomb attraction of the outgoing charged lepton and
charged ${\rm D}^*$.
This corresponds to a shift in $|V_{cb}|$ of about~$1\%$.

There are also virtual corrections from diagrams with $W$ and $Z$ 
bosons.
The leading parts of these effects are enhanced by the large logarithm
$\ln(M_Z/M_B)$, which arise from distances much shorter than the
QCD scale,
and  their net effect is the factor~$\eta_{\rm EW}^2$ in
Eq.~(\ref{eq:exclusiveVcb:dGdw}). One finds~[\ref{Sirlin:1981ie}]
\begin{equation}
    \eta_{\rm EW} = 1 + \frac{\alpha}{\pi}\ln(m_Z/\mu)
    \label{eq:exclusiveVcb:sirlin}
\end{equation}
where the scale $\mu$ separates weak and  strong effects.
It is natural to set $\mu=M_B$, in which case $\eta_{\rm EW}=1.0066$.
Should the accuracy of the QCD form factor $\cF$ fall 
below 1\%,  it might be necessary  to go beyond the leading log description of 
Eq.~(\ref{eq:exclusiveVcb:sirlin}), but that could require the introduction
of new form factors besides 
$\cF$, so a general treatment is difficult.

\vspace{2mm}

\boldmath 
\subsection{Semileptonic  B decays to a hadronic system heavier than 
D or D$^*$ }
\label{dssback}
\unboldmath 
Semileptonic B decays into $p$-wave charm mesons are the most
important sources of background polluting the measurement of the
 $\btodslnu$ decay rate. The  hadronic system heavier than $\rm D^{(*)}$
is commonly identified as `$\rm D^{**}$'.

In infinite quark mass limit, hadrons containing a single heavy quark
can be classified by their total spin $J$ and by the angular momentum $j$ of
their light degree of freedom. In this limit, heavy quark mesons
come in degenerate doublets with total spin $J=j\pm \frac{1}{2}$. 
Therefore, the four charm meson states `$ \rm D^{**}$' corresponding to the 
angular momentum $l=1$ are classified in two doublets:
$\rm D_0,{D^*}_1$ with $j=\frac{1}{2}$ and $J^P = (0^+,1^+)$, and
$\rm D_1,{D^*}_2$ with $j=\frac{3}{2}$ and $J^P = (1^+,2^+)$.
Both $\rm D_1$ and ${\rm D^*}_2$ are narrow states ($\Gamma \simeq 20$~MeV).
This small width is a consequence of their strong decay proceeding
through d-wave transitions.
The resonances of the other doublet are expected to be rather
broad, as they decay through s-wave pion emission.

 The existence of the narrow resonant states is well established
[\ref{ref:pdg02}] and a signal for a broad resonance has been seen by CLEO
[\ref{CB}], but the decay characteristics of these states in $b$-hadron
s.l.\  decays have large uncertainties. 
The average of ALEPH
[\ref{aleph-d2star}], CLEO [\ref{cleo-d2star}] and DELPHI
[\ref{delphi-d2star}] narrow state branching fractions show that
the ratio 
\begin{equation}
  R_{**} = \frac{{\cal B}(\rm \overline{B} \to D^*_2 \ell
\bar{\nu})} {{\cal B}(\rm \overline{B} \to D_1 \ell \bar{\nu})}
\end{equation}
 is smaller than one ($<0.6$ at 95\% C.L.[\ref{HFnote}]), in
disagreement with HQET calculations where an infinite quark mass is
assumed [\ref{wrong}], but in agreement with calculations which take into
account finite quark mass corrections [\ref{ligeti}].

To estimate the `$ {\rm D}^{**}$', the
LEP experiments use the treatment of narrow ${\rm D}^{**}$
proposed in [\ref{ligeti}] which accounts for ${\cal O}(1/m_c)$
corrections. 
Ref.~[\ref{ligeti}] provides several possible approximations of the
`$ {\rm D}^{**}$' form factors, that depend on  five different
expansion schemes  (A, A$_{inf}$, B$_{inf}$, B$_1$, B$_2$) 
and on three input parameters ($\eta_{ke}, t_{h1}, z_{h1}$).

Each proposed scheme is tested with the
relevant input parameters varied over a range consistent with the
experimental limit on $R_{**}$.
  The $\cF(1) V_{cb}$ analysis is repeated for each allowed
point of the scan and the systematic
error is the maximal difference from the central value obtained
in this way. 
Non-resonant terms may not be modelled correctly in this approach.

\boldmath 
\subsection{Review and future prospects for the exclusive 
determination of \vcb}
\unboldmath 

\subsubsection{\vcb\ from $\btodslnu$ decays}
\label{sec:vcbfrombtodst}

The decay $\btodslnu$ has been studied in
experiments performed at the $\y4s$ center of mass
energy and at the $\rm Z^0$ center of mass energy at LEP. At the
$\y4s$, experiments have the advantage that the $w$
resolution is good. However, they have more limited
statistics near $w=1$ in the decay $\rm \overline{B}^0\rightarrow
D^{* +} \ell^- \bar\nu_\ell$, because of the lower reconstruction
efficiency of the slow pion, from the $\rm D^{* +} \to \pi^{+}
D^0$ decay. The decay $\rm B^-\rightarrow D^{* 0} \ell^-
\bar{\nu}_\ell$ is not affected by this problem and CLEO
[\ref{cleo-vcb}] uses both channels. In addition, kinematic
constraints enable $\y4s$ experiments to identify the final state
without a large contamination from the poorly known s.l.\ 
B decays to  `$\rm  D^{**}$'.
 At LEP, B's are produced with a large
momentum (about ~30 GeV on average). This makes the determination of
$w$ dependent upon the neutrino four-momentum reconstruction, thus
giving a relatively poor resolution and limited physics background
rejection capabilities. The advantage that LEP experiments have is an
efficiency which is only mildly dependent upon $w$.

Experiments determine the product $(\cF(1)\cdot |V_{cb}|)^2$
by fitting the measured ${d\Gamma}/{dw}$ distribution.
Measurements at the $\y4s$ have been performed by CLEO
[\ref{cleo-vcb}]  and BELLE [\ref{belle-dslnu}].  At LEP data are
available from ALEPH [\ref{ALEPH_vcb}], DELPHI [\ref{DELPHI_vcb}] and
OPAL [\ref{OPAL_vcb}].  
At LEP, the
dominant source of systematic error is the uncertainty on the
contribution to ${d\Gamma}/{dw}$ from s.l.\  $\rm B\to D^{**}$
decays. The ``${\rm D}^{**}$'' includes 
both narrow orbitally excited charmed meson and non-resonant or broad
species.
The treatment of the ``${\rm D}^{**}$'' spectra is described in 
\ref{dssback}, while branching ratios of
the processes which affect the value of \vcb\ are taken from [\ref{HFnote}].
\begin{table}[htbp]
\begin{center}
\begin{tabular}{|l|c|c|c|c|} \hline
experiment & \fvcb\ $(\times 10^{3})$ & $\rhoAone^2$ & $\rm
Corr_{stat}$ & References\\ \hline
 ALEPH published &  31.9$\pm$  1.8$\pm$ 1.9 & 0.31$\pm$ 0.17$\pm$ 0.08 & 92\% 
&  [\ref{ALEPH_vcb}]\\
 ALEPH update    &  31.5$\pm$  2.1$\pm$ 1.3 & 0.58$\pm$ 0.25$\pm$ 0.11 & 94\% 
&[\ref{HFnote}] \\
 DELPHI          &  35.5$\pm$  1.4$\pm$ 2.4 & 1.34$\pm$ 0.14$\pm$ 0.23 & 94\% 
& [\ref{DELPHI_vcb}]\\
 OPAL            &  37.1$\pm$  1.0$\pm$ 2.0 & 1.21$\pm$ 0.12$\pm$ 0.20 & 90\% 
& [\ref{OPAL_vcb}]\\
 BELLE           &  35.8$\pm$  1.9$\pm$ 1.8 & 1.45$\pm$ 0.16$\pm$ 0.20 & 90\% 
&  [\ref{belle-dslnu}]\\
 CLEO            &  43.1$\pm$  1.3$\pm$ 1.8 & 1.61$\pm$ 0.09$\pm$ 0.21 & 86\% 
& [\ref{cleo-vcb}]\\
\hline
\end{tabular}
\end{center}

\vspace{-2mm}

\caption{\it Experimental results as published by the collaborations.
LEP numbers use theoretical predictions for R$_1$ and R$_2$. The
published ALEPH result is obtained using a linear fit and the old
ISGW model [\ref{isgw}] for ${\rm D}^{**}$. The updated
ALEPH numbers (used in our average) are obtained using the same
fit parameterization and ${\rm D}^{**}$ models as the other
LEP experiments [\ref{lepvcb}]. The BELLE result listed in the
Table uses R$_1$ and R$_2$ from CLEO data. \label{t:publ}}
\end{table}
\begin{table}[htbp]
\begin{center}
\begin{tabular}{|l|c|c|}
\hline
 Parameter & Value & Reference \\ \hline
 $\Rb=\Gamma(Z\to b\bar{b})/\Gamma(Z\to had)$ & (21.64 $\pm$ 0.07)\% & 
[\ref{EWWG}]  \\
 $\fd = {\cal B}(b\to {\rm B}_d)$ & (40.0  $\pm 1.1$)\% & [\ref{Bosc}] \\
$\tau(\rm B^0)$& (1.54   $\pm$ 0.015) ps & [\ref{life}] \\
${x_E}^{LEP}=E(\rm B~meson)/\sqrt{s}$  & 0.702  $\pm$ 0.008 & [\ref{EWWG}] \\
$ {\cal B}(\rm D^{* +}\to\Dz\pi^+$) & (67.7 $\pm$ 0.5) \% & [\ref{ref:pdg02}] \\
\hline
R$_1$ & $1.18 \pm 0.32$ & [\ref{r1r2cleo}] \\
R$_2$ & $0.71 \pm 0.23$ & [\ref{r1r2cleo}] \\ \hline
 ${\cal B}$(\Btau)           & (1.27  $\pm$ 0.21)\% &  [\ref{HFnote}] \\
 ${\cal B}(\rm B^- \to D^{* +} \pi^- \ell \bar{\nu}$) & (1.29$\pm0.16$) \% &
             [\ref{HFnote}]  \\
${\cal B}(\Bzerod~ \to~ \rm D^{* +} \pi^0 \ell \bar{\nu})$ &
$(0.61\pm0.08)$\% &
 [\ref{HFnote}] \\
${\cal B}({\rm B}_s \to \Dstars \rm K \ell \bar{\nu})$ & $(0.65\pm0.23)$\% &
[\ref{HFnote}]
\\ \hline
\end{tabular}
\end{center}

\vspace{-2mm}

\caption{\it Values of the most relevant parameters affecting the
measurement of $\vcb$. The three ${\rm D}^{**}$ production
rates are fully correlated.} \label{tab:summary}
\end{table}

Table~\ref{t:publ} summarizes all published data as quoted in the
original papers. 
To combine the published data, the central values and the errors
of $\cF(1) \vcb$ and $\rhoAone^2$ are re-scaled to the same set
of input parameters. These common
inputs are listed in Table~\ref{tab:summary}. The ${\cF}(1) \vcb$ 
values used for obtaining an average are extracted with
\begin{figure}[htbp]
\centering
\includegraphics*[width=4.7in]{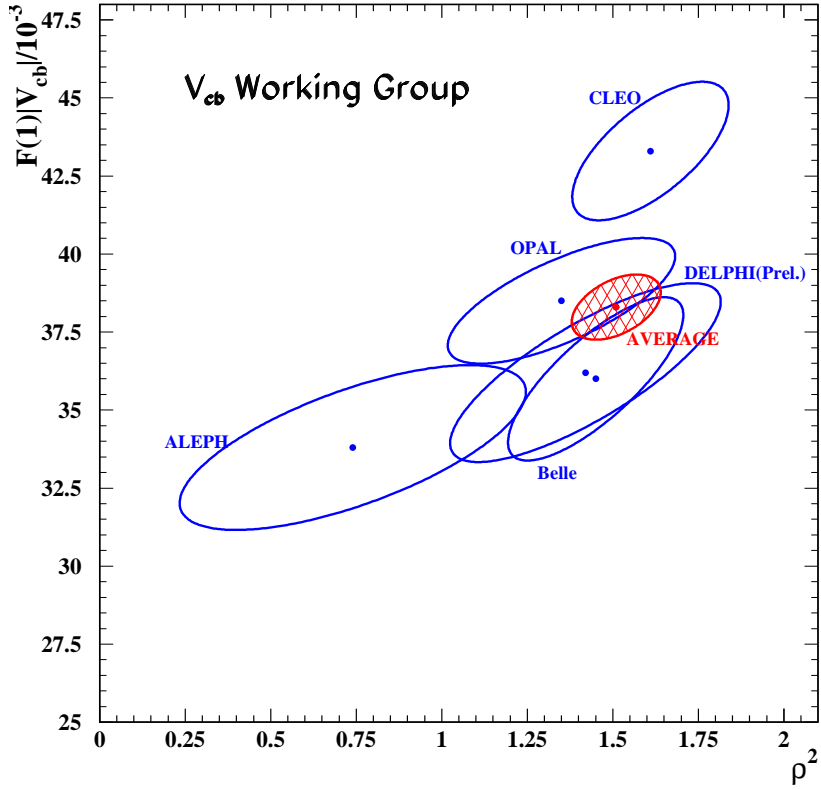}

\caption{\it The error ellipses for the corrected measurements and
world average for $\cF(1)\vcb$ vs $\rhoAone^2$. The ellipses
correspond to a 39\% C.L.} \label{fig:vcbell}
\end{figure}
the parametrization of \eq{A13rd}, taking on the experimental
determinations of the vector and axial form factor ratios $R_1$
and $R_2$ [\ref{r1r2cleo}]. The LEP data, which originally used
theoretical values for these ratios, are re-scaled accordingly
[\ref{lepvcb}]. Table~\ref{t:correxp} summarizes the corrected
data. The averaging procedure [\ref{lepvcb}] takes into account
statistical and systematic correlations between $\cF(1) \vcb$
and $\rhoAone^2$. Averaging the measurements in Table 1, we get:
$$\cF (1) \vcb = (38.3  \pm 1.0) \times 10^{-3}$$ and
$$\rhoAone^2 = 1.5 \pm 0.13$$ with a confidence level \footnote{
The $\chi^2$ per degree of freedom is less than 2, and we do not
scale the error.} of 5.1\%. The error ellipses for the corrected
measurements and for the world average are shown in
Fig.~\ref{fig:vcbell}.
\begin{table}
\begin{center}
\begin{tabular}{|l|ccc|} \hline
experiment   & \fvcb\ $(\times 10^{-3})$ & $\rhoAone^2$ & $\rm
Corr_{stat}$ \\ \hline
 ALEPH       &  33.8$\pm$  2.1$\pm$ 1.6 & 0.74$\pm$ 0.25$\pm$ 0.41  & 94\% \\
 DELPHI      &  36.1$\pm$  1.4$\pm$ 2.5 & 1.42$\pm$ 0.14$\pm$ 0.37 & 94\% \\
 OPAL        &  38.5$\pm$  0.9$\pm$ 1.8 & 1.35$\pm$ 0.12$\pm$ 0.31 &  89\% \\
 BELLE       &  36.0$\pm$  1.9$\pm$ 1.8 & 1.45$\pm$ 0.16$\pm$ 0.20 &  90\% \\
 CLEO        &  43.3$\pm$  1.3$\pm$ 1.8 & 1.61$\pm$ 0.09$\pm$ 0.21 &  86\% \\
\hline
World average&  38.3 $\pm$ 0.5$\pm$ 0.9 & 1.51$\pm$ 0.05$\pm$ 0.12 & 86\%\\
\hline 
\end{tabular}
\end{center}

\caption{\it Experimental results after the correction to common
inputs and world average. The LEP numbers are corrected to use
R$_1$ and R$_2$ from CLEO data. $\rhoAone^2$ is the slope parameter 
as defined in \eq{A13rd} at zero recoil. \label{t:correxp}}
\end{table}

The main contributions to the systematic error in $\cF (1) \vcb$ are
from the uncertainty on the $\rm B\dec D^{**}\ell \nu$ shape and on ${\cal
B}(b\to {\rm B}_d)$ ($0.57\times 10^{-3}$), fully correlated among the LEP
experiments, the branching fraction of D and $\rm D^{*}$ decays
($0.4\times 10^{-3}$), fully correlated among all the experiments, and
the slow pion reconstruction from BELLE and CLEO (0.28$\times
10^{-3}$), which are uncorrelated. The main contribution to the
systematic error on $\rhoAone^2$ is from the uncertainties in the
CLEO's measurement of R$_1$ and R$_2$ (0.12), fully correlated among
experiments. Because of the large contribution of this uncertainty to
the non-diagonal terms of the covariance matrix, the averaged
$\rhoAone^2$ is higher than one would naively expect.  This situation
will improve substantially in the next few years through a better
determination of $R_1$ and $R_2$, using the higher statistics samples
being accumulated at the B-factories, as well as through the full
exploration of the s.l.\ B decays to ${\rm D}^{**}$.

Using $\cF (1) = 0.91 \pm 0.04$, as given in \sec{sec:exclusiveVcb:F(1)} 
but with a symmetrized error, one gets 
\begin{equation}
\begin{array}{|l|}
\hline
        \vcb = (42.1 \pm 1.1_{exp} \pm 1.9_{th}) \times 10^{-3}.\\
\hline
\end{array}
        \label{eq:exclusiveVcb:Vcb}
\end{equation}
The dominant error is theoretical, but
there are good prospects to reduce it through improvements in lattice QCD 
calculations, particularly removing the quenched approximation.

\subsubsection{\vcb\ from $\btodlnu$ decays}
\label{sec:btodlnu}

The strategy to extract $\vcb \cG(1)$ is identical to
that used for $ \vcb \cF(1)$ in $\btodlnu$ decays. 
As discussed above, theoretical estimates of $\cG(1)$ are not,
at this time, as accurate.
This channel is much more challenging
also from the experimental point of view because 
${d\Gamma_D}/{dw}$ is more heavily suppressed near $w=1$ than
${d\Gamma_{D^*}}/{dw}$, due to the helicity mismatch between
initial and final states, and because
it is hard to isolate
from the dominant background, $\rm B\dec D^* \ell \nu$, as well
as from fake D-$\ell$ combinations.  Thus, the extraction of
$|V_{cb}|$ from this channel is less precise than the one from
the $\rm B\dec D^* \ell \nu$ decay. Nevertheless,  the $\rm B\dec
D \ell \nu$ channel provides a consistency check.

BELLE [\ref{belle-dplnu}] and ALEPH [\ref{ALEPH_vcb}] have studied the
$\rm \overline{B}^0\dec D^+ \ell^- \bar{\nu}$ channel, while CLEO
[\ref{cleo-dplnu}] has studied both $\rm B^+\dec D^0 \ell^+ \bar{\nu}$
and $\rm \overline{B}^0\dec D^+ \ell^- \bar{\nu}$ decays. The parametrization
used in these studies for the extrapolation to zero recoil is that of
\eq{V13rd}. The published results are shown in Table~\ref{t:D},
together with the results scaled to common inputs. Averaging the
latter according to the procedure of
[\ref{lepvcb}], we get $\cG(1) |V_{cb}| = (41.3 \pm 4.0)\times 10^{-3}$
and $\rhoG^2 = 1.19 \pm 0.19$, where $\rhoG^2$ is the slope parameter
of $\cG(w)$ at zero recoil. 

\begin{table}
\begin{center}
\begin{tabular}{|l|cc|c|} \hline
experiment   &  $\cG(1)|V_{cb}|(\times 10^{-3})$ & $\rhoG^2$ & References\\
\hline
Published values &  & & \\
 ALEPH       &  31.1$\pm$  9.9$\pm$ 8.6 & 0.20$\pm$ 0.98$\pm$ 0.50 & 
 [\ref{ALEPH_vcb}] \\
 BELLE       &  41.1$\pm$  4.4$\pm$ 5.2 & 1.12$\pm$ 0.22$\pm$ 0.14 & 
[\ref{belle-dplnu}] \\
 CLEO        &  44.4$\pm$  5.8$\pm$ 3.7 & 1.27$\pm$ 1.25$\pm$ 0.14 & 
[\ref{cleo-dplnu}] \\
\hline \hline
Scaled values & &  &\\
 ALEPH       &  37.7$\pm$  9.9$\pm$ 6.5 & 0.90 $\pm$ 0.98$\pm$ 0.38 &\\
 BELLE       &  41.2$\pm$  4.4$\pm$ 5.1 & 1.12$\pm$ 0.22$\pm$ 0.14  &\\
 CLEO        &  44.6$\pm$  5.8$\pm$ 3.5 & 1.27$\pm$ 0.25$\pm$ 0.14  &\\
\hline
World average &  41.3 $\pm$ 2.9$\pm$ 2.7 & 1.19$\pm$ 0.15$\pm$ 0.12 &\\
\hline
\end{tabular}
\end{center}
\caption{\it Experimental results before and after the correction to
common inputs and world average. $\rhoG^2$ is the slope parameter
as defined in \eq{V13rd}. \label{t:D}}
\end{table}


Using $\cG(1)=1.00\pm 0.07$, as given in \sec{sec:exclusiveVcb:G(1)}, 
we get 
\begin{equation}
        |V_{cb}| = (41.3 \pm 4.0_{exp} \pm 2.9_{theo}) \times 10^{-3},
\end{equation}
consistent with Eq.~(\ref{eq:exclusiveVcb:Vcb}) from $\btodslnu$ decay,
but with an uncertainty about twice as large.

Since \vcb ~drops out of the measured ratio $\cG(w)/\cF(w)$,
this can be compared to theoretical calculations independently 
of their basis.
In the heavy-quark limit, both form factors 
are given by the same function of $w$. A precise measurement of their 
ratio would provide information about the size of symmetry-breaking corrections
away from zero recoil. 
Some experiments have also looked at the differential decay rate
distribution to extract the ratio $\cG(w)/ \cF(w)$. 
However, data are not precise enough to measure the 
symmetry-breaking corrections away from zero recoil.
From the measured values of $\cG(1)\vcb$ and
$\cF(1)\vcb $, we get $\cG(1)/\cF(1) = 1.08 \pm 0.09,$
consistent with the form factor values that we used.

\boldmath 
\section{Exclusive determination of $\vub$}
\label{sec:vubexcl}
\def\fpbpi{f^+_{B\pi}}
\def\fzbpi{f^0_{B\pi}}
\def\qsqmax{q^2_{\mathrm{max}}}
\newcommand{\plus}{\makebox[15pt][c]{$+$}}
\newcommand{\minus}{\makebox[15pt][c]{$-$}}
\newcommand{\er}[2]
{\hskip-0.5em\raisebox{0.08em}{\scriptsize{$\;\begin{array}{@{}l@{}}
\plus\makebox[0.25em][r]{#1\hfill} \\[-0.12em]
\minus\makebox[0.25em][r]{#2\hfill} 
\end{array}$}}}
\newcommand{\err}[2]
{\hskip-0.5em\raisebox{0.08em}{\scriptsize{$\;\begin{array}{@{}l@{}}
\plus\makebox[0.50em][r]{#1\hfill} \\[-0.24em]
\minus\makebox[0.50em][r]{#2\hfill} 
\end{array}$}}}

\unboldmath 

As seen in \sec{sec:vubincl}, $|V_{ub}|$ can be measured from the inclusive
$b \rightarrow u l \nu$ rate --- blind to the particular decay mode.
Such measurements require, however, that kinematic selections be made
to discriminate against the dominant $b \rightarrow c l \nu$
background.  This introduces additional theoretical uncertainties
that can be significant.

An alternative route to measure $|V_{ub}|$ is the exclusive
reconstruction of particular $b \rightarrow u l \nu$ final states.
Experimentally this provides some extra kinematical constraints for
background suppression, and theoretically the uncertainties are of a
different nature. The extraction of $|V_{ub}|$ is complicated by the
fact that the quarks are not free, but bound inside mesons. The
probability that the final state quarks will form a given meson is
described by form factors. And unlike exclusive $b\to cl\nu$ decays,
heavy quark symmetry does not help to normalize these form factors at
particular kinematic points. A variety of calculations of these form
factors exists, based on lattice QCD, QCD sum rules, perturbative QCD,
or quark models. At present, none of these methods allows for a fully
model-independent determination of $|V_{ub}|$, though lattice
calculations should, in time, provide a means to reach this goal.  It
is thus very important to obtain a consistent measurement of
$|V_{ub}|$ with both the inclusive and exclusive approach and also to
find consistent results for the various exclusive modes. The simplest
mode theoretically is ${\rm B} \rightarrow \pi l \nu$, since a description
of its rate involves only one form factor in the limit of vanishing
lepton mass, instead of the three required for vector final states.

The differential rate for
${\rm B}^0\rightarrow\pi^-l^+\nu$ decays ($l=e$ or $\mu$) is given by
\begin{equation}
  \label{eq:differential_decay_rate}
  \frac{1}{|V_{ub}|^2}
  \frac{d\Gamma}{dq^2} =
  \frac{G_F^2}{24\pi^3}
  \l[(v\cdot k)^2-m_\pi^2\r]^{3/2}
  |\fpbpi(q^2)|^2,
\end{equation}
where the form factor $\fpbpi(q^2)$ is defined through
\begin{equation}
  \label{eq:f+f0}
  \langle\pi^-(k)|\bar{b}\gamma^{\mu}u|{\rm B}^0(p)\rangle
  = \fpbpi(q^2) 
  \left[ 
    (p+k)^{\mu} 
    - \frac{M_B^2-m_{\pi}^2}{q^2} q^{\mu}
  \right]
  + \fzbpi(q^2) \frac{M_B^2-m_{\pi}^2}{q^2} q^{\mu},
\end{equation}
with $q^2$ the momentum transfer squared,
$q^2=(p-k)^2=M_B^2+m_\pi^2-2M_B v\cdot k$, and $p=M_Bv$. In the
s.l.\  domain, $q^2$ takes values in the range from 0 to
$\qsqmax\equiv (M_B-m_\pi)^2$ which corresponds to $v\cdot k$ varying
from $M_B/2+m_\pi^2/(2M_B)$ to $m_\pi$. The form factor $\fzbpi(q^2)$
does not contribute to the rate in the limit of vanishing lepton mass.

\subsection{Lattice QCD determinations of semileptonic heavy-to-light 
form factors\protect}

Lattice QCD simulations potentially provide a means of calculating
heavy-to-light decay form factors from first 
principles\footnote{An introductory, though slightly dated, review of
some of the subjects covered in this section can be found 
in~[\ref{3Lellouch:1999dz}]}.  
These
calculations are model independent in the sense that they are based on
approximations of QCD that can be systematically improved to
arbitrarily high accuracy.  In practice, however, all calculations to
date have been performed in the quenched approximation, where the
effect of sea quarks is treated as a mean field. This introduces a
systematic error that is difficult to estimate {\em a priori}, though
experience shows that for many hadronic quantities, the deviations
induced by the quenched approximation are in the 10 to 15\% range.

Besides the quenched approximation, which will be lifted (at least
partially) in the near future, there are two major practical
limitations in the lattice calculation of heavy-to-light form factors.
One is that the spatial momenta of the initial and final state hadrons
are restricted to be less than about 2~GeV, to avoid large
discretization errors.  The other is that light-quark masses are much
larger than their physical value and the corresponding ``pion'' mass
is $m_\pi\gsim$ 400~MeV, so that an extrapolation to the physical
light quarks is needed (the so-called chiral extrapolation). As a
result, the available region for $q^2$ is limited to values above
about $\qsqmax/2$.

\subsubsection{Results for ${\rm B}^0\rightarrow\pi^-l^+\nu$ form factors}

In addition to the extrapolations in light-quark mass, an understanding
of the dependence of the form factors on heavy-quark mass is necessary.
For both these purposes, the
HQET motivated form factors~[\ref{Burdman:1993es}] 
$f_1(v\cdot k)$ and $f_2(v\cdot k)$
are useful. They are related to the form factors $\fpbpi$ and
$\fzbpi$ of \eq{eq:f+f0} through
\begin{eqnarray}
  \label{eq:f+_from_f1f2}
  \fpbpi(q^2) & = &
  \sqrt{M_B} \left\{
      \frac{f_2(v\cdot k)}{v\cdot k} +
      \frac{f_1(v\cdot k)}{M_B}
    \right\},
  \\
  \label{eq:f0_from_f1f2}
  \fzbpi(q^2) & = &
  \frac{2}{\sqrt{M_B}}
  \frac{M_B^2}{M_B^2-m_{\pi}^2}
  \Biggl\{
    \left[ 
      f_1(v\cdot k) + f_2(v\cdot k)
    \right]
  \nonumber\\
  & & 
  \left.
    -
    \frac{v\cdot k}{M_B}
    \left[ 
      f_1(v\cdot k) + 
      \frac{m_{\pi}^2}{(v\cdot k)^2}
      f_2(v\cdot k)
    \right]
  \right\}.
\end{eqnarray}
The HQET form factors are defined such that
the heavy quark scaling with $M_B\rightarrow\infty$
is manifest, namely, $f_{1,2}(v\cdot k)$ become
independent of $M_B$ up to logarithms coming from the
renormalization of the heavy-light current. 
The corrections due to finite $M_B$ are then described as a
power series in $1/M_B$. 
At leading order in the $1/M_B$ expansion,
$\fpbpi(q^2)$ is proportional to $f_2(v\cdot k)$, while 
$\fzbpi(q^2)$ is proportional to a linear combination
$f_1(v\cdot k)+f_2(v\cdot k)$.
Thus, the heavy quark scaling of $\fpbpi(q^2)$ and $\fzbpi(q^2)$
is given by, 
\begin{eqnarray}
  \label{eq:scaling_f+f0}
  \fpbpi(q^2) & \sim & \sqrt{M_B},\\
  \fzbpi(q^2) & \sim & \frac{1}{\sqrt{M_B}},
\end{eqnarray}
for fixed $v\cdot k$, up to logarithms and $1/M_B$ corrections.

\medskip

Recently four major lattice groups, UKQCD [\ref{Bowler:1999xn}], APE
[\ref{Abada:2000ty}], Fermilab [\ref{El-Khadra:2001rv}], and JLQCD
[\ref{Aoki:2001rd}], have performed quenched calculations of
${\rm B}\rightarrow\pi l\nu$ form factors.  The UKQCD [\ref{Bowler:1999xn}]
and APE [\ref{Abada:2000ty}] collaborations use 
non-perturbatively $O(a)$-improved Wilson fermions
[\ref{Sheikholeslami:1985ij},\ref{Luscher:1996sc},\ref{Luscher:1996jn}])
and treat heavy quarks relativistically.  In this formalism, the
leading discretization errors induced by the heavy-quark mass, $m_Q$,
are reduced from $am_Q$ to $(am_Q)^2$, with $a$ the lattice spacing.
To keep these errors under control with the lattice spacing $a\sim
1/2.7\,\gev$ available to them, they have to perform the calculations
for heavy-quark masses in the neighborhood of the charm-quark mass and
extrapolate to the bottom.  The drawback of this approach is that the
extrapolation can be significant and that discretization errors may be
amplified if this extrapolation is performed before a continuum limit
is taken. The Fermilab group
[\ref{El-Khadra:2001rv}], on the other hand, uses a 
formalism for heavy quarks in which correlation functions computed
with Wilson-type fermions are reinterpreted using HQET
[\ref{El-Khadra:1996mp},\ref{Harada:2001fi}].  
In this way, they can reach both the charm and
bottom quarks without extrapolation, and they investigate the
discretization errors using three lattice spacings ($\beta$ = 6.1, 5.9
and 5.7) covering $1/a\sim$ 1.2--2.6~GeV.  The JLQCD collaboration
[\ref{Aoki:2001rd}] employs a lattice NRQCD action
[\ref{Thacker:1990bm},\ref{Lepage:1992tx}] for heavy quarks so that the bottom
quark mass is covered by interpolation, and the calculation is done on
a coarse lattice, $1/a\sim$ 1.6~GeV ($\beta$ = 5.9).  Both the Fermilab
and NRQCD approach are based on expansions of QCD in powers of $1/m_Q$
and precision calculations at the physical $b$-quark mass require the
inclusion of corrections proportional to powers of $1/m_b$ which can
be difficult to compute accurately. In the case of NRQCD, one is also
confronted with the fact that the continuum limit cannot be taken.
All groups use an $O(a)$-improved Wilson action
[\ref{Sheikholeslami:1985ij}] for light quarks.
\begin{figure}[tbp]
  \centering
  \includegraphics*[width=10cm]{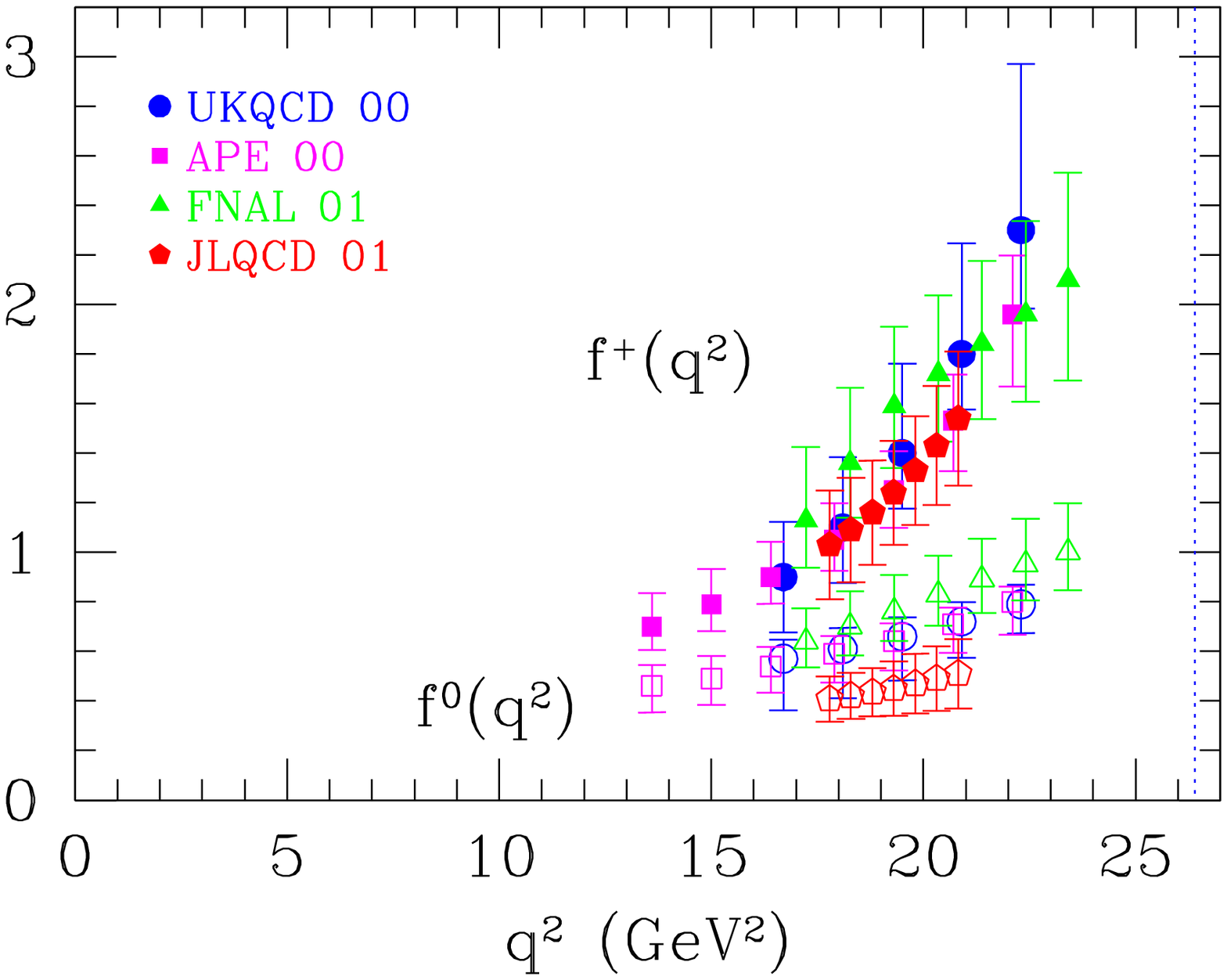} 
  \caption{\it Recent lattice results for ${\rm B}^0\rightarrow\pi^-
   l^+\nu$ form factors $\fpbpi(q^2)$ and $\fzbpi(q^2)$. Statistical
   and systematic errors are added in quadrature.
  \label{fig:fpf0_comparison}}
\end{figure}
Fig.~\ref{fig:fpf0_comparison} shows a comparison of
recent results for the 
${\rm B}^0\rightarrow\pi^- l^+\nu$ form factors $\fpbpi(q^2)$ and
$\fzbpi(q^2)$ from the four groups 
[\ref{Bowler:1999xn},\ref{Abada:2000ty},\ref{El-Khadra:2001rv},\ref{Aoki:2001rd}].
For convenience, the values of these form factors are also reported in
\tab{tab:diff_rate}. The lattice results are available only for the large
$q^2$ region (13~GeV$^2\gsim q^2\gsim$ 23~GeV$^2$) corresponding to
small spatial momenta of the initial B and final pion.

Good agreement is found amongst the different groups for
$\fpbpi(q^2)$, while the results for $\fzbpi(q^2)$ show a slight
disagreement.  To assess where these differences may come from and,
more generally, to estimate systematic errors, 
the heavy and light quark extrapolations, which form a core part
of the underlying analysis, are now briefly reviewed.
\begin{table}
  \begin{center}
    \begin{tabular}{|c|c|c|c|c|}
\hline
      Ref. & $q^2$ [GeV$^2$] &
      $\fzbpi(q^2)$ & $\fpbpi(q^2)$ & 
      $1/|V_{ub}|^2d\Gamma/dq^2$  [ps$^{-1}$GeV$^{-2}$]\\
      \hline
APE & { 13.6} & {${0.46(7)^{+5}_{-8}}$} & {${0.70(9)^{+10}_{-3}}$} & 0.33(9)\err{9}{3}\\
APE & { 15.0} & {${0.49(7)^{+6}_{-8}}$} & {${0.79(10)^{+10}_{-4}}$} & 0.31(8)\err{8}{3}\\
APE & { 16.4} & {${0.54(6)^{+5}_{-9}}$} & {${0.90(10)^{+10}_{-4}}$} & 0.28(6)\err{6}{3}\\
UKQCD & $16.7$ & $0.57^{+6}_{-6}\ {}^{+\ 5}_{-20}$ & $0.9^{+1}_{-2}\ {}^{+2}_{-1}$ & $0.29^{+10}_{-\ 9}\ {}^{+11}_{-\ 6}$\\
FNAL &  17.23  & 0.64\er{9}{3}\err{10}{10}   & 1.13\err{24}{9}\err{17}{17}  & 0.35\err{15}{6}(11)\\
JLQCD &      17.79 & 0.407(92) & 1.03(22) & 0.25(11) \\
APE & { 17.9} & {${0.59(6)^{+4}_{-10}}$} & {${1.05(11)^{+10}_{-6}}$} & 0.25(5)\err{5}{3}\\
UKQCD & $18.1$ & $0.61^{+6}_{-6}\ {}^{+\ 6}_{-19}$ & $1.1^{+2}_{-2}\ {}^{+2}_{-1}$ & $0.27^{+8}_{-7}\ {}^{+11}_{-\ 1}$ \\
FNAL &  18.27  & 0.70\er{9}{4}\err{11}{11}   & 1.36\err{23}{9}\err{20}{20}  & 0.37\err{13}{5}(11)\\
JLQCD &      18.29 & 0.421(92) & 1.09(21) & 0.240(94)\\
JLQCD &      18.80 & 0.435(98) & 1.16(21) & 0.231(84)\\
APE & { 19.3} & {${0.64(6)^{+4}_{-10}}$} & {${1.25(13)^{+9}_{-8}}$} & 0.22(5)\err{3}{3}\\
JLQCD &      19.30 & 0.45(11)  & 1.24(21) & 0.221(76)\\
FNAL &  19.31  & 0.76\err{10}{4}\err{11}{11} & 1.59\err{21}{7}\err{24}{24}  & 0.36\err{10}{3}(11)\\
UKQCD & $19.5$ & $0.66^{+5}_{-5}\ {}^{+\ 6}_{-17}$ & $1.4^{+2}_{-2}\ {}^{+3}_{-1}$ & $0.25^{+7}_{-6}\ {}^{+11}_{-\ 1}$\\
JLQCD &      19.81 & 0.47(12)  & 1.33(22) & 0.210(71)\\
JLQCD &      20.31 & 0.49(13)  & 1.43(24) & 0.199(68)\\
FNAL &  20.35  & 0.83\err{10}{4}\err{12}{12} & 1.72\err{18}{8}\err{26}{26}  & 0.28\err{6}{3}(9)\\
APE & { 20.7} & {${0.71(6)^{+3}_{-10}}$} & {${1.53(17)^{+8}_{-11}}$} & 0.19(4)\err{2}{3}\\
JLQCD &      20.82 & 0.51(14)  & 1.54(27) & 0.187(66)\\
UKQCD & $20.9$ & $0.72^{+5}_{-4}\ {}^{+\ 6}_{-14}$ &$1.8^{+2}_{-2}\ {}^{+4}_{-1}$ & $0.23^{+6}_{-5}\ {}^{+11}_{-\ 1}$ \\
FNAL &  21.38  & 0.89\err{10}{4}\err{13}{13} & 1.84\err{20}{14}\err{27}{27} & 0.20\err{4}{3}(6)\\
APE & { 22.1} & {${0.80(6)^{+1}_{-12}}$} & {${1.96(23)^{+6}_{-18}}$} & 0.16(4)\err{1}{3}\\
FNAL &  22.41  & 0.95\err{12}{3}\err{14}{14} & 1.96\err{24}{20}\err{29}{29} & 0.13\err{3}{3}(4)\\
FNAL &  23.41 & 1.00\err{13}{3}\err{15}{15} & 2.10\err{29}{25}\err{32}{32} & 0.09\err{2}{2}(2)\\
\hline
   \end{tabular} \end{center} \caption{ Form factors and differential
   rate for $B^0\to\pi^-l\nu$ decays from UKQCD [\ref{Bowler:1999xn}],
   APE [\ref{Abada:2000ty}], FNAL [\ref{El-Khadra:2001rv}] and JLQCD
   [\ref{Aoki:2001rd}]. The first set of errors is statistical and the
   second, systematic. In the case of JLQCD, these two sets of errors
   were combined quadratically. } \label{tab:diff_rate}
\end{table}

\subsubsection*{Heavy quark scaling}

\begin{figure}[htbp] 
  \centering
  \includegraphics*[width=10cm]{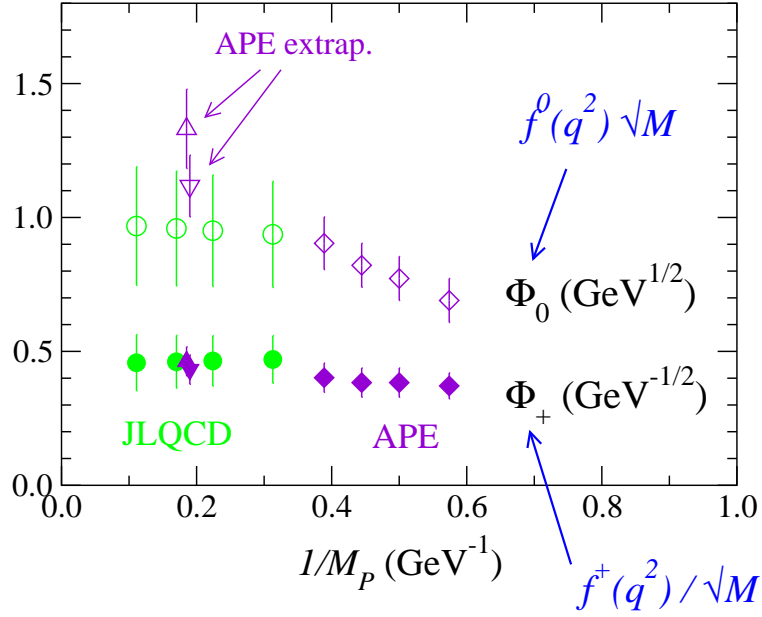}
  \caption{\it 
    $1/M_B$ scaling of the form factors 
    $\fpbpi(q^2)/\sqrt{M_B}$ (filled symbols) and 
    $\fzbpi(q^2)\sqrt{M_B}$ (open symbols) at  fixed $v\cdot k\sim 0.95\,\gev$.  
    Data from APE [\ref{Abada:2000ty}] (diamonds) and
    JLQCD [\ref{Aoki:2001rd}] (circles). 
  }
  \label{fig:phi_APE}
\end{figure}

At a fixed value of $v\cdot k$, the $1/M_B$ dependences of the form
factors $\fpbpi(q^2)/\sqrt{M_B}$ and $\fzbpi(q^2)\sqrt{M_B}$ from
JLQCD are compared to those of APE [\ref{Abada:2000ty}] in
Fig.~\ref{fig:phi_APE}.  Both collaborations agree that there is no
significant $1/M_B$ dependence in $\fpbpi(q^2)/\sqrt{M_B}$.  For
$\fzbpi(q^2)\sqrt{M_B}$, on the other hand, the APE
[\ref{Abada:2000ty}] result has a significant slope, which is
also supported by the Fermilab result
[\ref{El-Khadra:2001rv}] (not shown in the plot), 
while JLQCD do not see such dependence.
The reason for this disagreement is not clear, but it partly 
explains the smaller value of $\fzbpi(q^2)$ of JLQCD data in
Fig.~\ref{fig:fpf0_comparison}.

\vspace{-5mm}
\subsubsection*{Chiral extrapolation}

\begin{figure}[tbp]
  \centering \includegraphics*[width=10cm]{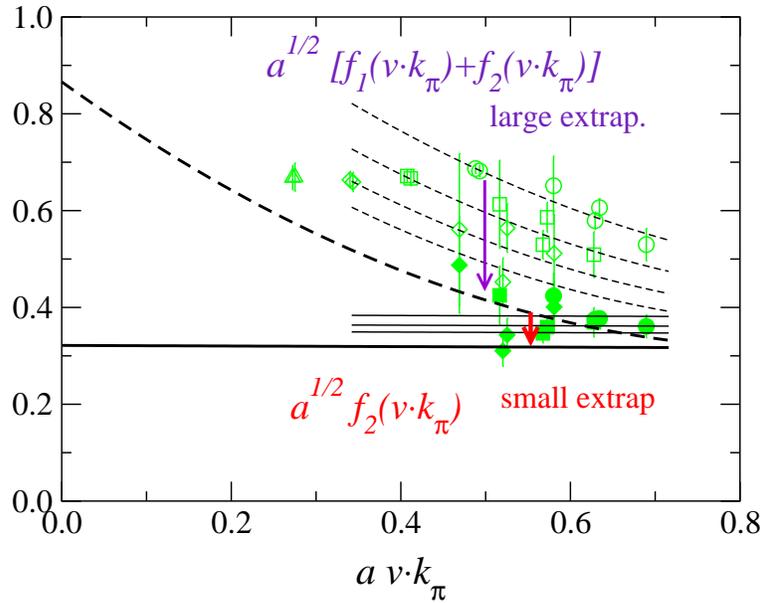}
  \caption{\it  The chiral extrapolation of the HQET form factors
  $f_1(v\cdot k)+f_2(v\cdot k)$ (open symbols) and $f_2(v\cdot k)$
  (filled symbols) is indicated by the vertical, downward-pointing
  arrows.  Data from JLQCD [\ref{Aoki:2001rd}].  
  \label{fig:f1f2_light_dependence}}
\end{figure}

The chiral extrapolation of the HQET form factors
$f_1(v\cdot k)+f_2(v\cdot k)$ and $f_2(v\cdot k)$ is
demonstrated in~Fig.~\ref{fig:f1f2_light_dependence}. This
extrapolation is performed at fixed $v\cdot k$ by fitting the
form factors to a power
series in the light quark mass, as suggested in [\ref{Bowler:1999xn}]. 
No attempt is made to account for chiral logarithms
because they are not correctly
reproduced in the quenched 
theory [\ref{Becirevic:2002sc},\ref{Fleischer:1992tn}]. 
The figure shows that the extrapolation is insignificant for
$f_2(v\cdot k)$ (or $\fpbpi(q^2)$), while a large extrapolation
is involved in $f_1(v\cdot k)+f_2(v\cdot k)$ 
(or $\fzbpi(q^2)$). 


\vspace{-5mm}
\subsubsection*{Summary of current status}

The current status of quenched lattice calculations of the
${\rm B}\rightarrow\pi l\nu$ form factors may be summarized as follows:
\begin{itemize}
\item 
  The physical form factor $\fpbpi(q^2)$ has small $1/M_B$ corrections
  in the range of recoils explored. As a result, neither the
  extrapolation from the charm-quark-mass region (in the UKQCD and APE
  results) nor the truncation of the $1/M_B$ expansion (in the Fermilab
  and JLQCD results) is a dominant source of systematic error.
  $\fzbpi(q^2)$ is more sensitive to $1/M_B$ corrections, and the
  agreement among different groups is poorer.
\item 
  The form factor $\fpbpi(q^2)$ is relatively insensitive to
  light-quark mass, and simple polynomial chiral extrapolations are
  stable. This is not the case for $\fzbpi(q^2)$, which displays significant
  light-quark-mass dependence.
\item
  The agreement amongst the four groups for $\fpbpi(q^2)$ as shown in
  Fig.~\ref{fig:fpf0_comparison} is remarkable, because the groups
  use different methods for modelling the $b$ quark, for matching the
  lattice current to the continuum one, for performing chiral
  extrapolations,  {\it etc.}  This agreement is probably due 
  to the fact that this form factor is relatively insensitive to heavy- and
  light-quark masses, as long as ${\qsqmax}$ is not approached too
  closely.
\end{itemize}
These observations allow us to conclude that the
systematic error is under control at the level of accuracy
shown in Fig.~\ref{fig:fpf0_comparison}.

On the other hand, the lattice calculations reviewed have
important drawbacks:
\begin{itemize}
\item
  They are performed in the quenched approximation.
\item 
  The available lattice results are restricted to the large $q^2$ region.
  They may be used to predict the partially integrated decay rate, but
  predictions for the total decay rate usually introduce some model
  dependence.
\item
  For the physical form factor $\fpbpi(q^2)$, the current error is of
  order 20\% for all groups and a significant reduction in error will 
  be challenging.
\end{itemize}

\subsubsection*{Strategies for determining $|V_{ub}|$}

With the quenched lattice results for ${\rm B}^0\to\pi^-l^+\nu$ decays presented
above, the only unknown in the expression of
\eq{eq:differential_decay_rate} for the differential decay rate is
$|V_{ub}|$. To illustrate this point, the results of 
the four collaborations for this rate are reproduced in
\tab{tab:diff_rate}. It is clear, then, that $|V_{ub}|$ can be
determined without assumptions about the $q^2$
dependence of form factors, once experiments measure the differential
or partially integrated rate in the range of $q^2$ values reached in
these calculations. Future lattice calculations in full, unquenched
QCD will permit completely model-independent determinations of
$|V_{ub}|$.

The total rate or the differential rate closer to $q^2=0$ can also be
used to extract $|V_{ub}|$, but then an extrapolation becomes
necessary. This extrapolation usually introduces model dependence and
the resulting $|V_{ub}|$ thus inherits a systematic error that is
difficult to quantify.

Pole dominance models suggest the following momentum
dependence for the form factors, 
\begin{equation}
  f^i_{B\pi}(q^2)=\frac{f_{B\pi}(0)}{(1-q^2/M_i^2)^{n_i}},
\label{eq:pdmodel}
\end{equation}
where $i=+,\ 0$, $n_i$ is an integer exponent and the kinematical
constraint $\fpbpi(0)=\fzbpi(0)$ has already been imposed.  Combining
this with the HQS scaling relations of \eq{eq:scaling_f+f0} implies
$n_+=n_0+1$. Light-cone sum rule scaling further suggests
$n_0=1$~[\ref{3DelDebbio:1997kr}]~\footnote{Pole/dipole behaviour for
$\fzbpi$ and $\fpbpi$ was also suggested in [\ref{3Charles:1998dr}].}.  
Another pole/dipole
parametrization for $\fzbpi$ and $\fpbpi$, which accounts for the
${\rm B}^*$ pole in $\fpbpi$ correctly, has been suggested by
Becirevic and Kaidalov (BK)~[\ref{BecirevicKaidalov}]:
\begin{eqnarray}
  \fpbpi(q^2)&=&\frac{f_{B\pi}(0)}
  {(1-q^2/m^2_{B^{\star}})(1-\alpha q^2/m^2_{B^{\star}})} \nonumber \\
  \fzbpi(q^2)&=&\frac{f_{B\pi}(0)}{(1- q^2/\beta m^2_{B^{\star}})}.      
\label{eq:bkparam}
\end{eqnarray}
Fitting this parametrization to the results of each of the four
collaborations yields the results summarized in
\tab{tab:bkfits}. Though uncertainties are still quite large,
consistency amongst the various lattice predictions, as well as with
the LCSR result, is good.

\begin{table}[t]
\begin{center}
\begin{tabular}{|c|c|c|c|}
\hline
Ref. & $f_{B\pi}(0)$ & $\alpha$ & $\beta$\\
\hline
UKQCD M-I[\ref{Bowler:1999xn}] & $0.30^{+6+4}_{-5-9}$ & $0.46^{+9+37}_{-10-5}$ & 
$1.27^{+14+4}_{-11-12}$\\
APE M-II [\ref{Abada:2000ty}] & $0.28(6)^{+5}_{-5}$ & $0.45(17)^{+6}_{-13}$ & $1.20(13)^{+15}$\\
APE M-I [\ref{Abada:2000ty}] & $0.26(5)^{+4}_{-4}$ & $0.40(15)^{+9}_{-9}$ & $1.22(14)^{+15}$\\
FNAL M-I[\ref{El-Khadra:2001rv}] & $0.33^{+2}_{-3}$ & $0.34^{+9}_{-3}$ & $1.31^{+3}_{-9}$\\
JLQCD M-II[\ref{Aoki:2001rd}] & $0.23^{+4}_{-3}$ &$ 0.58^{+12}_{-9}$ & $1.28^{+12}_{-20}$\\
\hline
LCSR [\ref{Khodjamirian:2000ds}] & 0.28(5) & $0.32^{+21}_{-7}$ & \\
\hline
\end{tabular}
\caption{Results of fits of the lattice results from the four groups to the BK parametrization
of \eq{eq:bkparam}. In the results of UKQCD and APE, the second set of
uncertainties corresponds to systematic errors. Method I (M-I)
consists in first extrapolating the form factors obtained from the
simulation in light-quark mass, heavy-quark mass etc.  and then
fitting to the BK parametrization.  Method II (M-II) corresponds to
first fitting the BK parametrization to the form factors obtained from
the simulation, before any chiral, heavy-quark, $\ldots$
extrapolations, and then peforming the extrapolations on the fit
parameters. The row entitled LCSR corresponds to a fit to light-cone
sum rule results.}
\label{tab:bkfits}
\end{center}
\end{table}

Using the results of these fits, UKQCD [\ref{Bowler:1999xn}] and APE 
[\ref{Abada:2000ty}], 
obtain the following total rate:
\begin{equation}
\label{eqn:my_decay_rate}
  \Gamma({\rm B}^0\to\pi^-l^+\nu)/|V_{\rm ub}|^2=\left\{\begin{array}{ll}
9^{+3}_{-2}{}^{+3}_{-4}\ {\rm ps}^{-1} & \mbox{UKQCD~[\ref{Bowler:1999xn}]}\\
7.0\pm 2.9\ {\rm ps}^{-1} & \mbox{APE~[\ref{Abada:2000ty}]}
\end{array}\right.
\ ,
\end{equation}
where the first error in the UKQCD result is statistical and the
second is the systematic error, which includes the difference between
the parametrizations of \eqs{eq:pdmodel}{eq:bkparam}. In the APE
result, where a fit to the pole/dipole parametrization is not
considered, the error includes statistical and systematic errors
summed in quadrature. Nevertheless, because of the model dependence of
these results, a larger systematic error cannot be excluded.

\subsubsection{Future directions}

In the following we discuss the directions that should be explored in the
near future to improve the accuracy in the determination of
$|V_{ub}|$.

\subsubsection*{Extension toward lower $q^2$}

\begin{figure}[tbp]
  \centering \includegraphics*[width=10cm]{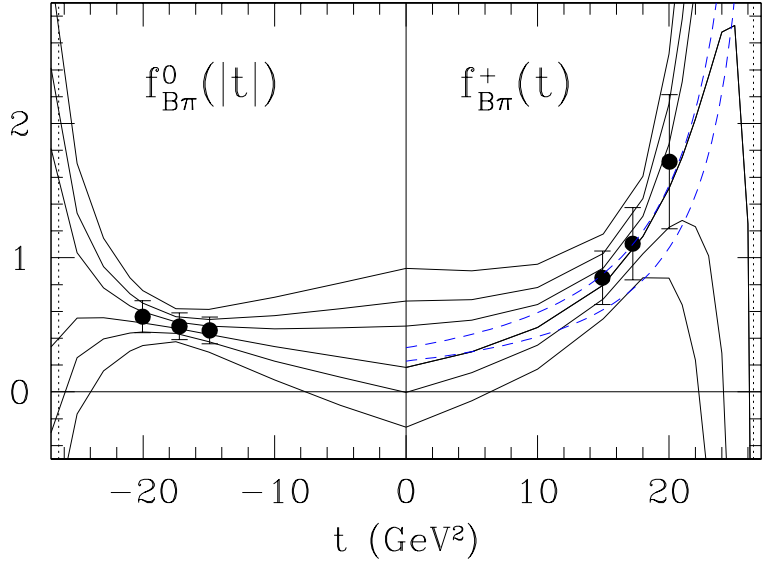} \caption{\it 
  Dispersive bounds for $f_0(|t|)$ and $f_+(t)$ in
  ${\rm B}^0\to\pi^-\ell^+\nu$ decays [\ref{Lellouch:1995yv}].  The points
  are the lattice results of [\ref{Burford:1995fc}] with added
  systematic errors. The pairs of fine curves are, from the outermost
  to the innermost, the 95\%, 70\% and 30\% bounds, where percentages
  represent the likelihood that the form factor take a value between
  the corresponding pair of curves at the given $t$. The dashed curves
  are the LCSR results of \eq{eq:KRWWY} [\ref{Khodjamirian:2000ds}]. 
Comparable results 
  are given in \eq{eq:Ball} [\ref{BallZ}].}  \label{fig:dispersive_bound}
\end{figure}

As already discussed, it is not straightforward to extend lattice
calculations of heavy-to-light form factors to the low $q^2$ region,
though finer lattices will eventually get us there. Extrapolations to
lower $q^2$ values can be performed using models which incorporate
many of the known constraints on the form factors, but this introduces
a model dependence which is difficult to quantify.  It has been
proposed, however, to use dispersion relations together with lattice
data to obtain model-independent bounds for the form factors over the
entire $q^2$ range
[\ref{Boyd:1994tt},\ref{Lellouch:1995yv},\ref{Boyd:1995tg}].  
These techniques are based on the same 
ingredients as those used to constrain the shape of the form factors
for ${\rm B}\to {\rm D}^{(*)}l\nu$ decays, briefly presented in
\sec{sec:exclusiveVcb:extrap_in_w}, though details of the implementation 
are quite different. An example is shown in
Fig.~\ref{fig:dispersive_bound}. The bounds in that figure were
obtained using the lattice results for ${\rm B}\to\pi l\nu$ form
factors from [\ref{Burford:1995fc}], the most complete set available at
the time.

Since then lattice calculations have improved significantly and it
would be interesting to derive new bounds by combining modern lattice
results for the form factors with the techniques developed in
[\ref{Lellouch:1995yv}]. It may also be advantageous to take into account
additional constraints on the form factors. Furthermore, other ways of
extending the range of lattice calculations to lower values of $q^2$
should be investigated.

\subsubsection*{Unquenching}

Lattice calculations have to be performed with dynamical sea quarks to
yield truly model independent results.  Some groups already have gauge
configurations for two flavours of sea quarks with degenerate masses
$\gsim m_s/2$ (instead of the two very light $u$ and $d$ quarks and
the lightish $s$ quark found in nature). The study of B meson
decays on these backgrounds presents no conceptual difficulty.

In practice, however, the chiral extrapolations required to reach
the $u$ and $d$ quark masses may be rather delicate as it is
not clear that the light-quark masses used in the simulations
are light enough to be sensitive to the so-called chiral
logarithms which are expected to dominate the small mass behaviour
of many physical quantities 
(see e.g. [\ref{Durr:2002zx}--\ref{3Lellouch:2002nj}] 
for recent discussions).
It will be very important to control this light-quark-mass behaviour to
obtain accuracies better than 10\%.

\subsubsection*{Using $\rm D\rightarrow\pi l\nu$ decays to
improve predictions for $\rm B\rightarrow\pi l\nu$ form factors}

In the heavy charm and bottom limit, heavy quark symmetry relates the
$\rm B\rightarrow\pi l\nu$ form factors to $\rm D\rightarrow\pi l\nu$. 
Burdman \textit{et al.} [\ref{Burdman:1993es}] proposed to 
consider the ratio 
\begin{equation}
\label{eq:bdratio}
\l.  \frac{
    d\Gamma({\rm B}^0\rightarrow\pi^- l^+\nu)/d(v\cdot k)}{
    d\Gamma({\rm D}^0\rightarrow\pi^- l^+\nu)/d(v\cdot k)
  }\r|_{\mathrm{same}\;v\cdot k}
  =
  \left|\frac{V_{ub}}{V_{cd}}\right|^2
  \left(\frac{M_B}{M_D}\right)^2
  \left|
    \frac{
      f^+_{B\pi}/\sqrt{M_B}}{
      f^+_{D\pi}/\sqrt{M_D}
    }
  \right|^2,
\end{equation}
from which one may extract the ratio of CKM matrix elements
$|V_{ub}/V_{cd}|$. In view of the high-precision measurements of D
decays promised by CLEO-$c$, such an approach to determining
$|V_{ub}|$ is becoming increasingly relevant.  

It is convenient to factorize the nearest pole contribution to
$\fpbpi(q^2)$, which is expected to dominate the $q^2$ behaviour of
this form factor in the heavy-quark limit, at least close to zero
recoil. 
 Thus, the breaking of heavy quark
symmetry may be parametrized as
\begin{equation}
  \frac{
    f^+_{B\pi}/\sqrt{M_B}}{
    f^+_{D\pi}/\sqrt{M_D}
  }
  =
  \frac{v\cdot k + \Delta_D}{v\cdot k + \Delta_B}
  R_{BD}(v\cdot k),
\end{equation}
where $\Delta_{B,D}\equiv m_{B^*,D^*}-m_{B,D}$ and $(R_{BD}(v\cdot k)-1)$
describes the $1/M_{D,B}$ corrections to be calculated on the lattice.
The question then becomes whether $R_{BD}(v\cdot k)$ can be calculated
more accurately on the lattice than $\fpbpi(q^2)$. The answer is
``yes'' as a number of uncertainties are expected to cancel in
the ratio. It is also encouraging that the heavy-quark-mass dependence of
$f_2(v\cdot k)$ appears to be mild, as discussed previously.

To reach a level of 5\% accuracy or better, the systematic errors
associated with the heavy quark have to be under good control for both
charm and bottom quarks. These errors should also be as similar as
possible in the two regimes in order for them to cancel
effectively. For these reasons, the relativistic and Fermilab
approaches seem to be preferable to the use of NRQCD. Indeed, NRQCD
involves an expansion of QCD in powers of $1/(am_Q)$ which requires
either the inclusion of high-orders or coarse lattices ($a^{-1}\ll
m_Q$) when $m_Q$ approaches the charm mass.  High orders are difficult
to implement in practice and coarse lattices imply large
discretization errors.

To reach such levels of accuracy, it is also important to study
carefully the extent to which uncertainties associated with the chiral
extrapolation of the form factors and with the presence of chiral
logarithms cancel in the ratio of bottom to charm amplitudes.

\subsubsection*{B to vector meson semileptonic decays}

The rate for $\rm B\rightarrow \rho l\nu$ is less strongly suppressed
kinematically near ${\qsqmax}$ than is the rate for ${\rm B}\rightarrow\pi
l\nu$ and it is larger overall.  Thus, the number of events will be
larger in the region where the lattice can compute the relevant
matrix elements reliably. In
[\ref{Flynn:1995dc}], the UKQCD collaboration suggested that $|V_{ub}|$
be obtained directly from a fit to the differential decay rate around
${\qsqmax}$, with the overall normalization of this rate, up to a
factor of $|V_{ub}|^2$, determined using lattice results. With their
lattice results, such a measurement would allow an extraction of
$|V_{ub}|$ with a 10\% statistical and a 12\% systematic error coming
from theory.~\footnote{Other early lattice work on ${\rm B}\rightarrow \rho
l\nu$ can be found in [\ref{Abada:1993dh},\ref{Allton:1994ui}].} A first
measurement of this differential rate has actually already been
performed by CLEO [\ref{Behrens:1999vv}]. 

\begin{figure}[tbp]
  \centering \includegraphics*[width=10.5cm]{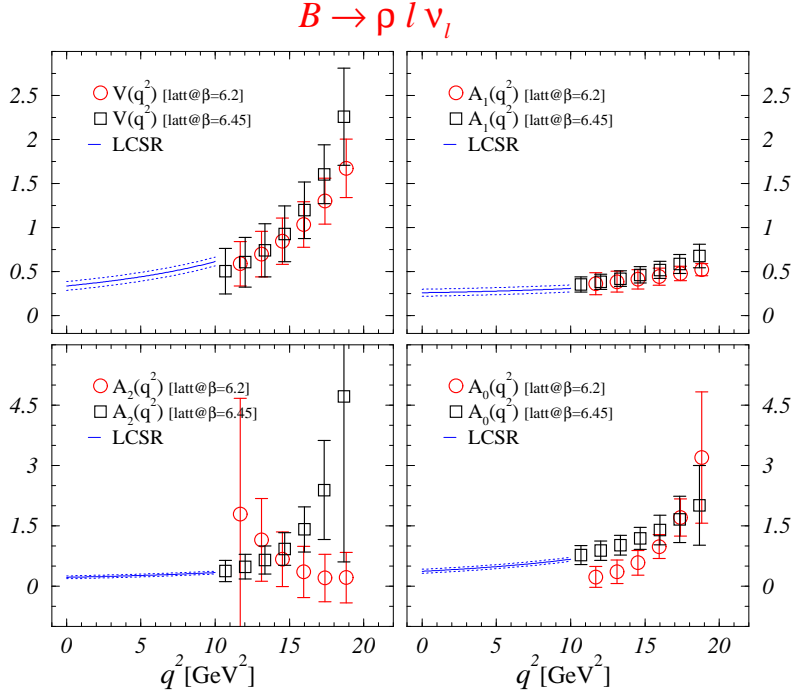} \caption{\it 
Example of quenched lattice results for ${\rm B}\rightarrow \rho l\nu$ 
form factors plotted as a function of $q^2$ [\ref{Abada:2002ie}]. These
results were obtained at two values of the inverse lattice spacing 
$1/a=3.7\,\gev$ and $2.7\,\gev$, corresponding to bare couplings
values $\beta=6.45$ and $6.2$ respectively. Also shown at low
$q^2$ are the light-cone sum rule results of [\ref{3Ball:1998kk}]. }
\label{fig:btorho}
\end{figure}

Very recently, two lattice collaborations (UKQCD [\ref{Gill:2001jp}]
and SPQcdR [\ref{Abada:2002ie}]) have begun revisiting ${\rm B}\rightarrow
\rho l\nu$ decays. Their calculations are performed in the 
quenched approximation and results are still 
preliminary. Shown in
\fig{fig:btorho} are 
the four independent form factors required to describe the
${\rm B}\rightarrow \rho$ s.l.\  matrix elements, as obtained by
SPQcdR [\ref{Abada:2002ie}] at two values of the lattice spacing. Also
shown are results from light-cone sum rule calculations
[\ref{3Ball:1998kk}] which are expected to be reliable for lower values
of $q^2$. These sum rule results look like very natural extensions of
the form factors obtained on the finer lattice. Combining the LCSR
results for $q^2\le 10\,\gev^2$ with the results on the finer lattice
for $q^2> 10\,\gev^2$ yields $\Gamma({\rm B}^0\to\rho^-l\nu)=(19\pm
4)|V_{ub}|^2\,\mathrm{ps}^{-1}$ [\ref{Abada:2002ie}]. It will be
interesting to see what the calculations of
[\ref{Gill:2001jp},\ref{Abada:2002ie}] give for the differential rate above
$10\,\gev^2$ once they are finalized.

As was the case for ${\rm B}\to\pi l\nu$ decays, derivation of the full
$q^2$ dependence of the form factors from lattice data involves a
large extrapolation from $q^2> 10\,\gev^2$ all the way down to
$q^2=0$. Here, the use of dispersion relations is complicated by the
singularity structure of the relevant correlation functions and form
factors. There exist, however, lattice-constrained parametrizations of
${\rm B}\rightarrow \rho l\nu$ form factors, which are consistent with
lattice results and heavy-quark scaling relations at large $q^2$, and
with kinematic constraints and light-cone sum rule scaling at $q^2=0$
[\ref{3DelDebbio:1997kr}].  These parametrizations provide simple,
few-parameter descriptions of s.l.\ form factors.~\footnote{Away from
$q^2=0$, these parametrizations are actually not fully consistent with
the large-recoil symmetry relations derived in
[\ref{3Charles:1998dr}] amongst the soft contributions to
the relevant form factors. For completeness, let us mention that the
$\alpha_s$ corrections to these symmetry relations were calculated in
[\ref{3Beneke:2000wa}] and corrections in powers of 
$1/m_b$ were investigated in 
[\ref{3Chay:2002vy},\ref{3Bauer:2002uv},\ref{3Beneke:2002ph}]} 
However, at values of the recoil
for which there are not lattice results (i.e. low $q^2$), they are not
predictions of (quenched) QCD.

\subsubsection{Summary}

Four groups have recently performed quenched lattice calculations of
${\rm B}\rightarrow\pi l\nu$ form factors for $q^2$ $\gsim$ $12\,\gev^2$ and
their results agree. Agreement is best for $\fpbpi$ which determines
the rate for these decays in the limit of vanishing lepton mass. The
error on this form factor is of order 20\%. The main sources of
remaining systematic errors are quenching and light-quark-mass
extrapolations for all the groups, and heavy-quark-mass extrapolations,
discretization, and perturbative matching, depending on the group.

A substantial reduction in the error (i.e. below 10\%) will be
difficult to achieve solely in lattice QCD.  This is where the use of
ratios of s.l.\  B and D rates, such as the one given in
\eq{eq:bdratio}, could be very helpful.

There is still a substantial number of improvements to be made to
present calculations. The list includes unquenching, the use of
dispersive bounds or other means of extending the kinematic reach of
lattice calculations, the determination of ratios of s.l.\  B
and D rates, and more investigations of ${\rm B}\rightarrow\rho l\nu$
decays.

\subsection{Heavy-to-light form factors from light-cone sum rules}

The QCD light-cone sum rules (LCSR) [\ref{lcsr},\ref{cz}]
provide estimates  of various 
heavy-to-light transition form factors. In particular, 
${\rm B}\to P,V$ form factors ($P=\pi,\rm K$ and
$V=\rho,\rm K^*,\phi$) have been calculated at small and 
intermediate momentum transfers, typically 
at $0<q^2 \leq m_b^2-2m_b\Lambda_{QCD}$. 
The upper part of this interval overlaps with the 
region accessible to the lattice calculations
of the same form factors, allowing one to compare the results 
of two methods. 
In what follows we will concentrate on the LCSR prediction 
for the ${\rm B}\to \pi$  form factor $f^+_{B\pi}$ 
[\ref{Bpi}--\ref{Ball1}].
Its accuracy has been recently improved in Ref.~[\ref{BallZ}].
For the LCSR ${\rm B}\to V$ form factors we refer to the NLO 
calculation in Ref.~[\ref{3Ball:1998kk}] and to the resulting parametrization
in Ref.~[\ref{AliB}]. 

The LCSR approach to calculate $f^+_{B\pi}$ 
employs a specially designed theoretical object, 
the  vacuum-to-pion correlation function 
\begin{eqnarray}
F_\mu (p,q)=
i\!\! \int \!d^4xe^{iqx}\langle \pi^+(p)\!\mid T\{\bar{u}\gamma_\mu b(x),
m_b\bar{b}i\gamma_5 d(0)\}\mid\! 0\rangle= F((p+q)^2,q^2)p_\mu+
O(q_\mu)\,,
\label{eq:lcsrcorr}
\end{eqnarray}
where the $b \to u$ weak current is correlated with  
the quark current which has the B meson 
quantum numbers, and $p^2=m_\pi^2$.
Writing down  the dispersion relation 
for the invariant amplitude $F$:
\begin{equation}
F((p+q)^2,q^2)=
\frac{2f_B f^+_{B\pi}(q^2)M_B^2}{M_B^2-(p+q)^2}
+\sum\limits_{B_h}\frac{2f_{B_h} f^+_{B_h\pi}(q^2)m_{B_h}^2}{m_{B_h}^2-(p+q)^2}~,
\label{eq:displcsr}
\end{equation}
one represents the correlation function (\ref{eq:lcsrcorr})
in terms of  hadronic degrees of freedom in  the B channel.
The ground-state contribution 
in Eq.~(\ref{eq:displcsr}) contains a product
of the B meson decay constant $f_B$ and the form factor 
$f^+_{B\pi}(q^2)$ we are interested in, whereas the 
sum over ${\rm B}_h$ accounts for  the contributions
of excited and continuum B states. 

The dispersion relation is then matched 
to the result of QCD calculation of $F((p+q)^2,q^2)$ 
at large virtualities, that is, at 
$\mid (p+q)^2-m_b^2 \mid \gg \Lambda_{QCD}^2$
and $q^2\ll m_b^2$. In this region the 
operator-product expansion (OPE) near the light-cone $x^2=0$
is employed:
\begin{equation}
F((p+q)^2,q^2)= \sum\limits_{t=2,3,4}
\int Du_i \sum\limits_{k=0,1}\left(\frac{\alpha_s}{\pi}\right)^k
~T^{(t)}_k((p+q)^2,q^2,u_i,m_b,\mu)
\varphi^{(t)}_\pi(u_i,\mu)\,. 
\label{eq:sr}
\end{equation}
This generic expression is a convolution 
of calculable short-distance
coefficient functions $T^{(t)}_k$ and universal pion light-cone 
distribution amplitudes (DA) $\varphi^{(t)}_\pi(u_i,\mu)$ of twist $t$. 
Here, $m_b$ is the one-loop $b$-quark pole mass, 
$\mu$  is the factorization scale and the integration 
goes over the pion momentum fractions $u_i={u_1,u_2,...}$
distributed among quarks and gluons, so that 
$Du_i\equiv du_1du_2...\delta(1-\sum_i u_i)$. 
In particular, $\varphi^{(2)}_\pi(u_1,u_2,\mu)=
f_\pi\varphi_\pi(u,\mu)$, ($u_1=u,u_2=1-u$)
where $\varphi_\pi$ is the lowest twist 2, 
quark-antiquark pion DA normalized to unity: 
\begin{equation}
\varphi_\pi(u,\mu)=6u(1-u)\left(1+\sum\limits_n a_{2n}(\mu)
C_{2n}^{3/2}(2u-1)\right)\,.
\label{eq:phipi}
\end{equation}
In the above, $C_{2n}$ are Gegenbauer polynomials and the oefficients
$a_n(\mu)$, that are suppressed logarithmically at large $\mu$,
determine the deviation of $\varphi_\pi(u)$ from its asymptotic form.
Importantly, the contributions to Eq.~(\ref{eq:sr}) corresponding to
higher twist and/or higher multiplicity pion DA are suppressed by
inverse powers of the $b$-quark virtuality $(m_b^2-(p+q)^2)$,
allowing one to retain a few low twist contributions in this
expansion.  Furthermore, one uses quark-hadron duality to approximate
the sum over ${\rm B}_h$ in Eq.~(\ref{eq:displcsr}) by a dispersion integral
over the quark-gluon spectral density, introducing a threshold
parameter $s_0^B$. The final step involves a Borel transformation
$(p+q)^2 \to M^2$, where the scale of the Borel parameter $M^2$
reflects the characteristic virtuality at which the correlation
function is calculated.

The resulting sum rule relation obtained by matching 
Eqs.~(\ref{eq:displcsr}) and (\ref{eq:sr})
can be cast in the following form:
\begin{eqnarray}
f_Bf_{B\pi}^+(q^2)=\frac{1}{M_B^2}
\exp\left (\frac{M_B^2}{M^2}\right)
\sum\limits_{t=2,3,4}~~\sum\limits_{k=0,1} \left(
\frac{\alpha_s}{\pi}\right)^{k}
{\cal F } ^{(t)}_{k}(q^2,M^2;m_b,s_0^B,\mu;\mbox{\{DA\}}^{(t)}) \,,
\label{eq:lcsrfBpi}
\end{eqnarray}
where the double expansion (in twists and in $\alpha_s$) 
and the dependence on the relevant parameters are made explicit.
In particular, $\{ DA \}^{(t)}$ denotes the non-perturbative 
normalization constant and non-asymptotic coefficients
for each given twist component, e.g., for $\varphi_\pi$:
$\{DA\}^{(2)}=\{f_\pi,a_i\}$.   
The sum rule (\ref{eq:lcsrfBpi}) includes all zeroth order in
$\alpha_s$, twist 2,3,4 contributions containing quark-antiquark and 
quark-antiquark-gluon DA of the pion. The perturbative expansion
has NLO accuracy, including the 
$O(\alpha_s)$ corrections to the twist 2 ~[\ref{Bpialphas}] 
and twist 3  coefficient functions, the latter recently calculated  
in Ref.~[\ref{BallZ}]. 
More details on the derivation
of LCSR (\ref{eq:lcsrfBpi}) and the explicit expressions  
can be found in the review papers~[\ref{KR}--\ref{CK}].

For the B meson decay constant entering LCSR (\ref{eq:lcsrfBpi})
one usually employs the conventional
SVZ sum rule [\ref{Shifman1}] for the two-point correlator
of $\bar{b}i\gamma_5 q$ currents with $O(\alpha_s)$ 
accuracy (a recent update of this sum rule [\ref{Jamin:2001fw}]
is discussed in Chapter 4 of the present document): 
\begin{equation}
f_B=
\sum\limits_{d=0,3 \div 6}~\sum\limits_{k=0,1}
\left(\frac{\alpha_s}{\pi}\right)^k
C_k^{(d)}(\overline{M}^2,m_b,s_0^B,\mu)\langle 0| \Omega_d(\mu)|0\rangle\,,
\label{eq:fBsr}
\end{equation}
where the expansion contains the perturbative term 
with dimension $d=0$ ($\Omega_0=1$), and, at $d\geq 3 $, goes over 
condensates, the vacuum averages of 
operators $\Omega_d=\bar{q}q, 
G^{a}_{\mu\nu}G^{a\mu\nu}, ....$, multiplied by calculable   
short-distance coefficients $C_k^{(d)}$. The Borel parameter 
$\overline{M}$ is correlated with $M$. 
The LCSR prediction for the ${\rm B}\to \pi$ form factor is finally obtained 
dividing Eq.~(\ref{eq:lcsrfBpi}) by Eq.~(\ref{eq:fBsr}):
\begin{equation}
f^+_{B\pi}(q^2)=(f_Bf^+_{B\pi}(q^2))_{LCSR}/(f_B)_{2ptSR}\,.
\label{eq:div}
\end{equation}
In order to demonstrate that the expansion in both twist and
$\alpha_s$ in this relation works well, we present the approximate
percentage of various contributions to the resulting form factor
(\ref{eq:div}): 

\vspace{2mm}

\begin{center}
\begin{tabular}{|cccc|}
\hline
twist& DA & LO& $O(\alpha_s)$ NLO\\
\hline
2   & $\bar{q}q$ & $\sim$  50 \% &  $\sim$ 5\%\\
\hline
 3   & $\bar{q}q$ & $\sim $ 40 \% &   $\sim$  1\%\\
\hline
\multicolumn{2}{|c}{
\begin{tabular}{cc}
 4   & $\bar{q}q$ \\ 
 3+4 & $\bar{q}qG$
\end{tabular} }
&  $\Big\} \sim$ 5 \%   & -\\
\hline
\end{tabular}
\end{center}

\vspace{2mm}

The input parameters used in the numerical analysis of  
the sum rules (\ref{eq:lcsrfBpi}) and (\ref{eq:fBsr}) have a limited accuracy.
The theoretical uncertainty is estimated  
by varying these inputs within the allowed regions and 
 adding up linearly the separate uncertainties induced by these
variations in the numerical prediction for $f^+_{B\pi}$. The resulting total
uncertainties are given below, together with  the parametrizations
of the form factor. A detailed theoretical error analysis can be found 
in Ref.~[\ref{Khodjamirian:2000ds}]. 
To summarize it briefly, one source of uncertainty 
is the value of the $b$-quark one-loop pole mass.
The two most recent LCSR analyses use 
$m_b= 4.7\pm 0.1 $ GeV~[\ref{Khodjamirian:2000ds}] and $m_b=4.6\pm 0.1$ GeV
[\ref{BallZ}]. In both studies, the threshold $s_0^B$ is not 
an independent parameter, being  determined 
by stabilizing $f_B$ calculated from Eq.~(\ref{eq:fBsr})
at a given $b$-quark mass. The uncertainty induced
by varying the factorization scale $\mu$ (adopted 
simultaneously as the normalization scale for $\alpha_s$) is very small,  
firstly, because the NLO approximation is implemented 
for both dominant twist 2 and 3 terms, and, secondly, because the 
relatively large $O(\alpha_s)$ corrections
to the twist 2 contribution and to the $f_B$ sum rule 
cancel in the ratio (\ref{eq:div}).
Another source of uncertainty is our limited knowledge of the 
non-asymptotic part in the pion DA
(determined by the coefficients $a_{2n}$ and the 
analogous coefficients in twist 3,4 DA). In Ref.~[\ref{Khodjamirian:2000ds}] 
these coefficients were varied  
from a certain non-asymptotic  ansatz 
of DA (motivated by QCD sum rules) 
to purely asymptotic DA. Such a substantial variation 
covers the existing constraints 
on non-asymptotic coefficients obtained from LCSR for pion 
form factors. The latter constraints have been used 
in Ref.~[\ref{BallZ}]. In fact, LCSR involve integration over 
normalized DA, therefore it is natural that 
the results only moderately depend on the  
non-asymptotic coefficients.
Finally, to assess the reliability of the LCSR procedure 
one has to comment on the use of quark-hadron duality, 
which is the most sensitive point in the sum rule approach. 
We expect that the sensitivity to 
the duality approximation is substantially reduced: 1) by 
restricting the Borel parameter at not too large values
and 2) by dividing out the $f_B$ sum rule which 
depends on the same threshold. The fact that the QCD sum rule 
prediction for $f_B$ (see Chapter 4 and [\ref{Jamin:2001fw}]) 
is in a good agreement with the lattice results 
indicates that quark-hadron duality is indeed valid in the B channel.

For a convenient use in the experimental analysis, 
the LCSR results for ${\rm B}\to \pi$ form factor 
are usually fitted to simple parametrizations. 
One of them, suggested in Ref.~ [\ref{Khodjamirian:2000ds}] employs
the ansatz [\ref{BecirevicKaidalov}] based
on the dispersion relation for $f_{B\pi}^+(q^2)$.
The latter is fitted to the LCSR predictions for $f^+_{B\pi}$ 
in its validity region $0< q^2 \leq 14-16$ GeV$^2$. For 
the  ${\rm B}^*$-pole term the ${\rm B}^*B\pi$  coupling [\ref{coupling}] is 
determined from the same correlation function (\ref{eq:sr}). The result is: 

\vspace{1mm}

\begin{equation}
f_{B\pi}^+(q^2)=\frac{0.23\div 0.33 }{(1-q^2/M_{B^*}^2)
(1-\alpha_{B\pi}q^2/M_{B^*}^2)}\,.
\label{eq:KRWWY}
\end{equation}

\vspace{1mm}

\noindent
where the values of the slope parameter correlated with the 
lower and upper limits of the interval for 
$f_{B\pi}^+(0)$ are almost equal:
$\alpha_{B\pi}=0.39\div 0.38$.
A different parametrization 
was suggested recently in~Ref.~[\ref{BallZ}]:
\begin{eqnarray}
f_{B\pi}^+(q^2)&&=\frac{0.26 \pm 0.06 \pm 0.05}{1-a(q^2/M_B^{2})+
b(q^2/M_B^2)^2},~~~ q^2<q_0^2,
\nonumber
\\
&&= \frac{c}{1- q^2/M_{B^*}^2}, ~~~q^2>q_0^2\,,
\label{eq:Ball}
\end{eqnarray}
where the LCSR result is extrapolated to large $q^2$ 
using the ${\rm B}^*$-pole form. In Eq.~(\ref{eq:Ball}) the ranges 
of fitted parameters, 
$a=2.34 \div 1.76$, $b=1.77\div 0.87 $, 
$c= 0.384\div 0.523$, and $q_0^2= 14.3\div 18.5$ GeV$^2$,
are correlated with the first error in $f_{B\pi}^+(0)$,
whereas the second error is attributed to the 
uncertainty of the quark-hadron duality  approximation.
Note that all values within the uncertainty intervals 
in Eqs.~(\ref{eq:KRWWY}) and (\ref{eq:Ball})
have to be considered as equally acceptable 
theoretical predictions, without any ``preferred central value''. 
The  numerical differences between the form factors
(\ref{eq:KRWWY}) and (\ref{eq:Ball}) are smaller than the 
estimated  uncertainties and are caused by slightly different inputs and
by the small  $O(\alpha_s)$ correction to the twist 3 term taken 
into account in Eq.~(\ref{eq:Ball}) but not in Eq.~(\ref{eq:KRWWY})
(where an additional uncertainty was attributed to this missing 
correction).

\begin{figure}[htbp] 
\centerline{
\epsfig{file=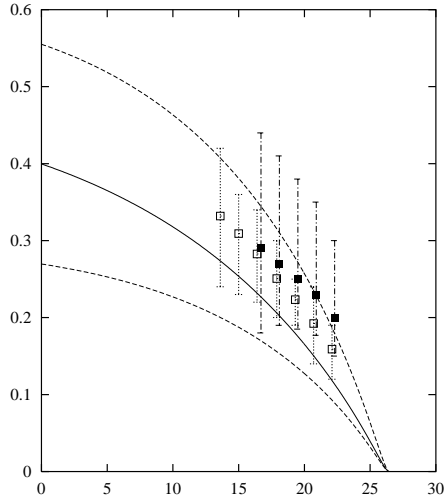,scale=0.45,%
clip=}}
\vspace{0.3cm}
\caption{{\it LCSR prediction for the ${\rm B}\to \pi l \nu$ 
decay distribution [\ref{Khodjamirian:2000ds}] at the nominal values of 
inputs (solid), with the interval of theoretical
uncertainties (dashed), compared with  some of the 
recent lattice calculations taken from Refs.~[\ref{UKQCD}] (solid points) and 
[\ref{Abada:2000ty}] (open points).}}
\label{fig:lcsr1}
\end{figure}

Having at hand the form factor, one can  predict 
the ${\rm B}\to \pi l \nu$ decay distribution using 
Eq.~(\ref{eq:differential_decay_rate}), as  is shown in Fig.~\ref{fig:lcsr1} in the case
of  the form factor 
(\ref{eq:KRWWY}). The corresponding 
integrated s.l.\  width 
is 
\begin{equation}
\Gamma(\rm B^0\to \pi^- l^+ \nu)=(7.3 \pm 2.5)|V_{ub}|^2 \mbox{ps}^{-1}\,,
\label{gamma}
\end{equation}
where the indicated error is mainly caused 
by the uncertainty of $f^+_{B\pi}(0)$, whereas 
the uncertainty of the form factor shape is insignificant.
This prediction [\ref{Khodjamirian:2000ds}] was recently used by BELLE 
in their preliminary analysis of ${\rm B}\to \pi l \nu$ decay
(see \sec{sec:bpi_expt}).
A similar estimate of $|V_{ub}|$ was obtained 
in Ref.~[\ref{Khodjamirian:2000ds}] using the older
CLEO measurement [\ref{cleopi}] of the  ${\rm B}\to \pi l \nu$ width.

The advantage of LCSR is that one can easily switch from 
the ${\rm B}\to \pi $ to $\rm D\to \pi $ form factor replacing $b$ quark by 
$c$ quark in the underlying correlation function (\ref{eq:sr}).
The LCSR prediction for $f_{D\pi}^+$ obtained 
in Ref.~[\ref{Khodjamirian:2000ds}] and 
parametrized in the form analogous to Eq.~(\ref{eq:KRWWY}) 
yields a  $\rm D^*$-pole dominance:
\begin{equation}
f_{D\pi}^+= \frac{0.65\pm 0.11}{1- q^2/m_{D^*}^2}\,,
\label{eq:KRWWYDpi}
\end{equation}
at $0<q^2<(M_D-m_\pi)^2$. The corresponding s.l.\  width
$\Gamma({\rm  D}^0\to\pi^- e^+ \nu_e)/|V_{cd}|^2=0.13 \pm 0.05 ~\mbox{ps}^{-1}$
calculated with the known value of $|V_{cd}|$ 
is, within errors,  in agreement with the experimental number 
$0.174 \pm 0.032~\mbox{ps}^{-1}$ [\ref{ref:pdg02}]. To make this comparison
more decisive it would be very important to have new, more accurate 
measurements of  the decay distribution and integrated  width of
${\rm D}^0\to\pi^- e^+ \nu_e$. 

Are further improvements of the LCSR result for $f^+_{B\pi}$ and 
$f^+_{D\pi}$ possible? As we have seen, the accuracy of  
OPE for the correlation function is quite sufficient.  
The  $O(\alpha_s^2)$ level recently
achieved in the sum rule for $f_B$ [\ref{Jamin:2001fw}] is certainly 
not an immediate task for LCSR, being also technically very difficult.
More important is to improve the accuracy of the input parameters by
1) narrowing the interval of the $b$ quark mass and 
2) gaining a better control over the parameters of pion DA.
For the latter, in particular, one needs more precise data on 
pion form factors,
especially on $\gamma^*\gamma \to \pi^0$ (the latter form factor 
can in principle be measured at the same $e^+e^-$ B-factories)  
and, eventually, lattice QCD simulations of $\varphi_\pi(u)$ and other DA.     
A better control over duality approximation in the B and D channels
can be achieved  if radially excited B and D states  
are accurately identified with their masses and widths. 
Optimistically, one may hope to reduce the overall 
uncertainty of the LCSR prediction for $f^+_{B\pi}$ and other  
heavy-to-light form factors 
to the level of $\pm 10 \%$, which is a natural limit for any 
QCD sum rule prediction.

In conclusion, we emphasize that, in addition to providing
estimates of the form factors, LCSR help in understanding important
physical aspects of the heavy-to-light transitions. First of all, LCSR
allow to {\em quantitatively} assess the role of the soft (end-point)
vs hard (perturbative gluon exchange) contributions to the form
factors, because both contributions are taken into account in this
approach. Secondly, using  LCSR one is able to predict [\ref{cz}] the
$m_b\to \infty $ limit, $f_{B\pi}^+(0)\sim 1/m_b^{3/2}$, which is used
in some lattice extrapolations.  Last but not least, LCSR can be
expanded in powers of $1/m_b$ and $1/E_\pi$ assessing the size of
$1/m_b$ and $1/E$ corrections to various relations predicted in
effective theories for heavy-to-light decays.

\subsection{Review and future prospects for the exclusive 
determination of $\vub$}

\subsubsection{Measurements of 
$BR({\rm B} \rightarrow \pi \ell \nu)$}
\label{sec:bpi_expt}

The first exclusive measurement of the mode ${\rm B} \rightarrow \pi \ell
\nu$ was presented by the CLEO collaboration in
1996~[\ref{cleopi}]. The neutrino momentum is inferred from the missing
momentum in the event, using the hermeticity of the detector. Events
with multiple charged leptons or a non-zero total charge are rejected,
resulting in a reduced efficiency in favour of an improved neutrino
momentum resolution. Isospin relations for the relative partial width
are used to combine the ${\rm B}^+$ and ${\rm B}^0$ modes.
A fit is performed using the variables $M_{cand}=\sqrt{E_{\rm beam}^2
  - |\vec{p}_\nu + \vec{p}_\ell + \vec{p}_{\rho,\omega,\pi}|^2}$ and
$\Delta E = (E_{\rho,\omega,\pi} + E_\ell + |\vec{p}_{\rm miss}|c) -
E_{\rm beam}$, where $E_{\rm beam}$ is the well known beam energy. The
modes ${\rm B} \rightarrow \rho \ell \nu$ (with $\rho^0$ and $\rho^-$) and
${\rm B} \rightarrow \omega \ell \nu$ are also included in the fit because
of cross-feed between these modes and ${\rm B} \rightarrow \pi \ell \nu$. The
$\rho$ ($\omega$) mode uses the invariant two (three) $\pi$ mass in
the fit to  distinguish better between resonant and non-resonant final
states. Backgrounds from continuum processes are subtracted using
off-resonance data. The shape of the five signal contributions, the $b
\rightarrow c$, and $b \rightarrow u$ backgrounds are provided by
Monte Carlo simulation. The final results for the branching ratio and
$|V_{ub}|$ are obtained by averaging over four separate form factor
calculations: two quark models (ISGW2~[\ref{isgw2}] and
Melikhov~[\ref{melikhov}]), a model by Wirbel Stech and
Bauer~[\ref{wsb}], and a hybrid model that uses a
dispersion-relation-based calculation of the $\pi\ell\nu$ form
factor~[\ref{burdman}] and combines lattice calculation of the
$\rho\ell\nu$ form factors~[\ref{flynn}] with predicted $\rho\ell\nu$
form factor relations~[\ref{stech}]. The dominant systematic
uncertainties arise from uncertainties in the detector simulation and
modelling of the $b \rightarrow u \ell \nu$ backgrounds. The result
using $2.66 \mbox{ fb}^{-1}$ on resonance data is
\begin{eqnarray}
{\cal B}({\rm B}^0\rightarrow~\pi^-~e^+~\nu) & = & (1.8~\pm~0.4~\pm~0.3~\pm~0.2)~\times 10^{-4}\mbox{ , and}\\
|V_{ub}| & = & (3.3~\pm~0.2~{}^{+0.3}_{-0.4}~\pm~0.78)~\times~10^{-3}\;.\label{eq:cleopi}
\end{eqnarray}
The errors given are statistical, systematic, and theoretical, in the
order shown. Note that the above value of $|V_{ub}|$ is extracted
using both the $\pi$ and $\rho$ modes.
At ICHEP 2002 BELLE presented a preliminary result using $60\mbox{
fb}^{-1}$ on-peak and $9\mbox{ fb}^{-1}$ off-peak
data~[\ref{bellepiICHEP}]. Results are quoted for the UKQCD
model~[\ref{3DelDebbio:1997kr}]
\begin{eqnarray}
{\cal B}({\rm B}^0\rightarrow~\pi^-~e^+~\nu) & = & (1.35~\pm~0.11~\pm~0.21)~\times 10^{-4}\mbox{ , and}\\
|V_{ub}| & = & (3.11~\pm~0.13~\pm~0.24~\pm~0.56)~\times~10^{-3}\;
\end{eqnarray}
and for the LCSR model~[\ref{Khodjamirian:2000ds}]
\begin{eqnarray}
{\cal B}({\rm B}^0\rightarrow~\pi^-~e^+~\nu) & = & (1.31~\pm~0.11~\pm~0.20)~\times 10^{-4}\mbox{ , and}\\
|V_{ub}| & = & (3.58~\pm~0.15~\pm~0.28~\pm~0.63)~\times~10^{-3}\;.
\end{eqnarray}

The CLEO collaboration submitted a preliminary updated
analysis~[\ref{cleovubICHEP}]
to ICHEP 2002 based on $9.7 \times 10^6$ ${\rm B}\overline{\rm B}$ pairs.  In
addition to more data compared to Ref.~[\ref{cleopi}], the analysis has
been improved in several ways: the signal rate is measured
differentially in three $q^2$ regions so as to minimize modelling
uncertainties arising from the $q^2$ dependence of the form factors
(this is the first time this has been done in the ${\rm B} \rightarrow \pi
\ell \nu$ mode); minimum requirements on the signal charged lepton
momentum were lowered for both the pseudoscalar and vector modes,
thereby increasing the acceptance and also reducing the model
dependence; and the $X_u \ell \nu$ feed-down modelling included a
simulation of the inclusive process using a parton-level calculation
by De Fazio and Neubert~[\ref{deFazioNeubert}],
its non-perturbative parameters measured in the CLEO analysis of the
${\rm B} \rightarrow X_s \gamma$ photon energy
spectrum~[\ref{shape2},\ref{bsgammacleo}],
with the ISGW2~[\ref{isgw2}] model used to describe a set of expected resonant
states\footnote{Note that the inclusive rate is reduced to allow for
that portion of the total rate that is treated exclusively by the ISGW2
model.}.  The preliminary CLEO result~[\ref{cleovubICHEP}] for the branching
fraction was
\begin{eqnarray}
{\cal B}({\rm B}^0\rightarrow~\pi^-~e^+~\nu) & = &
(1.376~\pm~0.180~^{+0.116}_{-0.135}~\pm~0.008~\pm~0.102~\pm~0.021)~\times 10^{-4},
\end{eqnarray}
where the uncertainties are statistical, experimental systematic, and
the estimated uncertainties from the $\pi\ell\nu$ form factor, the
$\rho\ell\nu$ form factors, and from modelling the other ${\rm B}\to X_u \ell
\nu$ feed-down decays, respectively.  By extracting rates independently
in three separate $q^2$ ranges, the CLEO analysis demonstrated a
significant reduction in the model dependence due to efficiency
variations as a function of $q^2$.

In a preliminary effort to reduce the impact of theoretical
uncertainties on the form factor normalization, the CLEO
collaboration~[\ref{cleovubICHEP}] used $q^2$-dependent partial
branching fractions to extract $|V_{ub}|$ using a $\pi\ell\nu$ form
factor from light cone sum rules in the range $q^2 < 16$~GeV$^2$ and
from lattice QCD calculations above this range to obtain the averaged
preliminary result
\begin{eqnarray}
|V_{ub}| & = & (3.32~\pm~0.21~\pm~^{+0.17}_{-0.19}~\pm~^{+0.55}_{-0.39}~\pm~0.12~\pm~0.07)~\times~10^{-3},
\end{eqnarray}
where the uncertainties represent the same quantities defined in the
branching-fraction expression above.  In addition, by performing
simple $\chi^2$ fits of $|V_{ub}|$ across the three $q^2$ ranges with
a given form factor model, the CLEO method can discriminate between
competing form factor model shapes on the basis of $\chi^2$
probabilities in the fits to the data.  The CLEO technique has been
used, for example, to demonstrate that the ISGW2~[\ref{isgw2}] model is
likely to be unreliable for the extraction of $|V_{ub}|$ from the
$\pi\ell\nu$ mode.

\subsubsection{Measurements of  
$BR({\rm B} \rightarrow \rho \ell \nu$)}

Analyses that are optimized for the modes ${\rm B} \rightarrow \rho \ell
\nu$ were performed by CLEO~[\ref{Behrens:1999vv}] and
BaBar~[\ref{babarrho}]. BELLE also presented a preliminary result at
ICHEP 2002~[\ref{bellerho}]. Again the modes \brholnu, \brhochglnu,
\bomegalnu, \bpilnu, and \bpichglnu\ (with $\rho^0 \rightarrow
\pi^+\pi^-$, $\rho^- \rightarrow \pi^0\pi^-$, and $\omega \rightarrow
\pi^0\pi^+\pi^-$) are fully reconstructed, the inclusion of charge
conjugate decays is implied throughout. The neutrino momentum is
inferred from the missing momentum in the event. The selection is
somewhat looser than for the other analysis (see above), resulting in
a higher efficiency but decreased $\Delta E$ resolution.
Off-resonance data, taken below the $\Upsilon(4S)$ resonance, are used
for continuum subtraction. The shape of the five signal contributions,
the $b \rightarrow c$, and $b \rightarrow u$ background are provided
by Monte Carlo simulation. A fit with the two variables $M_{\pi \pi
(\pi)}$ and $\Delta E$ is performed, simultaneously for the five decay
modes and for two (for CLEO three) lepton-energy regions. $M_{\pi \pi
(\pi)}$ is the invariant hadronic mass of the $\rho$ ($\omega$) meson
and $\Delta E$ is the difference between the reconstructed and the
expected B meson energy, $\Delta E \equiv E_{\rho,\omega,\pi} +
E_{\ell} + |{\vec{p}}_{{\rm miss}}|c - E_{\rm beam}$. These analyses
are most sensitive for lepton energies above $2.3\mbox{ GeV}$, below
that backgrounds from $b \rightarrow c \ell \nu$ decays
dominate. Isospin and quark model relations are again used to couple
the ${\rm B}^+$ and ${\rm B}^0$ and $\rho$ and $\omega$ modes. The dominant
systematic uncertainties arise from uncertainties in the detector
simulation and modelling of the $b \rightarrow u \ell \nu$
backgrounds.\\
The CLEO and BaBar analyses obtain their results for the branching
ratio and $|V_{ub}|$ by averaging over five separate form factor
calculations: two quark models (ISGW2~[\ref{isgw2}] and
Beyer/Melikhov~[\ref{beyer98}]), a lattice calculation
(UKQCD~[\ref{3DelDebbio:1997kr}]), a model based on light cone sum rules
(LCSR~[\ref{3Ball:1998kk}]), and a calculation based on heavy quark and $SU(3)$
symmetries by Ligeti and Wise~[\ref{ligetiwise}]. CLEO published the 
result~[\ref{Behrens:1999vv}]
\begin{eqnarray}
{\cal B}({\rm B}^0\rightarrow~\rho^-~e^+~\nu) & = & (2.69~\pm~0.41~{}^{+0.35}_{-0.40}\pm~0.50)~\times 10^{-4}\mbox{ , and}\\
|V_{ub}| & = & (3.23~\pm~0.24~{}^{+0.23}_{-0.26}~\pm~0.58)~\times~10^{-3}\;.
\end{eqnarray}
BaBar uses $50.5\mbox{ fb}^{-1}$ on resonance and $8.7\mbox{ fb}^{-1}$
off-resonance data and obtains the preliminary result~[\ref{babarrho}]
\begin{eqnarray}
{\cal B}({\rm B}^0\rightarrow~\rho^-~e^+~\nu) & = & (3.39~\pm~0.44~\pm~{0.52}\pm~0.60)~\times 10^{-4}\mbox{ , and}\\
|V_{ub}| & = & (3.69~\pm~0.23~\pm~0.27~{}^{+0.40}_{-0.59})~\times~10^{-3}\;.
\end{eqnarray}
BELLE quotes preliminary results only for the ISGW2 model (without theoretical error)
using $29\mbox{ fb}^{-1}$ on resonance and $3\mbox{ fb}^{-1}$ off-resonance data
\begin{eqnarray}
{\mathcal B}(\rm B^+ \rightarrow \rho^0 \ell^+ \nu) & = & (1.44 \pm 0.18 \pm 0.23) \times 10^{-4}\mbox{ , and}\\
|V_{ub}| & = & (3.50 \pm 0.20 \pm 0.28) \times 10^{-3}\,.
\end{eqnarray}
Another result was obtained by CLEO earlier 
(this analysis was described in the previous Section [\ref{cleopi}])
\begin{eqnarray}
{\cal B}({\rm B}^0\rightarrow~\rho^-~e^+~\nu) & = & (2.5~\pm~0.4~{}^{+0.5}_{-0.7}~\pm~0.5)~\times 10^{-4}\mbox{ , and}\\
|V_{ub}| & = & (3.3~\pm~0.2~{}^{+0.3}_{-0.4}~\pm~0.78)~\times~10^{-3}\;\mbox{ (as in Eq. \ref{eq:cleopi})}. \nonumber
\end{eqnarray}
Note that the above value of $|V_{ub}|$ is extracted using both the $\pi$ and $\rho$ modes.
CLEO quotes the following average result for the two analyses that were presented in 
Refs.~[\ref{cleopi},\ref{Behrens:1999vv}]:
\begin{equation}
|V_{ub}| = (3.25~\pm~0.14~{}^{+0.21}_{-0.29}~\pm~0.55)~\times~10^{-3}\;.
\end{equation}

More recently, the CLEO collaboration has presented a preliminary
analysis~[\ref{cleovubICHEP}] that uses the neutrino-reconstruction
technique to reconstruct the modes ${\rm B}\to \rho~\ell~\nu$ in a
self-consistent way along with the other experimentally accessible $b
\to u~\ell~\nu$ exclusive modes.  Whereas the analyses described in
Refs.~[\ref{Behrens:1999vv},\ref{babarrho},\ref{bellerho}] 
are principally sensitive to
the lepton endpoint region above 2.3~GeV, the improved CLEO
measurement~[\ref{cleovubICHEP}] imposes a charged-lepton momentum
criterion of 1.5~GeV/$c$ with a view to reducing the dominating
theoretical uncertainties.  Due to the large uncertainties in
$\rho\ell\nu$ from modelling the simulated feed-down ${\rm B}\to X_u\ell\nu$
backgrounds, at the time of ICHEP 2002 the $\rho\ell\nu$ mode was not
used by CLEO to determine a preliminary $|V_{ub}|$ value.

\subsubsection{Measurements of 
$BR({\rm B} \rightarrow \omega \ell \nu$)}

A first preliminary result was presented at ICHEP 2002 by the BELLE
collaboration~[\ref{belleomega}]. The analysis uses electrons with
$E>2.2$~GeV and is based on $60\mbox{ fb}^{-1}$ on resonance
and $6\mbox{ fb}^{-1}$ off-resonance data. Events are selected by
requiring that the missing mass is consistent with zero ($M_{\rm
miss}^2<3.0\mbox{ GeV}/c^2$), and that the Dalitz amplitude is $75\%$
of its maximum amplitude ($A =|p_{\pi^+} \times p_{\pi^-}|>0.75 \times
A_{\rm max}$). After subtraction of all backgrounds, $59\pm 15$ signal
events remain. The dominant systematic error is the background
estimation ($18\%$). The preliminary result using the ISGW2
form factors is
\begin{equation}
{\mathcal B}({\rm B}^+ \rightarrow \omega e^+ \nu)= (1.4 \pm 0.4 \pm 0.3) \times 10^{-4}\,.
\end{equation}
No value for $|V_{ub}|$ is given for this analysis.

\subsubsection{Measurements of 
$BR({\rm B} \rightarrow \eta \ell \nu$)}

As in the case of the ${\rm B} \rightarrow \pi \ell \nu$ mode, the decay 
$\eta \ell \nu$ is described
by only one form factor; however, the extraction of $|V_{ub}|$ is
complicated by the $\eta-\eta^\prime$ mixing. Experimentally, the
$\eta$ has a clear signal and, due to its large mass, one can study the
region of low $\eta$ momenta, where lattice calculations are most
reliable. ${\rm B} \rightarrow \eta \ell \nu$ decays can be related via
Heavy Quark Symmetry to $\rm D \rightarrow \eta \ell \nu$. It is
envisioned that future measurements of the latter mode by CLEO-c can
be used to calibrate the lattice calculations, and the B-factories can
then use the calibrated lattice to measure ${\rm B} \rightarrow \eta \ell
\nu$. A first preliminary result using approximately $9.7\times10^6$
${\rm B}\overline{\rm B}$ events was presented at DPF 2002 by the CLEO
collaboration~[\ref{cleoeta}]:
\begin{equation}
{\mathcal B}({\rm B}^+ \rightarrow \eta \ell^+ \nu)= 
( 0.39 {}^{+0.18}_{-0.16} {}^{+0.09}_{-0.08}) \times 10^{-4}\,.
\end{equation}
The separate CLEO global exclusive study~[\ref{cleovubICHEP}],
submitted to ICHEP 2002, also found evidence for the mode ${\rm B}^+
\rightarrow \eta \ell^+ \nu$ with a significance of 2.5$\sigma$.  No
value for $|V_{ub}|$ was determined from these analyses.

\subsubsection{Summary}
\begin{figure}[htbp] 
\begin{center}
\epsfig{file=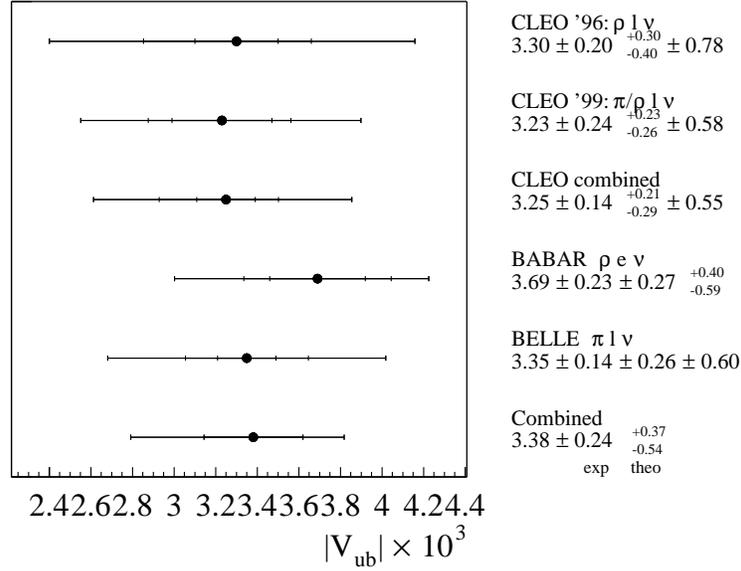, width=11cm}
\caption{\it Current s.l.\  exclusive measurements of
$|V_{ub}|$. The combined value is explained in the text.}
\label{fig:combined}
\end{center}
\end{figure}

Several mature measurements of the channels ${\rm B} \rightarrow \pi \ell
\nu$ and ${\rm B} \rightarrow \rho \ell \nu$ exist and can be used to extract
the value of $|V_{ub}|$. That these results are limited by
 the large theoretical uncertainties on the heavy-to-light form factor
 shapes and normalizations renders the exclusive approaches important
 to help clarify the non-perturbative QCD aspect of these decays, besides
  providing  an alternative avenue to $|V_{ub}|$.  With larger data
 samples, increased experimental acceptances, and improvements in our
 understanding of the background processes, the competing form factor
 models and calculations can now begin to be tested  through
 shape-sensitive comparisons with data.
A summary of some of the results is shown in
Fig.~\ref{fig:combined}. For the BELLE ${\rm B} \rightarrow \pi \ell \nu$
result the average of the two form factor model results is shown. 

A combined value (the last row in Fig.~\ref{fig:combined}) has been
calculated as weighted average of the combined CLEO result, the BaBar
${\rm B} \rightarrow \rho e \nu$ result and 
BELLE's ${\rm B} \rightarrow \pi \ell
\nu$ result. The weights are determined by the statistical error added
in quadrature with the uncorrelated part of the systematic
uncertainty. We assume that the systematic uncertainty is composed
quadratically out of an uncorrelated part and a correlated part of
about equal size, where the correlated part arises mainly from the
modelling of the $b\to u $ feed-down background. The experimental error of the
combined value includes this correlated contribution. The relative
theoretical error is similar for all measurements; we take the one
from the BaBar measurement. The result is


\begin{equation}
\begin{array}{|c|}\hline
|V_{ub}|_{excl} = (3.38~\pm~0.24_{exp}~{}^{+0.37}_{-0.54\;th})~\times~10^{-3}\;.\\ 
\hline
\end{array}
\end{equation}


\setcounter{footnote}{0}
\section{B hadron lifetimes and lifetime differences}
\label{sec:introlife}

Beside the direct determination of inclusive and exclusive 
s.l.\  decay widths, there are several other measurements of 
B meson properties which 
are  instrumental in testing some of the theoretical tools  
(OPE,  HQET, and lattice QCD) and are relevant in the precision 
determination of the CKM parameters.
For instance,  a precise evaluation of $\Delta M_d$ from the measurement of 
the time integrated $\Bd-\Bdb$ oscillation rate requires 
an accurate measurement of the $\Bdb$ meson lifetime. 
The accuracy of the $\Bdb$ lifetime and of the lifetime 
ratio of charged to neutral mesons are also a source of 
uncertainty in the extraction of  $\left| V_{cb} \right|$ with the exclusive 
method.
Measurements of B lifetimes test the decay dynamics, giving 
important information on non-perturbative QCD effects  induced by 
the spectator quark(s). 
Decay rates are expressed using the
OPE formalism, as an expansion  in $\Lambda_{QCD}/m_Q$. 
Spectator effects contribute at  $O(1/m_Q^3)$
and non-perturbative contributions  can be reliably evaluated, at least in principle, using 
lattice QCD calculations.

\begin{figure}[htbp] 
\begin{center}
\includegraphics[width=127mm]{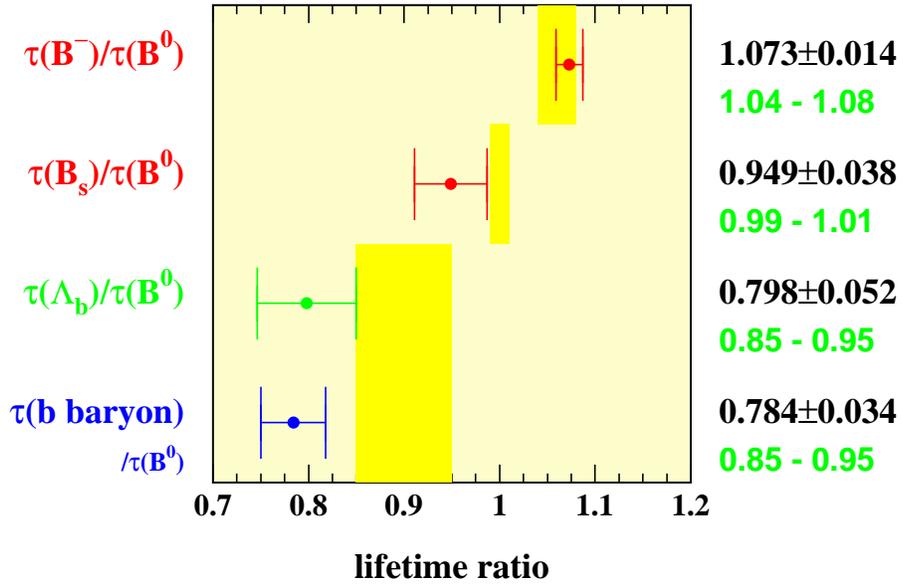}


\caption{\it Ratios of exclusive B hadrons lifetimes~[\ref{lifeWG}], 
compared with the theoretical predictions given in Secs.~\ref{sec:dgbstheo} 
and \ref{sec:blifetheory} and shown by the dark 
yellow bands.}\label{3fig:liferatio}
\end{center}
\end{figure}

Since the start of the data taking at LEP/SLC/Tevatron, an intense activity 
has been devoted to studies of inclusive and exclusive B hadron lifetimes. 
Most of the exclusive lifetime measurements are based on the 
reconstruction of the beauty hadron proper time by determining its decay 
length and momentum. The most accurate measurements are based on inclusive 
or partial reconstructions (such as topological reconstruction of B decay 
vertex and determination of its charge or reconstruction of 
${\rm B} \to \bar{{\rm D}^{(*)}} \ell^+ \nu X$). These techniques 
exploit the kinematics offered by $e^+e^-$ colliders at energies around the 
$Z^0$ peak, and also by hadron colliders, and the excellent tracking 
capabilities of the detectors. The accuracy  of the results for ${\rm B}_d$ 
and ${\rm B}_u$ mesons, where the samples of candidates 
are larger, are dominated by systematics, including backgrounds, $b$-quark 
fragmentation, branching fractions and modelling of the detector response.
In the case of ${\rm B}_s$ and $\Lambda_b$, the uncertainty is still 
statistical dominated. Final averages of the results obtained are given in 
Table~\ref{table:life} [\ref{lifeWG}]. The averages for the ${\rm B}^0_d$ 
and ${\rm B}^+$ lifetimes include also the recent 
very precise measurements by the B factories [\ref{lifeBfact}].
\begin{table}[htbp] 
\begin{center}
\begin{tabular}{|@{}ll|}
\hline
 ~B Hadrons                            &    ~~~~~~ Lifetime~[ps]                 \\ \hline 
~~$\tau({b})$                         & 1.573 $\pm$ 0.007 ~ (0.4~$\%$) \\ 
~~$\tau({{\rm B}^0_d})$                     & 1.540 $\pm$ 0.014 ~ (0.9~$\%$) \\ 
~~$\tau({{\rm B}^+})$                       & 1.656 $\pm$ 0.014 ~ (0.8~$\%$) \\
~~$\tau({{\rm B}^0_s})$                     & 1.461 $\pm$ 0.057 ~ (3.9~$\%$) \\
~~$\tau({\Lambda^0_b})$               & 1.208 $\pm$ 0.051 ~ (4.2~$\%$) \\ \hline
\multicolumn{2}{c}{$\tau({{\rm B}^+_u})/\tau({{\rm B}^0_d})$        ~~~~~=~ 1.073 $\pm$ 0.014}         \\
\multicolumn{2}{c}{$\tau({{\rm B}^0_s})/\tau({{\rm B}^0_d})$       ~~~~~~=~ 0.949 $\pm$ 0.038 }        \\
\multicolumn{2}{c}{$\tau({\Lambda^0_b})/\tau({{\rm B}^0_d})$ ~~~~~~=~ 0.798 $\pm$ 0.052 }        \\ 
\multicolumn{2}{c}{$\tau({b\,{\mathrm{baryon}}})/\tau({{\rm B}^0_d})$       ~=~ 0.784 $\pm$ 0.034 }        \\ 
\end{tabular}
\end{center}

\vspace{-2mm}

\caption{\it Summary of B hadron lifetime results provided by the Lifetime 
Working Group [\ref{lifeWG}].}\label{table:life}
\end{table}
Fig.~\ref{3fig:liferatio} gives the ratios of different B hadron lifetimes,
compared with theory predictions (dark yellow bands).
The achieved experimental precision of the hadron lifetimes -- 
from a fraction of percent to a few percent -- is quite remarkable. 
The phenomenological interpretation 
of these results in terms of exclusive lifetime ratios is discussed 
extensively in Sec~\ref{sec:blifetheory} 

The longer lifetime of charged B mesons as compared to
the neutral ones has been established at 5$\sigma$ level.
The ${\rm B}^0_d$ and ${\rm B}^0_s$ lifetimes are found to be equal 
within a $\simeq$4$\%$ accuracy. The lifetimes of b-baryons 
appear to be shorter than those of ${\rm B}^0_d$ mesons. Although this is in 
qualitative agreement with expectations, the magnitude of the lifetime 
ratio of beauty baryons to  mesons has been the subject of 
intense scrutiny, both by experiments and theorists, in view of a possible
discrepancy. Indeed, recent calculations of higher order terms have improved 
the agreement of $b$ baryon lifetime predictions with the present experimental 
results.
The most precise determinations of the $b$ baryon lifetimes 
come from two classes of partially reconstructed decays. The first has 
a $\Lambda_c^+$ baryon exclusively reconstructed in association with a 
lepton of opposite charge. The second uses more inclusive final states, 
where the enrichment in beauty baryons is obtained by requiring a proton 
or a $\Lambda^0$ to be tagged together with a lepton in the decay. 
These measurements are affected by uncertainties related to the $\Lambda_b$ 
polarization and to poorly known beauty baryon fragmentation functions and 
decay properties.

\newcommand{\dgs}{\Delta\Gamma_s}

Accessing the lifetime differences $\dgs$ offers also an independent
possibility of constraining the CKM unitarity triangle. This quantity is
sensitive to a  combination of CKM parameters very similar to the one entering 
$\Delta M_s$ (see Eq.~(\ref{eq:bec}) below), 
and an upper bound on $\dgs$ translates in a upper bound on $\Delta M_s$.
With future accurate determinations, this method 
can therefore provide, in conjunction with the determination of $\Delta M_d$,  
an extra constraint on the $\bar \rho$ and $\bar \eta$ parameters.

In the Standard Model the width difference $\left(\Delta\Gamma/\Gamma\right)$ 
of ${\rm B}_s$ mesons is expected to be rather large and within the reach of experiment
in the near future. Recent experimental studies already provide an 
interesting bound on this quantity as will be detailed in Sec.~\ref{sec:dgexp}
On the other hand, 
 the two mass eigenstates of the neutral ${\rm B}_d$ system 
 have in the SM only slightly different lifetimes.
This is because  the difference in the lifetimes is
CKM-suppressed with respect to that in the ${\rm B}_s$ system.
A rough estimate leads to
$\frac{\Delta \Gamma_d}{\Gamma_d} \, \sim \, 
\frac{\Delta \Gamma_s}{\Gamma_s} \cdot \lambda^2
\approx  0.5 \%~,$ 
where  
$\Delta \Gamma_s/\Gamma_s \approx 15\%$ [\ref{BBGLN},\ref{Ben}]. 

\boldmath 
\subsection{Theoretical description of the width difference 
of ${\rm B}_s$ mesons}
\label{sec:dgbstheo}
\unboldmath 

The starting point in the study of beauty hadron lifetimes is the construction 
of the effective weak Hamiltonian for the $\Delta B=1$ transitions, which is 
obtained after integrating out the heavy degrees of freedom of the $W$ and 
$Z^0$-bosons and of the top quark. 

Neglecting the Cabibbo suppressed contribution of $b \to u$ transitions 
and terms proportional to $\vert V_{td}\vert/\vert V_{ts}\vert$ 
 ($\sim \lambda$) in the penguin sector, 
the $\DB=1$ effective Hamiltonian can  be written 
as (cf.\ Eq.~(\ref{intro:b1}) of Chapter 1)
\begin{equation}
\label{eq:hdb1}
\Heff^{\DB=1}=
\frac{G_F}{\sqrt{2}} V^\ast_{cb} 
\sum_i C_i(\mu)Q_i + h.c. 
\end{equation}
The explicit expressions for the various operators 
can be found e.g.\ in [\ref{nlodb1}].
The Wilson coefficients $C_i(\mu)$ 
in the effective
Hamiltonian contain the information about the physics at short distances 
(large energies) and are obtained by matching the full (Standard Model) 
and the  effective theory ($\Heff^{\DB=1}$) at the scale $\mu \simeq M_W$. 
This matching, as well as the evolution from $M_W$ to the typical scale
 $\mu \simeq m_b$, 
are known at the next-to-leading order (NLO) in perturbation 
theory~[\ref{nlodb1}].

Through the optical 
theorem,  the width difference of ${\rm B}_s$ mesons can be related 
to the absorptive part of the forward scattering amplitude
\begin{equation}
\Delta \Gamma_{{\rm B}_s} =-\frac{1}{M_{{\rm B}_s}} \mathrm{Im} 
\langle \bar {\rm B}_s \vert {\cal T} \vert {\rm B}_s 
\rangle\, ,
\label{eq:master}
\end{equation}
where the transition operator ${\cal T}$ is written as
\begin{equation}
{\cal T} = i \int d^4x \; T \left( {\cal H}_{ eff}^{\Delta B=1}(x) 
{\cal H}_{ eff}^{\Delta B=1}(0) \right)\,,
\label{eq:T1}
\end{equation}
in terms of the $\Delta B=1$ effective Hamiltonian. 

\begin{figure}[htbp] 
\begin{center}
\epsfxsize=12.5cm
\epsfbox{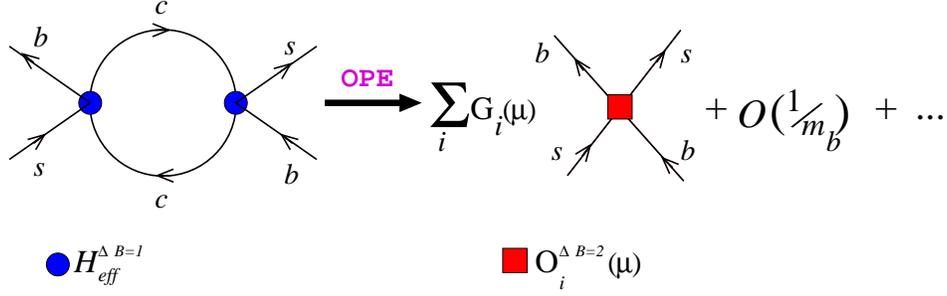}
\caption{\it \label{figA1} Heavy quark expansion: the non-local 
T-product of the l.h.s. (with the doubly inserted 
${\cal H}_{eff}^{\Delta B=1}$) is expanded in 
the series in $1/m_b$, each coefficient being the sum of local 
${\Delta B=2}$ operators.}
\end{center}
\end{figure}


Because of the large mass of the $b$-quark, it is possible to construct an OPE 
for the transition operator ${\cal T}$, which results in a sum of local 
operators of increasing dimension. The contributions of higher dimensional
operators are suppressed by higher powers of the $b$-quark mass. In the case of
the width difference $\dgamma$, the leading term in the expansion is 
parametrically of order $16 \pi^2 (\Lambda_{\rm QCD}/m_b)^3$. The result of 
this second OPE, which is illustrated in Fig.~\ref{figA1}, reads

\begin{equation}
 \label{eq3}
\Delta \Gamma_{{\rm B}_s} = \frac{G_F^2 m_b^2}{12 \pi M_{{\rm B}_s}} \vert V_{cb}^\ast 
V_{cs} \vert^2 \ \biggl\{ G_1(\mu) \langle 
\bar {\rm B}_s\vert O_1(\mu)\vert {\rm B}_s\rangle +  G_2(\mu) 
\langle \bar {\rm B}_s\vert O_2(\mu)\vert  {\rm B}_s\rangle + 
\delta_{1/m_b} \biggr\}\  ,
\end{equation}
where the $\Delta B=2$ operators on the r.h.s. are
\begin{eqnarray}
 & & O_1= \bar b \gamma_\mu (1-\gamma_5) s \ \bar b
 \gamma_\mu (1 - \gamma_5) s\,, \nonumber \\
& & O_2 = \bar b (1-\gamma_5) s \ \bar b  (1 - \gamma_5) s\,,
\end{eqnarray}

\noindent where a sum over 
repeated colour indices ($i,j$) is  understood;
 $\delta_{1/m_b}$ contains the $1/m_b$ correction [\ref{benekebd}]. 
Contributions 
proportional to $1/m_b^n$ ($n\geq 2$) are neglected. The short distance 
physics effects (above the scale $\mu$) are now encoded in the 
coefficient functions 
$G_{1,2}(\mu)$ which are combinations of the $\Delta B=1$ Wilson coefficients. 

The NLO corrections to the coefficients $G_{1,2}$ have been computed 
in Ref.~[\ref{BBGLN}]. They are large ($\sim 35\%
$) and  their inclusion is 
important. 
The long distance QCD dynamics is described in Eq.~(\ref{eq3}) by the matrix 
elements of the local operators $O_1$ and $O_2$, which are parametrized as
\vspace{0.5mm}
\begin{equation}
 \label{eq1}
\langle \bar {\rm B}_s\vert O_1(\mu)\vert {\rm B}_s\rangle  = \frac{8}{3} F_{{\rm B}_s}^2 
M_{{\rm B}_s}^2 B_1 (\mu)\,,\qquad \langle \bar {\rm B}_s\vert O_2(\mu)\vert  {\rm B}_s\rangle 
= -\frac{5}{3} \left(\frac{F_{{\rm B}_s} M_{{\rm B}_s}^2}{m_b(\mu)+ m_s(\mu)} \right)^2  
B_2(\mu) \,,
\end{equation} 
\vspace{0.5mm}
where the ${\rm B}$-parameters are equal to unity in the vacuum saturation
approximation (VSA). To measure the deviations from the VSA one should also 
include the non-factorizable (non-perturbative) QCD effects. For such a 
computation a suitable framework is provided by the lattice QCD simulations. 
In principle, the lattice QCD approach allows 
the fully non-perturbative estimate of the hadronic quantities to an arbitrary 
accuracy. In practice, however, several approximations need to be made which, 
besides the statistical, introduce also a systematic uncertainty in the 
final results. The steady progress in increasing the computational power, 
combined with various theoretical improvements, helps reducing ever more  
systematic uncertainties. 
Various approximate treatments of the heavy quark on the lattice, and thus 
various ways to compute the $B$-parameters of Eq.~(\ref{eq1}), have been used:

\vspace{1mm}

\begin{itemize}
\item {\underline{HQET}}: After discretizing the HQET lagrangian (to make it 
tractable for a lattice study), the matrix elements of Eq.~(\ref{eq1}) were 
computed in Ref.~[\ref{3GR}], but only in the static limit ($m_b \to \infty$). 
\item {\underline{NRQCD}}: A step beyond the static limit has been made in 
Ref.~[\ref{hiroshima}], where the $1/m_b$-corrections to the NRQCD lagrangian 
have been included, as well as a large part of $1/m_b$-corrections to the
matrix elements of the four-fermion operators. It is important to note, however, that
discretization errors associated with the light degrees of freedom
cannot be reduced by taking a continuum limit, $a\to 0$, since the
NRQCD expansion requires $a\sim 1/m_Q$. Instead, these errors are
reduced by including higher and higher dimension operators whose
coefficients are adjusted to improve the discretization. Such a
procedure is difficult to carry out beyond terms of $O(a)$ and one
must therefore show that the residual discretization and $1/m_b$
power-correction effects are small at finite $a$ [\ref{Aoki:2002bh}].
\item {\underline{Relativistic approach}}: In Ref.~[\ref{ape}], 
the matrix elements were computed 
by using an $O(a)$-improved action
in the region of masses close to the charm quark and then 
extrapolated to the $b$-quark sector by using the heavy quark scaling laws. 
However, this extrapolation can be significant and
discretization errors will be amplified to varying degrees depending on the quantity
studied, if it is performed before a
continuum limit is taken. A discussion of this amplification in the context of
neutral B meson mixing can be found in [\ref{Lellouch:2000tw}].
\end{itemize}

\vspace{1mm}

As of now, none of the above approaches is accurate enough on its own and all 
of them should be used to check the consistency of the obtained results.

\begin{figure}[t]
\begin{center}
\epsfxsize=8cm
\epsfbox{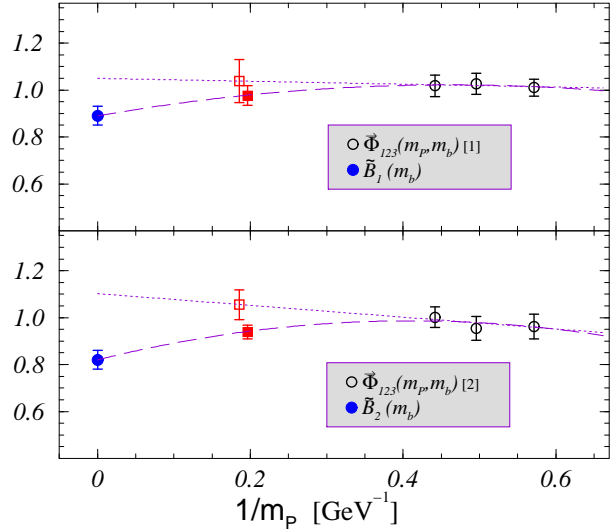}
\caption{\it \label{figB1} The lattice determination of $B_1(m_b)$ and  $B_2(m_b)$ 
obtained in QCD with three heavy--light mesons $m_P$ are combined with the 
static HQET result, $m_P \to \infty$. The result of the linear extrapolation to
$1/M_{{\rm B}_s}$ is marked by the empty squares, whereas the {\it interpolation}
 is denoted by the filled squares.}
\end{center}
\end{figure}

\begin{figure}[t]
\begin{center}
\vspace*{-1.0cm}
\epsfxsize=8cm
\epsfbox{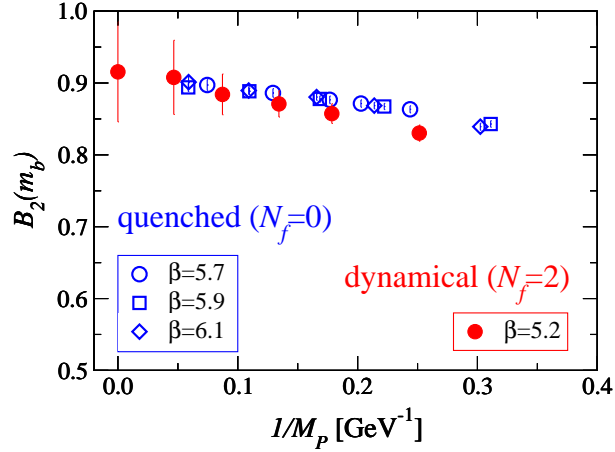}\vspace*{-.4cm}
\caption{\it \label{figC1} Results of the JLQCD 
collaboration~[\ref{norikazu}], 
showing that the effects of quenching are negligible.}
\end{center}
\end{figure}

A more accurate determination of the $B$-parameters 
relevant for $\dgamma$ has been recently obtained in Ref.~[\ref{damir}]. To 
reduce the systematics of the heavy quark extrapolation, the results obtained 
in the static limit of the HQET ~[\ref{3GR}] were combined with those of 
Ref.~[\ref{ape}], where lattice QCD is employed for three mesons of masses in 
the region of D$_s$-mesons. As a result,  one actually {\it interpolates} to the 
mass of the B$_s$-meson. This interpolation is shown in Fig.~\ref{figB1}. The 
resulting values from Ref.~[\ref{damir}], 
in the $\overline{\rm MS}$(NDR) scheme of Ref.~[\ref{BBGLN}], are
\begin{equation}
 \label{mi}
 B_1 (m_b) = 0.87(2)(5)\;, \hspace*{8mm}
 B_2 (m_b) = 0.84(2)(4)\;,
\end{equation}
where the first errors are statistical and the second include various sources 
of systematics. An important remark is that the above results are obtained in 
the quenched approximation ($n_f=0$), and the systematic error due to 
quenching 
could not be estimated. The effect of the inclusion of the dynamical 
quarks has 
been studied within the NRQCD approach. The authors of Ref.~[\ref{norikazu}] 
conclude that the $B$-parameters are essentially insensitive to the change from
$n_f=0$ to $n_f = 2$ (see Fig.~\ref{figC1}). From their (high statistics) 
unquenched simulation, they quote
\begin{eqnarray}
\label{oni}
&& B_1 (m_b)_{(n_f=2)} = 0.83(3)(8)\;,\hspace*{8mm} B_2 (m_b)_{(n_f=2)} = 0.84(6)(8)\; ,\nonumber \\
&& B_1 (m_b)_{(n_f=0)} = 0.86(2)(5)\;,\hspace*{8mm} B_2 (m_b)_{(n_f=0)} = 0.85(1)(5)\; ,
\end{eqnarray}
where, for comparison, we also display  their most recent 
results obtained in the quenched 
approximation~[\ref{JLQCD_new}]. The results of the two lattice approaches 
(Eqs.~(\ref{mi}) and (\ref{oni})) are in good agreement.

The theoretical estimate of $(\Delta\Gamma / \Gamma )_{{\rm B}_s}$ is obtained by 
combining the lattice calculations of the matrix elements with the Wilson 
coefficients. To that purpose two different formulas have been proposed which 
are both derived from Eq.~(\ref{eq3}):
\begin{itemize}  
\item 
In Ref.~[\ref{BBGLN}] the width difference has been  normalized by using the
s.l.\  branching ratio $BR({\rm B}_d \to X \ell \nu_\ell )$ which is 
experimentally determined. In this way one obtains the expression
\begin{equation}
\label{eq:ben}
\left(\frac{\Delta\Gamma}{\Gamma }\right)_{{\rm B}_s}=
\frac{128 \pi^2 BR({\rm B_d \to X \ell \nu_\ell}) }{3 \,m_b^3\, g_{SL}\,\eta_{QCD}} 
\vert V_{cs}\vert^2 F_{{\rm B}_s}^2 M_{{\rm B}_s} \, {\cal M}\,,
\end{equation}
where 
\begin{equation}
\label{eq:emme}
{\cal M} = G_1(z) B_1(m_b) + 
\frac{5}{8} \frac{M_{{\rm B}_s}^2}{\left( m_b(m_b) + m_s(m_b) \right)^2} G_2(z) 
B_2(m_b)  + \tilde \delta_{1/m}\,,
\end{equation}
with $z=m_c^2/m_b^2$, and  the phase space factor $g_{SL}=F(z)$ 
and $\eta_{QCD}=1-\frac23 \frac{\alpha_s}{\pi} f(z)$ are given  in Sec.~\ref{sec:vcbincl}

\item Alternatively, one can use the measured mass difference in the ${\rm B}_d$ 
neutral meson system to write~[\ref{ape}]:
\begin{equation}
\label{eq:bec} 
\left(\frac{\Delta\Gamma}{\Gamma }\right)_{{\rm B}_s} \ = \  
\frac{4\pi}{3} \frac{m_b^2}{M_W^2}\left|\frac{V_{cb} V_{cs}}{V_{td} V_{tb}}
\right|^2 \left( \tau_{{\rm B}_s} \Delta M_{B_d} \frac{M_{{\rm B}_s}}{M_{B_d}} \right)^
{\rm (exp)}\ \frac{B_1(m_b) \xi^2}{\eta_B(m_b) S_0(x_t)} \, {\cal M}  \ ,
\end{equation}
where $\xi$ is defined as $\xi= (F_{{\rm B}_s}\sqrt{\hat B_{{\rm B}_s}})/
(f_{B_d}\sqrt{\hat B_{B_d}})$, and $S_0(x_t)$ is defined in Sec.~\ref{subsec:epsformula}
of Chapter~4.
\end{itemize}
From the point of view of the hadronic parameters, the advantage of the second 
formula is that it is expressed in terms of the ratio $\xi$, in the evaluation 
of which many systematic uncertainties of the lattice calculations cancel. The 
estimate of $\xi$, however, is affected by an uncertainty due to the chiral 
extrapolation, which comes from the fact that in present 
lattice calculation it 
is not possible to simulate directly quark masses smaller than $\sim m_s/2$. 
Therefore an extrapolation to the physical $d$-quark mass is necessary. The 
first formula, instead, is expressed in terms of the  decay constant $F_{{\rm B}_s}$ 
whose determination does not require a chiral extrapolation. However,
other systematic uncertainties may be important in this case such as those 
coming from the value of the absolute lattice scale (inverse lattice spacing), 
the renormalization of the axial current and $1/m_b$-corrections.

In the numerical analysis, to derive a prediction for $\dgamma$, we use the 
values of parameters listed in Table~\ref{tab:A}. Notice that in the error 
for $\xi$ the uncertainty due to the chiral extrapolation is quoted separately 
(second error).

\begin{table}[ht]
\begin{center}
\begin{tabular}{|c|c||c|c|} 
\hline
\hspace{-4.mm}{\phantom{\huge{l}}}\raisebox{-.2cm}{\phantom{\Huge{j}}} 
{\sl Parameter} & {\sl Value and error} & {\sl Parameter} & 
{\sl Value and error} \\ \hline
$\alpha_s(m_b)$ & $0.22$ & $m_t$ & $165 \pm 5$ GeV \\
$M_W$ & 80.41 GeV & $m_b$ & $4.26 \pm 0.09$ GeV \\
$M_{{\rm B}_d}$ & 5.28 GeV & $m_c/m_b$ & $0.28 \pm 0.02$ \\
$M_{{\rm B}_s}$ & 5.37 GeV & $m_s $ & $ 105 \pm 25$ MeV \\
$\tau_{{\rm B}_s}$ &$ 1.461 \pm 0.057 \, ps$ & $\eta_B(m_b)$ & $0.85 \pm 0.02 $\\
$\vert V_{cb}\vert$ & $0.0395 \pm 0.0017$ & $F_{{\rm B}_s}$ & $ 238 \pm 31$ MeV\\
$\vert V_{ts}\vert$ & $0.0386 \pm 0.0013$ & $\xi$ & $  1.24\pm 0.03\pm 0.06$ \\
$\vert V_{cs}\vert  $ & $0.9756\pm 0.0005$ & $B_1 (m_b)$ &$ 0.86\pm 0.06$\\
$\vert V_{td }\vert  $ & $0.0080 \pm 0.0005$ & $B_2 (m_b)$ &$ 0.84\pm 0.05$\\
$\Delta M_{{\rm B}_d}$ & $0.503 \pm 0.006\,  ps^{-1}$ & $G(z)$ & $0.030\pm 0.007$ \\
$BR({\rm B}_d \to X l \nu_l)$ & $10.6 \pm 0.3$\% & $G_S(z)$ & $0.88 \pm 0.14$ \\ 
\hline
\end{tabular}
\caption{\it \label{tab:A} Average and errors of the main parameters 
used in the numerical analysis. When the error is negligible it has been 
omitted. The heavy-quark masses ($m_t$, $m_b$ and $m_c$) are the $\msbar$ 
masses renormalized at their own values, e.g. $m_t=\mtms$. The strange quark 
mass, $m_s=\msms$, is renormalized in $\msbar$ at  the scale $\mu=2$~GeV. 
The value for $F_{{\rm B}_s}$ and $\xi$ are taken 
from Ref.~[\ref{3Lellouch:2002nj}]. }
\end{center}
\end{table}  

The value of the $b$-quark mass deserves a more detailed discussion. The 
$b$-{\sl pole} mass, which corresponds at the NNLO to the $\overline{MS}$ mass 
$m_b = 4.26$ GeV quoted in Table~\ref{tab:A}, is $m_b^{pole}= 4.86$ GeV. Since 
the formulae for $\dgamma$ have been derived only at the NLO, however, it may 
be questionable whether to use $m_b^{pole}= 4.86$~GeV or $m_b^{pole}= 
4.64$~GeV, corresponding to $m_b \simeq  4.26$~GeV at the NLO. That difference 
is very important for  the value of $\tilde \delta_{1/m}$ which, computed in 
the VSA, varies between $-0.4$ and $-0.6$. In addition, 
a first principle non-perturbative estimates of the matrix elements entering 
the  quantity $\tilde \delta_{1/m}$ is still lacking.
For this reason we include $\pm 30\%
$ of uncertainty in the 
estimate of $\tilde \delta_{1/m}$ stemming from the  use of the VSA. 
We finally obtain the predictions:


\begin{equation}
\left(\frac{\Delta\Gamma}{\Gamma}\right)_{{\rm B}_s}^{Eq.~(\ref{eq:ben})} =
(8.5 \pm 2.8) \times 10^{-2}  \, , \qquad
\left(\frac{\Delta\Gamma}{\Gamma} \right)_{{\rm B}_s}^{Eq.~(\ref{eq:bec})} = 
(9.0 \pm 2.8) \times 10^{-2}   \, .
\label{eq:final}
\end{equation}

\begin{figure}[htbp] 
\begin{center}
\epsfxsize=7.5cm
\epsfbox{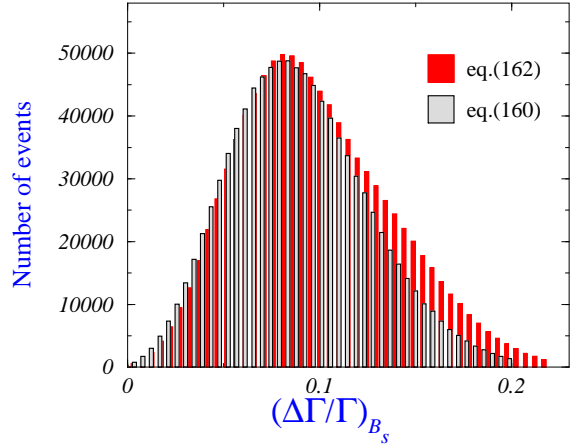}

\vspace{-2mm}

\caption{\label{figD1}{\it Probability density function (pdf) for $\dgamma$ 
using the formulas~\ref{eq:ben} and~\ref{eq:bec}. The pdf corresponding to 
the smaller value is the one obtained with Eq.~(\ref{eq:ben}). }}
\end{center}
\end{figure}

In Fig.~\ref{figD1} we show the corresponding probability distribution 
functions (pdf).

We see that the results obtained with the two formulas are in good agreement. 
From the pdfs we observe that $\dgamma$ can span a very large range of values,
say between 0.03 and
0.15: the theoretical uncertainty on this quantity is large. 
The main source of  uncertainty, besides the assumption of local quark-hadron 
duality, comes from the $1/m_b$ corrections parameterized by 
$\tilde \delta_{1/m}$. That uncertainty is enhanced by a rather large 
cancellation between the leading contributions (first two terms of 
Eq.~(\ref{eq:emme})) and it is very difficult to reduce, since it would 
require the non-perturbative estimate of many dimen\-sion-7, $\Delta B=2$, 
operators. Such a calculation is very challenging and most probably beyond 
the present capability of the  lattice QCD approach.
Given the present theoretical uncertainty on $(\Delta\Gamma/\Gamma )_{{\rm B}_s}$ it 
is unlikely that  signals of physics beyond the Standard Model may be 
detected from the measurement of this quantity.

\boldmath 
\subsection{Width difference of ${\rm B}_d$ mesons}
\label{sec:dgbdtheo}
\unboldmath 

The phenomenology of $\dg$ has been mostly neglected so far, in contrast to 
the lifetime difference in the ${\rm B}_s$ system, because the 
present data fall 
so short of the needed accuracy. However, in the prospect of experiments with 
high time resolution and large statistics, its study will become relevant. In fact,
it may affect a precise determination of the CKM phase $\beta$, and it also provides 
several opportunities for detecting New Physics.

The width difference $\dg/\Gamma_d$ has been  estimated 
in [\ref{Paper}]
including the $1/m_b$ contribution and part of the NLO QCD corrections. 
Adding the latter corrections decreases the value of $\dg/\Gamma_d$ 
computed at the leading order by a factor of almost $2$. 
This yields 
\begin{equation}
\Delta \Gamma_d / \Gamma_d = (2.6 ^{+1.2}_{-1.6}) \times 10^{-3} \, .
\label{dg2.6}
\end{equation}
Using another expansion of the partial NLO QCD corrections proposed 
in [\ref{breport}], one gets 
\begin{equation}
\Delta \Gamma_d / \Gamma_d = (3.0 ^{+0.9}_{-1.4}) \times 10^{-3}\, ,
\label{dg3.0}
\end{equation}
where  preliminary values 
for the bag factors from the JLQCD collaboration~[\ref{jlqcd}] are used.
The contributions to the error (in units of $10^{-3}$) are
$\pm 0.1$ each from the uncertainties in the values of the CKM parameters
and the parameter $x_d = (\Delta M_d / \Gamma )_d$, $\pm 0.5$ each from 
the bag parameters and the mass of 
the $b$ quark, $\pm 0.3$ from the assumption of naive factorization 
made for  the $1/m_b$ matrix elements, and $^{+0.5}_{-1.2}$ from the scale dependence.
The error due to the missing terms in the NLO contribution is 
estimated to be $\pm 0.8$ in the calculation of Eq.~(\ref{dg2.6}).
Although it is reduced in the calculation of Eq.~(\ref{dg3.0}),
a complete NLO calculation is definitely desirable for a more reliable result.

\boldmath 
\subsection{Relation between $\sin(2 \beta)$ and $\dg$}
\unboldmath 
\label{gold}
The time-dependent CP asymmetry measured through the 
`gold-plated' mode ${\rm B}_d \to J/\psi \rm K_S$ is
\begin{equation}
\aa_{CP}  = \frac{\Gamma[\overline{\rm B}_d(t) \to J/\psi {\rm K}_S] - 
\Gamma[{\rm B}_d(t) \to J/\psi {\rm K}_S]}
{\Gamma[\overline{\rm B}_d(t) \to J/\psi {\rm K}_S] +
\Gamma[{\rm B}_d(t) \to J/\psi {\rm K}_S]} \,
 \approx   \sin(\Delta M_d t) \sin(2\beta)~~,
\label{acp-approx}
\end{equation}
which is valid when the lifetime difference, the direct CP
violation, and the mixing in the neutral K mesons are 
neglected. As the accuracy of this measurement increases,
the corrections due to these factors will need to be taken 
into account. Using the effective parameter  $\bar{\epsilon}$ 
that absorbs several small effects and uncertainties, 
including penguin contributions
(see [\ref{Paper}] for a precise definition),
and keeping only linear terms in that effective parameter,  
the asymmetry becomes 
\begin{eqnarray}
\aa_{CP} & = & \sin(\Delta M_d t) 
        \sin(2\beta) \left[ 1 - \sinh \left( \frac{\dg t}{2} \right)
        \cos(2 \beta) \right] \\ \nonumber
&  & + 2 {\rm Re}(\bar{\epsilon}) 
\left[ -1 + \sin^2(2 \beta) \sin^2(\Delta M_d t) - \cos(\Delta M_d t) 
\right] \\ \nonumber
&  & + 2 {\rm Im}(\bar{\epsilon}) \cos(2\beta) \sin(\Delta M_d t) ~~.
\label{dg-corr}
\end{eqnarray}
The first term represents the  standard approximation 
of Eq.~(\ref{acp-approx}) 
together with  the correction due to the lifetime difference $\dg$.
The other terms include corrections due to CP 
violation in the ${\rm B}$--$\bar{\rm B}$ and $\rm K$--$\bar{\rm K}$ 
mixings.

Future experiments aim to measure $\beta$ with  an accuracy of 
0.005 [\ref{lhc}].
The corrections due to $\bar{\epsilon}$ and $\dg$ will become a large 
fraction of the 
systematic error. This error can be reduced by a  simultaneous fit of 
$\sin(2\beta), \dg$ and $\bar{\epsilon}$.
The BaBar Collaboration gives a bound on the coefficient of 
$\cos(\Delta M_d t)$ in 
Eq.~(\ref{dg-corr}), where other correction terms are 
neglected~[\ref{babar-direct}]. 
When measurements will become accurate enough to really constrain  the 
$\cos(\Delta M_d t)$ 
term, all the other terms in Eq.~(\ref{dg-corr})
would also be measurable. In this case, the complete expression 
for ${\cal A}_{CP}$  needs to be used.

\subsection{New Physics signals}
\label{contrast}
The lifetime difference in neutral ${\rm B}$ mesons can be written in the form
\begin{equation}
\Delta\Gamma_q = - 2 |\Gamma_{21}|_q  \cos(\Theta_q - \Phi_q)~~,
\label{theta-phi}
\end{equation}
where $\Theta_q \equiv {\rm Arg}(\Gamma_{21})_q , 
\Phi_q \equiv {\rm Arg}(M_{21})_q $,
and $q \in \{ d,s \}$ (see Sec.~\ref{subsec:BBformula}).
In the ${\rm B}_s$ system, the new physics effects can only decrease  
the value of $\Delta\Gamma_s$ with respect to the SM [\ref{grossman}]. In the ${\rm B}_d$ 
system, an upper bound for $\dg$ can be  given,
depending on the additional assumption of three-generation
CKM unitarity:
\begin{equation}
\Delta \Gamma_d \leq  \frac{\Delta \Gamma_d({\rm SM})}
{\cos[ {\rm Arg}(1 + \delta f) ]}~~,
\label{dgd-bd}
\end{equation}
where $\delta f$ depends on hadronic matrix elements.
The bound in Eq.~(\ref{dgd-bd}) can be calculated up to higher order corrections. 
In [\ref{Paper}], $|{\rm Arg}(1 + \delta f)| < 0.6$, so that   
$\Delta \Gamma_d < 1.2 ~ \Delta \Gamma_d({\rm SM})$.
A violation of this bound would indicate a violation of the unitarity of $3\times 3$ 
CKM matrix. A complete NLO calculation would  provide a stronger bound. 

The ratio of two effective lifetimes can be used to measure the quantity
$\Delta\Gamma_{obs(d)} \equiv \cos(2\beta) \Delta\Gamma_d/\Gamma_d$ 
(see Sec.~\ref{sec:measure}).
In the presence of new physics, this quantity is in fact
(see Eq.~(\ref{theta-phi}))
\begin{equation}
\Delta\Gamma_{obs(d)} =  - 2 (|\Gamma_{21}|_d/\Gamma_d)  
\cos(\Phi_d) \cos(\Theta_d - \Phi_d),
\end{equation}
where in the Standard Model 
\begin{equation}
\Delta\Gamma_{obs(d)}({\rm SM}) = 2 (|\Gamma_{21}|_d/\Gamma_d) 
\cos(2 \beta) \cos[ {\rm Arg}(1 + \delta f) ] 
\end{equation}
is predicted to be positive.
New physics is not expected to affect $\Theta_d$, but it may affect 
$\Phi_d$ in such a way that $\cos(\Phi_d) \cos(\Theta_d - \Phi_d)$ changes sign.
A negative sign of $\Delta\Gamma_{obs(d)}$ would  therefore be a clear signal for 
New Physics.
The time-dependent asymmetry in $J/\psi {\rm K}_S$ (or $J/\psi K_L$) measures  
${\cal A}_{CP} = - \sin(\Delta M_d t) \sin(\Phi_d)$, 
where $\Phi_d = -2\beta$ in the SM. 
The measurement of $\sin(\Phi_d)$ still allows for a discrete ambiguity
$\Phi_d \leftrightarrow \pi - \Phi_d$. 
If $\Theta_d$ can be determined independently of 
the mixing in the ${\rm B}_d$ system,  then the measurement of 
$\Delta\Gamma_{obs(d)}$ 
will in principle resolve the discrete ambiguity.

In conclusion, the measurement of $\dg$ and related quantities 
should become possible in a near future, providing further important 
informations on the flavour sector of the SM.

\boldmath 
\subsection{Experimental review and future prospects for 
$\Delta \Gamma$ measurements}
\label{sec:dgexp}
\unboldmath 

The width difference $\Delta\Gamma_s=\Gamma_{long}-\Gamma_{short}$ can 
be extracted from lifetime measurements of ${\rm B}_s$ decays. 
A first method is based on a double exponential lifetime fit to samples 
of events containing mixtures of CP eigenstates, like s.l.\  or 
$\rm D_s$-hadron ${\rm B}_s$ decays. 
A second approach consists in isolating samples of a single CP eigenstate,  
such as ${\rm B}_s\rightarrow J/\psi\phi$. 
The former method has a quadratic sensitivity to $\Delta\Gamma_s$, whereas 
the latter has a linear dependence and suffers from a much reduced statistics.
A third  method has been also proposed~[\ref{alephBR}]  and consists 
in measuring the branching fraction 
${\rm B}_s\rightarrow {\rm  D}_s^{\left(*\right)+} 
{\rm D}_s^{\left(*\right)-}$. 
More details on the different analyses performed are given in the following.

L3~[\ref{l3}] and DELPHI~[\ref{delphi1}] use inclusively reconstructed 
${\rm B}_s$ and ${\rm B}_s\rightarrow {\rm D}_s l\nu X$ events, respectively.
If those sample are fitted assuming a single exponential lifetime, then, 
assuming $\frac{\Delta\Gamma_s}{\Gamma_s}$ is small, 
the measured lifetime is given by:
\begin{equation}
\tau_{{\rm B}_s^{incl.}}= 
\frac{1}{\Gamma_s}\frac{1}{1-\left(
\frac{\Delta\Gamma_s}{2\Gamma_s}\right)^2} ~~({\rm incl.\, B}_s) \quad \quad ; 
\quad \quad
\tau_{{\rm B}_s^{semi.}}= \frac{1}{\Gamma_s}
\frac{1+\left(\frac{\Delta\Gamma_s}{2\Gamma_s}\right)^2}
{1-\left(\frac{\Delta\Gamma_s}{2\Gamma_s}\right)^2} ~~ ({\rm B}_s\rightarrow {\rm D}_s l\nu X)
\label{eq:dg_inclbs}
\end{equation}
The single lifetime fit is thus more sensitive to the effect of $\Delta \Gamma$
in the s.l.\ case than in the fully inclusive one.
The same method is used for the $\rm B_s$ world average lifetime 
(recomputed without the DELPHI measurement~[\ref{wal}]) obtained
by using only the s.l.\  decays.
The technique of reconstructing only decays at defined CP 
has been exploited by ALEPH, DELPHI and CDF.
ALEPH~[\ref{alephBR}], reconstructs the decay 
$\rm B_s\rightarrow D_s^{\left(*\right)+} 
D_s^{\left(*\right)-}\rightarrow \phi\phi X$ 
which is predominantly CP even. 
The proper time dependence of the ${\rm B}_s^0$ component is a simple
exponential and the lifetime is related to $\Delta\Gamma_s$ via 
\begin{equation}
\rm \frac{\Delta\Gamma_s}{\Gamma_s}=2\left(\frac{1}
{\Gamma_s}\frac{1}{\tau_{{\rm B}_s^{short}}}-1\right).
\label{eq:dg_phiphi}
\end{equation}
Another method consists in using the branching fraction, 
$\rm BR(B_s\rightarrow D_s^{\left(*\right)+} D_s^{\left(*\right)-})$. 
Under several theoretical assumptions [\ref{theoBR}] 
\begin{equation}
BR({\rm B}_s\rightarrow {\rm D}_s^{\left(*\right)+} 
{\rm D}_s^{\left(*\right)-}) = \frac{\Delta\Gamma_s}{\Gamma_s\left(1+\frac{\Delta\Gamma_s}
{2\Gamma_s}\right)}. 
\label{eq:dg_ratio}
\end{equation}
This is the only constraint on $\frac{\Delta\Gamma_s}{\Gamma_s}$ 
which does not rely on the measurement of the B$_s^0$(${\rm B}_d^0$) lifetime.
DELPHI~[\ref{delphi2}] uses a sample of $\rm B_s\rightarrow D_s-hadron$, 
which is expected to have an increased CP-even component as the contribution 
due to ${\rm D}_s^{\left(*\right)+} {\rm D}_s^{\left(*\right)-}$ events is enhanced by 
selection criteria.  
CDF~[\ref{cdfjpsi}] reconstructs ${\rm B}_s\rightarrow J/\psi\phi$ with 
$J/\psi \rightarrow \mu^+\mu^-$ and $\phi\rightarrow \rm K^+K^-$ 
where the CP even component is equal to $0.84\pm 0.16$ obtained by 
combining CLEO~[\ref{cleo}] measurement of CP even fraction in 
$\rm B_d\rightarrow J/\psi K^{*0}$ and possible $SU(3)$ symmetry  
correction. 
The results, summarized in Table~\ref{table:expsum}, 
are combined 
following the procedure described in [\ref{our}].  The log-likelihood of each 
measurement  are summed and normalized with respect to its minimum. 
\begin{table}[t]
\begin{center} 
 \begin{tabular}{|l|ll|}  \hline
Experiment        & $\rm B_s$ decays   &  $  \Delta\Gamma_s/ \Gamma_s$ \\ \hline   
DELPHI             & $\rm B_s\rightarrow D_sl\nu X$ & $<0.47$ \\
Other s.l.\ & $\rm B_s\rightarrow D_sl\nu X$ & $<0.31$ \\
ALEPH                 & $\rm B_s\rightarrow \phi\phi X$ & $0.43^{+0.81}_{-0.48}$ \\
ALEPH (BR method) & $\rm B_s\rightarrow \phi\phi X$ & $0.26^{+0.30}_{-0.15}$ \\
DELPHI              & $\rm B_s\rightarrow D_s-hadron$& $<0.70$ \\
CDF                   & $\rm B_s\rightarrow J/\psi\phi$ & $0.36^{+0.50}_{-0.42}$ \\
\hline
\end{tabular}   
\caption{\it Summary of the available measurements 
on $\Delta\Gamma_s/ \Gamma_s$ used to calculate the limit.}
\label{table:expsum}
\end{center}
\end{table}  
   Two measurements are excluded from the average for different reasons:
   \begin{itemize}
     \item L3 inclusive analysis: the likelihood is not available and it 
cannot be reconstructed from the numerical result;
     \item ALEPH branching ratio analysis: the theoretical assumptions 
in Eq.~(\ref{eq:dg_ratio}) are controversial and the systematic
          error due to these assumptions has not been estimated. 
   \end{itemize}
The $65\%$,  $95\%$ and $99\%$ confidence level contours are 
%
\begin{figure}[t]
\begin{minipage}[b]{.46\linewidth}
\includegraphics[width=7.5cm]{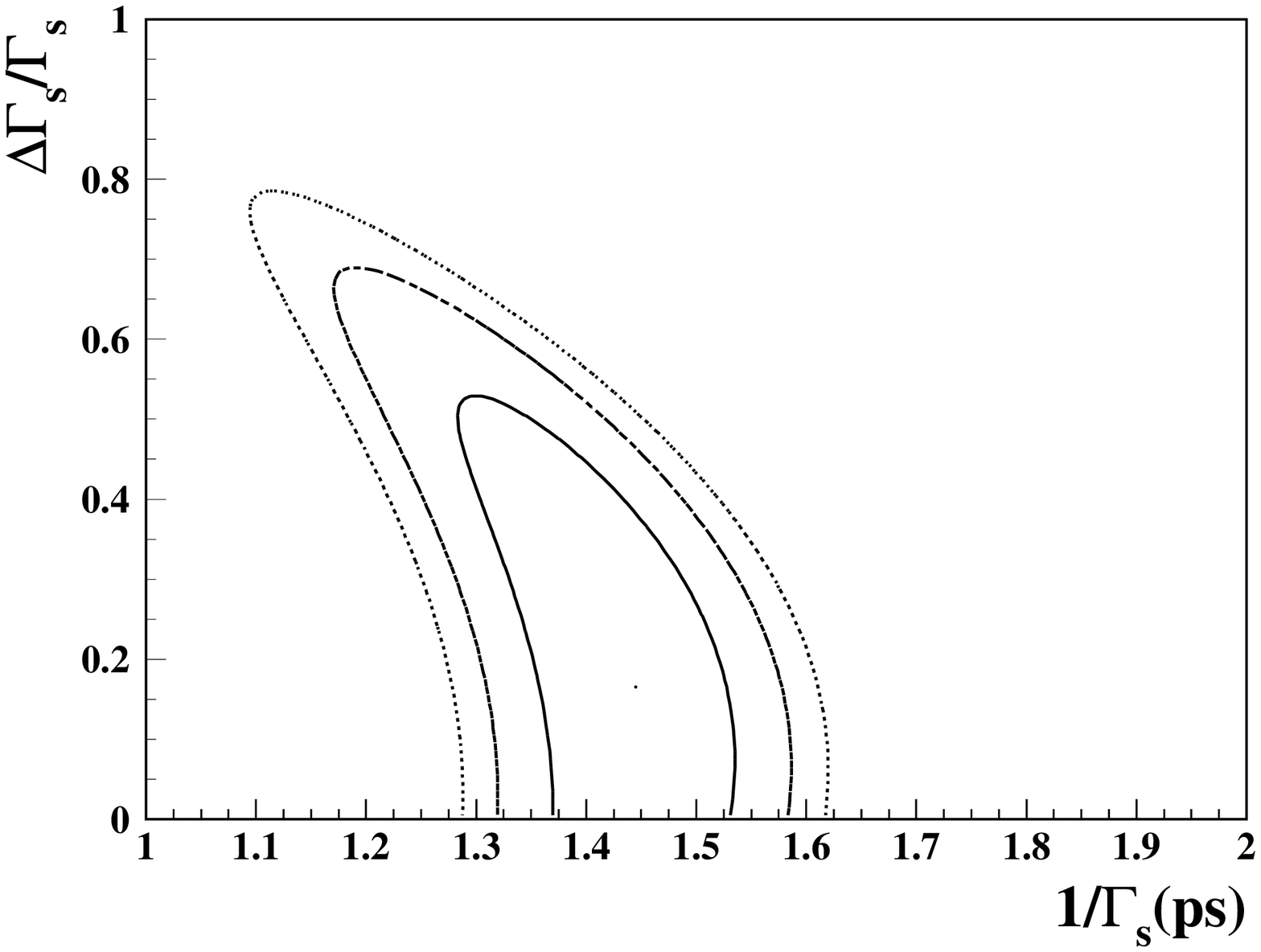}
\caption{\it $65\%$,  $95\%$ and $99\%$ C. L.\ contours of negative 
log-likelihood in the $\tau_{B_s}$, $\Delta\Gamma_s/\Gamma_s$ plane.}
\label{fig:contnocon}
\end{minipage} \hfill
\begin{minipage}[b]{.46\linewidth}
\includegraphics[width=7.5cm]{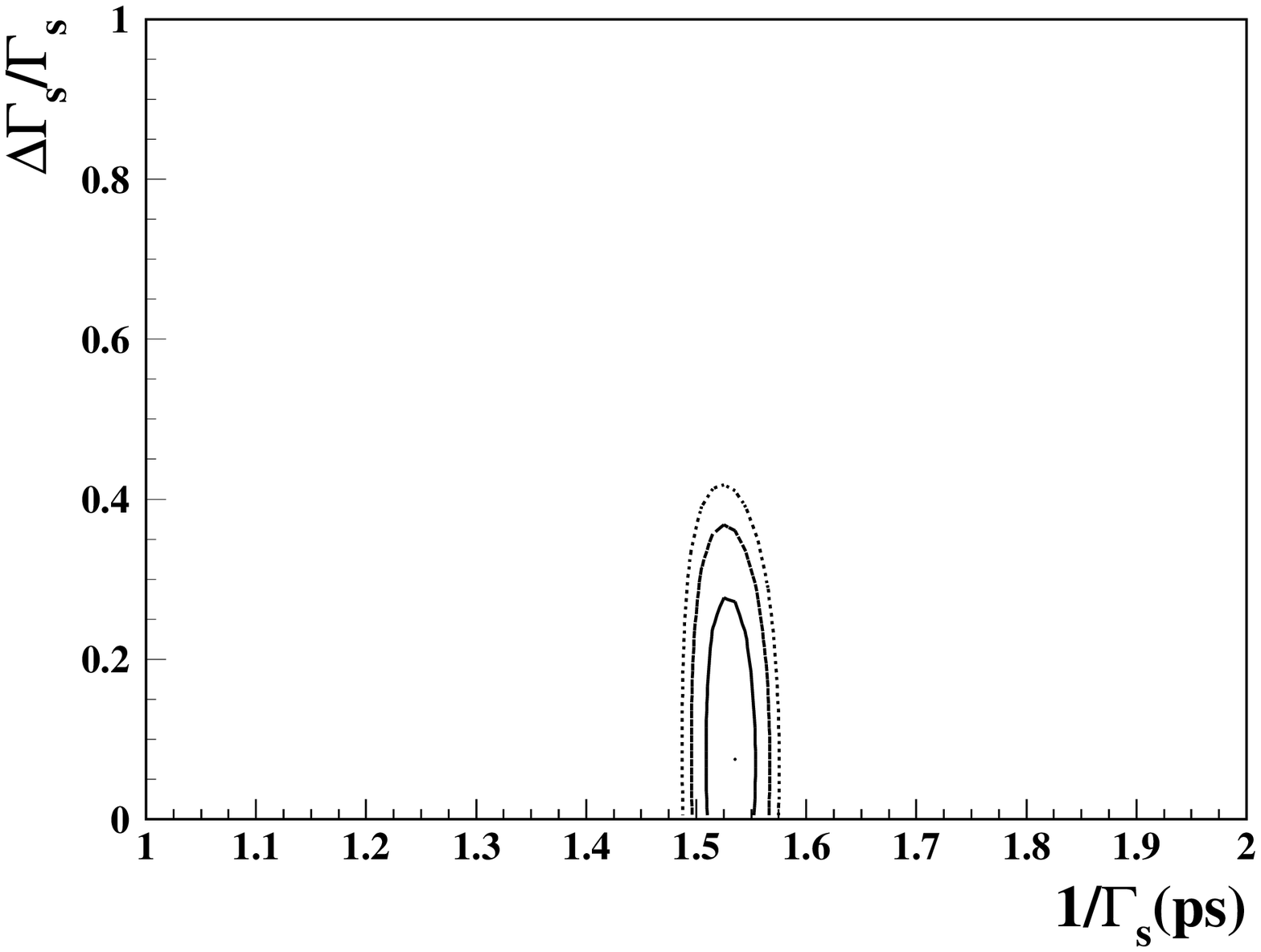}
\caption{\it Same as Fig.~\ref{fig:contnocon}  with the constraint 
$1/\Gamma_s~=~\tau_{{\rm B}_d}$.}
\label{fig:contcon}
\end{minipage}
\end{figure}
shown in Fig.~\ref{fig:contnocon}. The result is
\begin {center} 
$\Delta\Gamma_s/\Gamma_s= 0.16^{+0.15}_{-0.16}$\\
$\Delta\Gamma_s/\Gamma_s<0.54$ at $95\%$ C.L.
\end{center}
In order to improve the limit the constraint $1/\Gamma_s = \tau_{{\rm B}_d}$ can be imposed.
This is well motivated theoretically, as the total widths of the ${\rm B}_s^0$ and
the ${\rm B}_d^0$ mesons are expected to be equal within less than 1\%
(see Fig.~\ref{3fig:liferatio})
and that $\Delta \Gamma_{{\rm B}_d}$ is expected to be small.
It results in:
\begin {center} 
$\Delta\Gamma_s/\Gamma_s= 0.07^{+0.09}_{-0.07}$\\
$\Delta\Gamma_s/\Gamma_s<0.29$ at $95\%$ C.L.
\end{center}
The relative confidence level contours plot is shown in Fig.~\ref{fig:contcon}.

\subsubsection{Prospects for Tevatron experiments}
CDF measured the ${\rm B}_s\rightarrow J/\psi\,\phi$ 
lifetime~[\ref{cdfjpsiphilife}] 
and polarization~[\ref{cdfjpsiphipola}] separately. 
In the future the idea is to combine these two measurements by fitting both 
the lifetime and the transversity angle\footnote{The transversity angle is 
defined as the angle between the $\mu^+$ and the $z$ axis in the rest 
frame of the $J/\psi$, where the $z$ axis is orthogonal to the plane 
defined by the $\phi$ and K$^+$ direction.}. 
The use of the transversity allows to separate the CP even from the CP 
odd component. A study has been performed, by assuming similar performances
as those achieved during Run~I (mass resolution, background fractions, etc.) 
and improved proper time resolution ($18~\mu$m).
With an integrated luminosity of 2 fb$^{-1}$, corresponding to about 4000 
events, an accuracy on $\Delta \Gamma_s/\Gamma_s$ of 5\% could be reached
\footnote{In this results it is assumed that the CP$_{even}$ fraction is 
$0.77\pm 0.19$. If CP$_{even}=0.5 (1)$, the error becomes 
$\sigma\left( \Delta \Gamma_s/\Gamma_s\right)=0.08\left(0.035\right)$.}.
Using the same integrated luminosity and the impact parameter 
trigger~[\ref{svt}],
CDF could expect to reconstruct 2500 
${\rm B}_s\rightarrow {\rm D}_s^+{\rm D}_s^-$ events, 
with a signal-to-noise ratio of 1:1.5.
Using this sample the lifetime of the short component can be measured 
with an error of 0.044 ps, which corresponds to 
$\sigma\left( \Delta \Gamma_s/\Gamma_s\right) = 0.06$.
The ${\rm D}_s\pi$ and ${\rm D}_s3\pi$ decays could be also used. Those events are flavour-specific, 
thus they correspond to  well defined mixtures of ${\rm B}_s^{short}$ and 
${\rm B}_s^{long}$.
By using $\sim 75,000$ events $1/\Gamma_s$ can be measured with an error of 0.007~ps. 
Combining together the flavour specific measurement 
and the ${\rm D}_s^+{\rm D}_s^-$ analysis, CDF can reach an error 
$\sigma\left( \Delta \Gamma_s/\Gamma_s\right)=0.04$.

D{\O} has based its studies of B$_s$ lifetime difference measurements
on its strong dimuon trigger and the extensive coverage of the
calorimeter and the muon detector. It is expected that approximately
7000 $\rm B_s \rightarrow J/\psi \, \phi$ events will be reconstructed
with an integrated luminosity of 2 fb$^{-1}$. The sensitivity of the
measurements depends on two parameters: (a) the fraction of the
CP-even component of the $J/\psi \, \phi$ final state\footnote{The
CP$_{\rm even}$ fraction has been measured by CDF in Run-I:
(77$\pm$19)\%.}, and (b) the CP-violating phase 
$\phi$ in the mixing of the B$_s$
system\footnote{The CP-violating phase, defined by $\alpha_{CP}({\rm B}_s
\rightarrow J/\psi \, \phi) \sim \sin \phi$, is expected to be small in
the Standard Model.}. The methods discussed here invoke utilization of
CP eigenstates, therefore an angular analysis is needed to
disentangle the admixture of CP-even and CP-odd contributions. 

The $J/\psi \, \phi$ channel can be exploited in two ways: 
\begin{itemize}
\item {by comparison of the CP-eigenstate lifetimes: the
sensitivity in this measurement is proportional to
$\Delta \Gamma_s \, \cos \phi = \Delta \Gamma_{CP} \, \cos^2 \phi$.}

\item {by comparison of a CP-eigenstate lifetime to that of a
``50-50'' admixture, {\it e.g.}: $\Delta \Gamma_s = 2 \, \cos \phi \times [
\Gamma({\rm B}_s^{\rm CP~{\rm even}}) - \Gamma({\rm B}_s^{\rm CP~{50-50}}) ]$. About 1000
events of the B$_s \rightarrow {\rm D}_s \, \pi$ decay will be used for the
extraction of $\Gamma({\rm B}_s^{\rm CP~{50-50}})$.}

\end{itemize}
Additional decay channels may include B$_s \rightarrow J/\psi \, \eta$
and $J/\psi \, \eta ^\prime$ (both being CP-even states).
Combining all modes, D{\O} can achieve a measurement on $\Delta
\Gamma_s \, / \Gamma_s$ with precision between $\sigma = 0.04$
(CP$_{\rm even} = 100\%$) and $\sigma = 0.07$
(CP$_{\rm even} = 50\%$)

BTEV studied their $\Delta \Gamma_s/\Gamma_s$ reach in three 
different scenarios. 
Assuming a $b\bar b$ cross section of $100~\mu$b, the number of expected 
events, using 2 fb$^{-1}$ of integrated luminosity, are:

\vspace{2mm}

\begin{itemize} 
\item[{1.}] 91700 ${\rm B}_s\rightarrow {\rm D}_s\,\pi$ 
\item[{2.}] 1700 ${\rm B}_s\rightarrow J/\psi \,\eta$ and 6400 
 $\rm B_s\rightarrow J/\psi \,\eta\prime$, where 
            $\tau_{B_s^{short}}=1/\Gamma_s^{short}$ is measurable;
\item[{3.}] 41400 ${\rm B}_s\rightarrow J/\psi\, \phi$ where the lifetime, 
 $\tau_x=1/\Gamma_s^x$,
            is a mixture of a $\Gamma_s^{short}$ and a $\Gamma_s^{long}$ \\
 components. 
\end{itemize}

\vspace{2mm}

\noindent The analysis details are discussed in~[\ref{breport}].
The results are summarised in Table~\ref{table:btevres}, obtained 
under the assumption that $\Delta\Gamma_s/\Gamma_s=0.15$.
\begin{table}[t]
\begin{center} 
\begin{tabular}{|c|c|c|c|}  \hline
Decay Modes Used & \multicolumn{3}{|c|}{Error on $\Delta\Gamma_s/\Gamma_s$}\\ 
\hline 
   Integrated Luminosity in fb$^{-1}$                            &    2   &   10    & 20      \\
${\rm D}_s\pi$, $J/\psi \eta^{\left(\prime\right)}$                    & 0.0273 & 0.0135  & 0.0081  \\
${\rm D}_s\pi$, $ J/\psi \phi$                                         & 0.0349 & 0.0158  & 0.0082  \\
${\rm D}_s\pi$, $J/\psi \eta^{\left('\right)}$,$ J/\psi \phi$          & 0.0216 & 0.0095  & 0.0067  \\ \hline
\end{tabular}   
\caption{\it Projection for statistical error on $\Delta\Gamma_s/\Gamma_s$ which can be obtained by the
BTeV experiment.}
\label{table:btevres}
\end{center}
\end{table}  

\subsubsection{Prospects for LHC experiments}
The LHC experiments have investigated the measurement of $\DG$ in the exclusive \Bst decay
following the studies done in ~[\ref{DUNIETZ}].
In these analyses, $\DG$  and $\Gamma_{s}$  are fitted simultaneously 
with the weak phase  $\phi_{s}=arg (V_{cs}^{*} V_{cb} / V_{cs} V_{cb}^{*})$
and  the two helicity amplitude values, $ A_{||}$ and $A_{\perp}$, while 
the mixing parameter  $\xis=\Delta m_s/\Gamma$  is assumed to be known and kept fixed. 
The results summarised in Table~\ref{table:sumary} correspond to 3 (5) years running at a luminosity of
$10^{33}cm^{-2}s^{-1}$($2 \cdot 10^{32}cm^{-2}s^{-1}$) for ATLAS and CMS (LCHb).

\begin{table}\begin{center}
\begin{tabular}{|c|c|c|c|}             \hline 
                         &     LHCb & ATLAS & CMS    \\ \hline
 $\sigma (\frac  {\DG} {\Gamma_{s}}) /\frac  {\DG } {\Gamma_{s}} $   
&    8.4\%   &  11.3\% &   7.5\%  \\
$\sigma (\frac  {\DG} {\Gamma_{s}})  $   &    0.013   &  0.017 &   0.011  \\
 $\sigma (\Gamma_{s}) /\Gamma_{s} $            &  0.6\%   & 0.7\% & 0.5\%  \\
$\sigma (A_{||})/A_{||}$                &  0.7\%   & 0.8\% & 0.6\%  \\
$\sigma (A_{\perp})/A_{\perp}$             &    2\%   &   3\% &   2\%  \\ 
$\phi_{s}$ ($\xis = 20$) &   0.02   &  0.03 & 0.014  \\
$\phi_{s}$ ($\xis = 40$) &   0.03   &  0.05 & 0.03   \\
\hline 
 \end{tabular}
\end{center}\vspace*{-0.5cm}
\caption{\it Expected statistical uncertainties on 
\Bst\ parameters for each experiment under the assumptions presented 
in the text. The  value   $\frac  {\DG} {\Gamma_{s}} =0.15$ 
is used as input to the fit.     }\label{table:sumary}
\end{table}

\subsubsection{Measurement of $\Delta\Gamma_d/\Gamma_d$}
\label{sec:measure}
In the case of $\Delta\Gamma_d/\Gamma_d$, the time resolution is no longer 
a limiting factor in the accuracy of lifetime measurements.
At present, the only experimental limit comes 
from DELPHI~[\ref{delphideltaG}], 
which has been obtained by fitting a sample of inclusive B decays to determine 
the mass difference $\Delta M_d$ without neglecting 
the $\Delta\Gamma_d$ term. 
At $90\%$ C.L.\ $\Delta\Gamma_d/\Gamma_d <0.20$. 
Given the large number of ${\rm B}_d$ produced at LHC  and the 
proposed super B factories, it should be possible to
measure $\dg/\Gamma_d \sim 0.5$\% .
Using the time evolution of a single final state, however, is not 
sufficient as the time measurements of the decay of an untagged 
${\rm B}_d$ to a single final state can only be 
sensitive to quadratic terms in $\dg/\Gamma_d$,  [\ref{Paper}]. 
This problem can be circumvented by combining
the information from two different decay modes or by using 
angular distributions.
It is then possible to have observables linear in  $\dg/\Gamma_d$, 
which can provide  $\dg/\Gamma_d \sim 0.5$\% .
A viable option, perhaps the most efficient among those 
in~[\ref{Paper}], is to compare the measurements of the
{\it average untagged lifetimes}  
of the s.l.\  decay mode $\tau_{SL}$ and 
of the CP-specific decay modes $\tau_{CP_{\pm}}$.
The ratio between the two lifetimes is 
\begin{equation}
\frac{\tau_{SL}}{\tau_{CP\pm}} = 1 \pm \frac{\cos(2\beta)}{2} 
\frac{\dg}{\Gamma_d}
+ {\cal O} \left[ (\dg/\Gamma_d)^2 \right]~~.
\label{cp-sl}
\end{equation}
The measurement of these two lifetimes 
will give a value of $|\dg|$,
since $|\cos(2\beta)|$ will already be known with good accuracy
by that time. 

The LHC expects about $7 \times 10^5$ events of 
$J/\psi {\rm K}_S$ per year, whereas the number of 
s.l.\  decays at LHCb alone that will be directly useful in the 
lifetime measurements is expected to exceed $10^6$ per year. 
The s.l.\  data sample may be further increased by including 
self-tagging decay modes, such as ${\rm D}_s^{(*)+}{\rm D}^{(*)-}$.

At hadronic machines, the ${\rm B}_d/\overline{\rm B}_d$~production
asymmetry may be a stumbling block for the determination of the
average untagged lifetimes. This drawback is obviously absent at the
B~factories.  There, the most promising approach is to constrain
$\Delta\Gamma_d / \Gamma_d$ by using $\Upsilon(4 {\rm S})$~events
where one B~meson is fully reconstructed in a CP-specific decay mode,
and the decay point of the second B~meson is reconstructed using an
inclusive technique that relies predominantly on s.l. and other
self-tagging modes. 
For these events, only the {\it signed} difference
  of proper decay-times, $\Delta t = t_{\rm CP} - t_{\rm tag}$, i.e.\ not
  the decay times themselves, can be inferred since the production point
  cannot be reconstructed. The average value of~$\Delta t$ is given by
\begin{equation} \left<
\Delta t \right> = \eta_{\rm CP} \cos(2\beta) \, \tau_{{\rm B}_d}
\frac{\Delta\Gamma_d}{\Gamma_d} +{\cal O} \left[ (\Delta\Gamma_d /
\Gamma_d)^3 \right]
\end{equation}
where $\eta_{\rm CP}$~denotes the CP~eigenvalue of the CP-specific
final state considered. The BaBar potential has been studied using
$J/\psi \, {\rm K}_S$ and similar charmonium final states. The expected
statistical precision on~$\Delta\Gamma_d / \Gamma_d$ is determined
using the B~reconstruction efficiencies and the experimental $\Delta
t$~resolution determined from BaBar's first data. From extrapolations
based on published BaBar measurements of $\tau_{{\rm
B}_d}$~and~$\sin(2\beta)$, the precision on $\tau_{{\rm
B}_d}$~and~$\cos(2\beta)$ is expected to improve at the same time as
the precision on~$\left< \Delta t \right>$, and to remain good enough
to turn the $\left< \Delta t \right>$~measurement into an evaluation
of $\Delta\Gamma_d / \Gamma_d$. Using $\tau_{{\rm B}_d} = 1.55$~ps,
$\sin(2\beta) = 0.6$ and 30~fb$^{-1}$ of data one gets:
$\sigma\left(\Delta\Gamma_d / \Gamma_d \right) = 0.073$.  Using
300~fb$^{-1}$ of data $\sigma\left(\Delta\Gamma_d / \Gamma_d \right) =
0.023$ is expected, and for 500~fb$^{-1}$ $\sigma\left(\Delta\Gamma_d
/ \Gamma_d \right) = 0.018$.\\ At super B~factories, 50~ab$^{-1}$ of
data may be obtained. A statistical precision at the $0.2$~\%~level
could be achieved. Strategies to reduce the systematic uncertainties
to this level have not yet been studied in detail.

\begin{comment}{
At hadronic machines, the production asymmetry may be a stumbling
block for determining the average untagged lifetimes. This 
drawback is absent at the B factories. The BaBar potential have been 
studied exploiting s.l.\  and $J/\psi K^0_s$ B  decays.
The average 
$$\langle \Delta t\rangle=\langle t_{CP}-t_{Tag}\rangle= 
\eta_{CP}\cos\left(2\beta\right)\tau_{{\rm B}_d}\frac{\Delta\Gamma_d}{\Gamma_d}$$
 where $t_{CP}$ and $t_{Tag}$ are the reconstructed time for the 
$J/\psi {\rm K}^0_s$ and s.l.\  decays respectively, $\eta_{CP}$ is CP 
phase sign.  The precision on $\tau_{{\rm B}_d}$ and $\cos\left(2\beta\right)$ 
will be enough to turn the  $\langle \Delta t\rangle$ 
measurement into a $\Delta\Gamma_d/\Gamma_d$ evaluation. 
Using   $\tau_{{\rm B}_d} =1.55$ ps, 
$\sin\left(2\beta\right)=0.6$ and  30 fb$^{-1}$ of data 
one gets: 
$\sigma\left(\Delta\Gamma_d/\Gamma_d\right)=0.073$. Using 300 fb$^{-1}$ 
of data $\sigma\left(\Delta\Gamma_d/\Gamma_d\right)=0.023$ is expected 
and for 500 fb$^{-1}$ $\sigma\left(\Delta\Gamma_d/\Gamma_d\right)=0.018$.
At super B factories 50 ab$^{-1}$ of data may be obtained. 
If the systematic errors will decrease like the statistical, 
an accuracy of 0.2\% could be reached. 
}\end{comment}

\subsection{Theoretical description of $b$-hadron lifetimes and 
comparison with 
      experiment}
\label{sec:blifetheory}

The same theoretical tools used to study $\dgamma$ in Sec.~\ref{sec:dgbstheo} can also be 
applied to describe the lifetime ratios of hadrons containing a $b$-quark, 
such as $\rp$, $\rs$, $\rl$. The leading contributions in the heavy quark 
expansion (HQE) are represented, in the present case, by the dimension-3 
operator $\bar bb$ (${\cal O}(1)$) and the dimension-$5$ operator $\bar b 
\sigma_{\mu \nu} G^{\mu \nu} b$ (${\cal O}(1/m_b^2)$). The first term in the 
expansion reproduces the predictions of the na\"\i ve quark spectator model. 
At this order, the hadronic decay is described in terms of the free $b$-quark 
decay, and the lifetime ratios of beauty hadrons are all predicted to be 
unity.
The leading corrections, of ${\cal O}(1/m_b^2)$, describe the soft 
interactions 
of the spectator quark(s) inside the hadron, but  give a 
small contribution ($\simle 2\% $) to the lifetime ratios.

\begin{figure}[t] 
\begin{center}
\epsfxsize=10.7cm
\epsfbox{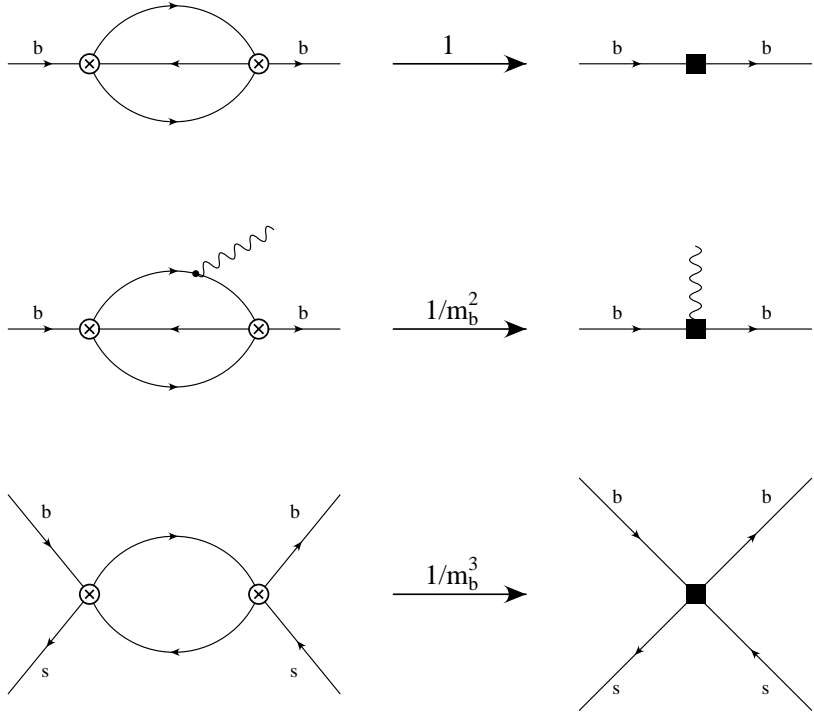}
\end{center}
\caption{\it \label{figA2}  Examples of LO contributions 
to the transition operator ${\cal T}$
(left) and to the corresponding local operator (right). The crossed circles 
represent the insertions of the $\DB=1$ effective Hamiltonian. The black 
squares represent the insertion of a $\DB=0$ operator.}
\end{figure}

The large lifetime difference of beauty hadrons which has been observed experimentally can be
explained by considering hard spectator effects, that appear at ${\cal O}(1/m_b^3)$. 
Although suppressed by an 
additional power of $1/m_b$, these effects are enhanced with respect to the 
leading contributions by a phase-space factor of $16\pi^2$, being $2\to 2$ 
processes instead of $1\to 3$ decays (see Fig.~\ref{figA2}). 
As in the case of the OPE for $\dgamma$, the starting point to describe the
beauty hadron lifetimes is the effective $\DB=1$ weak Hamiltonian, which enter
the transition operator 
\begin{equation}
{\cal T} = i 
\int d^4x \; T \left( \Heff^{\DB=1}(x) \Heff^{\DB=1}(0) \right)\,.
\label{eq:T2}
\end{equation}
From the forward matrix elements of this operator, and using the optical 
theorem,
one computes the inclusive decay width of a hadron $H_b$ containing a $b$ 
quark 
\begin{equation}
\Gamma(H_b) =
\frac{1}{M_{H_b}} 
\mathrm{Im} \langle H_b \vert {\cal T} \vert H_b \rangle\, .
\end{equation}
The result of the HQE, in this case, is expressed in terms of matrix 
elements of $\DB=0$ operators and it is given by   
\begin{equation}
\Gamma(H_b) =
\frac{G_F^2 |V_{cb}|^2 m_b^5}{192 \pi^3}\left[
c^{(3)} \frac{\langle H_b\vert \bar b b \vert H_b\rangle}{2 M_{H_b}} +
c^{(5)} \frac{g_s}{m_b^2} \frac{\langle H_b\vert \bar b\sigma_{\mu\nu}
G^{\mu\nu}b \vert H_b\rangle}{2 M_{H_b}} +
\frac{96\pi^2}{m_b^3} \dsum_{k} c^{(6)}_k 
\frac{\langle H_b\vert O_k^{(6)}\vert H_b\rangle}{2 M_{H_b}}\right]\, ,
\label{eq:gamma}
\end{equation}
where we have included all contributions up to ${\cal O}(1/m_b^2)$ and those 
$1/m_b^3$ corrections which are enhanced by the phase-space factor $16\pi^2$. 
The complete list of the dimension-6 operators $O_k^{(6)}$, which represent the
contribution of hard spectator effects, includes
\begin{equation}
\begin{array}{ll}
{\cal O}^q_1= (\bar b \,q)_{V-A} \, (\bar q \,b)_{V-A}\,, &
{\cal O}^q_2= (\bar b \,q)_{S-P} \, (\bar q \,b)_{S+P}\,, \\
{\cal O}^q_3= (\bar b \,t^{a} q)_{V-A} \, (\bar q \,t^{a} b)_{V-A}\,, &
{\cal O}^q_4= (\bar b \,t^{a} q)_{S-P} \, (\bar q \,t^{a} b)_{S+P}\, ,
\end{array}
\label{eq:opeff}
\end{equation}
with $q=u,d,s,c$, and the penguin operator
\begin{equation}
{\cal O}_P=(\bar b t^{a} b)_{V} \dsum\limits_{q=u,d,s,c}(\bar q t^{a} q)_{V}\,.
\label{eq:openg}
\end{equation}
It is important to emphasize that the symbols $b$ and $\bar b$ in the 
operators~(\ref{eq:opeff},\ref{eq:openg}) denote the heavy quark field HQET. 
The reason is that renormalized operators, in QCD, 
mix with operators  of lower dimension, with coefficients
 proportional to powers of the $b$-quark 
mass. Therefore, the dimensional ordering of the HQE, based on the assumption 
that contributions of higher dimensional operators are suppressed by 
increasing 
powers of the $b$-quark mass, would be lost in this case. In order to 
implement 
the expansion, the matrix elements of the local operators should be cut-off at 
a scale smaller than the $b$-quark mass, which is naturally realized in the 
HQET. The HQE can be expressed in terms of QCD operators in those cases in 
which, because of their specific quantum numbers, these operators cannot mix 
with lower lower dimensional operators. This is the case, for instance, 
for the 
leading contributions in the HQE of $\dgamma$ and of the lifetime ratio $\rp$.

The Wilson coefficients $c^{(3)}$ and $c^{(5)}$ in 
Eq.~(\ref{eq:gamma}) have been computed at the LO in Ref.~[\ref{Bigi:1992su}], 
while the NLO corrections to $c^{(3)}$ have been evaluated in 
[\ref{ref:niri},\ref{gamma1}--\ref{gamma4}]. 
The NLO corrections to $c^{(5)}$ are still 
unknown, but their impact on the lifetime ratio is expected to be negligible.
The coefficient functions $c^{(6)}_k$ of the current-current operators of 
dimension-6 have been computed at the LO in 
Refs.~[\ref{NS}--\ref{charmBB}]. At this 
order the coefficient of the penguin operator $c_P^{(6)}$  vanishes. The NLO 
correction to $c^{(6)}_k$  for the operators ${\cal O}^q_k$ with $q=u,d$ has 
been recently completed in Refs.~[\ref{Roma},\ref{Lenz}], 
and extended to $q=s$ in 
Ref.~[\ref{Roma}]. A complete list of these coefficients, calculated at NLO in 
the $\msb$ scheme of Ref.~[\ref{reyes}], is given in Table~\ref{tableone}. The 
operators containing the valence charm quark ($q=c$ in Eq.~(\ref{eq:opeff}))
are expected to give a negligible contribution to the non-charmed 
hadron decay 
rates. The calculation of the NLO corrections to these coefficient functions, 
as well as the NLO calculation of the coefficient function of the penguin 
operator, have not been performed yet. 
\begin{table}[htbp]
\renewcommand{\arraystretch}{1.2} 
\begin{center}
\begin{tabular}{|c|c c c|}
\hline
$ q$ & $u$ & $d$ & $s$ \\ \hline 
$ c^{\,q}_1$  & $-0.29^{+0.02}_{-0.04}$ & $-0.03^{-0.01}_{+0.01}$
              & $-0.03^{-0.01}_{+0.01}$ \\
$ c^{\,q}_2$  & $-0.02^{-0.01}_{+0.01}$ & $\phantom{-}0.03^{+0.01}_{-0.02}$ 
              & $\phantom{-}0.04^{+0.00}_{-0.02}$ \\
$ c^{\,q}_3$  & $\phantom{-}2.37^{+0.12}_{-0.10}$ & $-0.68^{-0.01}_{+0.01}$ 
              & $-0.58^{-0.00}_{+0.01}$ \\
$ c^{\,q}_4$  & $-0.05^{-0.01}_{+0.00} $  & $\phantom{-}0.68^{-0.00}_{+0.00}$  
              & $\phantom{-}0.65^{-0.00}_{+0.00}$ \\ \hline
\end{tabular}
\end{center}
\renewcommand{\arraystretch}{1.0} 

\vspace{-3mm}

\caption{\it \label{tableone} 
Wilson coefficients $c^q_k(\mu_0)$ computed in the HQET, at NLO, at the scale 
$\mu_0=m_b$. The coefficients also have a residual dependence on the
renormalization scale $\mu_1$ of the $\DB=1$ operators, which is a NNLO effect.
The uncertainty due to the variation of the scale $\mu_1$ is reflected in the 
error bars (central values are obtained  by using $\mu_1=m_b$, upper error for 
$\mu_1=m_b/2$ and the lower one for $\mu_1=2\, m_b$). 
In the evaluation we take 
$m_c/m_b=0.28$. All the coefficients remain unchanged under the variation of 
$m_c/m_b=0.28\pm 0.02$ except for $ c^{\,q}_3$, which changes by about $2\%$.}
\end{table}

The matrix elements of dimension-3 and dimension-5 operators, appearing in 
Eq.~(\ref{eq:gamma}), can be expressed in terms of the HQET parameters 
$\mu_\pi^2(H_b)$ and $\mu_G^2(H_b)$ as
\begin{eqnarray}
\langle H_b \vert \bar b b\vert H_b\rangle &=&2 M_{H_b} \left( 1-
\frac{\mu_\pi^2(H_b)-\mu_G^2(H_b)}{2 m_b^2}+{\cal O}(1/m_b^3)\right)\, ,\nn \\
\langle H_b \vert \bar b g_s \sigma_{\mu\nu}G^{\mu\nu} b\vert H_b \rangle &=& 
2M_{H_b}\left(2\mu^2_G(H_b)+{\cal O}(1/m_b)\right)\, .
\label{eq:hqet}
\end{eqnarray}
Using these expansions in the lifetime ratio of two beauty hadrons one 
finds
\begin{eqnarray}
\label{eq:ratio}
\frac{\tau(H_b)}{\tau(H_b^\prime)}&=&1+\frac{\mu_\pi^2(H_b)-
\mu_\pi^2(H_b^\prime)}{2 m_b^2}-\left(\frac{1}{2}+\frac{2 c^{(5)}}{c^{(3)}}
\right) \frac{\mu_G^2(H_b)-\mu_G^2(H_b^\prime)}{m_b^2}\nn\\
&&\quad-\frac{96\pi^2}{m_b^3\, c^{(3)}}\dsum\limits_{k}
c^{(6)}_k \left(\frac{\langle H_b^\prime \vert O^{(6)}_k\vert H_b \rangle}
{2M_{H_b}}-\frac{\langle
H_b^\prime \vert O_k^{(6)}\vert H_b^\prime\rangle }{2M_{H_b^\prime}}\right)\,.
\end{eqnarray}
From the heavy hadron spectroscopy one obtains
$\mu_\pi^2(\Lambda_b)-\mu_\pi^2({\rm B}) \approx 0.01(3)\ \gev^2$ and 
$\mu_\pi^2(\Lambda_b)-\mu_\pi^2({\rm B})\approx 0$. 
Therefore the impact of the 
second term in the above formula is completely negligible. On the other hand, 
$\mu_G^2({\rm B}_q)= 3(M_{{\rm B}_q^\ast}^2-M_{{\rm B}_q}^2)/4$, 
which gives $\mu_G^2({\rm B}_{u,d})
\approx 0.36~\gev^2$, $\mu_G^2({\rm B}_s)\approx 0.38~\gev^2$, while   
$\mu_G^2(\Lambda_b)= 0$. Therefore, only in the case $\rl$, the third term 
gives a contribution that is visibly different from zero. By using 
$(1/2 + 2c^{(5)}/c^{(3)})=-1.10(4)$, we thus obtain
\begin{equation}
\label{eq:del}
\frac{\tau({\rm B}^+)}{\tau({\rm B}_d)}= 1.00 -
\Delta^{{\rm B}^+}_{\rm\scriptstyle spec}\,,\quad
\frac{\tau({\rm B}_s)}{\tau({\rm B}_d)}= 1.00 -
\Delta^{{\rm B}_s}_{\rm\scriptstyle spec}\,,\quad
\frac{\tau(\Lambda_b)}{\tau({\rm B}_d)}= 0.98(1)-
\Delta^{\Lambda}_{\rm\scriptstyle spec} \,,
\end{equation}
where the $\Delta^{H_b}_{\rm spec}$ represent the $1/m_b^3$ contributions of 
hard spectator effects (second line in Eq.~(\ref{eq:ratio})).

The comparison of Eq.~(\ref{eq:del}) with the experimental results given in 
Table~\ref{tab:complr} shows that without inclusion of the spectator 
effects the experimental values could not be explained. 
\begin{table}[htbp]
\renewcommand{\arraystretch}{1.5} 
\begin{center}
\begin{tabular}{|l|c|c|}
\hline
                                    &  Theory Prediction & World Average \\ \hline
$\frac{\tau({\rm B}^+)}{\tau({\rm B}_d)}$       &  1.06$\pm$0.02 & 1.073$\pm$0.014 \\
$\frac{\tau({\rm B}_s)}{\tau({\rm B}_d)}$       &  1.00$\pm$0.01 & 0.949$\pm$0.038 \\
$\frac{\tau(\Lambda_b)}{\tau({\rm B}_d)}$ &  0.90$\pm$0.05 & 0.798$\pm$0.052 \\ \hline
\end{tabular}
\end{center}

\caption{\it Comparison of theoretical expectations and experimental results for the ratios 
of exclusive lifetimes.}
\label{tab:complr}
\end{table}

Beside the coefficient functions presented in Table~\ref{tableone}, the 
essential ingredients entering the corrections $\Delta^{H_b}_{\rm spec}$ are 
the hadronic matrix elements. We follow [\ref{noantri}] and parameterize 
the B meson matrix elements as follows

\vspace{2mm}

\begin{equation}
\begin{array}{ll}
\frac{\langle {\rm B}_q\vert {\cal O}^q_1 \vert {\rm B}_q\rangle}{2M_{{\rm B}_q}}=
\frac{F_{{\rm B}_q}^2M_{{\rm B}_q}}{2} \, \left( B_1^{\, q} +
\delta_1^{\, qq}\right) \, , \quad
\frac{\langle {\rm B}_q\vert {\cal O}^q_3 \vert {\rm B}_q\rangle}{2M_{{\rm B}_q}}=
\frac{F_{{\rm B}_q}^2M_{{\rm B}_q}}{2} \, \left( \ep_1^{\, q} +
\delta_3^{\, qq}\right) \, , \\
\frac{\langle {\rm B}_q\vert {\cal O}^q_2 \vert {\rm B}_q\rangle}{2M_{{\rm B}_q}}=
\frac{F_{{\rm B}_q}^2M_{{\rm B}_q}}{2} \, \left( B_2^{\, q} +
\delta_2^{\, qq}\right) \, , \quad
\frac{\langle {\rm B}_q\vert {\cal O}^q_4 \vert {\rm B}_q\rangle}{2M_{{\rm B}_q}}=
\frac{F_{{\rm B}_q}^2M_{{\rm B}_q}}{2} \, \left( \ep_2^{\, q} +
\delta_4^{\, qq}\right) \,,\\
\frac{\langle {\rm B}_{q}\vert {\cal O}^{q'}_k\vert {\rm B}_{q}\rangle}{2M_{{\rm B}_q}}=
\frac{F_{{\rm B}_q}^2M_{{\rm B}_q}}{2} \, \delta_k^{\,q'q} \quad 
\,, \quad\quad\quad
\frac{\langle {\rm B}_q\vert {\cal O}_P\vert {\rm B}_q\rangle}{2M_{{\rm B}_q}}=
\frac{F_{B}^2M_{B}}{2} \, P^{\, q} \,.
\end{array}
\label{eq:bpar}
\end{equation}

\vspace{2mm}

\noindent 
where the parameters $\delta_k^{\, qq}$ are defined as the $\delta_k^{\, qq'}$ 
in the limit of degenerate quark masses ($m_q=m_{q'}$). For the $\Lambda_b$
baryon we define
\begin{eqnarray}
\label{eq:bpar4}
&& \frac{\langle \Lambda_b\vert {\cal O}^q_1 \vert \Lambda_b\rangle}
{2M_{\Lambda_b}}= \frac{F_{B}^2M_{B}}{2} \, \left( L_1 +
\delta_1^{\,\Lambda q}\right) \quad {\rm for} \,\, q=u,d \, ,\nonumber \\
&& \frac{\langle \Lambda_b\vert {\cal O}^q_3 \vert \Lambda_b\rangle}
{2M_{\Lambda_b}}= \frac{F_{B}^2M_{B}}{2} \, \left( L_2 +
\delta_2^{\,\Lambda q}\right) \quad {\rm for} \,\, q=u,d  \, , \nonumber \\
&& \frac{\langle \Lambda_b\vert {\cal O}^q_1\vert \Lambda_b\rangle}
{2M_{\Lambda_b}}= \frac{F_{B}^2M_{B}}{2} \, \delta_1^{\, \Lambda q}
\quad {\rm for}\,\, q=s,c   \, ,\\
&& \frac{\langle \Lambda_b\vert {\cal O}^q_3\vert \Lambda_b\rangle}
{2M_{\Lambda_b}}= \frac{F_{B}^2M_{B}}{2} \, \delta_2^{\, \Lambda q}
\quad {\rm for}\,\, q=s,c   \, ,\nonumber \\
&& \frac{\langle \Lambda_b\vert {\cal O}_P\vert \Lambda_b\rangle}
{2M_{\Lambda_b}}=\frac{F_{B}^2M_{B}}{2} \, P^{\, \Lambda} \nonumber \,.
\end{eqnarray}
In addition, in the case of $\Lambda_b$, the following relation holds up to 
$1/m_b$ corrections:
\begin{equation}
\langle \Lambda_b\vert {\cal O}^q_1 \vert \Lambda_b\rangle = -2 \,
\langle \Lambda_b\vert {\cal O}^q_2 \vert \Lambda_b\rangle \, , \quad
\langle \Lambda_b\vert {\cal O}^q_3 \vert \Lambda_b\rangle = -2 \,
\langle \Lambda_b\vert {\cal O}^q_4 \vert \Lambda_b\rangle \, .
\end{equation}
In Eqs.(\ref{eq:bpar}) and (\ref{eq:bpar4}), $B_{1,2}$, $L_{1,2}$ and 
$\varepsilon_{1,2}$ are the ``standard" bag parameters, introduced in 
Ref.~[\ref{NS}]. Those parameters have already been computed in both the 
lattice 
QCD and QCD sum rule approaches. The parameters $\delta_k$ have been 
introduced 
in Ref.~[\ref{noantri}] to account for the corresponding penguin contractions. 
A non-perturbative lattice calculation of the $\delta_k$ parameters is 
possible, in principle. However, the difficult 
problem 
of subtractions of power-divergences has prevented their
calculation.

In terms of parameters introduced above, the spectator contributions 
to the 
lifetime ratios, $\Delta^{H_b}_{\rm\scriptstyle spec}$, are expressed in the 
form

\begin{eqnarray}
\label{eq:masterformula}
\Delta^{B^+}_{\rm\scriptstyle spec}&=& 48\pi^2\, 
\frac{F_B^2 M_B}{m_b^3 c^{(3)}}
\, \dsum_{k=1}^4 \left( c_k^{\,u}- c_k^{\,d} \right)
{\cal B}_k^{\, d} \, , \nn \\
\Delta^{{\rm B}_s}_{\rm\scriptstyle spec}&=&
48\pi^2\, \frac{F_B^2 M_B}{m_b^3 c^{(3)}} \, \left\{ \dsum_{k=1}^4
\left[r \,  c_k^{\,s} \, {\cal B}_k^{\, s} -  c_k^{\,d} \,
{\cal B}_k
^{\, d} + \left(  c_k^{\,u} +  c_k^{\,d} \right) \left(r\,
\delta_k^{\,ds} - \delta_k^{\,dd} \right) + \right. \right.  \nn \\ && \qquad
\qquad \quad \left. \left.
 c_k^{\,s} \left(r \, \delta_k^{\,ss} - \delta_k^{\,sd} \right) +
 c_k^{\,c}
\left(r \, \delta_k^{\,cs} - \delta_k^{\,cd} \right) \right] + c_P
\left(r P^{\, s} - P^{\, d} \right) \right\} \, ,\\
\Delta^{\Lambda}_{\rm\scriptstyle spec}&=&
48\pi^2\, \frac{F_B^2 M_B}{m_b^3 c^{(3)}} \, \left\{ \dsum_{k=1}^4 \left[\left(
  c_k^{\,u} +  c_k^{\,d} \right) {\cal L}_k^{\, \Lambda} -
  c_k^{\,d} \, {\cal B}_k^{\, d} + \left(  c_k^{\,u} +
 c_k^{\,d} \right) \left(\delta_k^{\,\Lambda d} - \delta_k^{\,dd} \right)
+ \right. \right.  \nn \\
&& \qquad \qquad \quad \left. \left.  c_k^{\,s}\left(\delta_k^{\,
\Lambda s}- \delta_k^{\,sd} \right) +  c_k^{\,c} \left(\delta_k^{\,
\Lambda c}- \delta_k^{\,cd} \right) \right] + c_P \left(P^{\,
\Lambda} - P^{\, d} \right)
\right\} \, . \nn
\end{eqnarray}
where $r$ denotes the ratio $(F_{{\rm B}_s}^2 M_{{\rm B}_s})/(F_B^2 M_B)$ and, in order to 
simplify the notation, we have defined the vectors of parameters

\begin{eqnarray}
&& \vec {\cal B}^q=\{B_1^q,B_2^q,\ep_1^q,\ep_1^q\} \, ,\nonumber \\[1mm]
&& \vec {\cal L}=\{L_1,-L_1/2,L_2,-L_2/2\} \, ,\\[1mm]
&& \vec \delta^{\Lambda q}=\{\delta^{\Lambda q}_1,-\delta^{\Lambda q}_1/2,
\delta^{\Lambda q}_2,-\delta^{\Lambda q}_2/2\} \nonumber \, .
\end{eqnarray}

An important result of Eq.~(\ref{eq:masterformula}) is that, because of 
the $SU(2)$ symmetry, the non-valence ($\delta$s) and penguin ($P$s)
contributions cancel out in the expressions of the lifetime ratio $\rp$. Thus,
the theoretical prediction of this ratio is at present the most accurate, 
since 
it depends only on the non-perturbative parameters actually computed by 
current 
lattice calculations. The prediction of the ratio $\rl$, instead, is affected 
by both the uncertainties on the values of the $\delta$ and $P$ parameters, 
and 
by the unknown expressions of the Wilson coefficients $c_k^{\, c}$ and $c_P$ 
at 
the NLO. For the ratio $\rs$ the same uncertainties exist, although their 
effect
is expected to be smaller, since the contributions of non-valence and penguin
operators cancel, in this case, in the limit of exact $SU(3)$ symmetry.

In the numerical analysis of the ratios $\rs$ and $\rl$, we will neglect the 
non-valence and penguin contributions ({\it i.e.} we set all $\delta=P=0$). 
The non-valence contributions vanish in the VSA, and present phenomenological 
estimates indicate that the corresponding matrix elements are suppressed, with 
respect to the valence contributions, by at least one order of 
magnitude~[\ref{Chernyak:1995cx},\ref{pirjol}]. On the other hand, the matrix 
elements 
of the penguin operators are not expected to be smaller than those of the 
valence operators. Since the coefficient function $c_P$ vanishes at the LO, 
this contribution is expected to have the size of a typical NLO corrections. 
Thus, from a theoretical point of view, a quantitative evaluation of the 
non-valence and penguin operator matrix elements would be of the greatest 
interest to improve the determination of the $\Lambda_B$ lifetime.

By neglecting the non valence and penguin contributions, and using for the
Wilson coefficients the NLO results collected in Table ~\ref{tableone}, one
obtains from Eq.~(\ref{eq:masterformula}) the following expressions
\begin{equation}
\label{eq:magic}
\renewcommand{\arraystretch}{1.2}
\begin{array}{ll}
\Delta^{B^+}_{\rm\scriptstyle spec} \, =&
-\, 0.06(2) \, B_1^d - 0.010(3) \, B_2^d + 0.7(2) \, 
\ep_1^d -0.18(5) \, \ep_2^d
\,,\\ \\
\Delta^{{\rm B}_s}_{\rm\scriptstyle spec} \, =&
-\, 0.010(2) \, B_1^s + 0.011(3) \, B_2^s - 0.16(4)\,
\ep_1^s + 0.18(5)\,\ep_2^s
\\ &
+\, 0.008(2)\, B_1^d - 0.008(2) \, B_2^d + 0.16(4) \, 
\ep_1^d -0.16(4)\, \ep_2^d
\,,\\ \\
\Delta^{\Lambda}_{\rm\scriptstyle spec} \, =&
-\, 0.08(2)\, L_1 + 0.33(8)\, L_2 \\ &
+\, 0.008(2)\, B_1^d - 0.008(2) \, B_2^d + 0.16(4) \, 
\ep_1^d -0.16(4)\, \ep_2^d
\,,\end{array}
\renewcommand{\arraystretch}{2.0}
\end{equation}
For the charm and bottom quark masses, and the B meson decay constants we have 
used the central values and errors given in Table~\ref{3tab:inputs}. The strong 
coupling constant has been fixed at the value $\as(m_Z)=0.118$. The parameter 
$c^{(3)}$ in Eq.~(\ref{eq:masterformula}) is a function of the ratio $m_c^2/
m_b^2$, and such a dependence has been consistently taken into account in the 
numerical analysis and in the estimates of the errors. For the range of masses 
given in Table~\ref{3tab:inputs}, $c^{(3)}$ varies in the interval $c^{(3)}=3.4
\div 4.2$~[\ref{gamma4}].
\begin{table}[htbp] 
\renewcommand{\arraystretch}{1.2}
\begin{center}
\begin{tabular}{|cc|}
\hline
$ B_1^d$ = $1.2 \pm 0.2$ & $ B_1^s$ = $1.0 \pm 0.2$  \\
$ B_2^d$ = $0.9 \pm 0.1$ & $ B_2^s$ = $0.8 \pm 0.1$ \\
$ \ep_1^d$ = $0.04 \pm 0.01$ & $ \ep_1^s$ = $0.03 \pm 0.01$ \\
$ \ep_2^d$ = $0.04 \pm 0.01$ & $ \ep_2^s$ = $0.03 \pm 0.01$ \\
\hline
$ L_1$ = $-0.2 \pm 0.1$ & $ L_2$ = $0.2 \pm 0.1$ \\
\hline
$m_b$ = $4.8\pm 0.1$ GeV & $m_b-m_c$ = $3.40\pm 0.06$ GeV \\
$F_B$ = $200\pm 25$ MeV & $F_{{\rm B}_s}/F_B$ = $1.16\pm 0.04$ \\
\hline
\end{tabular}
\end{center}
\caption{\it \label{3tab:inputs}
\it Central values and standard deviations of the 
input parameters used in the numerical analysis. The values of 
$m_b$ and $m_c$ refer to the pole mass definitions of these quantities.}
\end{table}

As discussed before, for the ratio $\rp$ the HQE can be also 
expressed in terms 
of operators defined in QCD. The corresponding coefficient functions can be 
evaluated by using the matching between QCD and HQET computed, at the NLO, in 
Ref.~[\ref{noantri}]. In this way, one obtains the expression
\begin{equation}
\label{eq:magic_qcd}
\Delta^{B^+}_{\rm\scriptstyle spec} =
-\, 0.05(1) \, \bar B_1^d - 0.007(2) \, \bar B_2^d + 0.7(2) \, \bar \ep_1^d 
-0.15(4) \, \bar \ep_2^d
\end{equation}
where the $\bar B$ and $\bar \ep$ parameters are now defined in 
terms of matrix elements of QCD operators. 

The errors quoted on the coefficients in Eq.~(\ref{eq:magic}) are strongly 
correlated, since they originate from the theoretical uncertainties on 
the same 
set of input parameters. For this reason, in order to evaluate the lifetime 
ratios, we have performed a Bayesian statistical analysis by implementing a 
short Monte Carlo calculation. The input parameters have been extracted with 
flat distributions, assuming as central values and standard deviations the 
values given in Table~\ref{3tab:inputs}. The results for the $B$-parameters are 
based on the lattice determinations of Refs.~[\ref{DiPierro98},\ref{APE}]
\footnote{For recent estimates of these matrix elements based on QCD 
sum rules, see Refs.~[\ref{Colangelo:1996ta}].}.
We have included in the errors an estimate of the uncertainties not taken into 
account in the original papers. The QCD results for the B meson 
$B$-parameters of Ref.~[\ref{APE}] have been converted to HQET at the NLO
[\ref{noantri}]\footnote{With respect to [\ref{noantri}], 
we use for the B meson 
$B$-parameters the results updated in [\ref{APE}].}.
The contributions of all the $\delta$ and $P$ parameters have been neglected.
\begin{figure}[t]
\begin{center}
\epsfxsize=12cm
\epsfbox{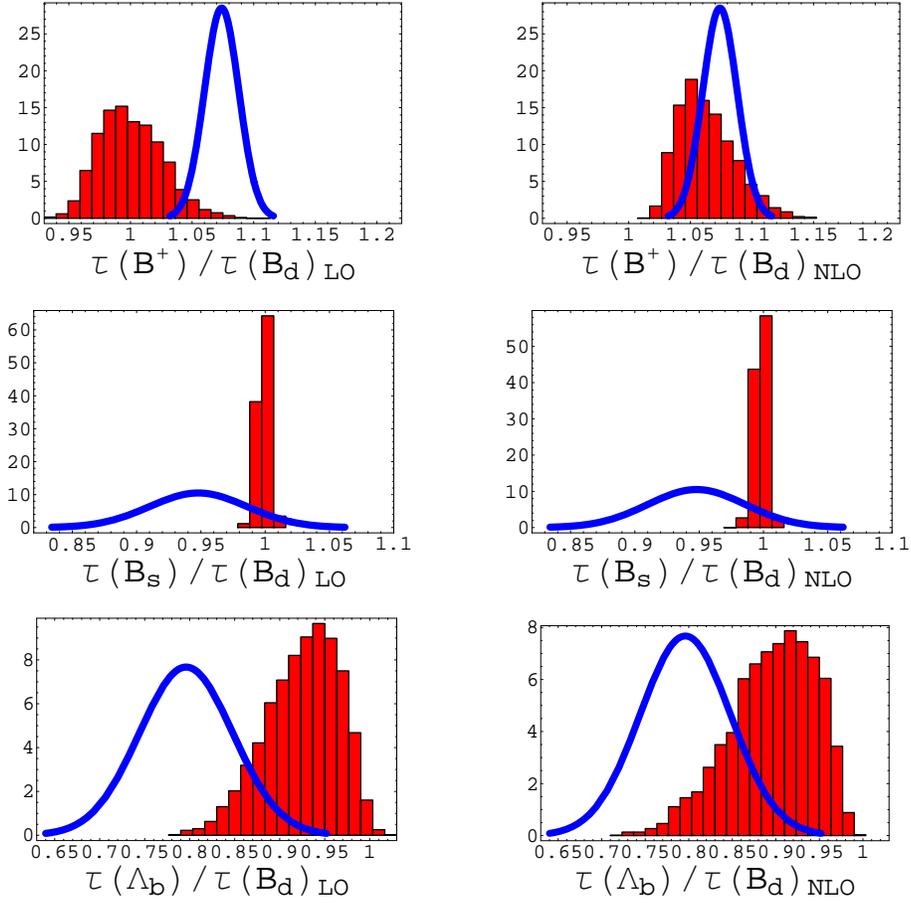}
\end{center}
\caption{\it Theoretical (histogram) vs experimental (solid line) 
distributions of lifetime ratios. The theoretical predictions are shown 
at the LO (left) and NLO (right).}
\label{figB2}
\end{figure}
In this way we obtain the final NLO predictions for the lifetimes ratios summarised in 
Table~\ref{tab:complr}.
The central values and uncertainties correspond to the average and the standard 
deviation of the theoretical distributions, shown in Fig.~\ref{figB2}, together with 
those from the experimental determinations. The uncertainty coming from the residual scale
dependence 
represents less than 20\% of the quoted errors. 

With the inclusion of the NLO corrections, the theoretical prediction for the ratio 
$\rp$ is in good agreement with the experimental measurement, also summarised 
in Table~\ref{tab:complr}.
The agreement is also good for the ratio $\rs$, with the difference between theoretical 
predictions and experimental determinations below $1\sigma$. A possible mismatch 
between the predicted and measured values for the ratio $\rl$ has been much debated 
in past years. Interpretation in terms of a breakdown of the HQE
framework and the appearance of a signal of quark-hadron duality violation have been 
claimed. The inclusion of higher order terms seems to reestablish a compatibility between 
predictions of the beauty baryon lifetime with the present experimental determinations.
However, this issue will require further scrutiny in view of new, more precise results 
expected from the Tevatron Run~II and from the fact that the theoretical predictions 
are less accurate in this case, since a reliable estimate of the contribution of the 
non-valence and penguin operators are not yet available. 

\boldmath 
\subsection{Future prospects for $b$-hadron lifetime measurements}
\label{sec:life}
\unboldmath 

\vspace{2mm}

The B factories are now providing new, accurate determinations of the lifetimes of the 
${\rm B}^0_d$ and ${\rm B}^+$ meson, which could decrease the relative error on to 
(0.4-0.5)\%. 
Results from the Tevatron Run~II are eagerly expected, since will provide precise 
measurements of the ${\rm B}^0_s$ and $\Lambda_b$ lifetimes and also results for the $\Xi_b$, 
$\Omega_b$ and the ${\rm B}_c$ beauty hadrons.
Further improvements are then expected from the LHC experiments, with special regard 
to ${\rm B}^0_s$ and baryon lifetimes.

CDF evaluated the lifetimes measurement capabilities exploiting separately 
the leptonic and the hadronic decay channels.  The leptonic decays considered 
are only to $J/\psi\rightarrow \mu\mu$, this means exclusive decays. The 
uncertainties shown in Table~\ref{table:lifecdf} are only statistical 
and are obtained by scaling by a factor 50 the Run I measurements. 
The systematic uncertainty is expected to be 
the same order as that for the Run~I analyses, at the level of $1\%$.
Since in Run~I there were no measurements based on hadronic decays, the 
Run~II estimations had to be based on Monte Carlo simulations. 
The major interest is in measuring the ${\rm B}_s$ and $\Lambda_b$ lifetime and the expected statistical 
errors are quoted in Table~\ref{table:lifecdf}.
With these measurements the ${\rm B}_s$/${\rm B}^0$ lifetime ratio to will have an uncertainty of 
$\sim 0.5\%$, which is of the same order of the predicted deviation from unity.
$\Lambda_b$ baryons, reconstructed in the $\Lambda_c\pi$, $pD^0\pi$, $p\pi$ 
and $pK$  decay channels, will allow a stringent test for the theoretical predictions of the 
lifetime ratio of $\Lambda_b$ to ${\rm B}^0$ if the signal to noise ratio of 1 can be obtained. 
\begin{table}[htbp] 
\renewcommand{\arraystretch}{1.5} 
\begin{center} 
 \begin{tabular}{|c|c|c|c|c|}  \hline
$\sigma\left(c\tau\right)/c\tau$ & ${\rm B}^{\pm}$ & ${\rm B}^{0}_d$  & ${\rm B}^{0}_s$  &$\Lambda_b$        \\ \hline
Run II leptonic triggers         & $0.6\%$   &  $0.6\%$   & $2\%$      & $3\%$             \\ 
 Run II hadronic trigger         &           &            & $0.5\%$    & $0.8\%$           \\ \hline
\end{tabular}  

\vspace{2mm}
 
\caption{\it CDF lifetime statistical error projections with leptonic 
and hadronic triggers for 2 fb$^-1$ of data. The systematic uncertainty 
is expected to be at the level of $1\%$.}
\label{table:lifecdf}
\end{center}
\end{table}  

The D{\O} experiment has concentrated its studies on the projection for the 
$\Lambda_b$ lifetime measurement. The preferred decay is $J/\psi \Lambda^0$ 
with $J/\psi\rightarrow \mu\mu$ and $\Lambda^0\rightarrow p\pi^-$. 
In 2~fb$^{-1}$ the expected number of reconstructed events is of order of 
15,000, corresponding to a relative lifetime accuracy of 9\%.

At LHC, lifetime measurements of different B hadron species will be based 
on even larger statistics, collected in individual exclusive channels. 
ATLAS~[\ref{sevelda},\ref{ATLASTDR}]
has performed a simulation for studying the statistical precision on the 
\Lamzerob lifetime using the \Lamzerobto ~decay channel. 
In three years of running at $10^{33}cm^{-2}s^{-1}$ luminosity, 75000 \Lamzerobto signal 
decays can be reconstructed (with 1500 background events, mostly $\rm J/\psi$ paired 
to a primary \Lamzero). Considering a proper time resolution of 0.073~ps, the estimated 
relative uncertainty on the \Lamzerob lifetime is 0.3$\%$.

\newpage

\addcontentsline{toc}{section}{References}
\section*{References}

\vspace{7mm}

\renewcommand{\labelenumi}{[\theenumi]}
\begin{enumerate}

\item \label{ope}
J. Chay, H. Georgi, and B. Grinstein, Phys.\  Lett.\  B~{\bf 247} (1990) 399; 
M. Voloshin and M.~Shifman, Sov.\  J.\  Nucl.\  Phys.\  {\bf 41} (1985) 120;
A.V. Manohar and M.B. Wise, Phys.\  Rev.\  D~{\bf 49} (1994) 1310;
B. Blok {\it et al.}, Phys.\  Rev.\  D~{\bf 49} (1994) 3356.

\vspace{3mm}

\item \label{opebigi}
I.I. Bigi {\it et al.}, Phys.\  Lett.\  B {\bf 293} (1992) 430; 
  Phys.\  Lett.\ B {\bf 297} (1993) 477 (E); 
Phys.\  Rev.\  Lett.\  {\bf 71} (1993) 496.

\vspace{3mm}

\item \label{Manohar:dt}
A.V.~Manohar and M.B.~Wise,
Cambridge Monogr.\ Part.\ Phys.\ Nucl.\ Phys.\ Cosmol.\  {\bf 10} (2000)~1;
I.I.~Bigi, M.A.~Shifman and N.G.~Uraltsev,
Ann.\ Rev.\ Nucl.\ Part.\ Sci.\  {\bf 47} (1997) 591
[hep-ph/9703290];
Z.~Ligeti,
eConf {\bf C020805} (2002) L02 [hep-ph/0302031].

\vspace{3mm}

\item \label{Uraltsev:2000qw}
N.G.~Uraltsev,
in \emph{At the Frontier of Particle Physics: Handbook of~QCD},
edited by M.~Shifman (World Scientific, Singapore, 2001)
[hep-ph/0010328].

\vspace{3mm}

\item \label{Shuryak:pg}
E.V.~Shuryak,
Phys.\ Lett.\ B {\bf 93} (1980) 134;
E.V.~Shuryak,
Nucl.\ Phys.\ B {\bf 198} (1982) 83;
J.E.~Paschalis and G.J.~Gounaris,
Nucl.\ Phys.\ B {\bf 222} (1983) 473;
F.E.~Close, G.J.~Gounaris and J.E.~Paschalis,
Phys.\ Lett.\ B {\bf 149} (1984) 209;
S.~Nussinov and W.~Wetzel,
Phys.\ Rev.\ D {\bf 36} (1987) 130.

\vspace{3mm}

\item \label{Shifman:sm}
M.A.~Shifman and M.B.~Voloshin,
Sov.\ J.\ Nucl.\ Phys.\  {\bf 45} (1987) 292
[Yad.\ Fiz.\  {\bf 45} (1987) 463].

\vspace{3mm}

\item \label{Shifman:1987rj}
M.A.~Shifman and M.B.~Voloshin,
Sov.\ J.\ Nucl.\ Phys.\  {\bf 47} (1988) 511
[Yad.\ Fiz.\  {\bf 47} (1988) 801].

\vspace{3mm}

\item \label{Isgur:vq}
N.~Isgur and M.B.~Wise,
Phys.\ Lett.\ B {\bf 232} (1989) 113.

\vspace{3mm}

\item \label{Isgur:ed}
N.~Isgur and M.B.~Wise,
Phys.\ Lett.\ B {\bf 237} (1990) 527.

\vspace{3mm}

\item \label{Eichten:1980mw}
E.~Eichten and F.~Feinberg,
Phys.\ Rev.\ D {\bf 23} (1981) 2724;
W.E.~Caswell and G.P.~Lepage,
Phys.\ Lett.\ B {\bf 167} (1986) 437;
E.~Eichten,
Nucl. Phys. B Proc. Suppl. {\bf 4} (1988) 170; 
G.P.~Lepage and B.A.~Thacker, Nucl. Phys. B Proc. Suppl. {\bf 4} (1988) 199; 
H.D.~Politzer and M.B.~Wise,
Phys.\ Lett.\ B {\bf 206} (1988) 681
and 
Phys.\ Lett.\ B {\bf 208} (1988) 504;
E.~Eichten and B.~Hill,
Phys.\ Lett.\ B {\bf 240} (1990) 193 and
Phys.\ Lett.\ B {\bf 243} (1990) 427;
B.~Grinstein,
Nucl.\ Phys.\ B {\bf 339} (1990) 253;
H.~Georgi,
Phys.\ Lett.\ B {\bf 240} (1990) 447;
A.F.~Falk, H.~Georgi, B.~Grinstein and M.B.~Wise,
Nucl.\ Phys.\ B {\bf 343} (1990) 1;
A.F.~Falk, B.~Grinstein and M.E.~Luke,
Nucl.\ Phys.\ B {\bf 357} (1991) 185;
T.~Mannel, W.~Roberts and Z.~Ryzak,
Nucl.\ Phys.\ B {\bf 368} (1992) 204;
J.G.~K\"orner and G.~Thompson,
Phys.\ Lett.\ B {\bf 264} (1991) 185;
S.~Balk, J.G.~K\"orner and D.~Pirjol,
Nucl.\ Phys.\ B {\bf 428} (1994) 499
[hep-ph/9307230].

\vspace{3mm}

\item \label{Flynn:1992fm}
J.M.~Flynn and N.~Isgur,
J.\ Phys.\ G {\bf 18} (1992) 1627
[hep-ph/9207223].

\vspace{3mm}

\item \label{Neubert:1993mb}
M.~Neubert,
Phys.\ Rept.\  {\bf 245} (1994) 259
[hep-ph/9306320].
 
\vspace{3mm}

\item \label{Neubert:2000hd}
M.~Neubert,
hep-ph/0001334.

\vspace{3mm}

\item \label{El-Khadra:2002wp}
A.H.~Hoang,
hep-ph/0204299; \\
A.~X.~El-Khadra and M.~Luke,
Ann.\ Rev.\ Nucl.\ Part.\ Sci.\  {\bf 52} (2002) 201
[hep-ph/0208114].

\vspace{3mm}

\item \label{Kronfeld1} 
R.~Tarrach,
{ Nucl.\ Phys.} B~{\bf 183} (1981)  384; \\
A.S.~Kronfeld,
{ Phys.\ Rev.} D~{\bf 58} (1998)  051501 
[hep-ph/9805215]; \\
P.~Gambino and P.A.~Grassi,
Phys.\ Rev.\ D {\bf 62} (2000) 076002
[hep-ph/9907254].

\newpage

\item \label{Bigi1}
I.I.~Bigi, M.A.~Shifman, N.G.~Uraltsev and A.I.~Vainshtein,
{ Phys.\ Rev.}  D~{\bf 50} (1994) 2234 
[hep-ph/9402360];
%
%
M.~Beneke and V.M.~Braun,
{ Nucl.\ Phys.}  B~{\bf 426} (1994) 301
[hep-ph/9402364].
%

\vspace{3mm}

\item \label{Broadhurst1}
N.~Gray, D.J.~Broadhurst, W.~Grafe and K.~Schilcher,
{ Z.\ Phys.}  C~{\bf 48} (1990) 673.
%

\vspace{3mm}

\item \label{Melnikov1}
K.~Melnikov and T.~v.~Ritbergen,
{ Phys.\ Lett.}  B~{\bf 482} (2000) 99 
[hep-ph/9912391]; \\
%
%
K.G.~Chetyrkin and M.~Steinhauser,
{ Nucl.\ Phys.}  B~{\bf 573} (2000) 617
[hep-ph/9911434].
%

\vspace{3mm}

\item \label{Hoang1}
A.H.~Hoang {\it et al.},
{ Eur.\ Phys.\ J.\ direct} C~{\bf 3} (2000) 1
[hep-ph/0001286].
%

\vspace{3mm}

\item \label{Bigi2}
I.I.~Bigi, M.A.~Shifman, N.G.~Uraltsev and A.I.~Vainshtein,
Phys.\ Rev.\ D {\bf 56} (1997) 4017 \break
[hep-ph/9704245].

\vspace{3mm}

\item \label{Bigi3}
I.I.~Bigi, M.A.~Shifman, N.G.~Uraltsev and A.I.~Vainshtein,
Phys.\ Rev.\ D {\bf 52} (1995) 196 \break
[hep-ph/9405410].
%

\vspace{3mm}

\item \label{Czarnecki5}
A.~Czarnecki, K.~Melnikov and N.G.~Uraltsev,
{ Phys.\ Rev.\ Lett.}  {\bf 80} (1998) 3189
[hep-ph/9708372].
%

\vspace{3mm}

\item \label{Beneke1}
M.~Beneke,
{ Phys.\ Lett.}  B~{\bf 434} (1998) 115
[hep-ph/9804241].
%

\vspace{3mm}

\item \label{Schroeder1}
Y.~Schroder,
{ Phys.\ Lett.}  B~{\bf 447} (1999) 321
[hep-ph/9812205]. \\
%
M.~Peter,
{ Phys.\ Rev.\ Lett.}  {\bf 78} (1997) 602
[hep-ph/9610209].
%
%

\vspace{3mm}

\item \label{Hoang2}
A.H.~Hoang, Z.~Ligeti and A.V.~Manohar,
{ Phys.\ Rev.\ Lett.}  {\bf 82} (1999) 277
[hep-ph/9809423];
%
{ Phys.\ Rev.}  D~{\bf 59} (1999) 074017
[hep-ph/9811239].
%

\vspace{3mm}

\item \label{Hoang3}
A.H.~Hoang and T.~Teubner,
{ Phys.\ Rev.} D~{\bf 60} (1999) 114027
[hep-ph/9904468].

\vspace{3mm}

\item \label{Pineda6}
A.~Pineda,
{ JHEP} {\bf 0106} (2001) 022
[hep-ph/0105008].

\vspace{3mm}

\item \label{Novikov1}
V.A.~Novikov, {\it et al.},
Phys.\ Rev.\ Lett.\  {\bf 38} (1977) 626
[Erratum-ibid.\  {\bf 38} (1977) 791]; \\
%
L.J.~Reinders, H.~Rubinstein and S.~Yazaki,
Phys.\ Rept.\  {\bf 127} (1985) 1.
%

\vspace{3mm}

\item \label{Voloshin2}
M.B.~Voloshin,
{ Int.\ J.\ Mod.\ Phys.}  A {\bf 10} (1995) 2865
[hep-ph/9502224].

\vspace{3mm}

\item \label{Kuhn1}
J.H.~K\"uhn, A.A.~Penin and A.A.~Pivovarov,
{ Nucl.\ Phys.}  B~{\bf 534} (1998) 356
[hep-ph/9801356].
%

\vspace{3mm}

\item \label{Penin1}
A.A.~Penin and A.A.~Pivovarov,
{ Phys.\ Lett.} B~{\bf 435} (1998) 413
[hep-ph/9803363];
%
{ Nucl.\ Phys.} B~{\bf 549} (1999) 217
[hep-ph/9807421].
%

\vspace{3mm}

\item \label{Hoang14}
A.H.~Hoang,
{ Phys.\ Rev.} D~{\bf 59} (1999) 014039
[hep-ph/9803454].
%

\vspace{3mm}

\item \label{Melnikov2}
K.~Melnikov and A.~Yelkhovsky,
{ Phys.\ Rev.} D~{\bf 59} (1999) 114009
[hep-ph/9805270].
%

\vspace{3mm}

\item \label{Jamin1}
M.~Jamin and A.~Pich,
{ Nucl.\ Phys.\ Proc.\ Suppl.}  {\bf 74} (1999) 300
[hep-ph/9810259].
%

\vspace{3mm}

\item \label{Hoang15}
A.H.~Hoang,
{ Phys.\ Rev.}  D~{\bf 61} (2000) 034005
[hep-ph/9905550].
%

\vspace{3mm}

\item \label{Beneke6}
M.~Beneke and A.~Signer,
{ Phys.\ Lett.}  B~{\bf 471} (1999) 233
[hep-ph/9906475].
%

\vspace{3mm}

\item \label{Hoang8}
A.H.~Hoang,
hep-ph/0008102.
%

\vspace{3mm}

\item \label{Kuhn2}
J.H.~Kuhn and M.~Steinhauser,
Nucl.\ Phys.\ B {\bf 619} (2001) 588
[Erratum-ibid.\ B {\bf 640} (2002) 415]
[hep-ph/0109084].

\vspace{3mm}

\item \label{Erler1}
J.~Erler and M.~x.~Luo, Phys.\ Lett.\ B {\bf 558} (2003) 125
[hep-ph/0207114].

\vspace{3mm}

\item \label{Eidemuller1}
M.~Eidemuller,
hep-ph/0207237.
%

\vspace{3mm}

\item \label{Bordes1}
J.~Bordes, J.~Penarrocha and K.~Schilcher,
hep-ph/0212083.

\vspace{3mm}

\item \label{Corcellamb}
A.H.~Hoang and G.~Corcella,
Phys.\ Lett.\ B {\bf 554} (2003) 133
[hep-ph/0212297].

%
\vspace{3mm}

\item \label{Pineda7}
A.~Pineda and F.J.~Yndurain,
{ Phys.\ Rev.} D~{\bf 58} (1998) 094022
[hep-ph/9711287].
%

\vspace{3mm}

\item \label{Hoang10}
A.H.~Hoang,
{ Nucl.\ Phys.\ Proc.\ Suppl.}  {\bf 86} (2000) 512
[hep-ph/9909356].
%

\vspace{3mm}

\item \label{Brambilla5}
N.~Brambilla, Y.~Sumino and A.~Vairo,
Phys.\ Rev.\ D {\bf 65} (2002) 034001
[hep-ph/0108084].
%

\vspace{3mm}

\item \label{Shifman1}
M.~A.~Shifman, A.~I.~Vainshtein and V.~I.~Zakharov,
{Nucl. Phys.} B  {\bf 147} (1979) 385 and  448.


\vspace{3mm}

\item \label{Eidemuller:2000rc}
M.~Eidemuller and M.~Jamin,
Phys.\ Lett.\ B {\bf 498} (2001) 203
[hep-ph/0010334]; \\
J.~Penarrocha and K.~Schilcher,
Phys.\ Lett.\ B {\bf 515} (2001) 291
[hep-ph/0105222].


\vspace{3mm}

\item \label{rimom}
G.~Martinelli, {\it et al.}, 
Nucl.\ Phys.\ B {\bf 445} (1995) 81
[hep-lat/9411010].

\vspace{3mm}

\item \label{sf}
K.~Jansen {\it et al.},
Phys.\ Lett.\ B {\bf 372} (1996) 275 [hep-lat/9512009].

\vspace{3mm}

\item \label{mc_ape}
D.~Becirevic, V.~Lubicz and G.~Martinelli,
Phys.\ Lett.\ B {\bf 524} (2002) 115 [hep-ph/0107124].

\vspace{3mm}

\item \label{mc_alpha}
J.~Rolf and S.~Sint  [ALPHA Collaboration],
JHEP {\bf 0212} (2002) 007 [hep-ph/0209255].

\vspace{3mm}

\item \label{Davies:1994pz}
C.T.~Davies {\it et al.},
Phys.\ Rev.\ Lett.\  {\bf 73} (1994) 2654
[hep-lat/9404012].

\vspace{3mm}

\item \label{mb_ape}
V.~Gimenez, L.~Giusti, G.~Martinelli and F.~Rapuano,
JHEP {\bf 0003} (2000) 018 
[hep-lat/0002007].

\vspace{3mm}

\item \label{martisach}
G.~Martinelli and C.T.~Sachrajda,
Nucl.\ Phys.\ B {\bf 559} (1999) 429
[hep-lat/9812001].

\vspace{3mm}

\item \label{mb_nrqcd}
S.~Collins,
hep-lat/0009040.

\vspace{3mm}

\item \label{direnzo}
F.~Di Renzo and L.~Scorzato,
JHEP {\bf 0102} (2001) 020 [hep-lat/0012011].

\vspace{3mm}

\item \label{mb_alpha}
J.~Heitger and R.~Sommer  [ALPHA Collaboration],
Nucl.\ Phys.\ Proc.\ Suppl.\  {\bf 106} (2002) 358 [hep-lat/0110016];
R.~Sommer,
hep-lat/0209162.

\vspace{3mm}


\item \label{Cronin-Hennessy:2001fk}
D.~Cronin-Hennessy {\it et al.}  [CLEO Collaboration],
Phys.\ Rev.\ Lett.\  {\bf 87} (2001) 251808 \break
[hep-ex/0108033].

\vspace{3mm}

\item \label{delphi_mx}
D.~Bloch {\it et al.} [DELPHI Collaboration], DELPHI 2002-070-CONF~604; \\
M.~Battaglia {\it et al.} [DELPHI Collaboration], DELPHI 2002-071-CONF-605;\\
M.~Calvi, hep-ex/0210046.

\vspace{3mm}

\item \label{babar9312} 
B. Aubert {\it et al.} [BaBar Collaboration], hep-ex/0207184.

\vspace{3mm}

\item \label{gremmetal}
M.~Gremm, A.~Kapustin, Z.~Ligeti and M.B.~Wise,
Phys.\ Rev.\ Lett.\  {\bf 77} (1996) 20
[hep-ph/9603314].

\vspace{3mm}

\item \label{voloshin0} M.B.~Voloshin,
Phys.\ Rev.\ D {\bf 51} (1995) 4934
[hep-ph/9411296].

\vspace{3mm}

\item \label{gremm-kap}
M.~Gremm and A.~Kapustin,
Phys.\ Rev.\ D {\bf 55} (1997) 6924
[hep-ph/9603448].

\vspace{3mm}

\item \label{chris}
C.W.~Bauer and M.~Trott,
Phys.\ Rev.\ D {\bf 67} (2003) 014021
[hep-ph/0205039].

\vspace{3mm}

\item \label{Bauer:2002sh}
C.W.~Bauer, Z.~Ligeti, M.~Luke and A.V.~Manohar,
Phys.\ Rev.\ D {\bf 67} (2003) 054012 
[hep-ph/0210027].

\vspace{3mm}

\item \label{kinmass} 
N.G.~Uraltsev, { Nucl.\,Phys.} B {\bf 491} (1997) 303.

\vspace{3mm}

\item \label{alphas} M.~Jezabek and J.H.~Kuhn,
Nucl.\ Phys.\ B {\bf 320} (1989) 20; \\
A.~Czarnecki, M.~Jezabek and J.H.~Kuhn,
Acta Phys.\ Polon.\ B {\bf 20} (1989) 961; \\
A.~Czarnecki and M.~Jezabek,
Nucl.\ Phys.\ B {\bf 427} (1994) 3
[hep-ph/9402326].

\vspace{3mm}

\item \label{alphas2b0}
M.~Gremm and I.~Stewart,
Phys.\ Rev.\ D {\bf 55} (1997) 1226
[hep-ph/9609341].

\vspace{3mm}

\item \label{FLS}
A.F.~Falk, M.E.~Luke and M.J.~Savage,
Phys.\ Rev.\ D {\bf 53} (1996) 2491
[hep-ph/9507284].

\vspace{3mm}

\item \label{Falk:1997jq}
A.F.~Falk and M.E.~Luke,
Phys.\ Rev.\ D {\bf 57} (1998) 424
[hep-ph/9708327].

\vspace{3mm}

\item \label{Battaglia:2002tm}
M.~Battaglia {\it et al.},
Phys.\ Lett.\ B {\bf 556} (2003) 41 [hep-ph/0210319].

\vspace{3mm}

\item \label{Uraltsev:2001ih}
N.G.~Uraltsev,
Phys.\ Lett.\ B {\bf 545} (2002) 337
[hep-ph/0111166]. 

\vspace{3mm}

\item \label{lattice}
A.S.~Kronfeld and J.N.~Simone,
Phys.\ Lett.\ B {\bf 490} (2000) 228
[Erratum-ibid.\ B {\bf 495} (2000) 441]
[hep-ph/0006345].

\vspace{3mm}

\item \label{bsgammacleo}
S.~Chen {\it et al.} [CLEO Collaboration], Phys. Rev. Lett. {\bf
87}, 251807 (2001).

\vspace{3mm}

\item \label{marina-lmom} R. Briere {\it et al.} [CLEO Collaboration], 
CLEO-CONF-02-10, hep-ex/0209024.

\vspace{3mm}

\item \label{h1}
A.H.~Hoang,  Nucl. Phys. B, Proc. Suppl. {\bf 86} (2000) 512
[hep-ph/9909356]; A.H.~Hoang and A.V.~Manohar, Phys. Lett. B
{\bf 483} (2000) 94 [hep-ph/9911461]; hep-ph/0008102; hep-ph/0102292.

\vspace{3mm}

\item \label{cleonew} A.H. Mahmood {\it et al.} [CLEO Collaboration],
CLNS 02/1810, CLEO 02-16, hep-ex/0212051.

\vspace{3mm}

\item \label{babarmom}
B. Aubert {\it et al.} [BaBar Collaboration], hep-ex/0207084.

\vspace{3mm}

\item \label{PQW76}
E.C.~Poggio, H.R.~Quinn and S.~Weinberg,
Phys.\ Rev.\ D {\bf 13} (1976) 1958.

\vspace{3mm}

\item \label{Greco:1978gy}
M.~Greco, G.~Penso and Y.~Srivastava,
Phys.\ Rev.\ D {\bf 21} (1980) 2520.

\vspace{3mm}

\item \label{shifman}
M.A.~Shifman, ``Quark-hadron duality,''
in B. Ioffe Festschrift
'At the Frontier of Particle Physics /
Handbook of QCD', ed. M. Shifman (World Scientific, Singapore, 2001), 
[hep-ph/0009131].

\vspace{3mm}

\item \label{BU2001}
I.I.~Bigi and N.G.~Uraltsev,
Int.\ J.\ Mod.\ Phys.\ A {\bf 16} (2001) 5201
[hep-ph/0106346].

\vspace{3mm}

\item \label{Boyd:1995ht}
C.G.~Boyd, B.~Grinstein and A.V.~Manohar,
Phys.\ Rev.\ D {\bf 54} (1996) 2081
[hep-ph/9511233].

\vspace{3mm}

\item \label{QCD2} R.F.~Lebed, N.G.~Uraltsev, 
Phys.\ Rev.\ D {\bf 62} (2000) 094011 [hep-ph/0006346], and refs.\ therein.

\vspace{3mm}

\item \label{chay}  J.~Chay, H.~Georgi, 
B.~Grinstein, Phys. Lett. B {\bf 247} (1990) 339.

\vspace{3mm}

\item \label{isgur} N.~Isgur, Phys.\ Lett.\ B~{\bf 448} (1999) 111 
[hep-ph/9811377].

\vspace{3mm}

\item \label{QCD} 
A. Le Yaouanc, {\it et al.}, 
Phys.\ Lett.\ B {\bf 480} (2000) 119 [hep-ph/0003087].

\vspace{3mm}

\item \label{articleNR} A. Le Yaouanc {\it et al.}, 
Phys.\ Rev.\ D~{\bf 62} (2000) 074007 [hep-ph/0004246].

\vspace{3mm}

\item \label{lettreNR} A. Le Yaouanc {\it et al.}, 
Phys.\ Lett.\ B~{\bf 517} (2001) 135  [hep-ph/010333].

\vspace{3mm}

\item \label{OH} A. Le Yaouanc {\it et al.}, 
 Phys.\ Lett.\ B~{\bf 488} (2000) 153  [hep-ph/0005039].


\vspace{3mm}

\item \label{ref:lukesav}
M. Luke, M.J. Savage and M.B. Wise, Phys. Lett. B~{\bf 345} (1995) 301.

\vspace{3mm}

\item \label{ref:niri}
Y. Nir, Phys. Lett. B~{\bf 221} (1989) 184.

\vspace{3mm}

\item \label{ref:cabi}
N. Cabibbo and L. Maiani, Phys. Lett. B~{\bf 79} (1978) 109;
the original QED calculation is published in R.E.
Behrends {\it et al.}, Phys. Rev. {\bf 101} (1956) 866.

\vspace{3mm}

\item \label{ref:melni3}
A. Czarnecki and K. Melnikov, Phys. Rev. D~{\bf 59} (1998) 014036.

\vspace{3mm}

\item \label{ref:bigim}
N.G.~Uraltsev, Int. J. Mod. Phys. A {\bf 11} (1996) 513.

\vspace{3mm}

\item \label{ref:order2}
M. Voloshin and M. Shifman, Sov. J. Nucl. Phys. {\bf 41} (1985) 120; \\
J. Chay, H. Georgi and B. Grinstein, Phys. Lett. B~{\bf 247} (1990) 399.

\vspace{3mm}

\item \label{vcb_new}  D.~Benson, I.I.~Bigi, Th.~Mannel, and
 N.G.~Uraltsev, hep-ph/0302262.

\vspace{3mm}

\item \label{uraltsevvcb} N.G.~Uraltsev, 
Mod.\ Phys.\ Lett.\ A~{\bf 17} (2002) 2317 
[hep-ph/0210413].

\vspace{3mm}

\item \label{ref:pdg00}
D.E. Groom {\it et al.}, Eur. Phys. J. C~{\bf 15} (2000) 1.

\vspace{3mm}

\item \label{ref:cleobad}
J. Bartelt {\it et al.} [CLEO Collaboration], CLEO CONF 98-21.

\vspace{3mm}

\item \label{ref:brandt}
Thorsten Brandt computed the average of $b$-hadron semileptonic
partial width inclusive measurements available for the Workshop.

\vspace{3mm}

\item \label{ref:lepbr}
The LEP Electroweak Working Group, {\tt http://lepewwg.web.cern.ch/LEPEWWG/},
CERN-EP-2001-098 and hep-ex/0112021.

\vspace{3mm}

\item \label{ref:babarbr}
B. Aubert {\it et al.} [BaBar Collaboration], hep-ex/0208018.

\vspace{3mm}

\item \label{ref:bellebr}
K. Abe {\it et al.} [BELLE Collaboration], hep-ex/0208033.

\vspace{3mm}

\item \label{ref:lepvcb}
The LEP Vcb Working group, see {\tt  http://lepvcb.web.cern.ch/LEPVCB/}.

\vspace{3mm}

\item \label{ref:pdg02}
K. Hagiwara {\it et al.}, Phys. Rev. D~{\bf 66} (2002) 010001.

\vspace{3mm}

\item \label{ref:marina}
M. Artuso and E. Barberio, hep-ph/0205163 
(a shorter version has been included 
in~[\ref{ref:pdg02}] as a minireview on $\vcb$). 

\vspace{3mm}

\item \label{vanRitbergen1}
T.~van Ritbergen,
{ Phys.\ Lett.} B~{\bf 454} (1999) 353
[hep-ph/9903226].

\vspace{3mm}

\item \label{uraltsevvub}
N.G.~Uraltsev,
Int.\ J.\ Mod.\ Phys.\ A {\bf 14} (1999) 4641
[hep-ph/9905520].

\vspace{3mm}

\item \label{Elcut}
R.~Fulton {\it et al.}  [CLEO Collaboration],
Phys.\ Rev.\ Lett.\  {\bf 64} (1990) 16; \\
J.~Bartelt {\it et al.}  [CLEO Collaboration],
Phys.\ Rev.\ Lett.\  {\bf 71} (1993) 411; \\
H.~Albrecht {\it et al.}  [ARGUS Collaboration],
Phys.\ Lett.\ B {\bf 255} (1991) 297.

\vspace{3mm}

\item \label{mXcut}
V.D.~Barger, C.S.~Kim and R.J.~Phillips,
Phys.\ Lett.\ B {\bf 251} (1990) 629; \\
A.F.~Falk, Z.~Ligeti and M.B.~Wise,
Phys.\ Lett.\ B {\bf 406} (1997) 225; \\
R.D.~Dikeman and N.G.~Uraltsev,
Nucl.\ Phys.\ B {\bf 509} (1998) 378; \\
I.I.~Bigi, R.D.~Dikeman and N.G.~Uraltsev,
Eur.\ Phys.\ J.\ C {\bf 4} (1998) 453.

\vspace{3mm}

\item \label{qsqcut}
C.W.~Bauer, Z.~Ligeti and M.E.~Luke,
Phys.\ Lett.\ B {\bf 479} (2000) 395 [hep-ph/0002161].

\newpage

\item \label{accmm}
G.~Altarelli, {\it et al.}, Nucl.\ Phys.\ B {\bf 208} (1982) 365; \\
M.~Wirbel, B.~Stech, M.~Bauer, Zeit.\ Phys.\ C~{\bf 29} (1985) 637.

\vspace{3mm}

\item \label{isgw} N. Isgur {\it et al.}, \Journal{\prd}{1989}{39}{799}.

\vspace{3mm}

\item \label{shape1}
M.~Neubert,
Phys.\ Rev.\ D {\bf 49} (1994) 3392; \\
T.~Mannel and M.~Neubert,
Phys.\ Rev.\ D {\bf 50} (1994) 2037.

\vspace{3mm}

\item \label{shapebigi}
I.I.~Bigi, M.A.~Shifman, N.G.~Uraltsev and A.I.~Vainshtein,
Int.\ J.\ Mod.\ Phys.\ A {\bf 9} (1994) 2467 [hep-ph/9312359].

\vspace{3mm}

\item \label{shape2}
M.~Neubert,
Phys.\ Rev.\ D {\bf 49} (1994) 4623;
A.L.~Kagan and M.~Neubert,
Eur.\ Phys.\ J.\ C {\bf 7} (1999) 5
[hep-ph/9805303].

\vspace{3mm}

\item \label{Bigi:2002qq}
I.I.~Bigi and N.G.~Uraltsev,
Int.\ J.\ Mod.\ Phys.\ A {\bf 17} (2002) 4709
[hep-ph/0202175].

\vspace{3mm}

\item \label{aglietti}
U.~Aglietti,
Phys.\ Lett.\ B {\bf 515} (2001) 308
[hep-ph/0103002] and 
Nucl.\ Phys.\ B {\bf 610} (2001) 293
[hep-ph/0104020].

\vspace{3mm}

\item \label{iraEl}
A.K.~Leibovich, I.~Low and I.Z.~Rothstein,
Phys.\ Rev.\ D {\bf 61} (2000) 053006 [hep-ph/9909404].

\vspace{3mm}

\item \label{BLM1}
C.W.~Bauer, M.E.~Luke and T.~Mannel,
hep-ph/0102089 and 
Phys.\ Lett.\ B {\bf 543} (2002) 261 \break
[hep-ph/0205150].

\vspace{3mm}

\item \label{Voloshin2001}
I.I.~Bigi and N.G.~Uraltsev,
Nucl.\ Phys.\ B {\bf 423} (1994) 33
[hep-ph/9310285]; \\
M.B.~Voloshin,
Phys.\ Lett.\ B {\bf 515} (2001) 74 
[hep-ph/0106040].

\vspace{3mm}

\item \label{Lig_Leib_Wise}
A.K.~Leibovich, Z.~Ligeti and M.B.~Wise,
Phys.\ Lett.\ B {\bf 539} (2002) 242
[hep-ph/0205148].

\vspace{3mm}

\item \label{NEUBERTVUB}
M.~Neubert,
Phys.\ Lett.\ B {\bf 543} (2002) 269
[hep-ph/0207002].

\vspace{3mm}

\item \label{cleovub}
A.~Bornheim {\it et al.}  [CLEO Collaboration],
Phys.\ Rev.\ Lett.\  {\bf 88} (2002) 231803
[hep-ex/0202019].

\vspace{3mm}

\item \label{delphivub} P.~Abreu {\it et al.} [DELPHI Collaboration], Phys.\ Lett.\,B~{\bf 478} (2000) 14.

\vspace{3mm}

\item \label{neubertq2}
M.~Neubert,
JHEP {\bf 0007} (2000) 022
[hep-ph/0006068].

\vspace{3mm}

\item \label{doublecut}
C.W.~Bauer, Z.~Ligeti and M.E.~Luke,
Phys.\ Rev.\ D {\bf 64} (2001) 113004
[hep-ph/0107074].

\vspace{3mm}

\item \label{cleovubmixed}
A.~Bornheim {\it et al.}  [CLEO Collaboration],
hep-ex/0207064.

\vspace{3mm}

\item \label{alephopalvub}
R.~Barate {\it et al.} [ALEPH Collaboration], 
Eur. Phys. J. C~{\bf 6} (1999) 555; \\
G.~Abbiendi {\it et al.} [OPAL Collaboration], 
Eur. Phys. J. C~{\bf 21} (2001) 399.

\vspace{3mm}

\item \label{l3vub}
M.~Acciarri {\it et al.} [L3 Collaboration], Phys. Lett. B436, 174 (1998).

\vspace{3mm}

\item \label{aglietti2}
U.~Aglietti, M.~Ciuchini and P.~Gambino,
Nucl.\ Phys.\ B {\bf 637} (2002) 427
[hep-ph/0204140].

\vspace{3mm}

\item \label{menke}
R.V.~Kowalewski and S.~Menke,
Phys.\ Lett.\ B {\bf 541} (2002) 29 [hep-ex/0205038].

\vspace{3mm}

\item \label{Neubert:td}
M.~Neubert,
Phys.\ Lett.\ B {\bf 264} (1991) 455.

\vspace{3mm}

\item \label{Neubert:1994vy}
M.~Neubert,
Phys.\ Lett.\ B {\bf 338} (1994) 84
[hep-ph/9408290].

\vspace{3mm}

\item \label{Boyd:1995sq}
C.G.~Boyd, B.~Grinstein and R.F.~Lebed,
Nucl.\ Phys.\ B {\bf 461} (1996) 493 [hep-ph/9508211].

\vspace{3mm}

\item \label{Boyd:1995tg}
C.G.~Boyd and R.F.~Lebed,
Nucl.\ Phys.\ B {\bf 485} (1997) 275
[hep-ph/9512363].

\vspace{3mm}

\item \label{Boyd:1997kz}
C.G.~Boyd, B.~Grinstein and R.F.~Lebed,
Phys.\ Rev.\ D {\bf 56} (1997) 6895
[hep-ph/9705252].

\vspace{3mm}

\item \label{Caprini:1997mu}
I.~Caprini, L.~Lellouch and M.~Neubert,
factors,''
Nucl.\ Phys.\ B {\bf 530} (1998) 153
[hep-ph/9712417].

\vspace{3mm}

\item \label{Neubert:1992wq}
M.~Neubert, Z.~Ligeti and Y.~Nir,
Phys.\ Lett.\ B {\bf 301} (1993) 101
[hep-ph/9209271].

\vspace{3mm}

\item \label{Neubert:1992pn}
M.~Neubert, Z.~Ligeti and Y.~Nir,
Phys.\ Rev.\ D {\bf 47} (1993) 5060
[hep-ph/9212266].

\vspace{3mm}

\item \label{Ligeti:1993hw}
Z.~Ligeti, Y.~Nir and M.~Neubert,
Phys.\ Rev.\ D {\bf 49} (1994) 1302
[hep-ph/9305304].

\vspace{3mm}

\item \label{bjorken}J. 
D. Bjorken, {\it in} Proceedings of the 4th Recontres
de Physique de la Vall\`{e}e d'Aoste, La Thuille, Italy, 1990,
ed. M. Greco (Editions Fronti\`{e}res, Gif-Sur-Yvette, 1990); \\
N. Isgur and M. B. Wise, Phys.\ Rev.\ D {\bf 43} (1991) 819.

\vspace{3mm}

\item \label{voloshin}M.B. Voloshin, Phys.\  Rev.\  D {\bf 46} (1992)  3062.

\vspace{3mm}

\item \label{alphasihqsr} 
A.G.~Grozin and G.P.~Korchemsky,
Phys. Rev. D~{\bf 53} (1996) 1378; \\
C.G.~Boyd, B.~Grinstein, and A.V.~Manohar, Phys.\ Rev.\
D {\bf 54} (1996) 2081; \\
C.G.~Boyd, Z.~Ligeti, I.Z.~Rothstein, and M.B.~Wise,
Phys.\ Rev.\ D {\bf 55} (1997) 3027.

\vspace{3mm}

\item \label{Uraltsev:2000ce}
N.G.~Uraltsev,
Phys.\ Lett.\ B {\bf 501} (2001) 86
[hep-ph/0011124].

\vspace{3mm}

\item \label{Luke:1990eg}
M.E.~Luke,
Phys.\ Lett.\ B {\bf 252} (1990) 447.

\vspace{3mm}

\item \label{Shifman:1995jh}
M.A.~Shifman, N.G.~Uraltsev, and A.I.~Vainshtein,
Phys.\ Rev.\ D {\bf 51} (1995) 2217 
[hep-ph/9405207];
[Erratum-ibid.\  {\bf 52} (1995) 3149].

\vspace{3mm}

\item \label{Falk:1993wt}
A.F. Falk and M. Neubert,
Phys.\ Rev.\ D {\bf 47} (1993) 2965 
[hep-ph/9209268].

\vspace{3mm}

\item \label{Kronfeld:2000ck}
A.S.~Kronfeld,
Phys.\ Rev.\ D {\bf 62}, 014505 (2000)
[hep-lat/0002008].

\vspace{3mm}

\item \label{Harada:2001fj}
J.~Harada, {\it et al.}, 
Phys.\ Rev.\ D {\bf 65}, 094514 (2002)
[hep-lat/0112045].

\vspace{3mm}

\item \label{Hashimoto:2001nb}
S.~Hashimoto, {\it et al.}, 
Phys.\ Rev.\ D {\bf 66} (2002) 014503
[hep-ph/0110253].

\vspace{3mm}

\item \label{Bigi:1997fj}
I.I.~Bigi, M.A.~Shifman and N.G.~Uraltsev,
Annu.\ Rev.\ Nucl.\ Part.\ Sci.\  {\bf 47} (1997) 591
[hep-ph/9703290].

\vspace{3mm}

\item \label{Czarnecki:1998wy}
A.~Czarnecki, K.~Melnikov and N.G.~Uraltsev,
Phys.\ Rev.\ D {\bf 57} (1998) 1769
[hep-ph/9706311].
%

%
\vspace{3mm}

\item \label{Czarnecki:1996gu}
A.~Czarnecki,
Phys.\ Rev.\ Lett.\  {\bf 76} (1996) 4124
[hep-ph/9603261].
%
\vspace{3mm}

\item \label{Czarnecki:1997cf}
A.~Czarnecki and K.~Melnikov,
Nucl.\ Phys.\ B {\bf 505} (1997) 65
[hep-ph/9703277].

%
\vspace{3mm}

\item \label{Uraltsev:1998bk}
N.G.~Uraltsev,
hep-ph/9804275.

\vspace{3mm}

\item \label{Mannel:1994wt}
T.~Mannel,
Phys.\ Rev.\ D {\bf 50} (1994) 428
[hep-ph/9403249].
%
\vspace{3mm}

\item \label{Ryan:2001ej}
S.~Ryan,
Nucl.\ Phys.\ B Proc.\ Suppl.\ {\bf 106} (2002) 86
[hep-lat/0111010].
%
\vspace{3mm}

\item \label{El-Khadra:1996mp}
A.X. El-Khadra, A.S. Kronfeld, and P.B. Mackenzie,
Phys.\ Rev.\ D {\bf 55} (1997) 3933
[hep-lat/9604004].

\vspace{3mm}

\item \label{Harada:2001fi}
J.~Harada, {\it et al.}, 
Phys.\ Rev.\ D {\bf 65} (2002) 094513 
[hep-lat/0112044].

\vspace{3mm}

\item \label{Hashimoto:2000yp}
S.~Hashimoto \emph{et al.},
Phys.\ Rev.\ D {\bf 61} (2000) 014502 
[hep-ph/9906376].

\vspace{3mm}

\item \label{Kronfeld:2002cc}
A.S.~Kronfeld, P.B.~Mackenzie, J.N.~Simone, S.~Hashimoto and
S.M.~Ryan,
in {\em Flavor Physics and $CP$ Violation}, edited by
R.G.C.~Oldemann,
hep-ph/0207122.

\vspace{3mm}

\item \label{Randall:1993qg}
L.~Randall and M.B.~Wise,
Phys.\ Lett.\ B {\bf 303} (1993) 135; \\
M.J.~Savage,
Phys.\ Rev.\ D {\bf 65} (2002) 034014 
[hep-ph/0109190].

\vspace{3mm}

\item \label{Arndt:2002ed}
D.~Arndt,
hep-lat/0210019.

\vspace{3mm}

\item \label{ScoraIsgurPRD5295}D. Scora and N. Isgur,
\Journal{\prd}{1995}{52}{2783}.

\vspace{3mm}

\item \label{Ligeti:1999yc}
Z.~Ligeti,
hep-ph/9908432.

\vspace{3mm}

\item \label{Lepage:2001ym}
G.P.~Lepage, {\it et al.},
Nucl.\ Phys.\ B Proc.\ Suppl.\  {\bf 106} (2002) 12
[hep-lat/0110175]; \\
C.~Morningstar,
\emph{ibid.}\ {\bf 109} (2002)  185 
[hep-lat/0112023].

\vspace{3mm}

\item \label{Atwood:1989em}
D.~Atwood and W.J.~Marciano,
Phys.\ Rev.\ D {\bf 41} (1990) 1736.

\vspace{3mm}

\item \label{Ginsberg:1968pz}
E.S.~Ginsberg,
Phys.\ Rev.\  {\bf 171} (1968) 1675;
{\bf 174} (1968) 2169(E);
{\bf 187} (1969) 2280(E).

\vspace{3mm}

\item \label{Sirlin:1981ie}
A.~Sirlin,
Nucl.\ Phys.\ B {\bf 196} (1982) 83.

\vspace{3mm}

\item \label{cleo-vcb} R.A. Briere  {\it et al.} [CLEO Collaboration],
\Journal{\prl}{2002}{89}{81803}; [hep-ex/0203032].

\vspace{3mm}

\item \label{belle-dslnu} K. Abe {\it et al.} [BELLE Collaboration],
\Journal{\plb}{2002}{526}{247}; [hep-ex/0111060].

\vspace{3mm}

\item \label{DELPHI_vcb} P. Abreu  {\it et al.} [DELPHI Collaboration],
\Journal{\plb}{2001}{510}{55} [hep-ex/0104026].

\vspace{3mm}

\item \label{ALEPH_vcb} D. Buskulic {\it et al.} [ALEPH Collaboration],
\Journal{\plb}{1997}{935}{373}.

\vspace{3mm}

\item \label{OPAL_vcb} G. Abbiendi  {\it et al.} [OPAL Collaboration],
\Journal{\plb}{2000}{482}{15}.

\vspace{3mm}

\item \label{CB} S. Anderson {\it et al.} [CLEO Collaboration],
{ Nucl.Phys. A}~{\bf 663} (2000) [hep-ex/9908009].

\vspace{3mm}

\item \label{aleph-d2star} D. Buskulic {\it et al.} [ALEPH Collaboration],
\Journal{\zpc}{1997}{73}{601}.

\vspace{3mm}

\item \label{cleo-d2star} A. Anastassov {\it et al.} [CLEO Collaboration],
\Journal{\prl}{1998}{80}{4127}.
%


\vspace{3mm}

\item \label{EWWG}
LEP/SLD Electroweak Heavy Flavor Group, 
results presented at the Winter 2001 Conferences, 
see {\tt http://lepewwg.web.cern.ch/LEPEWWG/heavy/}

\vspace{3mm}

\item \label{delphi-d2star} D. Block {\it et al.} [DELPHI Collaboration],
{contributed paper to ICHEP 2000}, DELPHI 2000-106 Conf. 45.

\vspace{3mm}

\item \label{HFnote}  ALEPH, CDF, DELPHI, L3, OPAL, SLD, CERN-EP/2001-050.

\vspace{3mm}

\item \label{wrong} V. Morenas {\it et al.}, 
{ Phys.\ Rev.\ D} {\bf 59} (1997) 5668
 [hep-ph/9706265]; \\
M.Q.~Huang, C.~Li and Y.B.~Dai, { Phys.\ Rev.\ D} {\bf 61} (2000) 54010
[hep-ph/9909307].

\vspace{3mm}

\item \label{ligeti} A.K. Leibovich, Z. Ligeti, I.W.~Stewart, M.B.~Wise
{ Phys.\ Rev.\ D} {\bf 57} (1998) 308 [hep-ph/9705467] and
                { Phys.\ Rev.\ Lett.} {\bf 78} (1997) 3995 [hep-ph/9703213].

\vspace{3mm}

\item \label{lepvcb} LEP $V_{cb}$ Working Group, Internal Note,
see   {\tt http://lepvcb.web.cern.ch/LEPVCB/}

\vspace{3mm}

\item \label{Bosc} CDF, LEP, SLD B Oscillations Working Group, 
Internal Note, see\\
{\tt http://lepbosc.web.cern.ch/LEPBOSC/}.  

\vspace{3mm}

\item \label{life} CDF, LEP, SLD B-hadron Lifetime Working Group, 
Internal Note, see\\
{\tt http://claires.home.cern.ch/claires/lepblife.html}.  

\vspace{3mm}

\item \label{r1r2cleo}J.E. Duboscq {\it et al.} [CLEO Collaboration], 
\Journal{\prl}{1996}{76}{3898}.

\vspace{3mm}

\item \label{Grinstein:2001yg}
B.~Grinstein and Z.~Ligeti,
Phys.\ Lett.\ B {\bf 526} (2002) 345
[hep-ph/0111392].

\vspace{3mm}

\item \label{belle-dplnu}K. Abe {\em et al.} [BELLE Collaboration],
\Journal{\plb}{2002}{526}{258}  [hep-ex/0111082].

\vspace{3mm}

\item \label{cleo-dplnu} J. Bartelt {\em et al.} [CLEO Collaboration],
\Journal{\prl}{1999}{82}{3746}. 

\vspace{3mm}

\item \label{3Lellouch:1999dz}
L.~Lellouch,
hep-ph/9912353.

\vspace{3mm}

\item \label{Bowler:1999xn}
K.`C.~Bowler {\it et al.}  [UKQCD Collaboration],
Phys.\ Lett.\ B {\bf 486} (2000) 111
[hep-lat/9911011].

\vspace{3mm}

\item \label{Abada:2000ty}
A.~Abada {\it et al.}, 
V.~Lubicz and F.~Mescia,
Nucl.\ Phys.\ B {\bf 619} (2001) 565
[hep-lat/0011065].

\vspace{3mm}

\item \label{El-Khadra:2001rv}
A.X.~El-Khadra {\it et al.}, 
Phys.\ Rev.\ D {\bf 64} (2001) 014502
[hep-ph/0101023].

\vspace{3mm}

\item \label{Aoki:2001rd}
S.~Aoki {\it et al.}  [JLQCD Collaboration],
Phys.\ Rev.\ D {\bf 64} (2001) 114505
[hep-lat/0106024].

\vspace{3mm}

\item \label{Sheikholeslami:1985ij}
B.~Sheikholeslami and R.~Wohlert,
Nucl.\ Phys.\ B {\bf 259} (1985) 572.

\vspace{3mm}

\item \label{Luscher:1996sc}
M.~Luscher, S.~Sint, R.~Sommer and P.~Weisz,
Nucl.\ Phys.\ B {\bf 478} (1996) 365
[hep-lat/9605038].

\vspace{3mm}

\item \label{Luscher:1996jn}
M.~Luscher, S.~Sint, R.~Sommer and H.~Wittig,
Nucl.\ Phys.\ B {\bf 491} (1997) 344
[hep-lat/9611015].

\vspace{3mm}

\item \label{Thacker:1990bm}
B.A.~Thacker and G.P.~Lepage,
Phys.\ Rev.\ D {\bf 43} (1991) 196.

\vspace{3mm}

\item \label{Lepage:1992tx}
G.P.~Lepage, {\it et al.}, 
Phys.\ Rev.\ D {\bf 46} (1992) 4052
[hep-lat/9205007].

\vspace{3mm}

\item \label{Burdman:1993es}
G.~Burdman, Z.~Ligeti, M.~Neubert and Y.~Nir,
Phys.\ Rev.\ D {\bf 49} (1994) 2331
[hep-ph/9309272].

\vspace{3mm}

\item \label{Burdman:gh}
G.~Burdman and J.F.~Donoghue,
Phys.\ Lett.\ B {\bf 280} (1992) 287.

\vspace{3mm}

\item \label{Wise:hn}
M.B.~Wise,
Phys.\ Rev.\ D {\bf 45} (1992) 2188.

\vspace{3mm}

\item \label{Yan:gz}
T.M.~Yan, {\it et al.}, 
Phys.\ Rev.\ D {\bf 46} (1992) 1148
[Erratum-ibid.\ D {\bf 55} (1997) 5851].

\vspace{3mm}

\item \label{Becirevic:2002sc}
D.~Becirevic, S.~Prelovsek and J.~Zupan,
Phys.\ Rev.\ D {\bf 67} (2003) 054010
[hep-lat/0210048].

\vspace{3mm}

\item \label{Fleischer:1992tn}
R.~Fleischer,
Phys.\ Lett.\ B {\bf 303} (1993) 147.

\vspace{3mm}

\item \label{Boyd:1994tt}
C.G.~Boyd, B.~Grinstein and R.F.~Lebed,
Phys.\ Rev.\ Lett.\  {\bf 74} (1995) 4603
[hep-ph/9412324].

\vspace{3mm}

\item \label{Lellouch:1995yv}
L.~Lellouch,
Nucl.\ Phys.\ B {\bf 479} (1996) 353
[hep-ph/9509358].

\vspace{3mm}

\item \label{Burford:1995fc}
D.R.~Burford, {\it et al.}
                  [UKQCD Collaboration],
Nucl.\ Phys.\ B {\bf 447} (1995) 425
[hep-lat/9503002].

\vspace{3mm}

\item \label{3DelDebbio:1997kr}
L.~Del Debbio, {\it et al.} 
  [UKQCD Collaboration],
Phys.\ Lett.\ B {\bf 416} (1998) 392
[hep-lat/9708008].

\vspace{3mm}

\item \label{3Charles:1998dr}
J.~Charles, {\it et al.}, 
Phys.\ Rev.\ D {\bf 60} (1999) 014001
[hep-ph/9812358].

\vspace{3mm}

\item \label{Khodjamirian:2000ds}
A.~Khodjamirian, {\it et al.}, 
Phys.\ Rev.\ D {\bf 62} (2000) 114002
[hep-ph/0001297].

\vspace{3mm}

\item \label{Durr:2002zx}
S.~D\"urr,
DESY-02-121, hep-lat/0208051.

\vspace{3mm}

\item \label{Bernard:2002yk}
C.~Bernard, {\it et al.}, 
hep-lat/0209086.

\vspace{3mm}

\item \label{Hashimoto:2002vi}
S.~Hashimoto {\it et al.}  [JLQCD Collaboration],
hep-lat/0209091.

\vspace{3mm}

\item \label{3Lellouch:2002nj}
L.~Lellouch, plenary talk presented at  
ICHEP-2002, Amsterdam, July 2002,
hep-ph/0211359.

\vspace{3mm}

\item \label{Flynn:1995dc}
J.~M.~Flynn {\it et al.}  [UKQCD Collaboration],
Nucl.\ Phys.\ B {\bf 461} (1996) 327
[hep-ph/9506398].

\vspace{3mm}

\item \label{Behrens:1999vv}
B.H.~Behrens {\it et al.}  [CLEO Collaboration],
Phys.\ Rev.\ D {\bf 61} (2000) 052001
[hep-ex/9905056].

\vspace{3mm}

\item \label{Abada:1993dh}
A.~Abada {\it et al.},
Nucl.\ Phys.\ B {\bf 416} (1994) 675
[hep-lat/9308007].

\vspace{3mm}

\item \label{Allton:1994ui}
C.R.~Allton {\it et al.}  [APE Collaboration],
Phys.\ Lett.\ B {\bf 345} (1995) 513
[hep-lat/9411011].

\vspace{3mm}

\item \label{Gill:2001jp}
J.~Gill  [UKQCD Collaboration],
Nucl.\ Phys.\ Proc.\ Suppl.\  {\bf 106} (2002) 391
[hep-lat/0109035].

\vspace{3mm}

\item \label{Abada:2002ie}
A.~Abada, {\it et al.} 
                  [SPQcdR Collaboration],
hep-lat/0209116.

\vspace{3mm}

\item \label{3Beneke:2000wa}
M.~Beneke and T.~Feldmann,
Nucl.\ Phys.\ B {\bf 592} (2001) 3
[hep-ph/0008255].

\vspace{3mm}

\item \label{3Chay:2002vy}
J.~Chay and C.~Kim,
Phys.\ Rev.\ D {\bf 65} (2002) 114016
[hep-ph/0201197].

\vspace{3mm}

\item \label{3Bauer:2002uv}
C.W.~Bauer, D.~Pirjol and I.W.~Stewart,
Phys.\ Rev.\ D {\bf 66} (2002) 054005
[hep-ph/0205289].

\vspace{3mm}

\item \label{3Beneke:2002ph}
M.~Beneke, {\it et al.}, 
Nucl.\ Phys.\ B {\bf 643} (2002) 431
[hep-ph/0206152].

\vspace{3mm}

\item \label{3Ball:1998kk}
P.~Ball and V.M.~Braun,
Phys.\ Rev.\ D {\bf 58} (1998) 094016
[hep-ph/9805422].

\vspace{3mm}

\item \label{Yamada:2001xp}
N.~Yamada {\it et al.}  [JLQCD Collaboration],
Nucl.\ Phys.\ Proc.\ Suppl.\  {\bf 106} (2002) 397
[hep-lat/0110087].

\vspace{3mm}

\item \label{lcsr}
I.I.~Balitsky, V.M.~Braun and A.V.~Kolesnichenko,
 Nucl.\ Phys.\  B~{\bf 312} (1989) 509; \\
V.M.~Braun and I.E.~Filyanov,
 Z.\ Phys.\  C~{\bf 44} (1989) 157.

\vspace{3mm}

\item \label{cz}
V.L.~Chernyak and I.R.~Zhitnitsky,
Nucl.\ Phys.\  B~{\bf 345} (1990) 137.

\vspace{3mm}

\item \label{Bpi} 
V.M.~Belyaev, A.~Khodjamirian and R.~R\"uckl,
Z.\ Phys.\  C~{\bf 60} (1993) 349.

\vspace{3mm}

\item \label{Bpialphas} 
A.~Khodjamirian, R.~R\"uckl, S.~Weinzierl and O.~Yakovlev,
Phys.\ Lett.\  B~{\bf 410} (1997) 275;\\
E.~Bagan, P.~Ball and V.M.~Braun,
 Phys.\ Lett.\  B~{\bf 417} (1998) 154.

\vspace{3mm}

\item \label{Ball1}
P.~Ball, 
 JHEP {\bf 9809} (1998) 005.

\vspace{3mm}

\item \label{BallZ}
P.~Ball and R.~Zwicky,
JHEP {\bf 0110} (2001) 019 [hep-ph/0110115].

\vspace{3mm}

\item \label{AliB}
A.~Ali, P.~Ball, L.T.~Handoko and G.~Hiller,
Phys.\ Rev.\ D {\bf 61} (2000) 074024.

\vspace{3mm}

\item \label{KR}
A.~Khodjamirian and R.~Ruckl,
in {\em Heavy Flavors}, 2nd edition, eds., A.J. Buras and M. Lindner,
World Scientific (1998), p. 345, hep-ph/9801443.

\vspace{3mm}

\item \label{Brauntalk}
V.M.~Braun,
hep-ph/9911206.

\newpage

\item \label{CK}
P.~Colangelo and A.~Khodjamirian,
 {\em Boris Ioffe Festschrift 
'At the Frontier of Particle Physics; Handbook of QCD'}, ed. M.
Shifman (World Scientific, Singapore, 2001), p.1495 [hep-ph/0010175].

\vspace{3mm}

\item \label{Jamin:2001fw}
M.~Jamin and B.O.~Lange,
Phys.\ Rev.\ D {\bf 65} (2002) 056005.

\vspace{3mm}

\item \label{BecirevicKaidalov}
D.~Becirevic and A.B.~Kaidalov,
Phys.\ Lett.\ B {\bf 478} (2000) 417 [hep-ph/9904490].

\vspace{3mm}

\item \label{coupling}
V.M.~Belyaev, V.M.~Braun, A.~Khodjamirian and R.~Ruckl,
Phys.\ Rev.\ D {\bf 51} (1995) 6177 [hep-ph/9410280];
A.~Khodjamirian, R.~Ruckl, S.~Weinzierl and O.~Yakovlev,
Phys.\ Lett.\ B {\bf 457} (1999) 245 [hep-ph/9903421].

\vspace{3mm}

\item \label{UKQCD}
K.C.~Bowler {\it et al.}  [UKQCD Collaboration],
Phys.\ Lett.\ B {\bf 486} (2000) 111.

\vspace{3mm}

\item \label{Belle}
H. Ishino [Belle Collaboration], talk at XXXVII Rencontres de Moriond,
March 2002.  

\vspace{3mm}

\item \label{cleopi}
J.P.~Alexander {\it et al.}  [CLEO Collaboration],
Phys.\ Rev.\ Lett.\  {\bf 77} (1996) 5000.

\vspace{3mm}

\item \label{isgw2} D. Scora and N. Isgur, { Phys. Rev.} D~{\bf 52} (1995)  2783.

\vspace{3mm}

\item \label{melikhov} D. Melikhov, { Phys. Rev.} D~{\bf 53} (1996) 2160.

\vspace{3mm}

\item \label{wsb} M. Wirbel, B. Stech and M. Bauer, 
Z. Phys. C~{\bf 29} (1985) 637.

\vspace{3mm}

\item \label{burdman} 
G.~Burdman and J.~Kambor,
Phys.\ Rev.\ D {\bf 55} (1997) 2817
[hep-ph/9602353].

\vspace{3mm}

\item \label{flynn} J.M. Flynn {\it et al.}, { Nucl. Phys.} 
B~{\bf 461} (1996) 327.

\vspace{3mm}

\item \label{stech} B. Stech, { Phys. Lett.} B~{\bf 354} (1995) 447.

\vspace{3mm}

\item \label{bellepi} BELLE Collaboration, BELLE-CONF 0124 (2001).

\vspace{3mm}

\item \label{bellepiICHEP} BELLE Collaboration, see 
{\tt  http://www.ichep02.nl/Transparencies/CP/CP-1/CP-1-4.kwon.pdf}, 
ICHEP (2002).

\vspace{3mm}

\item \label{babarrho} BaBar Collaboration,  hep-ex/0207080. 

\vspace{3mm}

\item \label{bellerho}  Y. Kwon [BELLE Collaboration],
ICHEP CP-1-4 (2002).

\vspace{3mm}

\item \label{beyer98} M. Beyer and D. Melikhov, { Phys. Lett.} B~{\bf 436} (1998) 344.

\vspace{3mm}

\item \label{cleovubICHEP}
 N.~Adam {\it et al.} [CLEO Collaboration], CLEO-CONF/02-09,
 ICHEP02-ABS931. 

\vspace{3mm}

\item \label{deFazioNeubert}
 F. De Fazio and M. Neubert, JHEP {\bf 9906} (1999) 017.

\vspace{3mm}

\item \label{ligetiwise} Z. Ligeti and M.B. Wise, { Phys. Rev.} D~{\bf 53} 
(1996) 4937.
.
\vspace{3mm}

\item \label{belleomega} BELLE Collaboration, BELLE-CONF 02/42, ICHEP ABS 732.

\vspace{3mm}

\item \label{cleoeta} CLEO Collaboration, talk at DPF-2002, see \\ 
{ \tt http://dpf2002.velopers.net/talks\_pdf/452talk.pdf}. 

\vspace{3mm}

\item \label{lifeWG}        LEP B Lifetime Working Group, see \\   
                    { \tt    http://lepbosc.web.cern.ch/LEPBOSC/lifetimes/ 
                        lepblife.html}   

\newpage

\item \label{lifeBfact} B. Aubert {\it et al.} [BaBar Collaboration], Phys.\
     Rev.\ Lett.\ {\bf 87} (2001) 201803; \\
K. Abe {\it et al.} [BELLE Collaboration], 
     Phys.\ Rev.\ Lett.\ {\bf 88} (2002) 171801.

\vspace{3mm}

\item \label{BBGLN}
M.~Beneke {\it et al.}, 
Phys.\ Lett.\ B {\bf 459} (1999) 631
[hep-ph/9808385].

\vspace{3mm}

\item \label{Ben}
M.~Beneke and A.~Lenz, J.~Phys.~G~{\bf G27} (2001) 1219
[hep-ph/0012222]; \\
D.~Becirevic,
hep-ph/0110124 and refs.\ therein.

\vspace{3mm}

\item \label{nlodb1}
A.J.~Buras, M.~Jamin, M.E.~Lautenbacher and P.H.~Weisz,
Nucl.\ Phys.\ B {\bf 400} (1993) 37
[hep-ph/9211304];
A.J.~Buras, M.~Jamin and M.E.~Lautenbacher,
Nucl.\ Phys.\ B {\bf 400} (1993) 75
[hep-ph/9211321];
M.~Ciuchini, E.~Franco, G.~Martinelli and L.~Reina,
Nucl.\ Phys.\ B {\bf 415} (1994) 403
[hep-ph/9304257].

\vspace{3mm}

\item \label{benekebd}
M.~Beneke, G.~Buchalla and I.~Dunietz,
Phys.\ Rev.\ D {\bf 54} (1996) 4419  
[hep-ph/9605259].

\vspace{3mm}

\item \label{3GR}
V.~Gimenez and J.~Reyes,
Nucl.\ Phys.\ Proc.\ Suppl.\  {\bf 93} (2001) 95 
[hep-lat/0009007].

\vspace{3mm}

\item \label{hiroshima}
S.~Hashimoto {\it et al.}, 
Phys.\ Rev.\ D {\bf 62} (2000) 114502  
[hep-lat/0004022].

\vspace{3mm}

\item \label{Aoki:2002bh}
S.~Aoki {\it et al.}  [JLQCD Collaboration],
hep-lat/0208038.

\vspace{3mm}

\item \label{ape}
D.~Becirevic {\it et al.}, 
Eur.\ Phys.\ J.\ C {\bf 18} (2000) 157  
[hep-ph/0006135].

\vspace{3mm}

\item \label{Lellouch:2000tw}
L.~Lellouch and C.J.~Lin  [UKQCD Collaboration],
Phys.\ Rev.\ D {\bf 64} (2001) 094501
[hep-ph/0011086]. 

\vspace{3mm}

\item \label{damir}
D.~Becirevic {\it et al.}, 
JHEP {\bf 0204} (2002) 025
[hep-lat/0110091].

\vspace{3mm}

\item \label{norikazu} 
S.~Hashimoto and N.~Yamada  [JLQCD Collaboration],
hep-ph/0104080.

\vspace{3mm}

\item \label{JLQCD_new}
S.~Aoki {\it et al.}  [JLQCD Collaboration],
hep-lat/0208038.

\vspace{3mm}

\item \label{Paper}
A.S.~Dighe, {\it et al.}, 
Nucl.\ Phys.\ B {\bf 624} (2002) 377 
[hep-ph/0109088] and hep-ph/0202070; \\
see also T.~Hurth {\it et al.},
J.\ Phys.\ G {\bf 27} (2001) 1277 [hep-ph/0102159].

\vspace{3mm}

\item \label{breport} K. Anikeev {\it et al.}, hep-ph/0201071. 

\vspace{3mm}

\item \label{jlqcd}
S.~Hashimoto and N.~Yamada  [JLQCD Collaboration],
hep-ph/0104080.

\vspace{3mm}

\item \label{lhc} P.~Ball {\it et al.},
CERN-TH-2000-101, hep-ph/0003238.

\vspace{3mm}

\item \label{babar-direct} 
B.~Aubert {\it et al.}  [BaBar Collaboration],
Phys.\ Rev.\ Lett.\  {\bf 87} (2001) 091801 [hep-ex/0107013].

\vspace{3mm}

\item \label{grossman}
Y.~Grossman,
Phys.\ Lett.\ B {\bf 380} (1996) 99
[hep-ph/9603244].

\vspace{3mm}

\item \label{Bigi:1992su}
I.I.~Bigi, N.G.~Uraltsev and A.I.~Vainshtein,
Phys.\ Lett.\ B {\bf 293} (1992) 430; Erratum {\bf 297} (1993) 477
[hep-ph/9207214].

\item \label{gamma1}
Q.~Ho-kim and X.-y.~Pham,
Phys.\ Lett.\ B {\bf 122} (1983) 297.

\vspace{3mm}

\item \label{gamma3}
E.~Bagan, P.~Ball, V.M.~Braun and P.~Gosdzinsky,
Nucl.\ Phys.\ B {\bf 432} (1994) 3
[hep-ph/9408306].

\vspace{3mm}

\item \label{gamma4}
E.~Bagan et al.,
Phys.\ Lett.\ B {\bf 342} (1995) 362; Erratum {\bf 374} (1996) 363
[hep-ph/9409440]; \\
E.~Bagan et al., 
Phys.\ Lett.\ B {\bf 351} (1995) 546
[hep-ph/9502338]; \\
A.F.~Falk et al., 
Phys.\ Lett.\ B {\bf 326} (1994) 145
[hep-ph/9401226].


\vspace{3mm}

\item \label{NS}
M.~Neubert and C.T.~Sachrajda,
Nucl.\ Phys.\ B {\bf 483}, 339 (1997)
[hep-ph/9603202].

\vspace{3mm}

\item \label{Ura}
N.G.~Uraltsev,
Phys.\ Lett.\ B {\bf 376} (1996) 303
[hep-ph/9602324].

\vspace{3mm}

\item \label{charmBB} 
M.~Beneke and G.~Buchalla,
Phys.\ Rev.\ D {\bf 53} (1996) 4991
[hep-ph/9601249].

\vspace{3mm}

\item \label{Roma}
E.~Franco {\it et al.}, 
Nucl.\ Phys.\ B {\bf 633} (2002) 212
[hep-ph/0203089].

\vspace{3mm}

\item \label{Lenz}
M.~Beneke {\it et al.}, 
hep-ph/0202106.

\vspace{3mm}

\item \label{reyes}
V.~Gimenez and J.~Reyes,
Nucl.\ Phys.\ B {\bf 545} (1999) 576
[hep-lat/9806023].

\vspace{3mm}

\item \label{noantri}
M.~Ciuchini, {\it et al.}, 
Nucl.\ Phys.\ B {\bf 625} (2002) 211
[hep-ph/0110375].

\vspace{3mm}

\item \label{Chernyak:1995cx}
V.~Chernyak,
Nucl.\ Phys.\ B {\bf 457} (1995) 96
[hep-ph/9503208].

\vspace{3mm}

\item \label{pirjol}
D.~Pirjol and N.G.~Uraltsev,
Phys.\ Rev.\ D {\bf 59} (1999) 034012
[hep-ph/9805488].

\vspace{3mm}

\item \label{DiPierro98}
M.~Di Pierro and C.T.~Sachrajda  [UKQCD Collaboration],
Nucl.\ Phys.\ B {\bf 534} (1998) 373 \break
[hep-lat/9805028].

\vspace{3mm}

\item \label{DiPierro:1999tb}
M.~Di Pierro, C.T.~Sachrajda and C.~Michael  [UKQCD Collaboration],
Phys.\ Lett.\ B {\bf 468} (1999) 143
[hep-lat/9906031];
M.~Di Pierro and C.T.~Sachrajda  [UKQCD Collaboration],
Nucl.\ Phys.\ Proc.\ Suppl.\  {\bf 73} (1999) 384
[hep-lat/9809083].

\vspace{3mm}

\item \label{APE} D.~Becirevic, update of hep-ph/0110124 for the present work.

\vspace{3mm}

\item \label{Colangelo:1996ta}
P.~Colangelo and F.~De Fazio,
Phys.\ Lett.\ B {\bf 387} (1996) 371
[hep-ph/9604425]; \\
M.S.~Baek, J.~Lee, C.~Liu and H.S.~Song,
Phys.\ Rev.\ D {\bf 57} (1998) 4091 [hep-ph/9709386]; \\
H.Y.~Cheng and K.C.~Yang,
Phys.\ Rev.\ D {\bf 59} (1999) 014011
[hep-ph/9805222]; \\
C.S.~Huang, C.~Liu and S.L.~Zhu,
Phys.\ Rev.\ D {\bf 61} (2000) 054004
[hep-ph/9906300].



\item \label{alephBR} ALEPH Collaboration, Phys. Lett. B~{\bf 486} (2000) 286.

\vspace{3mm}

\item \label{l3} L3 Collaboration, Phys. Lett. B~{\bf 438} (1998) 417.

\vspace{3mm}

\item \label{delphi1} DELPHI Collaboration, Eur. Phys. J. C~{\bf 18} 
(2000)  229.

\vspace{3mm}

\item \label{wal} ALEPH Collaboration, Phys. Lett. B~{\bf 377} (1996) 205; \\
CDF Collaboration, Phys. Rev. D~{\bf 59} (1999) 032004; \\
OPAL Collaboration, Phys. Lett. B~{\bf 426} (1998) 161.

\vspace{3mm}

\item \label{theoBR} R. Aleksan {\it et al.}, Phys. Lett. B~{\bf 316} (1993) 567.

\vspace{3mm}

\item \label{delphi2} DELPHI Collaboration, Eur. Phys. J. C~{\bf 16} (2000)  555.

\vspace{3mm}

\item \label{cdfjpsi} CDF Collaboration, Phys. Rev. D~{\bf 57} (1998) 5382.

\vspace{3mm}

\item \label{cleo} CLEO Collaboration, Phys. Rev. Lett. {\bf 79} (1997) 4533.

\vspace{3mm}

\item \label{our} ALEPH, CDF, DELPHI, L3, OPAL and SLD Collaborations, 
CERN-EP/2001-050. \\
See also {\tt http://www.cern.ch/LEPBOSC}.

\vspace{3mm}

\item \label{cdfjpsiphilife} CDF Collaboration, 
Phys. Rev. Lett. {\bf 77} (1996) 1945.

\vspace{3mm}

\item \label{cdfjpsiphipola} CDF Collaboration, 
Phys. Rev. Lett. {\bf 75} (1995) 3068 and
Phys. Rev. Lett. {\bf 85} (2000) 4668.

\newpage

\item \label{svt} W. Ashmanskas {\it et al.}, 
{\it Performance of the CDF Online Silicon Vertex Tracker}
published in Proceedings of the 2001 IEEE Nuclear Science Symposium (NSS) 
and Medical Imaging Conference
(MIC), San Diego, CA, November  2001.

\vspace{3mm}

\item \label{delphideltaG} T. Allmendinger {\it et al.} [DELPHI Collaboration] 
 DELPHI 2001-054 CONF 482. 

\vspace{3mm}

\item \label{DUNIETZ}A.~Dighe, I.~Dunietz, H.~Lipkin, J.L.~Rosner, 
Phys.\ Lett.~B~{\bf 369} (1996) 144.

\vspace{3mm}

\item \label{sevelda} M.~Sevelda, PhD Thesis, Charles University of Prague, 
1999. 

\vspace{3mm}

\item \label{ATLASTDR} ATLAS TDR, CERN-LHCC-99-015.

\end{enumerate}

\newpage

\thispagestyle{empty}
~

\newpage

\chapter{CKM ELEMENTS FROM K~AND~B~MESON~MIXING}
\label{chap:IV}

\noindent
{\it Conveners   : J.M.~Flynn, M.~Paulini, S.~Willocq. \\
Contributors: D.~Abbaneo, C.~Bozzi, A.J.~Buras, R. Forty, R.~Gupta, 
R.~Hawkings, A.~Hoecker, M.~Jamin, P.~Kluit, A.~Kronfeld, V.~Lacker, 
F.~Le~Diberder, L.~Lellouch, C.~Leonidopoulos, D.~Lin, V.~Lubicz, 
H.G.~Moser, U.~Nierste, J.~Ocariz, F.~Parodi, 
C.~Paus, P.~Roudeau, Y.~Sakai, O.~Schneider, A.~Stocchi, 
C.~Weiser, N.~Yamada.}

\section{Basic formulae for particle--antiparticle mixing}

\boldmath
\subsection{K sector: basic formula for $\eps_K$}
\unboldmath
\label{subsec:epsformula}

\begin{figure}
\hbox to\hsize{\hss
\includegraphics[width=0.6\hsize,bb = 3 3 408 152]{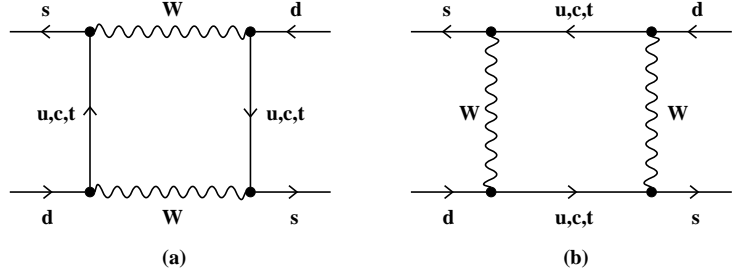}\hss}
\caption{\it Box diagrams contributing to ${\rm K}^0-\overline{\rm  K}^0$ mixing
in the SM.}
\label{fig:L9}
\end{figure}

In the $\rm K^0-\overline{\rm K}^0$ system, 
to lowest order in electroweak interactions $\Delta S=2$ transitions
are induced through the box diagrams of Fig.~\ref{fig:L9}. Including
leading and next-to-leading QCD corrections in renormalization group
improved perturbation theory the effective Hamiltonian for the $\Delta
S=2$ transitions for scales $\mu<\mu_c=\ord(m_c)$ is given by
\begin{eqnarray}\label{eq:hds2}
\mathcal{H}^{\Delta S=2}_{\rm eff}&=&\frac{G^2_{\rm F}}{16\pi^2}\mw^2
 \left[\lambda^2_c\eta_1 S_0(x_c)+\lambda^2_t \eta_2 S_0(x_t)+
 2\lambda_c\lambda_t \eta_3 S_0(x_c, x_t)\right] \times
\nonumber\\
& & \times \left[\as^{(3)}(\mu)\right]^{-2/9}\left[
  1 + \frac{\as^{(3)}(\mu)}{4\pi} J_3\right]  Q(\Delta S=2)
  + \mathrm{h.c.}
\end{eqnarray}
where $\lambda_i = V_{is}^* V_{id}^{}$, $\as^{(3)}$ is the strong
coupling constant in an effective three flavour theory and
$J_3 = 307/162 = 1.895$ in the NDR scheme [\ref{BJW90}]. In
(\ref{eq:hds2}), the relevant operator
\begin{equation}\label{qsdsd}
Q(\Delta S=2)=(\bar s\gamma_\mu(1-\gamma_5)d)(\bar s\gamma^\mu(1-\gamma_5)d),
\end{equation}
is multiplied by the corresponding Wilson coefficient function. This
function is decomposed into a charm-, a top- and a mixed charm-top
contribution. The functions $S_0(x_i)$ and $S_0(x_c, x_t)$ are given
by ($x_i=m^2_i/\mw^2$):
\begin{equation}\label{S0}
S_0(x_t)=\frac{4x_t-11x^2_t+x^3_t}{4(1-x_t)^2}-
 \frac{3x^3_t \ln x_t}{2(1-x_t)^3},\quad\quad S_0(x_c)=x_c, 
\end{equation}
\begin{equation}\label{BFF}
S_0(x_c, x_t)=x_c\left[\ln\frac{x_t}{x_c}-\frac{3x_t}{4(1-x_t)}-
 \frac{3 x^2_t\ln x_t}{4(1-x_t)^2}\right],
\end{equation}
where we keep only linear terms in $x_c\ll 1$, but of course all
orders in $x_t$. The exact expression can be found in [\ref{BSS}].

Short-distance QCD effects are described through the correction
factors $\eta_1$, $\eta_2$, $\eta_3$ and the explicitly
$\alpha_\mathrm{s}$-dependent terms in (\ref{eq:hds2}). The NLO values of
$\eta_i$ are given as follows 
[\ref{BJW90},\ref{HNa}--\ref{NIJA03}]:
%
\begin{equation}\label{etai}
\eta_1=(1.32\pm 0.32) \left(\frac{1.30\gev}{m_c(m_c)}\right)^{1.1},\qquad
\eta_2=0.57\pm 0.01,\qquad
  \eta_3=0.47\pm0.05~.
\end{equation}
It should be emphasized that the values of $\eta_i$ depend on the definition
of the quark masses $m_i$. The ones in (\ref{etai}) 
correspond to $m_t\equiv m_t(m_t)$ 
and $m_c\equiv m_c(m_c)$ . With this definition the dependences of $\eta_2$ 
on $m_t$ and of $\eta_3$ on $m_t$ and $m_c$ are fully negligible but 
the dependence of $\eta_1$ on $m_c$ turns out to be significant. It can
be well approximated by the formula in (\ref{etai}). The scale dependence in 
$m_t(\mu_t)$, where $\mu_t=\ord(m_t)$, present generally in the functions 
$S_0(x_t)$ and $S_0(x_t,x_c)$
is canceled to an excellent accuracy in the products  
$\eta_2 S_0(x_t)$ and $\eta_3 S_0(x_t,x_c)$. The corresponding scale 
dependence in $m_c(\mu_c)$, where $\mu_c=\ord(m_c)$, is cancelled to a 
large extent in the product 
$\eta_3 S_0(x_t,x_c)$ but remains still sizable in $\eta_1 S_0(x_c)$. 
As we use $m_c(m_c)$ and $m_t(m_t)$ we have included the
left-over scale uncertainties due to $\mu_c$ and $\mu_t$ present in 
(\ref{eq:hds2}) 
in the errors of $\eta_i$ that also include the uncertainties due to 
$\Lambda_{\overline{\mathrm{MS}}}$, the scale in the QCD running coupling.
The small changes in $\eta_1$ and $\eta_3$ relative to the original papers 
are due to changes in $\alpha_s(M_Z)$.

Now, $\eps_K$ is defined by
\begin{equation}\label{eq:ek}
\eps_K
={{A({\rm K}_{\rm L}\rightarrow(\pi\pi)_{I=0}})\over{A({\rm K}_{\rm 
S}\rightarrow(\pi\pi)_{I=0})}}
\end{equation} 
with $I$ denoting isospin. From (\ref{eq:ek}) one finds
\begin{equation}
\label{eq:ekxi}
\eps_K = \frac{\exp(i \pi/4)}{\sqrt{2} \Delta M_K} \,
\left( \IM M_{12} + 2 \bar\xi \RE M_{12} \right),
\quad\quad
\bar\xi = \frac{\IM A_0}{\RE A_0}
\label{eq:epsdef}
\end{equation}
with the off-diagonal element $M_{12}$ in
the neutral ${\rm K}$-meson mass matrix representing 
${\rm K}^0$-$\overline {\rm K}^0$ mixing  given~by
\begin{equation}
2 M_K M^*_{12} = 
\langle \overline{\rm  K}^0| \Heff(\Delta S=2) |\rm K^0\rangle\,.
\label{eq:M12Kdef}
\end{equation}
The factor $2 M_K$ reflects our normalization
of external states and $A_0$ is the isospin amplitude.
$\Delta M_K$ is the ${\rm K}_L-{\rm K}_S$ mass difference that is taken from experiment 
as it cannot be reliably calculated due to long distance
contributions. The expression in (\ref{eq:ekxi}) neglects higher order
CP-violating terms: see the discussion in the review article in
reference~[\ref{deRafaelTASI94}].

Defining the renormalization group invariant parameter $\hat B_K$ by
[\ref{BJW90}]
\begin{equation}
\hat B_K = B_K(\mu) \left[ \alpha_\mathrm{s}^{(3)}(\mu) \right]^{-2/9} \,
\left[ 1 + \frac{\alpha_\mathrm{s}^{(3)}(\mu)}{4\pi} J_3 \right]~,
\label{eq:BKrenorm}
\end{equation}
\begin{equation}
\langle \overline{\rm K}^0| Q(\Delta S=2) |{\rm K}^0\rangle
\equiv \frac{8}{3} B_K(\mu) F_K^2 M_K^2
\label{eq:KbarK}
\end{equation}
and using (\ref{eq:M12Kdef}) and (\ref{eq:hds2}) one finds
\begin{equation}
M_{12} = \frac{G_{\rm F}^2}{12 \pi^2} F_K^2 \hat B_K M_K \mw^2
\left[ {\lambda_c^*}^2 \eta_1 S_0(x_c) + {\lambda_t^*}^2 \eta_2 S_0(x_t) +
2 {\lambda_c^*} {\lambda_t^*} \eta_3 S_0(x_c, x_t) \right],
\label{eq:M12K}
\end{equation}
where $F_K=160~\mev$ is the ${\rm K}$-meson decay constant and $M_K$
the ${\rm K}$-meson mass. 

To proceed further we neglect the last term in (\ref{eq:epsdef}) as in the
standard CKM phase convention it
constitutes at most a $2\,\%$ correction to $\eps_K$. This is justified
in view of other uncertainties, in particular those connected with
$\hat B_K$. Inserting (\ref{eq:M12K}) into (\ref{eq:epsdef}) we find
\begin{equation}
\eps_K=C_{\eps} \hat B_K \IM\lambda_t \left\{
\RE\lambda_c \left[ \eta_1 S_0(x_c) - \eta_3 S_0(x_c, x_t) \right] -
\RE\lambda_t \eta_2 S_0(x_t) \right\} \exp(i \pi/4)\,,
\label{eq:epsformula}
\end{equation}
where we have used the unitarity relation $\IM\lambda_c^* = {\rm
Im}\lambda_t$ and  have neglected $\RE\lambda_t/\RE\lambda_c
 = \ord(\lambda^4)$ in evaluating $\IM(\lambda_c^* \lambda_t^*)$.
The numerical constant $C_\eps$ is given by
\begin{equation}
C_\eps = \frac{G_{\rm F}^2 F_K^2 M_K \mw^2}{6 \sqrt{2} \pi^2 \Delta M_K}
       = 3.837 \cdot 10^4 \, .
\label{eq:Ceps}
\end{equation}
To this end we have used the experimental value of 
$\Delta M_K= 3.837 \cdot 10^{-15}~\gev$
and $\mw=80.4~\gev$.
 
The main uncertainty in (\ref{eq:epsformula}) resides in the parameter
$\hat B_K$. The present status of $\hat B_K$ is discussed in
Sec.~\ref{sec:KKbar-mixing} Here we note only that when $\hat
B_K>0$, as found by all non-perturbative methods, the formula
(\ref{eq:epsformula}) combined with the experimental value for
$\eps_K$ implies $0<\delta<\pi$ in the standard parametrization
 or equivalently $\bar\eta>0$ in the Wolfenstein
parametrization.

\boldmath
\subsection{B sector: basic formulae for {$\Delta M_{d,s}$} 
oscillation frequencies}
\label{subsec:BBformula}
\unboldmath

The strengths of the ${\rm B}^0_{d,s}-\overline{\rm B}^0_{d,s}$ 
mixings are described by
the mass differences
\begin{equation}
\label{eq:deltamb}
\Delta M_{d,s}= M_H^{d,s}-M_L^{d,s}
\end{equation}
where the subscripts $H$ and $L$ denote the heavy and light mass
eigenstates respectively. The long distance contributions are
estimated to be very small, in contrast to the situation for $\Delta
M_K$, and $\Delta M_{d,s}$ are very well approximated by the relevant
box diagrams. Moreover, since $m_{u,c}\ll m_t$ only the top sector can
contribute significantly to $\Delta M_{d,s}$. The charm and mixed
top-charm contributions are entirely negligible.

$\Delta M_{d,s}$ can be expressed in terms of the off-diagonal element
in the neutral B-meson mass matrix as follows
\begin{equation}
\Delta M_q= 2 |M_{12}^{(q)}|, \qquad q=d,s
\label{eq:xdsdef}
\end{equation}
with $M_{12}$ given by a formula analogous to (\ref{eq:M12Kdef})
\begin{equation}
2 M_{B_q} |M_{12}^{(q)}| = 
|\langle \overline{\rm B}^0_q| \Heff(\Delta B=2) |{\rm B}^0_q\rangle|.
\label{eq:M12Bdef}
\end{equation}
In the case of ${\rm B}_d^0-\overline{\rm B}_d^0$ mixing
\begin{eqnarray}\label{hdb2}
\mathcal{H}^{\Delta B=2}_{\rm eff}&=&\frac{G^2_{\rm F}}{16\pi^2}M^2_W
 \left(V^\ast_{tb}V_{td}\right)^2 \eta_{B}
 S_0(x_t)\times
\nonumber\\
& &\times \left[\alpha^{(5)}_s(\mu_b)\right]^{-6/23}\left[
  1 + \frac{\alpha^{(5)}_s(\mu_b)}{4\pi} J_5\right]  Q(\Delta B=2) + h. c.
\end{eqnarray}
Here $\mu_b=\ord(m_b)$, $J_5 = 5165/3174 = 1.627$ in the NDR scheme
[\ref{BJW90}],
\begin{equation}\label{qbdbd}
Q(\Delta B=2)=
(\bar b\gamma_\mu(1-\gamma_5)d)(\bar b\gamma^\mu(1-\gamma_5)d)
\end{equation}
and 
\begin{equation}
\eta_B=0.55\pm0.01
\end{equation}
summarizes the NLO QCD corrections [\ref{BJW90},\ref{UKJS}]. In the case of
${\rm B}_s^0-\overline{\rm B}_s^0$ mixing one should simply replace $d\to s$ in
(\ref{hdb2}) and (\ref{qbdbd}) with all other quantities and numerical
values unchanged. Again $m_t\equiv m_t(m_t)$.

Defining the renormalization group invariant parameters $\hat B_{B_q}$
in analogy to (\ref{eq:BKrenorm}) and (\ref{eq:KbarK})
\begin{equation}
\hat B_{B_q} =
  B_{B_q}(\mu) \left[ \alpha_\mathrm{s}^{(5)}(\mu) \right]^{-6/23} \,
\left[ 1 + \frac{\alpha_\mathrm{s}^{(5)}(\mu)}{4\pi} J_5 \right]~,
\label{eq:BBrenorm}
\end{equation}
\begin{equation}
\langle \overline{\rm B}_q^0| Q(\Delta B=2) |{\rm B}_q^0\rangle
\equiv \frac{8}{3} B_{B_q}(\mu) F_{B_q}^2 M_{B_q}^2
\label{eq:BbarB}
\end{equation}
one finds
 using (\ref{hdb2}) 
\begin{equation}
\Delta M_q = \frac{G_{\rm F}^2}{6 \pi^2} \eta_B M_{B_q} 
(\hat B_{B_q} F_{B_q}^2 ) \mw^2 S_0(x_t) |V_{tq}|^2,
\label{eq:xds}
\end{equation}
where $F_{B_q}$ is the ${\rm B}_q$-meson decay constant. This implies two
approximate but rather accurate formulae
\begin{equation}\label{eq:DMD}
\Delta M_d=
0.50/{\rm ps}\cdot\left[ 
\frac{\sqrt{\hat B_{B_d}}F_{B_d}}{230\mev}\right]^2
\left[\frac{\mtb(\mt)}{167\gev}\right]^{1.52} 
\left[\frac{\vtd}{7.8\cdot10^{-3}} \right]^2 
\left[\frac{\eta_B}{0.55}\right]  
\end{equation}
and
\begin{equation}\label{eq:DMS}
\Delta M_{s}=
17.2/{\rm ps}\cdot\left[ 
\frac{\sqrt{\hat B_{B_s}}F_{B_s}}{260\mev}\right]^2
\left[\frac{\mtb(\mt)}{167\gev}\right]^{1.52} 
\left[\frac{\vts}{0.040} \right]^2
\left[\frac{\eta_B}{0.55}\right] \,.
\end{equation}
The main uncertainty here stems from the parameters $\fB{d,s}$ and
$\bbhat{d,s}$. The most recent lattice and QCD sum rule results are 
summarized in Sec.~\ref{subsec:fBxi}

\subsection{Basic formulae for B oscillation probabilities}

The probability $\mathcal{P}$ for a $\mbox{B}^0_q$ meson ($q=d,s$)
produced at time $t=0$ to decay as $\mbox{B}^0_q$ at proper time $t$ is
given as
\begin{equation}
{\cal P}({\rm B}^0_q\rightarrow {\rm B}^0_q) = \frac{1}{2}\,\Gamma_q\,
e^{- \Gamma_{q}t}\,
[\cosh (\frac{\Delta \Gamma_q}{2} t)~+~\cos (\Delta M_q t) ].
\end{equation}
Here we neglect effects from CP violation, while
$\Gamma_q = \frac{\Gamma_q^H + \Gamma_q^L}{2}$, $\Delta
 \Gamma_q = \Gamma_q^H-\Gamma_q^L$ and $\Delta M_q$ 
is defined in Eq.~(\ref{eq:deltamb}).
The Standard Model predicts $\Delta \Gamma_q \ll \Delta M_q$.
Neglecting a possible lifetime difference between the 
heavy and light mass eigenstates of the ${\rm B}^0_q$, 
the above expression simplifies to:
\begin{equation}
{\cal P}_{B_q^0}^{\rm unmix} = {\cal P}({\rm B}_q^0\rightarrow
{\rm B}_q^0)~=~\frac{1}{2}\, \Gamma_{q}\, e^{- \Gamma_q t}\,
 [ 1 + \cos ({\Delta M_q t} ) ]
\label{eq:proba}
\end{equation}
Similarly, the probability for the ${\rm B}^0_q$ to decay as $\overline{\rm B}^0_q$ is given by
\begin{equation}
{\cal P}_{B_q^0}^{\rm mix} = {\cal P}({\rm B}_q^0\rightarrow
\overline{\rm B}_q^0)~=~\frac{1}{2}\, \Gamma_{q}\, e^{- \Gamma_q t}
 [ 1 - \cos ({\Delta M_q t} ) ].
\label{eq:prob_mix}
\end{equation}
Thus, a measurement of 
the oscillation frequency gives a direct measurement of the mass difference
between 
the two physical B~meson states\footnote
{$\Delta {M}_q$ is usually given in ps$^{-1}$, where 1 ps$^{-1}$ corresponds
to 6.58 10$^{-4}$eV.}.

\begin{figure}
\centerline{
\includegraphics[width=0.48\hsize]{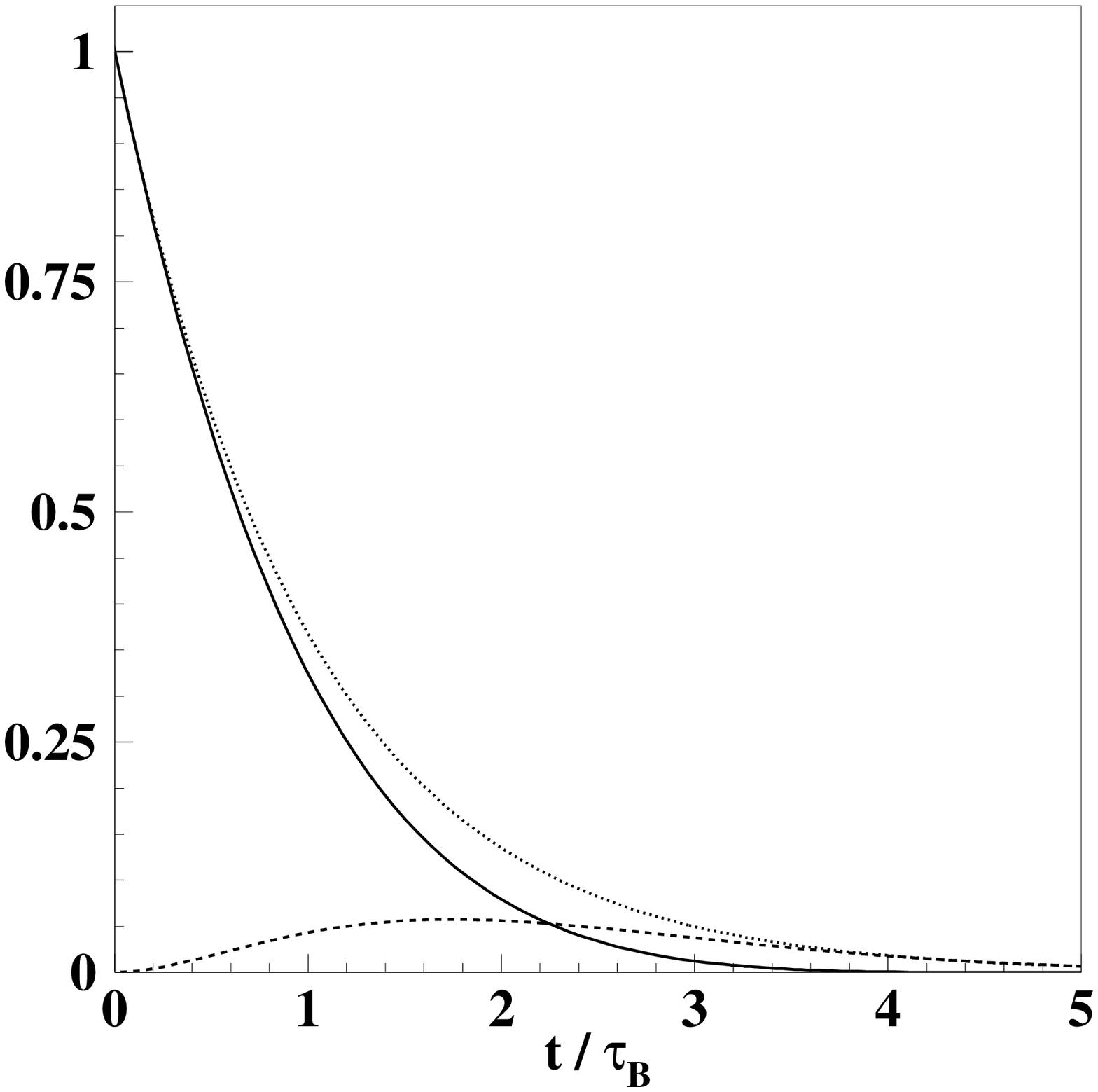}
\hfill
\includegraphics[width=0.48\hsize]{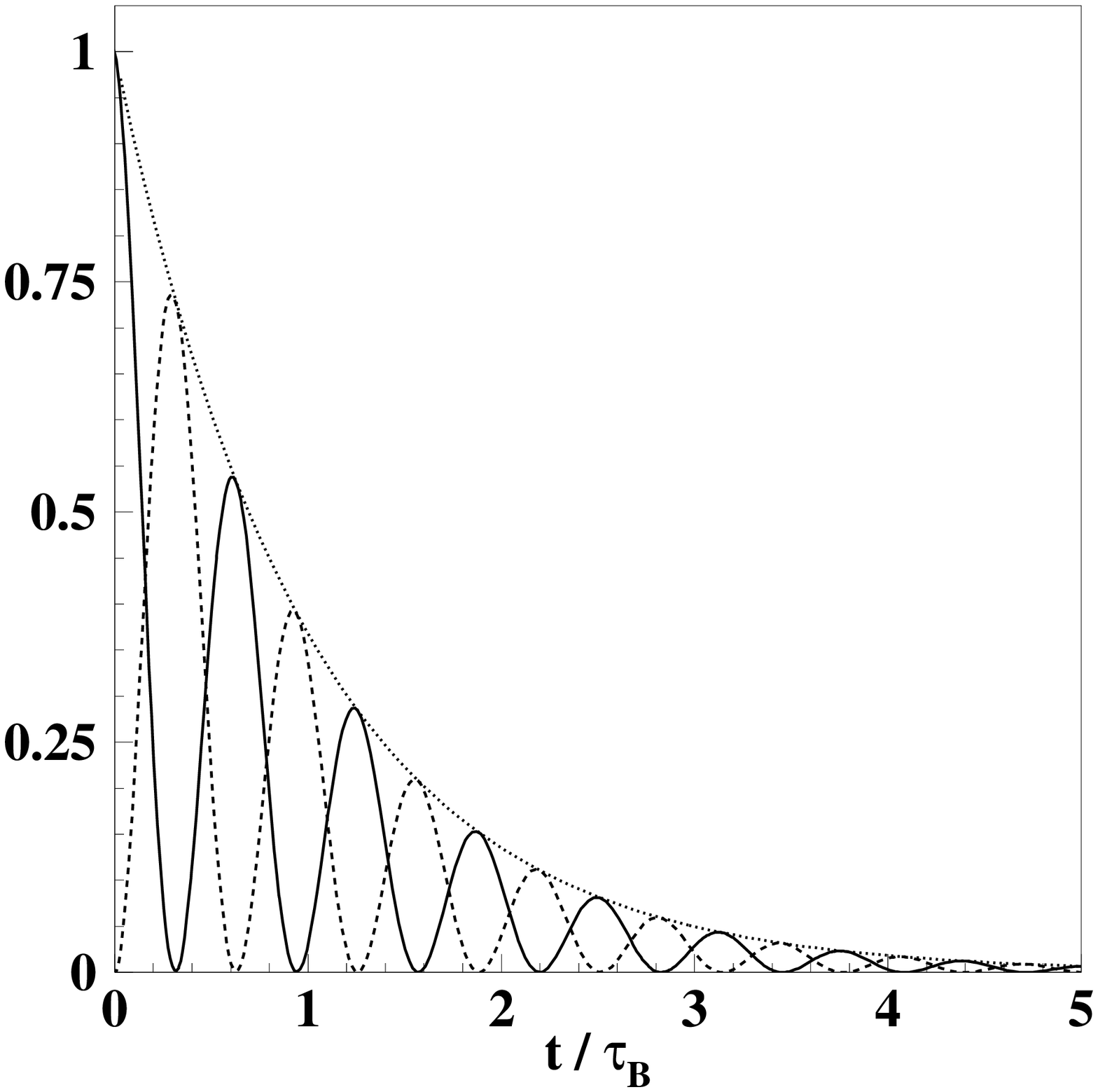}
\put(-275,190){\Large\bf (a)}
\put(-40,190){\Large\bf (b)}
}
\caption{\it Time evolution of ${\rm B^0}$--$\overline{\rm B}^0$ 
oscillations displaying the
unmixed (solid) and the mixed (dashed) contribution as well as the sum
of the two (dotted) for (a) slow and (b) fast oscillation frequencies 
$\Delta M_q$. 
}
\label{fig:bmix_box}
\end{figure}

Figure~\ref{fig:bmix_box} shows the time evolution of 
$B^0 - \overline{\rm B}^{0}$
oscillations displaying the unmixed (solid) and mixed (dashed)
contributions for two different oscillation frequencies $\Delta M$.
The sum of $\mathcal{P}^{\rm mix}$ and $\mathcal{P}^{\rm unmix}$ is just the
exponential particle decay $\Gamma_q\,{\rm e}^{-\Gamma_q t}$ and is
shown by the dotted line in Fig.~\ref{fig:bmix_box}.

The integral of the probability ${\mathcal P}_{B_q^0}^{\rm mix}$
defined in Eq.~(\ref{eq:prob_mix}) gives the mixing parameter:
\begin{eqnarray}
\chi_{q} = \frac{x^2_{q}}{2\,(1 + x^2_{q})}
\hspace*{1.0cm}{\rm with}
\hspace*{0.3cm}
x_{q} = \Delta M_{q}\,\tau_{B_{q}},
\label{eq:integosci}
\end{eqnarray}
where the lifetime $\tau_{B_{q}}=1/\Gamma_q$.

\section{Theoretical issues}

\subsection{Non-perturbative parameters for B meson mixing}
\label{subsec:fBxi}

From the discussion in Sec.~\ref{subsec:BBformula} above, the main
uncertainty in determining $\vtd$ from $\Delta M_d$ comes from the
factor $\fB d \sqrt{\bbhat d}$ in Eq.~\ref{eq:DMD}. In the standard
analysis of the Unitarity Triangle (see Chapter 5), $\Delta
M_s$ is used in a ratio with $\Delta M_d$, so that the important
quantity is $\xi$, that is crucial for the determination of
${\vtd}/{|V_{ts}|}$:
\begin{equation}
\frac{\vtd}{|V_{ts}|}= 
  \xi\sqrt{\frac{M_{{\rm B}_s}}{M_{{\rm B}_d}}}
  \sqrt{\frac{\Delta M_d}{\Delta M_s}},
\qquad
\xi = 
  \frac{F_{B_s} \sqrt{\hat B_{B_s}}}{F_{B_d} \sqrt{\hat B_{B_d}}}.
\label{eq:vtdvtsxi}
\end{equation}
Although the quantities $\fB q \sqrt{\bbhat q}$ for $q=d,s$ are needed
for UT fits, it is common to find $\fB q$ and $\bbhat q$ 
separately.

\subsubsection{{$F_{B_q}$ and $\xi$} from lattice QCD}

Lattice calculations are based on a first-principles evaluation of the
path integral for QCD on a discrete space-time lattice. They have
statistical errors arising from the stochastic (Monte Carlo)
techniques used to evaluate the integral. They also have systematic
errors from discretization effects, finite volume effects, the
treatment of heavy quarks, chiral extrapolations and quenching (or
partial quenching). We now briefly discuss these different sources of
error.

Statistical, discretization and finite volume errors can all be
addressed by brute-force improvement of numerical simulations. We can
also use improved discretization procedures (to reduce discretization
effects at a given lattice spacing) and understand (and even make use
of) the finite volume effects.

Lattice results need to be matched either directly to physical
quantities, or perhaps to quantities defined in some continuum
renormalization scheme. On the lattice side this can be done using
lattice perturbation theory, but with the development of
non-perturbative renormalization methods, the uncertainty from the
lattice can be systematically reduced. For a physical quantity (such
as the decay constant $\fB q$) this is the end of the story. If matching
is made to a quantity in a continuum scheme (such as $B_{B_q}$ in
$\overline\mathrm{MS}$), the remaining uncertainty comes from the
\emph{continuum} perturbation theory: see for example the discussion
in~[\ref{rome-paris-ckm-2000}].

There are a number of ways to treat the heavy $b$-quark on the
lattice. Results for ${\rm B}^0$--$\overline{\rm B}^0$ 
mixing obtained using different
approaches broadly agree, suggesting that the heavy quark mass
dependence is under control.

This leaves chiral extrapolations and quenching to consider. We will
start with quenching. Recall that the QCD path integral is over both
gauge and fermion fields. However, since the fermions appear
quadratically in the action, the fermion integral can be done exactly
to leave a determinant (actually a determinant for each flavour of
quark). The calculation of the determinant is extremely intensive
numerically, so the so-called \emph{quenched approximation} replaces
it with a constant, together with a shift in the bare couplings. This
is not a controlled approximation, but today more and more lattice
simulations are being done including the determinant for at least some
of the quarks. The first dynamical quark algorithms produced sea
quarks in degenerate pairs (in order to get a positive weight function
for the Monte Carlo generation of the gauge field ensemble) and
two-flavour ($N_f=2$) dynamical simulations are still the most
commonly encountered. However, methods are being developed to cope
with single flavours of dynamical quark and $N_f=2+1$ simulation
results, with two degenerate light flavours and one heavy flavour, are
beginning to appear, although there are still questions about the
validity of some steps in the algorithm.

Each quark whose determinant is evaluated is labeled as a `dynamical'
or `sea' quark in lattice parlance. A typical lattice calculation of a
hadronic correlation function (from which masses and/or matrix
elements may be extracted) involves an average over an ensemble of
gauge fields of a combination of quark propagators. These propagators
are evaluated on the background of each gauge field in the ensemble by
means of a matrix inversion. The set of masses used for the
propagators define the `valence' masses of the simulation, which may
or may not be the same as the dynamical masses which were incorporated
(via determinant factors) when generating the gauge field ensemble.
Usually the valence and sea masses are different and we talk of a
`partially quenched' calculation.

Results for $F_B$ from quenched calculations have remained stable for
a number of years. Numerical simulations using two flavours of
dynamical quarks show an increase in $F_B$ compared to quenched
results. The latest developments have seen the first $3$-flavour
dynamical results~[\ref{milc-fnal-nf3-lat2001},\ref{milc-fnal-nf3-lat2002}],
where two flavours are `light' and one is heavier, around the strange
quark mass. For the future, the development of more realistic
dynamical simulations will continue.

Another important (and related) issue is that of chiral
extrapolations, the subject of a panel
discussion~[\ref{chiextrap-lat2002}] at the Lattice 2002 conference. It
is difficult to simulate realistically light (valence or sea) quarks,
so that calculations of $\fB q$, say, are made for a a set of
(valence) quark masses $m_q$, typically in a range from about $m_s/2$
to $2 m_s$ and the results are interpolated or extrapolated as
required. Likewise, in partially quenched calculations, results from
simulations with a range of sea quark masses need to be extrapolated.
The control of these extrapolations is a serious issue for UT fits
because of their effect on the final values of $\fB d$ and $\fB s$ and
hence on the impact of the $\Delta M_s/\Delta M_d$ constraint. As far
back as late 1994 Booth noted the striking difference between the
quenched and QCD chiral logarithms~[\ref{booth-qchpt-hl-mesons}] and
posted a warning that $\fB s/\fB d$ in QCD would be larger than in the
quenched approximation. Recently, this issue has attracted much more
attention~[\ref{kronfeld-ryan-2002}--\ref{bfpz-2002}].

Consider an idealized lattice calculation of the decay constant of a
heavy-light pseudoscalar meson with valence content $Q\bar q$, where
$Q$ is the heavy quark and $\bar q$ a light quark. Imagine that the
simulation is performed either with or without the presence of $N_f$
flavours of (degenerate) sea quarks $f$ and let $\Delta \fB q$ be the
correction to $\fB q$ depending on the mass(es) of the valence ($q$)
and sea ($f$) quarks. With no sea quark effects included, the
calculation is quenched. When $m_q \neq m_f$ the calculation is
partially quenched and when $m_q=m_f$ it is QCD(-like). The
dependence of $\Delta \fB q$ on the valence and sea quark masses can
be calculated in quenched (Q), partially quenched (PQ) or ordinary
chiral perturbation theory, and shows up as dependence on the masses
$m_{qq}$, $m_{qf}$ and $m_{ff}$ of pseudoscalar mesons made from the
corresponding quarks~[\ref{sharpe-zhang-hl-qchpt-1996}]. The
expressions are as follows
\begin{eqnarray}
(\Delta \fB q)^\mathrm{QQCD} &=&
 {1\over(4\pi f)^2} ( X m_{qq}^2 + Y m_0^2 )
 \ln\Big( {m_{qq}^2\over\Lambda^2} \Big) \label{eq:DF1}\\
(\Delta \fB q)^\mathrm{PQQCD} &=&
 -{(1+3g^2)\over(4\pi f)^2}
  \left[
    {N_f\over2}m_{qf}^2\ln\Big( {m_{qf}^2\over\Lambda^2} \Big)
    +
    {(m_{f\!f}^2-2m_{qq}^2)\over2N_f}
    \ln\Big( {m_{qq}^2\over\Lambda^2} \Big)
  \right] \label{eq:DF2}\\
(\Delta \fB q)^\mathrm{QCD} &=&
 -{(1+3g^2)\over(4\pi f)^2} \Big({N_f\over2}-{1\over2N_f}\Big) m_{qq}^2
 \ln\Big( {m_{qq}^2\over\Lambda^2} \Big) \label{eq:DF3}
\end{eqnarray}
with $m_{qf}^2 = (m_{qq}^2 + m_{f\!f}^2)/2$ (at this order of
calculation). In the factor $1/(4\pi f)^2$, $f$ is equal to the common
light pseudoscalar meson decay constant at leading order, while $X$,
$Y$ and $m_0$ are also built from coefficients of the effective
Lagrangian. The dependence on the ultraviolet cutoff $\Lambda$ is
canceled by that of `analytic terms' not shown here. The coupling $g$
comes from the leading interaction term in the heavy meson chiral
Lagrangian (see the textbook by Manohar and Wise~[\ref{mw-hqp-2000}]
for details and original references) and fixes the $\rm B^*B\pi$ coupling
in the limit $M_B\to\infty$ by
\begin{equation}
g_{ B^*B\pi} = {2\, g \,M_B\over f}
\end{equation}
where
\begin{equation}
\langle {\rm B}^+(p)\pi^-(q) | 
{\rm B}^*(\epsilon,p') \rangle = g_{ B^* B\pi}
\epsilon\mathord{\cdot}q.
\end{equation}
The decay $\rm B^*\to B\pi$ is not kinematically allowed, but $g$ can be
estimated using CLEO results~[\ref{cleo-gamma-dstar-2001}] for $\rm D^*\to
D\pi$, or from a lattice QCD calculation of the matrix element of the
light-quark axial current between B and ${\rm B}^*$
mesons~[\ref{ukqcd-bstarbpi}] (or D and
$\rm D^*$~[\ref{abada-etal-gdstardpi-2002}]). The CLEO results lead to $g
=0.6$, consistent with the recent lattice
calculation~[\ref{abada-etal-gdstardpi-2002}].

The expressions in Eqs. (\ref{eq:DF1}), (\ref{eq:DF2}) and
(\ref{eq:DF3}) show that both the quenched and partially quenched
`chiral logarithms' diverge as the valence quark mass and hence
$m_{qq}$ vanishes while the sea quark mass is held fixed. In contrast,
there are no divergences when the sea quark masses vanish with the
valence masses held fixed. For the QCD-like case, things also remain
finite as the joint valence and sea quark mass vanishes. The problem
for lattice practitioners is how best to perform the chiral
extrapolations from results calculated with sets of $m_q$ and $m_f$
values, particularly since it is very difficult to make the masses
small enough to see the logarithmic dependence.

For $\fB d$ the situation is like the `QCD' case above where the
valence $d$ quark in the ${\rm B}_d$ meson and (some of) the sea quarks are
very light. For $\fB s$, the valence mass is fixed at $m_s$ and the
sea quark masses are extrapolated to small values (more like the
partially quenched case above). The JLQCD collaboration
find~[\ref{jlqcd-chiextrap-lat2002}] that these different
extrapolations tend to decrease the value of $\fB d$ relative to $\fB
s$, and therefore increase $\xi$. However, a number of caveats must be
kept in mind~[\ref{lellouch-lattphen-ichep2002}]. Although the data is
consistent with the chiral logarithmic forms, all the data points are
at masses beyond the region of strong variation in the logarithms.
Moreover, at these larger masses, higher order terms in the chiral
expansion may be required. Furthermore, in dynamical simulations the
lattice spacing changes as the sea quark mass changes at fixed lattice
coupling ($\beta$), so that care is needed not to interpret
lattice-spacing (and volume) dependence as sea-quark mass dependence.
An added twist is that JLQCD find that their results for $F_\pi$ are
\emph{not} consistent with the expected logarithmic behaviour.

The MILC collaboration have also estimated chiral logarithm effects as
part of their extensive analysis of $N_f=2$ simulations for
heavy-light decay constants~[\ref{milc-fnal-stat-nf2-2002}]. Their
method is based on extrapolation of the ratio of the light-light to
the heavy-light decay constant, where the chiral logarithmic terms
cancel to a large extent. MILC's conclusion is that these effects do
tend to increase the value of the ratio $\fB s/\fB d$ and MILC ascribe
a systematic error of $+0.04$ from chiral logarithms to a central
value of $1.16$ for $\fB s/\fB d$.

Kronfeld and Ryan (KR)~[\ref{kronfeld-ryan-2002}] consider the ratios
$\xi_f = \fB s/\fB q$ and $\xi_B = B_{B_s}/B_{B_q}$ as the mass of the
quark $q$ varies from the strange mass down to that of the light
quarks $u$ and $d$ and match ChPT to lattice data for $m_q$ not too
far from $m_s$. Their analysis gives $\xi=1.32(10)$. Another more
recent phenomenological analysis (BFPZ)~[\ref{bfpz-2002}] supports the
increase in $\xi$ coming from chiral logarithms and leads to a
consistent result $\xi=1.22(7)$. This value is extracted using the
double ratio
\begin{equation}
R = {(\fB s \sqrt{M_{B_s}})/(\fB d \sqrt{M_{B_d}})\over F_K/F_\pi}.
\end{equation}
An expression for $R$ in leading order heavy meson and pion chiral
perturbation theory (in full, $3$-flavour QCD) is combined with the
experimental ratio $(F_K/F_\pi)_\mathrm{expt} = 1.22(1)$ to extract
$\fB s/\fB d$. Systematic error in both analyses arises from the
uncertain values of parameters in the heavy meson and pion chiral
Lagrangian, namely the coupling $g$ in the leading interaction term,
already encountered above, together with sums of coefficients of
higher-order terms in the heavy meson chiral Lagrangian. In addition
the analysis using $R$ depends on $L_5$, the coefficient of a
higher-order term in the pion chiral Lagrangian through its use of the
ratio $F_K/F_\pi$.

In conclusion, lattice results for $F_B$ can show significant
light-quark mass dependence and more work is needed to understand to
what extent this dependence is physical. At present a reasonable
conservative view~[\ref{lellouch-lattphen-ichep2002}] is to allow a
\emph{decrease} of up to $-10\%$ in $\fB d$ with a negligible change
in $\fB s$ as added systematic errors. These are included in the final
estimates presented in Eq.~(\ref{eq:latt-final}).

\begin{figure}
\hbox to\hsize{%
\includegraphics[height=0.37\hsize]{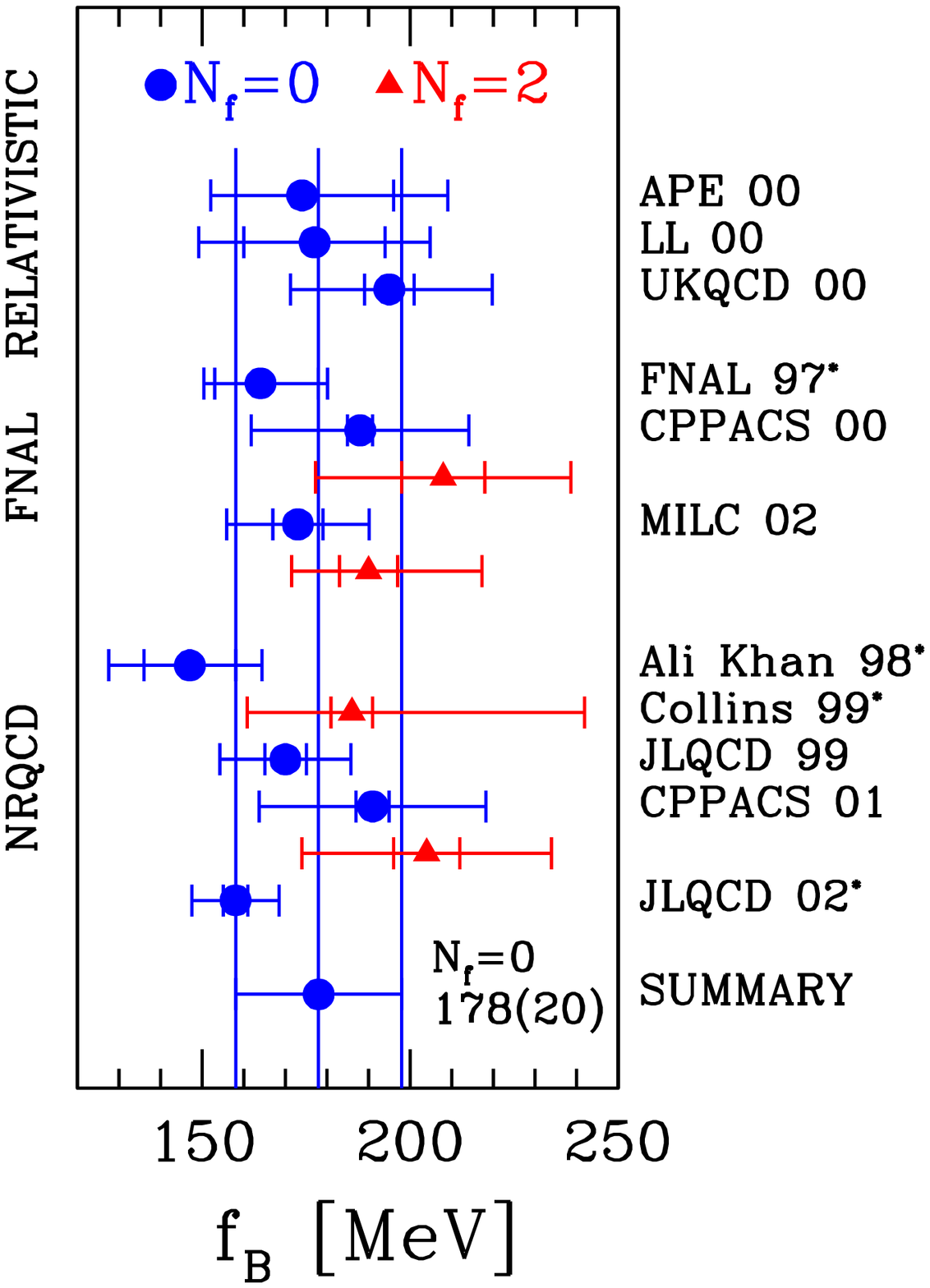}
\includegraphics[height=0.37\hsize]{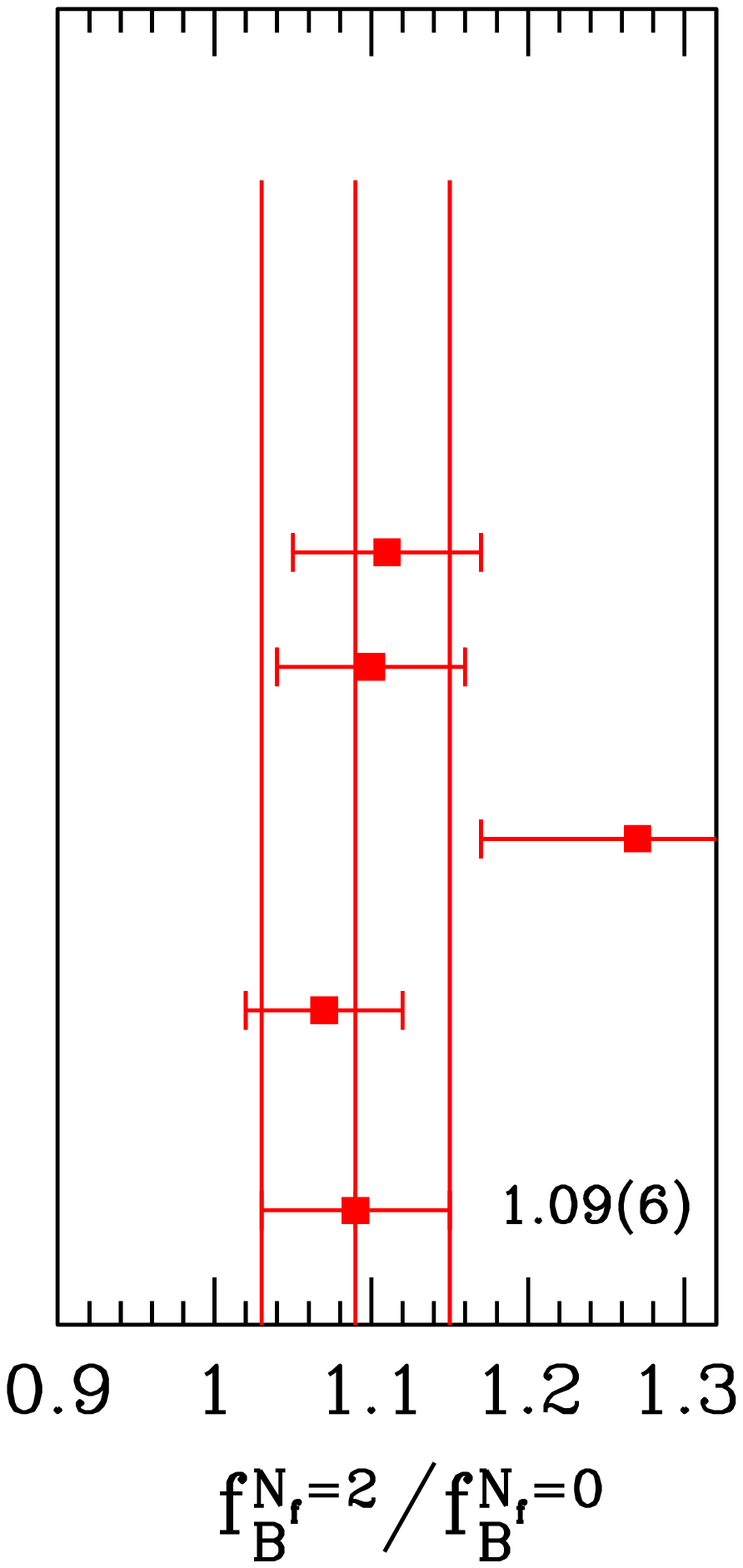}
\hss
\includegraphics[height=0.37\hsize]{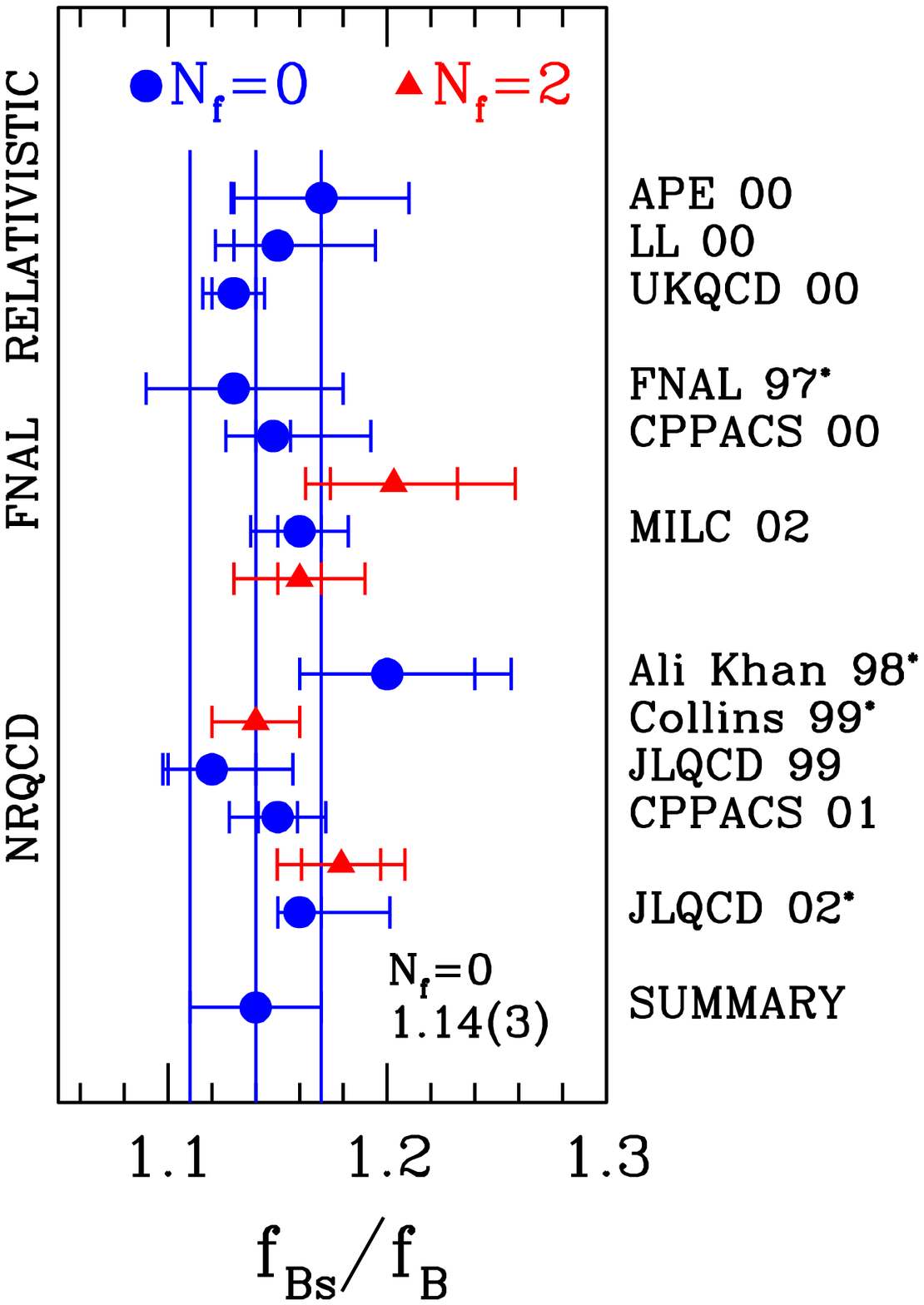}
\includegraphics[height=0.37\hsize]{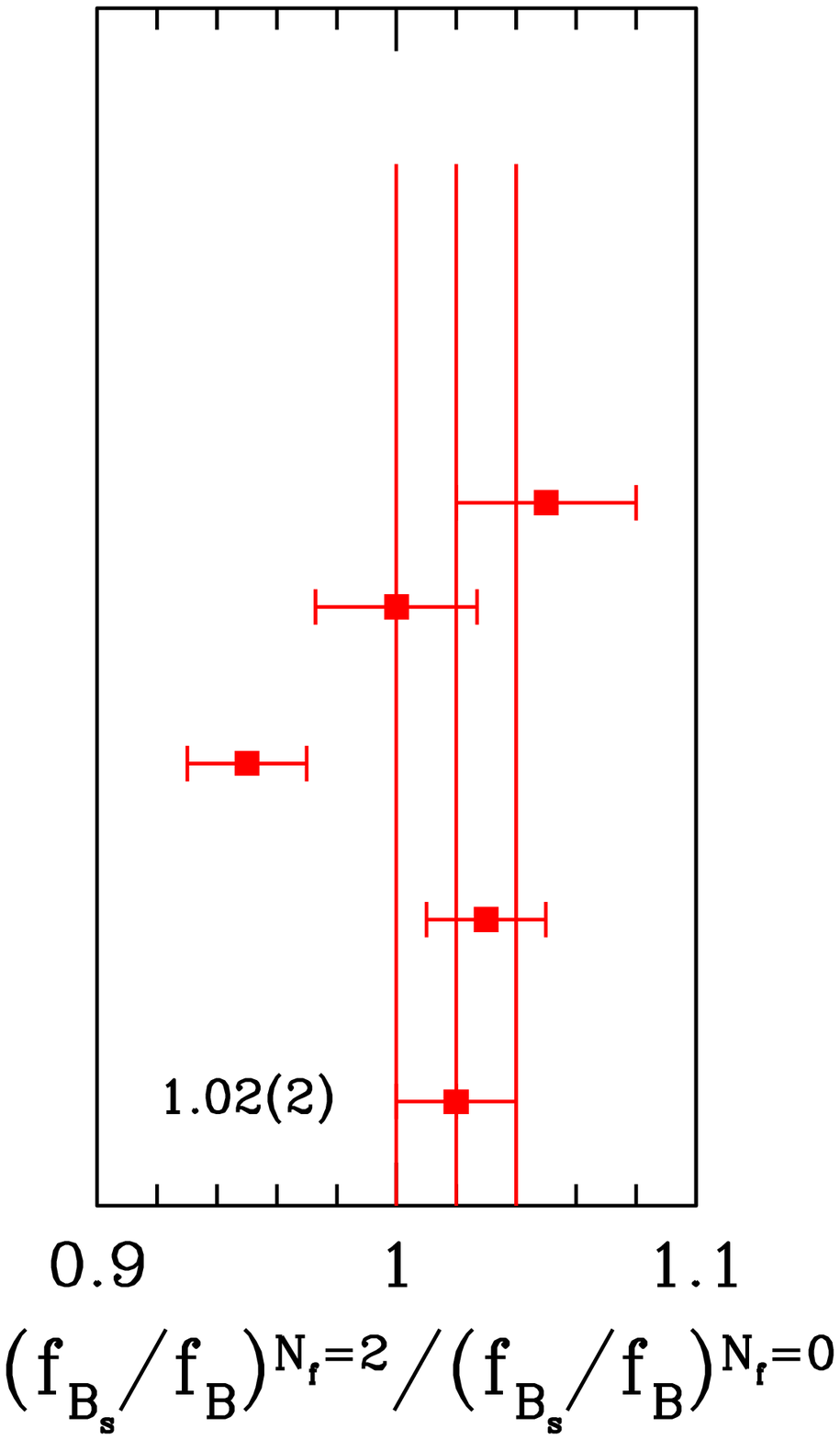}}
\caption{\it From left to right: lattice results published after 1996 for
(a) $\fB d$ in quenched ($N_f=0$) and two-flavour ($N_f=2$) QCD, (b)
the ratio $\fB d^{N_F=2}/\fB d^{N_f=0}$, (c) $\fB s/\fB d$ in quenched
($N_f=0$) and two-flavour ($N_f=2$) QCD, (d) the ratio $(\fB s/\fB
d)^{N_f=2}/(\fB s/\fB d)^{N_f=0}$. The results are grouped according
to the formulation used to treat the heavy quark and the references
are: APE 00~[\ref{beci-rel-quen-2001}], LL
00~[\ref{ukqcd-ll-rel-quen-2001}], UKQCD 00~[\ref{ukqcd-rel-quen-2001}],
FNAL 97~[\ref{fnal-fnal-quen-1998}], CPPACS
00~[\ref{cppacs-fnal-quen-nf2-2001}], MILC
02~[\ref{milc-fnal-stat-nf2-2002}], Ali Khan
98~[\ref{glok-nrqcd-quen-1998}], Collins 99~[\ref{glok-nrqcd-nf2-1999}],
JLQCD 99~[\ref{jlqcd-nrqcd-quen-2000}], CPPACS
01~[\ref{cppacs-nrqcd-nf2-2001}] and JLQCD
02~[\ref{yamada-hqreview-lat2002}]. Figs. taken
from~[\ref{lellouch-lattphen-ichep2002}].
\label{fig:fb-summary}}
\end{figure}

A summary of lattice calculations for the decay constants, published
after 1996, is given in Fig.~\ref{fig:fb-summary} (taken from the
review by Lellouch~[\ref{lellouch-lattphen-ichep2002}]), which shows
results for $\fB d$ and the ratio $\fB s/\fB d$. The `summary' numbers
at the bottom of the plots give quenched averages for $\fB d$ and $\fB
s/\fB d$, together with ratios of these quantities for $N_f=2$ and
$N_f=0$:
\begin{equation}
\label{eq:fbxi-latt-nf02}
\begin{array}{r@{\;}c@{\;}lr@{\;}c@{\;}l}
\fB d^{N_f=0} &=& 178(20)\mev &
{\fB d^{N_f=2}\over \fB d^{N_f=0}} &=& 1.09(6) \\[3.5ex]
(\fB s/\fB d)^{N_f=0} &=& 1.14(3) &
{(\fB s/\fB d)^{N_f=2}\over (\fB s/\fB d)^{N_f=0}} &=& 1.02(2)
\end{array}
\end{equation}

For the mixing parameter $B_{B_q}$, the situation with quenching and
chiral extrapolation looks more favourable. Very little variation is
observed between quenched ($N_f=0$) and $N_f=2$ results. The partially
quenched chiral logarithm for $B_{B_q}$ has a coefficient containing
$1-3g^2 \simeq -0.1$ compared to $1+3g^2 \simeq 2.1$ in the $\fB q$
case (using $g=0.6$ as discussed above) so the chiral extrapolation is
better-controlled and leads to a small error in $\bbhat s/\bbhat
d$~[\ref{kronfeld-ryan-2002},\ref{lellouch-lattphen-ichep2002},\ref{yamada-hqreview-lat2002}]. The heavy quark mass dependence is mild and
different formulations agree at the physical point for B-mesons.

There is, however, an issue concerning lattice results for $\xi$ which
are normally quoted by combining
results for $F_B$ and $\hat B_B$. Of course, it is also possible to
evaluate $\xi$ directly from the ratio of $\Delta B=2$ matrix
elements. In this case $\xi$ turns out to be larger, although with
large errors~[\ref{bbs-1998},\ref{ukqcd-ll-rel-quen-2001}]. Clearly the two
procedures should give consistent answers, so this issue will need to
be resolved.

\subsubsection{Summary on $F_{B_q}$ and $\xi$ from the lattice QCD}

Using the quenched averages as a starting point together with the
ratios of $N_f=2$ to $N_f=0$ results allows an extrapolation to
$N_f=3$~[\ref{lellouch-lattphen-ichep2002}]. An additional systematic
error equal to the shift from $2$ to $3$ flavours is added to account
for the uncertainty in this procedure\footnote{An alternative way to
quote the final answer would be to use the $N_f=2$ results extracted
from Eq.~(\ref{eq:fbxi-latt-nf02}) and add a systematic error for
the extrapolation to $N_f=3$. In this case, the final central value
for $\fB s/\fB d$ would be $1.16$. The value of $1.18$, however, is
consistent with the latest preliminary MILC results for $N_f=3$, which
give $(\fB s/\fB d)^{N_f=3}=1.18(1)\errp41$}. This leads to:
\begin{equation}
\begin{array}{|c|}\hline\left.
\begin{array}{r@{\;}c@{\;}lr@{\;}c@{\;}lr@{\;}c@{\;}l}
\fB d & = & 203(27)\errp0{20} \mev &
\fB s & = & 238(31) \mev &
\displaystyle {\fB s\over\fB d} & = & 1.18(4)\errp{12}0 \\
\bbhat d & = & 1.34(12) &
\bbhat s & = & 1.34(12) &
\displaystyle {\bbhat s\over\bbhat d} & = & 1.00(3) \\
\fB d\sqrt{\bbhat d} & = & 235(33)\errp0{24} \mev &
\fB s\sqrt{\bbhat s} & = & 276(38)\mev &
\xi & = & 1.18(4)\errp{12}0
\end{array}   \ .\right.
\\ \hline\end{array}
\label{eq:latt-final}
\end{equation}

Here, the last, asymmetric, error, where present, is due the
uncertainty in the chiral extrapolation discussed above. The first
error combines statistical and all other systematic errors. In UT
analyses, the value of $\xi$ given above should be understood as
\begin{equation}
\label{eq:xi-latt-final}
\xi=1.24(4)(6)
\end{equation}
and likewise for other quantities affected by this asymmetric error.
Note that this does not apply for $\fB s$ and $\bbhat s$, for which
the chiral logarithmic uncertainties appear small compared to other
systematic errors. The result for $\xi$ in
Eq.~(\ref{eq:xi-latt-final}) is consistent with the
KR~[\ref{kronfeld-ryan-2002}] and BFPZ~[\ref{bfpz-2002}] analyses
mentioned above.

\subsubsection{$\fB d$ and $\fB s$ from QCD sum rules}

Within the framework of QCD sum rules~[\ref{svz:79},\ref{svz:79b}], the decay
constants $\fB d$ and $\fB s$ can be calculated by equating
phenomenological and theoretical spectral functions for the
pseudoscalar ${\rm B}_d$ and ${\rm B}_s$ mesons, which leads to the relation
[\ref{ps:01}--\ref{nar:01}]\footnote{A review of the procedure and
further original references can be found in [\ref{ck:00}].}
\begin{equation}
\label{eq:fbsr}
M_B^4 \fB d^2 = \int\limits_0^{s_0} e^{(M_B^2-s)/u}\rho(s)\,ds
\end{equation}
for the ${\rm B}_d$ meson and analogously for ${\rm B}_s$. Eq.~(\ref{eq:fbsr})
is the central relation for the sum rule analysis. The theoretical spectral 
function $\rho(s)\equiv\Im\Psi(s)/\pi$ can be obtained by calculating
the two-point correlator of hadronic currents
\begin{equation}
\label{Psip2}
\Psi(p^2) \equiv i \int \! dx \, e^{ipx} \,
\big<0\vert\,T\{\,j_5(x)\,j_5(0)^\dagger\}\vert 0\big>
\end{equation}
in perturbative QCD, including corrections from the operator product
expansion. For the B meson, the pseudoscalar current $j_5(x)$ takes
the form
\begin{equation}
\label{j5ub}
j_5(x) = (m_b+m_u):\!\bar u(x)\,i\gamma_5 b(x)\!: \,.
\end{equation}
The parameter $s_0$ in Eq.~(\ref{eq:fbsr}) indicates the energy range
up to which experimental knowledge of the phenomenological spectral
function is available. This parameter will be further discussed below.

Substantial progress in determining the theoretical spectral function
has been achieved very recently through a calculation of the
perturbative three-loop order $\as^2$ corrections
[\ref{cs:01a},\ref{cs:01b}]. These are important because the size of
higher-order corrections depends on the renormalization scheme
employed for the quark masses. As can be inferred from
refs.~[\ref{cs:01a},\ref{cs:01b}], the $\as^2$ term turns out to be of
similar order to the leading contribution if pole quark masses are
used, whereas good convergence of the perturbative series emerges for
quark masses defined in the $\overline{\rm MS}$ scheme. Nevertheless,
these scheme dependences influence only the theoretical uncertainties,
since $\fB d$ and $\fB s$ are physical quantities which certainly should
not depend on the quark mass definitions. Higher-dimensional operator
corrections to the sum rule are known up to dimension six~[\ref{jl:01}]
and are also under good theoretical control.

\begin{figure}
\begin{center}
\includegraphics[angle=270, width=0.75\hsize]{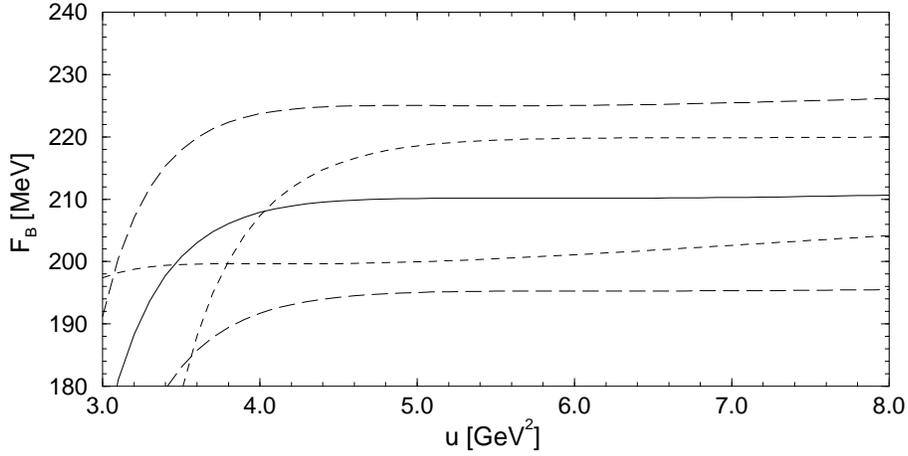}
\end{center}
\caption{\it \label{figfb}$\fB d$ as a function of the sum rule scale $u$
for different sets of input parameters. Solid line: central values of
Table~\ref{tabfb}; long-dashed line: $m_b(m_b)=4.16\;\gev$ (upper
line), $m_b(m_b)=4.26\;\gev$ (lower line); dashed line:
$\mu_m=3\;\gev$ (lower line), $\mu_m = 6\;\gev$ (upper line).}
\end{figure}
Figure~\ref{figfb} shows numerical results for $\fB d$ of Ref.~[\ref{jl:01}],
plotted as a function of the sum rule scale $u$, after evaluating the sum
rule of Eq.~(\ref{eq:fbsr}). Reliable values of $F_{B_q}$ can be extracted from
the sum rule if an energy region exists in which the physical quantity
is only weakly dependent on $u$. For $F_{B_d}$ this turns out to be the
case in the range $4\;\gev^2 \lsim u \lsim 6\;\gev^2$. Averaging
the results of refs.~[\ref{ps:01},\ref{jl:01}] in this energy range, one
extracts the central results $\fB d=208\;\mev$ and
$\fB s=242\;\mev$.\footnote{Owing to the criticism put forward in
Ref.~[\ref{jl:01}], the result of Ref.~[\ref{nar:01}] has not been
included in the average, despite the apparent agreement for the
numerical values.}

The dominant uncertainties in the sum rule determination of $\fB d$ and
$\fB s$ arise from the strong dependence on the value of the bottom
quark mass $m_b$ and correspondingly on the scale $\mu_m$ at which the
quark masses are renormalized. The ranges for the variation of these
parameters and the corresponding variations of $\fB d$ and $\fB s$ have
been collected in Tables~\ref{tabfb} and~\ref{tabfbs} respectively. The
reader should note that compared to Ref.~[\ref{jl:01}], the error on
$m_b(m_b)$ has been enlarged, in order to coincide with the value
employed throughout this report, although the larger uncertainty should
be considered very conservative. The Tables also list the values $u_0$ at
which the sum rule displays optimal stability, as well as the parameters
$s_0$ which can be determined consistently from an independent sum rule
for the ${\rm B}_d$ and ${\rm B}_s$ meson masses. 
Additional smaller uncertainties
are due to: variation of the strong coupling constant $\as$; higher order
QCD corrections; the value of the quark condensate $\uu$~[\ref{jam:02}]
which is the leading contribution from higher-dimensional operators;
the strange condensate $\langle\bar ss\rangle$ and the strange quark
mass $m_s$ in the case of $\fB s$. Ranges for these inputs together
with the variations of $\fB d$ and $\fB s$ are also collected in
Tables~\ref{tabfb} and~\ref{tabfbs}. For further details of the
numerical analysis, the reader is referred to Ref.~[\ref{jl:01}].
\begin{table}
\begin{center}
\def\arraycolsep{0.7em}
\def\arraystretch{1.1}
\begin{tabular}{|ccccc|}
\hline
Parameter & Value & $s_0$ & $u_0$ & $\Delta \fB d$ \\[0.5ex]
\hline
$m_b(m_b)$ & $4.21 \pm 0.08\;\gev$ & ${}^{32.8}_{34.6}$ & ${}^{6.5}_{5.0}$ &
$\mp 24$ \\
$\mu_m$ & $3.0-6.0\;\gev$ & ${}^{33.5}_{34.4}$ & ${}^{6.8}_{4.0}$ & $\pm10$ \\
$\uu(2\;\gev)$ & $-\,(267\pm17\;\mev)^3$ & ${}^{33.9}_{33.3}$ & ${}^{5.7}_{5.5}$
 & $\pm 6$ \\
$\mathcal{O}(\alpha_\mathrm{s}^2)$ & ${}^{2\times\mathcal{O}(\alpha_\mathrm{s}^2)}_{{\rm no}\;
\mathcal{O}(\alpha_\mathrm{s}^2)}$ & $33.6$ & $5.6$ & $\pm 2$ \\
$\alpha_\mathrm{s}(M_Z)$ & $0.1185 \pm 0.020$ & $33.6$ & $5.6$ & $\pm 1$ \\
\hline
\end{tabular}
\end{center}
\caption{\it Values for the dominant input parameters, continuum thresholds
$s_0\;[\gev^2]$, points of maximal stability $u_0\;[\gev^2]$, and corresponding
uncertainties for $\fB d\;[\mev]$.\label{tabfb}}
\end{table}

\begin{table}
\begin{center}
\def\arraycolsep{0.7em}
\def\arraystretch{1.1}
\begin{tabular}{|ccccc|}
\hline
Parameter & Value & $s_0$ & $u_0$ & $\Delta \fB s$ \\[0.5ex]
\hline
$m_b(m_b)$ & $4.21 \pm 0.08\;\gev$ & ${}^{34.3}_{36.9}$ & ${}^{5.8}_{4.6}$ &
$\mp 26$ \\
$\mu_m$ & $3.0-6.0\;\gev$ & ${}^{35.2}_{37.2}$ & ${}^{6.2}_{3.6}$ &
 ${}^{+8}_{-9}$ \\
$\langle\bar ss\rangle/\uu$ & $0.8 \pm 0.3$ & ${}^{35.9}_{35.2}$ &
${}^{5.3}_{4.7}$ & $\pm 8$ \\
$\uu(2\;\gev)$ & $-\,(267\pm17\;\mev)^3$ & ${}^{35.7}_{35.3}$ & ${}^{5.2}_{4.9}$
 & ${}^{+5}_{-4}$ \\
$m_s(2\;\gev)$ & $100 \pm 15\;\mev$ & $35.5$ & $5.1$ & $\pm 2$ \\
$\mathcal{O}(\alpha_\mathrm{s}^2)$ & ${}^{2\times\mathcal{O}(\alpha_\mathrm{s}^2)}_{{\rm no}\;
\mathcal{O}(\alpha_\mathrm{s}^2)}$ & $35.5$ & $5.1$ & $\pm 3$ \\
$\alpha_\mathrm{s}(M_Z)$ & $0.1185 \pm 0.020$ & $35.5$ & $5.1$ & $\pm 1$ \\
\hline
\end{tabular}
\end{center}
\caption{\it Values for the dominant input parameters, continuum thresholds
$s_0\;[\gev^2]$, points of maximal stability $u_0\;[\gev^2]$, and corresponding
uncertainties for $\fB s\;[\mev]$.\label{tabfbs}}
\end{table}

Adding all errors for the various input parameters in quadrature, the
final results for the ${\rm B}_d$ and ${\rm B}_s$ meson leptonic decay constants
from QCD sum rules are:
\begin{equation}
\label{eq:sumrulefbfbs}
\fB d = 208 \pm 27 ~\mev
\quad\qquad \mbox{and} \quad\qquad
\fB s = 242 \pm 29 ~\mev.
\end{equation}
Owing to the strong sensitivity of these results on the bottom quark
mass, one should note that for example using the very recent average
$m_b(m_b)=4.24~\gev$~[\ref{kl:02}], the resulting values for $\fB d$ and
$\fB s$ are lowered by almost $10~\mev$.

\subsubsection{${ B}_{B_d}$ and $B_{B_s}$ from QCD sum rules}

The status of the determination of the hadronic $B$-parameters
$B_{B_d}$ and $B_{B_s}$ from QCD sum rules is less satisfactory than
for the decay constants. In principle, the $B$-parameters can be
calculated from two different types of sum rules: namely three-point
function sum rules with the insertion of two pseudoscalar currents and
one four-quark operator~[\ref{op:88},\ref{ry:88}], or two-point function sum
rules with the insertion of two local four-quark
operators~[\ref{np:94},\ref{hnn:02}]. However, both approaches are plagued
with difficulties~\footnote{For a different approach see also 
Ref.~[\ref{Chernyak:1994cx}].}.

The first determinations of the hadronic
$B$-parameters~[\ref{op:88},\ref{ry:88}] employed three-point function sum
rules and found a value of $B_{B_d}(m_b)=0.95 \pm 0.10$, slightly
lower than the factorization approximation which results in $B_{B_d} =
1$. The dominant non-factorizable contribution due to the gluon
condensate turned out to be negative, thus lowering the $B$-parameter.
However, the perturbative part was only considered at the leading
order, and thus the scale and scheme dependences of $B_{B_d}$ were not
under control. Besides, the analytic structure of three-point function
sum rules is more delicate than for two-point correlators, and
therefore great care has to be taken to properly extract the quantity
in question~[\ref{ck:00}].

For the case of the two-point function sum rules, next-to-leading
order QCD corrections have been calculated in Ref.~[\ref{np:94}], which
provides better control over the renormalization dependence of $B_B$.
This analysis resulted in $B_{B_d}(m_b)=1.0 \pm 0.15$. However, here
the phenomenological parametrization of the spectral function is more
complicated, since contributions from intermediate states containing
B$^*$ mesons have to be taken into account in addition to the B
meson. Steps in this direction have recently been taken in
Ref.~[\ref{hnn:02}] were the value $B_{B_d}(m_b)=1.15 \pm 0.11$ was
obtained, now indicating a positive correction.

Although averaging the results of the two approaches might appear problematic,
we nevertheless decided to quote a common value for the B meson
$B$-parameter from QCD sum rules:
\begin{equation}
\label{BBav}
B_{B_d}(m_b) = 1.10 \pm 0.15
\quad\qquad \mbox{and} \quad\qquad
\bbhat{d}= 1.67 \pm 0.23 ,
\end{equation}
which covers the outcome of both methods within the uncertainties. On the
other hand, general agreement exists for the flavour dependence of the
$B$-parameter. In all present sum rule approaches it was found to be
negligible, thus yielding $B_{B_s}/B_{B_d} = 1$ to a good approximation.

\boldmath
\subsection{$\rm K^0$--${ \overline{ \rm K}}^0$ mixing: determination of $B_K$}
\label{sec:KKbar-mixing}
\unboldmath
\subsubsection{$B_K$ from lattice QCD}

\begin{figure}
\hbox to \hsize{\begin{small}
\hss\setlength{\unitlength}{0.0026\hsize}
\begin{picture}(288,101)
\put(0,0){\includegraphics[width=0.75\hsize]{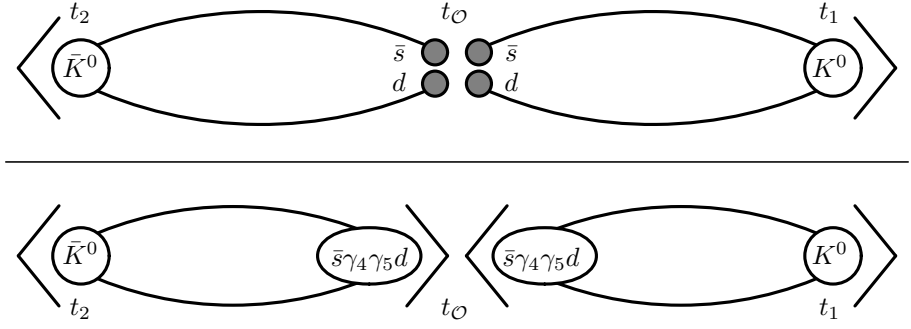}}
\point 24 21 {\bar K^0}
\point 24 81 {\bar K^0}
\point 264 21 {K^0}
\point 264 81 {K^0}
\point 24 98 {t_2}
\point 264 98 {t_1}
\point 144 98 {t_{\mathcal{O}}}
\point 24 3 {t_2}
\point 264 3 {t_1}
\point 144 3 {t_{\mathcal{O}}}
\point 116.5 19 {\bar s \gamma_4 \gamma_5 d}
\point 171.5 19 {\bar s \gamma_4 \gamma_5 d}
\point 126 85 {\bar s}
\point 162 85 {\bar s}
\point 126 76 d
\point 162 76 d
\end{picture}
\hss\end{small}}
\caption{\it Ratio of lattice correlation functions used to calculate
$B_K$.}
\label{fig:bkfig}
\end{figure}

The most commonly used method to calculate the matrix element $\langle
\overline{\rm  K}^0 \mathbin\vert Z\ (\bar s d)_{V-A} (\bar s d)_{V-A}(\mu)
\mathbin\vert {\rm K^0} \rangle$ is to evaluate the three point
correlation function shown in Fig.~\ref{fig:bkfig}. This corresponds
to creating a ${\rm K}^0$ at some time $t_1$ using a zero-momentum
source; allowing it to propagate for time $t_{\cal O}-t_1$ to isolate
the lowest state; inserting the four-fermion operator at time $t_{\cal
O}$ to convert the K$^0$ to a $\overline{\rm K}^0$; and finally allowing
the $\overline{\rm K}^0$ to propagate for long time $t_2 - t_{\cal O}$. To
cancel the K$^0$ ($\overline{\rm K}^0$) source normalization at times
$t_1$ and $t_2$ and the time evolution factors $e^{-E_K t}$ for times
$t_2 - t_{\cal O}$ and $t_{\cal O}-t_1$ it is customary to divide this
three-point function by the product of two 2-point functions as shown
in Fig 1. If, in the 2-point functions, the bilinear operator used to
annihilate (create) the ${\rm K^0}$ ($\overline{\rm K}^0$) at time $t_{\cal
O}$ is the axial density $\bar s \gamma_4 \gamma_5 d$, then the ratio
of the 3-point correlation function to the two 2-point functions is
$(8/3) B_K$.

$B_K$ is defined to be the value of the matrix element at the physical
kaon and normalized by the Vacuum Saturation Approximation value $ 8/3
M_K^2 F_K^2$
\begin{eqnarray*}
\langle {\rm K^0} \mathbin\vert Z\ (\bar s d)_{V-A} (\bar s d)_{V-A}(\mu)
                               \mathbin\vert \overline{\rm  K}^0   \rangle
                 &=&  {(8/3) B_K M_K^2 F_K^2} \,.
\end{eqnarray*}
The earliest calculations of $B_K$ were done using Wilson fermions and
showed significant deviations from this behaviour. It was soon
recognized that these lattice artifacts are due to the explicit
breaking of chiral symmetry in the Wilson formulation
[\ref{Cabibbo:Bk:PRL1984}--\ref{Gavela:Bk:1988bd}]. 
Until 1998, the only formulation that preserved sufficient chiral
symmetry to give the right chiral behaviour was Staggered fermions.
First calculations using this approach in 1989 gave the quenched
estimate $B_K (\mathrm{NDR}, 2\gev) = 0.70 \pm 0.01 \pm 0.03$. In
hindsight, the error estimates were highly optimistic, however, the
central value was only $10\%$ off the current best estimate, and most
of this difference was due to the unresolved $O(a^2)$ discretization
errors.

\begin{figure}
\hbox to \hsize{\hss
\includegraphics[width=0.6\hsize]{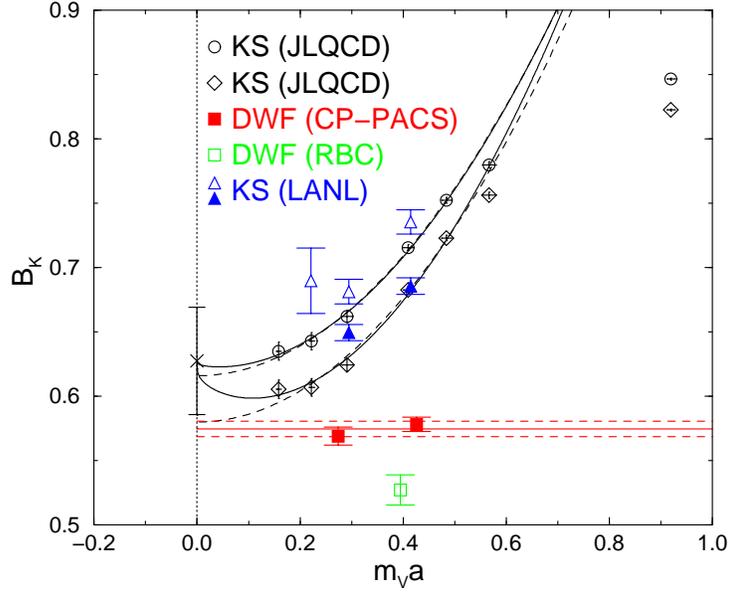}
\hss}
\caption{\it Published estimates of $B_K$ with fermion formulations that
respect chiral symmetry. All results are in the quenched approximation.}
\label{fig:bkestimates}
\end{figure}

In 1997, the staggered collaboration refined its calculation and
obtained $0.62(2)(2)$~[\ref{Kilcup:Bk:PRD1998}], again the error
estimate was optimistic as a number of systematic effects were not
fully included. The state-of-the-art quenched calculation using
Staggered fermions was done by the JLQCD collaboration in 1997 and
gave $B_K (2\gev) = 0.63 \pm 0.04$~[\ref{Aoki:Bk:1998nr}]. This
estimate was obtained using six values of the lattice spacing between
$0.15$ and $0.04$ fermi, thus allowing much better control over the
continuum extrapolation as shown in Fig.~\ref{fig:bkestimates} along
with other published results. This is still the benchmark against
which all results are evaluated and is the value exported to
phenomenologists. This result has three limitations: (i) It is in the
quenched approximation. (ii) All quenched calculations use kaons
composed of two quarks of roughly half the ``strange'' quark mass and
the final value is obtained by interpolation to a kaon made up of
$(m_s/2, m_s/2)$ instead of the physical point $(m_s, m_d)$. Thus,
SU(3) breaking effects ($m_s \neq m_d$) have not been incorporated.
(iii) There are large $O(a^2)$ discretization artifacts, both for a
given transcription of the $\Delta S=2$ operator on the lattice and
for different transcriptions at a given value of the lattice spacing,
so extrapolation to the continuum limit is not as robust as one would
like. These limitations are discussed after a brief summary of the
recent work.

\begin{table}
\begin{center}
 \begin{tabular}{|lccccc|c|}
\hline
\vrule height2.4ex width0pt depth0pt
Collaboration & year & $B_K(2\gev)$ & Formulation &
Renormalization & $a^{-1}$ ($\gev$) \\
\hline               
Staggered~[\ref{Kilcup:Bk:PRD1998}]  & 1997 & 0.62(2)(2)   & 
staggered       & 1-loop       &  $\infty$  \\[0.7ex] %
JLQCD~[\ref{Aoki:Bk:1998nr}]         & 1997 & 0.63(4)      & 
staggered       & 1-loop       &  $\infty$  \\[0.7ex]
Rome~[\ref{Becirevic:Bk:2002mm}]     & 2002 & 0.63(10)     & 
Improved Wilson & NP           & $\infty$    \\
Rome~[\ref{Becirevic:Bk:2002mm}]     & 2002 & 0.70(12)     & 
Improved Wilson & NP           & $\infty$    \\[0.7ex]
CP-PACS~[\ref{AliKhan:Bk:2001wr}]    & 2001 & 0.58(1)      & 
Domain Wall     & 1-loop       & $1.8$ GeV   \\
CP-PACS~[\ref{AliKhan:Bk:2001wr}]    & 2001 & 0.57(1)      & 
Domain Wall     & 1-loop       & $2.8$ GeV   \\
RBC~[\ref{Blum:Bk:2001xb}]           & 2002 & 0.53(1)      & 
Domain Wall     & NP           & $1.9$ GeV   \\[0.7ex]
DeGrand~[\ref{DeGrand:Bk:2002xe}]    & 2002 & 0.66(3)      & 
Overlap         & 1-loop       & $1.6$ GeV   \\[0.7ex] %
DeGrand~[\ref{DeGrand:Bk:2002xe}]    & 2002 & 0.66(4)      & 
Overlap         & 1-loop       & $2.2$ GeV   \\[0.7ex] %
GGHLR~[\ref{GGHLR:Bk:2002}]          & 2002 & 0.61(7)      & 
Overlap         & NP           & $2.1$ GeV  \\
\hline    
\end{tabular}
\end{center}
\caption{\it Quenched estimates for $B_K$ evaluated in the NDR scheme at
$2\gev$. The fermion formulation used in the calculation, the method
used for renormalizing the operators, and the lattice scale at which
the calculation was done are also given. NP indicates non-perturbative
renormalization using the RI/MOM scheme and $a^{-1}=\infty$ implies
that the quoted result is after a continuum extrapolation.}
\label{tab:lattices}
\end{table}
In the last four years a number of new methods have been developed and
the corresponding results are summarized in Table~\ref{tab:lattices}.
\begin{itemize}
\item
The Rome collaboration has shown that the correct chiral behaviour can
be obtained using $O(a)$ improved Wilson fermions provided
non-perturbative renormalization constants are used. Their latest
results, with two different ``operators'', are $B_K(2{\rm GeV}) =
0.63(10)$ and $0.70(12)$~[\ref{Becirevic:Bk:2002mm}]. These, while
demonstrating the efficacy of this method, do not supplant the
staggered result, as the continuum extrapolation is based on only
three points and the data have larger errors. The discretization
errors can be characterized as $B_K(a) = B_K(1 + a \Lambda$) with
$\Lambda \approx 400\mev$ and are similar in magnitude to those with
staggered fermions at $1/a=2$ GeV, as are the differences in estimates
with using different operators. In the staggered formulation, the
artifacts are, however, $O(a^2 \Lambda^2)$ and
$O(\alpha_\mathrm{s}^2)$ and the data suggest an unexpectedly large
$\Lambda \sim 900\mev$.
\item
Four collaborations have new results using domain wall and overlap
fermions as shown in Table~\ref{tab:lattices}
[\ref{Blum:Bk:1997mz},\ref{Blum:Bk:2001xb},\ref{AliKhan:Bk:2001wr},\ref{DeGrand:Bk:2002xe},\ref{GGHLR:Bk:2002}].
Both formulations have built in chiral symmetry at finite $a$ and
$O(a)$ improvement. Each of these collaborations have used slightly
different methodology, so they cannot be compared head on, or combined
to do a continuum extrapolation.  Thus, the results are quoted with
reference to the lattice spacing at which the calculation was done.
The differences reflect $O(a^2)$ (and $O(\alpha_s^2)$ in cases where
perturbative renormalization constants have been used) artifacts.
\item
Calculations are in progress~[\ref{twistedmass-Bk-lat2001}] using
another method with good chiral behaviour, twisted mass QCD.
\end{itemize}
Deriving an estimate for the physical $\hat B_K$, starting from the
current best quenched lattice estimate, the JLQCD staggered result
$B_K(2{\rm GeV})= 0.63(4)$, requires consideration of the following
issues.
\begin{itemize}
\item
The $O(a^2)$ errors in the staggered formulation are large.
Nevertheless, the error $0.04$ obtained by the JLQCD collaboration on
including both $O(a^2)$ and $O(\alpha_\mathrm{s}^2)$ terms in the
extrapolation is a reasonable $1\sigma$ estimate of both the
statistical and the extrapolation to continuum limit errors.
\item
A choice for $\alpha_\mathrm{s} $ and the number of flavours in the
perturbative expression has to be made to convert $B_K \to \hat B_K$.
It turns out that the result is insensitive to whether one uses
quenched or full QCD values. Using the 2-loop expression, the result
for the central value is $\hat B_K = 0.86(6)$.
\item
An estimate of the systematic uncertainty associated with the quenched
approximation and SU(3) breaking.  Preliminary numerical estimates
suggest that dynamical quarks would increase the value by about
$5\%$~[\ref{Ishizuka:Bk:1993ya},\ref{Kilcup:Bk:1993pa}]. Sharpe estimates,
using ChPT, that unquenching would increase $B_K$ by $1.05 \pm 0.15$,
and SU(3) breaking effects would also increase it by $1.05 \pm
0.05$~[\ref{sharpe-review-lat1996}].  This analysis of systematic
errors is not robust and, furthermore, the two uncertainties are not
totally independent. So one can take an aggressive and a conservative
approach when quoting the final result for $\hat B_K$.  In the
aggressive approach, the error estimate is given by combining in quadrature the
offset of the central values with respect to unity. This gives a $7\%$ uncertainty and
\begin{equation}
\hat B_K = 0.86 \pm 0.06 \pm 0.06 \,.
\end{equation}
In the conservative approach, advocated by
Sharpe~[\ref{sharpe-review-lat1996}], one combines the uncertainty in
quadrature to get a $16\%$ uncertainty. The final result in this case is 
\begin{equation}
\begin{array}{|c|}
\hline\left.
\hat B_K = 0.86 \pm 0.06 \pm 0.14 \,
\ \right.
\\ \hline\end{array}
\end{equation}

\end{itemize}

\vspace{2mm}

Given the lack of a robust determination of the systematic error, it
is important to decide how to fold these errors in a phenomenological
analysis. One recommendation is to assume a flat distribution for the
systematic error and add to it a Gaussian distribution with $\sigma =
0.06$ on either end, and do a separate analysis for the aggressive and
conservative estimates. In other words, a flat distribution between
$0.72$ and $1.0$ for a conservative estimate of $\hat B_K$ (or from
$0.80$ to $0.92$ for the aggressive estimate) to account for
systematic errors due to quenching and SU(3) breaking. Since this is
the largest uncertainty, current calculations are focused on reducing
it.

Finally, the reasons why the quenched lattice estimate of $B_K$ has
been stable over time and considered reliable within the error
estimates quoted above are worth reemphasizing:
\begin{itemize}
\item
The numerical signal is clean and accurate results are obtained with a 
statistical sample of even $50$ decorrelated lattices. 
\item
Finite size effects for quark masses $\ge m_s/2$ are insignificant
compared to statistical errors once the quenched lattices are larger
the $2$ fermi.
\item
In lattice formulations with chiral symmetry, the renormalization constant
connecting the lattice and continuum schemes is small ($< 15\%$), and
reasonably well estimated by one-loop perturbation theory.
\item
For degenerate quarks, the chiral expansion for the matrix element has
no singular quenched logarithms (they cancel between the $AA$ and $VV$
terms) that produce large artifacts at small quark masses in
observables like $M_\pi^2$, $f_\pi$, etc.  Also, the chiral expansions
have the same form in the quenched and full 
theories~[\ref{Bijnens:CPT:1984ec}--\ref{Golterman:QCL:1998st}].
\item
ChPT estimates of quenching and SU(3) breaking systematic errors are
at the $7$--$16\%$
level~[\ref{Sharpe:TASI:1994dc},\ref{Ishizuka:Bk:1993ya},\ref{Kilcup:Bk:1993pa}].
\end{itemize}

\subsubsection{$B_K$ from non-lattice approaches}

The parameter $B_K$ can also be calculated using other
non-perturbative approaches to QCD, like QCD sum rules, the
large-$N_c$ expansion or the chiral quark model. As for the parameter
$B_B$ in the ${\rm B}$-meson system, $B_K$ can be obtained from sum rules by
considering 
two-point~[\ref{praetal:91}--\ref{nar:95}] 
or three-point
[\ref{ry:87},\ref{bdg:88}] correlation functions. However, both methods
suffer from the same inadequacies as in the case of $B_B$. For the
two-point function sum rule, the phenomenological spectral function is
difficult to parametrise reliably, whereas for the three-point
function sum rule no next-to-leading order QCD corrections are
available and thus a proper matching with the Wilson coefficient
function is at present not possible. For these reasons, we shall
concentrate below on existing results in the large-$N_c$
expansion~[\ref{bbg:88}--\ref{bp:00}], 
which in our opinion are
developed furthest. After commenting on the large-$N_c$ approach in
more detail, the calculation of $B_K$ within the chiral quark model
[\ref{befl:98}] will also be briefly discussed.

Calculations of weak hadronic matrix elements in the framework of the
large-$N_c$ expansion were developed by Bardeen, Buras and G{\'e}rard
in the nineteen-eighties. For $B_K$, at the next-to-leading order in
$1/N_c$, this method resulted in $B_K=0.7\pm 0.1$ [\ref{bbg:88}], to be
compared with $B_K=0.75$ in the strict large-$N_c$ limit. However, at
that time the next-to-leading order correction to the Wilson
coefficient function~[\ref{bjw:90}] was not available, and anyhow it is
debatable whether the result of~[\ref{bbg:88}] can be properly matched
to the short distance coefficient. The proper matching of the scale
and scheme dependencies in matrix elements as well as Wilson
coefficients is, however, a crucial aspect for all approaches to weak
hadronic matrix elements.

In the approach of~[\ref{hks:99}] a significant dependence on the
matching scale is still present, resulting in sizable uncertainties
for $B_K$. Explicit cancellation of scale and scheme dependences was
demonstrated in Ref.~[\ref{pdr:00}] within the chiral limit, and, to a
lesser extent in Ref.~[\ref{bp:00}], also for a physical strange quark.
The main ingredients in the approaches of~[\ref{pdr:00},\ref{bp:00}] are:
the large-$N_c$ expansion; chiral perturbation theory to control the
low-energy end of the Green function required for the calculation of
the matrix elements; the operator product expansion to control the
higher-energy region of the Green function above roughly $1\,\gev$; a
model which connects the low- and high energy regimes. To this end,
in~[\ref{pdr:00}] the relevant Green function was saturated by the
lowest lying vector meson, the $\rho$, whereas in~[\ref{bp:00}] the
extended Nambu-Jona-Lasinio model was applied which, however, does
not display the correct QCD high-energy behaviour. The dependence on
these models constitutes the dominant uncertainty for the latter
approaches.

In the chiral limit, the findings $\hat B_K=0.38\pm 0.11$~[\ref{pdr:00}]
as well as $\hat B_K=0.32\pm 0.13$~[\ref{bp:00}] are in very good
agreement with the current algebra result $\hat B_K=0.33$~[\ref{dgh:82}],
obtained by relating $\hat B_K$ to the ${\rm K}^+\to\pi^+\pi^0$ decay rate.
In fact, this agreement could be interpreted as a successful description
of the ${\rm K}^+\to\pi^+\pi^0$ decay from large-$N_c$. The authors of
Ref.~[\ref{bp:00}] have also extended their calculation beyond the chiral
limit with the result $\hat B_K=0.77\pm 0.07$. The smaller error compared
to the chiral limit case is due to a reduced model dependence for a
physical strange quark. However, as is obvious from these results, the
chiral corrections amount to more than 100\%, and it remains to be seen
whether $\hat B_K$ of~[\ref{bp:00}] incorporates all such corrections.
Nevertheless, it is interesting to observe that the final result of
Ref.~[\ref{bp:00}] is again very close to the strict large-$N_c$
prediction, and is also in good agreement with the average from
lattice QCD quoted above.

An independent approach to hadronic matrix elements and to $B_K$ in
particular is the chiral quark model~[\ref{befl:98}]. The chiral quark
model provides a link between QCD and chiral perturbation theory and
bears some similarity to the extended Nambu-Jona-Lasinio model already
mentioned above. In this framework, the hadronic matrix elements
depend on the values of quark and gluon condensates, also present in
the QCD sum rule approach, as well as constituent masses for the
quarks. For values of these parameters which fit the $\Delta I=1/2$
rule for ${\rm K}\to\pi\pi$ decays, the authors of~[\ref{befl:98}] then
obtain $\hat B_K=1.1\pm 0.2$, where the error is dominated by the
variation of constituent quark mass and gluon condensate. However,
owing to a poor matching between long- and short-distance
contributions in the case of $B_K$, an additional systematic
uncertainty of the order of 15\% could be present in the result 
of~Ref.~[\ref{befl:98}].

\boldmath
\section{Experimental methods for the study of ${\rm B}^0$ and 
$\overline{\rm B}^0$ mixing}
\unboldmath
The system of neutral B mesons, ${\rm B}^0$ and $\overline{\rm B}^0$, can be
described in terms of states with well defined mass and lifetime
exhibiting the phenomenon of particle-antiparticle oscillations. The
frequency of $\Bd$ and $\Bs$ ~mixing can be described by the mass
difference $\Delta M_{d,s}$ as defined in Eq.~(\ref{eq:deltamb}). This
mass difference between the two mass eigenstates leads to a
time-dependent phase difference between the particle wave functions.
In the Standard Model, ${\rm B}^0$--$\overline{\rm B}^0$ 
mixing is described via
second order weak processes, as displayed for the case of 
${\rm K}^0$--$\overline{\rm K}^0$ mixing in Fig.~\ref{fig:L9}. The mass
difference $\Delta M_{d,s}$ can be determined by computing the
electroweak box diagram, where the dominant contribution is through
top quark exchange as can be seen in Eq.~(\ref{eq:xds}). A measurement of
$\Delta M_d$ or $\Delta M_s$ in principle allows the determination of the
Cabibbo-Kobayashi-Maskawa matrix elements $|V_{td}|$ or $|V_{ts}|$ as
indicated by the relations in Eq.~(\ref{eq:DMD}) and (\ref{eq:DMS}).
The main uncertainty in relating measurements of the mixing frequency
to the CKM matrix elements originates from the parameters $\fB{d,s}$
and $\bbhat{d,s}$ as discussed in Sec.~\ref{subsec:fBxi}. However,
in the ratio $\Delta M_d/\Delta M_s$ several of the theoretical uncertainties
cancel as is obvious from Eq.~(\ref{eq:vtdvtsxi}). 
Thus, the ratio $\Delta M_d/\Delta M_s$  is related to the ratio of CKM matrix
elements $|V_{td}|/|V_{ts}|$ and will ultimately determine one of the
sides of the CKM unitarity triangle.


\boldmath
\subsection{Time integrated oscillation analyses and determination of
{${\rm B}$}~hadron production rates}
\unboldmath

At the $\Upsilon(4S)$, only $\Bd$ and ${\rm B}^+$ mesons are produced,
whereas at high energy colliders $\Bs$ ~mesons and $b$-baryons are
also present. In the latter case, $\Bd$ and $\Bs$~mesons contribute to
time integrated mixing measurements with a weight proportional to
their relative production fractions:
\begin{equation}
\bar{\chi} = f_{\Bd}\ \chi_d\ +\ f_{\Bs}\ \chi_s.
\label{eq:chibar}
\end{equation}
Here, $f_{\Bd}$ and $f_{\Bs}$ are the production rates of $\Bd$
and $\Bs$ mesons in $b$ quark jets, while $\chi_{d,s}$ are the respective mixing parameters
defined in Eq. (\ref{eq:integosci}) \footnote{The world average for the 
time integrated mixing parameter is $\bar{\chi}$=0.1194 $\pm$ 0.0043 
[\ref{ewwg_chi}].}. 
The non-linear relation between $x$
and ${\chi}$ (see Eq.~(\ref{eq:integosci})\,) implies that ${\chi}$
becomes insensitive to $x$ for values greater than $x\sim5$. Thus, a
time dependent oscillation analysis is necessary to observe fast
oscillations as expected for $\Bs$~mesons. At the $\Upsilon(4S)$~resonance,
a measurement of $\chi_d$ allows to directly extract $x_d$ because
only slowly oscillating $\Bd$ mesons are produced.
A time integrated mixing analysis is, however, important to determine the 
hadron production fractions $f_{\Bd}$ and $f_{\Bs}$. 
For example, 
$f_{\Bd}$ is
an essential input for a measurement of $V_{cb}$ using 
$\overline{\rm B}^0_d \rightarrow D^{*+} \ell^- \bar{\nu}_{\ell}$ decays
and the source of an important 
systematic error in $\Delta M_d$ measurements at high energy colliders.
Furthermore, the sensitivity to $\Bs$--$\Bsb$ oscillations in inclusive
analyses depends on the ${\rm B}^0_s$~production rate $f_{\Bs}$.

The production rates of ${\rm B}$~hadrons in $b$~quark jets can be obtained
from the measured integrated oscillation rates of B mesons
(see Eq.~(\ref{eq:chibar})\,). When measuring the time integrated
oscillation parameter in a semileptonic sample, the mixing probability
can be written as
\begin{equation}
 \bar{\chi} = g_{\Bs} \chi_s + g_{\Bd} \chi_d,
\label{eq:chibarsl}
\end{equation}
where $g_{\Bs}$ and $g_{\Bd}$ are the fractions of $\Bd$ and $\Bs$~mesons in a
semileptonic sample. 
Assuming that the semileptonic width is the same for all 
${\rm B}$~hadrons, we obtain
\begin{equation}
  g_{B_{i}} = f_{B_{i}} \, R_i  \quad {\rm where } 
\quad R_i = \frac{\tau_i}{\tau_B}.
\end{equation}
This results in
\begin{eqnarray}
f_{\Bs} &=&\frac{1}{R_s}~\frac{(1+r)~\bar{\chi}-(1-f_{b{\rm -baryon}}
~R_{b{\rm -baryon}})~ \chi_d}
{(1+r)~ \chi_s - \chi_d}  \nonumber \\
f_{\Bd} &=& \frac{1}{R_d}~\frac{\bar{\chi}-(1-f_{b{\rm -baryon}} 
~{R_{b{\rm -baryon}}})~ \chi_s}
{\chi_d - (1+r)~\chi_s}
\label{eq:fs}
\end{eqnarray}
where $r=R_u/R_d = \tau(B^+)/\tau(\Bd)$. 
We assume $f_{\Bd}=f_{\Bu}$, $f_{\Bu}+f_{\Bd}+f_{\Bs}+f_{b{\rm -baryon}}=1$
and~$\chi_s=0.5$. 

From the previous expressions, the values of $f_{\Bs}$ and $f_{\Bd}$ are
determined 
and combined with those obtained from direct measurements 
(for more details see Ref.~[\ref{lepsldcdf}]).
The results are shown in Table \ref{tab:ratesstepa}. 
It is clear that $f_{\Bs}$ is essentially determined 
from the time integrated mixing measurement. 
The error on $f_{\Bs}$ is dominated by the uncertainty 
on the integrated oscillation parameter $\bar \chi$, which is not 
expected to improve substantially in the near future. 
Different uncertainties contribute to the error on $f_{\Bd}$. 
The most important one is the poor knowledge of the $b$-baryon 
production rates. It has to be noted that $f_{\Bd}$ is 
essentially determined by the DELPHI direct measurement [\ref{ref:fbddelphi}].

\begin{table}
\begin{center}
\begin{tabular}{|ccc|}
\hline
 $b$-hadron fractions & direct measurement  & direct plus mixing \\ \hline
 $f_{\Bs}$                & $(9.2 \pm2.4)$\% & $(9.3 \pm1.1)$\%   \\
 $f_{b-\mathrm{baryon}}$
                      & $(10.5\pm2.0)$\% & $(10.5\pm1.8)$\%  \\
 $f_{\Bd} = f_{\Bu}$          & $(40.1\pm1.3)$\% & $(40.1\pm1.1)$\%  \\
\hline
\end{tabular}
\end{center}
\caption{\it Average values of $b$-hadron production rates obtained from direct
  measurements and using time integrated mixing 
as of the ICHEP 2002 conference [\ref{ref:osciw}].
}
\label{tab:ratesstepa}
\end{table}

\subsection{Flavour tagging techniques}
\label{sec:flavourtag}

In general, a measurement of the time dependence of 
${\rm B}^0$--$\overline{\rm B}^0$ 
oscillations requires the knowledge of:
\begin{itemize}
\item 
the proper decay time $t$ of the ${\rm B}^0$~meson (see
Sec.~\ref{sec:description}),  
\item 
the flavour of the ${\rm B}$ or $\overline{\rm B}$~meson at both production and 
decay in order to determine whether 
the ${\rm B}^0$~meson has oscillated. 
\end{itemize}
Events are classified on the basis of the sign of the production
and decay tagging variables as mixed or unmixed. To accomplish this,
it is necessary to determine the $b$~quark content ($b$ or $\bar b$) 
of the ${\rm B}$~meson at production and at decay time. 
The figure of merit to compare 
different flavour tags is the so-called effective tagging
efficiency $\varepsilon (1-2\,p_W)^2$, 
where the efficiency $\varepsilon$ represents the fraction of events for
which a flavour tag exists
and $p_W$ is the mistag probability indicating the fraction of events 
with a wrong flavour tag. The mistag probability is related to the dilution 
$\cal D$, another quantity used to express the power of a flavour tag:  
\begin{equation}
  \mathcal{D} = 1 - 2\, p_W.
\end{equation}
The dilution $\cal D$ is defined as the
number of correctly tagged events $N_R$ minus the number of
incorrectly identified events $N_W$ divided by the sum:
\begin {equation}
  \mathcal{D} = \frac{N_R - N_W}{N_R + N_W}.
\end{equation}

\begin{figure}
\centerline{
\includegraphics[height=0.22\hsize]{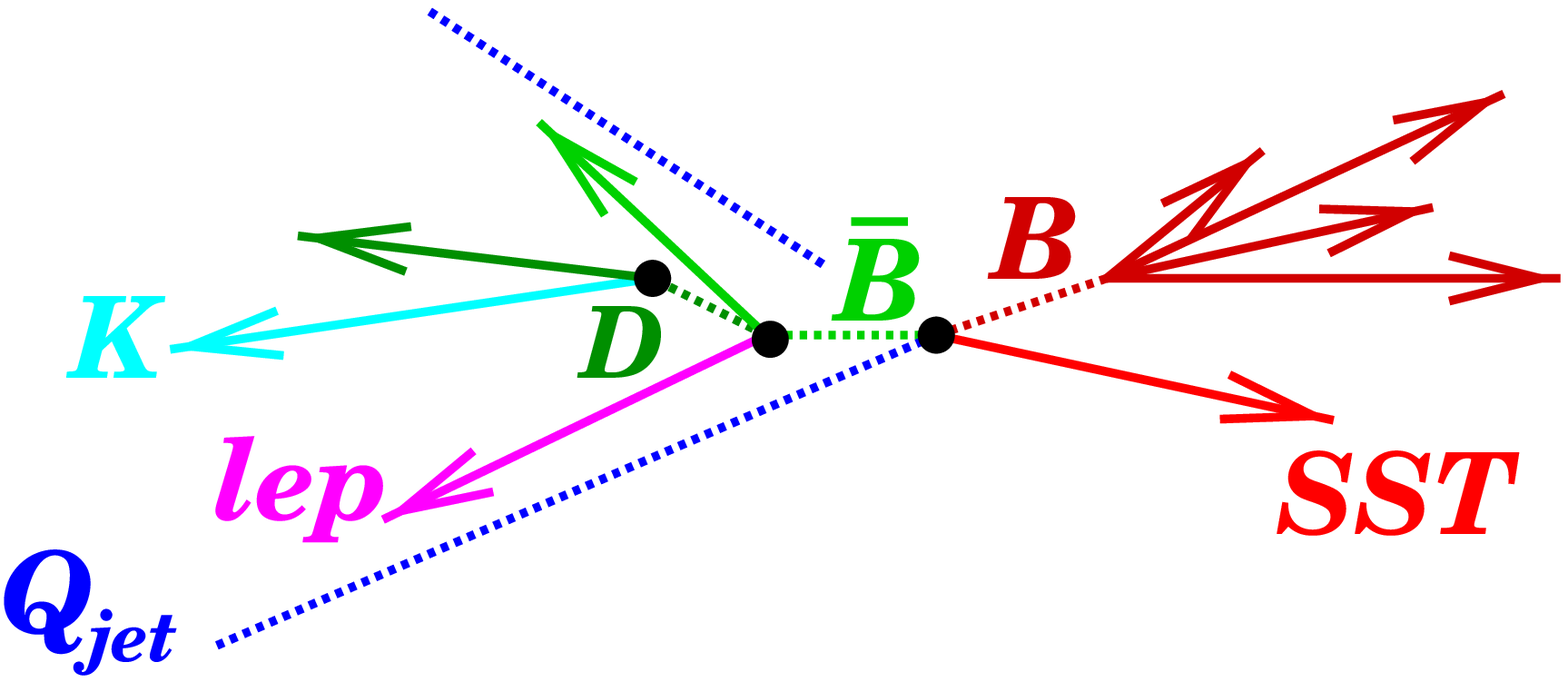}
\hfill
\includegraphics[height=0.22\hsize]{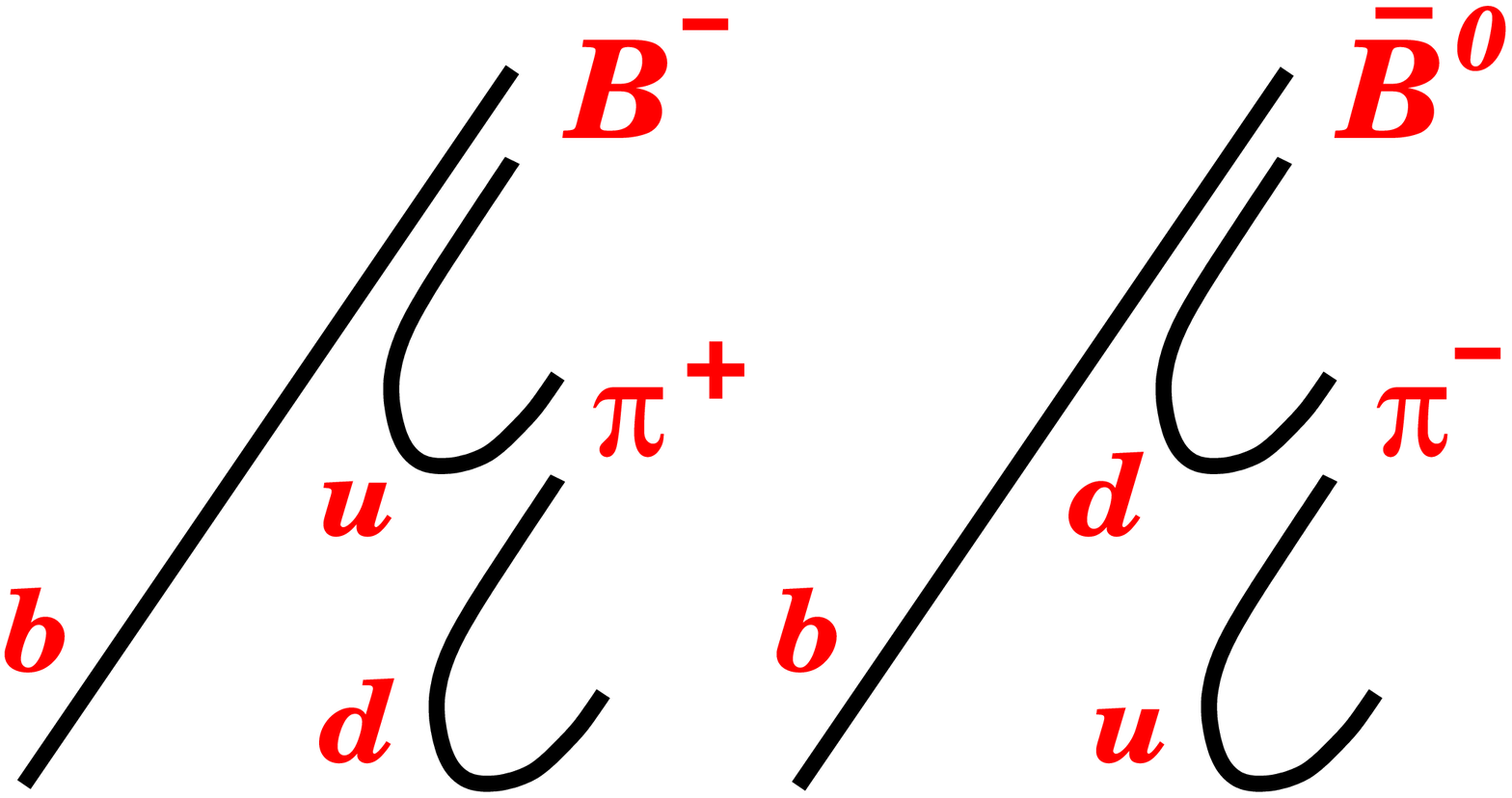}
\put(-445,85){\Large\bf (a)}
\put(-190,85){\Large\bf (b)}
}
\caption{\it 
(a) Schematic sketch of a typical ${\rm B}\overline{\rm B}$~event.
(b) A simplified picture of $b$~quark fragmentation into ${\rm B}$~mesons.
}
\label{fig:bmix_tag}
\end{figure}

Fig.~\ref{fig:bmix_tag}(a) is a sketch of a ${\rm B}\overline{\rm B}$~event
showing 
the ${\rm B}$ and $\overline{\rm B}$~mesons originating from the primary production
vertex and decaying at a secondary vertex indicating possible flavour
tags on the decay vertex side ($SST$) and opposite side
($\mathit{lep}$, ${\rm K}$, $Q_\mathit{jet}$).

\subsubsection{Decay flavour tagging}

Several techniques are used to determine the $b$~quark flavour at decay time.
The B flavour can be identified by
the charge of a lepton from a semileptonic ${\rm B}$~decay.
 In a prompt $b\rightarrow\ell^-$ decay, the charge of the lepton
reflects the $b$ flavour. 
However, other processes can also give a lepton in the final state such as  
cascade decays $b\rightarrow c\rightarrow \ell^+$
resulting in a wrong sign tag, right sign cascade decays $b \rightarrow W^- \rightarrow
\bar c \rightarrow \ell^-$, semileptonic $\tau$~decays $b\rightarrow W^- \rightarrow \tau^-
\rightarrow \ell^-$ or $b\rightarrow J/\psi X \rightarrow \ell^{\pm}$
decays giving both
sign leptons.  These processes resulting in wrong sign leptons can be suppressed 
by using the lepton momentum or transverse momentum with respect to the
$b$~jet axis. 

The $b$ quark flavour can also be inferred from the charge of a reconstructed charm meson
(${\rm D}^{*-}$ from $\Bd$ or ${\rm D}^-_s$ from $\Bs$) or that of a kaon assumed to
come from a $b\rightarrow c\rightarrow s$~transition.
In fully inclusive analyses, the $b$~flavour 
can be obtained from the jet
charge (see Eq.~(\ref{eq:jetcharge})\,), the charge of a reconstructed
dipole or from multitags as further detailed in
Sec.~\ref{sec:anamethods} 

\subsubsection{Production flavour tagging}

Methods to tag the production $b$ quark flavour differ somewhat
between high energy colliders (LEP, SLC, Tevatron) and the
${\rm B}$~factories. At high-energy colliders, the production flavour tags
can be divided into two groups, those that tag the initial charge
of the $b$~quark contained in the ${\rm B}$~candidate itself (same side tag)
and those that tag the initial charge of the other quark~($\bar b$)
produced in the same event (opposite side tag).

Same side tagging methods exploit correlations of the
${\rm B}$~flavour with the charge of 
particles produced in association with
the ${\rm B}$~meson. Such correlations are expected to arise from $b$~quark
hadronization and from ${\rm B}^{**}$~decays. It has been
suggested~[\ref{Rosner_sst}] that the electric charge of particles
produced near a ${\rm B}$~meson can be used to determine its initial
flavour. This can be understood in a simplified picture of $b$~quark
fragmentation as shown in Fig.~\ref{fig:bmix_tag}(b). For example, if
a $b$~quark combines with a $\bar u$~quark to form a ${\rm B}^-$~meson, the
remaining $u$ quark may combine with a $\bar d$~quark to form a
$\pi^+$. Similarly, if a $b$~quark hadronizes to form a $\overline{\rm B}^0$
meson, the associated pion would be a $\pi^-$. 
A similar charge correlation is expected for a charged kaon produced in association with a $\Bs$ meson. 
Decays of the orbitally excited ($L=1$) ${\rm B}^{**}$~mesons,
${\rm B}^{**0} \rightarrow \rm B^{(*)+} \pi^-$ or ${\rm B}^{**+} \rightarrow
\rm B^{(*)0} \pi^+$, also produce pions with the same charge correlation. 
This tagging method has
been successfully used for example at
CDF~[\ref{ref:bdmix_dstarl_sst_cdf},\ref{ref:bdmix_dstarl_sst_long_cdf}].

There are several methods of opposite side flavour tagging as
illustrated in Fig.~\ref{fig:bmix_tag}(a). The methods using a lepton
from the semileptonic decay of a $\rm B$~hadron, a kaon
or the presence of a charmed particle from the other $\bar B$~hadron in the
event, were already discussed above. 

The technique based on the jet charge exploits the
fact that the momentum weighted sum of the particle
charges of a $b$~jet is related to the $b$-quark charge.
In the most basic form, the jet charge can be defined as:
\begin{equation}
  Q_{\rm jet} = \frac{ \sum_i q_i \cdot (\vec{p}_i \cdot\hat{a}) }
                     { \sum_i \vec{p}_i \cdot\hat{a}  },
\label{eq:jetcharge} 
\end{equation}
where $q_i$ and $\vec{p}_i$ are the charge and momentum
of track $i$ in the jet and $\hat{a}$ is
a unit~vector defining the jet direction. 
On average, the sign of the jet charge is the same as the sign
of the $b$~quark charge that produced the jet.
More sophisticated weights (e.g. $(\vec{p}_i \cdot\hat{a}) ^\kappa$) or
track impact parameter
information  are often introduced 
to improve the $b$ flavour separation. 
The jet charge can also be used
as a same side tag, if tracks from primary vertex can be efficiently
distinguished with respect to those from secondary decay vertices.

Other tagging methods include the charge dipole method that  
aims of reconstructing the $b$~hadron decay chain topologically. This
method has been utilized at SLD taking advantage of the superb decay length
resolution of the SLD CCD pixel vertex detector to separate tracks from the
${\rm B}$~decay point from tertiary tracks emitted at the charm decay 
vertex~[\ref{ref:review_sld}].   
A charge dipole is defined as the distance between secondary and tertiary
vertices signed by the charge difference between them 
(see also Sec.~\ref{sec:anamethods}).

Another interesting production flavour tagging method is 
available at SLD. It exploits the large polarized 
forward-backward asymmetry in 
$Z\rightarrow b\bar b$~decays
[\ref{ref:bdmix_inclept_sld}--\ref{ref:bdmix_ktag_sld}]. 
This $b$~flavour production tag makes use of the large electron beam
polarization $P_e \sim 73\%$ at the SLC 
collider. A left- or right-handed incident electron tags
the quark produced in the forward hemisphere as a $b$~or $\bar b$~quark with
a mistag rate $p_W$ 
of 28\% at nearly 100\% efficiency~[\ref{ref:review_sld}].

At asymmetric $e^+e^-$~${\rm B}$~factories, $\Bd-\Bdb$~pairs are 
produced through the $\Upsilon(4S)$~resonance 
with a boost $\beta\gamma = 0.425$ and 0.55 at KEKB and 
PEP\,II, respectively.
The two neutral ${\rm B}_d$~mesons
produced from the $\Upsilon(4S)$~decay evolve in time in a coherent
$P$-wave state where they keep opposite flavours until one
of the ${\rm B}_d$~mesons decays. From this point in time onwards, 
the other ${\rm B}$~meson follows a time evolution according to 
the expression 
$\Gamma{\rm e}^{-\Gamma |\Delta t|}\,(1\pm\cos\Delta M\, \Delta t)$ where 
$\Delta t$ is the proper time difference between the two B~decays. 
Hence, the production flavour tag of one of the ${\rm B}$~mesons
can be taken as the decay flavour tag of the other.
The main flavour tagging methods
currently used at BaBar and Belle include $b\rightarrow\ell^-$ lepton tagging
and $b\rightarrow c\rightarrow s$ kaon tagging.

It is common to combine different production tags in an oscillation analysis
to achieve mistag probabilities of $p_W \sim 26\%$ at
LEP~[\ref{Barate:1998ua}--\ref{ref:bdmix_inclept_opal}]
or even 22\% for SLD~[\ref{Thom:2002fs}].
An equivalent figure for CDF in Run\,I
of the Tevatron is $p_W \sim 40\%$~[\ref{ref:bphys_run1_cdf}]. 
Effective mistag probabilities of $p_W \sim 24\%$ 
are achieved by the BaBar and Belle 
experiments~[\ref{Aubert:2001nu},\ref{Abe:2001xe}]. 
It is interesting to mention that the effect of $\Bd$ and $\Bs$~mixing 
substantially decreases the tagging power of opposite side tagging methods
at high-energy colliders while 
mixing of the other B meson
(i.e. the coherent mixing occurring before the first B decay) does
not contribute to a mistag probability at the $\Upsilon(4S)$.

\subsection{Analytical description of oscillation analyses}
\label{sec:description}

A physics function of the form $\Gamma{\rm e}^{-\Gamma t} \,(1\pm\cos\Delta M\, t)$  
is used to describe the signal in B oscillation analyses.
At high energy colliders such as LEP, SLC or the Tevatron, the ${\rm B}$~meson decay
proper time $t$ can be obtained from a measurement of the distance 
$L_B$ between the ${\rm B}$~production vertex and the ${\rm B}$~decay vertex. 
The proper time $t$ is related to the decay distance $L_B$ and to the boost
$\beta \gamma$ by 
\begin {equation}
c\,t =  \frac{L_B}{\beta \gamma} = L_B \, \frac{M_B}{p_B}.
\label{eq:tlxy1}
\end{equation}
At asymmetric $e^+e^-$~${\rm B}$~factories, the proper time difference $\Delta t$ 
between the two ${\rm B}$~candidate decays is the relevant measure. It is computed as: 
\begin {equation}
\Delta t = \Delta z/\beta\gamma c,
\label{eq:tlxy2}
\end{equation}
where $\Delta z$ is the spatial separation between the two ${\rm B}$~decay 
vertices along the boost direction. 

The uncertainty on the decay time $\sigma_t$ can be expressed in units
of the ${\rm B}$~lifetime $\tau_B$ as
\begin {equation}
  \frac{\sigma_t}{\tau_B} = \sqrt{\left( \frac{\sigma(L_B)}{L_B^0}\right)^2
   + \left(\frac{t}{\tau_B}\frac{\sigma(p_B)}{p}\right)^2}
  \hspace{1.0cm} {\rm where} \hspace{0.5cm}
  L_B^0 = c\tau_B \cdot p_B/M_B. 
\label{eq:resol_errors}
\end{equation}
The proper time resolution $\sigma_t$ depends on the uncertainty
$\sigma(L_B)$ to infer the decay length from the primary to the ${\rm B}$
decay vertex and on the ${\rm B}$~momentum resolution $\sigma(p_B)$.
Note that
the latter uncertainty scales with $t/\tau_B$, while the vertexing
resolution is independent of the proper time and only adds a constant error.

\begin{figure}
\centerline{
\includegraphics[width=0.33\hsize]{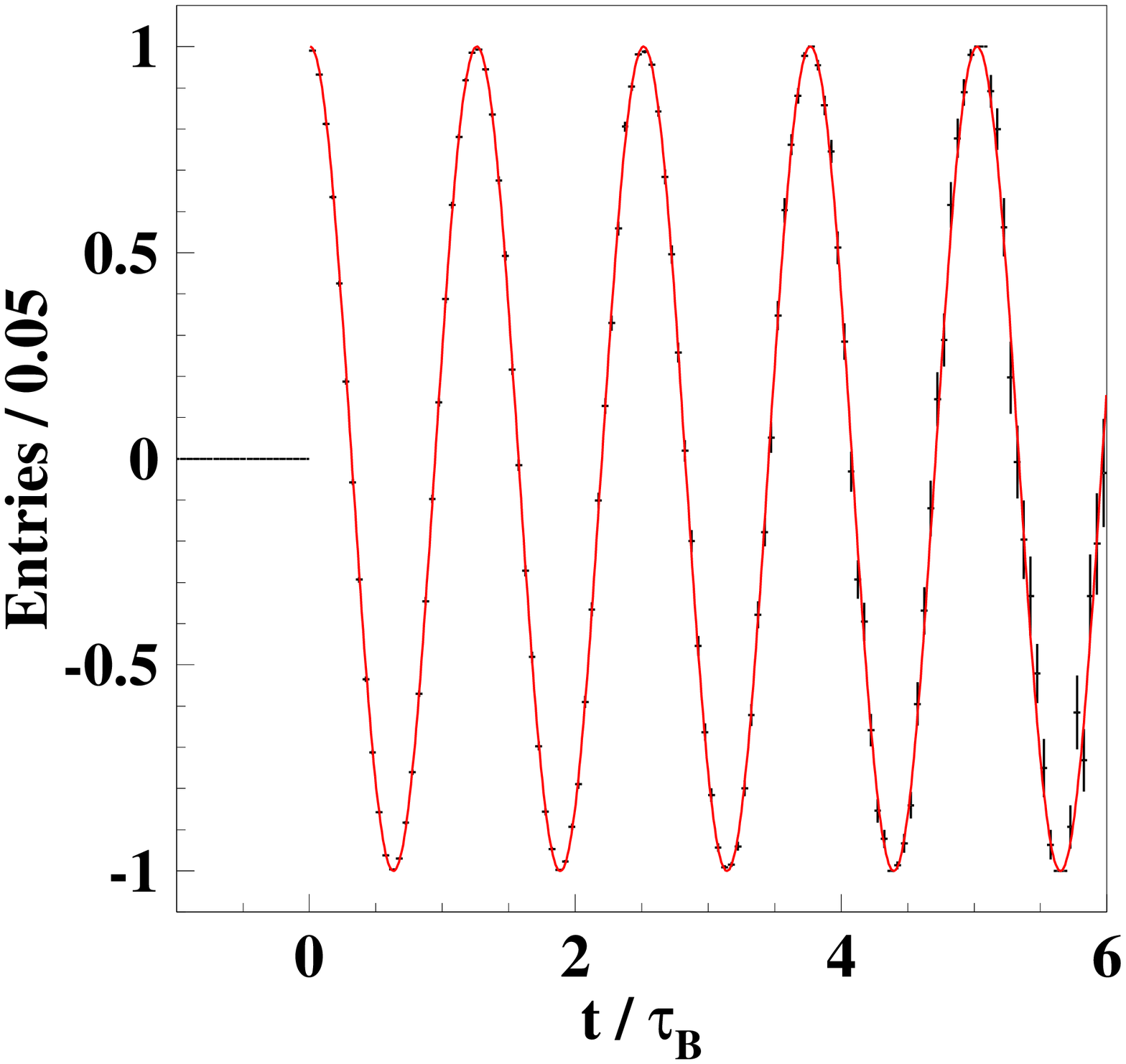}
\hfill
\includegraphics[width=0.33\hsize]{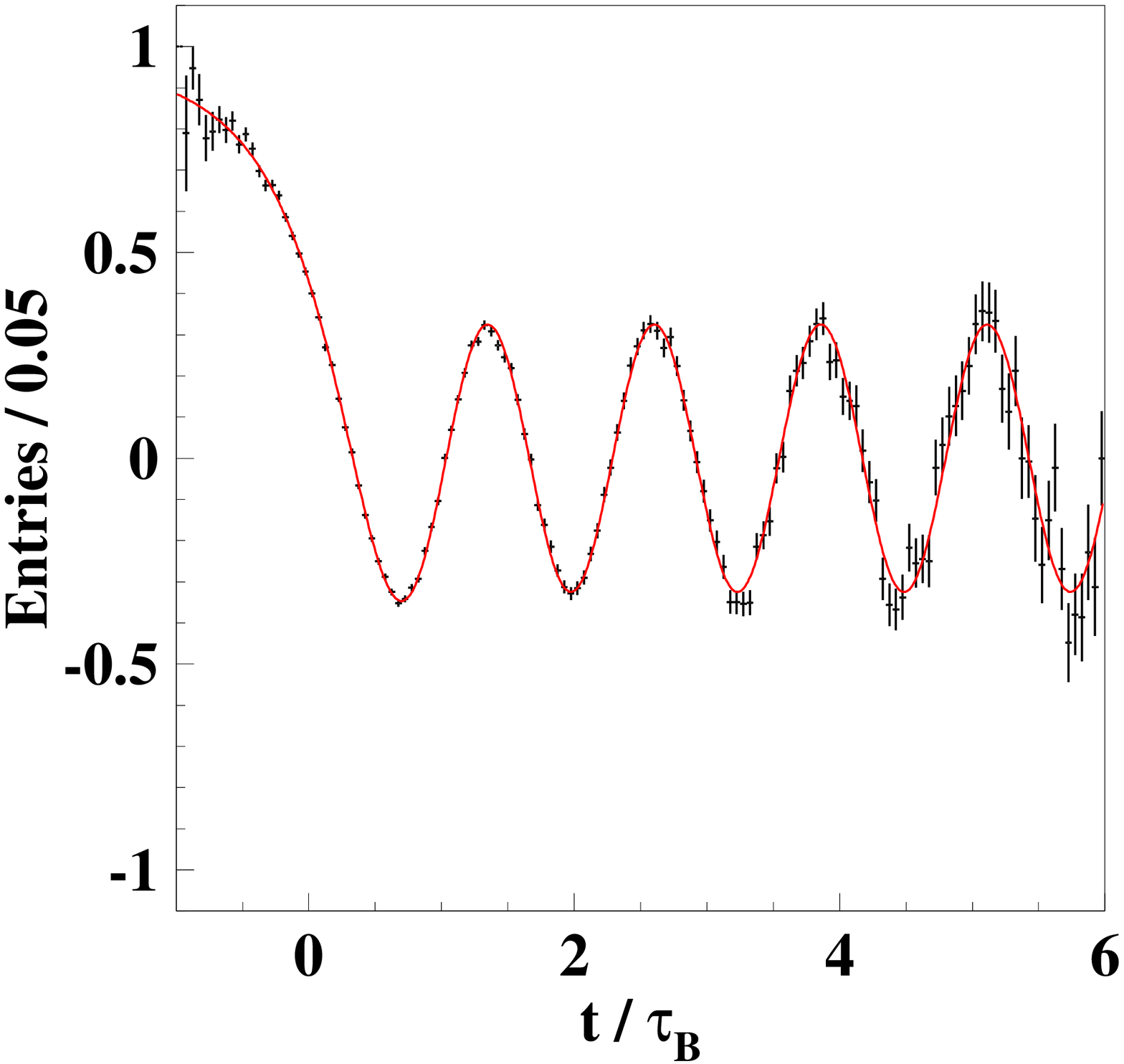}
\hfill
\includegraphics[width=0.33\hsize]{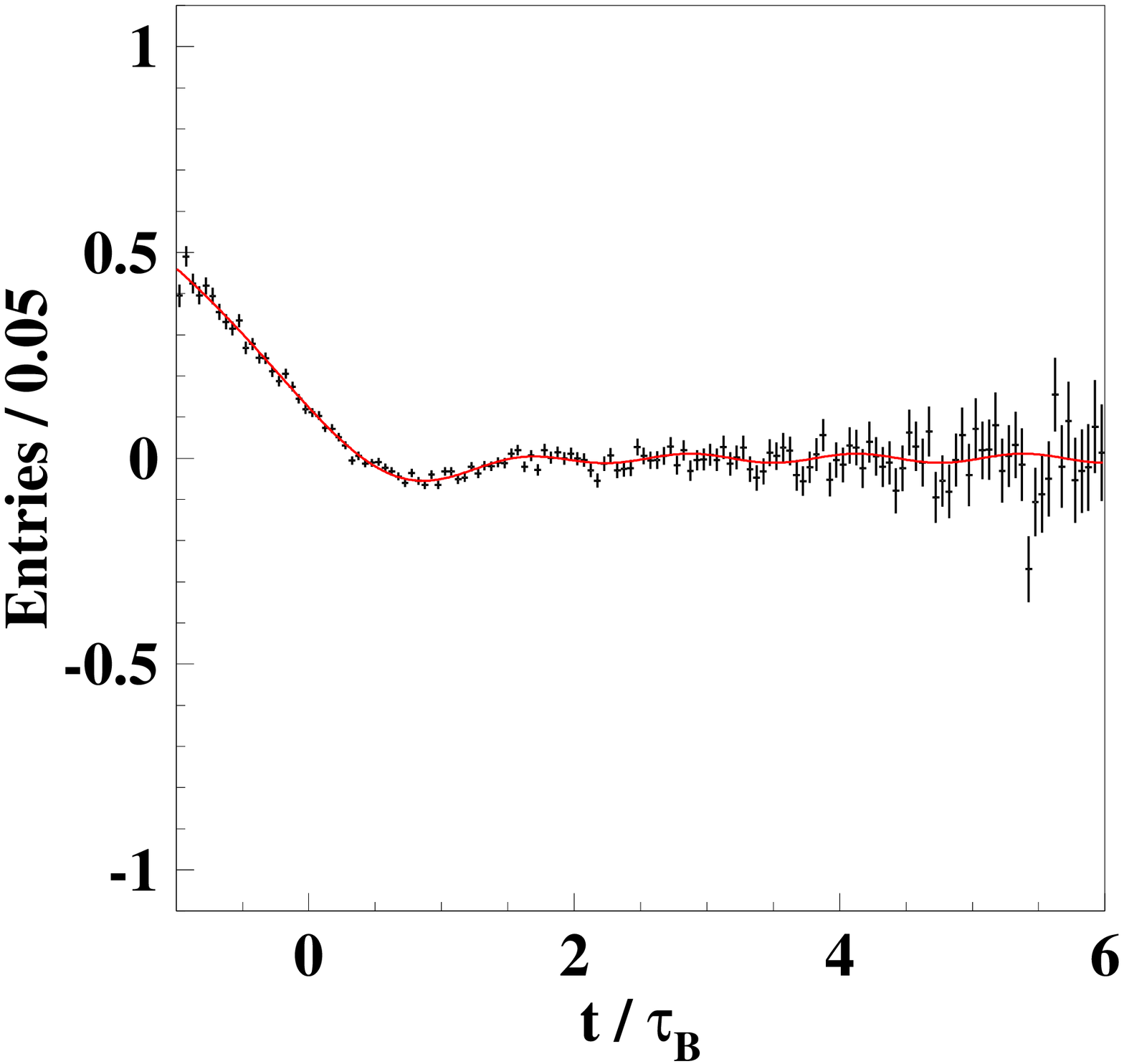}
\put(-425,30){\large\bf (a)}
\put(-271,30){\large\bf (b)}
\put(-120,30){\large\bf (c)}
}
\centerline{
\includegraphics[width=0.33\hsize]{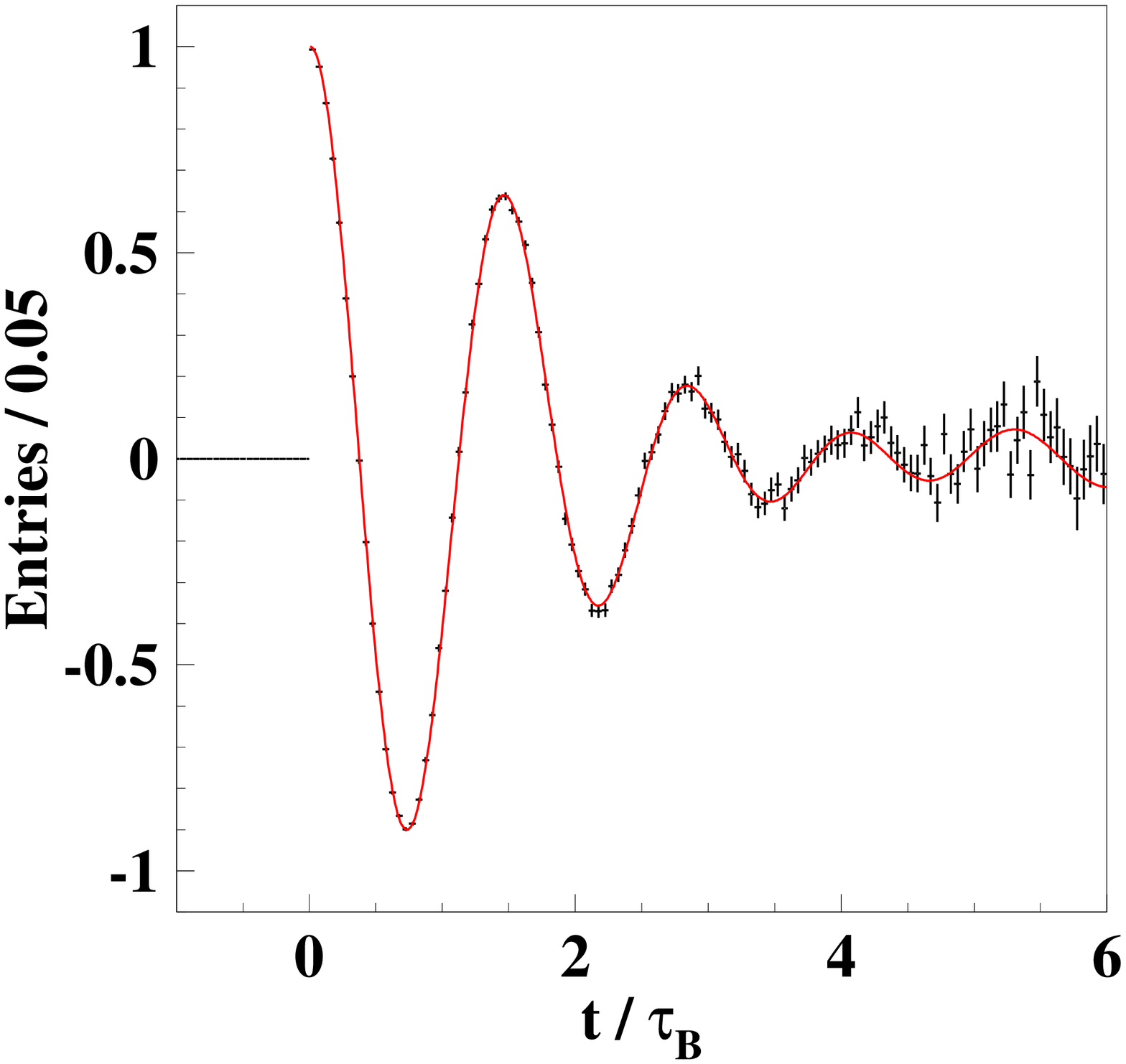}
\hfill
\includegraphics[width=0.33\hsize]{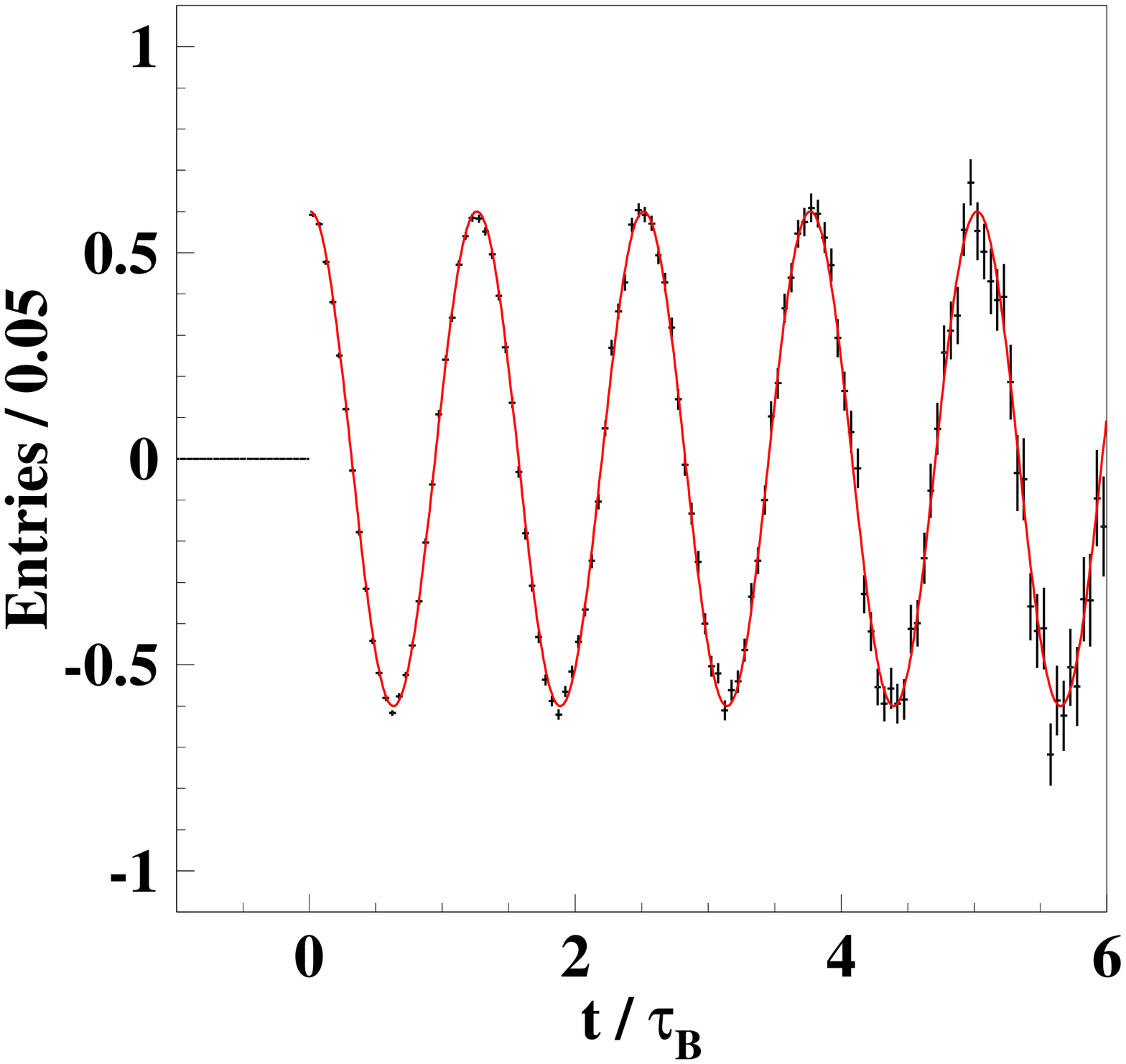}
\hfill
\includegraphics[width=0.33\hsize]{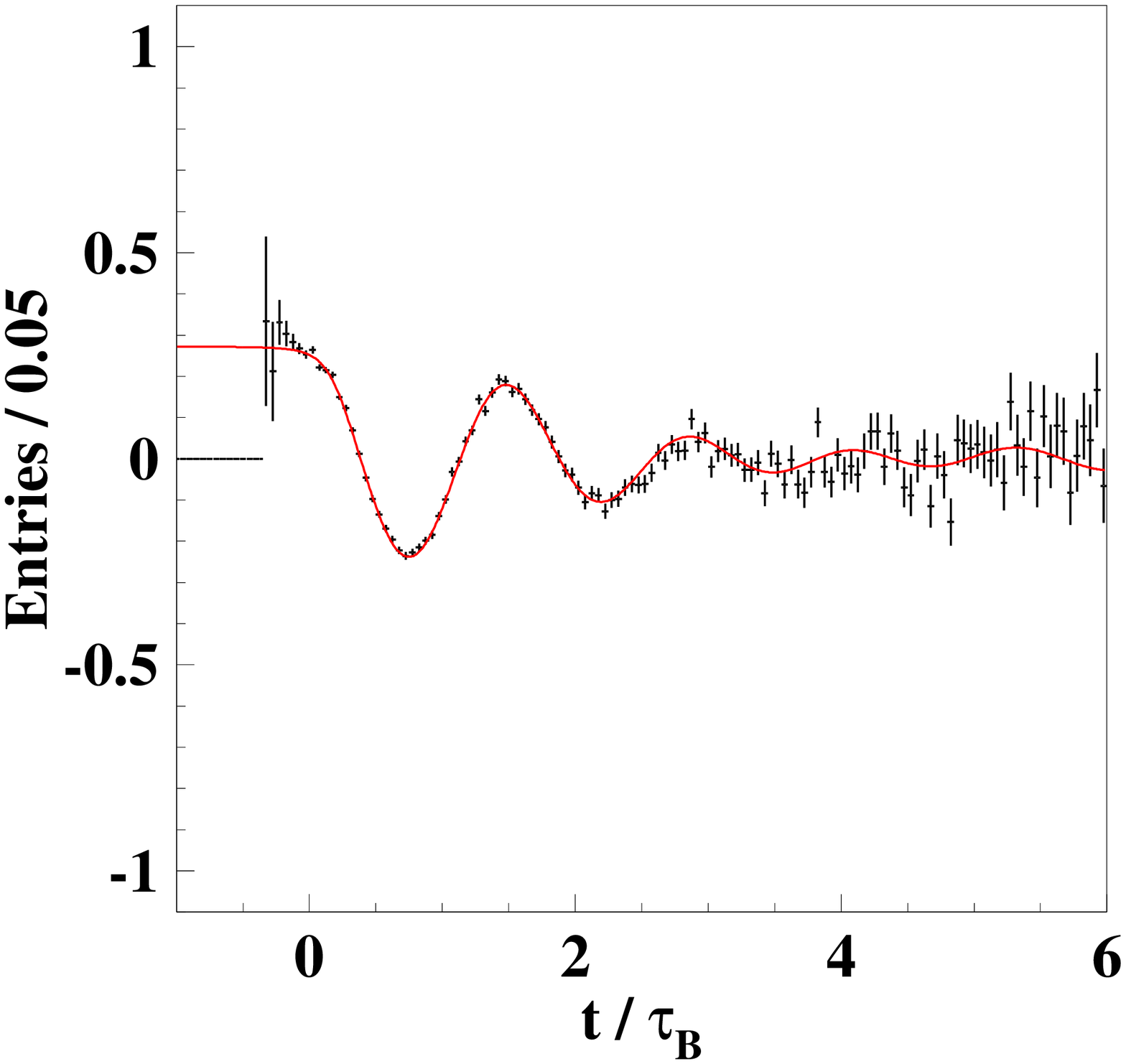}
\put(-425,30){\large\bf (d)}
\put(-271,30){\large\bf (e)}
\put(-120,30){\large\bf (f)}
}
\vspace{-.4cm}
\caption{\it 
Illustration of various detector and analysis effects on the mixing amplitude
${\mathcal A}_{\rm mix}$:  
(a) perfect resolution, 
(b) good decay length resolution,
(c) poor decay length resolution,
(d) finite momentum resolution,
(e) mistag probability and 
(f) decay length and momentum resolution plus mistag including background.
}
\label{fig:bmix_resol}
\end{figure}

The dependence of B~oscillations on the proper time resolution and other
detector effects is illustrated in Fig.~\ref{fig:bmix_resol}. Rather
than plotting the mixed and unmixed probabilities $\mathcal{P}_{\rm
unmix/mix}(t) = 1/2\,\,\Gamma{\rm e}^{-\Gamma t} \,(1\pm\cos\Delta M\,
t)$ as introduced in Eq.~(\ref{eq:prob_mix}) and
Eq.~(\ref{eq:proba}), it is customary in B~oscillation analyses
to either determine a mixing asymmetry $\mathcal{A}_{\rm mix}$ or to
calculate the fraction of mixed events $\mathcal{F}_{\rm mix}$
\begin{equation}
\mathcal{A}_{\rm mix} = \frac{\mathcal{P}_{\rm unmix} - \mathcal{P}_{\rm mix}} 
{\mathcal{P}_{\rm unmix} + \mathcal{P}_{\rm mix}} = \cos\Delta M\,t,
\hspace*{1.0cm}
\mathcal{F}_{\rm mix} = \frac{\mathcal{P}_{\rm mix}} 
{\mathcal{P}_{\rm unmix} + \mathcal{P}_{\rm mix}} = (1-\cos\Delta M\,t)/2.
\label{eq:mix_asym}
\end{equation}

As an example, Fig.~\ref{fig:bmix_resol}(a) shows the oscillation pattern
of ${\mathcal A}_{\rm mix}$ for $\Delta M = 5$~ps$^{-1}$ assuming
an ideal case with perfect tagging, ideal proper time resolution and no
background. 
The reduction of the amplitude due to  a finite decay length resolution 
is shown in Fig.~\ref{fig:bmix_resol}(b). 
Figure~\ref{fig:bmix_resol}(c)
indicates what happens when the resolution of the (silicon) vertex detector
is not sufficient to resolve the oscillations: 
${\mathcal A}_{\rm mix}$ is completely smeared out and oscillations are
no longer visible. 
The effect of a finite momentum resolution is displayed in
Fig.~\ref{fig:bmix_resol}(d). Since the uncertainty on the proper time
coming from the momentum resolution is linear in proper time $t$, as
seen in 
Eq.~(\ref{eq:resol_errors}), the rapid oscillation damps in time while
the first few ``wiggles'' can still be seen completely.
The oscillation amplitude is reduced if a  mistag probability is
introduced, as can be   
seen in Fig.~\ref{fig:bmix_resol}(e). 
Finally, in a real measurement, background will also be present
which additionally reduces the relative importance of the oscillation amplitude.
The effect of background on the mixing amplitude, in addition to a
finite decay length and momentum resolution, as well as a non-zero mistag
probability, is shown
in Fig.~\ref{fig:bmix_resol}(f).  Note, however, that this ``realistic''
distribution is based on half a million signal events. 
Imagine the corresponding error bars for a
measurement with 
a few hundred signal events and an oscillation frequency of $\Delta M =
20$~ps$^{-1}$. 

In a $\rm B^0$~mixing measurement, a value for $\Delta M$ is usually extracted
from the data using a maximum 
likelihood method. In the following, we illustrate some of the essential
steps for a $\Bd$~analysis determining $\Delta M_d$ in more detail.
We use the example of an analysis where like-sign (unlike-sign) events 
describe mixed (unmixed) events as would be the case, for example, in a
dilepton analysis.
The total probability to observe a like-sign tagged event at the reconstructed 
proper time $t_{rec}$ is:
\begin{eqnarray} 
 {\cal P}^{like}(t_{rec}) = f_{\bb} \sum_{q=d,s} f_{\rm B_q} p_W^{\rm B_q}
           {\cal P}^{\rm mix}_{rec. \rm B_q}(t_{rec}) +
           \mathnormal f_b \sum_{q=u,d,s,baryons}   f_{\rm B_q} (1- p_W^{\rm B_q}) 
           {\cal P}^{\rm unmix}_{rec. \rm B_q}(t_{rec}) &+ \nonumber \\   
           f_{bkg.} (1- p_W^{bkg.}) {\cal P}_{bkg.}(t_{rec})   
 \label{likes}
 \end{eqnarray} 
and correspondingly for an unlike-sign tagged event:
 \begin{eqnarray} 
 {\cal P}^{unlike}(t_{rec}) = f_{\bb} \sum_{q=d,s}  
                   f_{\rm B_q} (1- p_W^{\rm B_q}) {\cal P}^{\rm mix}_{rec. \rm B_q}(t_{rec}) +
                  \mathnormal     f_b \sum_{q=u,d,s,baryons}   f_{\rm B_q} p_W^{\rm B_q} 
                  {\cal P}^{\rm unmix}_{\rm B_q}(t_{rec}) &+ \nonumber \\   
                  f_{bkg.} p_W^{bkg.} {\cal P}_{bkg.} (t_{rec}).  
 \label{unlikes}
 \end{eqnarray} 
where $f_{\bb}$ is the fraction of $\bb$ events and $p_W^i$ are the mistag probabilities.
 The probability ${\cal P}^{\rm mix}_{rec. \rm B_q} (t_{rec})$
to observe the mixed $\mbox{B}^0_d$ or $\mbox{B}^0_s$ mesons 
at proper time  $t_{rec}$  is the result 
of a convolution of the oscillation probability function 
as given in Eq.~(\ref{eq:proba}) and Eq.~(\ref{eq:prob_mix}) with the 
detector resolution function $\cal{R}$ and weighted with an acceptance 
function $Acc(t)$ 
 \begin{equation} 
   {\cal P}^{\rm (un)mix}_{rec. \rm B_q} (t_{rec}) = \int_{0}^{\infty} 
  Acc(t) {\cal R}(t_{rec}-t,t) {\cal P}^{\rm (un)mix}_{\rm B_q} (t) dt .
 \end{equation} 

To extract the value $\Delta M$ of the oscillation frequency, the following
likelihood function is minimized :
 \begin{equation} 
 {\cal L} = - \sum_{like-sign} \ln({\cal P}^{like}(t_{rec}))  
  - \sum_{unlike-sign} \ln({\cal P}^{unlike}(t_{rec})) .
\label{eq:like}
 \end{equation}  

In order to fully exploit the
available statistics, more sophisticated mixing analyses make
use of those variables on an event-by-event basis, or often divide the event
sample into classes with e.g.~different tagging capabilities.

\subsubsection{The amplitude method}
\label{sec:amp_description}

For $\Delta M_s$ measurements, the amplitude method~[\ref{ref:moser}]
is used to set limits on $\Delta M_s$ and to combine
results from different analyses. For the mixed and unmixed $\mbox{B}_s^0$
events an amplitude $A$ is introduced in the expressions
describing the mixed and unmixed probabilities:
\begin{equation} 
{\cal P}_{\rm B_s^0}^{\rm unmix}~=
~\frac{1}{2} \Gamma_{\rm B_s} {\rm e}^{-\Gamma_{\rm B_s}t} 
[ 1 + A \cos {\Delta M_s t}  ]
\label{eq:amp1}
\end{equation}
and similarly:
\begin{equation} 
{\cal P}_{\rm B_s^0}^{\rm mix}~=
~\frac{1}{2} \Gamma_{\rm B_s} {\rm e}^{-\Gamma_{\rm B_s}t} 
[ 1 - A \cos {\Delta M_s t}  ]
\label{eq:amp2}
\end{equation}

The amplitude method works as follows. 
A $\Bs$ oscillation amplitude $A$ and its error $\sigma_{A}$ are extracted as
a function of  
a fixed test value of $\Delta M_s$ using a likelihood method in analogy to 
Eq.~(\ref{eq:like}) based on the physics functions defined in
Eq.~(\ref{eq:amp1}) and Eq.~(\ref{eq:amp2}).
To a very good approximation, the statistical uncertainty on $ A$ is Gaussian
and the experimental sensitivity is :

\begin {equation}
\mathcal{S} = \frac{1}{\sigma_A} \sim \sqrt{N/2}\,f_{\rm sig}\,(1-2p_w)\, 
{\mathrm e}^{-(\Delta M\,\sigma_t)^2/2}
\label{eq:bs_sensitivity}
\end{equation}
where $N$ and $f_{\rm sig}$ are the number of candidate events and the
fraction of signal in the selected sample, $p_W$ is the mistag probability
to incorrectly tag a decay as mixed or unmixed characterizing the effective
flavour tagging efficiency as discussed in Sec.~\ref{sec:flavourtag}, and
$\sigma_t$ is the resolution on proper time or proper time difference
in the case of the B~factories. The sensitivity $\cal S$ decreases
rapidly as $\Delta M$ increases. This dependence is controlled by
$\sigma_t$.

If $\Delta M_s$ equals its true
value $\dms^{\rm true}$, the amplitude method expects ${A} = 1$
within the total uncertainty $\sigma_{A}$. If $\Delta M_s$ is tested far
below its true value, a measurement consistent with ${A} = 0$ is
expected. A value of $\Delta M_s$ can be excluded at 95\%~C.L. if ${A} +
1.645\,\sigma_{A} \leq 1$. If the true $\Bs$ oscillation frequency
$\dms^{\rm true}$ is very large, far above the experimental
sensitivity, ${A} = 0$ is expected to be measured and all values
of $\Delta M_s$ such that $1.645\,\sigma_{A}(\Delta M_s) < 1$ are expected to
be excluded at 95\%~C.L. Because of proper time resolution, the
quantity $\sigma_{A}(\dms)$ is an increasing function of $\Delta M_s$. It
is therefore expected that individual values of $\Delta M_s$ can be excluded
up to $\dms^{\rm sens}$, where $\dms^{\rm sens}$ is called the
sensitivity of the analysis defined by $1.645\,\sigma_{A}(\dms^{\rm sens})
= 1$. 
The results from different analyses and experiments can be combined
by simple averaging different amplitude spectra.  

\subsection{Description of oscillation analyses}
\label{sec:anamethods}

Many different analysis methods have been devised to study $\Bd$ and
$\Bs$ mixing. These range from fully inclusive to fully exclusive
analyses and, thus, they differ significantly in terms of selection
efficiency, sample purity and mistag rates. Moreover, they make use of
various production and decay tags. The methods also differ in the
techniques used to reconstruct the ${\rm B}$ decay length and to estimate
the ${\rm B}$ momentum, and therefore have different proper time
resolutions. In the following, analysis methods developed to measure
$\Delta M_d$ are discussed first and those used in the search for
$\Bs$ oscillations are presented afterwards.

\subsubsection{${\rm B}_d^0$--$\overline{\rm B}_d^0$ oscillation analyses}

\subsubsection*{Exclusive methods}

The most straightforward and cleanest method relies on the exclusive
reconstruction of the \Bd\ decay chain. However, because of its low
efficiency, it has only recently become accessible with the advent of
$e^+ e^-$ asymmetric B factories. Using samples of $\sim$30M \BB\ events,
BaBar~[\ref{ref:bdmix_excl_babar}] and Belle~[\ref{ref:bdmix_excl_belle}] 
reconstruct the decays $\Bd \to \rm D^{(\ast) -} \pi^+$, 
${\rm D}^{(\ast) -} \rho^+$, ${\rm D}^{(\ast) -} a_1^+$, $J/\psi \rm K^{\ast 0}$ (BaBar), 
and $\Bd \to \rm D^{(\ast) -} \pi^+$, ${\rm D}^{\ast -} \rho^+$ (Belle),
where charmed mesons are fully reconstructed in 
several  ${\rm D}^{\ast -}$ and $\Dzb$ decay modes.
\begin{figure}
\begin{center}
\hbox to\hsize{%
\includegraphics[height=0.3\hsize]{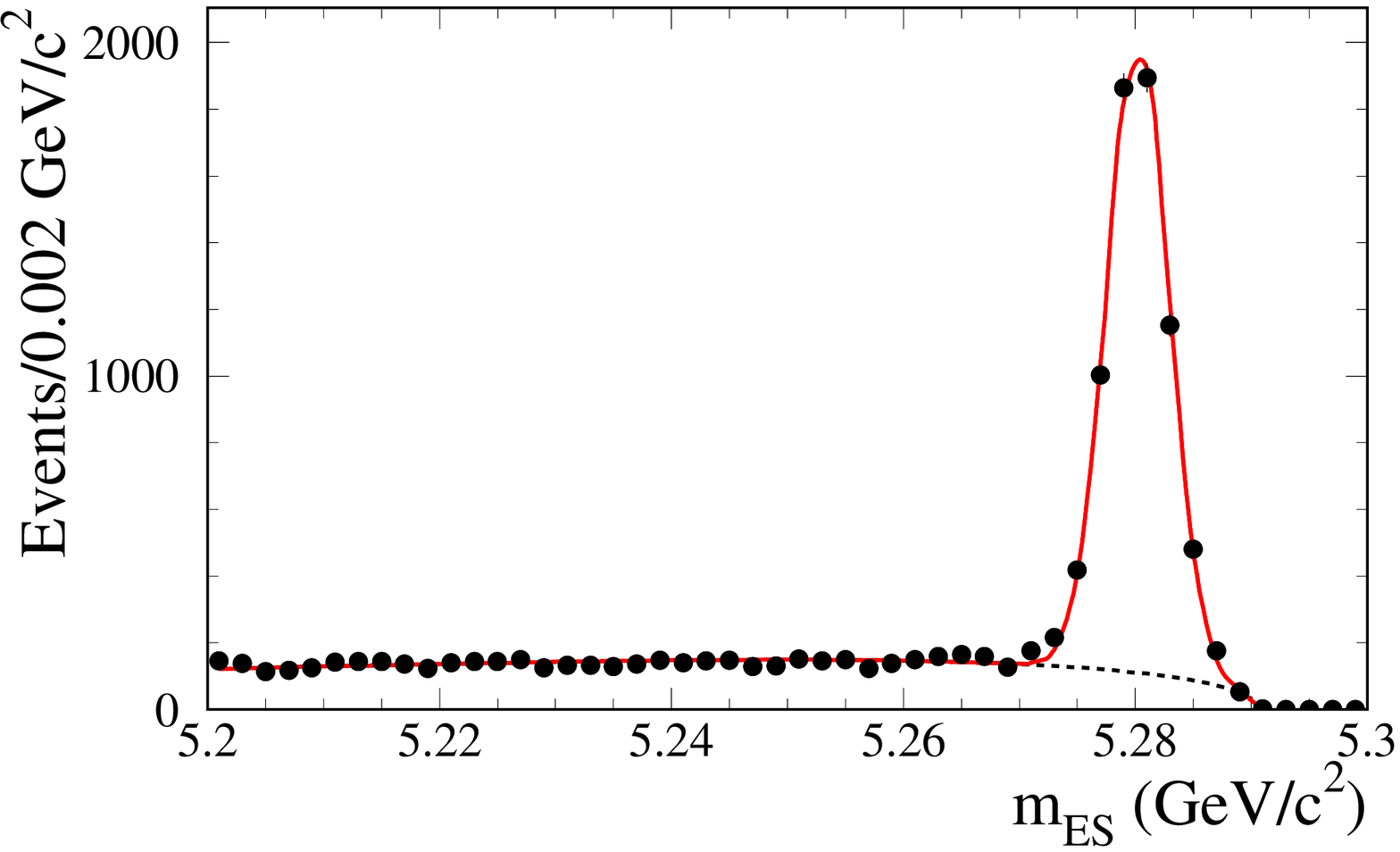}
\hfill
\includegraphics[height=0.3\hsize,bb=0 21 525 404]%
{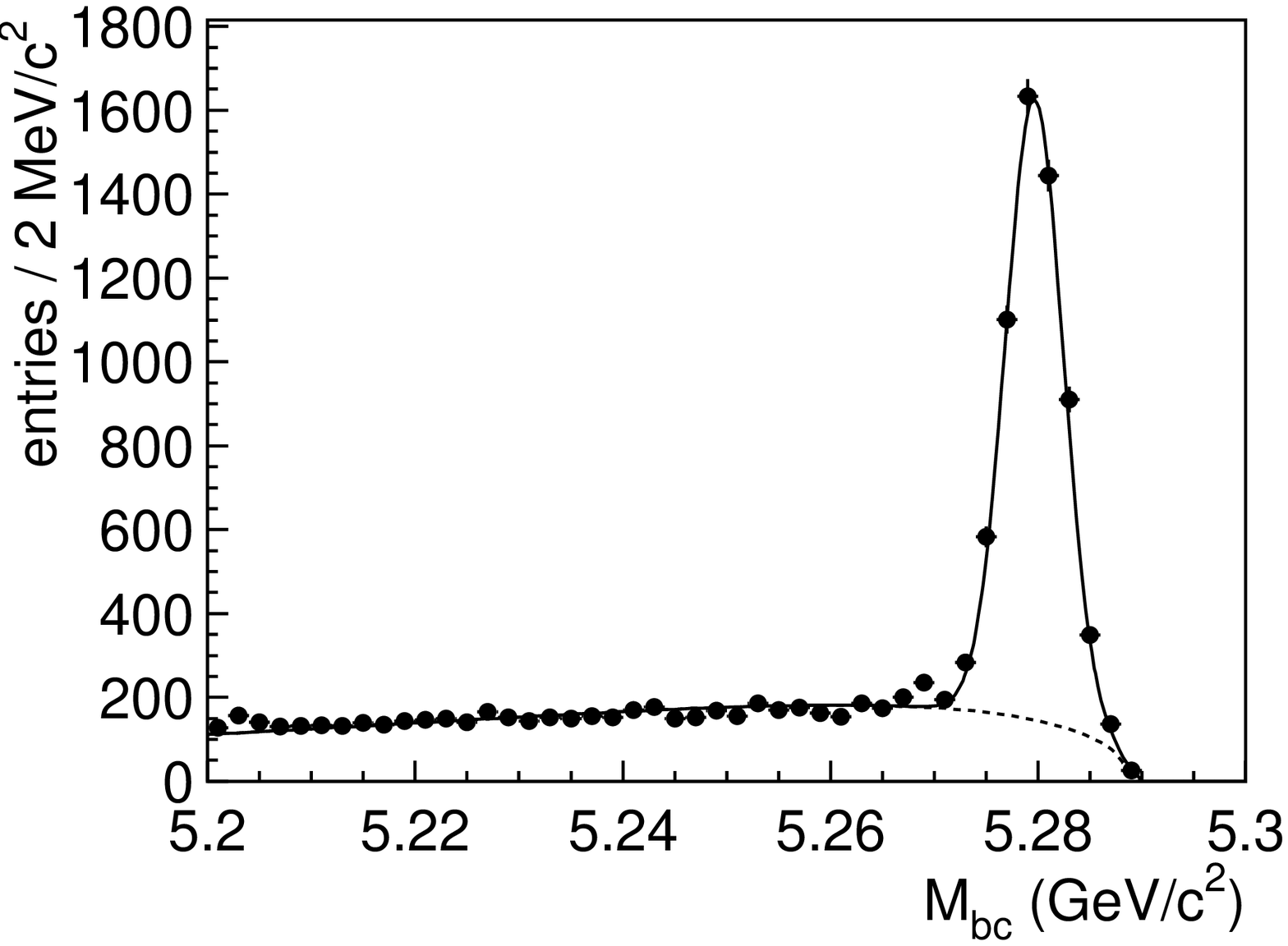}}
\caption{\it  {Distributions of beam-energy substituted mass for
 exclusively reconstructed $\Bd$ decays in the BaBar (left) and
 Belle (right) analyses.}}
\label{fig:mes_excl}
\end{center}
\end{figure}
Very clean signals are obtained, see Fig.~\ref{fig:mes_excl},
and the decay flavour is unambiguously determined by the charge of the
${\rm D}^{(\ast)}$ meson (or the charged kaon in case of the $J/\psi \rm K^{\ast0}$ decay).

The average separation of the two ${\rm B}$ decay points is $\Delta z =
255\;(200)\,\mu\mathrm{m}$ with $\sigma_z \simeq
180\;(140)\mu\mathrm{m}$ for Babar (Belle), which corresponds to a
resolution on $\Delta t$ (Eq. \ref{eq:tlxy2}) of about
$1.1\,\mathrm{ps}$. For a measurement of the \Bd\ oscillation
frequency it is therefore critical to have good control over the
resolution. Table~\ref{tab:bdmix_methods} summarizes the number of
events, signal mode purity and production flavour tag information for
these as well as all other analyses presented below.
\begin{table}[htbp]
\begin{center}
\begin{small}
\begin{tabular}{|llcll|}
\hline
Decay modes & Analysis & Events/Signal & $f_\mathrm{mode}$
  & Production flavour tag \\[0.5ex]
\hline
 $\Bd \to {\rm D}^{(\ast) -} h^{+a}$ &
 BaBar exclusive~[\ref{ref:bdmix_excl_babar}] &
 7380/6347 & 86\% & Multiple tags \\
 ~~~~~~~~~~~$J/\psi {\rm K}^{\ast 0}$&
 Belle exclusive~[\ref{ref:bdmix_excl_belle}] &
 8325/6660 & 80\% & Multiple tags \\
 $\Bd \to {\rm D}^{\ast -} \pi^+$ &
 Belle semi-incl.~[\ref{ref:bdmix_dstarpi_belle}] &
 4899/3433 & 70\% & Lepton \\
 $\Bd \to {\rm D}^{(\ast) -} X$ &
 ALEPH semi-excl.~[\ref{ref:bdmix_3methods_aleph}] &
 4059/2395 & 38?\% & Lepton+jet charge \\
 &
 CDF semi-excl.~[\ref{ref:bphys_run1_cdf}] &
 874/358 & 27\% & Lepton \\
 &
 DELPHI semi-excl.~[\ref{ref:bdmix_4methods_delphi}] &
 10030/4212~~ & 27?\% & Jet charge \\
 &
 OPAL semi-excl.~[\ref{ref:bdmix_dstar_opal}] &
 347/253 & 48\% & Lepton \\
 $\Bd \to {\rm D}^{(\ast) -} \ell^+ \nu$ &
 BaBar semi-excl.~[\ref{ref:bdmix_dstarl_babar}] &
 17506/14182 & 74\% & Multiple tags \\
 &
 Belle semi-excl.~[\ref{ref:bdmix_dstarl_belle}] &
 16397/15118 & 80\% & Multiple tags \\
 &
 CDF semi-excl.~[\ref{ref:bdmix_dstarl_cdf}] &
 888/530 &   & Lepton \\
 &
 CDF semi-excl.~[\ref{ref:bdmix_dstarl_sst_cdf}] &
 ~~~~~~~~/6266 &   & Same-side tag \\
 &
 OPAL semi-excl.~[\ref{ref:bdmix_dstar_opal}] &
 1200/926~~ & 65\% & Jet charge  \\
 &
 DELPHI semi-incl.~[\ref{ref:bdmix_4methods_delphi}] &
 5958/4135 & 59\% & Jet charge \\
 &
 OPAL semi-incl.~[\ref{ref:bdmix_pil_opal}] &
 ~~~~~~~~/7000 & 36\% & Multiple tags \\
 $\Bd \to X \ell^+ \nu$ &
 BaBar semi-incl.~[\ref{ref:bdmix_dilept_babar}] &
 99k/~~~~~~ & 37\% & Lepton \\
 &
 Belle semi-incl.~[\ref{ref:bdmix_dilept_belle}] &
 281k/~~~~~~~ &      & Lepton \\
 &
 ALEPH semi-incl.~[\ref{ref:bdmix_3methods_aleph}] &
 5957/~~~~~~~~ &      & Lepton \\
 &
 CDF semi-incl.~[\ref{ref:bdmix_dimu_cdf}] &
 5968/~~~~~~~~ & 39\% & Lepton ($\mu\mu$) \\
 &
 CDF semi-incl.~[\ref{ref:bphys_run1_cdf}] &
 10180/~~~~~~~~~~ &      & Lepton ($e \mu$) \\
 &
 DELPHI semi-incl.~[\ref{ref:bdmix_4methods_delphi}] &
 4778/~~~~~~~~ & 33\% & Lepton \\
 &
 L3 semi-incl.~[\ref{ref:bdmix_l3}] &
 1490/~~~~~~~~ &      & Lepton \\
 &
 L3 semi-incl.~[\ref{ref:bdmix_l3}] &
 2596/~~~~~~~~ & 34\% & Lepton (impact parameter) \\
 &
 OPAL semi-incl.~[\ref{ref:bdmix_dilept_opal}] &
 5357/~~~~~~~~ &      & Lepton \\
 &
 ALEPH semi-incl.~[\ref{ref:bdmix_3methods_aleph}] &
 62k/~~~~~~ &      & Jet charge \\
 &
 CDF semi-incl.~[\ref{ref:bdmix_inclept_cdf}] &
 13k/~~~~~~ &      & Lepton+jet charge \\
 &
 DELPHI semi-incl.~[\ref{ref:bdmix_4methods_delphi}] &
 60k/~~~~~~ & 29\% & Jet charge \\
 &
 OPAL semi-incl.~[\ref{ref:bdmix_inclept_opal}] &
 95k/~~~~~~ & 30\% & Jet charge \\
 &
 L3 semi-incl.~[\ref{ref:bdmix_l3}] &
 8707/~~~~~~~~ &      & Jet charge \\
 &
 SLD semi-incl.~[\ref{ref:bdmix_lepd_sld}] &
 581/~~~~~~ & 51\% & Polarization+jet charge\\
 &
 SLD semi-incl.~[\ref{ref:bdmix_inclept_sld}] &
 2609/~~~~~~~~ & 31\% & Polarization+jet charge\\
 $\Bd \to {\rm all}$ &
 ALEPH inclusive~[\ref{ref:bdmix_vtx_aleph}] &
 423k/~~~~~~~~ & 35\% & Jet charge\\
 &
 DELPHI inclusive~[\ref{ref:bdmix_vtx_delphi}] &
 770k/~~~~~~~~ & 40\% & Multiple tags\\
 &
 SLD inclusive~[\ref{ref:bdmix_kdipo_sld}] &
 3291/~~~~~~~~ & 60\% & Polarization+jet charge;\\
 & & & & Charge dipole decay tag\\
 &
 SLD inclusive~[\ref{ref:bdmix_kdipo_sld}] &
 5694/~~~~~~~~ & 60\% & Polarization+jet charge;\\
 & & & & Kaon decay tag 1993--95\\
 &
 SLD inclusive~[\ref{ref:bdmix_ktag_sld}] &
 7844/~~~~~~~~ & 60\% & Multiple tags;\\
 & & & & Kaon decay tag 1996--98\\
\hline
 \multicolumn{5}{l}{$^{a}$ $h^+$ stands for $\pi^+, \rho^+, a_1^+$.}
\end{tabular}
\end{small}
\end{center}
\caption{\it Summary of $\Bd$ mixing analyses showing the signal decay
modes, analysis method, total number of selected events and estimated
signal, fraction of signal decay mode in the selected sample
($f_\mathrm{mode}$), and production flavour tag.
  \label{tab:bdmix_methods}}
\end{table}

\subsubsection*{Semi-exclusive methods}
Several analyses have combined an identified lepton with a fully
reconstructed charmed hadron. 
Generally, the presence of a $\rm D^{(*)-}$, with charge opposite that of the 
lepton, tags the decay of a \Bd\ meson.
This simple picture is complicated by
decays of the type $\Bu \to \overline{\rm D}^{\ast\ast 0} \ell^+ \nu$,
where the $\overline{\rm D}^{\ast\ast 0}$ decays into a
$\rm D^{(\ast)-}$ meson.

Measurements have been performed at B
factories by BaBar~[\ref{ref:bdmix_dstarl_babar}] and
Belle~[\ref{ref:bdmix_dstarl_belle}] and at high energy colliders by
CDF~[\ref{ref:bdmix_dstarl_cdf},\ref{ref:bdmix_dstarl_sst_cdf}] and
OPAL~[\ref{ref:bdmix_dstar_opal}]. 
$\Bd$ mesons are partially
reconstructed in the mode $\Bd \to \rm D^{(\ast) -} \ell^+ \nu$, where the
${\rm D}^{\ast -}$ or ${\rm D}^-$ meson is fully reconstructed.
The selection relies on the kinematical properties
of $\Bd$ and ${\rm D}^{(\ast) -}$ decays. In particular, the low $Q$ value of
the decay ${\rm D}^{\ast -} \to \Dzb \pi^-$ is exploited to identify
${\rm D}^{\ast -}$ mesons efficiently and cleanly.
\begin{figure}
\hbox to\hsize{%
  \includegraphics[height=0.4\hsize,bb=6 9 554 436]%
{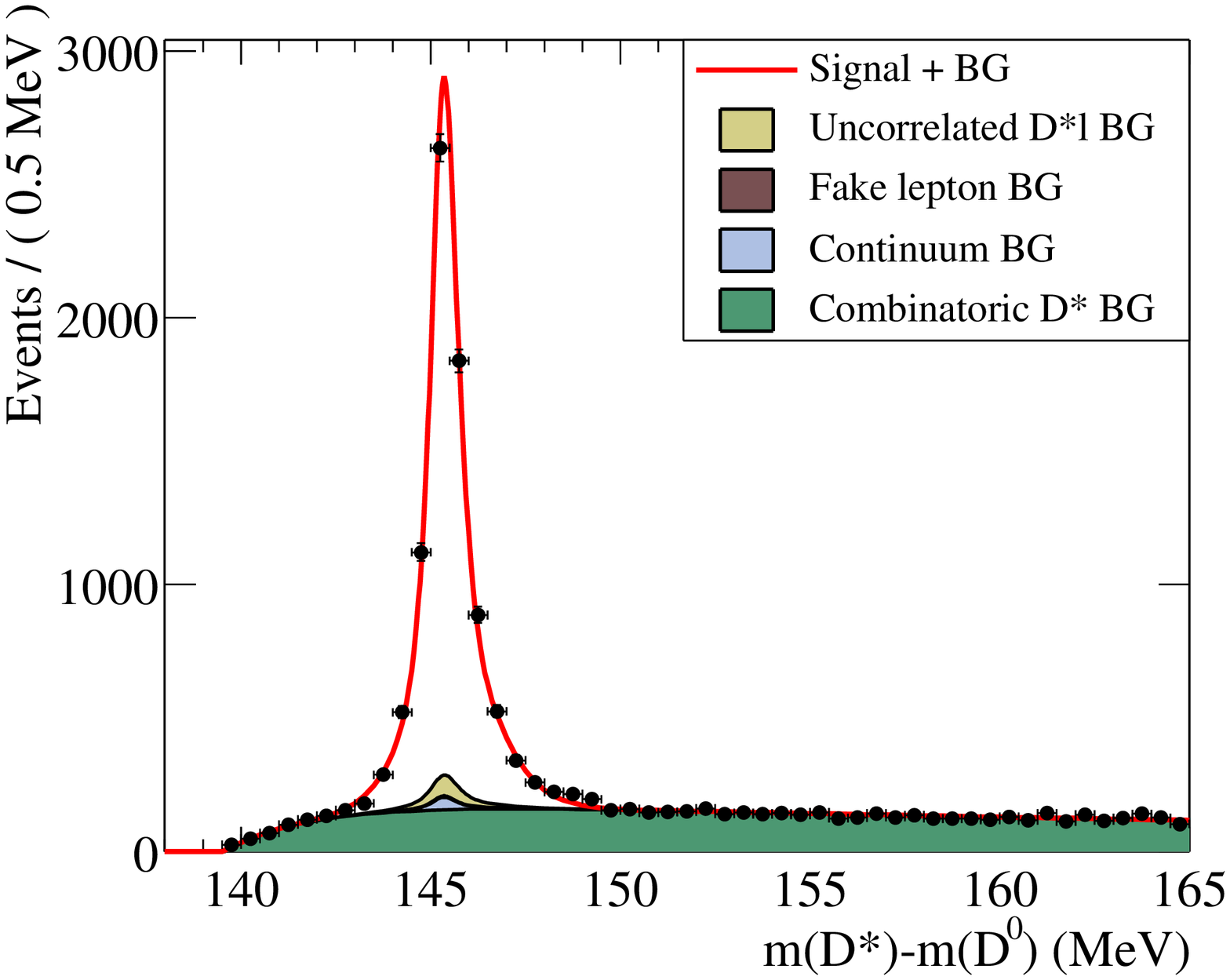}
\hfill
  \includegraphics[height=0.4\hsize,bb=5 15 454 456]%
{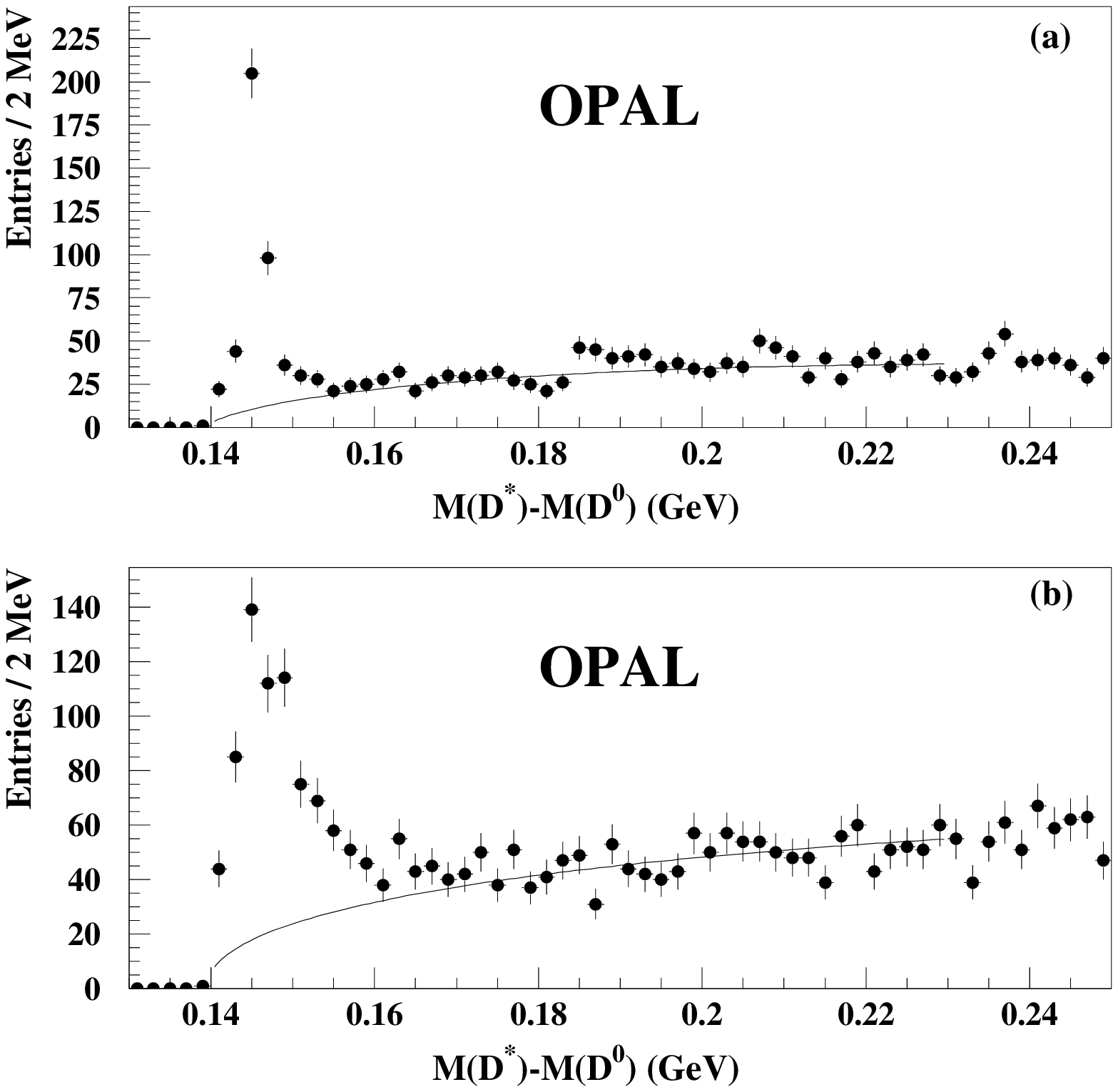}}
\caption{\it  {Distributions of the mass difference between
 ${\rm D}^{\ast -}$ and $\Dzb$ candidates for BaBar (left) and OPAL (right).
 The BaBar distribution is shown for ${\rm D}^{\ast -} e^+ \nu$ candidates.
 The distributions on the right correspond to the modes
 $\Dzb \to \rm K^+ \pi^-$ (top) and $\Dzb \to \rm K^+ \pi^- \pi^0$ (bottom),
 where the $\pi^0$ is not reconstructed.}}
\label{fig:deltam_babar_opal}
\end{figure}
Fig.~\ref{fig:deltam_babar_opal} shows the mass difference
$\Delta M = M(\rm D^{\ast -}) - M(\Dzb)$ in the BaBar and OPAL analyses.
Signal purities range from $\sim$45\% to
$\sim$90\% for the different experiments, depending mostly on the
\Dzb\ decay mode. 

In order to increase the selection efficiency, analyses by
ALEPH~[\ref{ref:bdmix_3methods_aleph}], CDF~[\ref{ref:bphys_run1_cdf}],
DELPHI~[\ref{ref:bdmix_4methods_delphi}], and
OPAL~[\ref{ref:bdmix_dstar_opal}] select $\Bd \to {\rm D}^{(\ast) -} X$ decays,
where the ${\rm D}^{(\ast) -}$ meson is also fully reconstructed.
Despite the more inclusive nature of this
method, the identification of a ${\rm D}^{(\ast) -}$ decay
guarantees that the $\Bd$ purity remains high.
However, \bbbar\ tagging is generally needed
to suppress the significant number of ${\rm D}^{\ast -}$ produced
in \ccbar\ events.

\subsubsection*{Semi-inclusive methods}

One of the semi-inclusive methods selects $\Bd \to \rm D^{\ast -}
\ell^+ \nu$ decays without attempting to fully reconstruct the \Dzb\ 
meson but only the lepton and the slow $\pi^-$ from the ${\rm D}^{\ast -} \to \Dzb \pi^-$ decay. 
This partial reconstruction method yields much larger data
samples than obtained with the exclusive reconstruction but suffers
from higher background. It has been applied by
DELPHI~[\ref{ref:bdmix_4methods_delphi}] and
OPAL~[\ref{ref:bdmix_pil_opal}]. 
The combinatorial background can be studied with same-sign
lepton-pion pairs and $\Delta M$ side bands. The $\Bu \to
\overline{\rm D}^{\ast\ast 0} \ell^+ \nu$ component needs to be estimated
from the simulation.

A similar technique is used by Belle~[\ref{ref:bdmix_dstarpi_belle}] 
to reconstruct $\Bd \to \rm D^{\ast-} \pi^+$ decays.
In this analysis, only the fast $\pi^+$ and the slow $\pi^-$ are reconstructed.
This information is sufficient to compute the \Dzb\ missing mass,
assuming that the $\Bd$ meson is at rest in the \fours\ rest frame and using
energy and momentum conservation.
The event is required to contain a high-momentum lepton
to tag the other B meson flavour and to suppress the large non-\BB\ background.
This method is only possible at the \fours\ where sufficient kinematical 
constraints are available.

The most widely used method relies on the inclusive reconstruction
of semileptonic decays. At high energy colliders, it has been employed
by ALEPH~[\ref{ref:bdmix_3methods_aleph}],
CDF~[\ref{ref:bdmix_dimu_cdf},\ref{ref:bdmix_inclept_cdf},\ref{ref:bphys_run1_cdf}], DELPHI~[\ref{ref:bdmix_4methods_delphi}],
L3~[\ref{ref:bdmix_l3}], OPAL~[\ref{ref:bdmix_dilept_opal},\ref{ref:bdmix_inclept_opal}], and SLD~[\ref{ref:bdmix_lepd_sld},\ref{ref:bdmix_inclept_sld}]. 
This method is efficient since the decay rate
for $\Bd \to X \ell^+ \nu$ is approximately 20\% (using electrons and
muons) and the decay flavour tag is excellent. A high-$p$ and
high-$p_T$ lepton is selected to suppress the contribution
from cascade leptons (from $b \to c \to \ell^+$ transitions)
and the accompanying charmed hadron
(denoted ``${\rm D}$'' in the following) is reconstructed inclusively using
charged tracks in the jet containing the lepton.
The position of the ${\rm B}$ decay vertex and the ${\rm B}$ momentum are obtained
using algorithms 
that aim to classify tracks as coming from either primary or secondary vertices.
The ${\rm B}$ decay vertex is then obtained 
by intersecting the trajectories of the lepton and that of a ${\rm D}$ candidate. 

  The analyses are combined with a variety of different production flavour tags
and are thus referred to as 
``dilepton'', ``lepton-jet charge'' and ``Multiple tags'' analyses 
(see Table \ref{tab:bdmix_methods}). 

  Dilepton analyses have also been performed by both
BaBar~[\ref{ref:bdmix_dilept_babar}] and
Belle~[\ref{ref:bdmix_dilept_belle}]. Here, there is no
attempt to reconstruct the ${\rm D}$ decay and the time difference is
extracted directly from the separation $\Delta z$ between the
intersections of the two leptons with the beam axis. Momentum and
angular cuts are applied to reduce the wrong-sign background from
cascade leptons.
In the BaBar analysis, the main background consists of $\Bu \Bub$ events and is
determined to be $\sim$55\% and the main source of mistag originates from
events containing one direct lepton and one cascade lepton,
amounting to 13\% of the total sample.

\subsubsection*{Inclusive methods}

A few analyses rely on fully inclusive techniques to select large
samples of $\Bd$ decays. These techniques aim to capture most decays
by using topological vertexing. As for the semi-inclusive methods, the
selection algorithms generally do not provide any enhancement in the
$\Bd$ purity. The primary issue here is the decay flavour tag.

  SLD uses two different decay tags: the charge of a kaon coming from the
${\rm B}$ decay chain~[\ref{ref:bdmix_kdipo_sld},\ref{ref:bdmix_ktag_sld}] 
or the charge dipole of the secondary vertex~[\ref{ref:bdmix_kdipo_sld}].
These analyses require the net charge of all tracks associated with
the decay to be zero to enhance the $\Bd$ fraction from $\sim$40\% to
$\sim$60\%.
The kaon decay tag is more efficient than the lepton decay tag
but has a worse mistag rate of $\sim$20\%.
The charge dipole technique takes advantage of the
$\Bd \to {\rm D}^- X^+$ dipole structure and
the fact that the $\Bd$ and
${\rm D}^-$ vertices are separated along the $\Bd$ line of flight due to the
finite charm lifetime. For the $\Bd$ analyses the charge dipole
is defined  as the difference between the weighted mean location of the positive tracks
and of the negative tracks along the axis joining the
primary and secondary vertices. The track
weights account for the uncertainty in determining the location of
each track. A positive (negative) charge dipole tags the decay flavour of the 
\Bdb\ (\Bd) meson.

  At LEP, DELPHI~[\ref{ref:bdmix_vtx_delphi}] also
developed a fully inclusive method based on the charge dipole tag.
The vertex algorithm uses topological and kinematical
information to separate primary and secondary tracks. A secondary lepton is found in a
subset of the vertices and provides the decay flavour tag (these
leptons are referred to as ``soft'' leptons since decays with high
$p$ and $p_T$ are used in other DELPHI analyses).
For the remainder of the sample, the ${\rm B}$ decay products are boosted back into
the ${\rm B}$ meson rest frame and a charge dipole is formed between the
forward and backward hemispheres (as defined by the thrust axis).
Given that the forward (backward) hemisphere contains mostly tracks
from the ${\rm D}$ (${\rm B}$) decay vertex, one expects a $\pm 2$ charge
difference between the two hemispheres.
The ALEPH inclusive analysis~[\ref{ref:bdmix_vtx_aleph}] reconstructs
topological vertices in both event hemispheres
as in the inclusive semileptonic analysis. The flavour tagging is performed by
computing the product of the jet charges in the two hemispheres of each
event. This product thus combines production and decay flavour tags
and is sensitive to whether mixing occurred or not.

Table~\ref{tab:bdmix_methods} summarizes the different $\Bd$ mixing
analyses. It should be noted
that this Table provides only an approximate representation
of the performance of each analysis. The reader is referred
to the specific papers for more detailed comparisons.

\subsubsection{${\rm B}_s^0$ and $\overline{\rm B}_s^0$ oscillation analyses}
\label{subsec:bsosc} 
  The study of time dependent  $\Bs$ oscillations 
has been performed with a wide range of analysis techniques at high
energy colliders. The study of $\Bs$ oscillations is more challenging
than that of $\Bd$ oscillations due to two main differences.
Only about 10\% of $b$ quarks hadronize into $\Bs$ mesons, as compared
to about 40\% into $\Bd$ mesons. The $\Bs$ oscillation frequency
is expected to be at least a factor of 20 larger than that for $\Bd$
oscillations. To address this, sophisticated analyses have been
developed with an emphasis on lowering the mistag rate, increasing the
$\Bs$ purity and, especially, improving the proper time resolution, all
of which affect the sensitivity to $\Bs$ oscillations.

\subsubsection*{Exclusive methods}

Fully exclusive analyses have been performed by
ALEPH~[\ref{ref:bsmix_aleph}] and DELPHI~[\ref{ref:bsmix_excl_delphi}]
via the (all charged particles) modes $\Bs \to {\rm D}_s^-\pi^+$, ${\rm D}_s^-
a_1^+$, $\overline{{\rm D}}^0 {\rm K}^- \pi^+$, $\overline{{\rm D}}^0 {\rm K}^- a_1^+$ (last
two for DELPHI only), where the ${\rm D}_s^-$ and $\overline{\rm D}^0$ 
are fully
reconstructed in several decay modes. The decays $\Bs \to
{\rm D}_s^{\ast-}\pi^+$, ${\rm D}_s^{\ast-}a_1^+$ and ${\rm D}_s^{(\ast)-}\rho^+$ are
also reconstructed by adding one or more photons to the above final
states (ALEPH only) or by considering the events falling into the
``satellite'' mass region below the $\Bs$ mass peak.

The number of selected signal decays is 
small~(see Table~\ref{tab:bsmix_methods}) 
but the method provides excellent proper time resolution for two reasons.
As there is no missing particle in the decay (at least for events in the main peak), 
the $\Bs$ momentum is known with good precision and therefore the contribution 
of the momentum uncertainty to the proper time resolution is small.
As a result, unlike all other methods, $\sigma_t$ does not grow
significantly when increasing the proper time $t$.
In addition, the reconstructed channels are two-body or quasi two-body 
decays, with an opening angle of their decay products which is on average 
larger than that in multi-body final states; this results in a better accuracy 
on the B decay length.
Despite the limited statistics, this method contributes to the study of $\Bs$ oscillations
at the highest values of $\Delta M_s$. As detailed in
Sec.~\ref{sec:futureprospects}, this is the preferred method for future studies of $\Bs$
oscillations at hadron colliders.
\subsubsection*{Semi-exclusive methods}

\vspace{2mm}

  Many analyses have been developed with semi-exclusive methods. $\Bs$
decays are partially reconstructed in the modes $\Bs \to {\rm D}_s^- \ell^+ \nu_{\ell} X$ 
and $\Bs \to {\rm D}_s^- h^+ X$, where $h$ represents any charged hadron
(or system of several hadrons) and the ${\rm D}_s^-$ meson decay is either
fully or partially reconstructed in the modes ${\rm D}_s^- \to \phi\pi^-$,
${\rm K}^{\ast 0} {\rm K}^-$, $\Ks {\rm K}^-$, $\phi\rho^-$, ${\rm K}^{\ast 0} {\rm K}^{\ast -}$,
$\phi\pi^-\pi^+\pi^-$, $\phi \ell^- \overline{\nu}$,
$\phi h^- X$.
Partial reconstruction in ${\rm D}_s^- h^+$ modes has the benefit of
larger statistics but the ${\rm D}_s^- \ell^+ \nu_{\ell} X$ channel has the advantage 
of a considerably higher $\Bs$ purity, lower mistag rate and 
higher proper time resolution.

  Analyses in the mode $\Bs \to {\rm D}_s^- \ell^+ \nu$ have been performed by
ALEPH~[\ref{ref:bsmix_aleph}], CDF~[\ref{ref:bsmix_phil_cdf}], 
DELPHI~[\ref{ref:bsmix_semil_delphi}] 
and OPAL~[\ref{ref:bsmix_dsl_opal}]. Selection of ${\rm D}_s^-$ decays proceeds
as described above. CDF only uses a partial reconstruction of the mode
${\rm D}_s^- \to \phi \pi^- X$.
Some background suppression (especially from ${\rm B} \to
{\rm D}_s {\rm D} X$) is achieved by requiring that the lepton and the ${\rm D}_s^-$
comes from the same vertex.

The hadronic channel $\Bs \to {\rm D}_s^- h^+ X$ has been used by
DELPHI~[\ref{ref:bsmix_excl_delphi}] and SLD~[\ref{ref:bsmix_dstracks_sld}].
Fully reconstructed ${\rm D}_s^-$ decays are selected only in the modes
${\rm D}_s^- \to \phi\pi^-$ and ${\rm K}^{\ast 0} {\rm K}^-$ because of their lower
background level. ${\rm D}_s^-$ candidates are then combined with one or
more secondary tracks to form $\Bs$ decay candidates.
Among $\Bs$ decays contributing to the
${\rm D}_s^-$ signal, approximately 10\% have the wrong decay flavour tag
due to the process $W^+ \to {\rm D}_s^+$ ($b \to c \bar{c} s$ transition).
This source of mistag is essentially absent in the
semileptonic analyses.
Despite lower statistics, the SLD analysis contributes to the $\Bs$
oscillation sensitivity at large $\Delta M_s$ thanks to its excellent 
decay length resolution (see Table~\ref{tab:bsmix_methods}).

\subsubsection*{Semi-inclusive methods}

\vspace{2mm}

The semi-inclusive lepton method, based on the process  
$\Bs \to X \ell^+ \nu_{\ell}$,
is the most sensitive method at LEP and has been used by
ALEPH~[\ref{ref:bsmix_aleph}], DELPHI~[\ref{ref:bsmix_semil_delphi}],
OPAL~[\ref{ref:bsmix_inclept_opal}] and SLD~[\ref{ref:bsmix_lepd_sld}].
The principle of the method (see the discussion above in the case of
$\Bd$ mixing) is to reconstruct the ${\rm D}_s^{-}$ inclusively by relying on
topological vertexing and kinematical information.
Fairly loose criteria are applied to select large event samples,
see Table~\ref{tab:bsmix_methods}.

For this method, it is important to reduce the contribution from cascade decays
and to increase the \Bs\ purity of the sample (\Bs\ mesons represent about
10\% of all $b$-hadrons produced, see Table~\ref{tab:ratesstepa}).
To enrich the sample in direct \Bs\ semileptonic decays,
the following quantities are used:
momentum and transverse momentum of the lepton, impact 
parameters of all tracks in the opposite hemisphere relative
to the main event vertex, kaons at primary or
secondary vertices in the same hemisphere, and charge of the secondary vertex.
Those variables are usually combined in a global discriminant variable.
The result of this procedure is to increase the \Bs\ purity by
about 30$\%$; the corresponding mistag rate at decay is $\sim$10\% or less. 
The above information, as well as the proper time resolution,
is then used on an event-by-event basis.  
\begin{figure}
\hbox to\hsize{\hss
  \includegraphics[width=0.91\hsize]{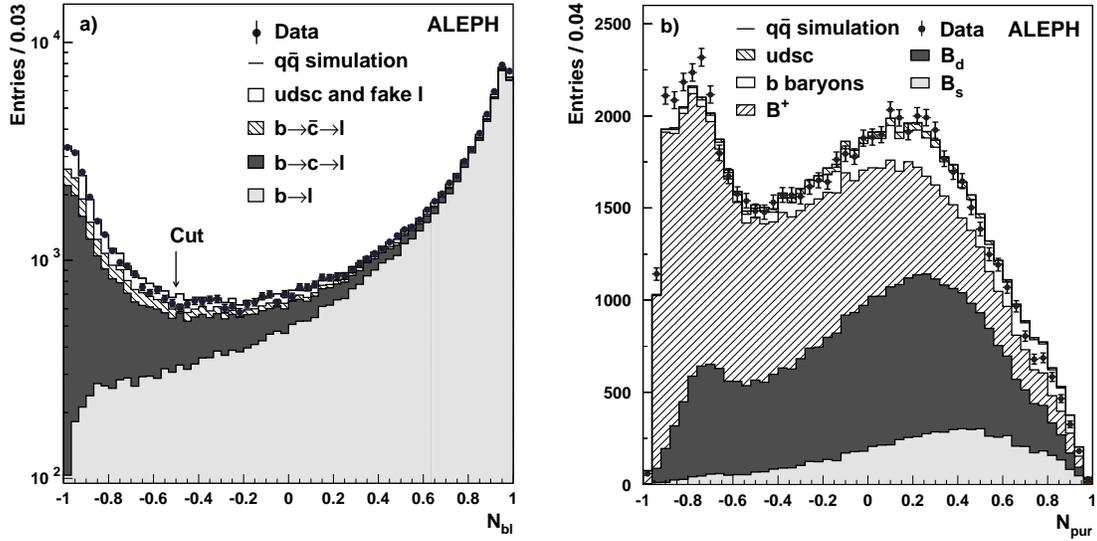}
\hss}
  \caption{\it \label{fig:purity_inclept_aleph}
  Distribution of (a) the $b \to \ell^-$ tagging variable and
  (b) the $\Bs$ purity variable for
  data (points) and Monte Carlo simulation (histograms)
  in the ALEPH inclusive lepton analysis.}
\end{figure}
As an example, Fig.~\ref{fig:purity_inclept_aleph} shows
the neural network output distributions sensitive to the
$b \to \ell^-$ fraction and the $\Bs$ purities in the ALEPH data.
The decay length resolution is somewhat worse than in the
case of semi-exclusive analyses due to missing or mis-assigned
tracks.

\subsubsection*{Inclusive methods}

\vspace{2mm}

\begin{figure}[htbp] %
\hbox to \hsize{\hss
\includegraphics[width=0.492\hsize,bb=21 129 540 624]{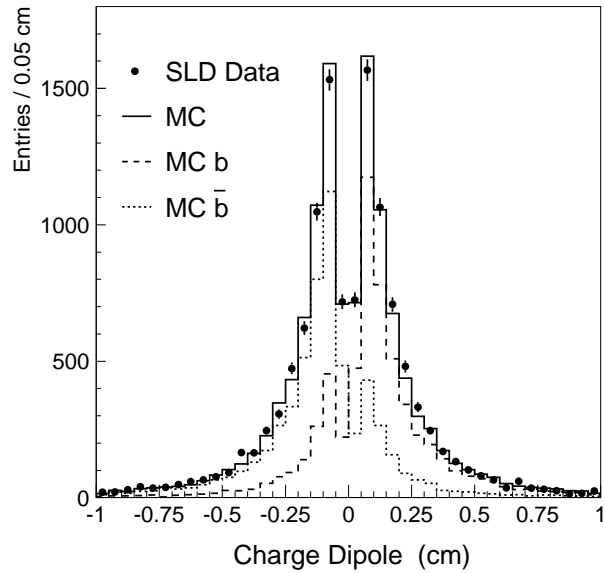}
\hss}
  \caption{\it \label{fig:dipole_sld}
  Distribution of the charge dipole for
  SLD data (points) and Monte Carlo (solid histogram).
  Also shown are the contributions from hadrons containing
  a $b$ quark (dashed histogram) or a $\bar{b}$ quark (dotted histogram).}
\end{figure}

Fully inclusive methods are sensitive to most ${\rm B}$ decay modes and,
thus, have high efficiency. Such techniques have been developed by
DELPHI~[\ref{ref:bdmix_vtx_delphi}] and
SLD~[\ref{ref:bsmix_dipole_sld}]. The analyses rely on inclusive
topological vertexing to select ${\rm B}$ decay products and to reconstruct
the ${\rm B}$ decay vertex. The DELPHI analysis is the same as the one
described earlier for $\Bd$ mixing. A very large data sample is
obtained but the mistag rates are high (see
Table~\ref{tab:bsmix_methods}).
SLD is able to exploit the excellent 3D
spatial resolution of its CCD-pixel vertex detector to cleanly
separate the charged decay products from secondary (originating
directly from the ${\rm B}$ decay) and tertiary (originating from cascade
${\rm D}$ decays) vertices. The decay flavour is determined from the
charge dipole $\delta Q$ defined as the distance between secondary and
tertiary vertices signed by the charge difference between them.
Positive (negative) values of $\delta Q$ tag $\overline{\rm B^0}$ (${\rm B}^0$) decays
as shown in Fig.~\ref{fig:dipole_sld}.

Table~\ref{tab:bsmix_methods} summarizes the different $\Bs$ mixing analyses.
It should be noted that the Table presents only the average performance
of the analyses and that most analyses substantially increase their
sensitivity by relying on event-by-event information.

\begin{table}
\begin{center}
\begin{small}
\begin{tabular}{|@{}llcrrrl@{}|}
\hline
Decay modes & Analysis & Events/Signal & $f_\mathrm{mode}$ & $p_W$
 & $\sigma_L$ & $\sigma_p/p$ \\[0.5ex]
\hline
 $\Bs \to {\rm D}_s^{(\ast)-} h^{+a}$ &
 ALEPH~[\ref{ref:bsmix_aleph}] &
 80/29 & 36\% &  0  & 180 $\mu$m & 0.005 (peak) \\
 & ~~~exclusive &
   &  &  &  & 0.03 (satellite) \\
 $\Bs \to {\rm D}_s^{(\ast)-} h^{+a}$ &
 DELPHI~[\ref{ref:bsmix_excl_delphi}] &
 44/23 & 52\% &  0  & 117 $\mu$m (58\%) & $^{b}$ \\
 ~~~~~~~~~~~$\Dzb {\rm K}^- h^{\prime +}$&
 ~~~exclusive &
   &  &  & 216 $\mu$m (42\%) &  \\
 $\Bs \to {\rm D}_s^- X$ &
 DELPHI~[\ref{ref:bsmix_excl_delphi}] &
 3079/1266 & 50\% & 10\% & 260 $\mu$m (77\%) & 0.10 (77\%)$^{c}$ \\
 & ~~~semi-excl. &
           &      &      & 304 $\mu$m (13\%) & 0.26 (23\%) \\
 &  &
           &      &      & 650 $\mu$m (10\%) &  \\
 &
 SLD~[\ref{ref:bsmix_dstracks_sld}] &
 361/174 & 55\% & 10\% &  50 $\mu$m (60\%) & 0.08 (60\%) \\
 & ~~~semi-excl. &
         &      &      & 151 $\mu$m (40\%) & 0.18 (40\%) \\
 $\Bs \to {\rm D}_s^- \ell^+ \nu$ &
 ALEPH~[\ref{ref:bsmix_aleph}] &
 333/156 & 47\% &   & 240 $\mu$m & 0.11 \\
 & ~~~semi-excl. &
         &      &   &            &      \\
 &
 CDF~[\ref{ref:bsmix_phil_cdf}] &
 ~~~~~~~~/1068 & 61\% &  &  & \\
 & ~~~semi-excl. &
         &      &   &            &      \\
 &
 DELPHI~[\ref{ref:bsmix_semil_delphi}] &
 ~~~~~~/436 & 53\% &  & 200 $\mu$m (82\%) & 0.07 (82\%) \\
 & ~~~semi-excl. &
        &        &  & 740 $\mu$m (16\%) & 0.16 (16\%) \\
 &
 OPAL~[\ref{ref:bsmix_dsl_opal}] &
 244/116 & 48\% &  & 500 $\mu$m & 0.10 \\
 & ~~~semi-excl. &
         &      &   &            &      \\
 $\Bs \to X \ell^+ \nu$ &
 ALEPH~[\ref{ref:bsmix_aleph}] &
 74k/~~~~~~ & 10\% & 13\%$^{d}$ & 251 $\mu$m (75\%) & 0.064 (60\%) \\
 & ~~~semi-incl. &
         &      &      & 718 $\mu$m (25\%) & 0.020 (40\%) \\
 &
 DELPHI~[\ref{ref:bsmix_semil_delphi}] &
 68k/~~~~~~ & 10\% & 8-18\% &  &  \\
 & ~~~semi-incl. &
         &      &  &  &  \\
 &
 OPAL~[\ref{ref:bsmix_inclept_opal}] &
 53k/~~~~~~ & 8\% & 12\%$^{d}$ &  &  \\
 & ~~~semi-incl. &
         &        &        &  &  \\
 &
 SLD~[\ref{ref:bsmix_lepd_sld}] &
 2k/~~~~ & 16\% & 4\% &  55 $\mu$m (60\%) & 0.06 (60\%) \\
 & ~~~semi-incl. &
       &      &     & 217 $\mu$m (40\%) & 0.18 (40\%) \\
 $\Bs \to {\rm all}$ &
 DELPHI~[\ref{ref:bdmix_vtx_delphi}] &
 770k/~~~~~~~~ & 10\% & 43\%$^{e}$ & 400 $\mu$m & 0.15 \\
 & ~~~inclusive &
           &        & 33\%$^{f}$ &   &   \\
 &
 SLD~[\ref{ref:bsmix_dipole_sld}] &
 11k/~~~~~~ & 16\% & 22\% &  78 $\mu$m (60\%) & 0.07 (60\%) \\
 & ~~~inclusive &
         &      &      & 304 $\mu$m (40\%) & 0.21 (40\%) \\
\hline
 \multicolumn{7}{l}{$^{a}$ $h^+$ stands for $\pi^+, \rho^+, a_1^+$ and $h^{\prime +}$
  stands for $\pi^+, a_1^+$.} \\
 \multicolumn{7}{l}{$^{b}$ For the best data subset ($\Bs$ peak and 1994-95 data).}\\
 \multicolumn{7}{l}{$^{c}$ Evaluated at $t = 1$ ps for the best subset of data.}\\
 \multicolumn{7}{l}{$^{d}$ Fraction of non-$(b \to \ell^-)$ decays.}\\
 \multicolumn{7}{l}{$^{e}$ For 615k vertices with charge dipole tag.}\\
 \multicolumn{7}{l}{$^{f}$ For 155k vertices with soft lepton tag.}
\end{tabular}
\end{small}
\end{center}
\caption{\it Summary of $\Bs$ mixing analyses showing the signal decay
modes, analysis method, total number of selected events and estimated
signal, fraction of signal mode in the selected sample
$f_\mathrm{mode}$, decay flavour mistag rate $p_W$ for $\Bs$ decays,
decay length and momentum resolutions. For semi-exclusive analyses,
the number of signal events corresponds to the number of ${\rm D}_s^-$
signal decays (not the number of signal events in the selected decay
mode) and $f_\mathrm{mode}$ represents the fraction of $\Bs$ in the
${\rm D}_s^-$ signal. \label{tab:bsmix_methods}}
\end{table}

\boldmath
\subsection{$\Bd$ oscillation results. Measurement of the $\Delta M_d$ frequency}
\label{sec:osciresults_bd}
\unboldmath
As detailed in Sec.~\ref{sec:anamethods}, many methods  
and channels have been used to study
$\Bd$--$\Bdb$~oscillations.
These analyses have been performed by the
ALEPH~[\ref{ref:bdmix_3methods_aleph},\ref{ref:bdmix_vtx_aleph}], 
BaBar~[\ref{ref:bdmix_excl_babar},\ref{ref:bdmix_dstarl_babar},\ref{ref:bdmix_dilept_babar}],
Belle~[\ref{ref:bdmix_excl_belle},\ref{ref:bdmix_dstarl_belle},\ref{ref:bdmix_dstarpi_belle},\ref{ref:bdmix_dilept_belle}], 
CDF~[\ref{ref:bdmix_dstarl_sst_cdf},\ref{ref:bdmix_dstarl_sst_long_cdf},\ref{ref:bphys_run1_cdf},\ref{ref:bdmix_dstarl_cdf},\ref{ref:bdmix_dimu_cdf},\ref{ref:bdmix_inclept_cdf}],
DELPHI~[\ref{ref:bsmix_excl_delphi},\ref{ref:bdmix_4methods_delphi},\ref{ref:bdmix_vtx_delphi}],
L3~[\ref{ref:bdmix_l3}],
OPAL~[\ref{ref:bdmix_dstar_opal},\ref{ref:bdmix_pil_opal},\ref{ref:bdmix_dilept_opal},\ref{ref:bdmix_inclept_opal}] and 
SLD~[\ref{ref:bdmix_inclept_sld}--\ref{ref:bdmix_ktag_sld}] collaborations. 

\begin{figure}
\centerline{
\includegraphics[height=0.34\hsize]{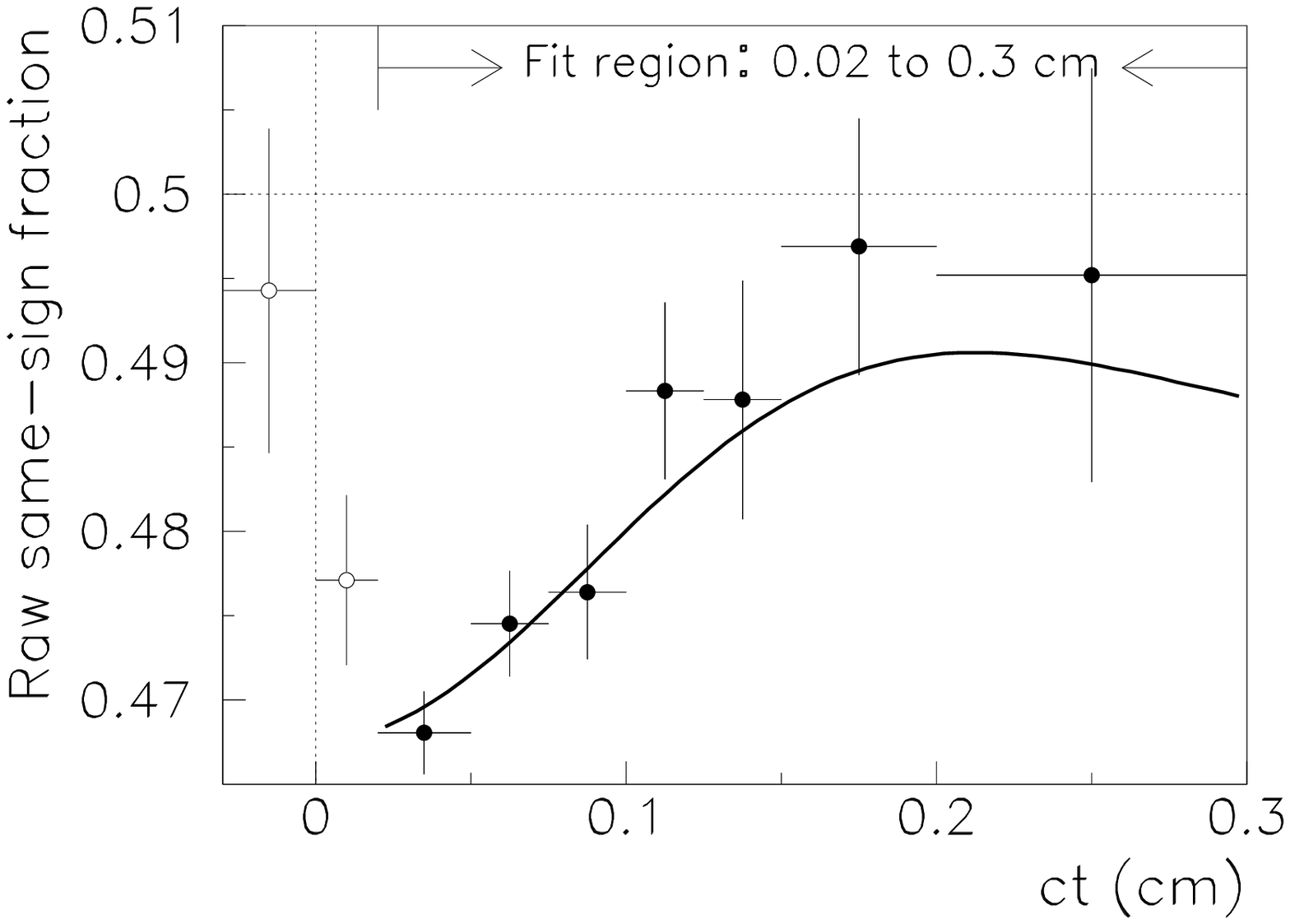}
\hfill
\includegraphics[height=0.34\hsize]{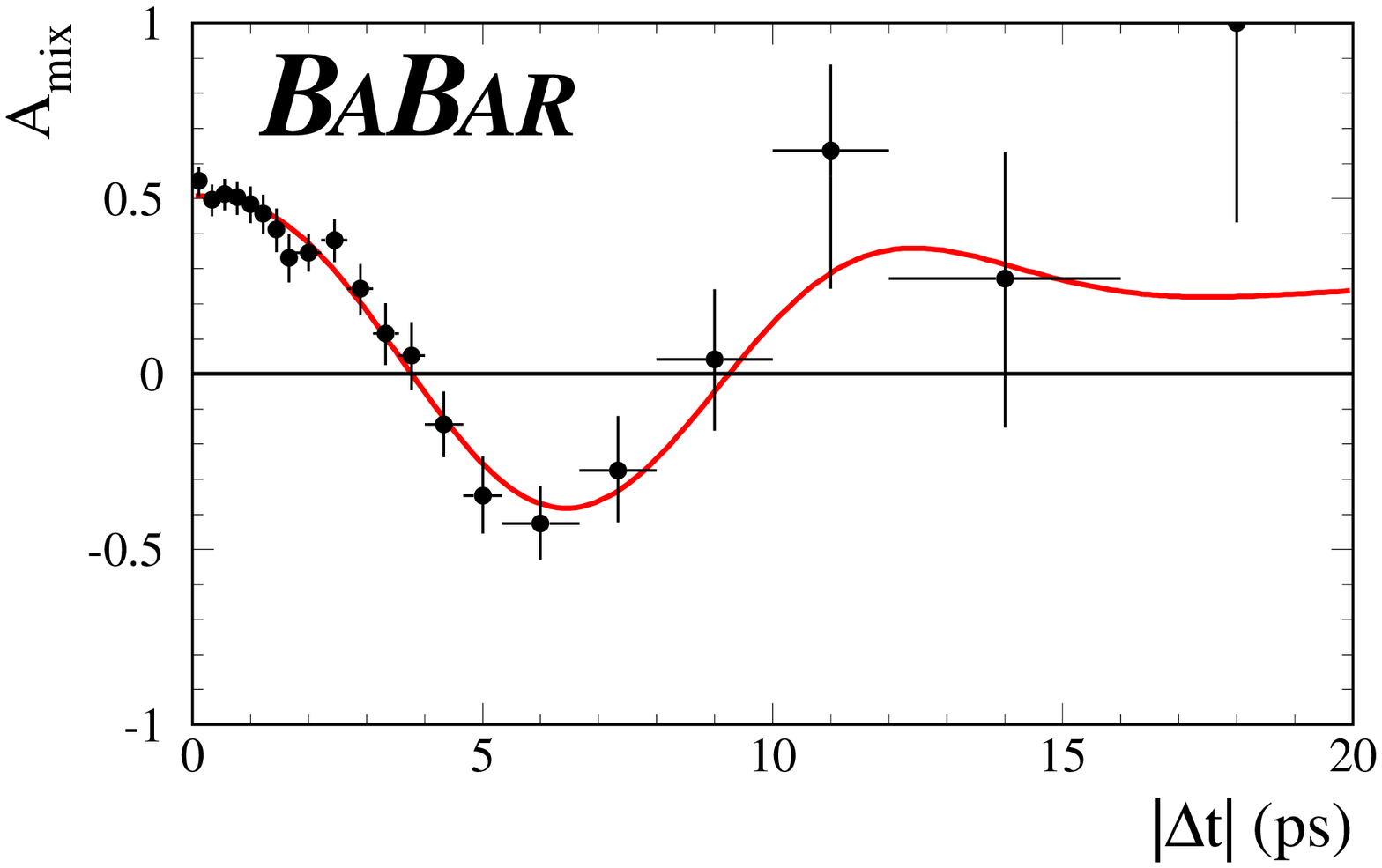}
\put(-285,40){\Large\bf (a)}
\put(-35,40){\Large\bf (b)}
}
\caption{\it Examples of $\Delta M_d$ results from
(a) CDF (Ref.~[\ref{ref:bdmix_inclept_cdf}])
and 
(b) BaBar (Ref.~[\ref{ref:bdmix_dilept_babar}]). 
See text for details.
}
\label{fig:bdmix_example1}
\end{figure}

In the following, we will discuss the results of a 
few representative measurements of $\Delta M_d$. 
Fig.~\ref{fig:bdmix_example1}(a) showss the fraction of mixed events as
a function of proper decay length for a semi-inclusive analysis at CDF
using a lepton sample with an inclusively reconstructed vertex combined, on
the opposite side, with a lepton and jet charge tag to
infer the production flavour [\ref{ref:bdmix_inclept_cdf}]. Although this
analysis is based on about 240,000 events, the total height of the oscillation
amplitude is small ($\sim 0.05$) due to an effective tagging efficiency of
$\varepsilon (1-2 p_W)^2 \sim 1\%$ for each tag yielding a value
of $\Delta M_d = (0.500 \pm 0.052 \pm 0.043)$~ps$^{-1}$. 
In this analysis, a large mistag rate $p_W$ resulting in $(1-2 p_W)$ being small is
compensated by the number of events $N$ being large (see
Eq.~(\ref{eq:bs_sensitivity})\,).
This result can be compared to
a measurement from BaBar~[\ref{ref:bdmix_dilept_babar}] based on about $\sim
6300$ neutral ${\rm B}$~mesons 
fully reconstructed in multihadronic modes (mainly 
${\rm B}_d^0\rightarrow {\rm \bar{D}}^{(*)} X$).
An opposite lepton and kaon tag with low mistag fractions of $p_W \sim 8\%$
and $\sim 16\%$, respectively, are the reason for an oscillation amplitude
of $\sim 0.5$ in the mixed asymmetry as shown in Fig.~\ref{fig:bdmix_example1}(b). 
Note the statistical error on the
$\Delta M_d$ value obtained by BaBar for this analysis: 
$\Delta M_d = (0.516 \pm 0.016 \pm 0.010)$ ps$^{-1}$.
From this example we can see the trade-off between a poor tagging power in
high statistics ${\rm B}$~samples produced for example in a hadronic 
$p\bar p$~environment at 
the Tevatron and lower statistics analyses with superior tagging and low
mistag probabilities in an $e^+e^-$~environment for example at the
B~factories. In addition, compared to 
inclusive methods, analyses with fully reconstructed ${\rm B}$~mesons have a higher 
sample purity.

\begin{figure}
\centerline{
\includegraphics[width=0.63\hsize]{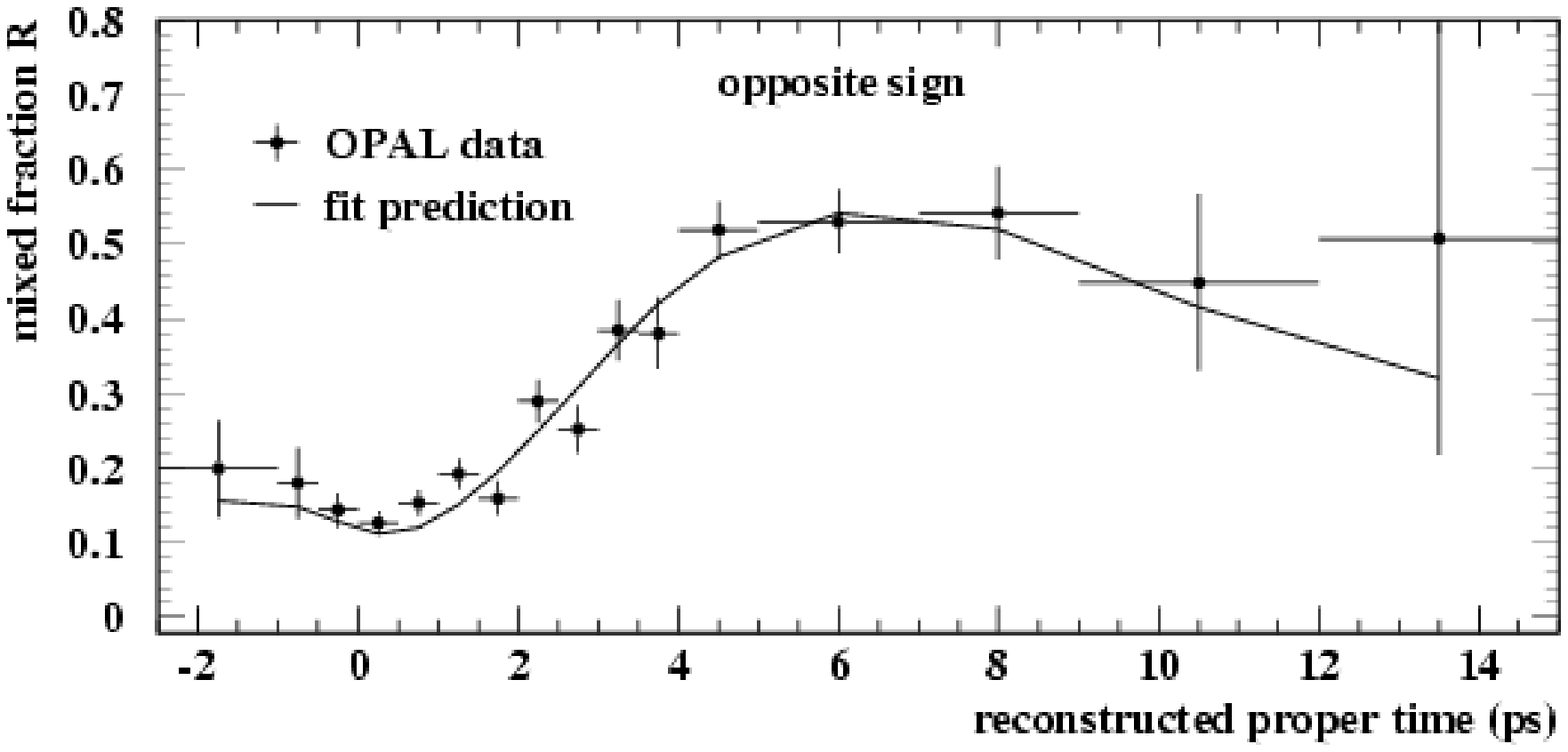}
\hfill
\includegraphics[width=0.37\hsize]{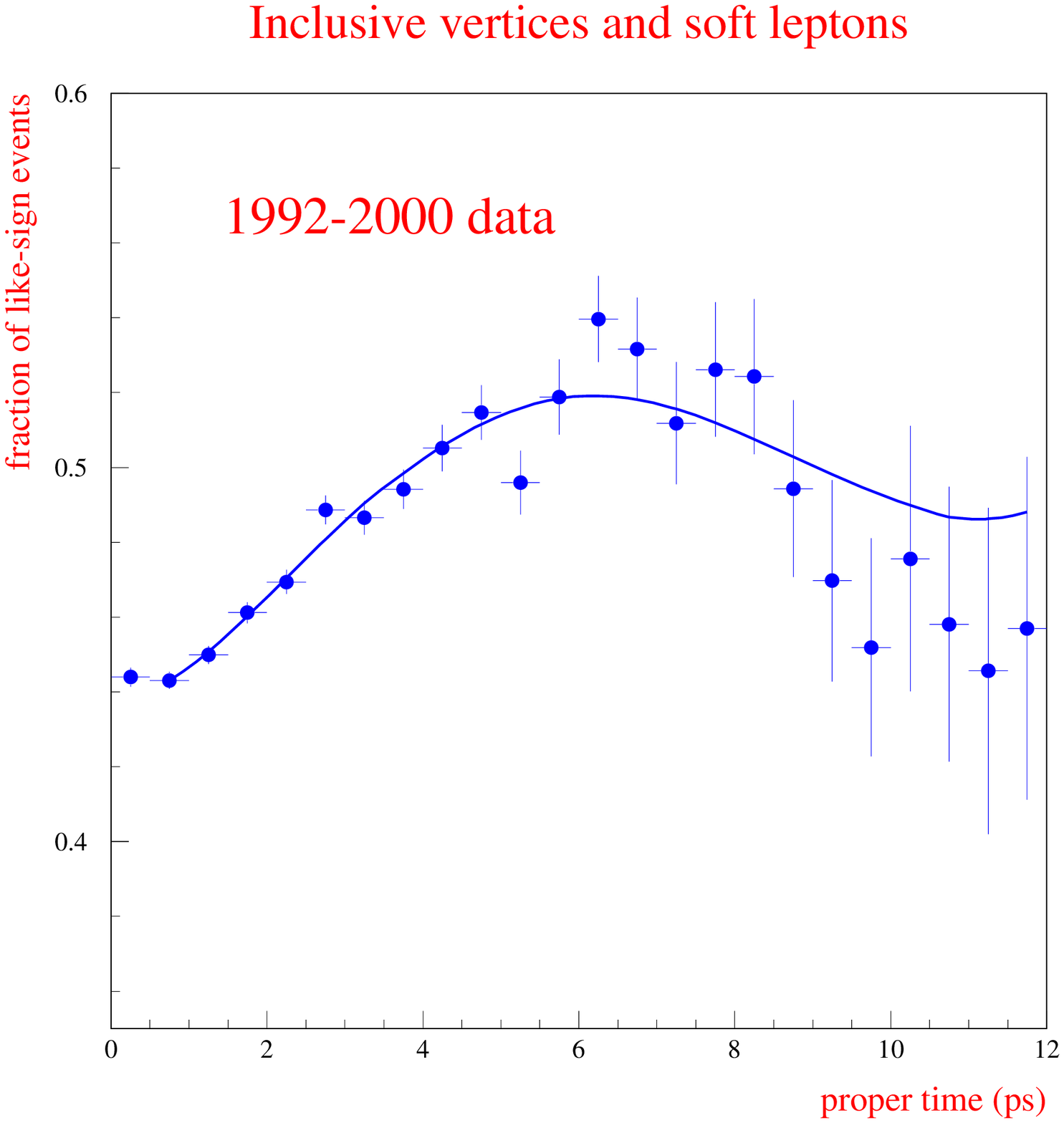}
\put(-210,35){\Large\bf (a)}
\put(-35,25){\Large\bf (b)}
}
\caption{\it 
Examples of $\Delta M_d$ results from
(a) OPAL (Ref.~[\ref{ref:bdmix_pil_opal}]) and 
(b) DELPHI (Ref.~[\ref{ref:bdmix_vtx_delphi}]).
See text for details.
}
\label{fig:bdmix_example2}
\end{figure}

Fig.~\ref{fig:bdmix_example2} shows the result of two other $\Delta M_d$
analyses.  
One of the most precise single measurements performed at the
$Z^0$~resonance is an inclusive ${\rm D}^*$~analysis by
OPAL~[\ref{ref:bdmix_pil_opal}] using 
${\rm B}^0\rightarrow {\rm D}^{*-}\ell^+\nu$~decays.    
High statistics  ${\rm D}^{*-}\rightarrow \bar {\rm D}^0\pi^-$~decays were
reconstructed using the 
slow $\pi^-$ from the ${\rm D}^{*-}$~decay while inferring the $\overline {\rm D}^0$ with an 
inclusive technique. Same-sign lepton-pion pairs serve to constrain
the combinatorial background in the opposite sign lepton-pion pair
signature. A clear oscillation signal is observed in the fraction of
mixed events as can be seen in
Fig.~\ref{fig:bdmix_example2}(a). A value of 
$\Delta M_d = (0.497\pm0.024\pm0.025)$~ps$^{-1}$ is extracted.
Another example of a precise $\Delta M_d$ analysis at the $Z^0$~pole by DELPHI 
is shown in Fig.~\ref{fig:bdmix_example2}(b).
A sample of 770,000 events with an inclusively reconstructed vertex 
has been selected. Tags based on several separating variables such as 
the jet charge, dipole charge and the transverse momentum of the (soft)
lepton have been combined into a probability to 
determine the fraction of like-sign events as displayed in  
Fig.~\ref{fig:bdmix_example2}(b). DELPHI obtains a value of 
$\Delta M_d = (0.531 \pm 0.025 \pm 0.007)$~ps$^{-1}$.

\begin{figure}
\centerline{
\includegraphics[width=0.99\hsize]{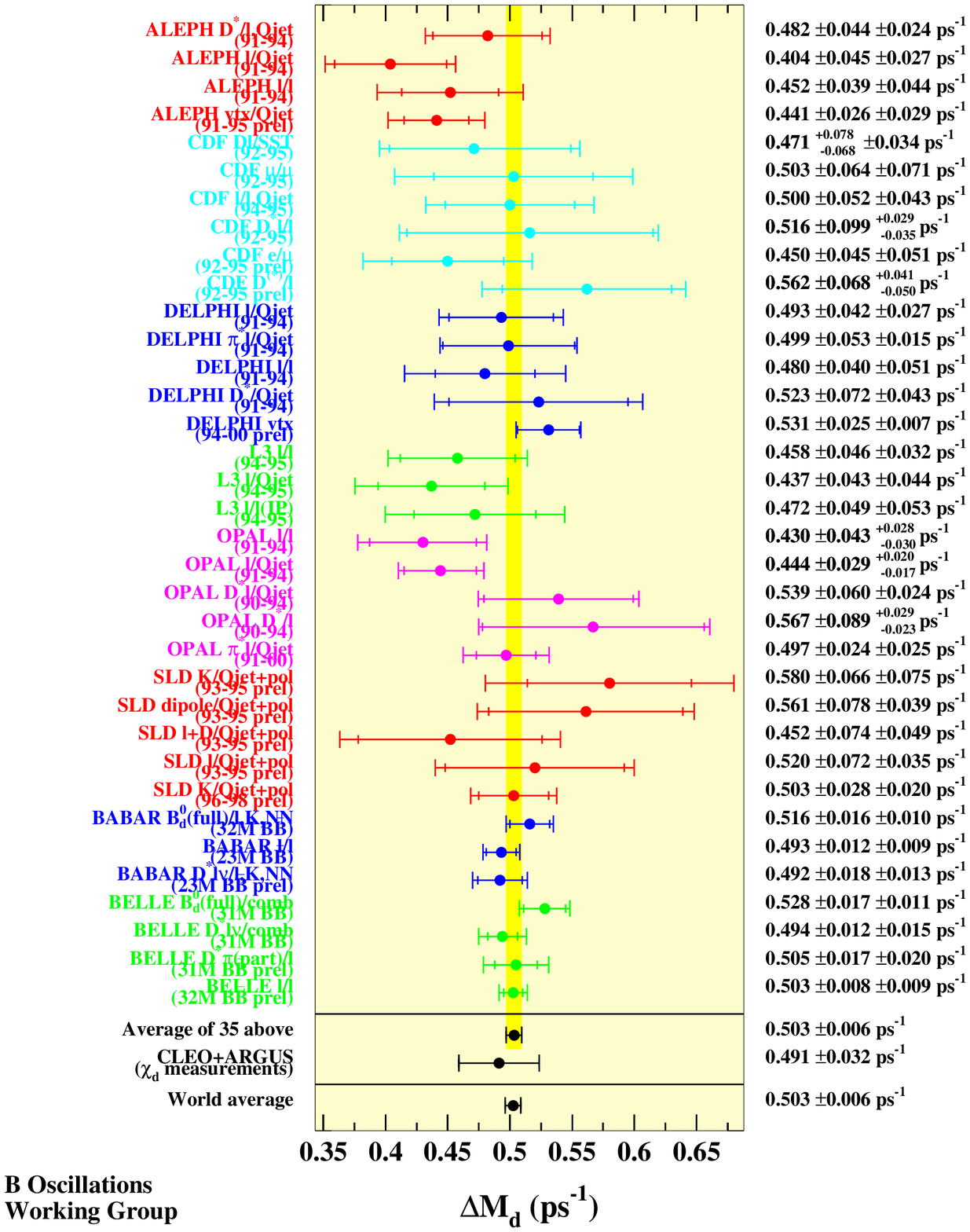}
}
\caption{\it 
Individual and combined measurements of $\Delta M_d$ at ${\rm B}$
factories, LEP, SLD and CDF as of the ICHEP 2002 conference~[\ref{ref:osciw}]. 
The quoted world average, at the bottom, also includes 
$\chi_d$ measurements performed by ARGUS and CLEO. 
}
\label{fig:dmd_comp_all}
\end{figure}

In order to combine all individual $\Delta M_d$ results to obtain a world average value, 
possible statistical correlations 
between individual analyses have to be taken into account and also
the systematic errors which are often not negligible have to be combined
properly. The main sources of systematic uncertainties are determinations of
sample compositions, mistag probabilities, $b$~hadron production fractions
and contributions from $b$~hadron lifetimes. 
Before being combined, the measurements are adjusted on the basis of a
common set of input values.
Details of the averaging  procedure
are described in Ref.~[\ref{lepsldcdf}].

\begin{figure}
\centerline{
\includegraphics[height=0.6\hsize]{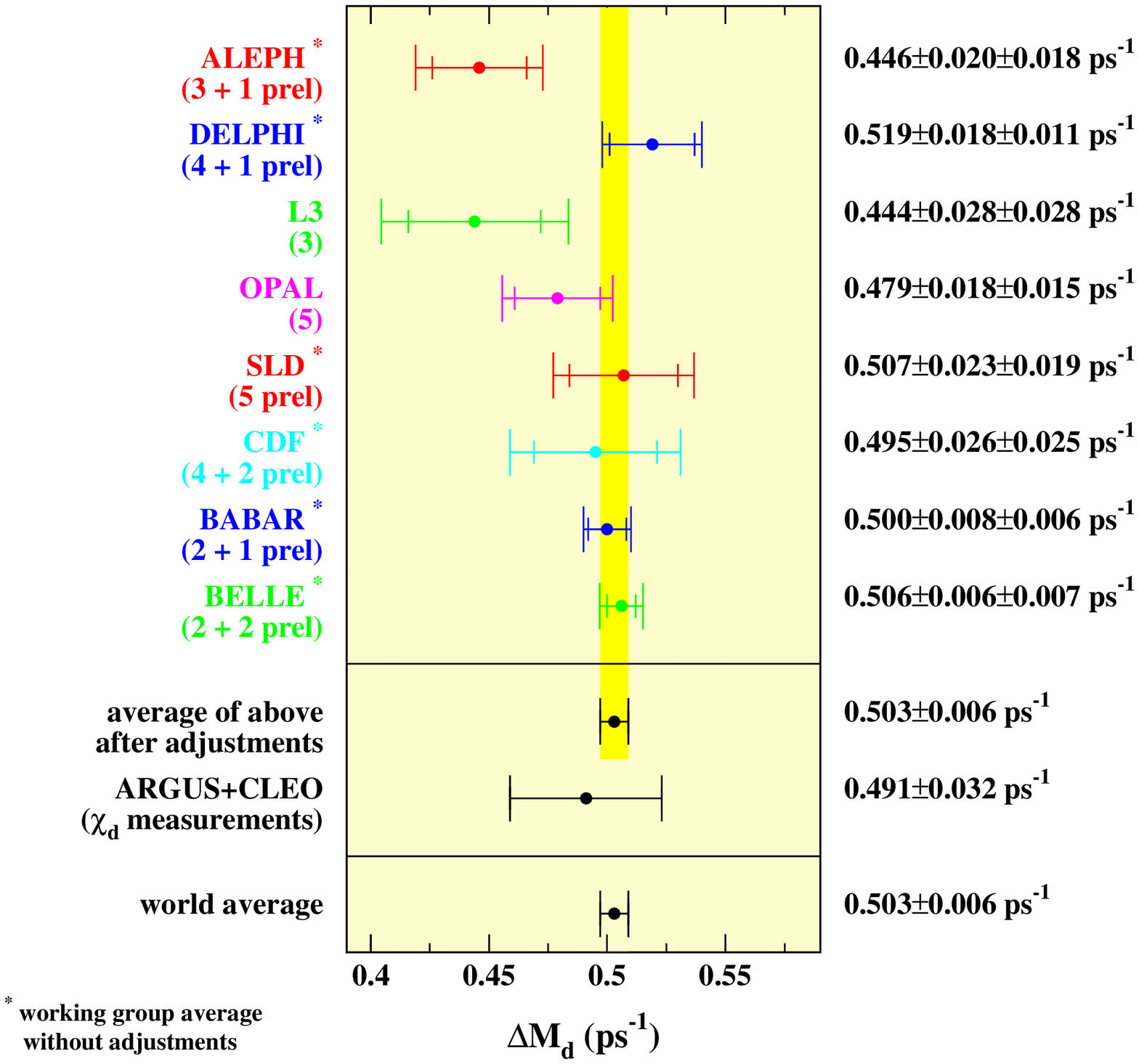}
}
\caption{\it 
Combined measurements of $\Delta M_d$ averaged by experiment 
as of the ICHEP 2002 conference [\ref{ref:osciw}].  
The quoted world average, at the bottom, also includes
$\chi_d$~measurements by ARGUS and CLEO. 
}
\label{fig:dmd_comp_exp}
\end{figure}

A compilation of all $\Delta M_d$ measurements available as of the 2002
ICHEP conference, 
can be found in Fig.~\ref{fig:dmd_comp_all}.
The individual results from each experiment are combined and averaged using
the procedure described above. 
There exist also time-integrated measurements of $\Bd$ mixing from the 
ARGUS~[\ref{Albrecht:1992yd},\ref{Albrecht:1994gr}]
and CLEO~[\ref{Bartelt:1993cf},\ref{Behrens:2000qu}]
collaborations which can be converted into a value for $\Delta M_d$ 
assuming the width difference $\Delta \Gamma_d$ in the $\Bd$~system to be
zero and no $CP$~violation in $\Bd$~mixing.
The quoted world average, at the bottom of Fig.~\ref{fig:dmd_comp_all},
also includes $\chi_d$~measurements by ARGUS and CLEO. 
The $\Delta M_d$~averages per experiment are
displayed in Fig.~\ref{fig:dmd_comp_exp}. 

The different results from the combination procedure are [\ref{ref:osciw}]:

\vspace{1mm}

\begin{equation}
\begin{array}{|c|}\hline\left.
\Delta M_d =
  \cases{(0.491 \pm 0.041)~\ps^{-1} & Argus-CLEO (from $\chi_d$)\cr
         (0.498 \pm 0.013)~\ps^{-1} & LEP-SLD-CDF\cr
         (0.503 \pm 0.007)~\ps^{-1} & Belle-BaBar\cr
         (0.503 \pm 0.006)~\ps^{-1} & world average\cr }          
\ .\right.
\\ \hline\end{array}
\label{eq:dmdcom}
\end{equation}

\vspace{3mm}

At the end of the LEP-CDF-SLD era, $\Delta M_d$ has been determined
with a relative precision of about $2.6\%$. The LEP-CDF-SLD results are
in excellent agreement with the Belle-BaBar measurements. After the
inclusion of the results from ${\rm B}$~factories, the precision on
$\Delta M_d$ is 
improved by a factor of two. The world average $\Bd$~mixing frequency is
now dominated by the results of ${\rm B}$~factories.

\boldmath
\subsection{Results on $\Bs$ oscillations. 
Limits on the $\Delta M_s$ frequency} 
\label{sec:osciresults_bs}
\unboldmath

\vspace{2mm}

$\Bs$--$\Bsb$~oscillations have also been the subject of many studies by
ALEPH~[\ref{Barate:1998ua},\ref{Barate:1998rv},\ref{ref:bsmix_aleph}],
CDF~[\ref{ref:bsmix_phil_cdf}], 
DELPHI~[\ref{ref:bsmix_excl_delphi},\ref{Abreu:2000sh},\ref{ref:bsmix_semil_delphi},\ref{Adam:1997pv}], 
OPAL~[\ref{ref:bsmix_dsl_opal},\ref{ref:bsmix_inclept_opal}] and 
SLD~[\ref{Thom:2002fs},\ref{ref:bsmix_dstracks_sld},\ref{ref:bsmix_lepd_sld},\ref{ref:bsmix_dipole_sld}]. 
No oscillation signal has been observed to date. To set lower
limits on the oscillation frequency $\Delta M_s$, all $\Bs$ mixing analyses
use the amplitude method~[\ref{ref:moser}] described in 
Sec.~\ref{sec:amp_description}  

\begin{figure}
\centerline{
\includegraphics[width=0.495\hsize]{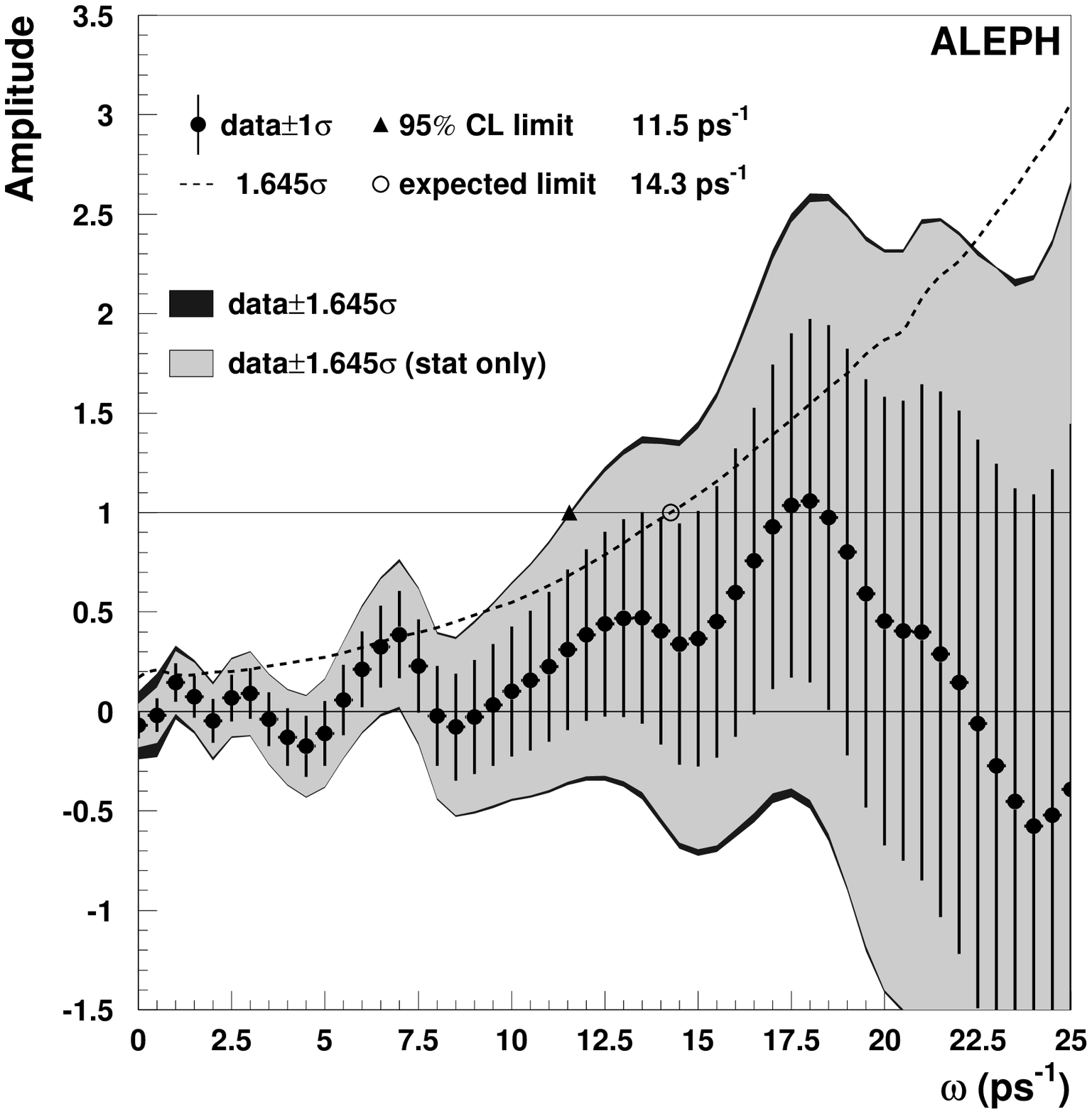}
\hfill
\includegraphics[width=0.505\hsize]{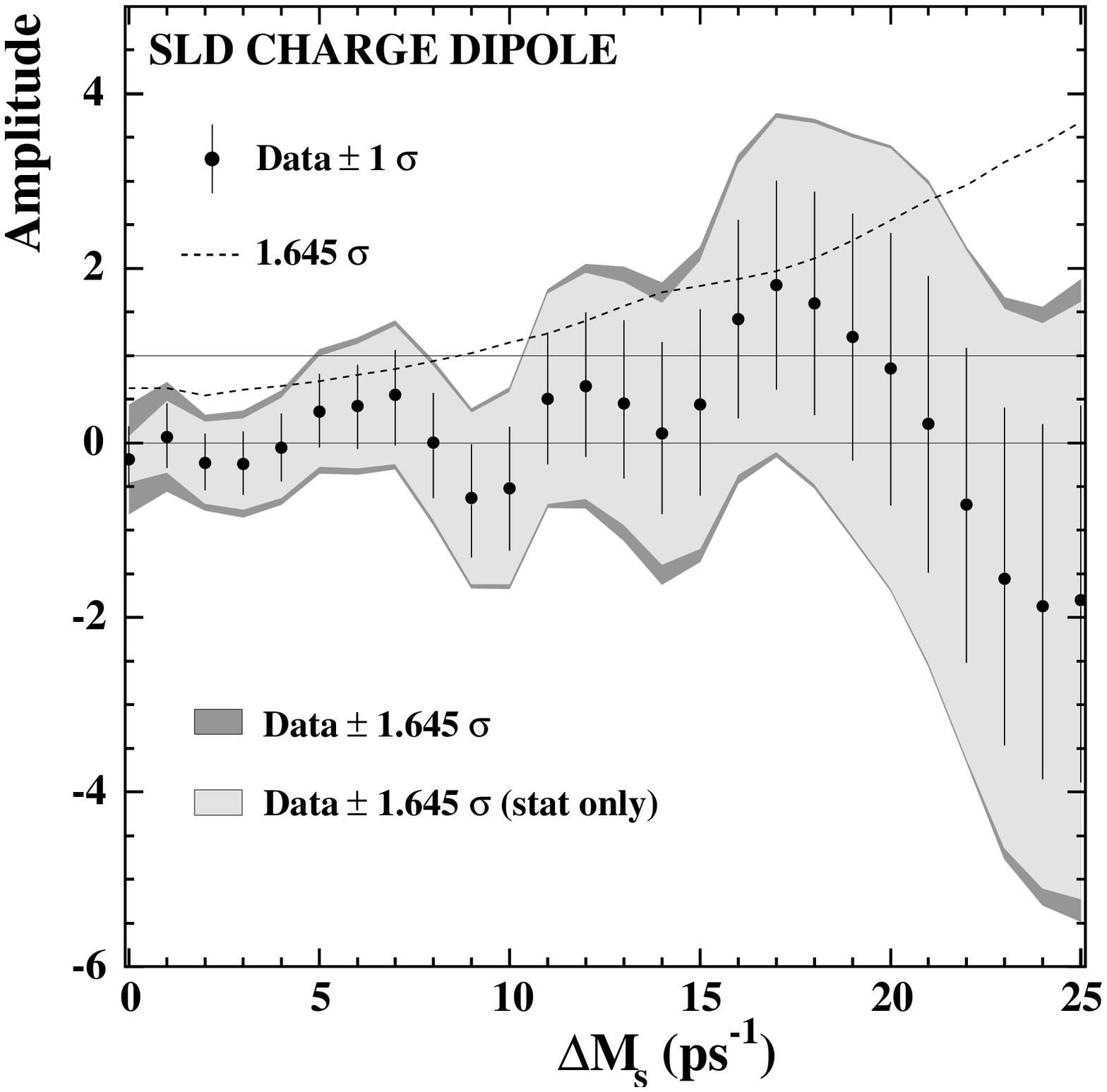}
\put(-415,35){\Large\bf (a)}
\put(-190,37){\Large\bf (b)}
}
\caption{\it 
Examples of measured $\Bs$ oscillation amplitudes as a function of the
mixing frequency $\Delta M_s$ from 
(a) ALEPH (Ref.~[\ref{ref:bsmix_aleph}]) and
(b) SLD (Ref.~[\ref{ref:bsmix_dipole_sld}]).
See text for details.
}
\label{fig:dms_comp_AS}
\end{figure}

\newpage

Two examples of measured $\Bs$ oscillation amplitudes as a function of the
mixing 
frequency $\Delta M_s$ are shown in Fig.~\ref{fig:dms_comp_AS}. 
The ALEPH collaboration recently presented an improved search for
$\Bs$ 
oscillations combining three analyses based on different 
final states~[\ref{ref:bsmix_aleph}]. 
First, fully reconstructed decays of $\Bs$ mesons yield a
small sample of $\Bs$ candidates with excellent decay length and
momentum resolution. Semileptonic decays with a reconstructed \Ds~meson
provide a second sample with larger statistics, high 
$\Bs$ purity but with a poorer
momentum and decay length resolution due to the partial decay
reconstruction. Finally, semileptonic ${\rm B}$~hadron decays are inclusively
selected and yield the data sample with the highest sensitivity to
$\Bs$ oscillations since the higher statistics compensates for the low
average $\Bs$ purity and the poorer proper time resolution. 
Fig.~\ref{fig:dms_comp_AS}(a) shows the fitted amplitude spectrum as a
function of $\Delta M_s$ for the third sample. From this
inclusive semileptonic sample alone, ALEPH excludes all frequencies below
11.5~ps$^{-1}$, while the combined 95\% C.L. limit from all three analyses
yields $\dms > 10.9$~ps$^{-1}$. 

Fig.~\ref{fig:dms_comp_AS}(b) shows the amplitude spectrum from an
analysis by SLD~[\ref{ref:bsmix_dipole_sld}].
This analysis determined the ${\rm B}$~flavour at production time by exploiting the
large forward-backward asymmetry of polarized $Z^0\rightarrow b\bar b$~decays and
uses additional information from the hemisphere opposite to that of the
reconstructed ${\rm B}$~decay such as the jet charge, the lepton and kaon tags. The
${\rm B}$~flavour at decay is tagged by a charge 
dipole method as explained in Sec.~\ref{subsec:bsosc}
 Although this analysis is based on only 11,000 decays, it
reaches a sensitivity of 8.8~ps$^{-1}$ because of the slower rise of the
uncertainty on the amplitude due to the excellent proper time resolution. 

\begin{figure}
\centerline{
\includegraphics[width=0.5\hsize]{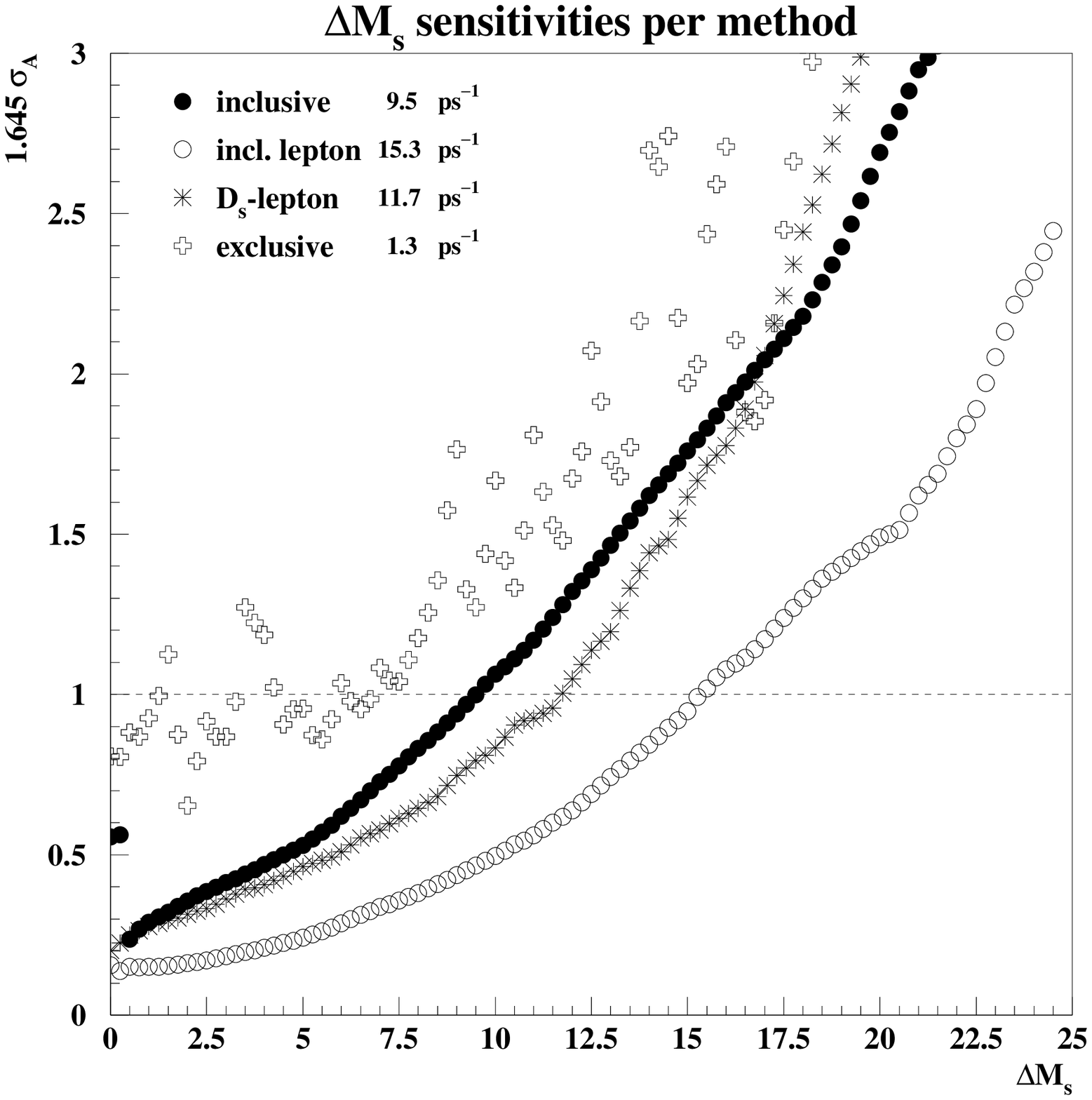}
\hfill
\includegraphics[width=0.5\hsize]{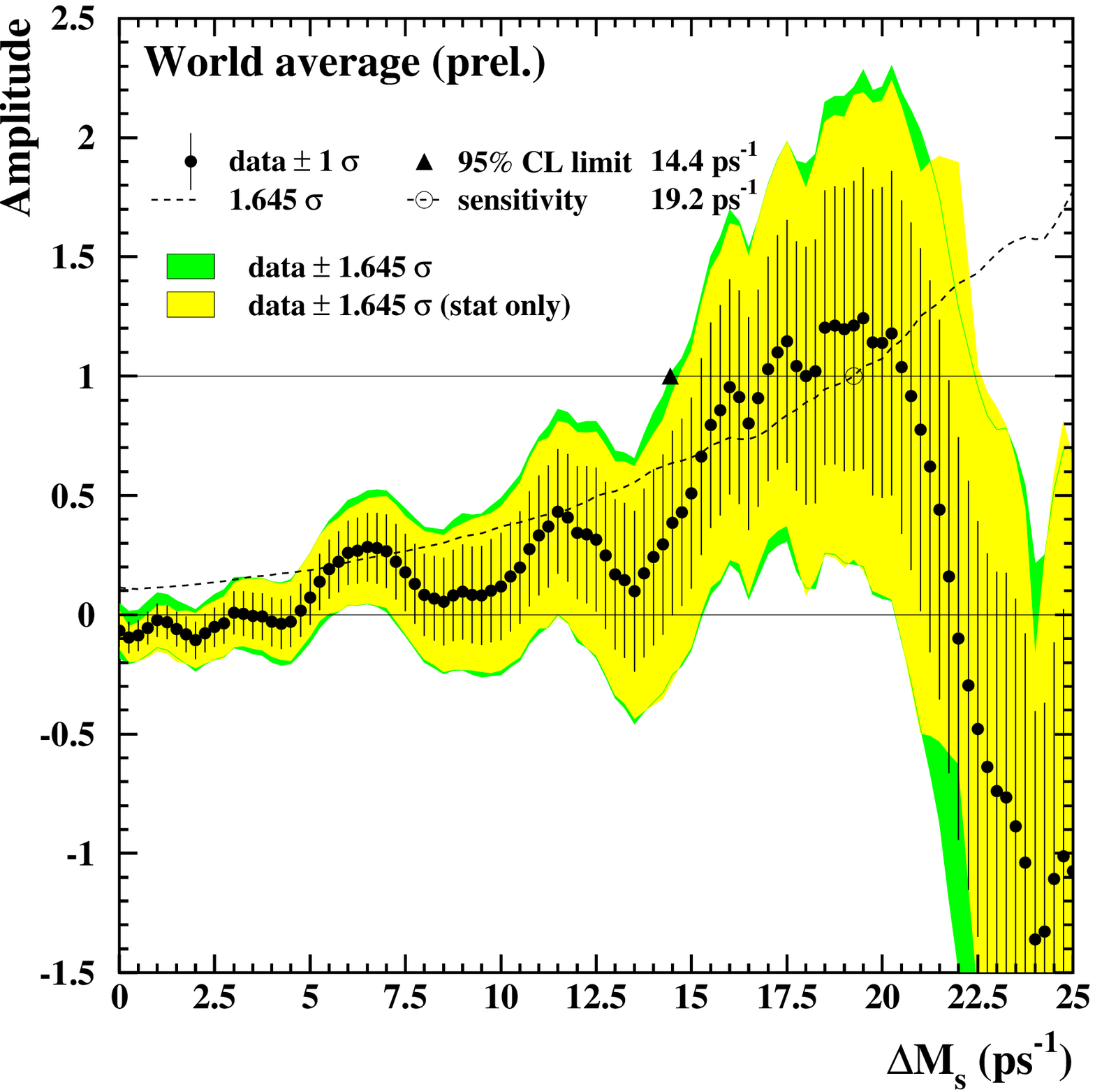}
\put(-265,35){\Large\bf (a)}
\put(-190,39){\Large\bf (b)}
}
\caption{\it 
(a) Uncertainty ($1.645\,\sigma_{A}$) on the amplitude $ A$ as a
function of $\Delta M_s$ for the various $\Bs$ oscillation analyses. The
dashed line corresponds to the condition $1.645\,\sigma_{A} = 1$ used to
define the 95\% C.L.~sensitivity. 
(b) Combined measurements of the $\Bs$ oscillation amplitude as a function of
$\Delta M_s$, including all results as of the ICHEP 2002 conference.
Neighbouring points are statistically correlated.
}
\label{fig:dms_comp_all}
\end{figure}

No $\Bs$ oscillation signal has been seen so far. 
The most sensitive analyses are the ones
based on the inclusive lepton samples at LEP. Because of better proper
time resolution, smaller data samples of inclusive decays analyzed at
SLD as well as measurements using only a few fully reconstructed
$\Bs$ decays at LEP, turn out to be very useful to explore the high
$\Delta M_s$ region. This point is illustrated in
Fig.~\ref{fig:dms_comp_all}(a) showing the $\Delta M_s$ sensitivities for the
different $\Bs$ oscillation analysis methods. The uncertainty
on the amplitude $A$ (actually $1.645\,\sigma_{A}$) is plotted
as a function of $\Delta M_s$ combining the existing results of the various
$\Bs$ analyses methods from different experiments.  
The combination of all fully inclusive methods crosses the
dashed line corresponding to the condition $1.645\,\sigma_{A} = 1$ used to
define the 95\% C.L.~sensitivity at about 9.5~ps$^{-1}$. 
This represents
the combined sensitivity of all inclusive methods from the various
experiments. 
Due to the combination of high statistics and adequate
vertexing resolution, the inclusive lepton methods give currently the most
sensitive results. The \Ds-lepton samples also reach a high sensitivity while the
exclusive methods that attempt to fully reconstruct hadronic $\Bs$
decays have a lower sensitivity because of the small number of $\Bs$
candidates that have been exclusively reconstructed to date. However,
the slow growth of the amplitude error for the exclusive method can be
inferred from Fig.~\ref{fig:dms_comp_all}(a).
Note, the visible scattering of points for the exclusive method which results
from the small number of events contributing in these analyses.

All available results on $\Delta M_s$~oscillations can be combined into a world
average exclusion limit using the amplitude method. All data on the
measurements of $\Bs$~oscillation amplitudes versus $\Delta M_s$, as
provided by the experiments,  
are averaged to yield a combined amplitude $ A$ as a function of $\Delta M_s$
as shown in Fig.~\ref{fig:dms_comp_all}(b). 
The individual results have been adjusted to common physics inputs and all
known correlations have been accounted for. The sensitivities of the
inclusive analyses which depend on the assumed fraction, $f_{\Bs}$, of $\Bs$ 
mesons have been re-scaled to a common average of 
$f_{\Bs} = 0.093 \pm 0.011$ (see Table~\ref{tab:ratesstepa}).
Figure~\ref{fig:dms_comp_all}(b) includes all results as of the ICHEP 2002 
conference. The measurements are dominated by statistical uncertainties.
Neighbouring points are statistically correlated. 
The combined result is [\ref{ref:osciw}]:
\begin{equation}
\begin{array}{|c|}
\hline
\dms > 14.4 ~ {\rm ps}^{-1}~{\rm at\  95\% \rm C.L.}  \\
{\rm~with~ a~ sensitivity~ of~ \dms = 19.2~{\rm ps}^{-1}}
\\ \hline\end{array}
\end{equation}
Values between $14.4\,\mathrm{ps}^{-1}$ and $\sim 22\,\mathrm{ps}^{-1}$ 
cannot be excluded because the data appear to be
compatible with a signal in this region.
The amplitude plot presents a deviation from $A=0$ at about 
$\dms \sim 17.5$~ps$^{-1}$ for which a significance of $\sim 2.2~\sigma$ can be 
derived.
This means that there is not enough sensitivity for the observation of
a $\Bs$--$\overline\Bs$ signal at this frequency.

\begin{figure}
\vspace{-2mm}
\centerline{
\includegraphics[width=0.7\hsize]{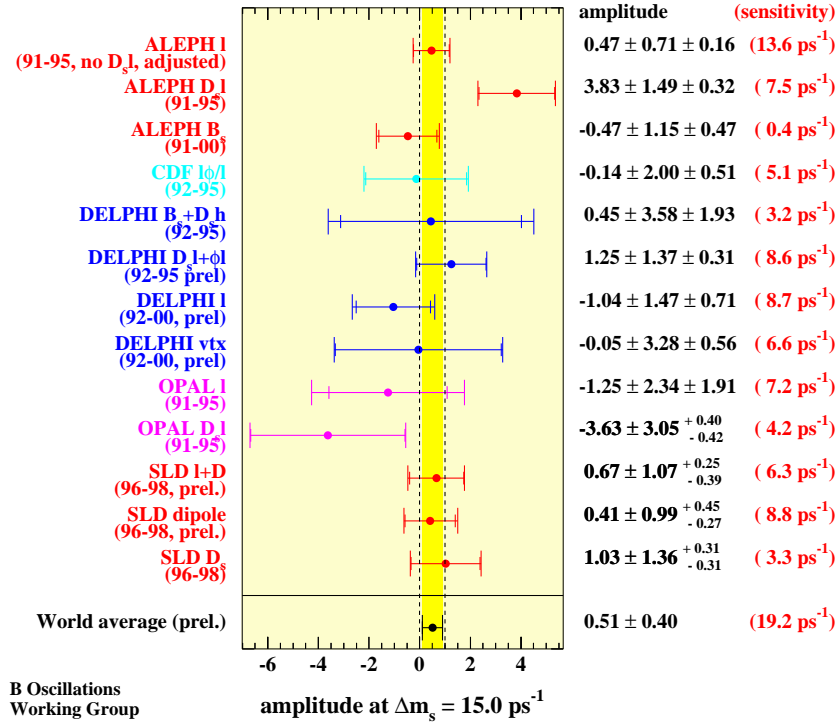}
}
\vspace{-.5cm}
\caption{\it 
Measurements of the $\Bs$ oscillation amplitude as of the ICHEP 2002 
conference. The amplitudes are given at 
$\dms = 15$~ps$^{-1}$ along with the relevant statistical and systematic
errors. The exclusion sensitivities are indicated on the right, within in parentheses,
The shaded area indicates the $\pm$ 1 $\sigma$ region on the average, 
and the dashed lines correspond to the values 0 and 1.}
\label{fig:dms_comp_exp}
\end{figure}

The different measurements of the $\Bs$ oscillation amplitude as of the
ICHEP 2002 conference are shown in Fig.~\ref{fig:dms_comp_exp}, where
the amplitudes for the various analyses are given at $\dms = 15$~ps$^{-1}$ 
along with the relevant statistic and systematic errors. 
The exclusion sensitivities are also indicated
Fig.~\ref{fig:dms_comp_exp} shows which analyses contribute most in the
high $\Delta M_s$ region.  
Note that the individual measurements are quoted as in the original
publications, but the averages include the effects of adjustments to a
common set of input parameters. 

Although all $\Bs$ mixing results are presently limited by statistics, a
discussion of systematic uncertainties in these analyses is relevant
for a future measurement of $\Bs$ oscillations. 
Critical analysis parameters
($\sigma_L$, $\sigma_p$ and $p_W$)
are extracted from detailed Monte Carlo simulation and are subject
to modelling uncertainties.
A first level of control is typically achieved with detailed comparisons
between data and MC.
In addition, measurements from calibration samples are performed
to cross-check the parameters directly from the data but not
all critical parameters can be tested in this manner.
Of particular importance to the sensitivity at large $\dms$ values
is the proper time resolution and, in particular, the
decay length resolution. The latter has been
tested with a variety of techniques: fit to the decay length
distribution of $\tau$ decays, fit for the primary vertex
in $Z^0$ decays to light-flavour quarks, study of tracks
with negative impact parameter.
These studies find that the decay length resolution is typically understood
at the 10\% level or better.

\boldmath
\subsection{Future prospects for \dmd\ and $\dms$ determination}
\label{sec:futureprospects}
\unboldmath
The current world average $\Bd$~oscillation frequency
constitutes a measurement at about 1$\%$ precision. It is dominated by the results of the
${\rm B}$~factories which will further improve the precision on $\Delta m_d$.
The uncertainty on the $\Bd$~lifetime starts to become a main contributor
to the systematic error on future measurements of $\Delta M_d$. A
simultaneous fit of the ${\rm B}$~lifetime and $\Delta M_d$ will improve this
situation as demonstrated in Ref.~[\ref{ref:bdmix_excl_babar}]. For a data
sample of 300~fb$^{-1}$, the BaBar and Belle experiments 
expect to improve the $\Bd$~oscillation frequency by a factor two, down 
to a precision of about 0.4\% 

The future interest in B~mixing clearly lies in a measurement of 
$\Bs$~oscillations. Some of the still preliminary analyses from LEP and SLD are
in the process of being finalized for publication while no new measurements
or improved limits are to be expected. Since no $\Bs$~mesons are produced
at the ${\rm B}$~factories and running at the $\Upsilon(5S)$~resonance as a
source of $\Bs$~mesons is not foreseen in the near future, the hopes of the
heavy flavour community focus on the Tevatron Collider experiments CDF and
D\O\ to measure $\Bs$~oscillations. 
For such a measurement it is important that
the resolution of the vertexing device is good enough to resolve the expected
(rapid) oscillations while a small boost correction will prevent the measured
oscillations to damp out with proper time. The path to measure
$\Bs$~oscillations is therefore to use fully reconstructed $\Bs$~mesons
rather than higher statistics samples of partially reconstructed
$\Bs$~candidates from e.g.~semileptonic decays. 

A measurement of $\dms$ will be the next crucial test of the Standard Model
probing whether the obtained result will fit to the current constraints on
the CKM triangle which are all in beautiful agreement (see results in Chapter 
\ref{chap:fits}). 
It is noteworthy to mention that physics with $\Bs$~mesons
is unique to the Tevatron until the start of the LHC in 2007.

\subsubsection{CDF and D\O\ detector upgrades in Run\,II at Tevatron}

The Fermilab accelerator complex has undergone a major upgrade since
the end of Run\,I in 1996. The centre-of-mass energy has been increased to
1.96~TeV and the Main Injector, a new 150~GeV
proton storage ring, has replaced the Main Ring as injector of protons and
anti-protons into the Tevatron. The Main Injector also 
provides higher proton intensity onto the anti-proton production target,
with the goal to allow for more than an order of magnitude higher
luminosities. Run\,II officially started in March 2001. 
The design luminosity during the first phase of Run\,II (Run\,IIa) is 
5-8$\cdot 10^{31}$~cm$^{-2}$s$^{-1}$ for a final integrated luminosity of
$\sim 2$~fb$^{-1}$ by the end of Run\,IIa. 

Since 1996, the CDF and D\O~detectors have also undergone major
upgrades~[\ref{Blair:1996kx},\ref{Ellison:2000hj}] 
to allow operation at high luminosities and bunch spacing of up to 132~ns.
Details of the D\O~detector upgrade can be found
elsewhere~[\ref{Ellison:2000hj}].  
The main upgrade for D\O\ is the installation of a tracking system contained
in a 2T~superconducting solenoid surrounded by a scintillator preshower
detector. The tracking upgrade includes a silicon microstrip tracker which
consists of six barrel segments with disks in between plus three more disks
located at each end of the tracker. In addition, there are two large
disks placed at the end of the silicon tracker to increase the
pseudorapidity coverage. The silicon 
system is enclosed within a central fiber tracker providing momentum
resolution at the level of $\sigma(p_T)/p_T = 0.02$-0.05 for low-$p_T$
tracks with high tracking efficiency for charged particles with
pseudo-rapidity $\eta<2.5$. Vertex reconstruction is expected with a resolution of
15-30~$\mu$m in the $r\phi$-plane and about 80~$\mu$m in the $rz$-plane. 
A major upgrade of the muon system together with central and forward
scintillators will allow D\O\ to trigger and reconstruct muon tracks. 
The ${\rm B}$~physics triggers at D\O\ allow to trigger on muons and electrons
while a new Level~1 tracking trigger and a Level~2 silicon trigger are
under construction.

The CDF detector improvements for Run\,II~[\ref{Blair:1996kx}] 
were motivated by the shorter 
accelerator bunch spacing of up to 132~ns and the increase in luminosity by
an order of magnitude. All front-end and trigger electronics has been
significantly redesigned and replaced. A DAQ upgrade allows the operation
of a pipelined trigger system. CDF's tracking devices 
were completely replaced. They consist of  
a new Central Outer Tracker (COT) with 30,200 sense wires
arranged in 96 layers combined into four axial and four stereo
superlayers. It also provides d$E$/d$x$ information for particle
identification. 
The Run\,II silicon vertex detector, covering a total radial area from
1.5-28~cm, consists of seven double sided layers and one single sided layer  
mounted on the beampipe. The
silicon vertex detector covers the full Tevatron luminous 
region which has a RMS spread of about 30~cm along the beamline and allows
for standalone silicon tracking up to a pseudo-rapidity $|\eta|$ of 2. The
forward calorimeters have been replaced by a new scintillator tile based
plug calorimeter which gives good electron identification up to 
$|\eta|=2$.
The upgrades to the muon system almost double the central
muon coverage and extent it up to $|\eta|\sim1.5$.

\subsubsection{Prospects for $\Bs$ mixing at CDF}

The most important improvements for ${\rm B}$~physics at CDF are a 
Silicon Vertex Trigger (SVT)
and a Time-of-Flight (ToF) system with a resolution of about
100~ps. The later employs 216 three-meter-long
scintillator bars located between the outer radius of the COT
and the superconducting solenoid.
More details about the CDF\,II Time-of-Flight detector and its performance
can be found in Ref.~[\ref{Cabrera:2002vp},\ref{Grozis:2002nr}].
The ToF system will be most beneficiary for the identification
of kaons with a 2\,$\sigma$-separation between $\pi$ and ${\rm K}$ for
$p<1.6$~GeV/$c$. This will enable CDF to make use of opposite side kaon
tagging and allows to identify same side fragmentation
kaons accompanying $\Bs$~mesons~[\ref{Cabrera:2002vp},\ref{Grozis:2002nr}].

In Run\,I, all ${\rm B}$~physics triggers at CDF were based on leptons including
single and dilepton triggers. A newly implemented Silicon Vertex Trigger
gives CDF access to purely hadronic ${\rm B}$~decays and makes CDF's ${\rm B}$~physics
program fully competitive with the one at the
$e^+e^-$~${\rm B}$~factories. The hadronic track trigger
is the first of its kind operating successfully at a hadron collider. It
works as follows: with a
fast track trigger at Level~1, CDF finds track pairs in the COT
with $p_T>1.5$~GeV/$c$. At Level~2, these tracks are
linked into the silicon vertex detector and cuts on the track impact
parameter (e.g.~$d > 100~\mu$m) are applied. 
The original motivation for CDF's hadronic track trigger was to select
the two tracks from the rare decay ${\rm B}^0 \rightarrow \pi\pi$ but it will play a
major role in collecting hadronic $\Bs$~decays for the measurement of 
$\Bs$~oscillations. 
Since the beginning of Run\,II, much work has gone
into commissioning the CDF detector. 
The Silicon Vertex Trigger was fully operational at the beginning of 2002.
A detailed discussion of the SVT and its
initial performance can be found 
elsewhere~[\ref{Cerri:2002ss},\ref{Lucchesi:2002tw}].

\begin{figure}
\centerline{
\includegraphics[width=0.5\hsize]{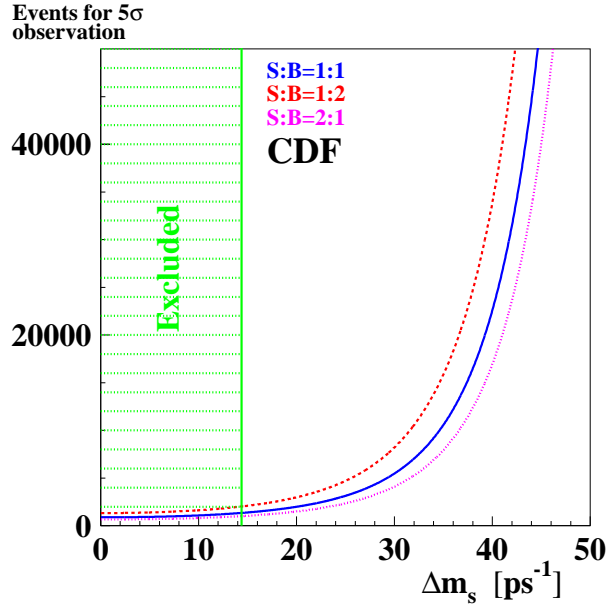}
}
\caption{\it 
Expected event yield of fully reconstructed $\Bs$~decays at CDF necessary
for a $5\,\sigma$-observation of $\Bs$~oscillations as a function of
$\dms$ for different signal-to-background ratios.
}
\label{fig:dms_exp_cdf}
\end{figure}

The CDF detector upgrades described above play an important role in CDF's
prospects for measuring $\Bs$~mixing.  
The inner layer of silicon mounted on the beampipe
improves the time resolution for measuring the $\Bs$~decay length 
from originally $\sigma_t = 0.060$~ps 
to 0.045~ps. This will be important if 
$\dms$ is unexpectedly large. The Time-of-Flight
system will enhance the effectiveness of ${\rm B}$~flavour tagging, especially
through same side tagging with kaons and opposite side kaon tagging, to a
total expected $\varepsilon {\cal D}^2 \sim 11.3\%$~[\ref{Anikeev:2001rk},\ref{Grozis:2002nr}]. 

Fig.~\ref{fig:dms_exp_cdf} shows the expected 
event yield of fully reconstructed $\Bs$~decays necessary
for a 5\,$\sigma$~observation of $\Bs$~oscillations
as a function of the mixing frequency $\dms$ for different
signal-to-background ratios. 
If the $\Bs$~mixing frequency is around
the current Standard Model expectation of $\dms \sim
18\ps^{-1}$ (see discussion in Sec.~\ref{sec:final} of Chapter \ref{chap:fits}), 
Fig.~\ref{fig:dms_exp_cdf} indicates that CDF 
would only need a few thousand fully reconstructed $\Bs$~mesons to discover
$\Bs$~flavour oscillations. 
Originally, CDF estimated to fully reconstruct a signal of about 75,000 
$\Bsb \rightarrow \Ds \pi^-$ and $\Bsb \rightarrow \Ds \pi^- \pi^+\pi^-$
events from the two-track hadronic trigger 
in 2~fb$^{-1}$~[\ref{Anikeev:2001rk}]. 
This assumes all detector components and triggers work as
expected. 
Although with the beginning of 2002, the CDF
detector is in stable running conditions operating with reliable physics
triggers, including the hadronic two-track trigger,  
there appear to be indications that the projected event yield
might be overestimated. Given this and the 
small amount of data delivered by the Tevatron and recorded by CDF to date
(about 100~pb$^{-1}$ by the end of 2002)
it will take some time until CDF
can present first results on $\Bs$~mixing~[\ref{Paulini:2003jk}].

\subsubsection{Prospects for $\Bs$ mixing at D\O}

The major difference for a search of $\Bs$~oscillations at D\O\ is the
collection of $\Bs$~candidate events. D\O\ currently does not operate a
hadronic track trigger. However, it will be able to collect $\Bs$~candidate
events using lepton triggers. Various $\Bs$~decay modes such as
$\Bsb \to \Ds \pi^-$, $\Bsb \to \Ds \pi^- \pi^+\pi^-$ and
$\Bsb \to \Ds \ell^- \nu$
are under investigation by the D\O~collaboration. 
The fully hadronic decay modes can be collected by single lepton triggers
where the trigger lepton serves as an opposite side lepton tag and the
$\Bs$~meson is reconstructed on the other side. In this case the event
yield is suppressed leaving D\O\ with a few thousand events of this type in
a data sample of 2~fb$^{-1}$. If the $\Bs$~oscillation frequency is small
enough, semileptonic $\Bs$~decays can be used utilizing D\O's lepton trigger
data. But due to the escaping neutrino, the boost resolution is reduced
limiting the $\dms$~reach. D\O\ expects to collect about 40,000 events in
the semileptonic channel in 2~fb$^{-1}$. Monte Carlo studies indicate that
D\O\ will be able to measure $\Bs$~oscillations in this mode up to a
mixing frequency of $\dms\sim 20$~ps$^{-1}$.  

\section{Use of the amplitude spectrum for CKM fits}

In this Section we discuss how to include $\dms$ information in
CKM fits starting from the amplitude spectrum given by the LEP 
Oscillation Working Group [\ref{ref:osciw}].

The 95\% C.L. limit and the sensitivity (see definition in
Eq.~(\ref{eq:bs_sensitivity})\,), are useful to summarize the results of the
analysis. However to include $\dms$ in a CKM fit and to determine
probability regions for the Unitarity Triangle parameters, continuous
information about the degree of exclusion of a given value of $\dms$
is needed. We describe how to include this information in both
Bayesian and frequentist approaches.
The requirements for an optimal method are:
\begin{itemize}
\item the method should be independent of the significance of the
      signal: this criterion is important to avoid switching from one
      method to another because of the presence (absence) of a
      significant signal (whose definition is arbitrary);
\item the probability regions derived should have correct coverage.
\end{itemize}
For the discussion in this Section we use the World Average computed
by the LEP Oscillation Working Group~[\ref{ref:osciw}] and presented a the 
CKM-Workshop, corresponding to a 95$\%$ C.L. lower limit at
15.0 ps$^{-1}$ and to a sensitivity at 18.0 ps$^{-1}$.

In Sec.~\ref{sec:now} we review and analyse how to include $\dms$
information for the CKM fits.
Sec.~\ref{sec:freqch4} describes the newly-proposed
frequentist method for including $\dms$ information in CKM fits.

\subsection{Review of the available methods. The likelihood ratio method}
\label{sec:now}

\subsubsection*{Modified $\chi^2$ method}

The first CKM fits~[\ref{ref:par_1}--\ref{ref:par_2}] used the
$\chi^2$ of the complete amplitude spectrum w.r.t. 1:
\begin{eqnarray}
  \chi^2 = \left( {{1-A} \over \sigma_A} \right )^2
\end{eqnarray}
The main drawback of this method is that the sign of the deviation of
the amplitude with respect to the value $A = 1$ is not used.
A signal might manifest itself by giving an amplitude value
simultaneously compatible with $A = 1$ and incompatible with
$A = 0$; in fact, with this method, values of $A>1$ (but still compatible with $A=1$) 
are disfavoured w.r.t. $A=1$, while it is expected that, because of
statistical fluctuations, the amplitude value corresponding to the
``true'' $\dms$ value could be higher than $1$. This problem was
solved, in the early days of using $\dms$ in CKM fits, by taking
$A=1$ whenever it was in fact higher.

A modified $\chi^2$ has been introduced in [\ref{ref:hock}] to solve
the second problem:
\begin{eqnarray}
\label{eq:chi2mod}
\chi^2 = 2\cdot
  \left[
   \mathrm{Erfc}^{-1}
    \left(\frac{1}{2}\mathrm{Erfc}
     \left({1-A}\over{\sqrt{2}\sigma_A}\right)
    \right)
  \right]^2
\end{eqnarray}

\subsubsection*{Relation between the log-likelihood and the Amplitude}

\begin{figure}
\centerline{
\includegraphics[width=0.5\hsize]{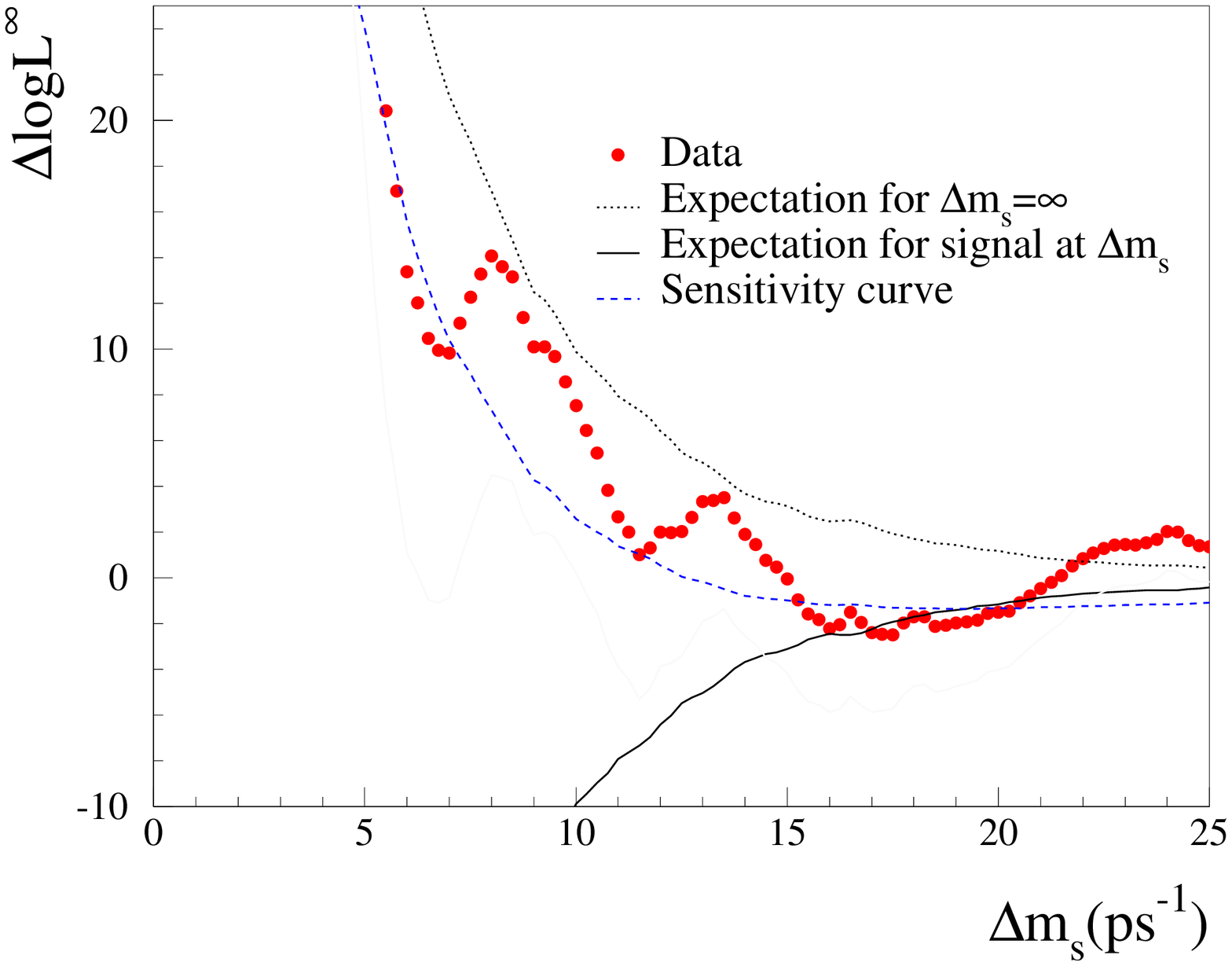}
}
\caption{\it World average amplitude analysis: 
$\Delta \log {\cal L}^{\infty}(\dms)$.}
\label{fig:amp_avg}
\end{figure}

The log-likelihood values can be
easily deduced from $ A$ and $\sigma_{A}$ using the expressions given
in~[\ref{ref:moser}]:
\begin{eqnarray}
\Delta \log\mathcal{L}^{\infty}(\dms) &=&
 \frac12\,
 \left[ \left(\frac{A-1}{\sigma_A}\right)^2-
        \left(\frac{A}{\sigma_A}\right)^2
 \right]
 = \left (\frac12-{A}\right)\frac1{\sigma_A^2} \ , 
\label{dms_right1}\\
\Delta \log{\cal L}^{\infty}({\dms})_\mathrm{mix}   &=&
   -\frac{1}{2}\frac{1}{\sigma_{A}^2} 
   \label{dms_right2} \ ,  \\
\Delta \log{\cal L}^{\infty}({\dms})_\mathrm{nomix} &=&
  \frac{1}{2}\frac{1}{\sigma_{A}^2 }\ . \label{dms_right3}
\end{eqnarray} 
The last two equations give the average log-likelihood value for
$\Delta M_s$ corresponding to the true oscillation frequency (mixing case) 
and for $\Delta M_s$ being far from the oscillation
frequency ($|\dms-\dms^\mathrm{true}| \gg \Gamma/2$, no-mixing
case). $\Gamma$ is here the full width at half maximum of the amplitude
distribution in case of a signal; typically $\Gamma \simeq
{1}/{\tau_{\Bs}}$. Fig.~\ref{fig:amp_avg} shows the variation of
$\Delta\mathcal{L}^\infty(\dms)$ corresponding to the amplitude
spectrum of Fig.~\ref{fig:dms_comp_all}(b).

\subsubsection*{Likelihood ratio method $R$}

Instead of the $\chi^2$ or the modified $\chi^2$ methods, 
the log-likelihood function $\Delta \log\mathcal{L}^{\infty}(\dms)$ 
can be used: this is the
log-likelihood referenced to its value obtained for
$\dms=\infty$~[\ref{ref:checchia1},\ref{ref:likr}]. The log-likelihood values
can easily be deduced from $A$ and $\sigma_A$, in
the Gaussian approximation, by using the expressions given
in Eqs.(\ref{dms_right1}), (\ref{dms_right2}), (\ref{dms_right3}).  
The Likelihood Ratio $R$, defined as,
\begin{equation}
\mathrm{R}(\dms) = e^{\textstyle {-\Delta \log
                     \mathcal{L}^{\infty}(\dms)}}
                 =  \frac{\mathcal{L}(\dms)}{\mathcal{L}
                     (\dms=\infty)} \ ,
\label{R_eq}
\end{equation}
has been adopted in~[\ref{ref:likr}] to incorporate the $\dms$ constraint.

\subsubsection*{Comparison between the two methods using the world
average amplitude spectrum}

\begin{figure}
\begin{center}
\includegraphics[width=0.75\hsize]{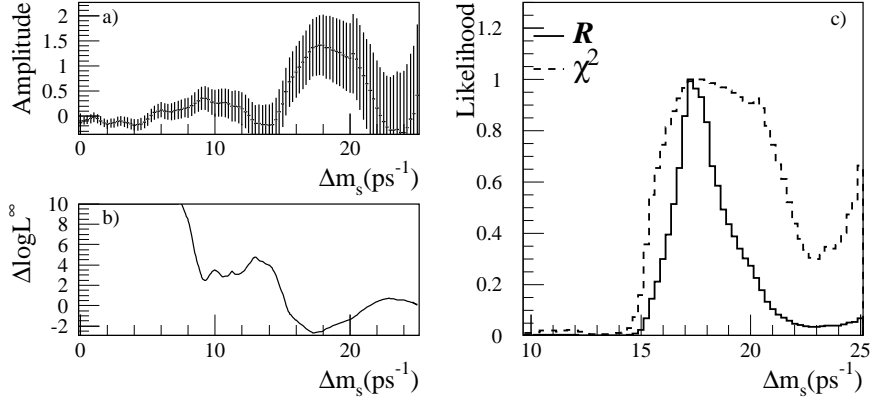}
\end{center}
\caption{\it  {World Average amplitude analysis:
         (a) amplitude spectrum, (b) $\Delta \log
         \mathcal{L}^{\infty}(\dms)$, (c) comparison between the
         Likelihood Ratio method ($R$) and the Modified $\chi^2$
         method ($\chi^2$). The information in (b) and in the solid
         histogram in (c) is identical.}}
\label{fig:realcase}
\end{figure}

The variation of the amplitude as a function of $\dms$ and the
corresponding $\Delta \log\mathcal{L}^{\infty}({\dms})$ value are
shown in Fig.~\ref{fig:realcase}-(a) and (b). The constraints
obtained using the Likelihood Ratio method (${R}$) and the
Modified $\chi^2$ method ($\chi^2$) are shown in Fig.
\ref{fig:realcase}-(c). In this comparison the Modified $\chi^2$
has been converted to a likelihood using $\mathcal{L}
\propto \exp(-\chi^2/2)$.
It is clear that the two methods ($R$ and $\chi^2$) give very
different constraints. In particular the Modified $\chi^2$ method,
with the present World Average, corresponds to a looser constraint for
CKM fits (and in particular for the determination of the $\bar{\rho}$
and $\gamma$ parameters).

\subsubsection*{The toy Monte Carlo}
 
In order to test and compare the statistical properties of the two methods it is necessary to generate several
experiments having similar characteristics as the data used for 
the World Average. We will call equivalent those
experiments having the same dependence of $\sigma_A$ as a function of $\dms$.

\begin{figure}
\begin{center}
\includegraphics[width=0.6\hsize,bb=10 0 284 193]{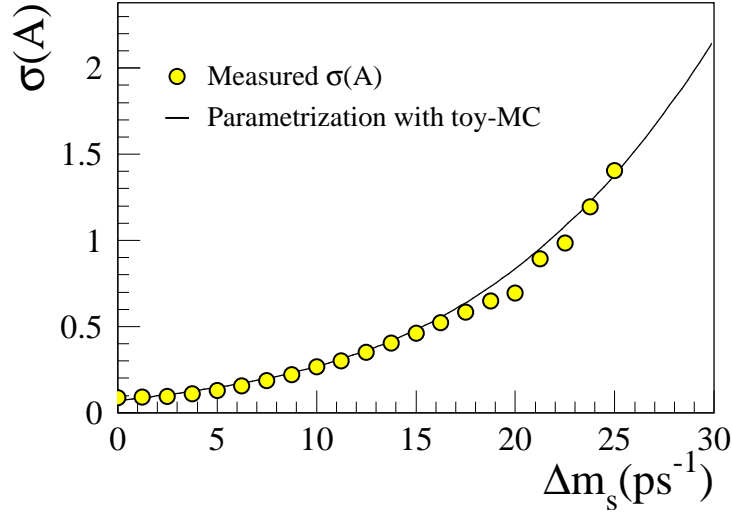}
\end{center}
\caption{\it Comparison between the error distribution computed with the toy-MC
(solid line) and the measured amplitude errors (circles).}
\label{fig:extra_dms}
\end{figure}

The dependence of $\sigma_A$ on $\dms$ can be reproduced by
tuning the parameters of a fast simulation (toy-MC). The method used
here is similar to the one presented in~[\ref{ref:boix}]. The error on
the amplitude can be written as:
\begin{eqnarray}
\sigma^{-1}_A= \sqrt{N}\, \eta_{{\Bs}}
\, (2\epsilon_d-1) \, (2\epsilon_p-1) \, W(\sigma_L,\sigma_P,\dms)
\end{eqnarray}        
where $N$ is the total number of events, $\eta_{{\Bs}}$ the ${\Bs}$ purity of
the sample, $\epsilon_{d(p)}$ the tagging purity at
the decay (production) time, $\sigma_L$ the uncertainty  on the ${\Bs}$ flight length
and $\sigma_p$ the relative uncertainty in the ${\Bs}$  momentum.
$W$ is the function that accounts for the damping of the oscillation
due to the finite proper time resolution. The parameters $\sigma_L$,
$\sigma_p$ and the global factor that multiplies the $W$ function are
obtained by adjusting the simulated error distribution to the one
measured with real events. Figure~\ref{fig:extra_dms} shows the
agreement between the toy-MC calculation and the real data up to
$\dms=25\,\mathrm{ps}^{-1}$ (the upper value of $\dms$ at which
amplitudes are given).
An additional problem is that, in principle,
one would like to define the likelihood within the interval
$[0,\infty]$ whereas the amplitude spectrum is measured only up to a
certain value. For the present World Average the value is
$25\,\mathrm{ps}^{-1}$. A procedure has to be introduced to continue
$\sigma_A$ and $A$.

The continuation for $\sigma_A$ is shown in Fig.\ref{fig:extra_dms}.
The continuation of $A$ is more delicate. In particular it
is more sensitive to the real amplitude spectrum. Nevertheless if
$\dms^\mathrm{sens} << \dms^\mathrm{last}$, the significance $S$
($S=A/\sigma_A$) is approximately constant. It is
then a good approximation to continue using: 
\begin{equation}
A(\dms) = 
   \frac{A(\dms^\mathrm{last})}
        {\sigma_A(\dms^\mathrm{last})}
   \sigma_A(\dms).
\end{equation}
Although this procedure is reasonable, it should be stressed that it
is very desirable to have all the amplitudes (with errors) up to the
$\Delta M_s$ value where the significance remains stable.

\subsubsection*{Comparison of the methods in case of an oscillation signal}

In this Section we compare the two methods in the
presence of a clear $\dms$ oscillation signal. We perform several
$\dms$ toy-MC analyses with the same $\sigma_A$ versus
$\dms$ behaviour as the World Average analysis. For this study we have
generated a $\dms$ signal at $17\,\mathrm{ps}^{-1}$. This value corresponds to
the value where there is the bump in the World Average amplitude
spectrum. The statistics of the virtual experiments is much larger than the 
registered data, at present, so that clear oscillation signals are expected.

\begin{figure}
\begin{center}
\begin{tabular}{cc}
\includegraphics[width=0.47\hsize]{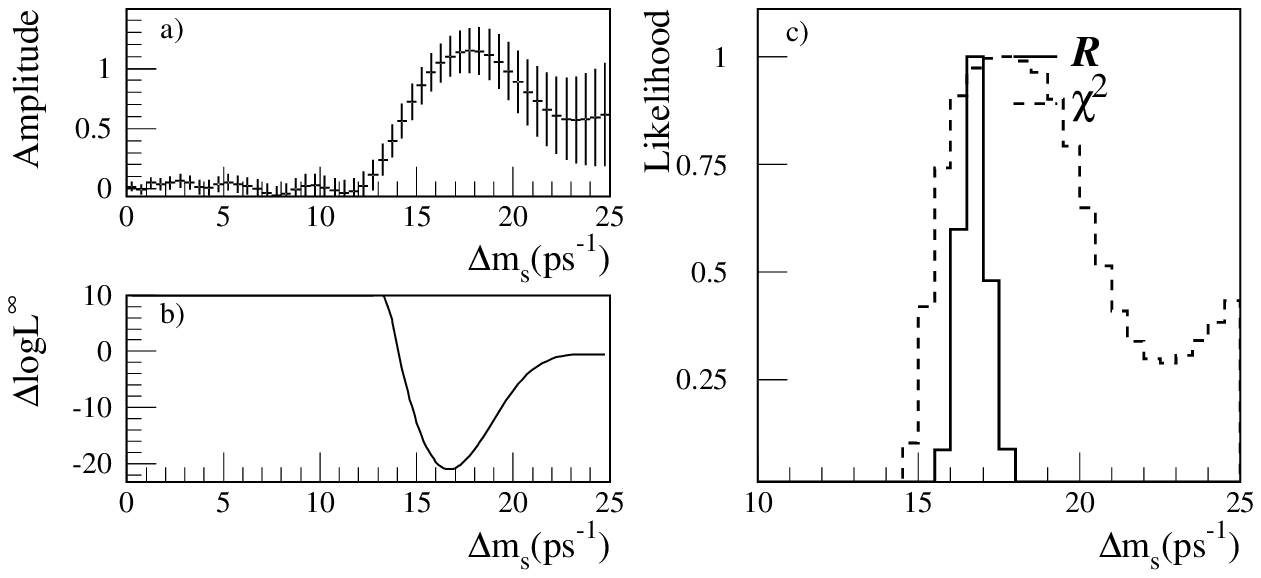} &
\includegraphics[width=0.47\hsize]{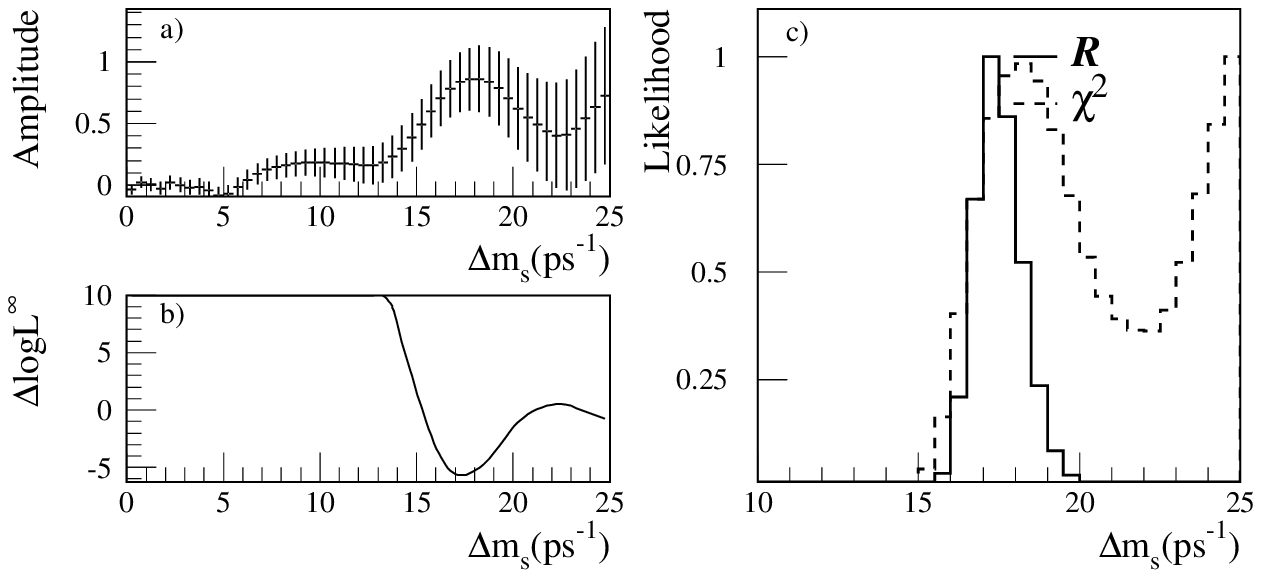} \\
\includegraphics[width=0.47\hsize]{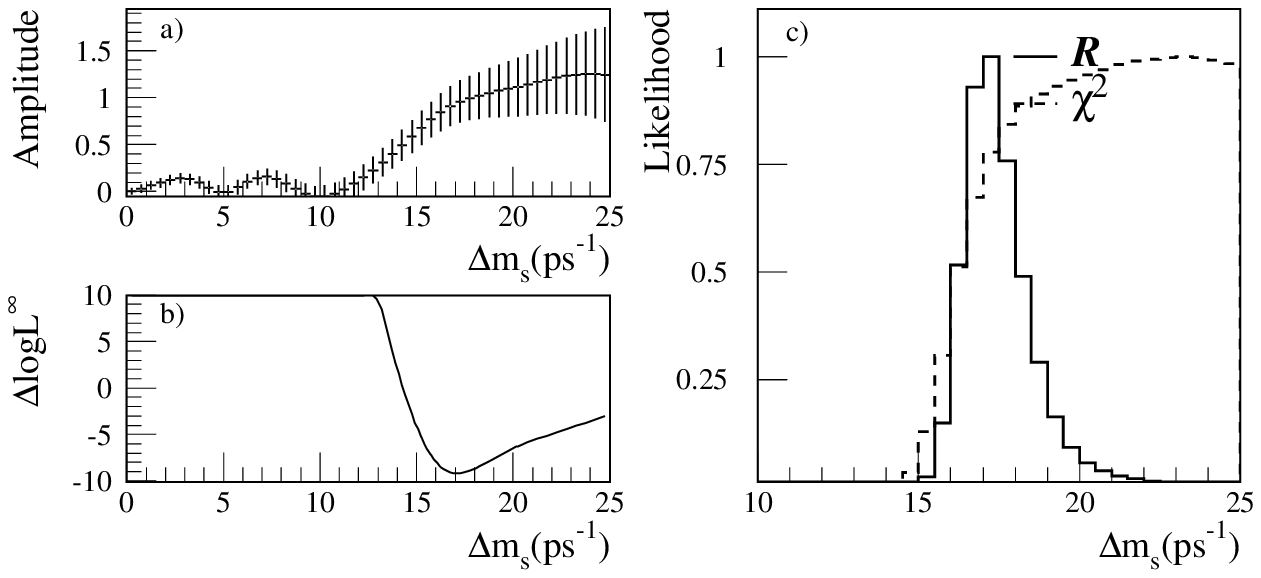} &
\includegraphics[width=0.47\hsize]{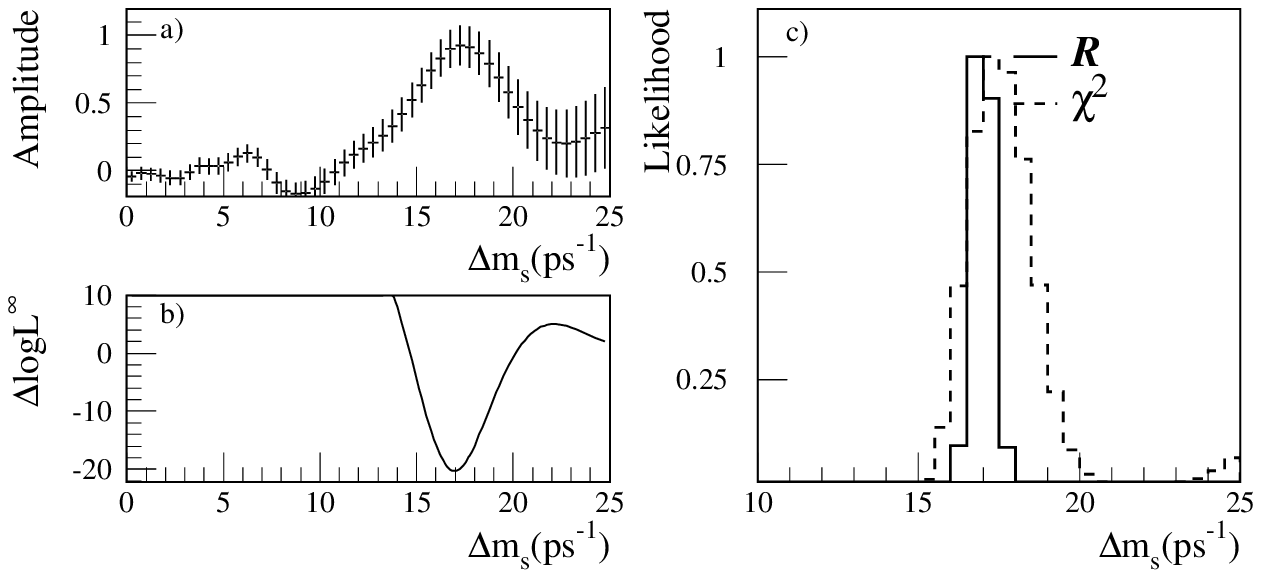}
\end{tabular}
\caption{\it Toy-MC analyses with $\dms$ generated at
$17\,\mathrm{ps}^{-1}$ corresponding  to four virtual experiments.
Each experiment is summarized in three plots:
   (a) amplitude spectrum,
   (b) $\Delta \log \mathcal{L}^{\infty}(\dms)$,
   (c) comparison between the Likelihood Ratio method ($R$) 
       and the Modified $\chi^2$ method ($\chi^2$).}
\label{fig:casesignal}
\end{center}
\end{figure}

The results in Fig.~\ref{fig:casesignal} show that only the
Likelihood Ratio method is able to see the signal at the correct
$\dms$ value. The same exercise has been repeated for different
generated values of $\dms$, always giving the same result.

\subsubsection*{Test of the coverage of the two methods applied to CKM fits}

In the absence of a clear ${\Bs}$ oscillation signal, the Likelihood Ratio method results
in a $\dms$ range which extends to infinity at any C.L. A criticism
was made in~[\ref{ref:hock}] that it is then dangerous to use this 
information in a CKM fit. The best way to answer this objection 
is to test the coverage of the probability regions (68\%, 95\% and 99\%) computed by
the fit by performing a Monte Carlo simulation.

To do this we have prepared a simplified CKM fit where we measure the
quantity $R_t$ (see Chapter~1),
using only the $\Delta M_d$ and the $\Delta M_d/\Delta M_s$
constraints. The set of constraints on the quantity~$R_t$~is:
\begin{eqnarray}
  \Delta M_d              &=& a^2R_t^2  \\
  \Delta M_d/\Delta M_s   &=& b^2 R_t^2 \qquad
  \mbox{(or $\Delta M_s = a^2/b^2$)}
\end{eqnarray}
where $a$ and $b$ are Gaussian distributed parameters with errors
$\sigma_a=20\%$ and $\sigma_b=10\%$, thus taking into account the theoretical
uncertainties.

Several experiments have been generated, each of them characterized
by the following set of parameters:
\begin{center}
\begin{tabular}{ll}
  $R_t$ & \\
  $a_\mathrm{theo}$     & extracted from the $a$ distribution\\
  $b_\mathrm{theo}$     & extracted from the $b$ distribution\\
  $\Delta M_d(\mathrm{theo})$ & computed from $R_t$ and $a$\\
  $\Delta M_s(\mathrm{theo})$ & computed from $R_t$ and $b$\\
  $\Delta M_d(\mathrm{exp})$  & from $\Delta M_d(\mathrm{theo})$
                          smeared by the experimental resolution \\
  Amplitude spectrum    & from a toy-experiment generated with
                          $\Delta M_s(\mathrm{theo})$ 
\end{tabular}
\end{center}
For each experiment the best-fit value for $R_t$ was determined and 
it was counted how many times it fell 
inside the $68\%$, $95\%$ and $99\%$ probability regions defined by
the Likelihood Ratio and by the Modified $\chi^2$ methods. This
exercise was repeated 1000 times.
The measured frequencies for the three probability regions using the Likelihood 
Ratio or the Modified $\chi^2$ method are given in Table~\ref{tab:likr} and
\ref{tab:mchi} respectively.

For the Likelihood Ratio method the measured frequencies correspond to the 
confidence level intervals and the coverage is close to correct.
This is not the case for the Modified $\chi^2$ method where the
confidence levels are significantly underestimated for the true value
of $\Delta M_s$. The effect stems from the fact that the $\chi^2$
defined in Eq.~\ref{eq:chi2mod} reaches its minimum systematically
above the true value of $\Delta M_s$. 
\begin{table}
\begin{center}
\begin{tabular}{|llll|}
\hline
            &      68\%    &      95\%     &      99\%      \\ \hline
$\dms=10$   & $67.5\pm1.5$ & $93.1\pm 0.8$ & $98.1\pm0.4$   \\
$\dms=18.2$ & $71.4\pm1.4$ & $96.1\pm 0.6$ & $99.6\pm0.2$   \\
$\dms=25$   & $69.5\pm1.5$ & $96.4\pm 0.6$ & $99.3\pm0.3$   \\ \hline
\end{tabular}
\end{center}
\caption{\it Results obtained with the Likelihood Ratio method.
For three different values of generated $\dms$ (left column) we
indicate the percentage of ``experiments'' for which the generated
true value of $R_t$ falls inside the 68\%, 95\% and 99\% probability
interval.}
\label{tab:likr}
\end{table}
\begin{table}
\begin{center}
\begin{tabular}{|llll|}
 \hline
  & 68\% & 95\% & 99 \% \\ \hline
$\dms=10$ & $48.6\pm1.6$ & $83.8\pm 1.2$ & $94.3\pm0.7$ \\
$\dms=18.2$   & $64.6\pm1.5$ & $93.0\pm 0.8$ & $99.2\pm0.3$ \\
$\dms=25$   & $77.5\pm1.5$ & $98.2\pm 0.4$ & $99.7\pm0.2$ \\ \hline
\end{tabular}
\end{center}
\caption{\it As for Table~\ref{tab:likr}, but for the Modified $\chi^2$
method.} 
\label{tab:mchi}
\end{table}

\subsubsection*{Some conclusions}

In this first part we have studied the problem of including in CKM
fits the $\dms$ World Average amplitude spectrum. 
We have tested two different methods and compared the results in case 
of an oscillation signal. MC simulations also were performed for a CKM fit to
test the coverage of the two methods. The conclusion
is that the Likelihood Ratio method, proposed in
[\ref{ref:checchia1},\ref{ref:likr}], is optimal because it gives probability 
intervals with correct coverage and, in case of a signal, it also
gives the correct value of $\dms$.

\subsection{Use of the amplitude spectrum in a frequentist approach}
\label{sec:freqch4}

The aim of this Section is to describe the frequentist method for
incorporating experimental constraints derived from the amplitude
spectrum as a function of the $\Bs$ oscillation frequency ($\dms$)
into a global CKM fit. In other words, we address the questions: what
is the pdf of a likelihood measurement of $\Delta M_s$, and what is
the confidence level (CL) as a function of $\Delta M_s$ to be
associated with an observation obtained with a given level of
sensitivity?

\subsubsection*{Infinite statistics}
\label{ParabolicBehaviorSection}

We assume that the $\xs$ measurement is performed using the
log-likelihood. 
The measured value of $\xs$
($\xsmes$) is defined to be the one maximizing $\Lik(\xs)$: the
outcome of one experiment $\xsmes$ is a random number. For infinite
statistics, the $\xsmes=\Delta M_s\tau_b$ random number follows a
(leading-order: lo) Gaussian probability density function:
\begin{equation}
\label{KeyFormula}
\Philo^{\xs}(\xsmes)={1\over\sqrt{2\pi}\Sigma(\xs)}
\exp{
\left(
-{1\over 2}
\left({\xsmes-\xs\over\Sigma(\xs)}\right)^2
\right)}
\end{equation}
where the standard deviation $\Sigma(\xs)$ is given by 
the second derivative of the expected $\Lik$, through the integral $\A$
\begin{eqnarray}
\label{Aintegral}
(\sqrt{N}\Sigma(\xs))^{-2}&=&\int\limits_{-\infty}^{+\infty} 
\left({(\Pdot_-)^2\over\P_-}+{(\Pdot_+)^2\over\P_+}\right)\dtmes
\equiv \A(\xs)
\\
\Pdot_\pm&=&{\partial\P_\pm\over\partial\xs}
  =
\mp\fs{1\over 2}d\ t\sin(\xs t)e^{-t}\otimes\Gt
\end{eqnarray}
$N$ is the total number of mixed and unmixed events and the integrals
are performed using the {true} value of $\xs$, not the measured one.
It follows from Eq.~(\ref{KeyFormula}) that one may set a confidence
level $\CLlo(\xshyp)$ on a given hypothetical value $\xshyp$ using the
$\chi^2$ law:
\begin{eqnarray}
\label{CLGauss}
\CLlo(\xshyp)&=&\int\limits_<\Philo^\xshyp(\xsmes^\prime)\d
\xsmes^\prime = \Prob(\chi^2,1)
\\
\label{chi2true}
\chi\equiv\chi^\xshyp(\xsmes)&=&{\xsmes-\xshyp\over\Sigma(\xshyp)}
\end{eqnarray}
where the integral is performed over the $\xsmes'$ domain where
$\Philo^{\xshyp}(\xsmes^\prime)<\Philo^{\xshyp}(\xsmes)$, that is to
say where $|\chi^\xshyp(\xsmes^\prime)|>|\chi^\xshyp(\xsmes)|$.

If the log-likelihood is parabolic near its maximum, as is the case
for infinite statistics, then, in the vicinity of $\xsmes$,
$\Sigma(\xshyp)\simeq \mathrm{cst}=\Sigma(\xsmes)$, and one can evaluate
$\Sigma$ as the second derivative of the experimental log-likelihood,
taken at the measured value $\xsmes$. In effect:
\begin{eqnarray}
{\partial^2\Lik\over\partial\xs^2}_{\mid\xs=\xsmes}
&=&\sum_-\left({\Pdotdot_-\P_- - (\Pdot_-)^2\over \P_-^2}\right)^2
     +\sum_+\left({\Pdotdot_+\P_+ - (\Pdot_+)^2\over \P_+^2}\right)^2
\\
 &\stackrel{N\rightarrow\infty}{=}& -N\A(\xs)=-\Sigma^{-2}    
\end{eqnarray}
where $\Pdotdot_\pm$ denotes the second derivative with respect to $\xs$:
\begin{equation}
\Pdotdot_\pm 
={\partial^2\P_\pm\over\partial\xs^2}
=
\mp\fs{1\over 2}d\ t^2\cos(\xs t)e^{-t}\otimes\Gt
\end{equation}
which does not appear in the final expression thanks to the
normalization of the probability density function, and assuming that
$\xsmes=\xs$ (which is true for infinite statistics).

Equivalently, one can evaluate $\Sigma$ by locating the value of
$\xshyp$ which yields a drop of $-1/2$ in the log-likelihood, for the
experiment at hand, or one can compute the $\chi^2$ directly using the
approximation
\begin{equation} 
\label{chi2Lik}
\chi^2(\xshyp)=\left({\xsmes-\xshyp\over\Sigma(\xshyp)}\right)^2
\simeq 2(\Lik(\xsmes)-\Lik(\xshyp))\equiv\tilde\chi^2(\xshyp)
\end{equation}

\subsubsection*{Finite statistics}

For large enough $\xshyp$, the approximation $\Sigma(\xshyp)\simeq
\Sigma(\xsmes)$ breaks down since the sensitivity of the experiment
vanishes: $\Sigma(\xshyp)\rightarrow\infty$ for $\xshyp \rightarrow
\infty$ . It follows that the likelihood is not parabolic for large
enough $\xshyp$, however large the statistics.

The vanishing sensitivity makes $\chi^2$, as defined by
Eq.~(\ref{chi2true}), a poor test statistic to probe for large $\xs$
values. Furthermore, it is not a straightforward task to infer the
correct $\CL(\xshyp)$ from the $\chi^2$ value: Eq.~(\ref{CLGauss})
does not apply (i.e., it is not a true $\chi^2$) because
Eq.~(\ref{KeyFormula}) is a poor approximation\footnote{The
redefinition of the $\chi^2$ using the right-hand side of
Eq.~(\ref{chi2Lik}) provides a test statistic more appropriate for
large values of $\xshyp$. Although Eq.~(\ref{CLGauss}) does not apply,
$\tilde\chi^2$ is capable of ruling out $\xshyp$ values lying beyond
the sensitivity reach (if $\Lik(\xsmes)$ is large enough) provided one
computes the CL using:
\[
\CL(\xshyp)=\int\limits_{\tilde\chi^2(\xshyp)}^\infty\Psi^\xshyp(\tilde\chi^{2\prime})\ 
\d\tilde\chi^{2\prime}
\]
where $\Psi^\xshyp$ is the probability density function of the
$\tilde\chi^2$ test statistic, for $\xs=\xshyp$, obtained using a toy
Monte Carlo. The rejection of $\xshyp$ values beyond the sensitivity
reach is not a paradox: it uses the fact that large values are
unlikely to yield an indication of a clear signal, especially at low
values of $\xs$. Such a treatment, as well as others (e.g., the
minimum value of the likelihood could be used to define another test
statistics) are satisfactory. We prefer here to use $x_s^{\rm mes}$,
and only this quantity, because an analytical expression for its
probability density function is available (Eq.~\ref{KeyNLOFormula})
and thus the computation of the CL can be carried out in practice.
This is nothing but the standard choice made when dealing with better
defined measurements.}.

In the realistic case of finite statistics, the next-to-leading order
statistical analysis of a likelihood measurement~[\ref{Bob}] is used
here to obtain the key-formula expressing the probability density
function of the random number $\xsmes$ beyond the Gaussian
approximation:
\begin{eqnarray}
\label{KeyNLOFormula}
\Phinlo^{\xs}(\xsmes)&=&
\Philo^{\xs}(\xsmes)\ e^{-\at^{\xs}\chi^3}
(1+\ao^{\xs}\chi)
\\
\label{derivativeofSigma}
\ao^{\xs}&=&
{2\mathrm{B}-\C\over 2\A}{1\over\sqrt{N \A}}
=-\dot\Sigma
\\
\at^{\xs}&=&
{3\mathrm{B}-\C\over 6\A}{1\over\sqrt{N \A}}
\end{eqnarray}
where $\A(\xs)$ is the integral defined in Eq.~(\ref{Aintegral}), 
$\B(\xs)$ and $\C(\xs)$ being two new integrals:
\begin{eqnarray}
B(\xs)&=&\int\limits_{-\infty}^{+\infty} 
\left({\Pdot_-\Pdotdot_-\over\P_-}
     +{\Pdot_+\Pdotdot_+\over\P_+}\right)\dtmes
\\
C(\xs)&=&\int\limits_{-\infty}^{+\infty} \left({(\Pdot_-)^3\over\P_-^2}
              +{(\Pdot_+)^3\over\P_+^2}\right)\dtmes          
\end{eqnarray}
The integral $\C$ tends to be small because, on the one hand the two
contributions have opposite signs, and on the other hand the
denominator is of order two: it follows that $\at\simeq\ao/2$. The
right hand side of Eq.~(\ref{derivativeofSigma}) links the
next-to-leading order correction terms $\ao$ and $\at$ to the
dependence on $\xs$ of $\Sigma$. When $\Sigma$ depends significantly
on $\xs$, not only is the standard treatment of 
Sec.~\ref{ParabolicBehaviorSection} invalid, but the well-known formula
Eq.~(\ref{CLGauss}) itself becomes incorrect, even if one uses the
correct $\Sigma(\xs)$.
\begin{figure}[t]
\begin{center}
\includegraphics[width=0.4\hsize,bb=30 17 517 512]{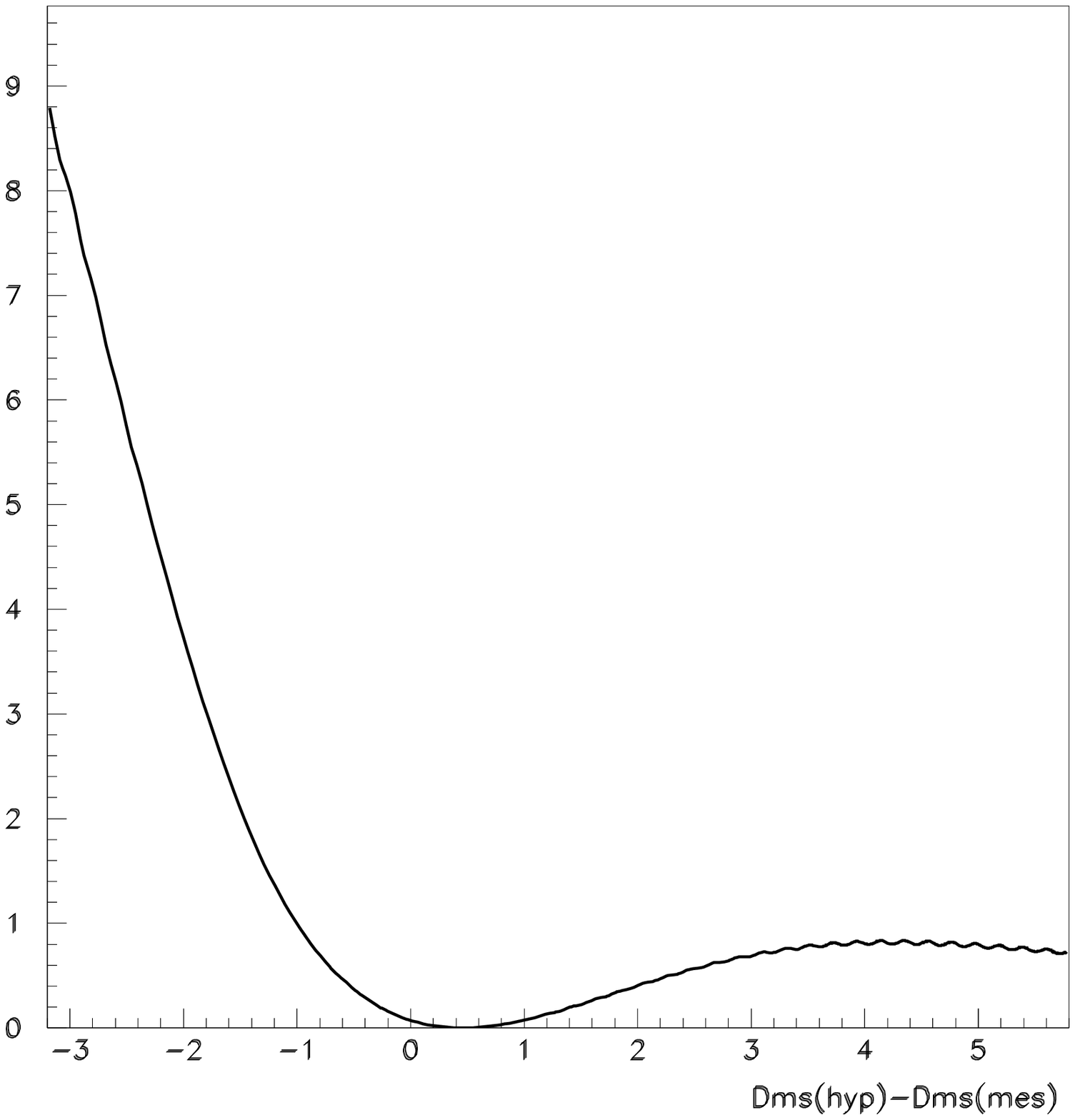}
\quad
\includegraphics[width=0.4\hsize,bb=30 17 517 512]{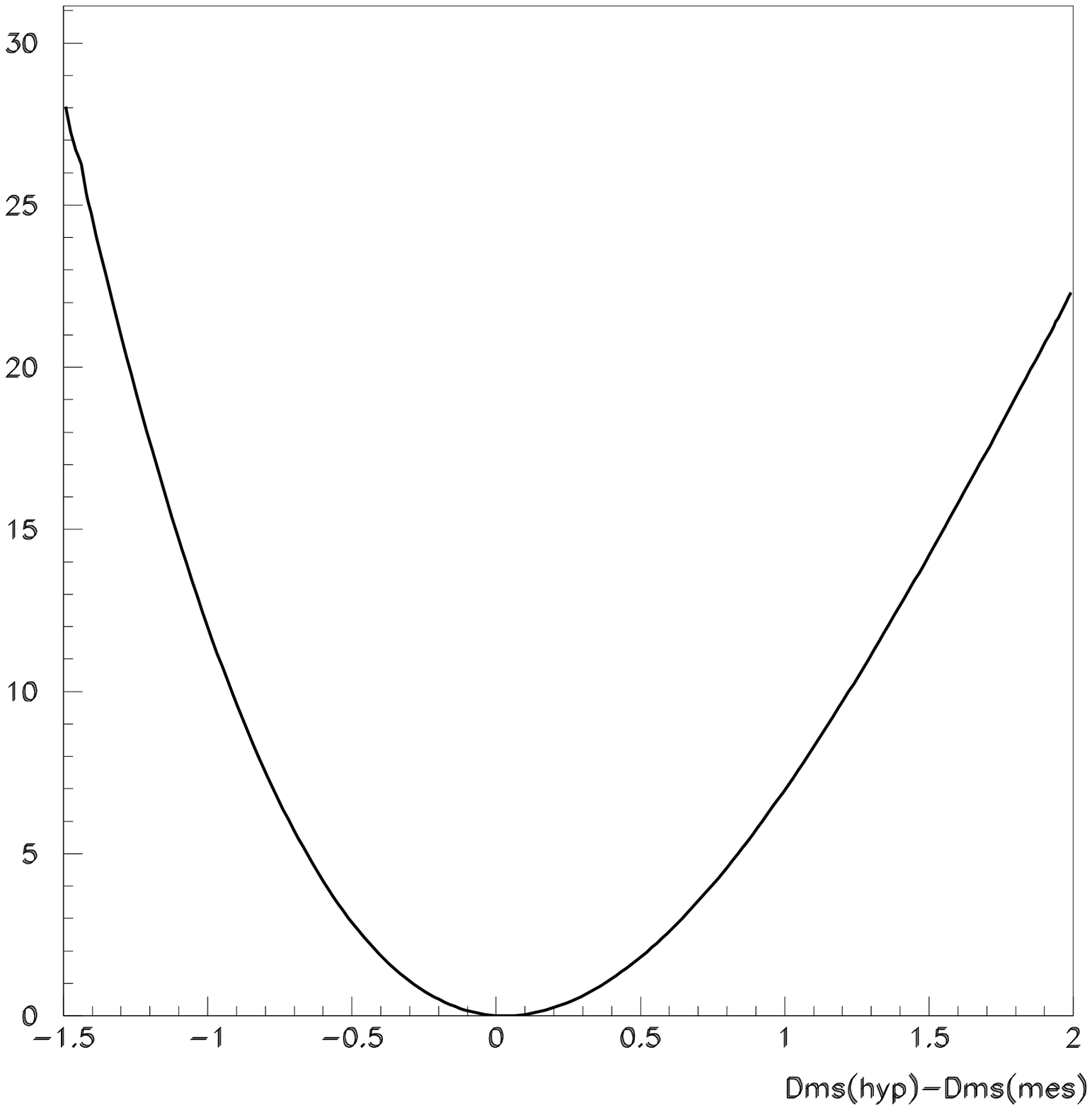}
\end{center}
\caption{\it \label{chi2CL}Left Plot: 
The equivalent $\chi^2$ (cf. Eq.~(\ref{CLGauss})\,) expressing the
confidence levels computed using the next-to-leading order expression
Eq.~(\ref{KeyNLOFormula}) in the actual situation where the maximum
value of the likelihood is reached for $\Delta M_s(\mathrm{mes})=17.2\,
\mathrm{ps}^{-1}$. The horizontal axis is the difference $\Delta
M_s(\mathrm{hypothetical})-\Delta M_s(\mathrm{mes})$. The minimum
value of the equivalent $\chi^2$ is not reached for $\Delta
M_s(\mathrm{hypothetical})=\Delta M_s(\mathrm{mes})$ because the
maximum of the next-to-leading order probability density function is
slightly shifted below the true $\Delta M_s$ value. The left hand side
of the plot is nearly parabolic and resembles closely the one that a
simplistic interpretation of the likelihood curve provides. The right
hand side of the plot states that there is almost no constraint on
high values of $\Delta M_s$. One is far from dealing with a
measurement in the usual (Gaussian) sense. Right Plot: The equivalent
$\chi^2$ in the would-be situation where the maximum value of the
likelihood is reached for $\Delta M_s(\mathrm{mes})=10\,
\mathrm{ps}^{-1}$. Although the equivalent $\chi^2$ is not truly
parabolic, the Gaussian limit is almost reached: one is close to
dealing with a measurement in the usual (Gaussian) sense.}
\end{figure}

The expression Eq.~(\ref{KeyNLOFormula}) is identical to
Eq.~(\ref{KeyFormula}) for small $\chi$ values. Although it extends
the range of validity to larger $\chi$ values, it cannot be trusted
too far away from the origin, where higher order corrections start to
play a role. In particular, $\Phinlo$ becomes negative (hence
meaningless) for $\chi>-\ao^{-1}$ ($\ao$~is negative since it is equal
to minus the derivative of $\Sigma$ with respect to $\xs$). Since
$\Phi$ is sizable only when $\chi\sim \mathcal{O}(1)$ the
next-to-leading order terms, when relevant, are of the form
$N^{-{1\over2}}\times ({\rm ratio\ of\ integrals})$: they are
negligible for large enough $N$ and for small enough ratio of
integrals. The double-sided CL is computed as in Eq.~(\ref{CLGauss}),
replacing $\Philo$ by the next-to-leading order approximation. Using
the right hand side of Eq.~(\ref{CLGauss}) to translate the confidence
level thus obtained into a more familiar equivalent\footnote{In the
CKMfitter package, it is this equivalent $\chi^2$ which is added to
the overall $\chi^2$.} $\chi^2$, one obtains the results shown in
Fig.~(\ref{chi2CL}) in two cases: first for the actual situation using
the parametrization of the world average likelihood as described in
Sec.~\ref{sec:now} where the maximum of the likelihood is reached
at the boundary of the experimental sensitivity; second for a
hypothetical situation where the maximum of the likelihood would be
reached well within the sensitivity region.

In conclusion, we have presented a frequentist analysis of the ${\rm B}_s$ 
oscillation.
Its domain of validity extends to the level of sensitivity reached by LEP
and SLD. The treatment presented here provides, in a frequentist
approach, a practical means to incorporate into a CKM fit the
information on $\Delta M_s$ contained in the data, both present and
future.

\section*{References}
\addcontentsline{toc}{section}{References}

\vspace{7mm}

\renewcommand{\labelenumi}{[\theenumi]}
\begin{enumerate}

\item \label{BJW90}
A.J. Buras, M.~Jamin and P.H. Weisz, Nucl. Phys. B~\textbf{347} (1990) 491.

\vspace{3mm}

\item \label{BSS}
A.J. Buras, W.~Slominski and H.~Steger, Nucl. Phys. B~\textbf{245} (1984) 369.

\vspace{3mm}

\item \label{HNa}
S.~Herrlich and U.~Nierste, Nucl. Phys. B~\textbf{419} (1994) 292,
[hep-ph/9310311].

\vspace{3mm}

\item \label{HNb}
S.~Herrlich and U.~Nierste, Phys. Rev. D~\textbf{52} (1995) 6505,
[hep-ph/9507262].

\vspace{3mm}

\item \label{HNc}
S.~Herrlich and U.~Nierste, Nucl. Phys. B~\textbf{476} (1996) 27,
[hep-ph/9604330].

\vspace{3mm}

\item \label{NIJA03}
M.~Jamin and U.~Nierste (2003). Recent update for this document.

\vspace{3mm}

\item \label{deRafaelTASI94}
E.~de~Rafael, in Proceedings of TASI 1994 (1995), hep-ph/9502254.

\vspace{3mm}

\item \label{UKJS}
J.~Urban, F.~Krauss, U.~Jentschura and G.~Soff, Nucl. Phys. B~\textbf{523}
  (1998) 40, [hep-ph/9710245].

\vspace{3mm}

\item \label{rome-paris-ckm-2000}
M.~Ciuchini {\it et al.}, JHEP \textbf{07} (2001) 013, [hep-ph/0012308].

\vspace{3mm}

\item \label{milc-fnal-nf3-lat2001}
C.~Bernard {\it et al.}, Nucl. Phys. Proc. Suppl. \textbf{106} (2002) 412,
  [hep-lat/0110072].

\vspace{3mm}

\item \label{milc-fnal-nf3-lat2002}
MILC Collaboration, C.~Bernard {\it et al.}, hep-lat/0209163.

\vspace{3mm}

\item \label{chiextrap-lat2002}
C.~Bernard {\it et al.}, in Proceedings of Lattice 2002 (2002) hep-lat/0209086.

\vspace{3mm}

\item \label{booth-qchpt-hl-mesons}
M.J. Booth, Phys. Rev. D~\textbf{51} (1995) 2338, [hep-ph/9411433].

\vspace{3mm}

\item \label{kronfeld-ryan-2002}
A.S. Kronfeld and S.M. Ryan, Phys. Lett. B~\textbf{543} (2002) 59,
[hep-ph/0206058].

\vspace{3mm}

\item \label{jlqcd-nrqcd-nf2-lat2001}
JLQCD Collaboration, N.~Yamada {\it et al.}, 
Nucl. Phys. Proc. Suppl. \textbf{106} (2002) 397, \break
[hep-lat/0110087].

\vspace{3mm}

\item \label{yamada-hqreview-lat2002}
N.~Yamada, in Proceedings of 20th Int. Symp. on Lattice Field Theory, Lattice
  2002, Boston, June 2002 (2002) hep-lat/0210035.

\vspace{3mm}

\item \label{milc-fnal-stat-nf2-2002}
MILC Collaboration, C.~Bernard {\it et al.}, Phys. Rev. D~\textbf{66} (2002) 094501,
  [hep-lat/0206016].

\vspace{3mm}

\item \label{bfpz-2002}
D.~Becirevic, S.~Fajfer, S.~Prelovsek and J.~Zupan, hep-ph/0211271.

\vspace{3mm}

\item \label{sharpe-zhang-hl-qchpt-1996}
S.R. Sharpe and Y.~Zhang, Phys. Rev. D~\textbf{53} (1996) 5125, [hep-lat/9510037].

\vspace{3mm}

\item \label{mw-hqp-2000}
A.V. Manohar and M.B. Wise, Heavy Quark Physics, vol.~10 of Cambridge
  Monographs on Particle Physics, Nuclear Physics and Cosmology (Cambridge
  University Press, 2000).

\vspace{3mm}

\item \label{cleo-gamma-dstar-2001}
CLEO Collaboration, S.~Ahmed {\it et al.}, Phys. Rev. Lett. \textbf{87} (2001)
  251801, [hep-ex/0108013].

\vspace{3mm}

\item \label{ukqcd-bstarbpi}
UKQCD Collaboration, G.M. de~Divitiis {\it et al.}, JHEP \textbf{10} (1998) 010,
[hep-lat/9807032].

\vspace{3mm}

\item \label{abada-etal-gdstardpi-2002}
A.~Abada {\it et al.}, hep-lat/0209092.

\vspace{3mm}

\item \label{jlqcd-chiextrap-lat2002}
JLQCD Collaboration, S.~Hashimoto {\it et al.}, hep-lat/0209091.

\vspace{3mm}

\item \label{lellouch-lattphen-ichep2002}
L.~Lellouch, in Proceedings of ICHEP 2002, 31st Int. Conf. on High Energy
  Physics, Amsterdam, July 2002, hep-ph/0211359.

\vspace{3mm}

\item \label{beci-rel-quen-2001}
D.~Becirevic {\it et al.}, Nucl. Phys. B~\textbf{618} (2001) 241, 
[hep-lat/0002025].

\vspace{3mm}

\item \label{ukqcd-ll-rel-quen-2001}
UKQCD Collaboration, L.~Lellouch and C.J.D. Lin, 
Phys. Rev. D~\textbf{64} (2001) 094501, \break
[hep-ph/0011086].

\vspace{3mm}

\item \label{ukqcd-rel-quen-2001}
UKQCD Collaboration, K.C. Bowler {\it et al.}, 
Nucl. Phys. B~\textbf{619} (2001) 507,
  [hep-lat/0007020].

\vspace{3mm}

\item \label{fnal-fnal-quen-1998}
A.X. El-Khadra {\it et al.}, Phys. Rev. D~\textbf{58} (1998) 014506, [hep-ph/9711426].

\vspace{3mm}

\item \label{cppacs-fnal-quen-nf2-2001}
CP-PACS Collaboration, A.~Ali~Khan {\it et al.}, Phys. Rev. D~\textbf{64} (2001)
  034505, [hep-lat/0010009].

\vspace{3mm}

\item \label{glok-nrqcd-quen-1998}
A.~Ali~Khan {\it et al.}, Phys. Lett. B~\textbf{427} (1998) 132, [hep-lat/9801038].

\vspace{3mm}

\item \label{glok-nrqcd-nf2-1999}
S.~Collins {\it et al.}, Phys. Rev. D~\textbf{60} (1999) 074504, [hep-lat/9901001].

\vspace{3mm}

\item \label{jlqcd-nrqcd-quen-2000}
JLQCD Collaboration, K.I. Ishikawa {\it et al.}, Phys. Rev. D~\textbf{61} (2000)
  074501, [hep-lat/9905036].

\vspace{3mm}

\item \label{cppacs-nrqcd-nf2-2001}
CP-PACS Collaboration, A.~Ali~Khan {\it et al.}, Phys. Rev. D~\textbf{64} (2001)
  054504, [hep-lat/0103020].

\vspace{3mm}

\item \label{bbs-1998}
C.W. Bernard, T.~Blum and A.~Soni, Phys. Rev. D~\textbf{58} (1998) 014501,
 [hep-lat/9801039].

\vspace{3mm}

\item \label{svz:79}
M.A. Shifman, A.I. Vainshtein and V.I. Zakharov, Nucl. Phys. B~\textbf{147}
  (1979) 385.

\vspace{3mm}

\item \label{svz:79b}
M.A. Shifman, A.I. Vainshtein and V.I. Zakharov, Nucl. Phys. B~\textbf{147}
  (1979) 448.

\vspace{3mm}

\item \label{ps:01}
A.A. Penin and M.~Steinhauser, Phys. Rev. D~\textbf{65} (2002) 054006,
  [hep-ph/0108110].

\vspace{3mm}

\item \label{jl:01}
M.~Jamin and B.O. Lange, Phys. Rev. D~\textbf{65} (2002) 056005, [hep-ph/0108135].

\vspace{3mm}

\item \label{nar:01}
S.~Narison, Phys. Lett. B~\textbf{520} (2001) 115, [hep-ph/0108242].

\vspace{3mm}

\item \label{ck:00}
P.~Colangelo and A.~Khodjamirian, hep-ph/0010175.

\vspace{3mm}

\item \label{cs:01a}
K.G. Chetyrkin and M.~Steinhauser, Phys. Lett. B~\textbf{502} (2001) 104,
  [hep-ph/0012002].

\vspace{3mm}

\item \label{cs:01b}
K.G. Chetyrkin and M.~Steinhauser, Eur. Phys. J. C~\textbf{21} (2001) 319,
  [hep-ph/0108017].

\vspace{3mm}

\item \label{jam:02}
M.~Jamin, Phys. Lett. B~\textbf{538} (2002) 71, [hep-ph/0201174].

\vspace{3mm}

\item \label{kl:02}
A.X. El-Khadra and M.~Luke, hep-ph/0208114.

\vspace{3mm}

\item \label{op:88}
A.A. Ovchinnikov and A.A. Pivovarov, Phys. Lett. B~\textbf{207} (1988) 333.

\vspace{3mm}

\item \label{ry:88}
L.J. Reinders and S.~Yazaki, Phys. Lett. B~\textbf{212} (1988) 245.

\vspace{3mm}

\item \label{np:94}
S.~Narison and A.A. Pivovarov, Phys. Lett. B~\textbf{327} (1994) 341,
  [hep-ph/9403225].

\vspace{3mm}

\item \label{hnn:02}
K.~Hagiwara, S.~Narison and D.~Nomura, Phys. Lett. B~\textbf{540} (2002) 233,
  [hep-ph/0205092].

\vspace{3mm}

\item \label{Chernyak:1994cx}
V.~Chernyak,
Nucl.\ Phys.\ B {\bf 457} (1995) 96
[hep-ph/9503208].

\vspace{3mm}

\item \label{Cabibbo:Bk:PRL1984}
N.~Cabibbo, G.~Martinelli and R.~Petronzio, Nucl. Phys. B~\textbf{244} (1984)
  381.

\vspace{3mm}

\item \label{Brower:Bk:PRL1984}
R.C. Brower, G.~Maturana, M.~Belen~Gavela and R.~Gupta, Phys. Rev. Lett.
  \textbf{53} (1984) 1318.

\vspace{3mm}

\item \label{Bernard:1984bb}
C.W. Bernard, in Proceedings of Gauge Theory On A Lattice, Argonne (1984),
  p.~85.

\vspace{3mm}

\item \label{Bernard:Bk:1985tm}
C.W. Bernard {\it et al.}, Phys. Rev. Lett. \textbf{55} (1985) 2770.

\vspace{3mm}

\item \label{Gavela:Bk:1988bd}
M.B. Gavela {\it et al.}, Nucl. Phys. B~\textbf{306} (1988) 677.

\vspace{3mm}

\item \label{Kilcup:Bk:PRD1998}
Staggered Collaboration, G.~Kilcup, R.~Gupta and S.R. Sharpe, Phys. Rev.
  D~\textbf{57} (1998) 1654, [hep-lat/9707006].

\vspace{3mm}

\item \label{Aoki:Bk:1998nr}
JLQCD Collaboration, S.~Aoki {\it et al.}, Phys. Rev. Lett. \textbf{80} (1998) 5271,
  [hep-lat/9710073].

\vspace{3mm}

\item \label{Becirevic:Bk:2002mm}
SPQCDR Collaboration, D.~Becirevic {\it et al.}, [hep-lat/0209136].

\vspace{3mm}

\item \label{AliKhan:Bk:2001wr}
CP-PACS Collaboration, A.~Ali~Khan {\it et al.}, Phys. Rev. D~\textbf{64} (2001)
  114506, [hep-lat/0105020].

\vspace{3mm}

\item \label{Blum:Bk:2001xb}
RBC Collaboration, T.~Blum {\it et al.}, hep-lat/0110075.

\vspace{3mm}

\item \label{DeGrand:Bk:2002xe}
MILC Collaboration, T.~DeGrand, hep-lat/0208054.

\vspace{3mm}

\item \label{GGHLR:Bk:2002}
GGHLR Collaboration, N.~Garron {\it et al.}, hep-lat/0212015.

\vspace{3mm}

\item \label{Blum:Bk:1997mz}
T.~Blum and A.~Soni, Phys. Rev. Lett. \textbf{79} (1997) 3595, [hep-lat/9706023].

\vspace{3mm}

\item \label{twistedmass-Bk-lat2001}
M.~Guagnelli {\it et al.}, hep-lat/0110097.

\vspace{3mm}

\item \label{Ishizuka:Bk:1993ya}
N.~Ishizuka {\it et al.}, Phys. Rev. Lett. \textbf{71} (1993) 24.

\vspace{3mm}

\item \label{Kilcup:Bk:1993pa}
G.~Kilcup, Phys. Rev. Lett. \textbf{71} (1993) 1677.

\vspace{3mm}

\item \label{sharpe-review-lat1996}
S.~Sharpe, Nucl. Phys. Proc. Suppl. \textbf{53} (1997) 181, [hep-lat/9609029].

\vspace{3mm}

\item \label{Bijnens:CPT:1984ec}
J.~Bijnens, H.~Sonoda and M.B. Wise, Phys. Rev. Lett. \textbf{53} (1984) 2367.

\vspace{3mm}

\item \label{Sharpe:QCL:1992ft}
S.R. Sharpe, Phys. Rev. D~\textbf{46} (1992) 3146, [hep-lat/9205020].

\vspace{3mm}

\item \label{Sharpe:TASI:1994dc}
S.R. Sharpe, hep-ph/9412243.

\vspace{3mm}

\item \label{Golterman:QCL:1998st}
M.F.L. Golterman and K.C. Leung, Phys. Rev. D~\textbf{57} (1998) 5703,
  [hep-lat/9711033].

\vspace{3mm}

\item \label{praetal:91}
J.~Prades {\it et al.}, Z. Phys. C~\textbf{51} (1991) 287.

\vspace{3mm}

\item \label{jp:94}
M.~Jamin and A.~Pich, Nucl. Phys. B~\textbf{425} (1994) 15, [hep-ph/9402363].

\vspace{3mm}

\item \label{nar:95}
S.~Narison, Phys. Lett. B~\textbf{351} (1995) 369, [hep-ph/9409428].

\vspace{3mm}

\item \label{ry:87}
L.J. Reinders and S.~Yazaki, Nucl. Phys. B~\textbf{288} (1987) 789.

\vspace{3mm}

\item \label{bdg:88}
N.~Bilic, C.A. Dominguez and B.~Guberina, Z. Phys. C~\textbf{39} (1988) 351.

\vspace{3mm}

\item \label{bbg:88}
W.~Bardeen, A.~Buras and J.M. G{\'e}rard, Phys. Lett. \textbf{211} (1988) 343.

\vspace{3mm}

\item \label{hks:99}
T.~Hambye, G.O. K{\"o}hler and P.H. Soldan, Eur. Phys. J. C~\textbf{10} (1999)
  271, [hep-ph/9902334].

\vspace{3mm}

\item \label{pdr:00}
S.~Peris and E.~{de Rafael}, Phys. Lett. B~\textbf{490} (2000) 213,
  [hep-ph/0006146].

\vspace{3mm}

\item \label{bp:00}
J.~Bijnens and J.~Prades, J. High Energy Phys. \textbf{01} (2000) 002,
  [hep-ph/9909244].

\vspace{3mm}

\item \label{befl:98}
S.~Bertolini, J.O. Eeg, M.~Fabbrichesi and E.I. Lashin, Nucl. Phys.
  B~\textbf{514} (1998) 63, \break
[hep-ph/9705244].

\vspace{3mm}

\item \label{bjw:90}
A.J. Buras, M.~Jamin and P.H. Weisz, Nucl. Phys. B~\textbf{~347} (1990) 491.

\vspace{3mm}

\item \label{dgh:82}
J.F. Donoghue, E.~Golowich and B.R. Holstein, Phys. Lett. B~\textbf{119} (1982)
  412.

\vspace{3mm}

\item \label{ewwg_chi} LEP/SLD Electroweak Working Group, CERN-EP/2001-021.

\vspace{3mm}

\item \label{lepsldcdf}
LEP/SLD/CDF Collaborations, CERN-EP/2001-050.

\vspace{3mm}

\item \label{ref:fbddelphi} DELPHI Collaborations, EPS-HEP99, 
contributed paper 5.515.

\vspace{3mm}

\item \label{ref:osciw}
{The LEP B Oscillation Working Group}, {http://www.cern.ch/LEPBOSC/}. 

\vspace{3mm}

\item \label{Rosner_sst}
M.~Gronau, A.~Nippe and J.L. Rosner, Phys. Rev. D~\textbf{47} (1993) 1988,
  [hep-ph/9211311].

\vspace{3mm}

\item \label{ref:bdmix_dstarl_sst_cdf}
CDF Collaboration, F.~Abe {\it et al.}, Phys. Rev. Lett. \textbf{80} (1998) 2057,
  [hep-ex/9712004].

\vspace{3mm}

\item \label{ref:bdmix_dstarl_sst_long_cdf}
CDF Collaboration, F.~Abe {\it et al.}, Phys. Rev. D~\textbf{59} (1999) 032001,
  [hep-ex/9806026].

\vspace{3mm}

\item \label{ref:review_sld}
P.C. Rowson, D.~Su and S.~Willocq, Ann. Rev. Nucl. Part. Sci. \textbf{51}
  (2001) 345, [hep-ph/0110168].

\vspace{3mm}

\item \label{ref:bdmix_inclept_sld}
SLD Collaboration  (1996). SLAC-PUB-7228, contribution to ICHEP96 Warsaw.

\vspace{3mm}

\item \label{ref:bdmix_lepd_sld}
SLD Collaboration, K.~Abe {\it et al.}, SLAC-PUB-7229, contribution to
  ICHEP96 Warsaw.

\vspace{3mm}

\item \label{ref:bdmix_kdipo_sld}
SLD Collaboration, SLAC-PUB-7230, contribution to ICHEP96 Warsaw.

\vspace{3mm}

\item \label{ref:bdmix_ktag_sld}
J.L. Wittlin  (2001). Ph.D. thesis, SLAC-R-582.

\vspace{3mm}

\item \label{Barate:1998ua}
ALEPH Collaboration, R.~Barate {\it et al.}, Eur. Phys. J. C~\textbf{4} (1998) 367.

\vspace{3mm}

\item \label{Barate:1998rv}
ALEPH Collaboration, R.~Barate {\it et al.}, Eur. Phys. J. C~\textbf{7} (1999) 553,
  [hep-ex/9811018].

\vspace{3mm}

\item \label{ref:bsmix_excl_delphi}
DELPHI Collaboration, P.~Abreu {\it et al.}, Eur. Phys. J. C~\textbf{18} (2000) 229,
  [hep-ex/0105077].

\vspace{3mm}

\item \label{Abreu:2000sh}
DELPHI Collaboration, P.~Abreu {\it et al.}, Eur. Phys. J. C~\textbf{16} (2000) 555,
  [hep-ex/0107077].

\vspace{3mm}

\item \label{ref:bdmix_dilept_opal}
OPAL Collaboration, K.~Ackerstaff {\it et al.}, Z. Phys. C~\textbf{76} (1997) 417,
  [hep-ex/9707010].

\vspace{3mm}

\item \label{ref:bdmix_inclept_opal}
OPAL Collaboration, K.~Ackerstaff {\it et al.}, Z. Phys. C~\textbf{76} (1997) 401,
  [hep-ex/9707009].

\vspace{3mm}

\item \label{Thom:2002fs}
J.~Thom, SLAC-R-585 (2002).

\vspace{3mm}

\item \label{ref:bphys_run1_cdf}
CDF Collaboration, M.~Paulini, Int. J. Mod. Phys. A~\textbf{14} (1999) 2791,
  [hep-ex/9903002].

\vspace{3mm}

\item \label{Aubert:2001nu}
BABAR Collaboration, B.~Aubert {\it et al.}, Phys. Rev. Lett. \textbf{87} (2001)
  091801, [hep-ex/0107013].

\vspace{3mm}

\item \label{Abe:2001xe}
Belle Collaboration, K.~Abe {\it et al.}, Phys. Rev. Lett. \textbf{87} (2001) 091802,
  [hep-ex/0107061].

\vspace{3mm}

\item \label{ref:moser}
H.G. Moser and A.~Roussarie, Nucl. Instrum. Meth. A~\textbf{384} (1997) 491.

\vspace{3mm}

\item \label{ref:bdmix_excl_babar}
BABAR Collaboration, B.~Aubert {\it et al.}, Phys. Rev. Lett. \textbf{88} (2002)
  221802, [hep-ex/0112044].

\vspace{3mm}

\item \label{ref:bdmix_excl_belle}
Belle Collaboration, T.~Tomura {\it et al.}, Phys. Lett. B~\textbf{542} (2002) 207,
  [hep-ex/0207022].

\vspace{3mm}

\item \label{ref:bdmix_dstarl_babar}
BABAR Collaboration, B.~Aubert {\it et al.}, [hep-ex/0212017].

\vspace{3mm}

\item \label{ref:bdmix_dstarl_belle}
Belle Collaboration, K.~Hara {\it et al.}  Phys. Rev. Lett. \textbf{89} (2002) 251803, 
 [hep-ex/0207045].

\vspace{3mm}

\item \label{ref:bdmix_dstarl_cdf}
CDF Collaboration, T.~Affolder {\it et al.}, Phys. Rev. D~\textbf{60} (1999) 112004,
  [hep-ex/9907053].

\vspace{3mm}

\item \label{ref:bdmix_dstar_opal}
OPAL Collaboration, G.~Alexander {\it et al.}, Z. Phys. C~\textbf{72} (1996) 377.

\vspace{3mm}

\item \label{ref:bdmix_3methods_aleph}
ALEPH Collaboration, D.~Buskulic {\it et al.}, Z. Phys. C~\textbf{75} (1997) 397.

\vspace{3mm}

\item \label{ref:bdmix_4methods_delphi}
DELPHI Collaboration, P.~Abreu {\it et al.}, Z. Phys. C~\textbf{76} (1997) 579.

\vspace{3mm}

\item \label{ref:bdmix_pil_opal}
OPAL Collaboration, G.~Abbiendi {\it et al.}, Phys. Lett. B~\textbf{493} (2000) 266,
  [hep-ex/0010013].

\vspace{3mm}

\item \label{ref:bdmix_dstarpi_belle}
Belle Collaboration, Y.~Zheng {\it et al.}, to appear in Phys. ReV. D, [hep-ex/0211065].

\vspace{3mm}

\item \label{ref:bdmix_dimu_cdf}
CDF Collaboration, F.~Abe {\it et al.}, Phys. Rev. D~\textbf{60} (1999) 051101.

\vspace{3mm}

\item \label{ref:bdmix_inclept_cdf}
CDF Collaboration, F.~Abe {\it et al.}, Phys. Rev. D~\textbf{60} (1999) 072003,
  [hep-ex/9903011].

\vspace{3mm}

\item \label{ref:bdmix_l3}
L3 Collaboration, M.~Acciarri {\it et al.}, Eur. Phys. J. C~\textbf{5} (1998) 195.

\vspace{3mm}

\item \label{ref:bdmix_dilept_babar}
BABAR Collaboration, B.~Aubert {\it et al.}, Phys. Rev. Lett. \textbf{88} (2002)
  221803, [hep-ex/0112045].

\vspace{3mm}

\item \label{ref:bdmix_dilept_belle}
Belle Collaboration, N.~Hastings {\it et al.},to appear in Phys. Rev. D, 
[hep-ex/0212033].

\vspace{3mm}

\item \label{ref:bdmix_vtx_delphi}
DELPHI Collaboration, J.~Abdallah {\it et~al.}, CERN-EP-2002-078.

\vspace{3mm}

\item \label{ref:bdmix_vtx_aleph}
ALEPH Collaboration, D.~Buskulic {\it et~al.}, ALEPH-97/027, contribution to
  EPS-HEP97 Jerusalem.

\vspace{3mm}

\item \label{ref:bsmix_aleph}
ALEPH Collaboration, A.~Heister {\it et~al.}, CERN-EP-2002-016.

\vspace{3mm}

\item \label{ref:bsmix_phil_cdf}
CDF Collaboration, F.~Abe {\it et al.}, Phys. Rev. Lett. \textbf{82} (1999) 3576.

\vspace{3mm}

\item \label{ref:bsmix_semil_delphi}
DELPHI Collaboration, J.~Abdallah {\it et~al.} DELPHI 2002-073, contribution
  to ICHEP 2002 Amsterdam.

\vspace{3mm}

\item \label{ref:bsmix_dsl_opal}
OPAL Collaboration, G.~Abbiendi {\it et al.}, Eur. Phys. J. C~\textbf{19} (2001) 241,
  [hep-ex/0011052].

\vspace{3mm}

\item \label{ref:bsmix_dstracks_sld}
SLD Collaboration, K.~Abe {\it et al.}, Phys. Rev. D~\textbf{66} (2002) 032009,
  [hep-ex/0207048].

\vspace{3mm}

\item \label{ref:bsmix_inclept_opal}
OPAL Collaboration, G.~Abbiendi {\it et al.}, Eur. Phys. J. C~\textbf{11} (1999) 587,
  [hep-ex/9907061].

\vspace{3mm}

\item \label{ref:bsmix_lepd_sld}
SLD Collaboration, K.~Abe {\it et al.}, SLAC-PUB-8568, [hep-ex/0012043]. 

\vspace{3mm}

\item \label{ref:bsmix_dipole_sld}
SLD Collaboration, K.~Abe {\it et al.}, Phys. Rev. D~\textbf{67} (2003) 012006,
  [hep-ex/0209002].

\vspace{3mm}

\item \label{Albrecht:1992yd}
ARGUS Collaboration, H.~Albrecht {\it et al.}, Z. Phys. C~\textbf{55} (1992) 357.

\vspace{3mm}

\item \label{Albrecht:1994gr}
ARGUS Collaboration, H.~Albrecht {\it et al.}, Phys. Lett. B~\textbf{324} (1994) 249.

\vspace{3mm}

\item \label{Bartelt:1993cf}
CLEO Collaboration, J.~Bartelt {\it et al.}, Phys. Rev. Lett. \textbf{71} (1993)
  1680.

\vspace{3mm}

\item \label{Behrens:2000qu}
CLEO Collaboration, B.H. Behrens {\it et al.}, Phys. Lett. B~\textbf{490} (2000) 36,
  [hep-ex/0005013].

\vspace{3mm}

\item \label{Adam:1997pv}
DELPHI Collaboration, W.~Adam {\it et al.}, 
Phys. Lett. B~\textbf{414} (1997) 382.

\vspace{3mm}

\item \label{Hagiwara:2002fs}
Particle Data Group Collaboration, K.~Hagiwara {\it et al.},
 Phys. Rev. D~\textbf{66}
  (2002) 010001.

\vspace{3mm}

\item \label{Blair:1996kx}
CDF-II Collaboration, R.~Blair {\it et al.}, FERMILAB-PUB-96-390-E.

\vspace{3mm}

\item \label{Ellison:2000hj}
D0 Collaboration, J.~Ellison, prepared for 15th International Workshop
  on High-Energy Physics and Quantum Field Theory (QFTHEP 2000), Tver, Russia,
  14-20 Sep 2000.

\vspace{3mm}

\item \label{Cabrera:2002vp}
S.~Cabrera {\it et al.}, Nucl. Instrum. Meth. A~\textbf{494} (2002) 416.

\vspace{3mm}

\item \label{Grozis:2002nr}
C.~Grozis {\it et al.}, hep-ex/0209027.

\vspace{3mm}

\item \label{Cerri:2002ss}
CDF Collaboration, A.~Cerri, presented at 31st International Conference
  on High Energy Physics (ICHEP 2002), Amsterdam, The Netherlands, 24-31 Jul
  2002.

\vspace{3mm}

\item \label{Lucchesi:2002tw}
CDF Collaboration, D.~Lucchesi, presented at Beauty 2002: 8th
  International Conference on B Physics at Hadron machines, Santiago de
  Compostela, Spain, 17-21 Jun 2002.

\vspace{3mm}

\item \label{Anikeev:2001rk}
K.~Anikeev {\it et al.}, hep-ph/0201071.

\vspace{3mm}

\item \label{Paulini:2003jk}
CDF Collaboration, M.~Paulini, hep-ex/0302016.

\vspace{3mm}

\item \label{ref:par_1}
P.~Paganini, F.~Parodi, P.~Roudeau and A.~Stocchi, Phys. Scripta \textbf{58}
  (1998) 556, [hep-ph/9711261].

\vspace{3mm}

\item \label{ref:sch}
Y.~Grossman, Y.~Nir, S.~Plaszczynski and M.H. Schune, 
Nucl. Phys. B~\textbf{511} (1998) 69, \break
[hep-ph/9709288].

\vspace{3mm}

\item \label{ref:par_2}
F.~Parodi, P.~Roudeau and A.~Stocchi, Nuovo Cim. A~\textbf{112} (1999) 833,
  [hep-ex/9903063].

\vspace{3mm}

\item \label{ref:hock}
A.~Hoecker, H.~Lacker, S.~Laplace and F.~Le~Diberder, Eur. Phys. J.
  C~\textbf{21} (2001) 225, \break
[hep-ph/0104062].

\vspace{3mm}

\item \label{ref:checchia1}
P.~Checchia, E.~Piotto and F.~Simonetto  (1999), hep-ph/9907300.

\vspace{3mm}

\item \label{ref:likr}
M.~Ciuchini {\it et al.}, JHEP \textbf{07} (2001) 013, [hep-ph/0012308].

\vspace{3mm}

\item \label{ref:boix}
G.~Boix and D.~Abbaneo, JHEP \textbf{08} (1999) 004, [hep-ex/9909033].

\vspace{3mm}

\item \label{Bob}
R.N. Cahn, Private communication.

\end{enumerate}

\newpage

\thispagestyle{empty}
~

\newpage


\chapter{FIT OF THE UNITARITY TRIANGLE PARAMETERS}
\label{chap:fits}

{\it Conveners : A.J.~Buras,  F.~Parodi. \\
     Contributors : 
 M.~Ciuchini,
 G.~Dubois-Felsmann,
 G.~Eigen,
 P.~Faccioli,
 E.~Franco,
 A.~Hocker,
 D.~Hitlin,
 H.~Lacker,
 S.~Laplace,
 F.~LeDiberder,
 V.~Lubicz,
 G.~Martinelli,
 F.~Porter,
 P.~Roudeau,
 L.~Silvestrini,
 A.~Stocchi,
 M.~Villa
}
\section{Introduction}
\setcounter{equation}{0}
In this Chapter  we will discuss the determination of the 
Unitarity Triangle (UT)
using as input the values of $\vus$, $\vcb$, and $|V_{ub}|$ from Chapters 
\ref{chap:II} and \ref{chap:III} and the constraints from $\varepsilon_K$ 
and $\Delta M_{d,s}$ with the values of the non-perturbative 
parameters $\hat B_K$, $F_{B_d}\sqrt{\hat B_{B_d}}$, 
$F_{B_s}\sqrt{\hat B_{B_s}}$ and $\xi$ determined in Chapter \ref{chap:IV}. 
We will also include in this analysis the most recent results for 
the CP asymmetry in ${\rm B}_d\to J/\psi {\rm K}_S$ that allows 
to determine the angle $\beta$ of the UT essentially without any 
theoretical uncertainty.
The list of the common quantities which have been used for the analyses
performed in this Chapter are summarised in Table \ref{tab:inputs}.

\begin{table}[!ht]
\begin{center}
\begin{tabular}{|c|c|c|c|c|}
\hline
                          Parameter                & Value  &   Gaussian   &  Theory       \\
                                                   &        &   $\sigma$   & uncertainty    \\ \hline
                          $\lambda$                & 0.2240(0.2210) &    0.0036 ~({\it 0.0020})    &     -        \\
 \hline
$\left | V_{cb} \right | (\times 10^{-3})$ (excl.) &  42.1  &      2.1     &     -         \\
$\left | V_{cb} \right | (\times 10^{-3})$ (incl.) &  41.4~({\it 40.4})  &      0.7     &     0.6({\it 0.8})       \\ \hline 
$\left | V_{ub} \right | (\times 10^{-4})$ (excl.) &  33.0({\it 32.5})  &     2.4({\it 2.9})     &     4.6({\it 5.5})       \\
$\left | V_{ub} \right | (\times 10^{-4})$ (incl.) &  40.9  &      4.6     &     3.6       \\ \hline
                  $\Delta M_d~(\mbox{ps}^{-1})$      &  0.503~({\it 0.494}) &      0.006~({\it 0.007})   &      -        \\
                  $\Delta M_s~(\mbox{ps}^{-1})$      & $>$ 14.4~({\it 14.9}) at 95\% C.L. & \multicolumn{2}{|c|}
                                                                           {sensitivity 19.2~({\it 19.3})}  \\
                             $m_t$ (GeV)           &  167   &       5      &      -        \\
                             $m_c$ (GeV)           &  1.3  &       -      &      0.1        \\
          $F_{B_d} \sqrt{\hat B_{B_d}}$(MeV)       &  223~({\it 230})   &      33~({\it 30})      & 12~({\it 15})  \\
$\xi=\frac{ F_{B_s}\sqrt{\hat B_{B_s}}}
                    { F_{B_d}\sqrt{\hat B_{B_d}}}$ & 1.24(1.18)   &     0.04~({\it 0.03})     & 0.06~({\it 0.04}) \\ \hline
                         $\hat B_K$                &  0.86  &     0.06     &      0.14      \\ \hline
                        sin 2$\beta$               & 0.734~({\it 0.762})   &     0.054~({\it 0.064})   &        -       \\ \hline
\end{tabular} 
\caption[]{ \it { Latest values of the relevant quantities entering 
into the expressions of 
$\epsilon_K$, $\Delta M_d$ and $\Delta M_s$. In the third and fourth columns 
the Gaussian and the flat part of the uncertainty are given, respectively.
The values within parentheses are the ones available at the 
time of the Workshop and  used when comparing
different fitting procedures. In case of asymmetric theoretical errors, 
like for $\vub$ exclusive,
the central values have been shifted to make them symmetric.
 \label{tab:inputs}}}
\end{center}
\end{table}

A very important issue in constraining the apex $(\bar\varrho,\bar\eta)$ 
of the UT is the treatment of the experimental and especially the theoretical
uncertainties. 
In the literature five different approaches can be found:  
Gaussian approach [\ref{Gaus}], 
Bayesian approach [\ref{C00}], frequentist approach [\ref{FREQ}], 
$95\%$ C.L. scan method [\ref{SCAN95}] and the simple 
(naive) scanning within one standard deviation.
Moreover the fact that different authors often use different input 
parameters makes the comparison of various analyses very difficult.

This situation is clearly unsatisfactory as different analyses give generally 
different allowed ranges for $(\bar\varrho,\bar\eta)$. While all these 
analyses find presently the SM consistent with all the available data, 
the situation may change in  
the future when the experimental and theoretical uncertainties will be 
reduced and additional decays relevant for the determination of the UT will be 
available.

It is then conceivable that some approaches will find the SM
consistent with the data whereas other will claim an indication for new
physics contributions. This clearly would be rather unfortunate.
However, even in the absence of new physics contributions, the increasing
accuracy of the data and the expected reduction of theoretical
uncertainties calls for an extensive comparison of
the different methods to gain the best understanding on the UT.

Another important issue is the sensitivity of the UT analysis to theoretical 
uncertainties. Some theoretical parameters have more impact on this 
analysis than others and it is important to identify those for which the 
reduction of errors through improved non-perturbative calculations can 
contribute to the quality of the determination of the UT most efficiently.

The  goals of this Chapter are:
\begin{itemize}
\item
to describe in some detail two of the most developed methods: the 
Bayesian approach and the frequentist approach,
\item
to compare the resulting allowed regions for $(\bar\varrho,\bar\eta)$ 
obtained from the Bayesian and frequentist approaches for the same input 
parameters,
\item
to identify those non-perturbative parameters for which the 
reduction of the uncertainty is most urgent.
\end{itemize}

This Chapter is organized as follows. 
In Section \ref{sec:consts} we express the 
constraints from $|V_{ub}/V_{cb}|$, $\varepsilon_K$ and $\Delta M_{d,s}$
in terms of Wolfenstein parameters [\ref{WO5}] including the generalization 
of [\ref{BLO5}]. The Bayesian method and the frequentist 
methods are discussed in 
Sections \ref{sec:bayes} and \ref{sec:freq}, respectively. 
The discussion in the frequentist 
case includes the \rfit~  and the scanning methods. In 
Section \ref{sec:qcdpar} the impact of the uncertainties of 
theoretical parameters on
the determination of the UT is discussed in detail using both the 
Bayesian approach and the scanning method. Finally in Section \ref{p:comp} we 
compare the Bayesian and \rfit~ methods and draw conclusions. 
In Section  \ref{sec:final}
we show some important results obtained in testing the consistency of the CKM
picture of the Standard Model.

\section{Constraints on the Unitarity Triangle parameters}
\label{sec:consts}
\begin{figure}[hbt]
\includegraphics[width=14cm]{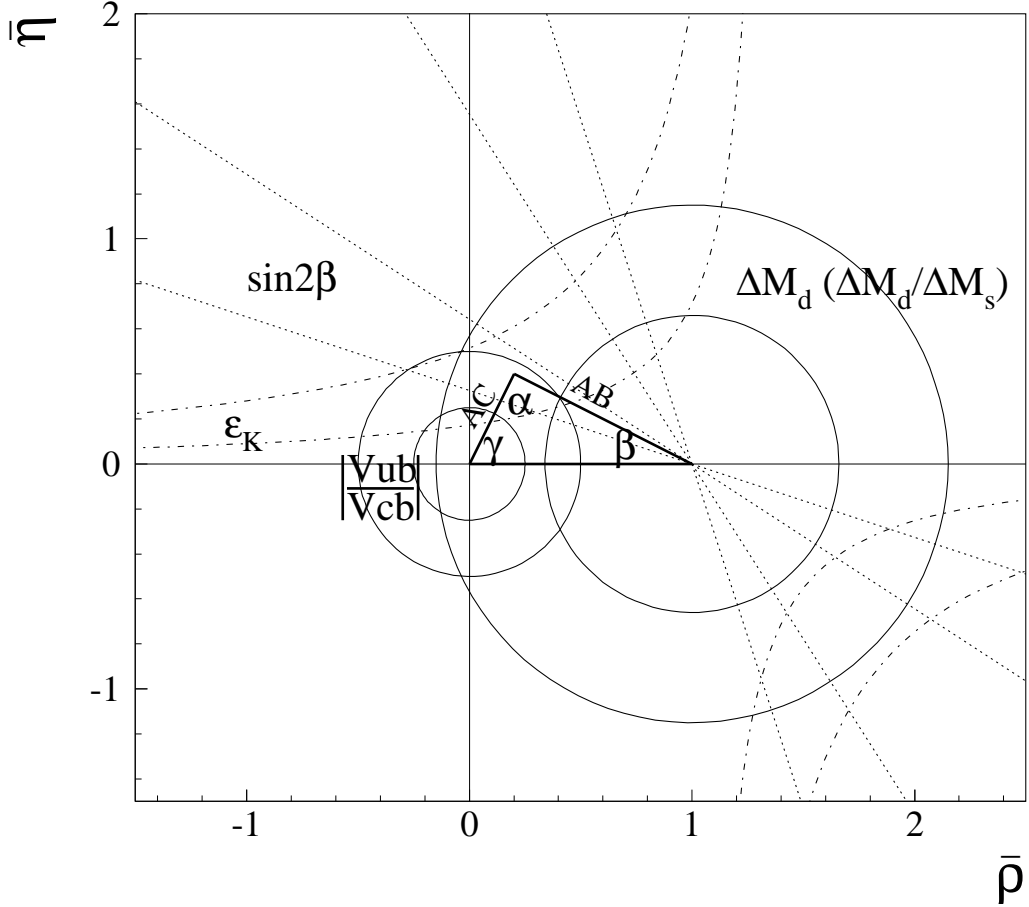}
\caption[]{\it Constraints which are contributing to the Unitarity Triangle parameter
determination.} 
\label{fig:utriangle}
\end{figure}

\setcounter{equation}{0}
Five measurements restrict at present the range of $(\bar\varrho,\bar\eta)$ 
within the SM: 
\begin{itemize}
\item
The $\vub$ constraint: \\
The length of the side AC of the UT (see Fig.~\ref{fig:utriangle})
is determined from
\begin{equation}\label{RB}
R_b = \sqrt{\bar\varrho^2 +\bar\eta^2}
= (1-\frac{\lambda^2}{2})\frac{1}{\lambda}
\left| \frac{V_{ub}}{V_{cb}} \right|.
\end{equation}
The constraint in the $(\bar\varrho,\bar\eta)$ plane resulting from
(\ref{RB}) is represented by a circle of radius 
$R_b$ that is centered at $(\bar\varrho,\bar\eta)=(0,0)$
( for the visualisation of this and following constraints  
see Fig.~\ref{fig:utriangle}).
\item 
The $\varepsilon_K$--constraint:
\begin{equation}\label{100}
\bar\eta \left[(1-\bar\varrho) A^2 \eta_2 S(x_t)
+ P_c(\varepsilon) \right] A^2 \hat B_K = 0.187~
\left(\frac{0.224}{\lambda}\right)^{10}~,
\end{equation}
that follows from the experimental value for $\varepsilon_K$
and the formula (Eq.~(\ref{eq:epsformula}) of Chapter \ref{chap:IV}). 
Here
\begin{equation}\label{102}
P_c(\varepsilon) = 
\left[ \eta_3 S_0(x_c,x_t) - \eta_1 x_c \right] \frac{1}{\lambda^4},
\qquad
x_t=\frac{\mt^2}{\mw^2}.
\end{equation}
 $P_c(\varepsilon)=0.29\pm0.07$ [\ref{HNab}] summarizes the contributions
of box diagrams with two charm quark exchanges and the mixed 
charm-top exchanges. We observe a very strong dependence of the
r.h.s. in (\ref{100}) on the parameter $\lambda=\vus$. However, this
dependence is cancelled to a large extent by the $\lambda$ dependence
of $P_c(\varepsilon)$ and of $A=\vcb/\lambda^2$ that enter the l.h.s
of (\ref{100}).
The main uncertainties in the constraint (\ref{100}) reside then in
$\hat B_K$ and to some extent in the factor $A^4$ or equivalently $\vcb^4$
which multiplies the 
dominant term. The status of $\hat B_K$ has been reviewed in Chapter \ref{chap:IV}.
Eq.~(\ref{100}) specifies 
an hyperbola in the $(\bar \varrho, \bar\eta)$ plane.
This hyperbola intersects the circle found from the $\vub$ constraint
in two points which correspond to two solutions for
the angle $\gamma$.
\item
The $\Delta M_d$--constraint: \\ 
The length $R_t$ of the side AB of the UT (see Fig.~\ref{fig:utriangle})
can be determined from the observed ${\rm B}^0_d-\overline{\rm B}^0_d$ mixing, parametrized 
by $\Delta M_d$ and given in Eq.~\ref{eq:xds} (in Chapter \ref{chap:IV}), with the result
\begin{equation}\label{106}
 R_t= \sqrt{(1-\bar\varrho)^2 +\bar\eta^2}= 
\frac{1}{\lambda}\frac{|V_{td}|}{\vcb} = 0.85 \cdot
\left[\frac{|V_{td}|}{7.8\cdot 10^{-3}} \right] 
\left[ \frac{0.041}{\vcb} \right]
\end{equation}
where
\begin{equation}\label{VT}
\vtd=
7.8\cdot 10^{-3}\left[ 
\frac{230\mev}{\sqrt{F_{B_d}\hat B_{B_d}}}\right]
\sqrt{\frac{\Delta M_d}{0.50/{\rm ps}}} 
\sqrt{\frac{0.55}{\eta_B}}\sqrt{\frac{2.34}{S_0(x_t)}}~.
\end{equation}
Since $\mt$, $\Delta M_d$ and $\eta_B$ are already rather precisely
known, the main uncertainty in the determination of $R_t$ and $\vtd$ from
${\rm B}_d^0-\overline{\rm B}_d^0$ mixing comes from $F_{B_d}\sqrt{\hat B_{B_d}}$.
Its theoretical status has been reviewed in Chapter \ref{chap:IV}.
$R_t$ suffers from additional uncertainty in $\vcb$.
The constraint in the $(\bar\varrho,\bar\eta)$ plane resulting from
(\ref{106}) is represented by a circle of radius 
$R_t$ that is centered at $(\bar\varrho,\bar\eta)=(1,0)$.
\item 
The $\Delta M_d/\Delta M_s$--constraint:\\
The measurement of ${\rm B}^0_s-\overline{\rm B}^0_s$ 
mixing parametrized by $\Delta M_s$
together with $\Delta M_d$  allows to determine $R_t$ in a different
manner:
\be\label{Rt}
R_t=\frac{1}{\lambda}\xi\sqrt{\frac{M_{{\rm B}_s}}{M_{{\rm B}_d}}} 
\sqrt{\frac{\Delta M_s}{\Delta M_d}} 
(1-\frac{\lambda^2}{2}+\bar\varrho\lambda^2),
\qquad \xi=\frac{F_{B_s}\sqrt{\hat B_s}}{F_{B_d}\sqrt{\hat B_d}}~.
\ee
This constraint follows from Eq.~(\ref{eq:xds}) 
(in Chapter \ref{chap:IV}) with the factor 
$(1-{\lambda^2}/{2}+\bar\varrho\lambda^2)$ representing 
the departure of $|V_{ts}/V_{cb}|$ from unity. For 
$0\le \bar\varrho\le 0.5$ this factor deviates from unity by less 
than~$2\%$. Neglecting this correction gives ($\lambda=0.224$) 
\be
R_t=0.86~\sqrt{\frac{\Delta M_d}{0.50/{\rm ps}}}
\sqrt{\frac{18.4/{\rm ps}}{\Delta M_s}}\left[\frac{\xi}{1.18}\right].
\ee
The advantage of determining $R_t$ by means of the ratio 
$\Delta M_d/\Delta M_s$ with respect to the $\Delta M_d$ constraint are 
smaller hadronic uncertainties in $\xi$ than in ${F_{B_d}\sqrt{\hat B_d}}$ 
and the absence of $m_t$ and $\vcb$ dependence. 
The present status of $\xi$ has been 
reviewed in Chapter \ref{chap:IV}.
\item
The $a(\psi K_S)$--constraint:\\ 
The mixing induced CP asymmetry $a_{\psi K_S}$ in $\rm B\to \psi K_S$ allows to 
determine 
the angle $\beta$ of the UT essentially without any hadronic 
uncertainties through
\be
(\sin 2\beta)_{\psi K_S}=0.734\pm 0.054~.
\label{ga}
\ee
 The value given in (\ref{ga}) is the world 
average from [\ref{NIR02}] and is dominated by the results of the 
BaBar [\ref{BaBar5}] and Belle [\ref{Belle5}] Collaborations. 
\end{itemize}

\section{Statistical methods for CKM fits}
In this Section we describe the basic ingredients for the different statistical 
approaches.
The plots and the results presented here have to be taken as illustrations of the
methods. Quantitative results and comparisons are given in the next Sections.

\subsection{Bayesian methods}
\label{sec:bayes}

In this Section we describe the basic ingredients of the Bayesian approach
and discuss the role of the systematic and theoretical uncertainties in deriving 
probability intervals for the relevant parameters.

Each of Eqs.~(\ref{RB}, \ref{100}, \ref{106}, \ref{Rt}, \ref{ga})
relates a  constraint $c_j$
to the parameters $\rhobar$ and $\etabar$, via 
the set of additional parameters ${\mathbf x}$, 
where ${\mathbf x} =\{x_1, x_2, \ldots, x_N\}$ stand for all 
experimentally determined or theoretically calculated 
quantities  on which the various $c_j$ depend ($m_t, ~\xi$ ...)
\begin{equation}
c_j=c_j(\rhobar,\etabar; {\mathbf x}). 
\label{eq:c_j}
\end{equation} 
In an ideal case of  exact knowledge of  $c_j$ and ${\mathbf x}$,
each of the  constraints  provides a curve in the 
$(\rhobar,\etabar)$ plane.  In a realistic  case, the analysis
suffers from  several uncertainties on the quantities $c_j$ and ${\mathbf x}$.
This means that, instead of a  single curve (\ref{eq:c_j}) in the $(\rhobar,\etabar)$ 
plane, we have a family  of curves which depends on the 
distribution of  the set $\{c_j,{\mathbf x}\}$.
As a result, the points  in the $(\rhobar,\etabar)$ plane get  
different weights (even if they were taken to be  equally probable 
{\it a priori}) and the {\it confidence} on the values of 
$\rhobar$ and $\etabar$ clusters in a region of the plane. 

The above arguments can be  formalized
by using the so called Bayesian approach (see~[\ref{ref:YR9903}]
for an introduction). In this approach, 
the uncertainty  is described in terms of 
a probability density function  (pdf)
which quantifies the confidence on the values of a given quantity. 
Applying Bayes Theorem in the case of a single  constraint
we obtain
\begin{eqnarray}
f(\rhobar,\etabar, c_j,  {\mathbf x }\,|\,\hat{c}_j) & \propto &  
f(\hat{c}_j\,|\,c_j,\rhobar,\etabar,{\mathbf x})\cdot 
f(c_j,\rhobar,\etabar,{\mathbf x}) \label{eq:bayes1}\\
& \propto & f(\hat{c}_j\,|\,c_j)\cdot f(c_j\,|\,\rhobar,\etabar,{\mathbf x})
      \cdot f({\mathbf x},\rhobar,\etabar) \label{eq:bayes3}\\
&\propto & f(\hat{c}_j\,|\,c_j)\cdot\delta(c_j-c_j(\rhobar,\etabar,{\mathbf x}))
 \cdot f({\mathbf x})\cdot f_\circ(\rhobar,\etabar)\,,
\label{eq:bayes33}
\end{eqnarray} 
where $\hat{c}_j$ is the experimental best  estimate of $c_j$ 
and $f_\circ(\rhobar,\etabar)$ denotes the {\it prior} distribution.

The various steps follow from  probability rules,
by assuming  the independence of the different quantities and 
 by noting that 
$\hat{c}_j$ depends on ($\rhobar,\etabar,{\mathbf x}$) only 
via $c_j$. This is true since  $c_j$ is  
unambiguously determined, within the Standard Model, from the values of 
 $\rhobar$, $\etabar$ and ${\mathbf x}$.

The extension of the formalism to several 
constraints is straightforward. 
 We can rewrite Eq.~(\ref{eq:bayes1})~as 
 \begin{eqnarray}
f(\rhobar,\etabar,{\mathbf x}\,|\,\hat c_1,...,\hat c_{\rm M})
&\propto &
\prod_{j=1,{\rm M}}f_j(\hat{c}_j\,|\,\rhobar,\etabar,{\mathbf x})
\times \prod_{i=1,{\rm N}}f_i(x_i) \times f_\circ(\rhobar,\etabar) \ .
 \label{eq:bayes_f}
\end{eqnarray} 
M and N run over the constraints and the parameters respectively.
In the derivation of (\ref{eq:bayes_f}),
we have  used  the  independence of the different quantities.  
By integrating Eq.~(\ref{eq:bayes_f}) over ${\mathbf x}$ 
we obtain 
\begin{equation}
f({\rhobar},{\etabar} \, | \, {\mathbf {\hat{c}}} , {\mathbf f} )
\propto {\cal L}({\mathbf {\hat{c}}} \, | \, \rhobar,\etabar, {\mathbf f})
\times f_\circ(\rhobar,\etabar)\, ,
\end{equation} 
where $\mathbf {\hat{c}}$ stands for the set of measured constraints,
 and
\begin{equation}
{\cal L}({\mathbf {\hat{c}}}\,|\,\rhobar,\etabar,{\mathbf f})
= \int 
\prod_{j=1,{\rm M}}f_j(\hat{c}_j\,|\,\rhobar,\etabar,{\mathbf x})
\prod_{i=1,{\rm N}}f_i(x_i)\, \mbox{d}{ x_i}
\label{eq:lik_int}
\end{equation}
is the effective overall likelihood which takes into account 
all possible values of $x_j$, properly weighted. We have
written explicitly that the overall likelihood depends 
on the best knowledge of all $x_i$, described by $f({\mathbf x})$. 
Whereas {\it a priori} all values for  $\rhobar$ 
and $\etabar$ are considered equally likely ($f_\circ(\rhobar,\etabar)$=cst), 
{\it a posteriori}  the probability clusters 
around the point which maximizes the  likelihood. 

In conclusion, the final (unnormalized) pdf obtained 
starting from a uniform pdf for $\rhobar$ and $\etabar$ is
\begin{equation}
f(\rhobar,\etabar) \propto 
 \int 
\prod_{j=1,{\rm M}}f_j(\hat{c}_j\,|\,\rhobar,\etabar,{\mathbf x})
\prod_{i=1,{\rm N}}f_i(x_i)\,\mbox{d}{x_i}\, .
\label{eq:flat_inf}
\end{equation}   
The integration can be performed by Monte Carlo methods and the 
normalization is trivial. 
Starting from the pdf for $\rhobar$ and $\etabar$, probability regions $P(w)$
are defined by the conditions:
\begin{eqnarray*}
\begin{array}{l}
   (\rhobar,\etabar) \in P(w) \, \mbox{ if $f(\rhobar,\etabar)~>~z_w$} \\
   \int_{P(w)} f(\rhobar,\etabar) d\rhobar d\etabar = w
\end{array}
\end{eqnarray*}

\begin{figure}[hbtp]
\begin{center}
\includegraphics[width=13.5cm]{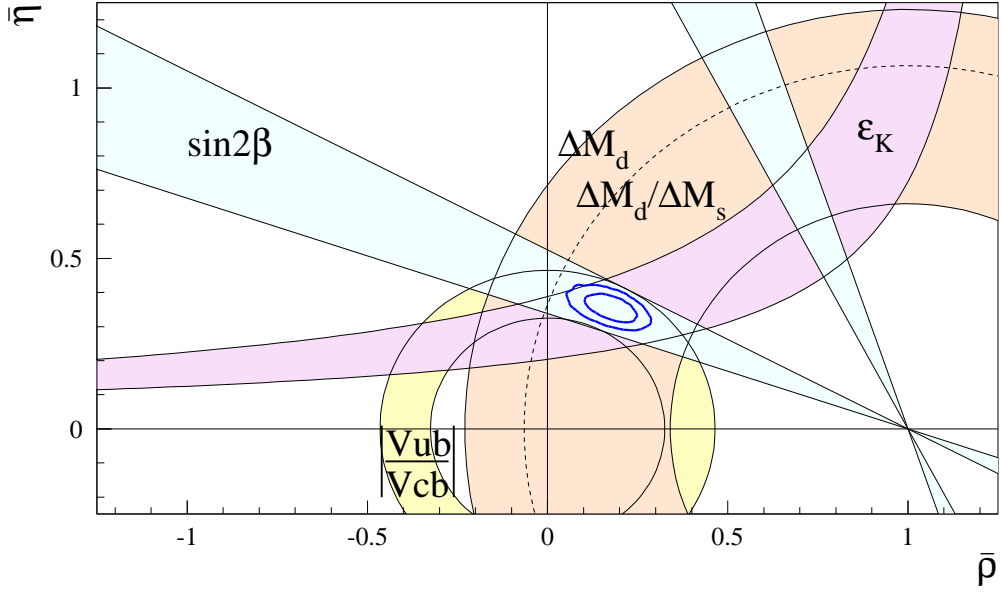}
\caption{{ \it The contours at 68\%, 95\% probability regions in
$\bar{\rho}$ and $\bar{\eta}$ as obtained using the 
Bayesian method, superimposed to the experimental constraints.}}
\label{fig:bande}
\end{center}
\end{figure}

An example of the typical output of this fit approach is shown in
Fig.~\ref{fig:bande} where the probability regions at 68$\%$ and
95$\%$ are shown together with the experimental constraints.

It is important to note that this method does not make any
distinction on whether the individual likelihood associated
with some  constraint is different from zero only in a narrow
region (and we usually refer to this case as ``measurement"), 
or if it goes to zero only on one side 
(e.g. when  $c_j\rightarrow \infty$ or~$0$). In the latter case,  
the data only provide an upper/lower bound to the value of the constraint. 
This is precisely what happens, at present, with $\Delta M_s$. Therefore, 
the experimental information about this constraint enters naturally
in the analysis.

One of the feature of the Bayesian approach
is that there is no conceptual distinction between 
the uncertainty due to random fluctuations, 
which might have occurred in the measuring process, 
the uncertainty about the parameters of the theory,
and the uncertainty about systematics of 
not-exactly-known value, which can be both of experimental or theoretical origin 
(in the Bayesian jargon there are often indicated as influence parameters).

We can simply extend the notation to include in ${\mathbf x}$
these influence parameters responsible for the systematic uncertainty, 
and use Eq.~(\ref{eq:lik_int}) in  an extended way.
Irrespectively of the assumptions
made on the pdf of  ${\mathbf x}$,
the overall likelihoods $f(\hat{c}_j)$ are approximately 
Gaussian because of a mechanism similar to the central limit 
theorem (i.e. just a matter of combinatorics). 
This makes the results largely stable against variations within 
choices of the distributions used to describe 
the uncertainties due to theory or systematics. For this reason 
we simplify the problem, by reducing the choice to only two possibilities.
We choose a Gaussian model when the uncertainty is 
dominated by statistical effects, or there are many
comparable  contributions to the systematic errors, so that the 
central limit theorem applies (${\cal{G}}(x-x_{0})$).
We choose a uniform pdf if the parameter is believed 
to be (almost) certainly in a given interval, and the 
points inside this interval are considered as equally probable.
The second model is applied to some theoretical uncertainties.
${\cal{U}}(x) = 1/2 \sigma_{\rm theo}$ for 
$x \in [x_{0}-\sigma_{\rm theo},x_{0}+\sigma_{\rm theo}]$
and ${\cal{U}}(x) = 0$ elsewhere. 
The combined pdf $\cal{P}$ is then obtained by convoluting 
the Gaussian pdf $\cal{G}$ with the uniform pdf 
$\cal{U}$: $\cal{P} = \cal{G} \otimes \cal{U}$.
When several determinations of the same quantity are available, 
the final p.d.f, in the Bayesian approach, is obtained by the 
product of the single pdfs.

An important point is how to evaluate the compatibility
among individual constraints.
In the CKM fits based on $\chi^2$ minimization, a conventional evaluation 
of compatibility stems automatically from 
the  value of the $\chi^2$ at its minimum. 

The compatibility between constraints in the Bayesian method is
evaluated by comparing partial pdfs
obtained when removing each constraint at a time. 
The integral over the overlap between the pdf with and without a given constraint
quantifies the compatibility.
In case of poor overlap the difference $\Delta_j$ between the two pdfs
can be determined, for each constraint $c_j$, by substituting
\begin{equation}
    c_j \rightarrow c_j(1+\Delta_j).
\label{eq:compa}
\end{equation}
Further investigation (based on physics and not on statistics) 
will be necessary to tell if the difference $\Delta_j$ comes 
from an incorrect evaluation of the input parameters or from new physics.


\subsection {Frequentist methods}
\label{sec:freq} 

As said in the introduction theoretical quantities play an important role 
in the formulae relating the measured quantities to the UT
parameters. These quantities are often inferred from 
theoretical calculations with uncertainties which can be 
associated to approximations. 
Uncertainties due to approximations are often estimated from 
more or less educated guesswork.  For example, we recall that  i) The quenched 
approximation in Lattice QCD calculations; ii) Model calculations 
of form factors where model parameters are varied within some 
range; iii) Higher order terms neglected in a power series for 
which the error is estimated from the ``natural size'' of the 
expansion parameter or a scale dependence in a perturbative 
series where the error is estimated by varying the scale within 
some {\it reasonable} range. 
This has driven the developments of statistical approaches 
based on a frequentist understanding of systematic theoretical 
uncertainties, which cannot be treated as statistically distributed quantities.

In this framework two approaches are presented: 
the \rfit~method and the Scanning method. In both methods, 
the ``theoretically allowed'' values for some theoretical 
parameters are ``scanned'', {\it i.e.} no statistical weight 
is assigned to these parameters as long as 
their values are inside a ``theoretically allowed'' range.

The \rfit~method starts by choosing 
a point in a parameter subspace of interest, {\it e.g.} a point 
in the $\rhobar$-$\etabar$ plane, and ask for the best 
set of theoretical parameters for this given point. 
This set is determined by minimizing a $\chi^2$ function 
with respect to all model parameters, except $\rhobar$ 
and $\etabar$. The theoretical parameters are free 
to vary inside their theoretically allowed range without 
obtaining any statistical weight. In this way, upper limits
of confidence levels in the parameter subspace of interest 
can be determined.

The basic idea of the Scanning method is to choose a 
possible set of values for the 
theoretical parameters and to ask whether this 
particular model leads to an acceptable fit for the 
given data set. If so, a confidence contour is drawn 
in a parameter subspace of interest, {\it e.g.} the 
$\rhobar$-$\etabar$ plane, representing the constraints 
obtained for this particular set of model parameters. 
This procedure is repeated for a large number of possible 
theoretical models by scanning allowed ranges of 
the non-perturbative parameters. The single confidence 
level contours cannot be compared from a statistical 
point of view. This method has been extended to facilitate
an analysis of the relative influence of experimental 
and theoretical uncertainties in determining the 
consistency of the measurements with theory.

\subsubsection{The {\rm \rfit}~approach}
\label{sec:rfit}

The CKM analysis using the \rfit~method \footnote{ 
The \rfit~method is implemented in the software package 
\CKMfitter~[\ref{CkmFitter}]. More details can be found
in [\ref{FREQ}].} is performed in three 
steps: 
1.~Testing the overall consistency between data 
and the theoretical framework, here the SM. 
2.~If data and the SM are found to be in reasonable 
agreement, confidence levels (CL) in parameter subspaces are 
determined. 
3.~Testing extensions of the SM.

The quantity $\chi^2 = -2 \ln{{\cal L}(y_{\rm mod})}$ is minimized 
in the fit, where the likelihood function is defined by 
${\cal L}(y_{\rm mod}) = 
{\cal L}_{\rm exp}(x_{\rm exp}-x_{\rm theo}(y_{\rm mod})) 
\cdot {\cal L}_{\rm theo}(y_{\rm QCD})$. 
The experimental part, ${\cal L}_{\rm exp}$, depends on measurements, 
$x_{\rm exp}$, and theoretical predictions, $x_{\rm theo}$, which are 
functions of model parameters, $y_{\rm mod}$. The theoretical part, 
${\cal L}_{\rm theo}$, describes our ``knowledge'' of the theoretical 
parameters, $y_{\rm QCD} \in\{ y_{\rm mod} \}$. We set 
${\cal L}_{\rm theo}=1$ within an ``allowed range'' ${\cal R}$ provided 
by a theoretical estimate, and ${\cal L}_{\rm theo}=0$ outside ${\cal R}$. 
That is, the $y_{\rm QCD}$ are free to vary within ${\cal R}$ without 
changing the ${\cal L}_{\rm theo}$ part of the $\chi^2$ function. It 
should be kept in mind that the choice of ${\cal R}$ is statistically 
not well-defined and reflects an intrinsic problem of all statistical 
analyses when dealing with theoretical uncertainties. 

It is worthwhile to emphasize that a uniform likelihood function is not 
identical to a uniform pdf. Whereas a 
uniform likelihood means that the theoretical parameter is free to vary 
within ${\cal R}$, a uniform pdf states that each value within ${\cal R}$ 
has equal probability and hence introduces a statistical weight. This 
has important consequences if more than one theoretical parameter enter 
a constraint or if the constraint depends on a nonlinear function of a 
theoretical parameter. For example, the $\eps_{K}$ constraint depends 
on the product $P=\hat B_{K} \cdot |V_{cb}|^4$. The theoretical likelihood 
for $|V_{cb}|^4$ reads ${\cal L}_{\rm theo}(|V_{cb}|^4)=1$ for all 
theoretically allowed $|V_{cb}|$ values given by 
${\cal L}_{\rm theo}(|V_{cb}|)=1$. The theoretical likelihood for the 
product $P$ reads ${\cal L}_{\rm theo}(P)=1$ for any value of $\hat B_{K}$ 
and $|V_{cb}|$ given by ${\cal L}_{\rm theo}(\hat B_{K})=1$ and 
${\cal L}_{\rm theo}(|V_{cb}|)=1$, respectively. That is, in \rfit,  
no statistical weight is introduced for any value of $P$, independent 
of the fact whether the single theoretical parameters are bound or 
unbound and independent of the particular parametrization chosen. On 
the contrary, the pdf for the theoretical part of $|V_{cb}|^4$ is 
proportional to $({|V_{cb}|)^{-3/4}}$ if the theoretical pdf for 
$|V_{cb}|$ is chosen to be uniform. The pdf for the product $P$ 
would be proportional to $-\log{|P|}$ in leading order if the pdfs 
for $\hat B_{K}$ and $|V_{cb}|^4$ were chosen to be uniform~[\ref{FREQ}].

\begin{figure}[t] 
\centerline{
\resizebox{0.62\textwidth}{!}{
  \includegraphics{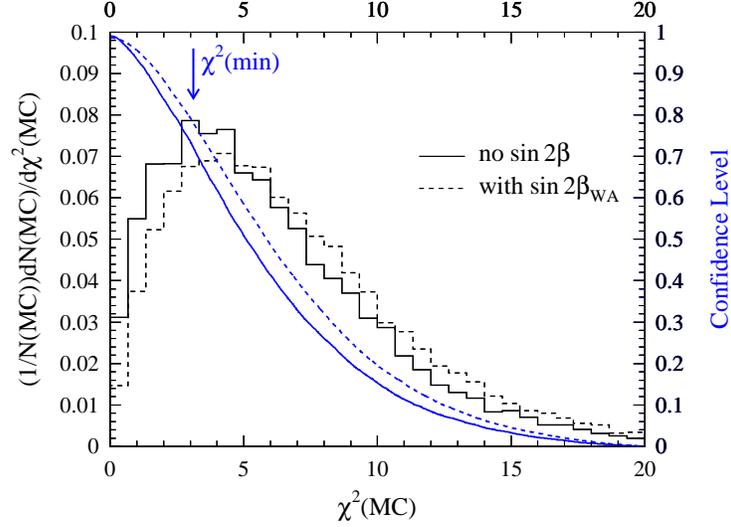}
}
}
\vspace{-0.0cm}       
\caption{\it
   Determination of CL(SM). The histograms show the test 
   statistics $F(\chi^2)$ built from the Monte Carlo technique 
   as described in the text for two different fits, including 
   or excluding the $\sin{2\beta}$ measurement. Integration 
   of the distributions above $\chi^{2}_{min}$ provides CL(SM).
}
\label{chi2min}      
\end{figure}

The agreement between data and the SM is gauged by the global minimum 
$\chi^2_{{\rm min};y_{\rm mod}}$, determined by varying freely all model 
parameters $y_{\rm mod}$. For $\chi^2_{{\rm min};y_{\rm mod}}$, a 
confidence level CL(SM) is computed by means of a Monte Carlo simulation. 
For the optimal set of model parameters $y_{\rm mod}$, a large number of 
pseudo-measurements is generated using the experimental likelihood function 
${\cal L}_{\rm exp}$. For each set of pseudo-measurements, the minimum 
$\chi^2_{{\rm min}}$ is determined and used to build a test statistics 
$F(\chi^2)$. The CL is then calculated as 
${\rm CL(SM)} = \int_{\chi^2_{{\rm min};y_{\rm mod}}}^{\infty} 
F(\chi^{2}) d\chi^{2}$
as illustrated in Fig.~\ref{chi2min}. If there is a hint of an
incompatibility between data and the SM one has to investigate
in detail which constraint leads to a small value for ${\rm CL(SM)}$.

If the hypothesis ``the CKM picture of the SM is correct'' is accepted, 
CLs in parameter subspaces $a$, 
{\it e.g.} $a=(\bar{\rho},\bar{\eta}),\sin{2\beta}, ...$, are evaluated. 
For a given point in $a$, one determines the best agreement between data 
and theory. One calculates 
$\Delta \chi^2(a) = \chi^2_{\rm min; \mu}(a) - \chi^2_{{\rm min};y_{\rm mod}}$,
by varying freely all model parameters $\mu$ (including $y_{\rm QCD}$) 
with the exception of $a$. The corresponding CL is obtained from 
${\rm CL}(a) = {\rm Prob}(\Delta \chi^2(a),N_{\rm dof})$ (see {\rm e.g.} 
Fig.~\ref{chi2min}) where $N_{\rm dof}$ is the number of degrees of freedom, 
in general the dimension of the subspace $a$. It has to be stressed that 
${\rm CL}(a)$ depends on the choice of ${\cal R}$. The usage of 
${\rm Prob}(\Delta \chi^2(a),N_{\rm dof})$ assumes Gaussian shapes for 
${\cal L}_{\rm exp}$. The CL obtained has been verified for several examples 
using a Monte Carlo simulation similar to the one described in the last section.

If the SM cannot accommodate the data, the analysis has to 
be repeated within extensions of the SM. Even in the case 
of a good agreement between data and the SM, it is worthwhile 
to perform the analysis for possible extensions of the SM 
in order to constrain New Physics parameters, see {\it e.g.} 
Ref.~[\ref{laplace}], or to determine the precision needed 
to study or exclude certain models.

\begin{figure}[!ht]
\centerline{
\resizebox{0.6\textwidth}{!}{
  \includegraphics{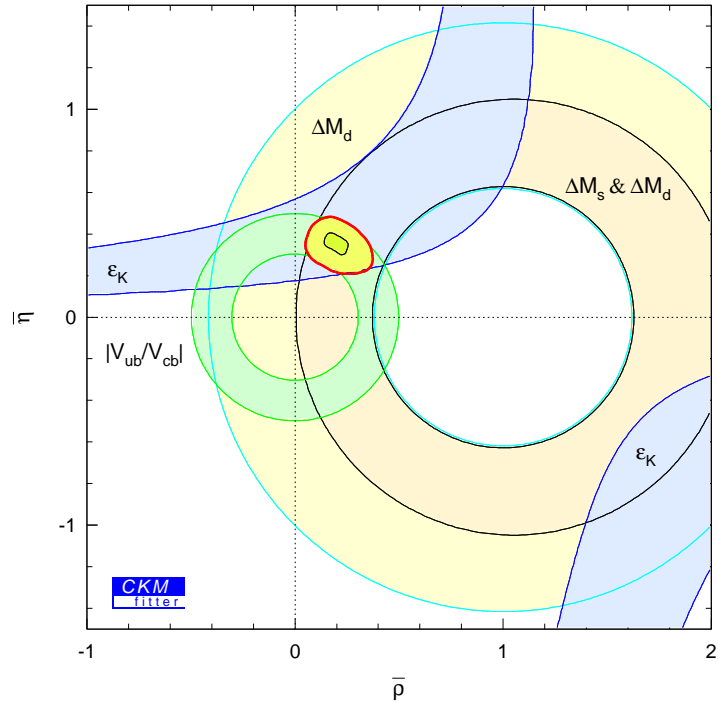}
}
}
\vspace{-0.6cm}       
\caption{\it
   Different single constraints in the $\bar{\rho}-\bar{\eta}$ plane 
   shown at $95~\%$ CL contours. The $95~\%$ and $10~\%$ CL  
   contours for the combined fit are also shown.
}
\label{rhoeta}      
\end{figure}


\subsubsection{The Scanning Method}

In the scanning method the following procedure is set to deal with
``not-statistically'' distributed quantities: we select a
specific set of theoretical parameters called a
``model'', 
\begin{equation}
{\cal M} \equiv \bigl \{{\cal F}_{D*}(1),\  \tilde \Gamma^c_{incl}, \ 
\tilde \Gamma^u_{excl},\tilde \Gamma^u_{incl}, \ 
F_{B_d} \sqrt{\hat B_{B_d}},\hat B_K,\ \xi, \ \eta_1, \eta_2, \ \eta_3, \ 
\eta_B \bigr \},
\end{equation}
where  ${\cal F}_{D*}(1)$ is the Isgur-Wise function of $\rm B \rightarrow D^*
\ell \nu$ at zero recoil corrected for finite $b$-quark mass,
$\tilde \Gamma^c_{incl}$ denotes the reduced decay rate for 
$b \rightarrow c \ell \nu$, $\tilde \Gamma^u_{excl}$ 
($\tilde \Gamma^u_{incl}$) represents
the reduced decay rate for $\rm B \rightarrow \rho \ell \nu$ 
($b \rightarrow u \ell \nu$), $F_{B_d} \ (F_{B_s})$ is the ${\rm B}^0_d~({\rm B}^0_s)$ 
decay constant, $B_{B_d},\ B_{B_s}$ and $\hat B_K$ 
parameterize the 
values of the hadronic matrix elements appearing in ${\rm B}^0_d- \overline{\rm B}^0_d$ mixing,
 ${\rm B}^0_s -\overline{\rm B}^0_s$ mixing and $\rm K^0 -\overline{K}^0$ mixing, respectively, 
$\xi = F_{B_s}/ F_{B_d} \sqrt{(\hat B_{B_s}/\hat B_{B_d})}$, and 
$\eta_1, \ \eta_2, \ \eta_3$, and $\eta_B$ denote QCD parameters.
Such a set of theoretical parameters carries by definition 
non-probabilistic uncertainties but still may
involve probabilistic errors. 
By choosing many different ``models'' we map out the allowed ranges 
of these theoretical parameters.

For each ``model'' ${\cal M}$ we construct and minimize the function
\begin{equation}
\chi^2_{{\cal M}}(A, \bar \rho, \bar \eta) = \sum_{i}
 \left[ \frac{E_i  - 
{\cal E}_i(A, \bar \rho, \bar \eta; C_k; {\cal M})}{\sigma_{E_i}} 
\right] ^2,
\end{equation}
where the $E_i$ are observables based on measured quantities, 
${\cal E}_i(A, \bar \rho, \bar \eta; C_k; {\cal M})$ 
is their parameterization in terms of 
$A$, $\bar \rho$, and $\bar \eta$, $C_k$ denotes measured quantities 
that possess experimentally derived or other probabilistic uncertainties,
such as masses and lifetimes, and the $\sigma_{E_i}$ denote all
measurement uncertainties contributing to both $E_i$ and
${\cal E}_i (A, \bar \rho, \bar \eta; C_k; {\cal M})$. 
This includes all uncertainties on the theoretical parameters that are
statistical in nature. 

The inputs used are those given in Table~\ref{tab:inputs}. 
To incorporate results on $\Delta M_{s}$ searches we include a $\chi^2$-term
defined as the maximum between the log-likelhood ratio used in [\ref{C00}] 
and 0:
\begin{equation} 
-2 ln {\cal L}_\infty (\Delta M_{s}) = 
\max \left( \frac{(1 - 2 \cal{A})}{\sigma_{{\cal A}}^2}, 0 \right)
\end{equation}
${\cal A}$ is the amplitude spectrum as function of $\dms$.

The minimization solution $(A, \bar \rho, \bar \eta)_{{\cal M}}$ for a  
particular ``model'' ${\cal M}$ 
incorporates no prior distribution for non-probabilistic uncertainties
of the theoretical parameters and meets the frequency interpretation. 
All uncertainties  depend only
on measurement errors and other probabilistic uncertainties
including any probabilistic component of the uncertainties on the 
theoretical parameters relevant to each particular measurement. 
At the moment, for practical reasons, we have treated the comparatively
small uncertainties arising from $\eta_1, \eta_2, \eta_3$, and
$\eta_B$ as probabilistic. 
The effects of this simplification will be explored in future fits.

A ``model'' ${\cal M}$ and its best-fit solution are kept only
if the probability of the fit satisfies
${\cal P}(\chi^2_{\cal M}) > {\cal P}_{min}$, which is typically chosen to 
be $5\%$. For each ``model'' ${\cal M}$ accepted, we draw a 95\%~C.L. contour 
in the $(\bar \rho, \bar \eta)$ plane. 
The fit is repeated for other ``models'' ${\cal M}$ by scanning through 
the complete parameter space specified by the theoretical uncertainties. 
This procedure derives from the technique originally described 
in~[\ref{bib:BABAR}].

The $\chi^2$ minimization thus serves three purposes:

\begin{itemize}

\item[1.] If a ``model'' ${\cal M}$ is consistent with the data, 
we obtain the best estimates for the 
three CKM parameters, and 95\%~C.L. contours are determined.

\item[2.] If a ``model'' ${\cal M}$ is inconsistent with the data the 
probability ${\cal P}(\chi^2_{\cal M})$ will be low. Thus, the
requirement of ${\cal P}(\chi^2_{\cal M})_{min} > 5\%$ provides 
a test of compatibility between data and its theoretical 
description.

\item[3.] By varying the theoretical parameters beyond their specified range
we can determine correlations on them imposed by the measurements. The first
results of this study are shown in Section~\ref{sec:partheo_freq}
\end{itemize}

If no ``model'' were to survive we would have evidence of an inconsistency 
between data and theory, independent of the calculations of the
theoretical parameters or the choices of their uncertainties.
Since the goal of the CKM parameter fits is to look for inconsistencies 
of the different measurements within the Standard Model, it is important
to be insensitive to artificially produced effects and  to 
separate the non-probabilistic uncertainties from 
Gaussian-distributed errors.

\begin{figure}[t]
\begin{tabular}{cc}
\includegraphics[width=7.9cm]{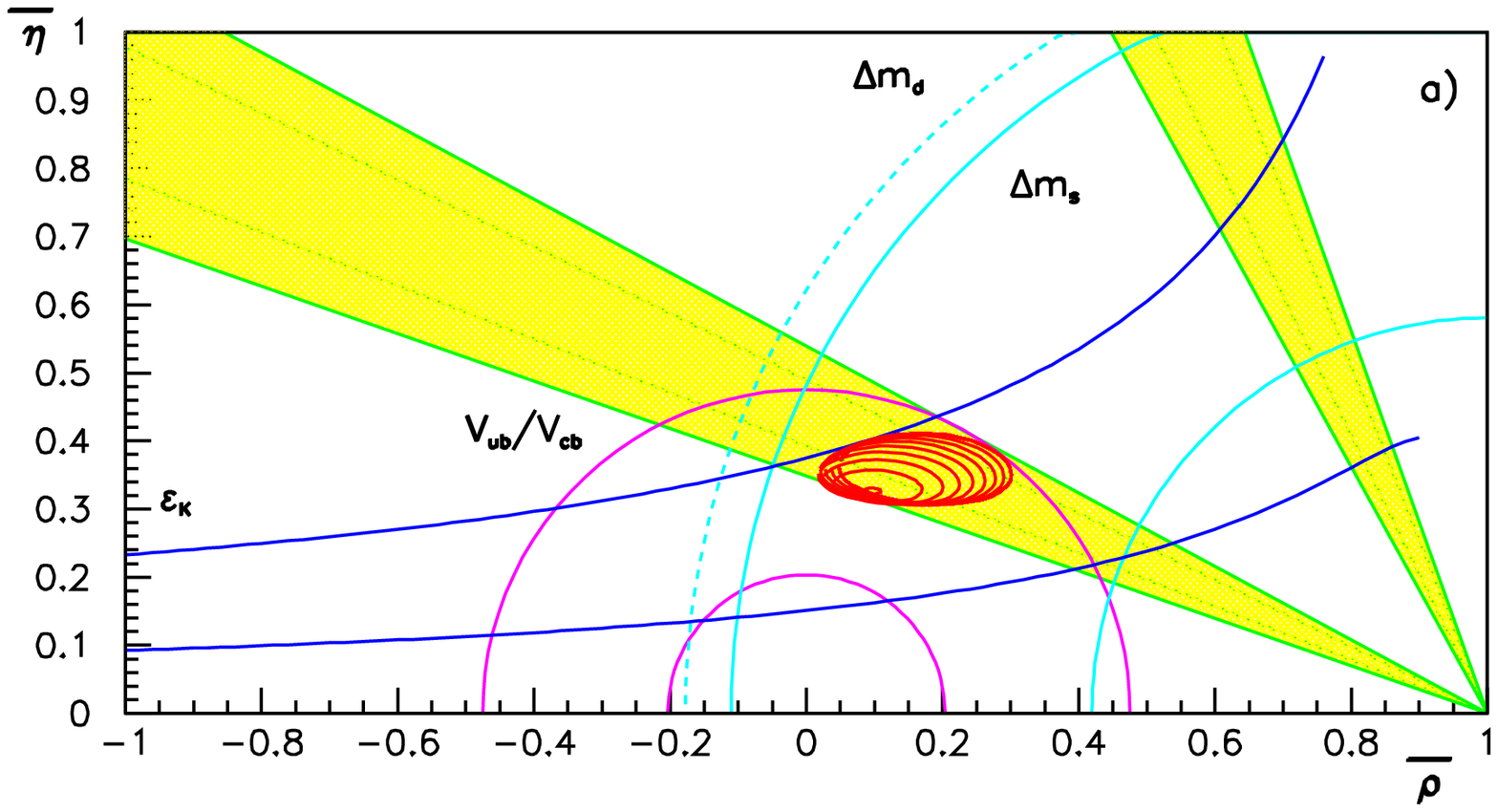} &
\includegraphics[width=7.9cm]{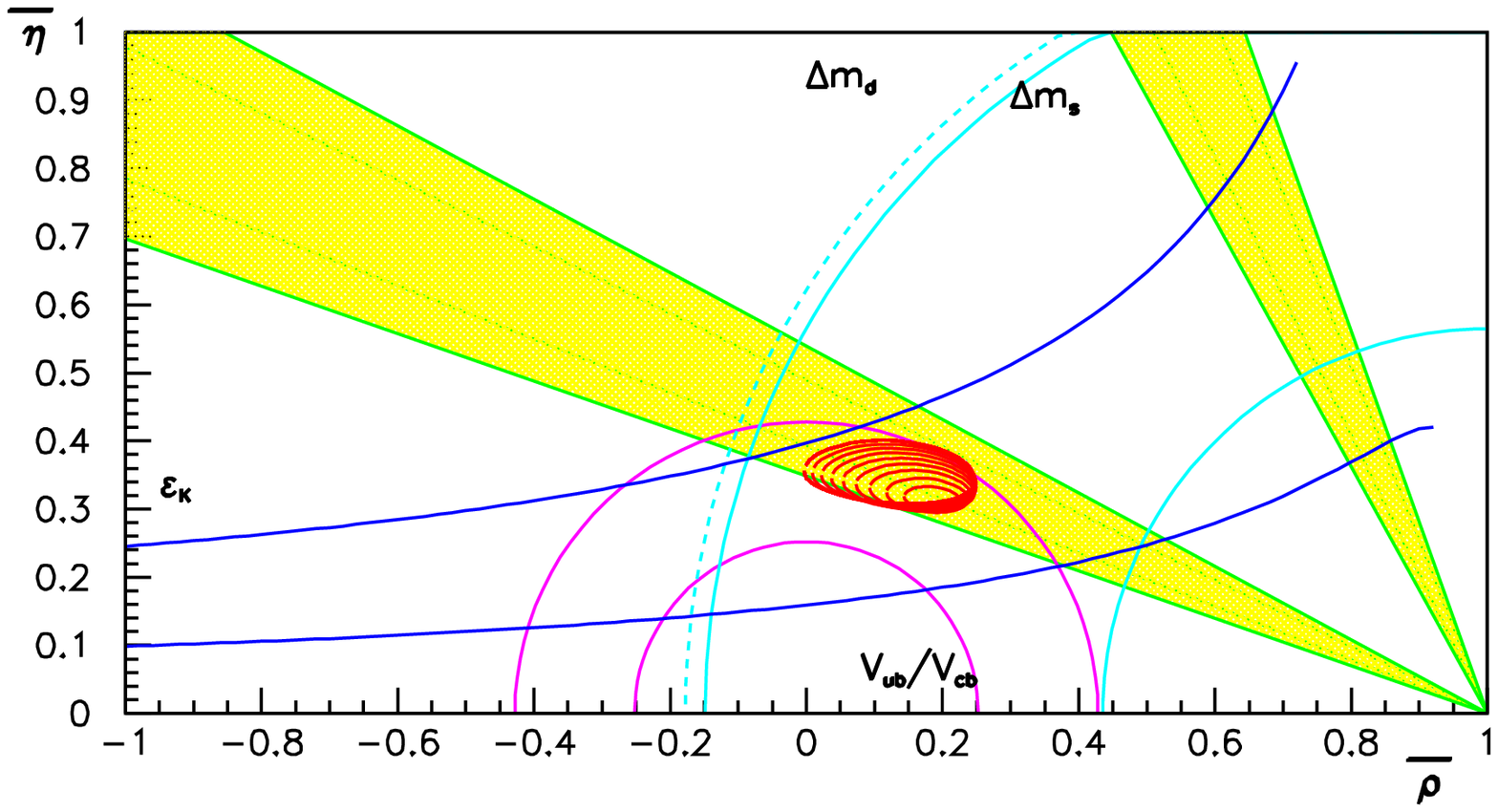} \\
\includegraphics[width=7.9cm]{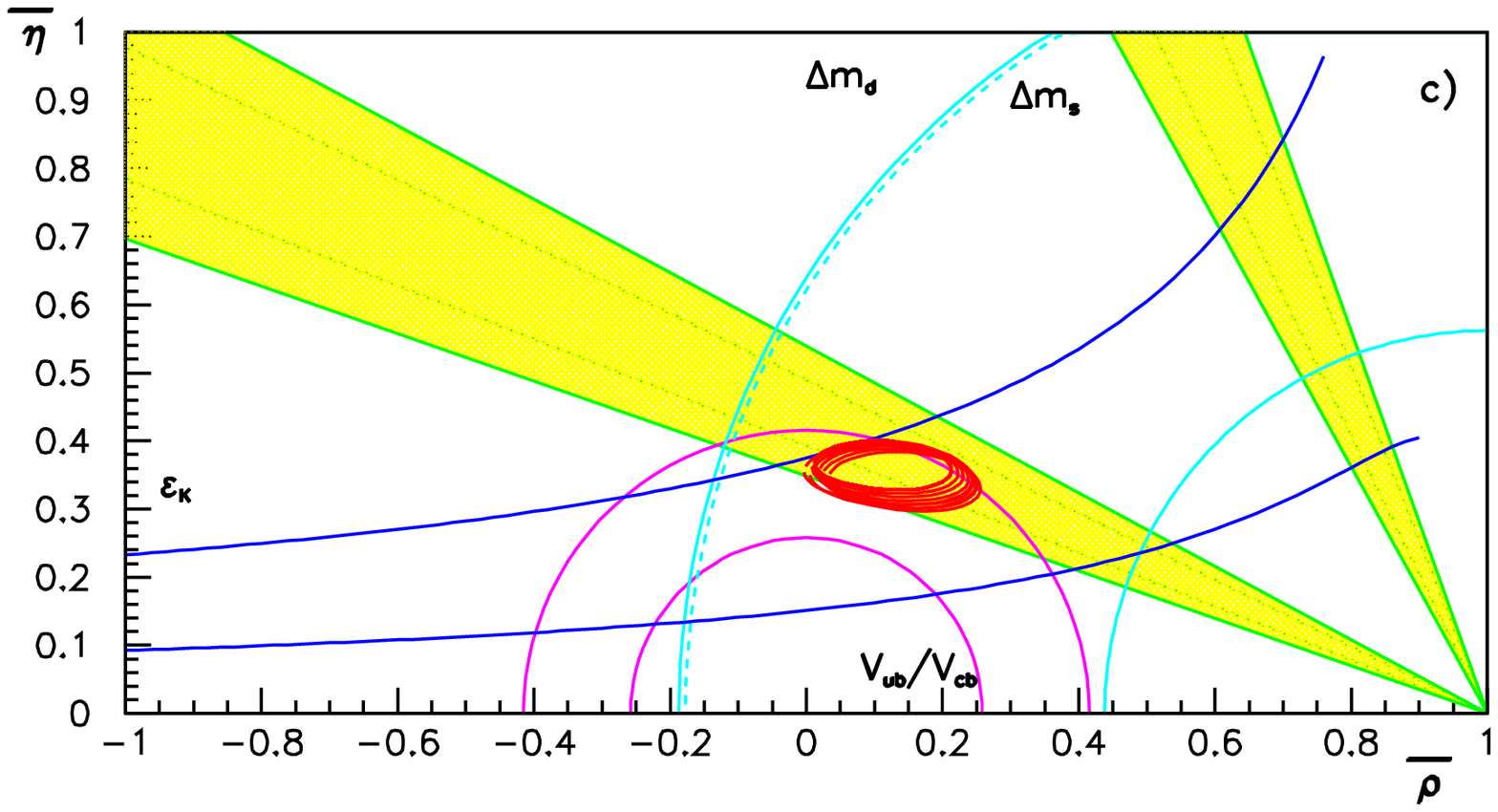} &
\includegraphics[width=7.9cm]{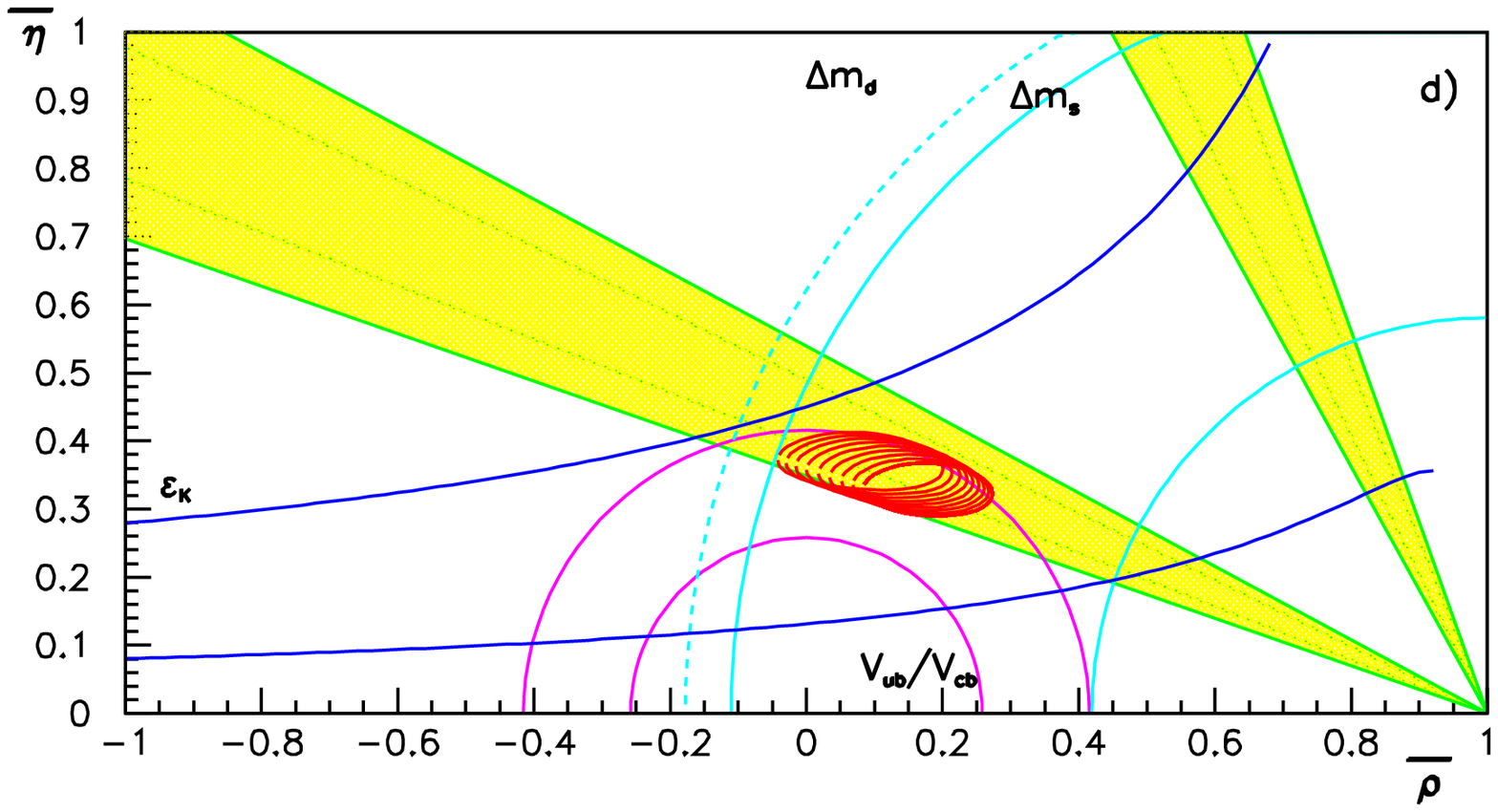}  \\
\includegraphics[width=7.9cm]{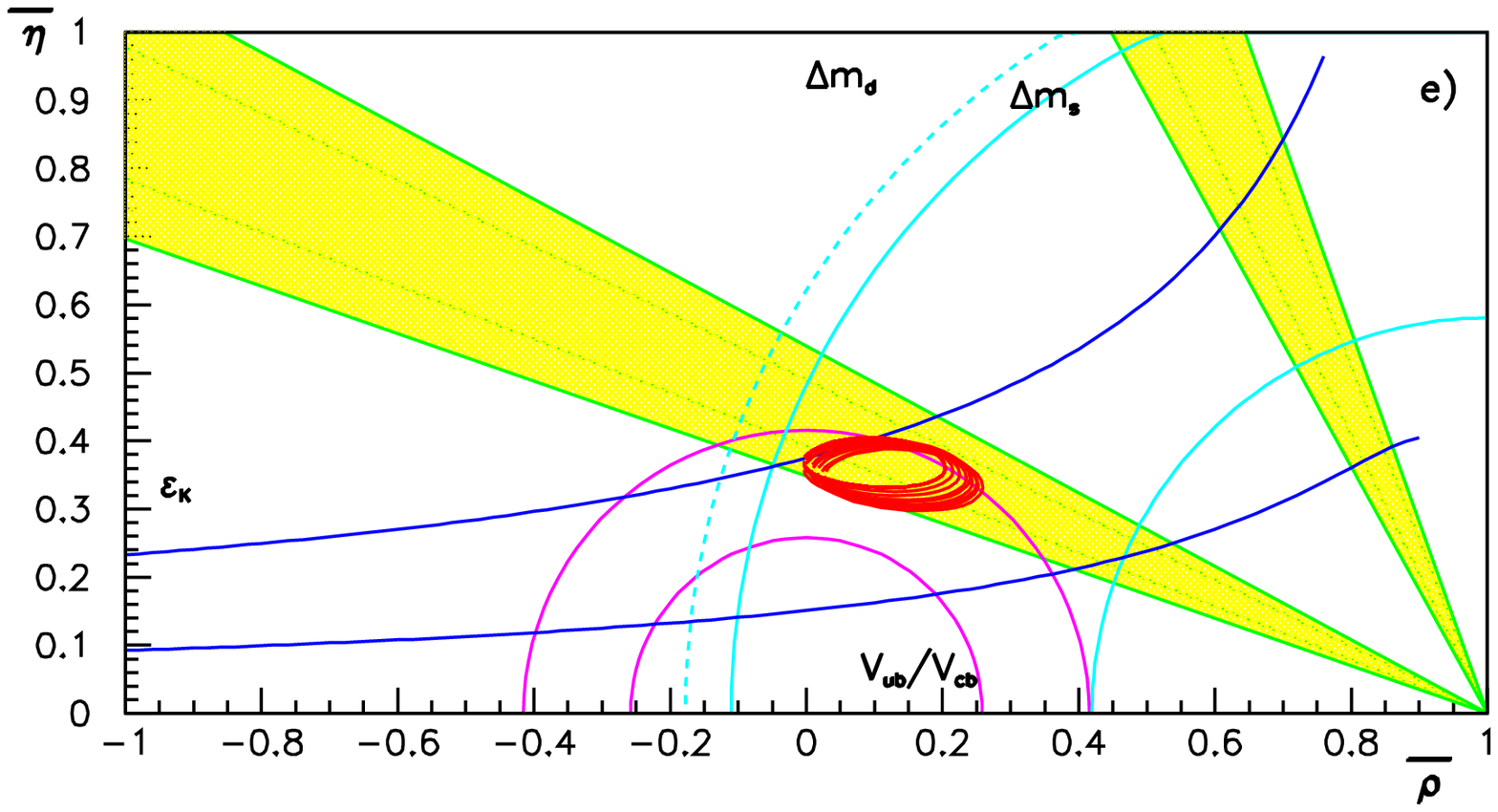}  &
\includegraphics[width=7.9cm]{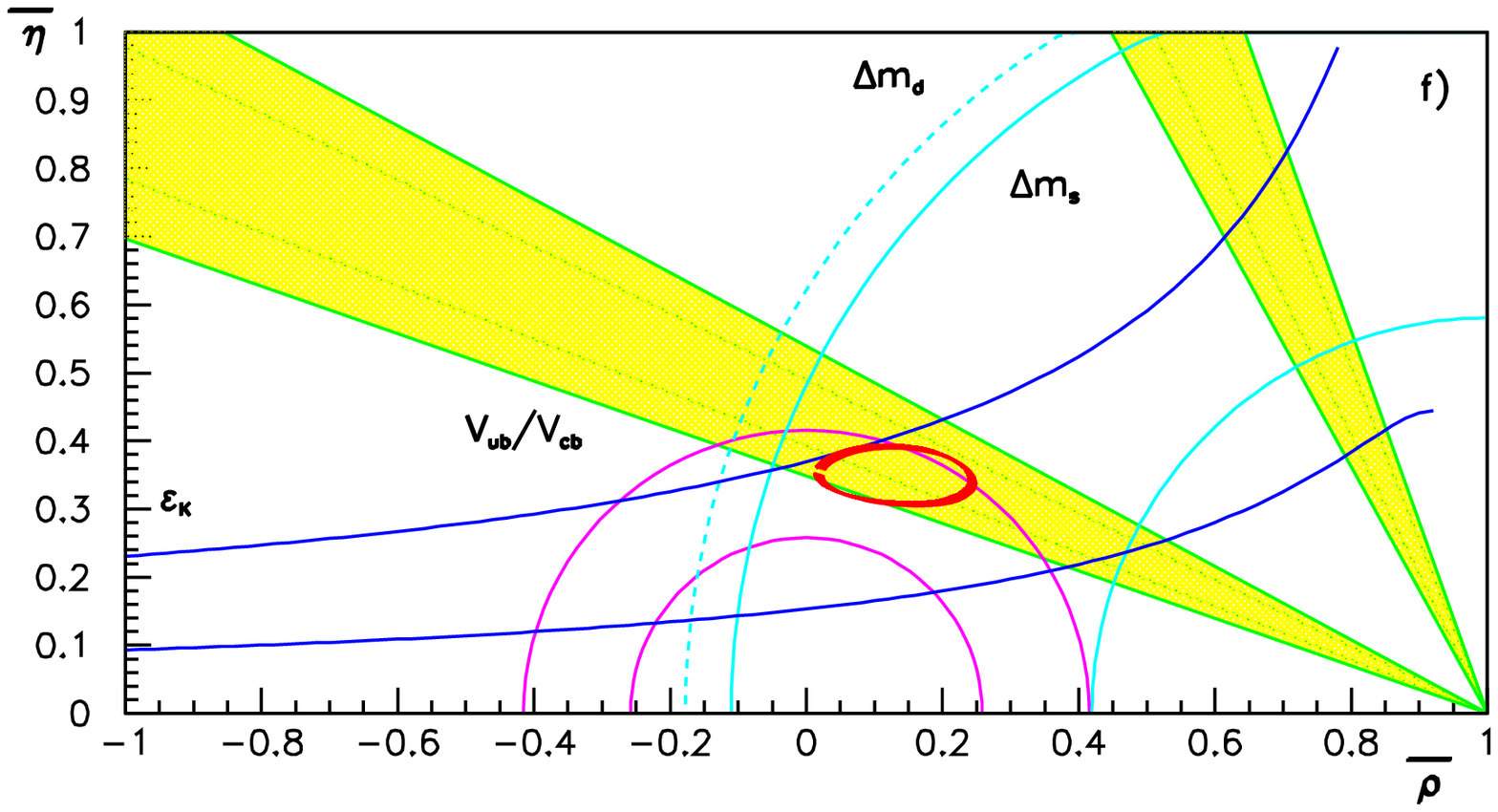}
\end{tabular}
\caption{\it Contours of different models in the $(\bar \rho, \bar \eta)$
plane, by varying only one theoretical parameter at a time, 
a) $\tilde\Gamma_{excl}$,
b) ${\cal F}_{D^*}(1)$, c)  $F_{B_d} \sqrt{\hat B_{B_d}}$, d)  $\hat B_K$,
e) $\xi$ (where $\Delta M_{s}$ is included in the fit), and f)
$\eta_1$. In each plot nine different models are considered by varying
the theoretically-allowed range from the minimum value to the maximum value.
The figures are arranged with a) in the upper left, b) in the upper right, 
{\it etc..} } 
\label{fig:ut_par}
\end{figure}

In order to demonstrate the impact of the different theoretical quantities 
on the fit results in the $(\bar \rho, \bar \eta)$ plane, 
Figs.~\ref{fig:ut_par}a--f show contours for
fits in which only one parameter was scanned while the others were kept at 
their central values. 
These plots demonstrate the impact of the model dependence in
$|V_{ub}|$ and $|V_{cb}|$ as well as that of $F_{B_d} \sqrt{\hat B_{B_d}}$ and 
$\hat B_K$, $\xi$, and $\eta_1$, respectively.
For each parameter we consider nine different models which span its range
equidistantly, starting with the smallest 
allowed value. Since these plots just serve illustrative purposes we use only
the measurements of $|V_{ub}|$, $|V_{cb}|$, $\Delta M_{d}$, and 
$\epsilon_K$ in the fits, except for Fig.~\ref{fig:ut_par}e 
where the information of  $\Delta M_{s}$ has been included in addition.
To guide the eye we show the boundaries of the
three bands for $|V_{ub}/V_{cb}|$,  $|V_{td}/V_{cb}|$, and $\epsilon_K$.
Since the theoretical parameters are kept at their central values except for
the one being varied, the bands corresponding to the other parameters 
reflect only experimental uncertainties.  

\begin{figure}[htb!]
\begin{center}
\includegraphics[width=13cm]{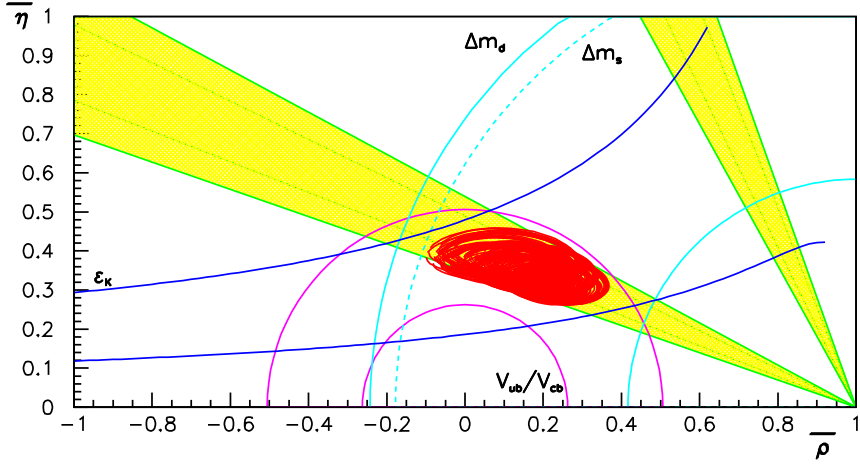}

\vspace*{-5mm}

\caption{\it
Contours in the $(\bar \rho, \bar \eta)$ plane for 
different models,
scanning theoretical parameters $\tilde\Gamma_{excl}$, ${\cal F}_{D^*}(1)$,
$F_{B_d} \sqrt{\hat B_{B_d}}$, $\hat B_K$, and $\xi$,
based on measurements of $|V_{ub}|$, $|V_{cb}|$, $\Delta M_{d}$, 
$\epsilon_K$, the amplitude for $\Delta M_{s}$, and
$\sin 2 \beta$.
}
\label{fig:xut_nodms}
\end{center}

\vspace*{-5mm}

\end{figure}

We now turn to scanning all parameters simultaneously within their
theoretically ``allowed'' ranges.
Figure~\ref{fig:xut_nodms} shows the resulting contours for a set of 
representative ``models'', when all available constraints are included.
Note that there is no frequency interpretation for comparing which models are
to be ``preferred'', other than the statement that at most one model is 
correct. In this analysis we cannot, and do not, give any relative 
probabilistic weighting among the contours, or their overlap regions.
Indeed, the entire purpose of the scanning method is to make clear
the relative importance of measurable
experimental errors and a-priori unknown
theoretical uncertainties. 

\vspace*{-3mm}

\section{Impact of the uncertainties on theoretical quantities in CKM Fits}
\label{sec:qcdpar}

As described in the previous sections, the ``correct'' way
to treat the theoretical parameters is not unambiguously defined but depends on 
the adopted statistical approach.
In this section we will not discuss the problems and the virtues of
the different statistical approaches on this point,
and concentrate  on the impact of the uncertainties on 
theoretical parameters in constraining $\rhobar~{\rm and}~\etabar$.

Two numerical analyses will be presented: one in the Bayesian
framework and one in the frequentist framework.
In the first analysis we study the effect on the UT fit from a modification 
(or a removal) of some theoretical parameter used as input parameter.
The second analysis introduces a graphical method to represent, in the space 
of the theoretical parameters, the goodness of the UT fit and to 
evaluate the relevance of the knowledge on these parameters.

\vspace*{-3mm}

\subsection{Bayesian analysis}

In the framework of the Bayesian method the input knowledge is expressed in 
terms of pdfs
for all quantities (theoretical and experimental parameters).
Following the procedure described in Section ~\ref{sec:bayes},
the output pdf\ can be computed for $\rhobar$, $\etabar$ and for
any other quantity of interest.

The impact of the uncertainty on a given quantity, which enters as a
parameter in a given constraint, is naturally evaluated by comparing the 
results obtained excluding the corresponding constraint or by varying the error 
of the input parameter. When the information on a certain quantity is excluded the
corresponding input pdf\  is taken as uniform.
The common set of inputs used for this analysis are the ones available at the time 
of the Workshop (Table~\ref{tab:inputs}).

\vspace*{-3mm}

\subsubsection{Determination of $F_{B_d}\sqrt{\hat B_{B_d}}$}
\begin{figure}[htbp] 
\begin{minipage}{0.49\hsize}{
\epsfig{figure=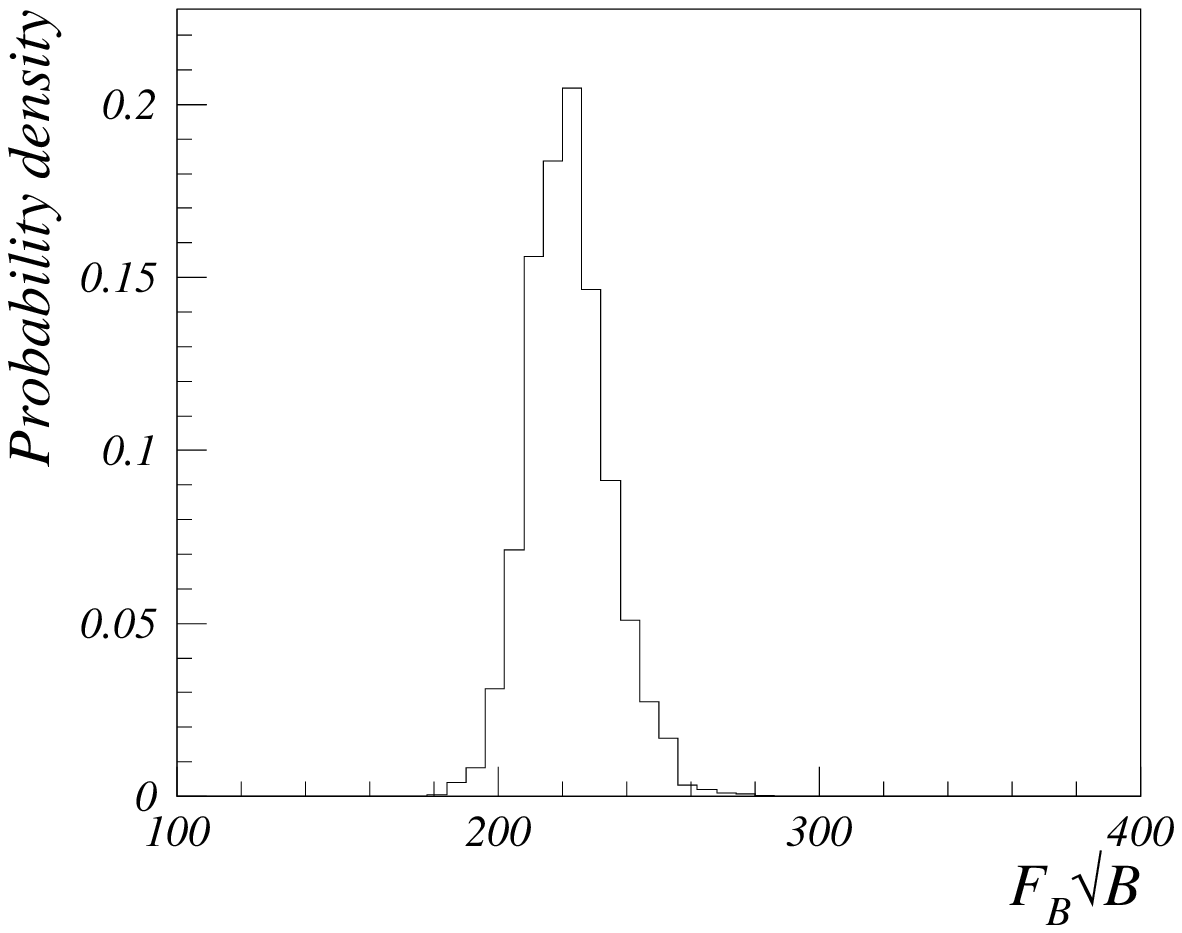,width=0.9\textwidth} 
}\end{minipage}\begin{minipage}{0.5\hsize}{
\begin{center}
\begin{tabular}{|c|c|c|}
\hline
$\sigma \pm \Delta/2$ & $\rhobar$ & $\etabar$      \\ \hline
$\pm 16 \pm 6$   & $0.183 \pm 0.040$ & $0.355 \pm 0.027$ \\
$\pm 33 \pm 12$  & $0.173 \pm 0.046$ & $0.357 \pm 0.027$ \\
$\pm 66 \pm 24$  & $0.173 \pm 0.046$ & $0.355 \pm 0.027$ \\
$\infty$         & $0.175 \pm 0.049$ & $0.355 \pm 0.027$ \\\hline 
\end{tabular}
\end{center}
}\end{minipage}
\caption{\it Left: output distribution for 
$F_{B_d}\sqrt{\hat B_{B_d}}$ assuming a flat input distribution.
Right: table reporting the results of the UT fit for
$\rhobar$ and $\etabar$ assuming different input values for the errors on 
$F_{B_d}\sqrt{\hat B_{B_d}}$ (in MeV)
($\sigma$ is the Gaussian error and $\Delta/2$ is the half-width of the 
systematic range). The last
column (``infinite error'') is obtained with a uniform input distribution.
}
\label{tab:fbb}
\end{figure}

First we consider the $F_{B_d}\sqrt{\hat B_{B_d}}$ parameter.
Quite remarkably the remaining constraints determine precisely this 
quantity and, from the output distribution shown in the left 
part of Fig.~\ref{tab:fbb},
we get
\be
F_{B_d}\sqrt{\hat B_{B_d}} = (223 \pm 12)~\rm MeV
\ee
This is in perfect agreement with the results from lattice calculation 
(see Table~\ref{tab:inputs}) and has a significantly smaller error. 
This suggests that, unless the lattice error
on $F_B\sqrt{\hat B_B}$ does not become smaller than 12~MeV, the theoretical
knowledge of this quantity is not quite relevant in UT fits.
The Table in Fig.~\ref{tab:fbb} quantifies the effect of changing the
uncertainty on $F_{B_d}\sqrt{\hat B_{B_d}}$ (keeping the same central value).

\subsubsection{Determination of $\hat B_K$}
\begin{figure}[htbp]
\begin{minipage}{0.49\hsize}{
\epsfig{figure=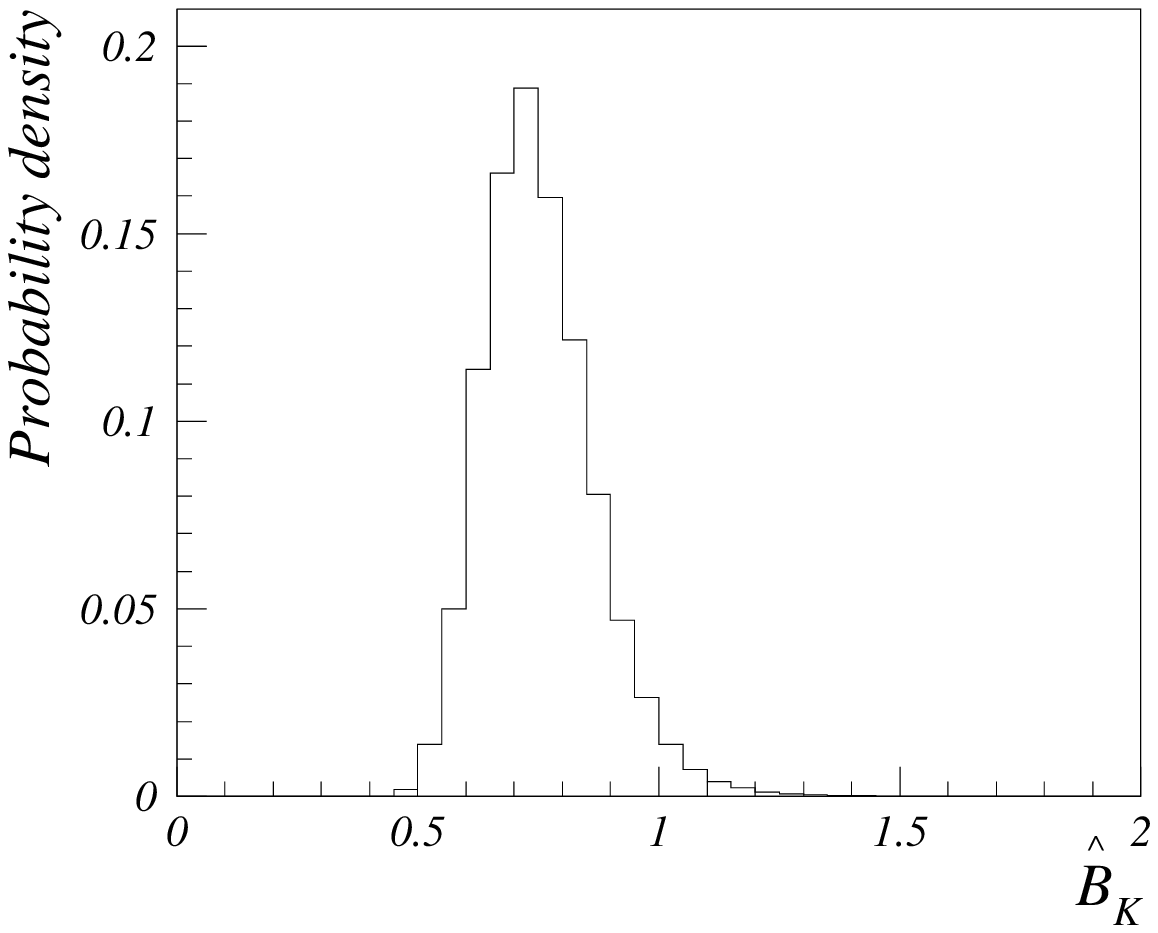,width=0.9\textwidth} 
}\end{minipage}\begin{minipage}{0.5\hsize}{
\begin{tabular}{|c|c|c|}
\hline
$\sigma \pm \Delta/2$ & $\rhobar$ & $\etabar$ \\ \hline
$\pm 0.03\pm0.065$ & $0.181 \pm 0.040$ & $0.349 \pm 0.025$ \\
$\pm 0.06\pm0.13$  & $0.173 \pm 0.046$ & $0.357 \pm 0.027$ \\
$\pm 0.12\pm0.26$  & $0.163 \pm 0.052$ & $0.365 \pm 0.030$ \\
$\infty$           & $0.161 \pm 0.055$ & $0.361 \pm 0.042$ \\ \hline
\end{tabular}
}\end{minipage}
\caption{\it 
Left: output distribution for $\hat B_K$
assuming a flat input distribution.
Right: table reporting the results of the CKM Fits for
$\rhobar$ and $\etabar$ assuming
different input values for the errors on $\hat B_K$ 
 ($\sigma$ is the Gaussian error and $\Delta/2$ is the half-width of the systematic range). The last
column ("infinite error") is obtained with a flat input distribution.}
\label{fig:bk}
\end{figure}

Here the same exercise has been repeated with the parameter $\hat B_K$.
Assuming for $\hat B_K$ a uniform input distribution between 0 and 2, from the
output distribution shown in Fig.~\ref{fig:bk} we obtain
\be
\hat B_K = 0.73 ^{+0.13}_{-0.07}
\ee
The fitted value is again in perfect agreement with the lattice value
(see Table \ref{tab:inputs}), but
in this case the fitted (output) uncertainty is similar to the 
theoretical (input) one. We then expect that ``lattice information'' plays a non
negligible role, in particular in the determination of $\etabar$
(because of simple geometrical arguments).
Table in Fig.~\ref{fig:bk} shows that, in fact, removing the information coming
from Lattice QCD (last row) the error on $\etabar$ increases by 50\%.

\subsubsection{Determination of $\xi$}
\begin{figure}[htbp]
\begin{minipage}{0.43\hsize}{
\epsfig{figure=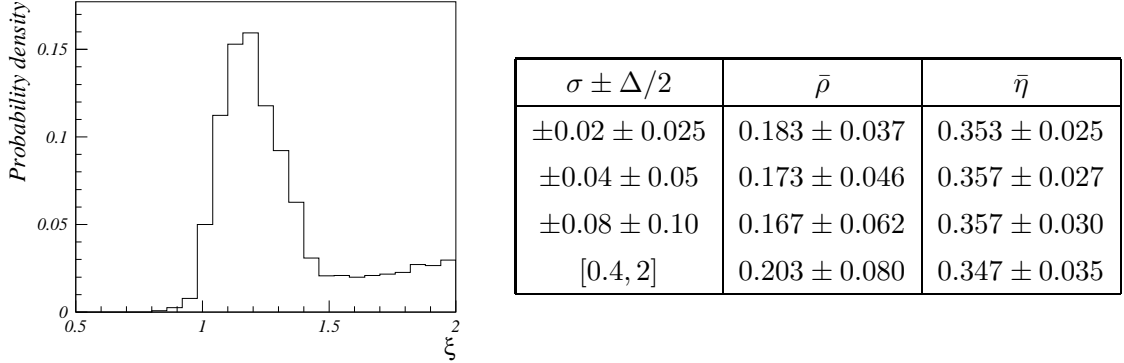,width=0.9\textwidth} 
}\end{minipage}\begin{minipage}{0.5\hsize}{
\begin{tabular}{|c|c|c|}
\hline
$\sigma \pm \Delta/2$ & $\rhobar$ & $\etabar$ \\ \hline
$\pm 0.02\pm0.025$ & $0.183 \pm 0.037$ & $0.353 \pm 0.025$ \\
$\pm 0.04\pm0.05$  & $0.173 \pm 0.046$ & $0.357 \pm 0.027$ \\
$\pm 0.08\pm0.10$  & $0.167 \pm 0.062$ & $0.357 \pm 0.030$ \\
$[0.4,2]$          & $0.203 \pm 0.080$ & $0.347 \pm 0.035$ \\ \hline
\end{tabular}
}\end{minipage}
\caption{\it
Left: output distribution for $\xi$
assuming a flat input distribution between 0.6 and 2.
Right: table reporting the results of the CKM Fits for
$\rhobar$ and $\etabar$ assuming
different input values for the errors on $\xi$ ($\sigma$ is the Gaussian error 
and $\Delta/2$ is the half-width of the systematic range).
The last column is obtained with a flat input distribution 
between 0.6 and 2.}
\label{fig:xi}
\end{figure}

Since $\dms$ has not yet been measured, $\xi$ cannot be determined
by the data.
Assuming a uniform distribution between 0.6 and 2 (the upper bound 
is obviously arbitrary),
the output distribution shown in Fig.~\ref{fig:xi} is obtained.
The tail on the right part of the plot shows that, at present, $\xi$ 
is only weakly constrained by experimental data;
for this reason the information on the $\xi$ parameter is very important, in particular
in the determination of $\rhobar$, as shown in the table in
Fig.~\ref{fig:xi}.

\subsection{Scan analysis}
\label{sec:partheo_freq}
In the study of the sensitivity of the UT fit to a given theoretical 
parameter one should define how to treat the remaining parameters. 
In the Bayesian approach (described in the previous section) the remaining 
parameters are integrated using their input pdf
while in the standard frequentist approach the confidence level for a parameter
is computed irrespectively of the values of all the remaining parameters
(logical ``OR'' over the values of the parameters).

The technique presented here aims at studying and visualizing the 
sensitivity of UT fits to theoretical uncertainties, in the theoretical 
parameters space (T), minimizing a priori inputs and intermediate 
combinations of parameters. The method tries to represent pairs or 
triplets of theoretical parameters while keeping
some information on the remaining (undisplayed) parameters.
The input knowledge on a theoretical parameter is described by a
"nominal central value" and a "theoretical preferred range".
In two dimensions the procedure is as follows:

\vspace{1mm}

\begin{itemize}

\item
Pick two of the parameters $T$ for display.  Call these the {\it
primary parameters}, $T_1$ and $T_2$.

\item
Pick a third $T$ parameter, the {\it secondary parameter} $T_s$.  This
parameter is singled out for special attention to the effects of
projecting over it.

\item
Call all the other T parameters the {\it undisplayed parameters}, $T_X$.

\item
For each point P in the grid of scanned values of $T_1 \otimes T_2$, 
a number of fits will have been attempted, 
covering all the scanned values of $T_s$ and $T_X$.  
For each P, evaluate the following hierarchy of
criteria against the ensemble of results of these fits, deriving 
for the point a value, we call it the ``Level'', which is an 
integer between 0 and 5 inclusive:

\vspace{5mm}

\begin{enumerate}

\item[1.] 
Define a minimum acceptable value for $P(\chi^2)$.
Did any of the fits for P pass this cut?  If not, assign Level~=~0 and
stop; otherwise assign Level~=~1 and continue.

\item[2.]
Did any of the remaining fits lie within the "theoretically preferred 
region" for {\it all} the undisplayed parameters \({T_X}\)?  If not, stop;
if yes, assign Level~=~2 and continue.

\item[3.]
Did any of the remaining fits have the secondary parameter \({T_s}\) within its
"theoretically preferred region"?
If not, stop;
if yes, assign Level~=~3 and continue.

\item[4.]
Did any of the remaining fits have \({T_s}\) equal to its "nominal central 
value"?  (That value must have been included in the scan grid for this to
make sense.)  If not, stop; if yes, assign Level~=~4 and continue.

\item[5.]
Did any of the remaining fits have {\it all} the undisplayed parameters 
\({T_X}\) also at their "nominal central values"?    If not, stop; 
if yes, assign Level~=~5 and stop.

\end{enumerate}

\vspace{3mm}

\item
Now display contours of the quantity Level over the grid in the
\({T_1}\otimes {T_2}\) plane. Assign a unique color to each
parameter T, so the contours for \({T_s}\) at Level~=~3,4 are
drawn in the color corresponding to that parameter.  The contours for
Level~=~4,5, which correspond to restrictions of parameters
exactly to their central values, are also drawn distinctively, with
dashing.

The Level~3 contour (solid, colored), in particular, displays the
allowed region, at the selected confidence level, for \({T_1}\) and
\({T_2}\), based on the experimental data and on limiting all other
theoretical parameters to their preferred ranges.
Study of the relative spacing of the Level~2, 3, and~4 contours readily 
reveals the effects of the application of the \({T_s}\) bounds on the 
fit results.

\item
Overlay the contours with straight lines showing the theoretically
preferred ranges and nominal central values for \({T_1}\) and
\({T_2}\), in their respective unique colors, again with dashing for the
central value.  This allows the theoretical bounds on \({T_1}\) and
\({T_2}\) to be evaluated directly for consistency against all other 
available data, yet avoiding any convoluted use of priors for these two
parameters.  

Comparison of these theoretical bounds for \({T_1}\) and \({T_2}\)
with the Level~3 contour that shows the experimental information, 
constrained by the application of the theoretical bounds on
\({T_s}\) and the \({T_x}\), allows a direct visual evaluation of the
consistency of all available information, with the effects of the 
application of all theoretical bounds manifest, not obscured by
convolutions performed in the fit itself.
\end{itemize}

\begin{figure}[htbp] 
\begin{center}
\epsfig{figure=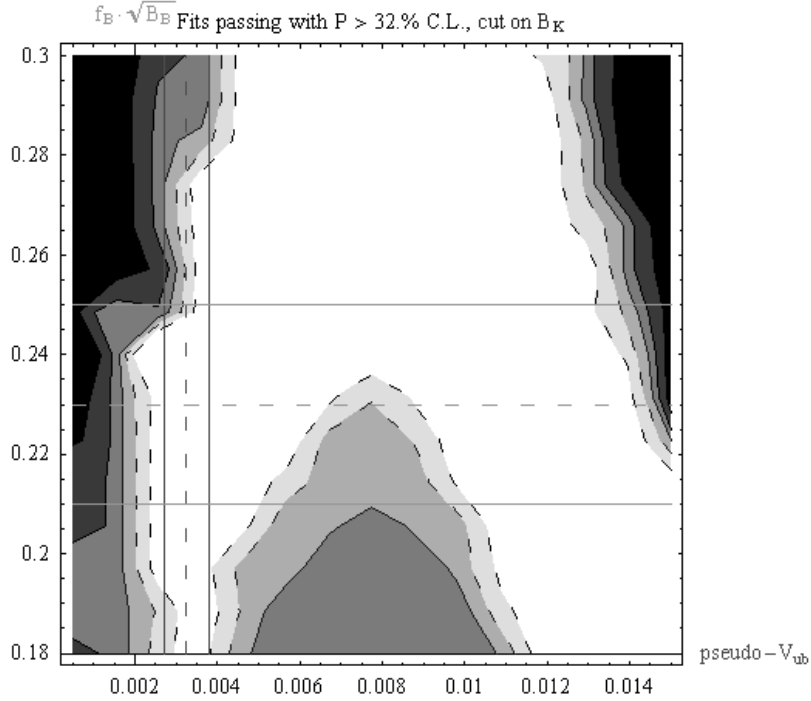,width=0.67\textwidth}
\end{center}
\vspace{-1cm}
\caption{\it Results of the procedure described in the text, using
$F_{B_d}\sqrt{\hat B_{B_d}}$ and $V_{ub}$ as primary parameters, $\hat B_K$ as 
secondary parameter, and $\xi$ as undisplayed parameter.}
\label{fig:greg}
\end{figure}

Figure~\ref{fig:greg} shows the results of the previous procedure, using
$F_B\sqrt{\hat B_B}$ and $V_{ub}$ as primary parameters, $\hat B_K$ as 
secondary parameter, while the undisplayed parameter (there is just one in
this case) is $\xi$.
What can be seen immediately is that the entire theoretically allowed region
for the primary parameters, shown by the crossing of the solid lines,
is consistent with all the other data, including the theoretical bound on
$\hat B_K$, and that even when all parameters are constrained to their
central values the resulting fit (there can be only one at that point)
is fully consistent. Changing the role of primary, secondary,
 and undisplayed parameters in many
different ways, helps to understand the role of these parameters~in~the~fit. 

These plots can  be extended in three dimensions by drawing the three 
bi-dimensional projections of the allowed region. Several three 
dimensional plots and further details can be found in ~[\ref{greg}].


      
\section{Fit comparison}
\label{p:comp}
In this section we compare the results on the CKM quantities 
obtained following two approaches: Bayesian and \rfit. The common set of inputs
are the ones available at the time of the Workshop (Table~\ref{tab:inputs}).
The Scan method has not been included in the comparison because it does
not evaluate overall allowed regions for the CKM parameters.
As explained in the previous sections, the main difference between 
the Bayesian and the \rfit~ analyses originates from the computation of the 
Likelihood functions (identified with pdfs in the Bayesian case) 
for each parameter and in the treatment of the Likelihood fit.

\subsection{Input likelihoods and combination of likelihoods}
\label{sec:inputlike}

In general a determination of a given quantity is characterized by a 
central value, a statistical 
error and a systematical error. Starting from such a splitting of the 
errors Bayesian and frequentist approaches may describe this quantity 
according to different likelihood functions. 
%

In the Bayesian approach, the basic assumption is that the value of any quantity 
is distributed according to a pdf. 
The final pdf\ of a certain quantity is obtained by convoluting the pdfs corresponding 
to the different uncertainties affecting the quantity.
In particular, the uncertainty on a quantity is usually splitted in two 
parts: a statistical part which can be described by a Gaussian pdf, 
${\cal{G}}(x-x_{0})$ (this part may contain many sources of
uncertainties which have been already combined into a single pdf) and another 
part which is usually of theoretical origin and is often related to 
uncertainties due to theoretical parameters. In the following we will denote it as 
theoretical systematics. It is often described using an uniform pdf: 
${\cal{U}}(x) = 1/2 \sigma_{\rm theo}$ for 
$x \in [x_{0}-\sigma_{\rm theo},x_{0}+\sigma_{\rm theo}]$
and ${\cal{U}}(x) = 0$ elsewhere. 
The combined pdf\ $\cal{P}$ is then obtained by convoluting 
the Gaussian pdf\ $\cal{G}$ with the uniform pdf\ 
$\cal{U}$: $\cal{P} = \cal{G} \otimes \cal{U}$.

In the frequentist analysis, no statistical meaning is attributed to the uncertainties
related to theoretical systematics. The likelihood function ${\cal{L}}$ 
for the quantity $x$ contains a statistical part, ${\cal{L}}_{\rm exp}(x-x'_{0})$, 
described by a Gaussian with mean value $x'_{0}$, 
and a ``not-statistical'' part, ${\cal{L}}_{\rm theo}(x'_{0})$.
The function ${\cal{L}}_{\rm theo}(x'_{0})$,
as denoted {\it R}fit likelihood, is a 
uniform function ${\cal{L}}_{\rm theo}(x'_{0})=1$ for     
$x'_{0} \in [x_{0}-\sigma_{\rm theo},x_{0}+\sigma_{\rm theo}]$ 
and ${\cal{L}}_{\rm theo}(x'_{0})=0$ elsewhere.
The final likelihood is given by the product 
${\cal{L}}={\cal{L}}_{\rm exp}(x-x'_{0}) \cdot {\cal{L}}_{\rm theo}(x'_{0})$. 
In conclusion, when a quantity contains an uncertainty to which the frequentists
do not attribute any statistical meaning, the likelihood which describes
 this quantity is obtained as a product between this uncertainty and the statistical one.

When several determinations of the same quantity are available 
one may combine them to obtain a single input for a quantity 
entering the fit (these considerations apply for example to 
the determinations of $|V_{ub}|$ and $|V_{cb}|$).
We suppose, in the following, that 
these determinations are not correlated. In addition, it is 
assumed that the various determinations of these quantities 
are compatible.
Then, for the combination, the Bayesian approach calculates 
the product of the single pdfs, whereas the frequentist 
approach calculates the product of the individual likelihoods. 
Hence, the mathematical concepts for the combination procedure 
of the two statistical approaches are identical.

\subsection{Distributions for the relevant quantities in the CKM fits}
\label{sec:inputpara}

\begin{figure}[t]
\begin{center}
\begin{tabular}{ll}
\includegraphics[width=6cm]{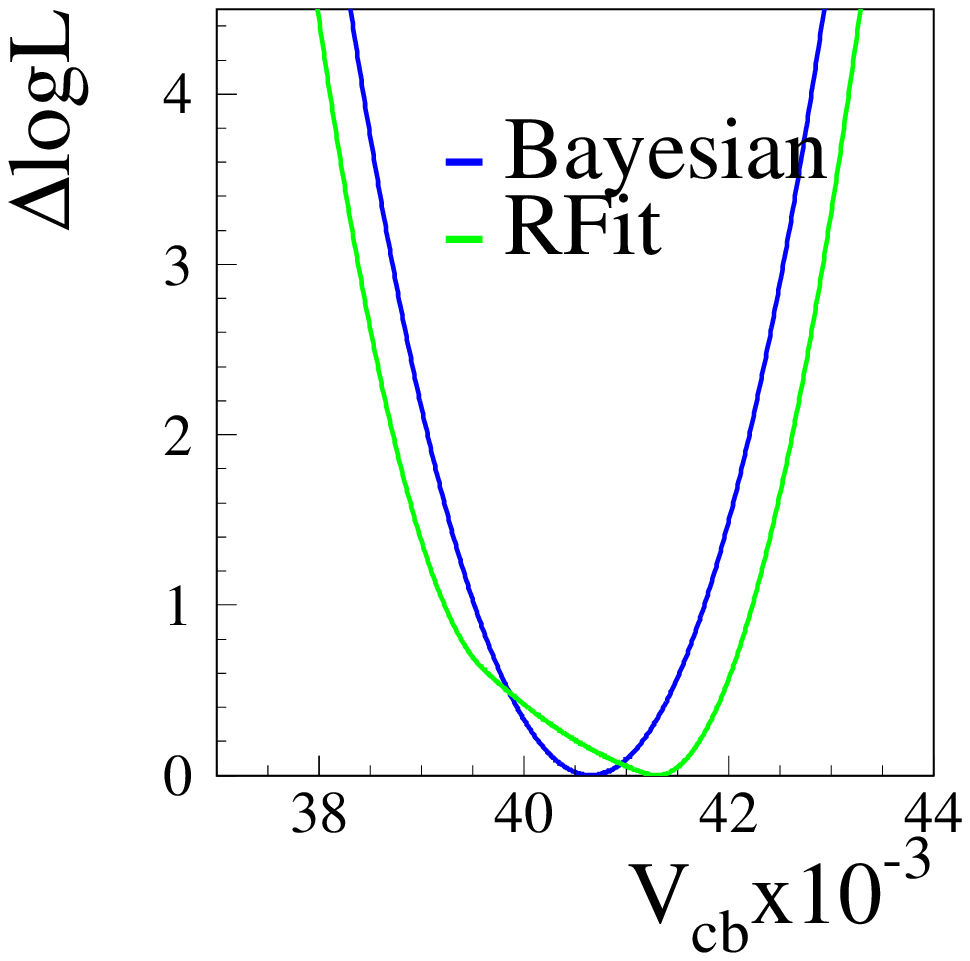} &
\includegraphics[width=6cm]{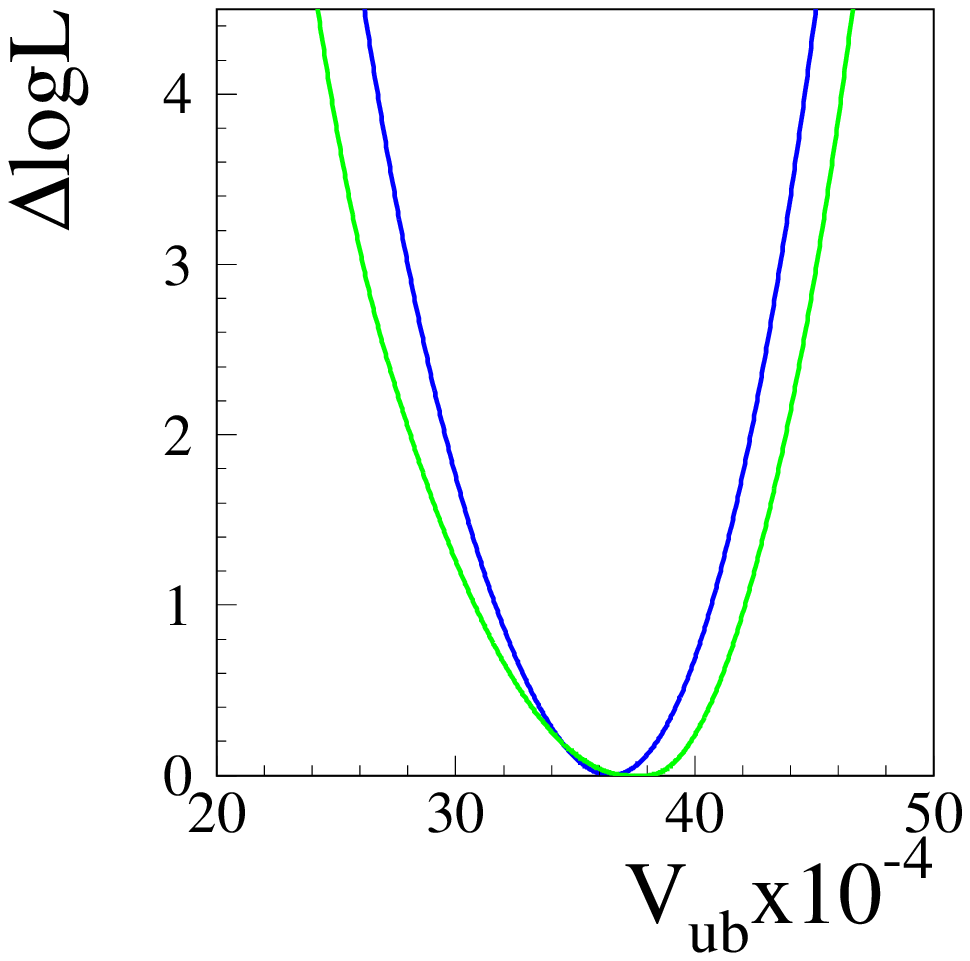}
\end{tabular}
\caption[]{\it{ The $\Delta$Likelihood for $|V_{cb}|$  and $|V_{ub}|$  
using the Bayesian and frequentist approaches when combining 
the inclusive and the exclusive 
determinations. }\label{fig:fig1}}
\end{center}
\end{figure}

\begin{figure}[t]
\begin{center}
\begin{tabular}{lll}
\includegraphics[width=5cm]{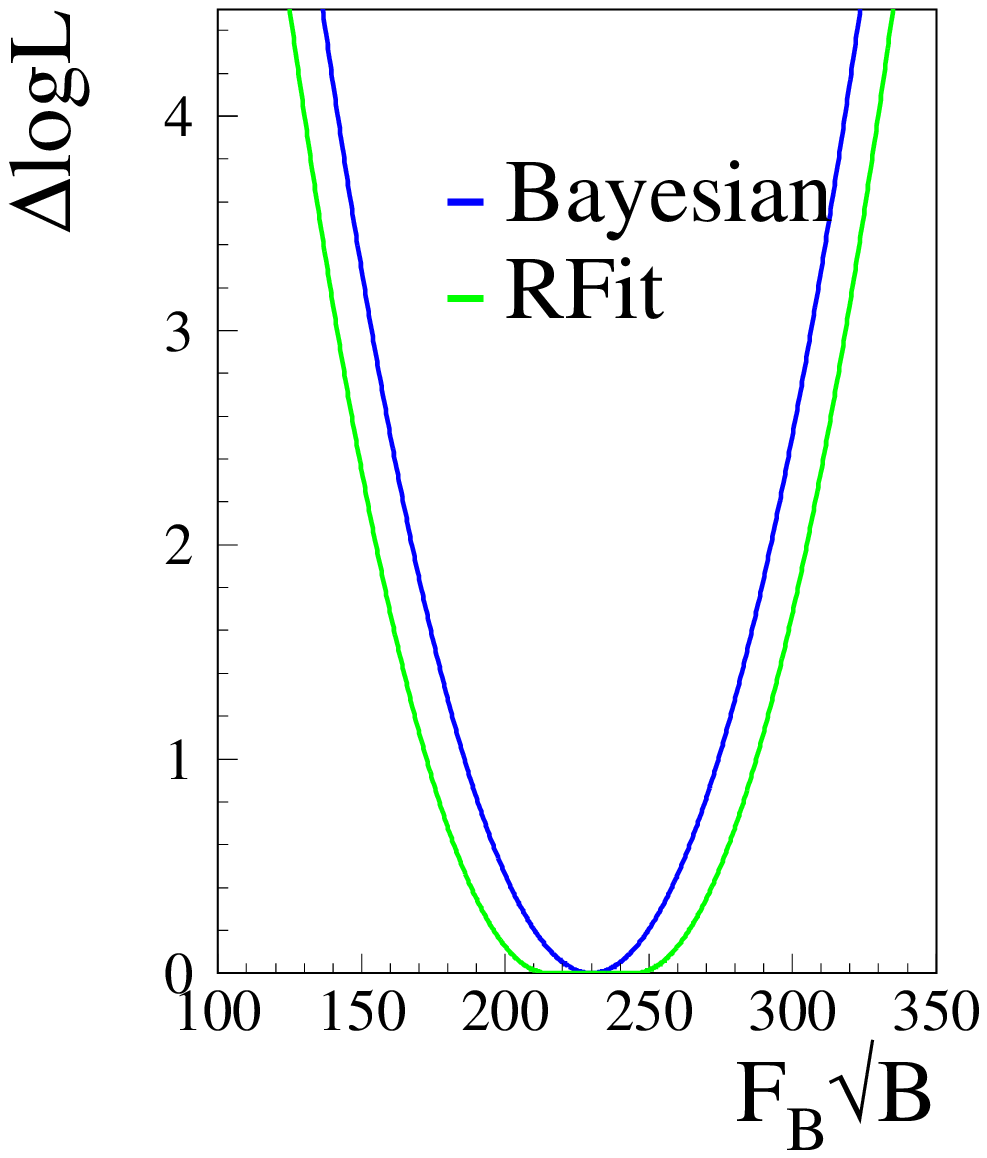} &
\includegraphics[width=5cm]{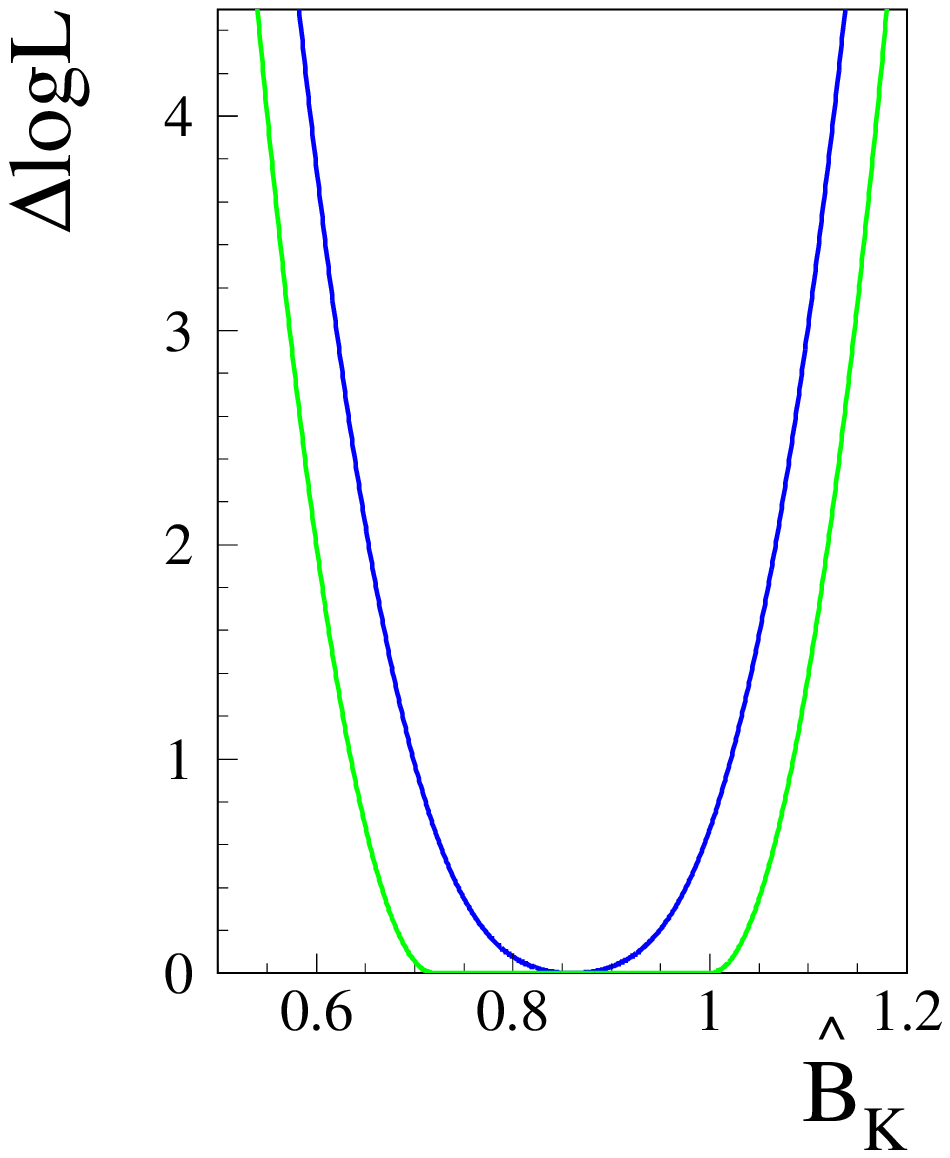} &
\includegraphics[width=5cm]{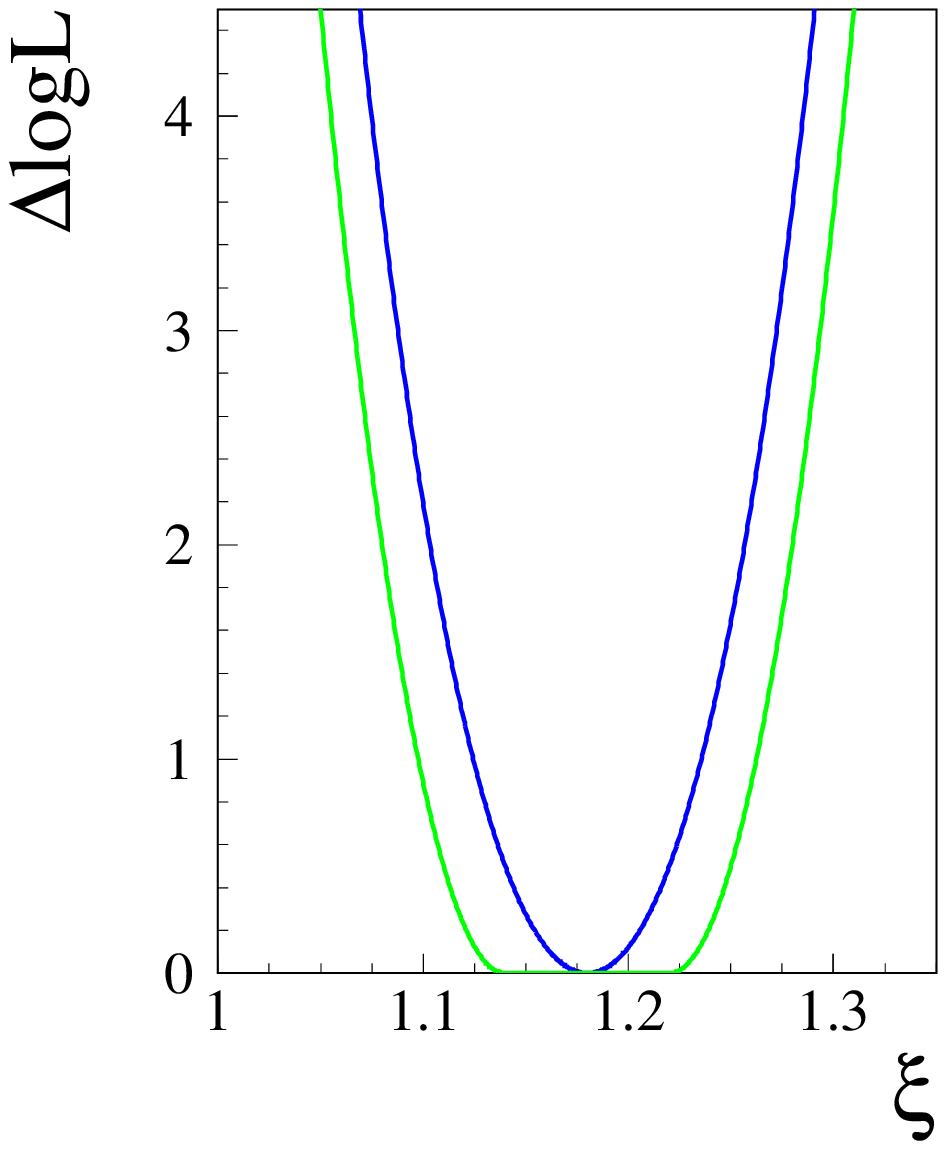}
\end{tabular}
\caption[]{\it { The $\Delta$Likelihood for $F_{B_d} \sqrt{\hat B_{B_d}}$, 
$\hat B_K$ and $\xi$ using the Bayesian and 
frequentist approaches.\label{fig:fig2} }}
\end{center}
\end{figure}

The relevant quantities entering the fit are
 summarized in Table~\ref{tab:inputs}
given at the beginning of this Chapter.
Figures.~\ref{fig:fig1} and \ref{fig:fig2} show the $\Delta$ Log(Likelihood) 
for $|V_{cb}|$ ,$|V_{ub}|$ and for the non-perturbative QCD parameters, 
$F_{B_d} \sqrt{\hat B_{B_d}}$ , $\xi$ and $\hat B_K$ 
as obtained following the Bayesian and the frequentist methods.
To be more explicit, in Table~\ref{tab:bkex} we show the 68$\%$ and 95$\%$ 
ranges as obtained following the Bayesian and the \rfit~ methods.
It can be noticed that differences on the input quantities between the two approaches
can be important and depend upon the chosen splitting of the errors. 
In the Bayesian approach the splitting of the total error in two errors 
is not really important, since, the two errors 
are often, already, the results of the convolution of many different source of errors. 
It has been also shown that the choice of the shape of the pdf to be attributed 
to the error has a moderate impact on the final 
results [\ref{ref:parodietal}], 
once the central value and the r.m.s. of the pdf\ has been fixed.
In the \rfit~ this splitting is crucial and a careful breakdown of the sources of the errors which
contribute to it should be done. 
For this comparison we have decided to keep this splitting and to classify certain errors
as a flat pdf
and ``non statistical'' for the Bayesian and \rfit~ approaches, respectively. 
\begin{table}[h]
\begin{center}
\begin{tabular}{|c|c|c|}
\hline
         Parameters                 &     68$\%$ range             &     95$\%$ range         \\ \hline
$\hat B_K$ \rfit~ (Bayes) [ratio R/B] & 0.68-1.06 (0.76-0.98) [1.70] & 0.62-1.12 (0.67-1.06) [1.25] \\
\hline
\end{tabular} 
\caption[]{ \it {68$\%$ and 95$\%$ ranges for some relevant quantities used in the CKM fits in 
the \rfit~ and Bayes approaches.}}
\end{center}
\label{tab:bkex}
\end{table}

\subsection{Results and Comparison}
\begin{table}[h]
  \begin{center}
    \begin{tabular}{cccc}
\hline
      \multicolumn{4}{|c|} {\rfit~ Method}                               \\ 
\hline
    Parameter    &$\leq 5\%$ CL & $\leq 1\%$ CL   &$\leq 0.1\%$ CL     \\ 
\hline
$\bar{\rho}$     &  0.091 - 0.317 & 0.071  - 0.351   &  0.042 - 0.379  \\ 
$\bar{\eta}$     &  0.273 - 0.408 & 0.257  - 0.423   &  0.242 - 0.442  \\
$\sin{2\beta}$   &  0.632 - 0.813 & 0.594  - 0.834   &  0.554 - 0.855  \\
$\gamma^{\circ}$ &   42.1 - 75.7  &  38.6  - 78.7    &   36.0 - 83.5   \\
    \hline 
                 &              &                &                     \\ 
\hline
      \multicolumn{4}{|c|} {Bayesian Method}                           \\ 
\hline
    Parameter    &    $5\%$ CL   &    $1\%$ CL    &    $0.1\%$ CL  \\ 
\hline
$\bar{\rho}$     &  0.137 - 0.295 & 0.108 - 0.317   &  0.045 - 0.347 \\
$\bar{\eta}$     &  0.295 - 0.409 & 0.278 - 0.427   &  0.259 - 0.449 \\
$\sin{2\beta}$   &  0.665 - 0.820 & 0.637 - 0.841   &  0.604 - 0.863 \\
$\gamma^{\circ}$ &   47.0 - 70.0  &  44.0 - 74.4    &  40.0 - 83.6   \\
     \hline 
                 &              &                &                 \\ 
\hline
      \multicolumn{4}{|c|} {Ratio \rfit/Bayesian Method}             \\ 
\hline
    Parameter    &    $5\%$ CL   &    $1\%$ CL    &    $0.1\%$ CL  \\ 
    \hline 
$\bar{\rho}$     &   1.43        &   1.34         &   1.12         \\
$\bar{\eta}$     &   1.18        &   1.12         &   1.05         \\
$\sin{2\beta}$   &   1.17        &   1.18         &   1.16         \\
$\gamma^{\circ}$ &   1.46        &   1.31         &   1.09         \\
\hline
    \end{tabular}
  \caption{ \it Ranges at difference C.L for $\bar{\rho}$, $\bar{\eta}$, 
           sin 2$\beta$ and $\gamma$. The measurements of  
           $\left | V_{ub} \right |/\left | V_{cb} \right |$ and $\Delta M_d$, the amplitude 
           spectrum for including the information from the ${\rm B}^0_s-\overline{{\rm B}}^0_s$oscillations, 
           $\epsilonk$  and the measurement of sin 2$\beta$ have been used. }
  \end{center}
  \label{tab:test1}
\end{table}

\begin{figure}[!ht]
\begin{center}
\begin{tabular}{cc}
\includegraphics[width=7.5cm]{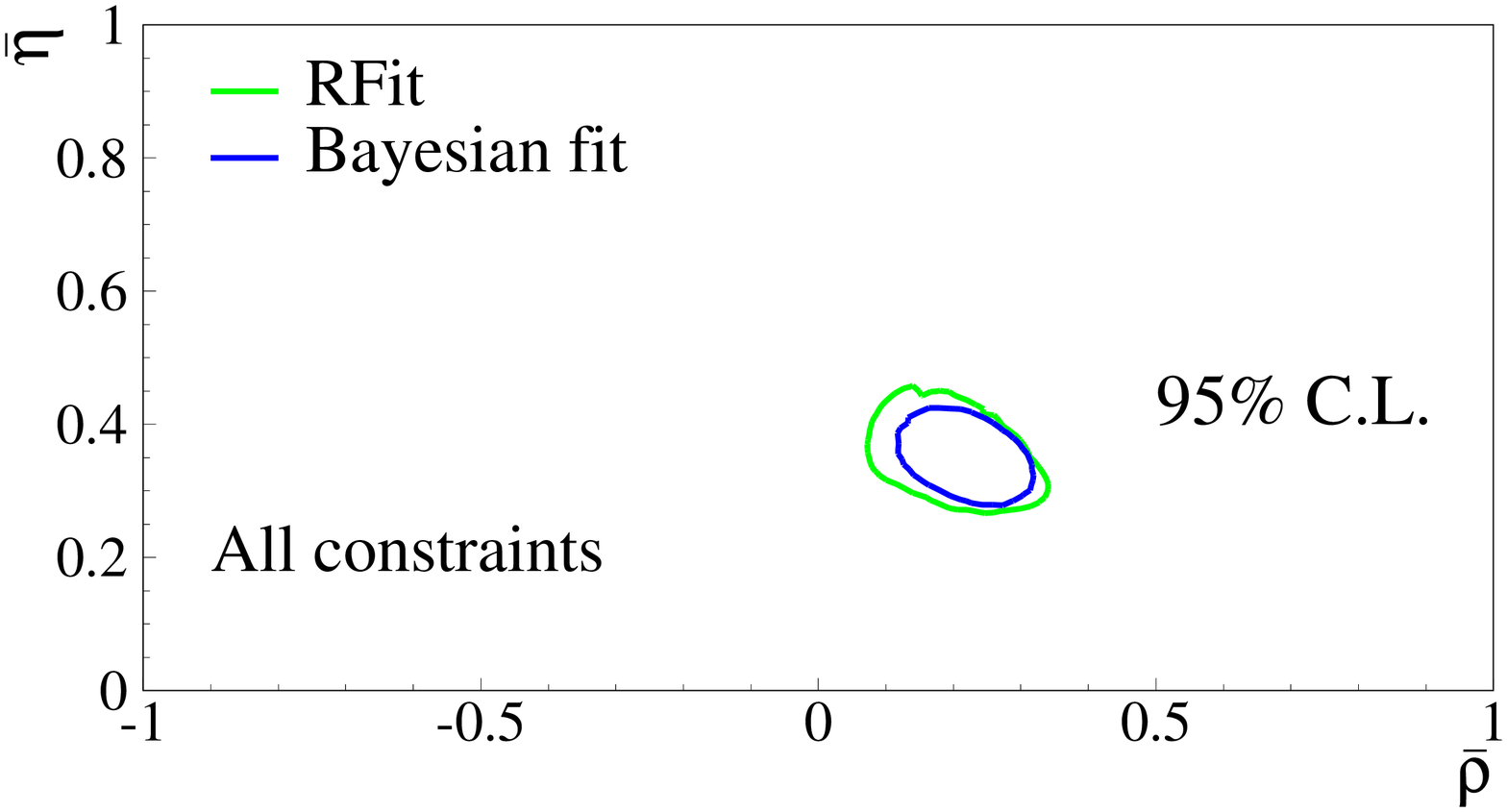} & 
\includegraphics[width=7.5cm]{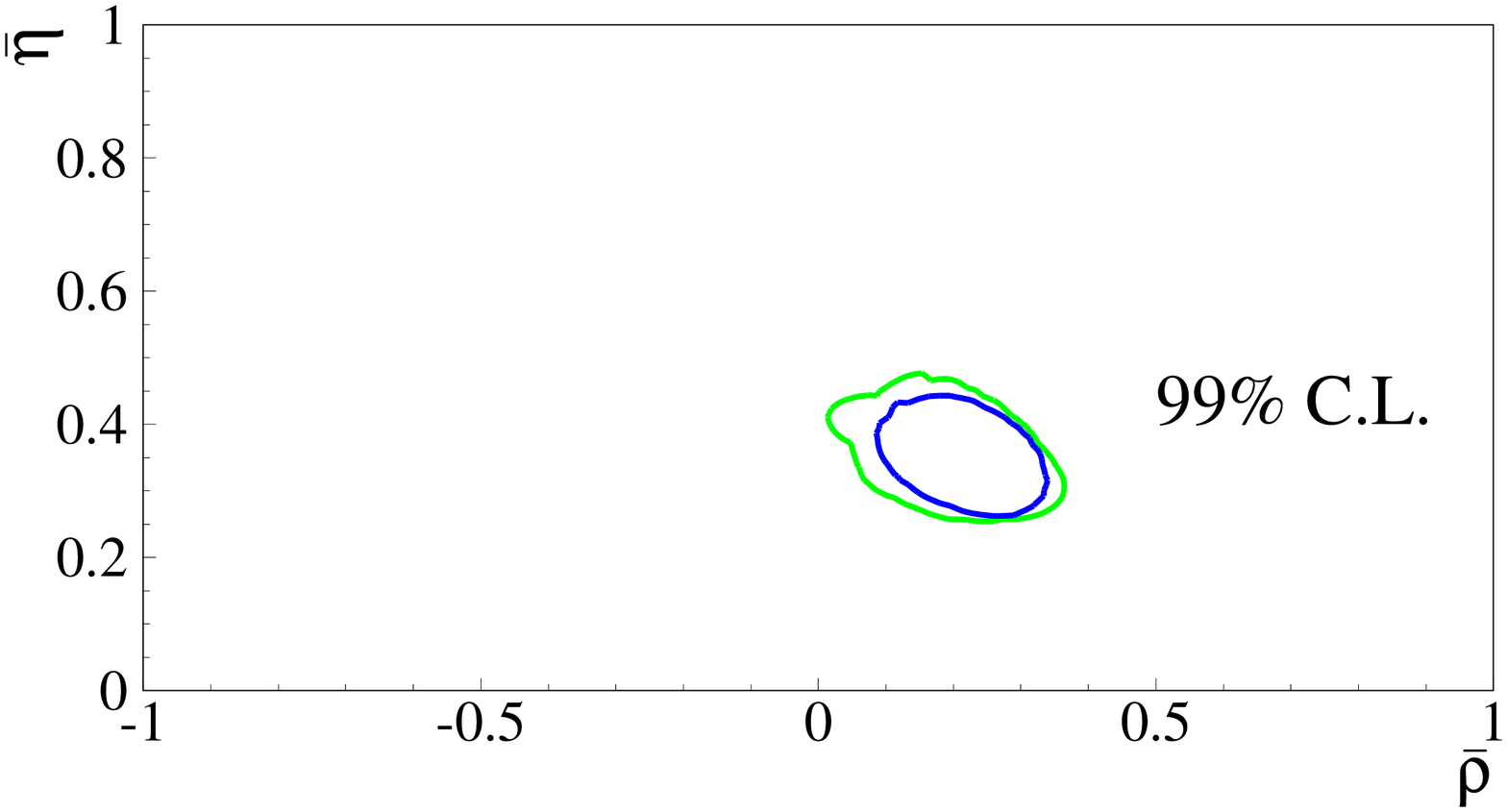}
\end{tabular}
\caption{ {\it Comparison Bayesian/\rfit~ Methods.
Allowed regions for $\bar{\rho}$ and $\bar{\eta}$ at 
95\% (left plot) and 99\% (right plot)
using the measurements of  
$\left | V_{ub} \right |/\left | V_{cb} \right |$, $\Delta M_d$,
the amplitude spectrum for including the information 
from the ${\rm B}^0_s-\overline{{\rm B}}^0_s$ oscillations,
$\epsilonk$ and the measurement of sin 2$\beta$.} \label{fig:test1}}
\end{center}
\end{figure}

\begin{figure}[!hbt]
\begin{center}
\begin{tabular}{cc}
\includegraphics[width=7.5cm]{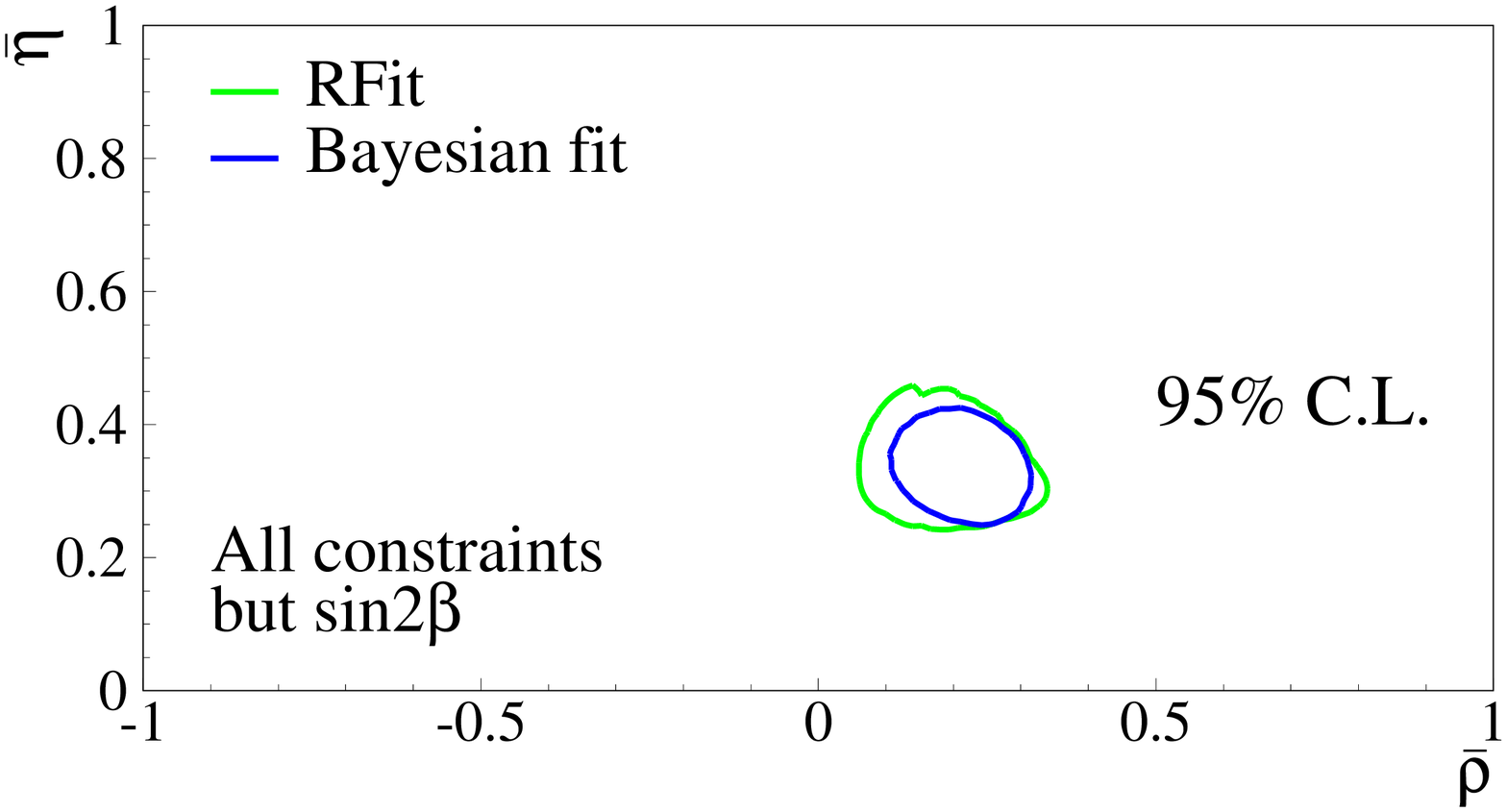} &
\includegraphics[width=7.5cm]{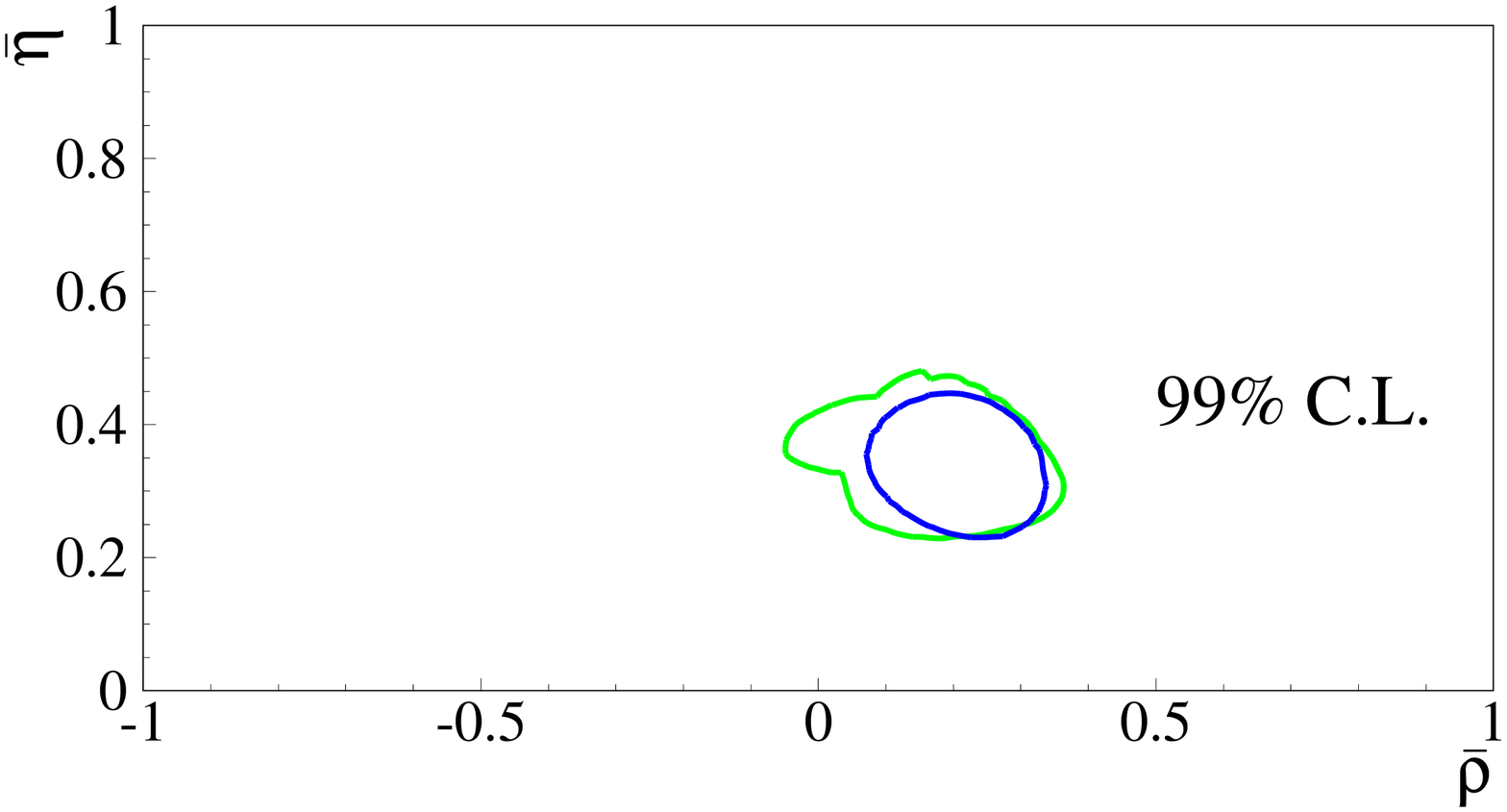} 
\end{tabular}
\caption{{\it Comparison Bayesian/\rfit~ Methods.
Allowed regions for $\bar{\rho}$ and $\bar{\eta}$ at 95\% 
(left plot) and 99\% (right plot)  
using the measurements of  
$\left | V_{ub} \right |/\left | V_{cb} \right |$, $\Delta M_d$,
the amplitude spectrum for including the information 
from the ${\rm B}^0_s-\overline{{\rm B}}^0_s$ oscillations
and~$\epsilonk$ }}
\label{fig:test1_sin_sin2b}

\vspace{-3mm}

\end{center}
\end{figure}

For the comparison of the results of the fit we use 
$\bar{\rho}$, $\bar{\eta}$, sin 2$\beta$ and $\gamma$. Those quantities 
are compared at the 95$\%$, 99$\%$ and 99.9$\%$~C.L. It has to be stressed 
that in the frequentist approach those confidence levels correspond to 
$\geq$95$\%$, $\geq$99$\%$ and $\geq$99.9$\%$.
All the available constraints have been used: the measurements of  
$\left | V_{ub} \right |/\left | V_{cb} \right |$, $\Delta M_d$, 
the amplitude spectrum 
for including the information from the ${\rm B}^0_s-\overline{{\rm B}}^0_s$ 
oscillations, $\epsilonk$ and the measurement of sin 2$\beta$. 
It has to be stressed once more that the inputs used are the same 
in the two approaches (in term of Gaussian and uniform uncertainties), 
but they correspond to different input likelihoods, for 
$|V_{cb}|$, $|V_{ub}|$, $\fbdsqbd$, $\hat B_K$ and $\xi$ as shown in 
the previous figures.
Figure~\ref{fig:test1} shows the comparison on the 
($\bar{\rho}$, $\bar{\eta}$) plane.
The numerical results are given in Table~\ref{tab:test1} 
Figure~\ref{fig:test1_sin_sin2b} 
shows the comparison between the allowed regions 
obtained using Bayesian or \rfit~ 
methods if the constraint from the direct measurement of sin2$\beta$ 
is removed from the fit. 

\vspace*{-12mm}
\subsubsection{Further comparisons}

\begin{figure}[hbtp]
\begin{center}
\begin{tabular}{cc}
\includegraphics[width=7.5cm]{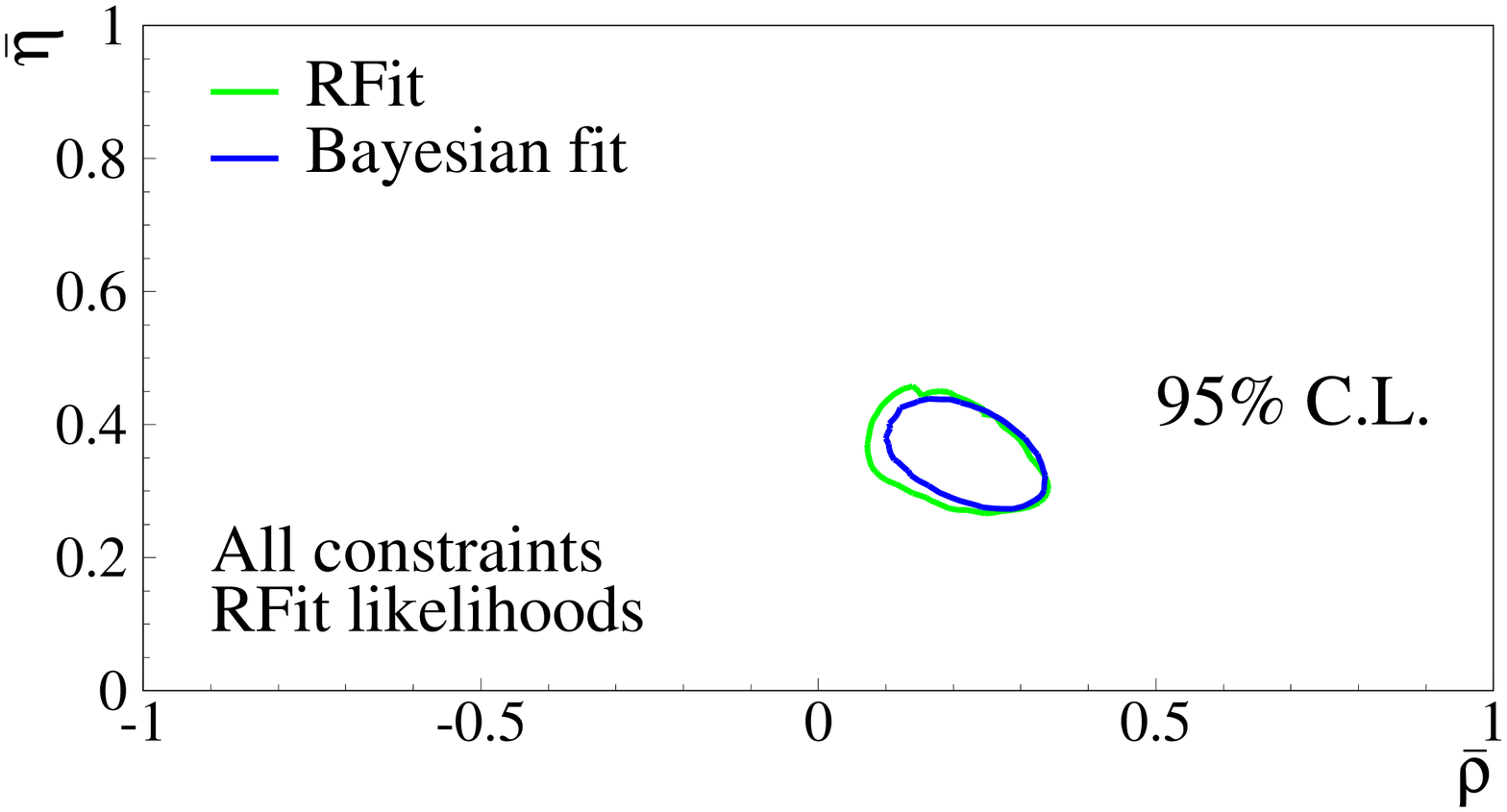} &
\includegraphics[width=7.5cm]{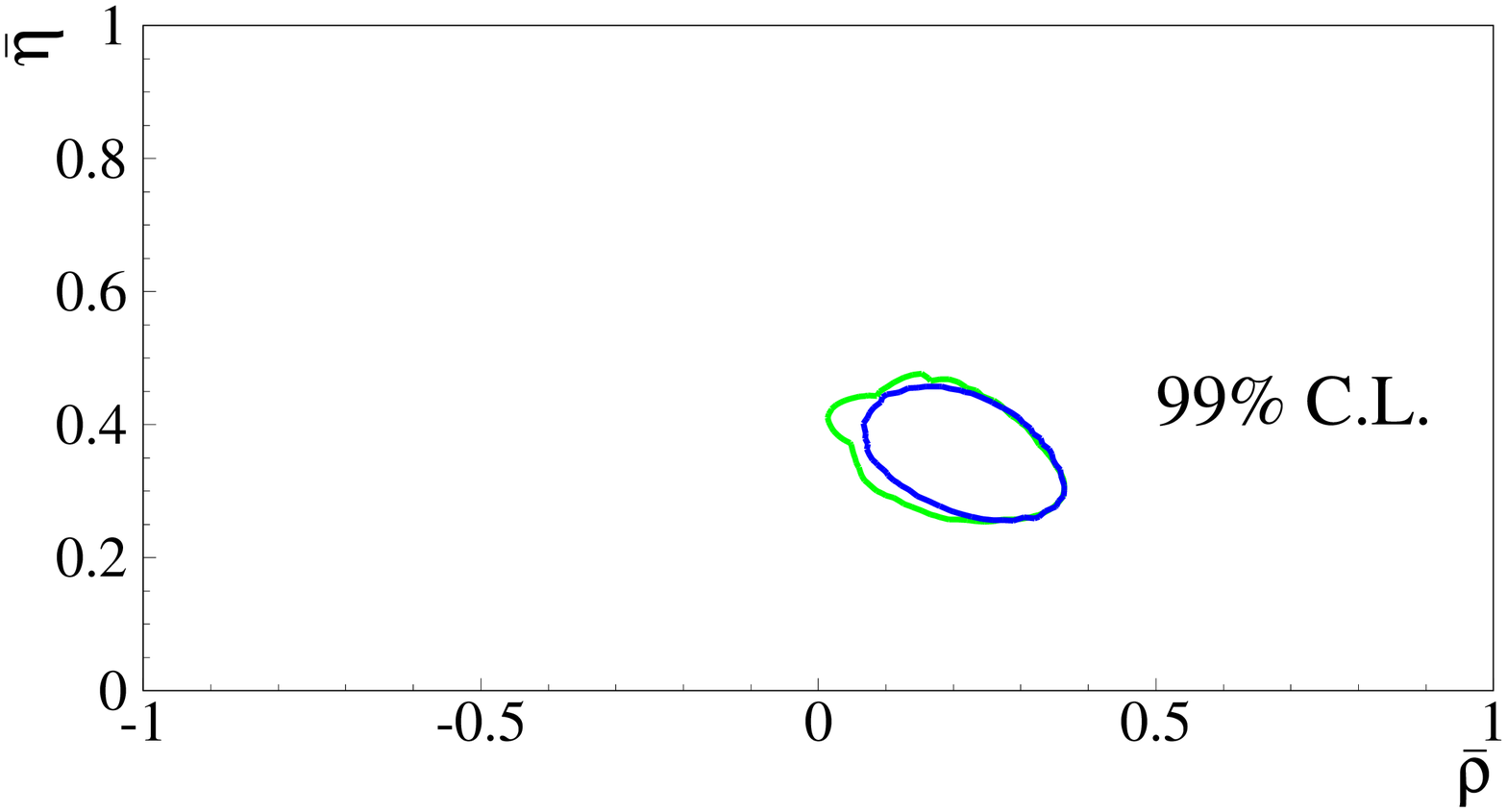} 
\end{tabular}
\caption{ {\it Comparison Bayesian/\rfit~ Methods using the distributions as obtained 
from \rfit~ to account for the information on input quantities.
Allowed regions for $\bar{\rho}$ and $\bar{\eta}$ at 95\% (left plot) and 99\%(right plot) 
using the measurements of $\left | V_{ub} \right |/\left | V_{cb} \right |$ and $\Delta M_d$
the amplitude spectrum for including the information from the ${\rm B}^0_s-\overline{{\rm B}}^0_s$ oscillations,
$\epsilonk$  and the measurement of sin 2$\beta$. }\label{fig:test2}}
\end{center}
\end{figure}

To further the orogin of the residual difference between the two methods, 
we have performed the following test: both methods use the distributions 
as obtained from \rfit~ or from the Bayesian method to account for the 
information on input quantities.
The results of the comparison using the input distributions
as obtained from \rfit~ are shown in Figs.~\ref{fig:test2} (Table~\ref{tab:test2}).
In some cases (0.1$\%$ C.L.) the ranges selected by the Bayesian approach are wider.
The comparison using the input distributions, as obtained from the Bayesian method,
give a maximal difference of 5\%.
These two tests show that, if same input likelihood are used, the results on the output 
quantities are very similar. The main origin of the residual difference on the output quantities, 
between the Bayesian and the \rfit~ method comes from the likelihood associated to the 
input quantities.

\vspace{-3mm}

\begin{table}[t] 
  \begin{center}
    \begin{tabular}{|cccc|}
    \hline
    Parameter    & $5\%$ CL & $1\%$ CL & $0.1\%$ CL  \\
    \hline 
$\bar{\rho}$     &   1.20    &   1.13    &   0.96      \\
$\bar{\eta}$     &   1.03    &   0.99    &   0.94      \\
$\sin{2\beta}$   &   1.07    &   1.08    &   1.07      \\
$\gamma^{\circ}$ &   1.24    &   1.12    &   0.95      \\
     \hline 
    \end{tabular}
  \caption{\it Comparison. Ratio for confidence levels \rfit/Bayesian using
             the distributions as obtained from \rfit~ to account for the information on input quantities }
\label{tab:test2} 
  \end{center}
\end{table}

\vspace{-2mm}

\subsubsection{Some conclusions on the fit comparison}
The Bayesian and the \rfit~ methods are compared in an agreed framework in terms 
of input and output quantities. For the input quantities the total error has 
been splitted in two errors. 
The splitting and the p.d.f distribution associated to any of the errors 
is not really important in the Bayesian approach. It becomes central in 
the \rfit~ approach where the systematic errors are treated as ``non statistical'' 
errors. The result is that, even if the same central values and errors 
are used in the two methods, the likelihood associated to the input parameters, 
which are entering in the fitting procedure, can be different. 
The output results ($\bar{\rho}$,$\bar{\eta}$, sin2 $\beta$, $\gamma$) differ 
by 15$\%$-45$\%$, 10$\%$-35$\%$ and 5-15$\%$ if the 
95$\%$, 99$\%$ and 99.9$\%$ confidence regions are compared, respectively,
with ranges from the frequentist method being wider.
If the same likelihoods are used the output results are very similar.

\section{Test of the CKM picture in the Standard Model}
\label{sec:final}
After comparing  different statistical methods, in this final Section we show
how the present data can be used to test the CKM picture of the Standard Model.
The results presented here have been obtained with a Bayesian fit to the latest inputs
of Table~\ref{tab:inputs}. The central values,  errors and 
 $68\%$ ($95\%$) [and $99\%$]  C.L. ranges obtained for various quantities of
interest are collected in Table \ref{tab:ckm_fit_final}.
\begin{table}[t]
\begin{center}
\begin{tabular}{|c|c|}
\hline
  $|V_{cb}|\times 10^3$     &     41.5  $\pm$ 0.8  \quad   (39.9,43.1) 
[39.1,43.9]       \\
  $\bar {\eta}$             &     0.341 $\pm$ 0.028 
                            \quad    (0.288,0.397) [0.271,0.415] \\
  $\bar {\rho}$             &     0.178 $\pm$ 0.046 
                            \quad    (0.085,0.265) [0.052,0.297] \\
  $\sin 2\beta$             &     0.705 $\pm$ 0.037 
                            \quad    (0.636,0.779) [0.612,0.799] \\
  $\sin 2\alpha$ &        --0.19 $\pm$ 0.25 
                 \quad    (--0.62,0.33) [--0.75,0.47] \\
  $\gamma$(degrees)         &         61.5 $\pm$ 7.0 
                            \quad   (49.0,77.0)   [44.3,82.1]\\
  $\Delta M_s$($ps^{-1}$)   &         18.3 $\pm$ 1.7 
                            \quad   (15.6,22.2)   [15.1,27.0]\\
\hline
\end{tabular}
\caption[]{ \it Values and errors for different quantities using the present 
knowledge summarized in Table \ref{tab:inputs}.
Within parentheses and brackets the 95$\%$ and 99$\%$ probability regions are, respectively, given.}
\label{tab:ckm_fit_final}
\end{center}
\end{table}

\begin{figure}[htb!]
\begin{center}
\includegraphics[width=12cm]{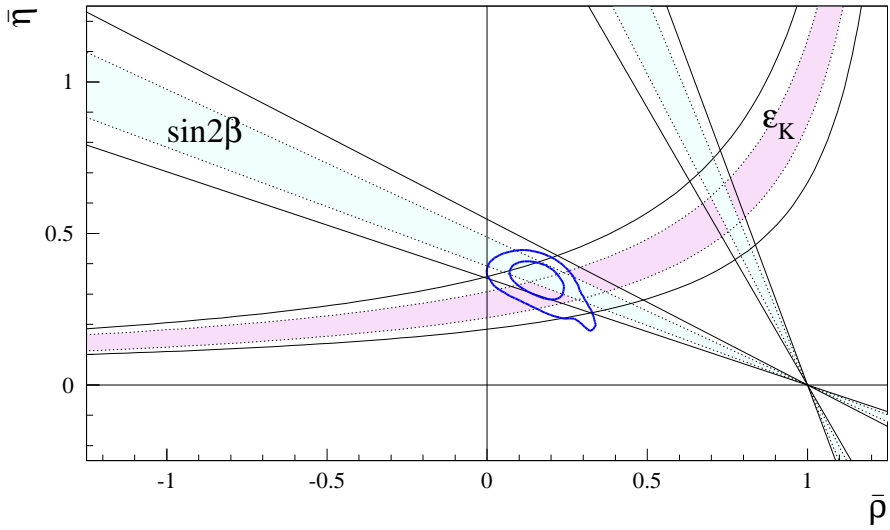}
\caption{\it The allowed regions for $\overline{\rho}$ and $\overline{\eta}$
(contours at 68\%, 95\%) as selected by the measurements of $\left | V_{ub} \right |/\left | V_{cb} \right |$, 
$\Delta {M_d}$, and by the limit on $\Delta {M_s}/\Delta {M_d} $ are compared with the bands (at 68\% and 95\% C.L.)
from CP violation in the kaon ($\epsilon_K$) and in the B (sin2$\beta$) sectors.}
\label{fig:testcp}
\end{center}
\end{figure}

The most crucial test is the comparison between the UT parameters  
determined with  quantities sensitive to the sides of the UT
(semileptonic B decays and oscillations) with the measurement of  CP violation in the 
kaon sector ($\epsilonk$) and, also with the one in the B (sin2$\beta$) sector. 
This test is shown in Fig.~\ref{fig:testcp}. 
It can be translated quantitatively into a comparison between the values of 
sin2$\beta$ obtained from the measurement of the CP asymmetry in the $J/\psi \rm K^0_S$ decays and the one 
determined from ``sides`` measurements:
\begin{eqnarray}
\begin{array}{|ccc|}\hline
\sin 2 \beta = & 0.685 \pm 0.052 ~(0.581,0.789)  & \rm {indirect\ -\ sides\ only}     \nonumber \\
\sin 2 \beta = & 0.734 \pm 0.054 ~(0.628,0.840)  & \rm \quad \rm B^0 \rightarrow J/\psi K^0_S 
\ ,
\\ \hline\end{array}
\label{eq:sin2beta}
\end{eqnarray}
where, within parentheses, we give also the 95$\%$ probability region.
The spectacular agreement between these values shows the consistency of the Standard 
Model in describing the CP violation phenomena in terms of one single complex parameter $\eta$.
Conversely, assuming the validity of the SM, this is 
also  an important test of the OPE, HQET and LQCD theories which have been used to extract the
CKM parameters. It has to be noted that the test is significant provided that the errors on sin 2$\beta$ 
from the two determinations are comparable. Presently, the accuracy of both 
is at the 10$\%$ level.
It is also of interest to explicitly make predictions for quantities which will be measured in 
the next future. We concentrate on $\dms$ which will be soon measured at Tevatron. 
The results obtained by excluding (or including) the information from the 
${\rm B}^0_s-\bar{{\rm B}}^0_s$ analyses are:
\begin{eqnarray}
\begin{array}{|cc|}\hline
\dms ({\rm with}~\dms~ {\rm included}) = & 18.3 \pm 1.7~(15.6,22.2)~[15.1,27.0]~\mbox{ps}^{-1} \nonumber \\
\dms ({\rm without}~ \dms~)   =  & 20.6 \pm 3.5~(14.2,28.1)~[12.8,30.7]~\mbox{ps}^{-1}
\ .
\\ \hline\end{array}
\label{eq:dms}
\end{eqnarray}
where, within parentheses, we give the 95$\%$ and the 99$\%$ regions.

It will be interesting to compare these results with future
measurements  with the goal of identifying new physics
contributions. Moreover a precise measurement of $\Delta M_s$ will
reduce significantly the uncertainties in the output quantities in 
Table~\ref{tab:ckm_fit_final}.

\newpage

\section*{References}
\addcontentsline{toc}{section}{References}

\vspace{7mm}

\renewcommand{\labelenumi}{[\theenumi]}
\begin{enumerate}

\item \label{Gaus}
A.~Ali and D.~London, Eur. Phys. J. C {\bf 18} (2001) 665; \\
S.~Mele, Phys. Rev. D {\bf 59} (1999) 113011; \\
D.~Atwood and A.~Soni, Phys. Lett. B {\bf 508} (2001) 17.

\vspace{3mm}

\item \label{C00}
M.~Ciuchini, G.~D'Agostini, E.~Franco, V.~Lubicz, G.~Martinelli, 
F.~Parodi, P.~Roudeau, A.~Stocchi,
JHEP 0107 (2001) 013, [hep-ph/0012308].

\vspace{3mm}

\item \label{FREQ}
A.~H\"ocker, H.~Lacker, S.~Laplace, F.~Le~Diberder, 
Eur.\ Phys.\ J.\ C {\bf 21} (2001) 225,
[hep-ph/0104062].

\vspace{3mm}

\item \label{SCAN95}
Y.~Grossman, Y.~Nir, S.~Plaszczynski and M.-H.~Schune,
Nucl. Phys. B {\bf 511} (1998) 69; \\
S.~Plaszczynski and M.-H.~Schune, hep-ph/9911280.

\vspace{3mm}

\item \label{WO5}
L.~Wolfenstein, Phys. Rev. Lett. {\bf 51} (1983) 1945.

\vspace{3mm}

\item \label{BLO5} A.J.~Buras, M.E.~Lautenbacher and G.~Ostermaier, Phys.
Rev. D {\bf 50} (1994) 3433.

\vspace{3mm}

\item \label{HNab}
S.~Herrlich and U.~Nierste, Nucl. Phys. B {\bf 419} (1994) 192, 
Phys. Rev. D {\bf 52} (1995) 6505 and recent update by M.~Jamin and 
U.~Nierste.

\vspace{3mm}

\item \label{NIR02} Y.~Nir, hep-ph/0208080.

\vspace{3mm}

\item \label{BaBar5}
B.~Aubert {\it et al.}, BaBar Collaboration, hep-ex/0207042. 

\vspace{3mm}

\item \label{Belle5}
K.~Abe {\it et al.}, Belle Collaboration, hep-ex/0208025.

\vspace{3mm}

\item \label{ref:YR9903} G.~D'Agostini, CERN Report 99--03.

\vspace{3mm}

\item \label{CkmFitter} ``\CKMfitter: code, numerical results and plots'', 
http://ckmfitter.in2p3.fr

\vspace{3mm}

\item \label{laplace} S.~Laplace, Z.~Ligeti, Y.~Nir, G.~Perez, Phys. Rev. D
{\bf 65}:094040 (2002)

\vspace{3mm}

\item \label{bib:BABAR}
The \babar\ Physics Book, \babar\ Collaboration, 
(P.F.~Harrison and H.R.~Quinn, eds.). 
SLAC-R-0504 (1998).

\vspace{3mm}

\item \label{greg}
G.P.~Dubois-Felsmann D.G.~Hitlin, F.C.~Porter and G.~Eigen,
CALT 68-2396 (2002).

\vspace{3mm}

\item \label{ref:parodietal} 
F.~Parodi, P.~Roudeau and A.~Stocchi, Nuovo Cim. {\bf 112A} (1999) 833.

\end{enumerate}

\chapter{FUTURE DIRECTIONS}
\label{chap:VI}

\vspace{2mm}

This chapter contains  a few contributions related to future possibilities for 
the extraction of the CKM elements and of the CP violating phases.
They include general strategies for the determination of the 
CKM matrix elements, 
radiative rare B decays, weak phase determination from hadronic 
B decays and rare ${\rm K}\to \pi\nu\bar\nu$ decays. Since these topics have 
not been the subject of a dedicated working group at this meeting of the 
Workshop, we present them in the form of collected papers under individual 
authorship.

\section{General strategies for the CKM matrix}
\label{sec:intro}
\setcounter{equation}{0}

{\it A.J. Buras, F. Parodi and A. Stocchi}

\vspace{2mm}

During the last two decades
several strategies have been proposed that should allow one to determine 
the CKM matrix and the related unitarity triangle (UT). We have already 
discussed a number of processes that can be used for the determination 
of the CKM parameters in Chapters 2--4. Additional processes and the 
related strategies will be discussed in this part. They are also 
reviewed in~[\ref{BABAR}--\ref{Erice}].
In this first opening section we want to address the determination of the 
CKM matrix and of the UT in general terms leaving the discussion of specific 
strategies to the following sections. 

To be specific let us first choose as the independent parameters
\begin{equation}\label{I1}
\vus, \qquad \vcb, \qquad \bar\varrho, \qquad \bar\eta~.
\end{equation}

\vspace{1mm}

\noindent
The best place to determine $\vus$ and $\vcb$, as discussed already in detail
in Chapters 2 and 3,  are the semi-leptonic K and 
B decays, respectively. The question that we want address  
here is the determination of the remaining two parameters 
$(\bar\varrho,\bar\eta)$.

There are many ways to determine $(\bar\varrho,\bar\eta)$. As the length 
of  one side
of the rescaled unitarity triangle is fixed to unity, we have to our disposal
two sides, $R_b$ and $R_t$ and three angles, $\alpha$, $\beta$ and $\gamma$.
These five quantities can be measured by means of rare K and B 
decays and in particular by studying CP-violating observables. While until
recently only a handful of strategies could be realized, the present decade
should allow several independent determinations of $(\bar\varrho,\bar\eta)$ 
that will test the KM picture of CP violation and possibly indicate
the physics beyond the Standard Model (SM).

The determination of $(\bar\varrho,\bar\eta)$ in a given strategy is subject 
to experimental and theoretical errors and it is important to identify 
those strategies that are experimentally feasible and in which
hadronic uncertainties are as much as possible under control. 
Such strategies are reviewed 
in~[\ref{BABAR}--\ref{Erice}] and
in the following sections below.

Here we want to address a different question. The determination of 
$(\bar\varrho,\bar\eta)$ requires at least two independent measurements. 
In most cases these
are the measurements of two sides of the UT, of one side and one angle or
the measurements of two angles. Sometimes $\bar\eta$ can be directly 
measured and combining it with the knowledge of one angle or one side 
of the UT, $\bar\varrho$ can be found. Analogous comments apply to 
measurements in which $\bar\varrho$ is directly measured. Finally
in more complicated strategies one measures various linear combinations 
of angles, sides or $\bar\varrho$ and $\bar\eta$.

Restricting first our attention to measurements in which sides and angles 
of the UT can be measured independently of each other, we end up with 
ten different pairs of measurements that allow the determination of
$(\bar\varrho,\bar\eta)$. The question then arises which of the pairs 
in question is most efficient in the determination of the UT?
That is, given the same relative errors on $R_b$, $R_t$, $\alpha$, 
$\beta$ and $\gamma$, we want to find which of the pairs gives the most 
accurate determination of $(\bar\varrho,\bar\eta)$.

 The answer to this question depends necessarily on
the values of $R_b$, $R_t$, $\alpha$, $\beta$ and $\gamma$ but as we will see
below just the requirement of the consistency of $R_b$ with the measured
value of $|V_{ub}/V_{cb}| $ implies
a hierarchy within the ten strategies mentioned above.

During the 1970's and 1980's $\alpha_{QED}$, the Fermi 
constant $G_F$ and the sine of the Weinberg angle ($\sin\theta_W$) 
measured in the $\nu$-$N$ scattering were 
the fundamental parameters in terms of which the electroweak tests of 
the SM were performed. After the $Z^0$ boson was  discovered
and its mass precisely measured at LEP-I, $\sin\theta_W$ has been replaced
by $M_Z$ and the fundamental set used in the electroweak precision studies 
in the 1990's has been $(\alpha_{QED},G_F,M_Z)$. It is to be expected that
when $M_W$ will be measured precisely this set will be changed to 
$(\alpha_{QED},M_W,M_Z)$ or ($G_F,M_W,M_Z)$.

We anticipate that an analogous development will happen in this decade 
in connection with the CKM matrix. While the set (\ref{I1}) has clearly 
many virtues and has been used extensively in the literature, one should
emphasize that presently no direct independent measurements of $\bar\eta$
and $\bar\varrho$ are available. $|\bar\eta|$ can be measured cleanly in
the decay ${\rm K}_L\to\pi^0\nu\bar\nu$. On the other hand to our knowledge
there does not exist any strategy for a clean independent measurement 
of $\bar\varrho$. 

Taking into account the experimental feasibility of various measurements
and their theoretical cleanness, the most obvious candidate for the 
fundamental set 
in the quark flavour physics for the coming years 
appears to be~[\ref{BUPAST}]
\begin{equation}\label{I2}
\vus, \qquad \vcb, \qquad R_t, \qquad \beta
\end{equation}

\vspace{2mm}

\noindent
with the last two variables describing the $V_{td}$ coupling that can 
be measured by means of the ${\rm B}^0-\bar B^0$ 
mixing ratio $\Delta M_d/\Delta M_s$ and the CP-asymmetry $a_{\psi {\rm K}_S}$, 
respectively.
In this context, we investigate~[\ref{BUPAST}], 
in analogy to the $(\bar\varrho,\bar\eta)$ 
plane and the planes $(\sin 2\beta,\sin 2\alpha)$~[\ref{betaalpha}] and 
$(\gamma,\sin 2\beta)$~[\ref{Lacker}]
considered in the past, the $(R_t,\beta)$
plane for the exhibition of various constraints on the CKM matrix. 
We also provide the parametrization of the CKM matrix given directly in
terms of the variables (\ref{I2}).

While the set (\ref{I2}) appears to be the best choice for the 
coming years, our analysis shows that in the long run other choices 
could turn out to be preferable.  
In this context it should be emphasized that
several of the results and formulae presented here are not entirely new 
and have been already discussed by us and other authors in the 
past. In particular in~[\ref{BBSIN}] it has been pointed out that only a 
moderately precise measurement of $\sin 2\alpha$ can be as useful for 
the UT as a precise measurement of the angle $\beta$. This has been recently 
reemphasized in~[\ref{Beneke:2001ev}--\ref{Luo:2001ek}], 
see contribution in this Chapter, Sec.~\ref{sec:benekesec}. 
Similarly the measurement of the pair 
$(\alpha,\beta)$ has been found to be a very efficient tool for
the determination of the UT~[\ref{BLO},\ref{B95}] and the construction of 
the full CKM matrix from the angles of various unitarity triangles has 
been presented in~[\ref{Kayser}]. Next, the importance of 
the pair $(R_t,\sin 2\beta)$ has been emphasized recently in a number of 
papers~[\ref{UUT}--\ref{AI01}]. 
Many useful relations relevant for the unitarity triangle
 can also be found in~[\ref{Branco},\ref{BOBRNERE}].
Finally, in a recent paper~[\ref{BUPAST}] we have presented a
systematic classification of the strategies in question and their 
comparison. In fact the results of this paper constitute the main 
part of this section.

\subsection{Basic formulae}
Let us begin our presentation by listing the formulae for $\bar\varrho$ 
and $\bar\eta$ in 
the strategies in question  that are labeled by the two measured quantities as
discussed above.  \\
\noindent
\underline{\boldmath{$R_t$} and  \boldmath{$\beta$}} 
\be\label{S1}
\bar\varrho=1-R_t\cos\beta,\qquad \bar\eta=R_t\sin\beta~.
\ee

\noindent
\underline{\boldmath{$R_b$} and \boldmath{$\gamma$}} 
\be\label{S2}
\bar\varrho=R_b\cos\gamma,\qquad \bar\eta=R_b\sin\gamma~.
\ee

\noindent
\underline{\boldmath{$R_b$} and \boldmath{$R_t$}} 
\be\label{S3}
\bar\varrho = \frac{1}{2} (1+R^2_b-R^2_t),
\qquad \bar\eta=\sqrt{R_b^2-\bar\varrho^2}
\ee
where $\bar\eta>0$ has been assumed.

\noindent
\underline{\boldmath{$R_t$} and \boldmath{$\gamma$}} \\
This strategy uses (\ref{S2}) with
\be\label{S4}
R_b=\cos\gamma\pm \sqrt{R^2_t-\sin^2\gamma}~.
\ee
The two possibilities can be distinguished by the measured 
value of $R_b$.\\

\noindent
\underline{\boldmath{$R_b$} and \boldmath{$\beta$}} \\
This strategy uses (\ref{S1}) and
\be\label{S5}
R_t=\cos\beta\pm \sqrt{R^2_b-\sin^2\beta}~.
\ee
The two possibilities can be distinguished by the measured 
value of $R_t$.\\

\noindent
\underline{\boldmath{$R_t$} and \boldmath{$\alpha$}} 
\be\label{S6a}
\bar\varrho=1-R_t^2\sin^2\alpha+R_t\cos\alpha\sqrt{1-R_t^2\sin^2\alpha},
\ee
\be\label{S6b}
\bar\eta=R_t\sin\alpha\left[R_t\cos\alpha+\sqrt{1-R_t^2\sin^2\alpha}\right] 
\ee
where $\cos\gamma>0$ has been assumed. For $\cos\gamma<0$ the signs in front
of the square roots should be reversed.\\

\noindent
\underline{\boldmath{$R_b$} and \boldmath{$\alpha$}} 
\be\label{S7a}
\bar\varrho=R_b^2\sin^2\alpha-R_b\cos\alpha\sqrt{1-R_b^2\sin^2\alpha},
\ee
\be\label{S7b}
\bar\eta=R_b\sin\alpha\left[R_b\cos\alpha+\sqrt{1-R_b^2\sin^2\alpha}\right]
\ee
where $\cos\beta>0$ has been assumed.\\

\noindent
\underline{\boldmath{$\beta$} and \boldmath{$\gamma$}} 
\be\label{S8}
R_t=\frac{\sin\gamma}{\sin(\beta+\gamma)},
\qquad R_b=\frac{\sin\beta}{\sin(\beta+\gamma)}
\ee
and (\ref{S3}).\\

\noindent
\underline{\boldmath{$\alpha$} and \boldmath{$\gamma$}} 
\be\label{S9}
R_t=\frac{\sin\gamma}{\sin\alpha},
\qquad R_b=\frac{\sin(\alpha+\gamma)}{\sin\alpha}
\ee
and (\ref{S3}).\\

\noindent
\underline{\boldmath{$\alpha$} and \boldmath{$\beta$}} 
\be\label{S10}
R_t=\frac{\sin(\alpha+\beta)}{\sin\alpha},
\qquad R_b=\frac{\sin\beta}{\sin\alpha}
\ee
and (\ref{S3}).

Finally we give the formulae for the strategies in which $\bar\eta$ is 
directly measured and the strategy allows to determine $\bar\varrho$.\\
\noindent
\underline{\boldmath{$\bar\eta$} and \boldmath{$R_t$} or \boldmath{$R_b$}} 
\be\label{S11}
\bar\varrho=1-\sqrt{R_t^2-\bar\eta^2}, \qquad
\bar\varrho=\pm\sqrt{R_b^2-\bar\eta^2}\,,
\ee

\vspace{2mm}

where in the first case we have excluded the + solution in view of 
 $R_b\le 0.5$ as extracted from the experimental data on $\vub$. \\
\noindent
\underline{\boldmath{$\bar\eta$} and \boldmath{$\beta$} or \boldmath{$\gamma$}}
\be\label{S12}
\bar\varrho=1-\frac{\bar\eta}{\tan\beta}, \qquad
\bar\varrho=\frac{\bar\eta}{\tan\gamma}~.
\ee

\vspace{3mm}

\subsection{CKM matrix and the fundamental variables}
It is useful for phenomenological purposes to express the CKM matrix 
directly in terms of the parameters selected in a given strategy.
This can be easily done by inserting the formulae for $\bar\varrho$ 
and $\bar\eta$ presented here into the known expressions for 
the CKM elements in terms of these variables~[\ref{WO},\ref{BLO}] as given 
in~Chapter~1.

Here we give explicit result only for the set (\ref{I2}). In order to 
simplify the notation we use $\lambda$ instead of $\vus$ as
$V_{us}=\lambda+\ord(\lambda^7)$. We find 

\vspace{2mm}

\be
V_{ud}=1-\frac{1}{2}\lambda^2-\frac{1}{8}\lambda^4 +\ord(\lambda^6),
\qquad
V_{ub}=\frac{\lambda}{1-\lambda^2/2}\vcb \left[1-R_t e^{i\beta}\right],
\ee
\vspace{1mm}
\be
V_{cd}=-\lambda+\frac{1}{2} \lambda \vcb^2 -
\lambda\vcb^2 \left[1-R_t e^{-i\beta}\right] +
\ord(\lambda^7),
\ee
\vspace{1mm}
\be
V_{cs}= 1-\frac{1}{2}\lambda^2-\frac{1}{8}\lambda^4 -\frac{1}{2} \vcb^2
 +\ord(\lambda^6),
\ee
\vspace{1mm}
\be
V_{tb}=1-\frac{1}{2} \vcb^2+\ord(\lambda^6),
\qquad
V_{td}=\lambda\vcb R_t e^{-i\beta}
+\ord (\lambda^7),
\ee
\vspace{1mm}
\begin{equation}\label{2.83d}
 V_{ts}= -\vcb +\frac{1}{2} \lambda^2 \vcb - 
\lambda^2 \vcb \left[1-R_t e^{-i\beta}\right]
  +\ord(\lambda^6)~.
\end{equation}

\vspace{2mm}

\subsection{Hierarchies of the various strategies}
\label{sec:Hierarchies}
\begin{figure}[htbp] 
\begin{center}
{\epsfig{figure=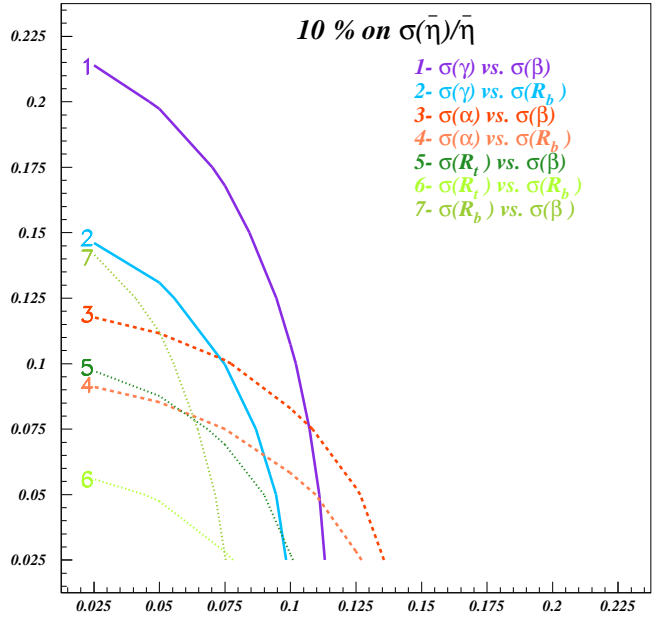,height=9.8cm}}

\vspace{-2mm}

\caption[]{\it The plot shows the curves of the 10$\%$ relative
precision on $\bar\eta$ as a function of the precision on the variables
of the given strategy.}
\label{fig:cl_eta}
\end{center}
\end{figure}

\begin{figure}[htbp] 
\begin{center}
{\epsfig{figure=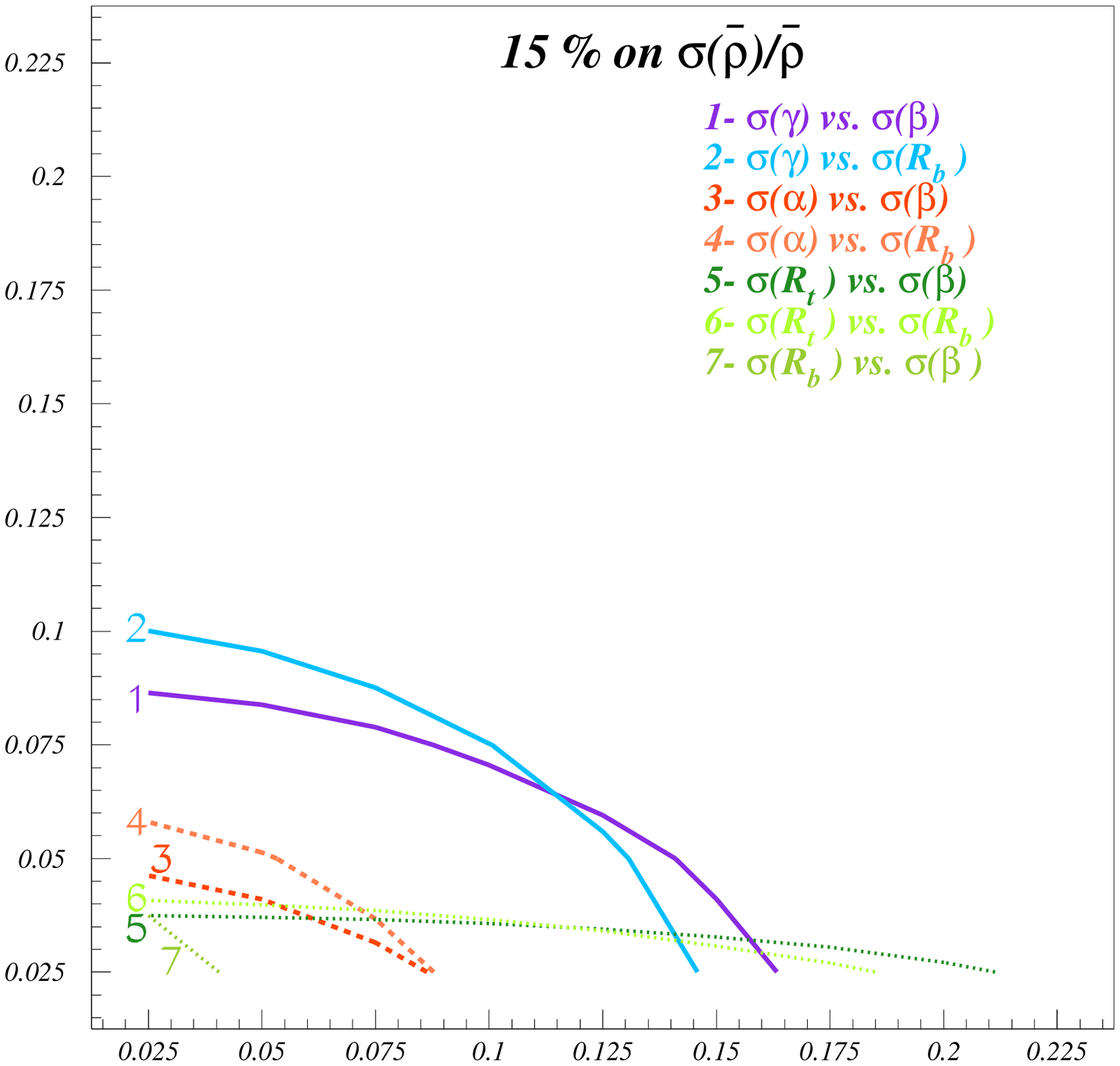,height=9.8cm}}

\vspace{-2mm}

\caption[]{\it{The plot shows the curves of the 15$\%$ relative
precision on $\bar\varrho$ as a function of the precision on the variables
of the given strategy.}}
\label{fig:cl_rho}
\end{center}
\end{figure}

\begin{figure}[htbp]
\begin{center}
{\epsfig{figure=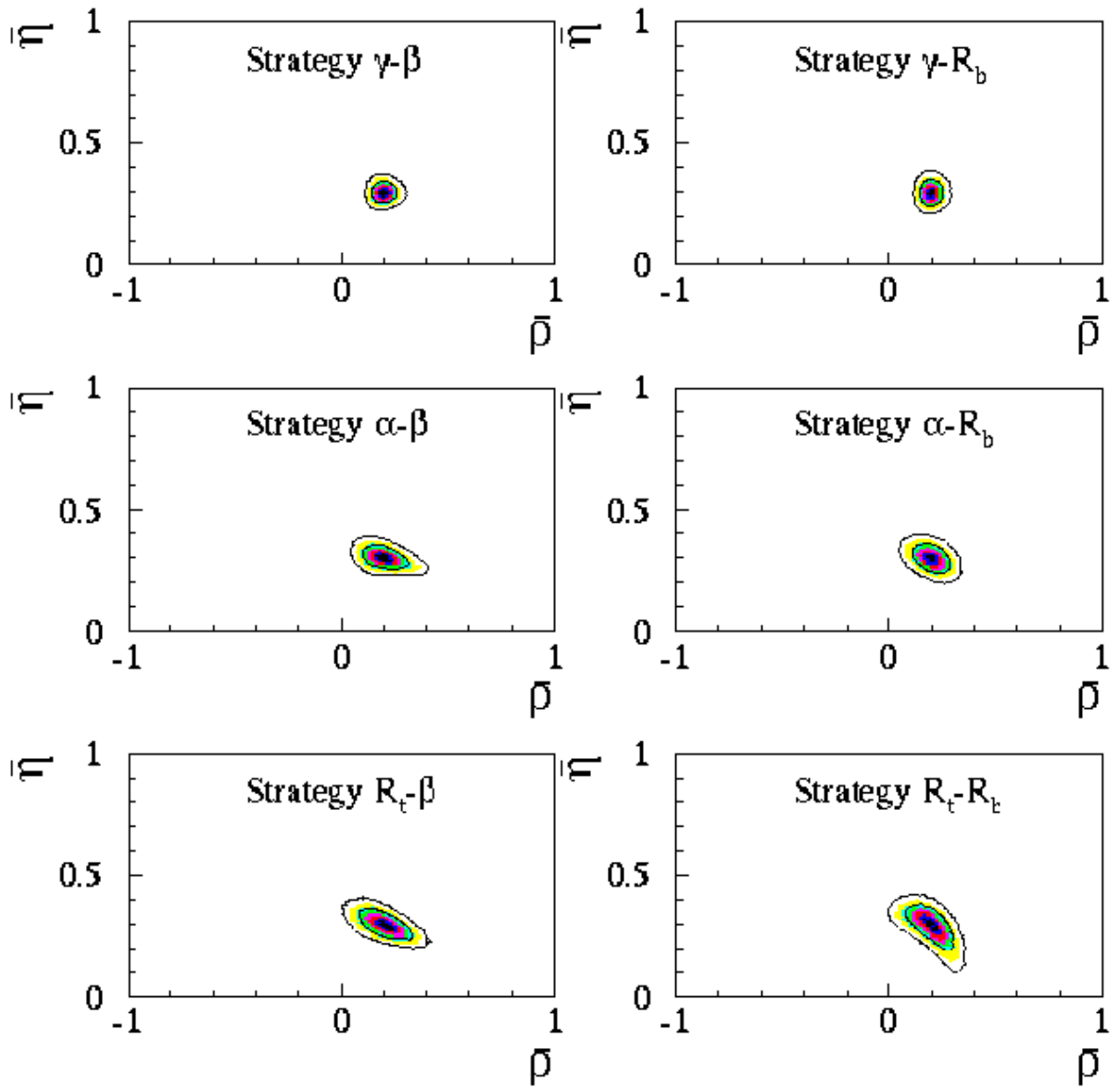,height=16cm}}
\caption[]{\it The plots show the allowed regions {(68$\%$ and 95$\%$)} 
in the $(\bar\varrho,\bar\eta)$ plane
obtained from the leading strategies assuming that each variable is measured
with $10\%$ accuracy.}
\label{fig:6plots}
\end{center}
\end{figure}

\begin{figure}[htbp]
\begin{center}
{\epsfig{figure=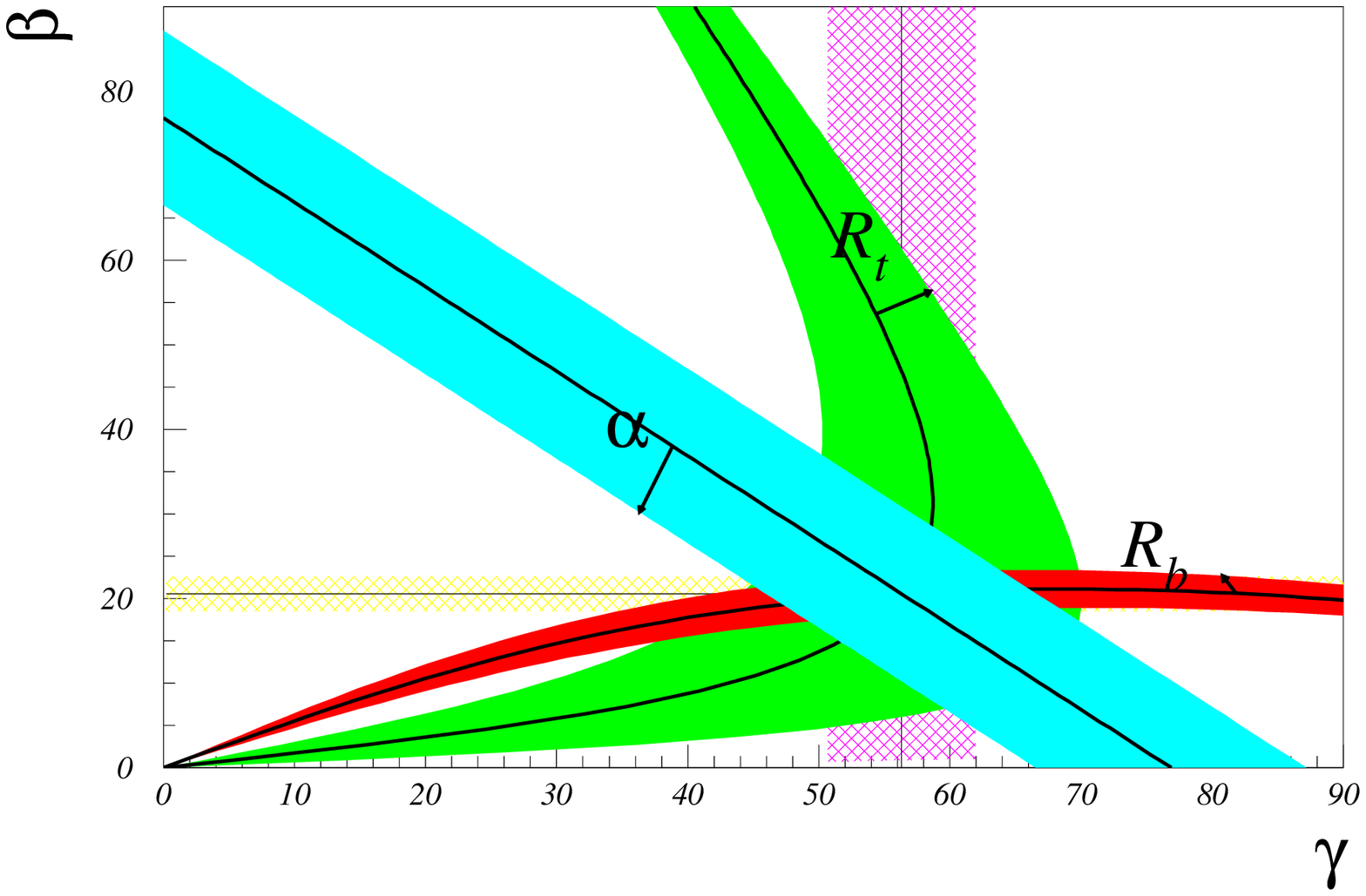,height=5.5cm}}
{\epsfig{figure=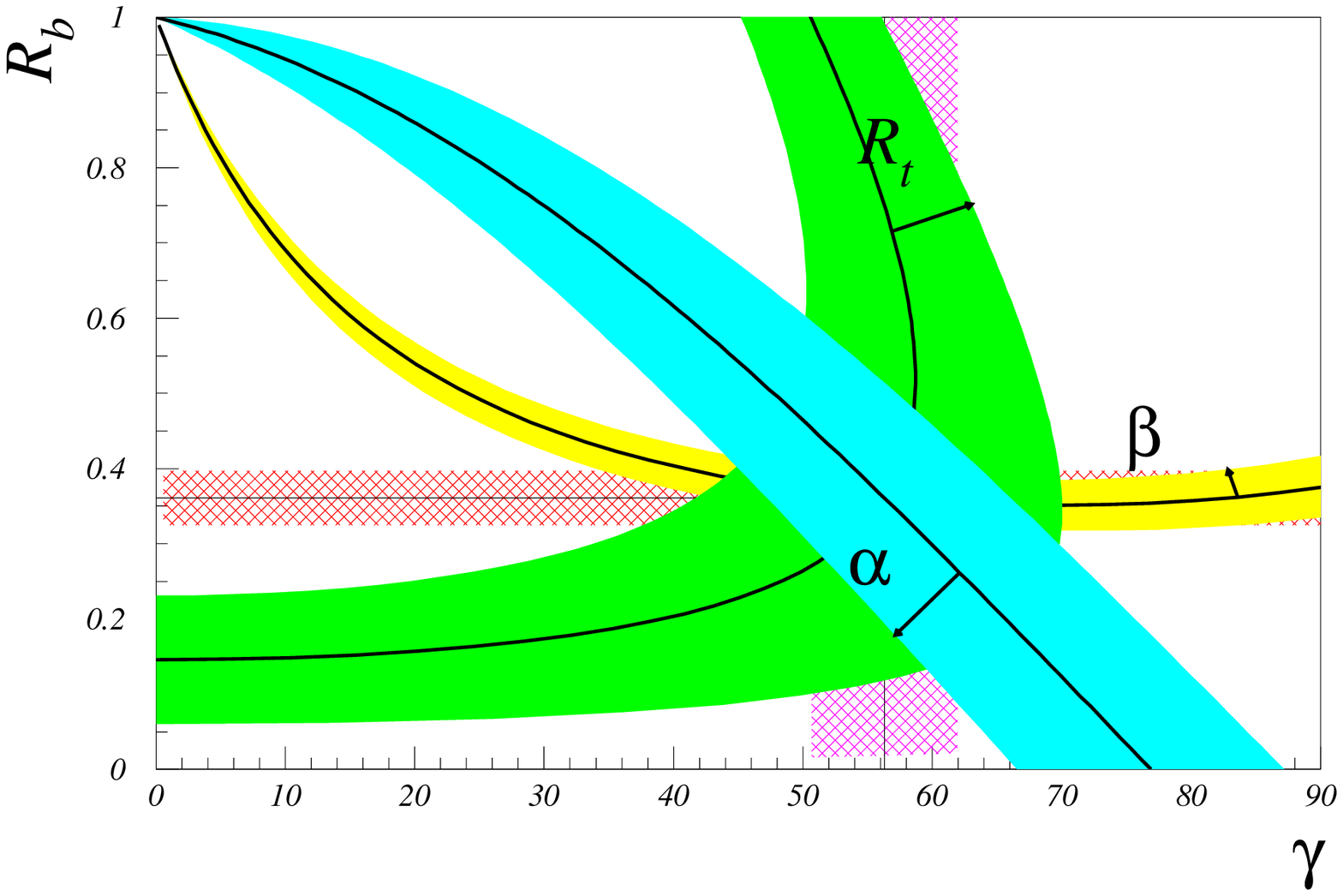  ,height=5.5cm}}\\
{\epsfig{figure=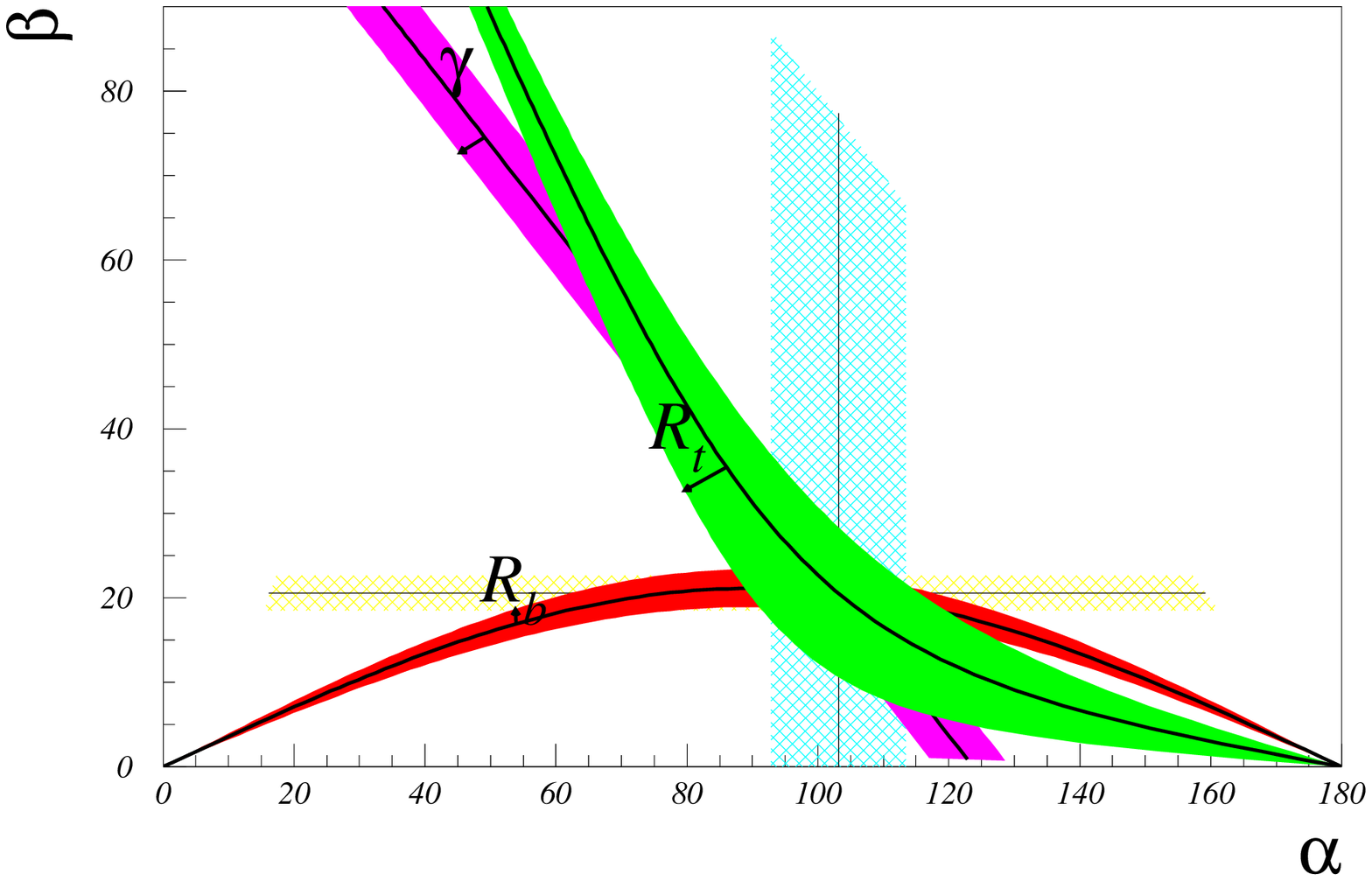,height=5.5cm}}
{\epsfig{figure=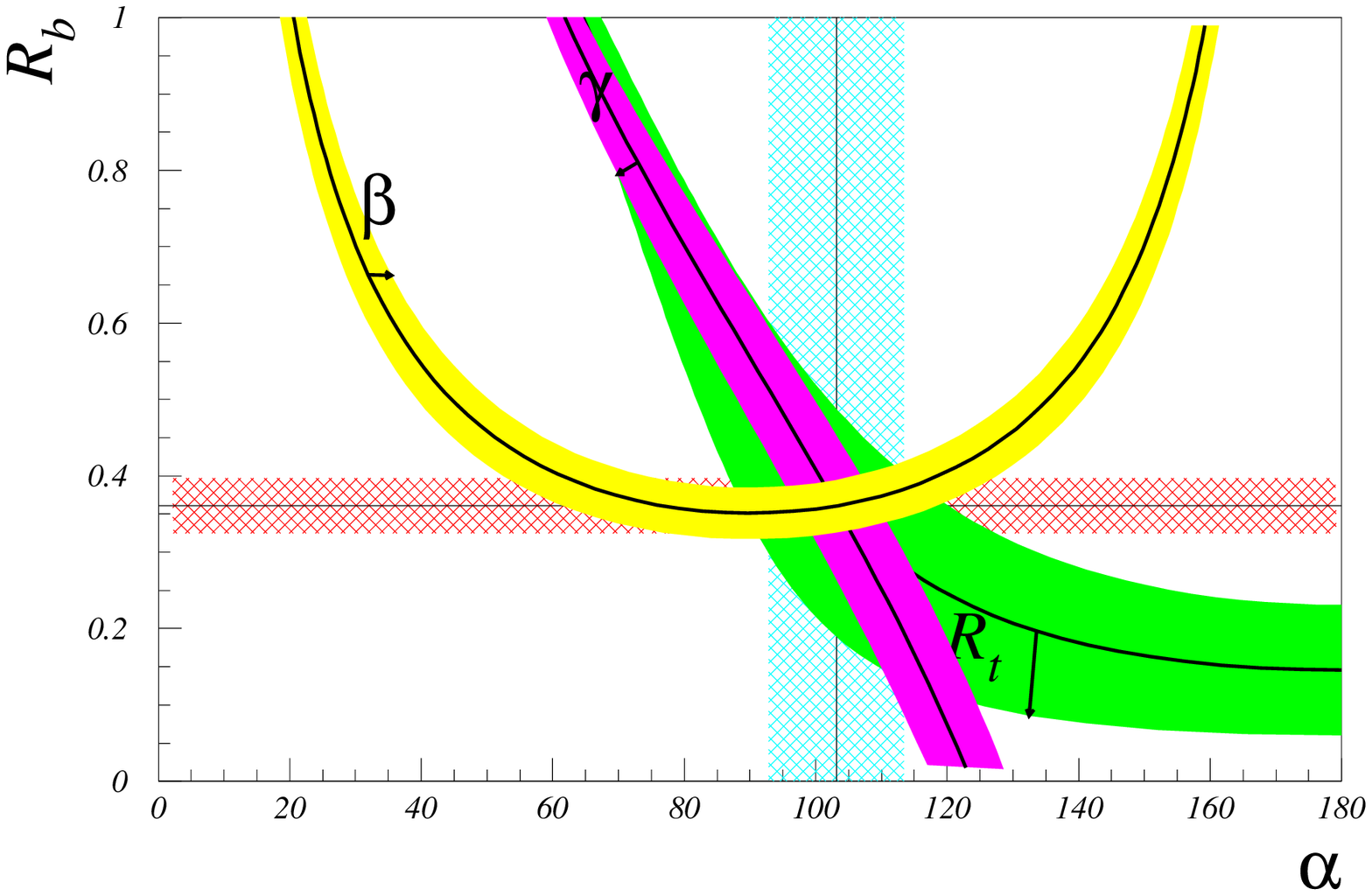  ,height=5.5cm}}\\
{\epsfig{figure=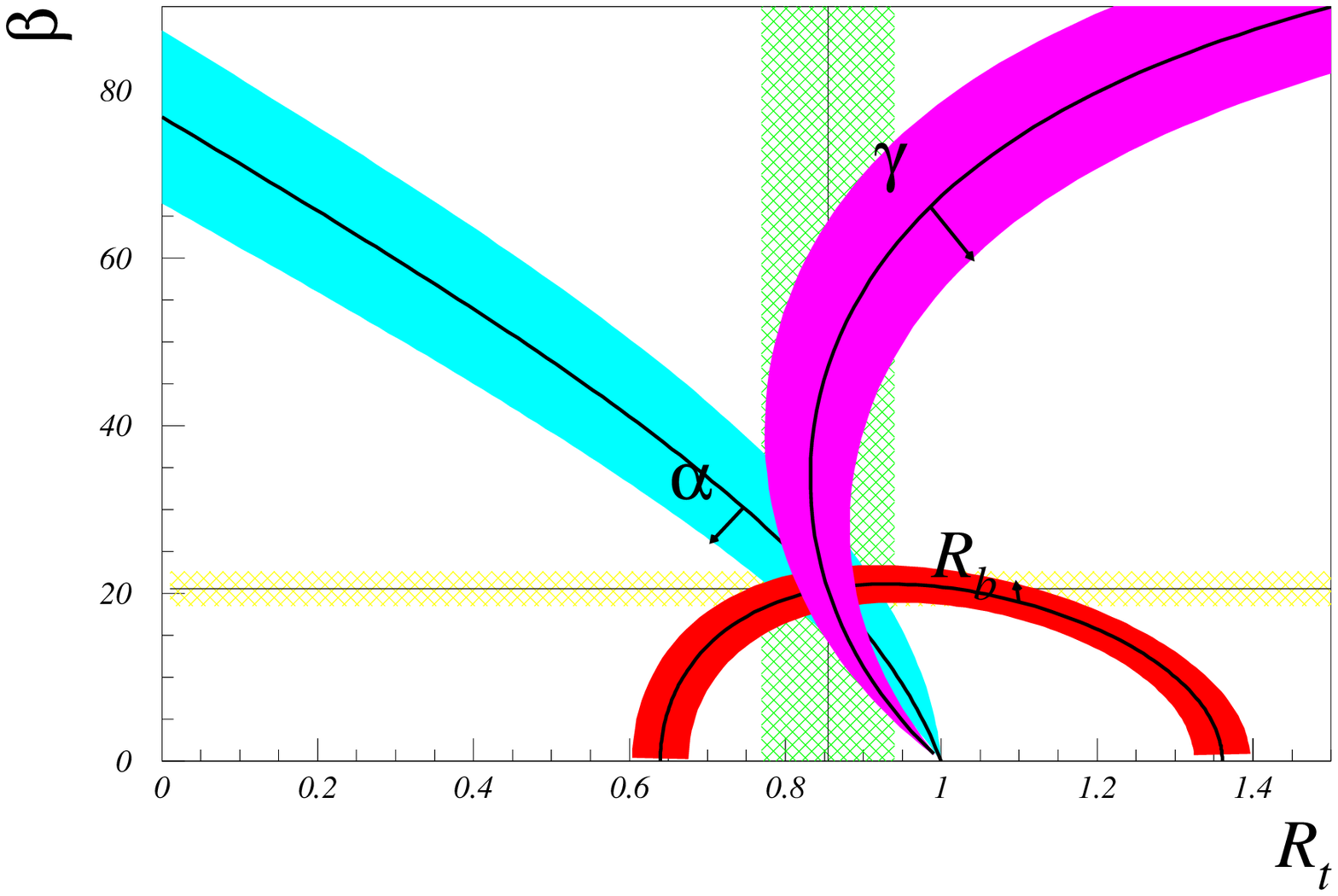   ,height=5.5cm}}
{\epsfig{figure=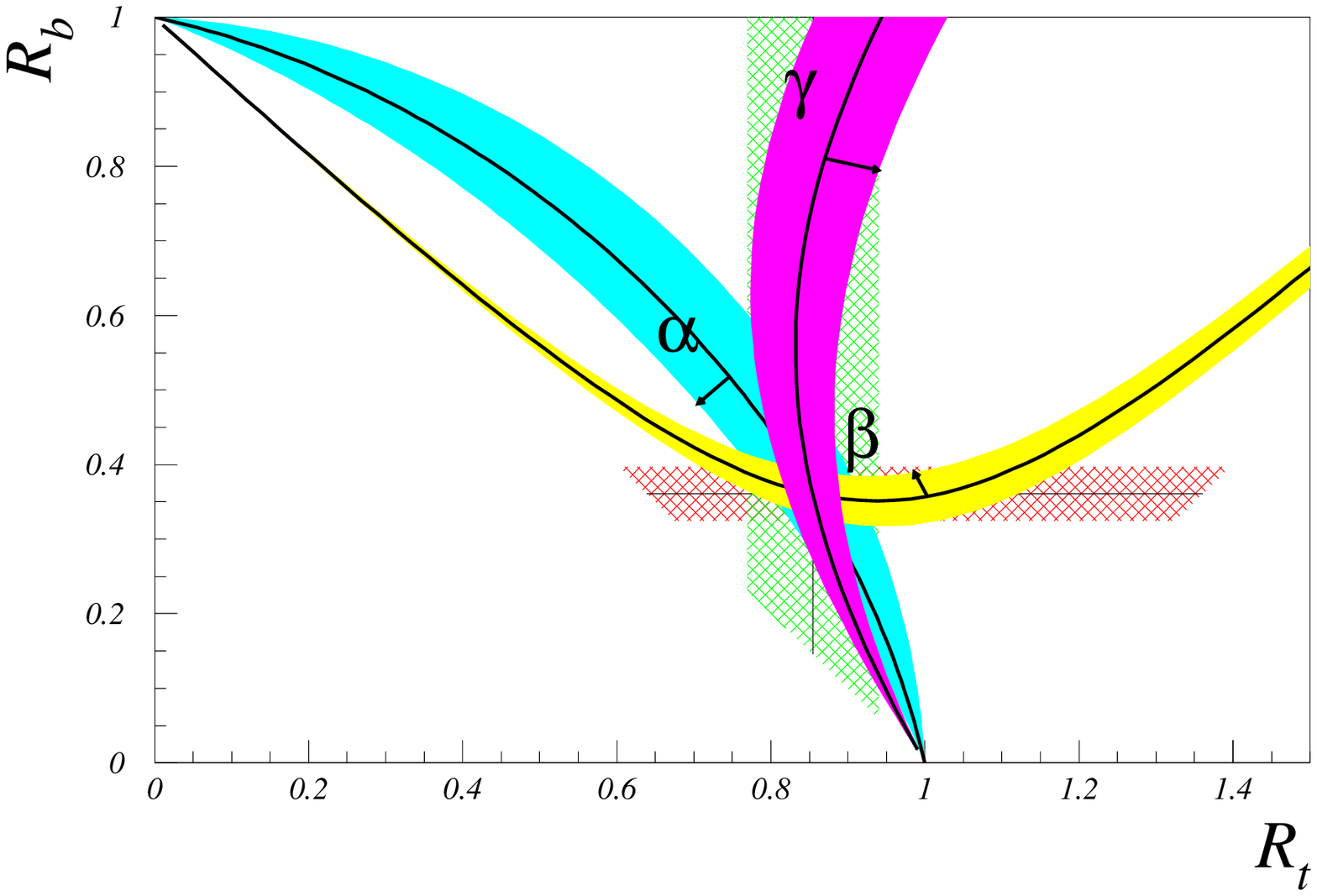     ,height=5.5cm}}
\caption[]{\it The plots show the different constraints 
(assuming a relative error of 10$\%$) in the different planes 
corresponding to the leading stategies of equation \ref{ranking1}. 
The small arrow indicates the range corresponding to an 
increase of 10$\%$ of the corresponding quantity.}
\label{fig:stra}
\end{center}
\end{figure}

The numerical analysis  of various strategies listed above
was performed using
the Bayesian approach as described in the previous Chapter.
The main results of this analysis are depicted in 
Figs.~\ref{fig:cl_eta},~\ref{fig:cl_rho},~\ref{fig:6plots}~and \ref{fig:stra}. 
In Figs.~\ref{fig:cl_eta} and \ref{fig:cl_rho} we plot the 
correlation between the precisions on the variables relevant for a given 
strategy required to reach the assumed precision on $\bar\eta$ and
$\bar\varrho$, respectively. For this exercise we have used, for
$\bar \eta$ and $\bar \rho$, 
the central values obtained in the previous Chapter.
Obviously strategies described by curves in
Figs. \ref{fig:cl_eta} and \ref{fig:cl_rho} that lie far from the origin 
are more effective in the determination of the unitarity  triangle than
those corresponding to curves placed close to the origin.

\vspace{-6mm}
%


Figures~\ref{fig:cl_eta} and \ref{fig:cl_rho}
reveal certain hierarchies within the strategies in question. In order to 
find these hierarchies and to eliminate the weakest ones not shown in these 
figures we divided first the five variables under consideration 
into two groups:

\vspace{1mm}

\be\label{group}
(R_t,\alpha,\gamma),\qquad (R_b,\beta)~.
\ee

\vspace{1mm}

It turned out then that the four strategies $(R_t,\alpha)$, $(R_t,\gamma)$,
$(\alpha,\gamma)$ and $(R_b,\beta)$ which involve pairs of variables
belonging to the same group are not particularly useful in the determination
of $(\bar\varrho,\bar\eta)$.  In the case of $(R_b,\beta)$ this is related to
the existence of two possible solutions as stated above. If one of these 
solutions can easily be excluded on the basis of $R_t$, then the 
effectiveness of this strategy can be increased. 
We have therefore included this strategy in our 
numerical analysis. The strategy $(R_t,\gamma)$ turns out to be less
useful in this respect. Similarly the strategy $(\gamma,\alpha)$ is
not particularly useful due to strong correlation between the variables
in question as discussed previously by many authors in the literature.

The remaining six strategies that involve pairs of variables belonging 
to different groups in (\ref{group}) are all interesting.
While some strategies are better suited for the determination of $\bar\eta$ 
and the other for $\bar\varrho$, as clearly seen in 
Figs. \ref{fig:cl_eta} and \ref{fig:cl_rho}, 
on the whole a clear ranking of
strategies seems to emerge from our analysis.

If we assume the same relative error on $\alpha$, $\beta$, $\gamma$,
$R_b$ and $R_t$ we find the following hierarchy:

\vspace{0.5mm}

\be\label{ranking1}
1)~(\gamma,\beta), \quad (\gamma,R_b) \qquad 
2)~(\alpha,\beta), \quad (\alpha,R_b) \qquad 
3)~(R_t,\beta), \quad (R_t,R_b), \quad (R_b,\beta).
\ee

We observe that in particular the strategies involving $R_b$ and $\gamma$ are
very high on this ranking list. This is related to the fact that 
$R_b<0.5<R_t$ and consequently the action in the $(\bar\varrho,\bar\eta)$
plane takes place closer to the origin of this plane than to the corner 
of the UT involving the angle $\beta$. Consequently the accuracy on 
$R_b$ and $\gamma$ does not have to be as high as for $R_t$ and $\beta$
in order to obtain the same accuracy for $(\bar\varrho,\bar\eta)$.
This is clearly seen in Figs. \ref{fig:cl_eta} and \ref{fig:cl_rho}.

This analysis shows how important is the determination of $R_b$ and $\gamma$ 
in addition to $\beta$ that is already well known.
On the other hand the strategy involving $R_t$ and $\beta$ will be 
most probably the cleanest one before the LHC experiments unless the 
error on angle $\gamma$ from B factories and Tevatron can be significantly 
decreased below $10\%$ and the accuracy on $R_b$ considerably 
improved. The explicit strategies for the determination of $\gamma$ 
are discussed in the following sections.

The strategies involving $\alpha$ are in our second best class. 
However, it has to be noticed that in order to get 10$\%$(15$\%$) relative 
precision on $\bar\eta$($\bar\rho$) it is necessary 
(see Figs. \ref{fig:cl_eta} and \ref{fig:cl_rho}) 
to determine $\alpha$ with better than 10$\%$ relative precision. 
If $\sin 2\alpha$ could be directly measured this could be soon achieved 
due to the high sensitivity of $\sin 2\alpha$ to $\alpha$ for 
$\alpha$ in the ball park of $90^\circ$ as obtained from the standard 
analysis of the unitarity triangle.
However, from the present perspective this appears to be very difficult in view
of the penguin pollution that could be substantial as indicated by the most 
recent data from Belle~[\ref{Bellealpha}]. On the other hand, as the BaBar  
data~[\ref{BaBaralpha}] do not indicate this pollution, 
the situation is unclear 
at present. These issues are discussed in detail in the following sections.

We have also performed a numerical analysis for the strategies in which
$|\bar \eta|$ can be directly measured. The relevant formulae are given in 
(\ref{S11}) and (\ref{S12}). It turns out that the strategy 
$(\gamma,\bar\eta)$ can be put in the 
first best class in (\ref{ranking1}) together with the strategies 
$(\gamma,\beta)$ and $(\gamma,R_b)$.

In Fig.~\ref{fig:6plots} we show the resulting regions in the 
$(\bar\varrho,\bar\eta)$
plane obtained from leading strategies assuming that each variable is measured
with $10\%$ accuracy. This figure is complementary to 
Figs. \ref{fig:cl_eta} and \ref{fig:cl_rho} and demonstrates clearly the 
ranking given in (\ref{ranking1}).

While at present the set (\ref{I2}) appears to be the leading candidate for 
the fundamental parameter set in the quark flavour physics for the coming 
years, it is not clear which set will be most convenient in the second half 
of this decade when the B-factories and Tevatron will improve considerably 
their measurements and LHC will start its operation. Therefore it is of 
interest to investigate how the measurements of three variables out of 
$\alpha,~\beta,~\gamma~,R_b$ and $R_t$ will determine the allowed values for 
the remaining two variables. We illustrate this in~Fig.~\ref{fig:stra} 
assuming a relative error of $10\%$ for the constraints used in each plot. 
While this figure is self explanatory a striking feature consistent with 
the hierarchical structure in (\ref{ranking1}) can be observed. While the
measurements of $(\alpha,R_t,R_b)$ and $(\alpha,\beta,R_t)$ as seen in 
the first two plots do not appreciably constrain the parameters of the two 
leading strategies $(\beta,\gamma)$ and $(R_b,\gamma)$, respectively, 
the opposite is true in the last two plots. There the measurements of 
$(R_b,\gamma,\alpha)$ and $(\beta,\gamma,\alpha)$ give strong constraints 
in the $(\beta,R_t)$ and $(R_b,R_t)$ plane, respectively.  
The last two plots illustrate also clearly that 
measuring only $\alpha$ and $\gamma$ does not provide a strong constraint on 
the unitarity triangle.

\subsection{Results for the presently available strategies}
\label{sec:present}

At present the concrete results can be obtained only for the strategies
($R_t,\beta$), ($R_b,R_t$) and ($R_b,\beta$)  as no direct measurements of
$\gamma$ and $\alpha$ are available. 

\begin{figure}[htbp] 
\begin{center}
{\epsfig{figure=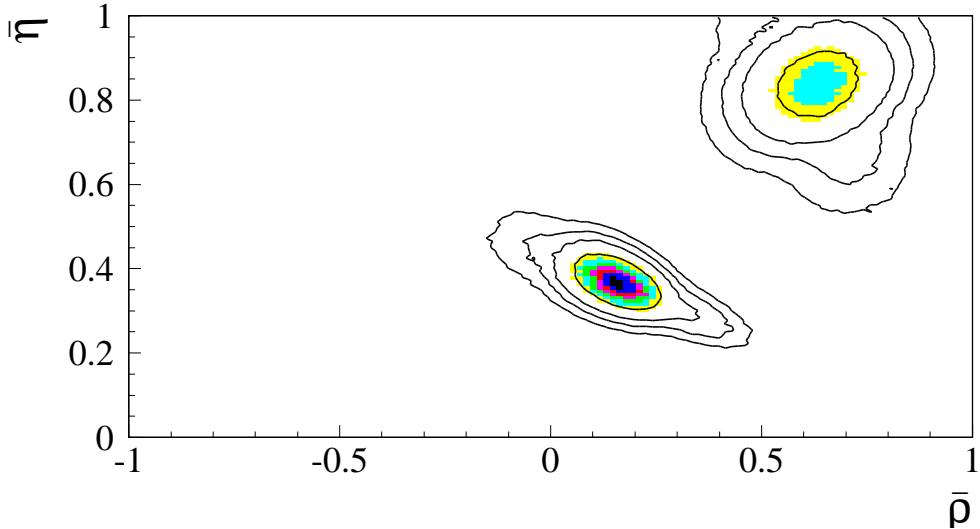,height=8cm}}
\caption[]{\it The plot shows the presently allowed regions 
{(68$\%$,95$\%$,99$\%$ and 99.9$\%$)} in the 
($\bar\rho,\bar\eta$) plane using the $(R_t,\beta)$ strategy: the 
direct measurement of sin 2$\beta$ and $R_t$ from $\Delta M_d$ 
and ${\Delta M_d}/{\Delta M_s}$.}
\label{fig:rtbeta_real}
\end{center}
\end{figure}

\begin{figure}[htbp]
\begin{center}
{\epsfig{figure=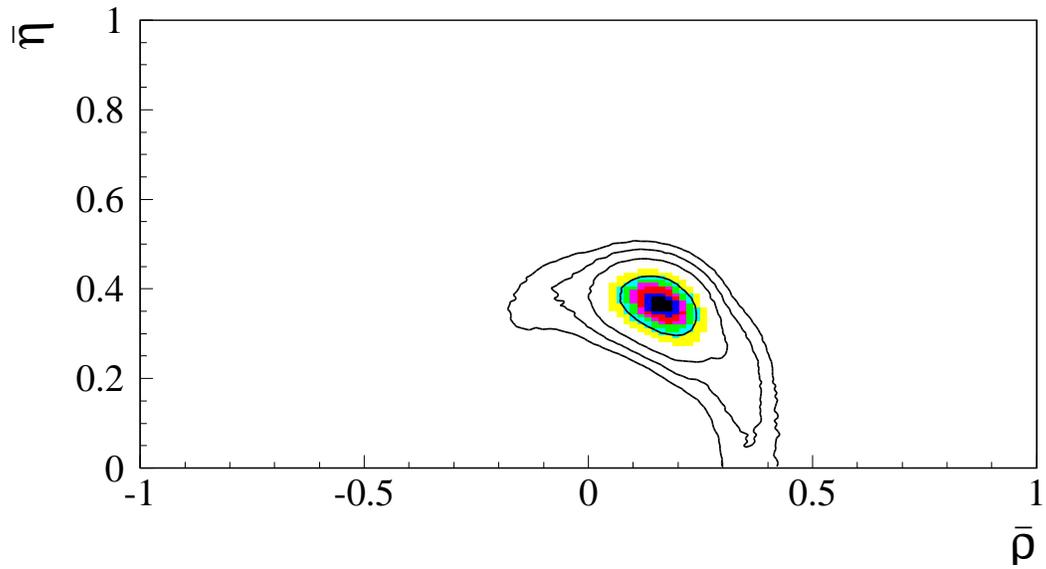,height=8.5cm}}
\caption{\it The plot shows the allowed regions 
{(68$\%$,95$\%$,99$\%$ and 99.9$\%$)} in the 
($\bar \rho,\bar \eta$) plane using the $(R_t,R_b)$ strategy: 
$R_t$ from $\Delta M_d$ and ${\Delta M_d}/{\Delta M_s}$ and $R_b$ from 
$\vub$.}
\label{fig:rtrb_real}
\end{center}
\end{figure}

\begin{figure}[htbp]
\begin{center}
{\epsfig{figure=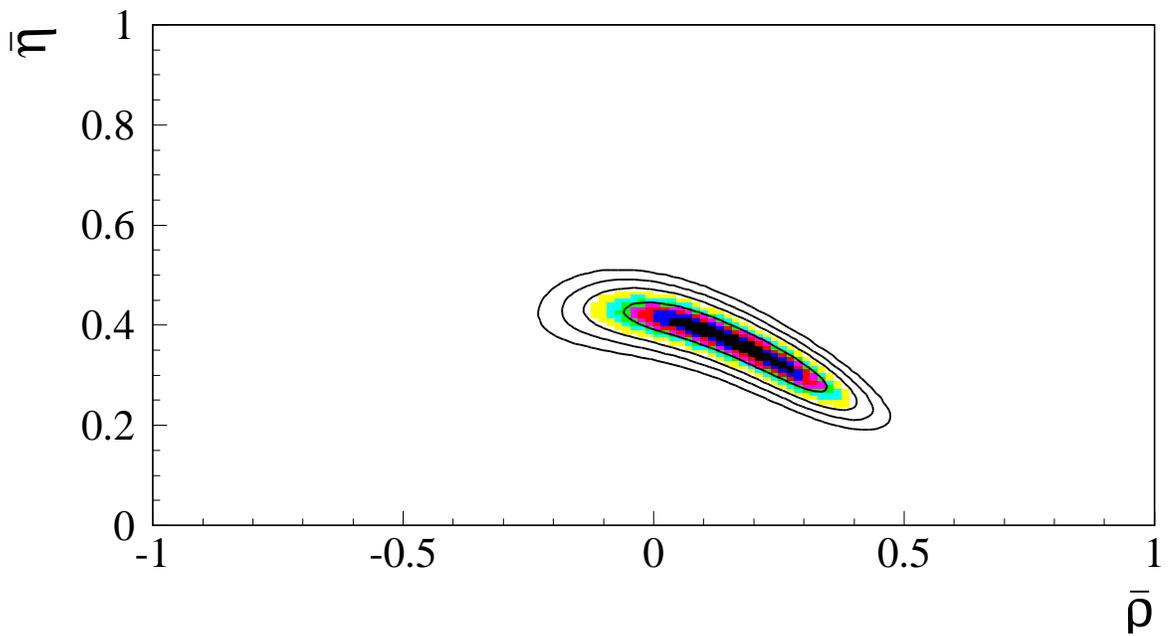,height=9.5cm}}
\caption{\it The plot shows the allowed regions {(68$\%$,95$\%$,99$\%$ 
and 99.9$\%$)} in the 
($\bar \rho,\bar \eta$) plane using the $(R_b,\beta)$ strategy: 
direct measurement of sin 2$\beta$ and $R_b$ from 
$\left|{V_{ub}}/{V_{cb}}\right|$.}
\label{fig:rbbeta_real}
\end{center}
\end{figure}

The results for $\bar \rho$ and $\bar \eta$ for the three strategies in
question are presented in Table \ref{results} and in Figs. 
\ref{fig:rtbeta_real}, \ref{fig:rtrb_real} and \ref{fig:rbbeta_real}. 
To obtain these results we have used the direct measurement of 
sin 2$\beta$~[\ref{ref:sin2b}], $R_t$ as extracted from
$\Delta M_d$ and ${\Delta M_d}/{\Delta M_s}$ by means of the formulae 
in~[\ref{Erice},\ref{ref:haricot}] and $R_b$ as extracted from~$\vub$.

\begin{table}[htbp] 
\begin{center}
\begin{tabular}{|c|c|c|}
\hline
  Strategy       &        $\bar \rho$            &      $\bar \eta$          \\ \hline
  ($R_t,\beta$)  &    0.157 $^{+0.056}_{-0.054}$ &    0.367 $^{+0.036}_{-0.034}$ \\
                 &       [0.047-0.276]           &        [0.298-0.439]          \\
  ($R_t,R_b$)    &    0.161$^{+0.055}_{-0.057}$  &   0.361 $^{+0.041}_{-0.045}$  \\  
                 &       [0.043-0.288]           &        [0.250-0.438]          \\
  ($R_b,\beta$)  &  0.137 $^{+0.135}_{-0.135}$   &   0.373 $^{+0.049}_{-0.063}$  \\   
                 &       [-0.095-0.357]          &        [0.259-0.456]          \\
\hline
\end{tabular}
\caption[]{ \it Results for $\bar \rho$ and $\bar \eta$ for the three
indicated strategies using the present knowledge summarized 
in Table \ref{tab:inputs} in Chapter \ref{chap:fits}. 
For the strategy ($R_t,\beta$), the solution compatible with 
the region selected by the $R_b$ constraint has been considered. In squared 
brackets the 95$\%$ probability regions are also given.}
\label{results}
\end{center}
\end{table}
\vspace{-12mm}
%

%

\begin{figure}[htbp]
\begin{center}
{\epsfig{figure=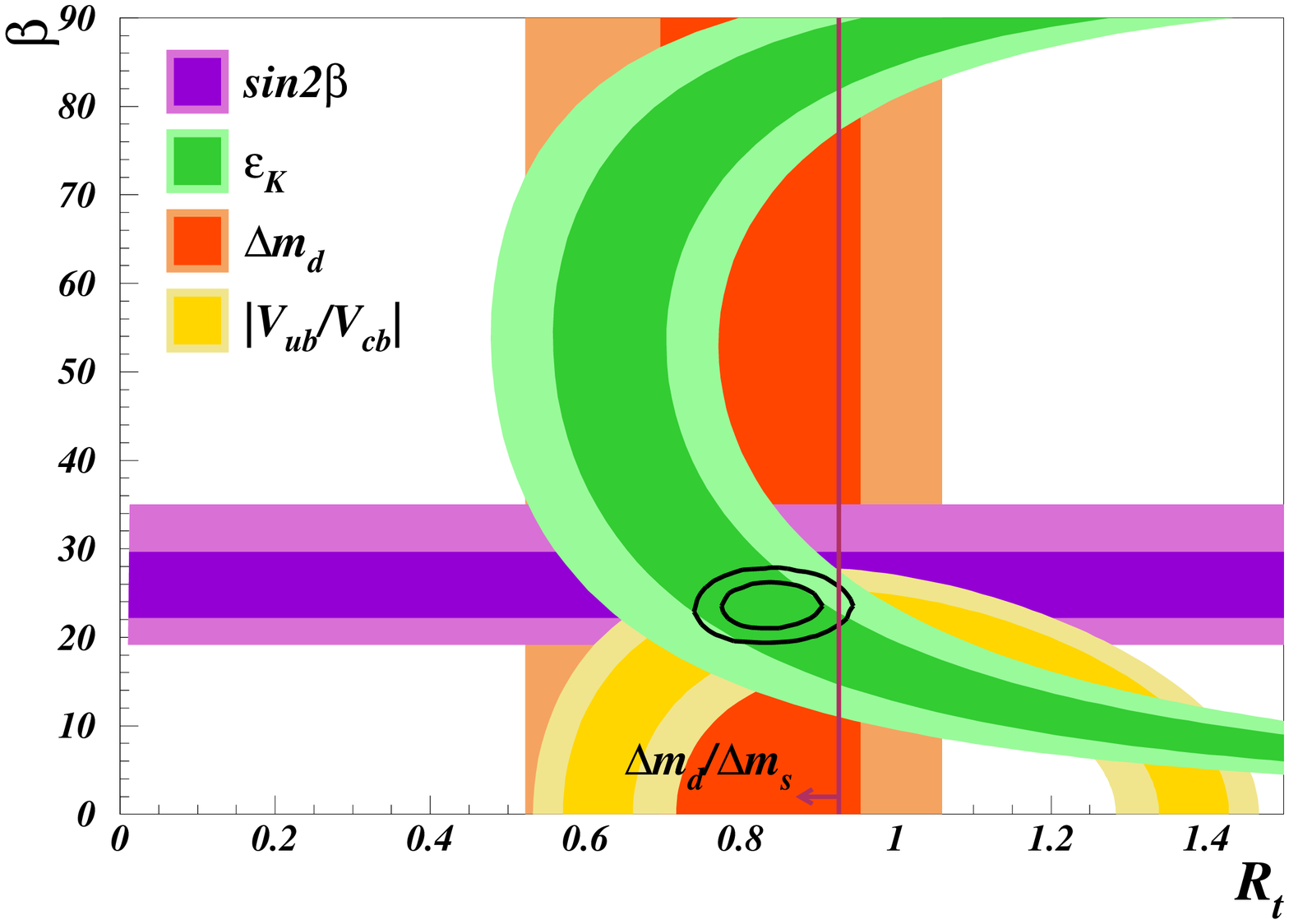,height=9.5cm}}
\caption[]{\it  {The plot shows the allowed regions {(68$\%$ and 95$\%$)} 
in the ($R_t,\beta$) plane. Different constraints are also shown.}}
\label{fig:rtbeta}
\end{center}
\end{figure}

The experimental and theoretical inputs are summarized in Chapter 4
It should be emphasized that these three presently available strategies are 
the weakest among the leading strategies listed in (\ref{ranking1}). 
Among them $(R_t,\beta)$ and $(R_t,R_b)$ appear to be superior to 
$(R_b,\beta)$ at present. We expect that once $\Delta M_s$ has been measured 
and the error on $\sin 2\beta$ reduced, the strategy $(R_t,\beta)$ will be 
leading among these three. Therefore in Fig.~\ref{fig:rtbeta} we show how 
the presently available 
constraints look like in the $(R_t,\beta)$ plane.

\subsection{Summary}
\label{sec:conclusions}
We have presented a numerical analysis of the unitarity triangle 
from a different point of view, that emphasizes the role of different 
strategies in the precise determination of the unitarity triangle parameters. 
While we have found that the pairs $(\gamma,\beta)$,
$(\gamma,R_b)$ and $(\gamma,\bar\eta)$ are most efficient in determining 
$(\bar\varrho,\bar\eta)$, we expect that the pair $(R_t,\beta)$ 
will play the leading role in the UT fits in the coming years, in particular,
when $\Delta M_s$ will be measured and the theoretical error on $\xi$ 
decreased. For this reason we have proposed 
to plot available constraints on the CKM matrix in the $(R_t,\beta)$ plane. 

It will be interesting to compare in the future the allowed ranges 
for $(\bar\varrho,\bar\eta)$ resulting from different strategies in order 
to see whether they are compatible with each other. Any discrepancies 
will signal the physics beyond the SM. We expect that the strategies 
involving $\gamma$ will play a very important role in this comparison.

For the fundamental set of parameters in the quark flavour physics given in 
(\ref{I2}) we find within the~SM
\begin{eqnarray}
\vus=0.2240\pm 0.0036,~\vcb=(41.3\pm 0.7) 10^{-3},~  
R_t=0.91\pm0.05,~\beta=(22.4 \pm 1.4)^\circ , \nonumber 
\label{eq:lastlast}
\end{eqnarray}
where the errors represent one standard deviations and the result  
for $\beta$ corresponds to $\sin 2\beta=0.705\pm 0.035$. 

A complete analysis of the usefulness of a given 
strategy should also include the discussion of its experimental feasibility 
and theoretical cleanness. Extensive studies of these two issues can be found
in~[\ref{BABAR}--\ref{Erice}] 
and in these proceedings. Again among various strategies, 
the $(R_t,\beta)$ strategy is 
exceptional as the theoretical uncertainties in the determination of these 
two variables are small and the corresponding experiments are presently 
feasible. In the long run, 
when $\gamma$ will be cleanly measured in ${\rm B}_d\to \rm D\pi$ and 
${\rm B}_s\to {\rm D}_s{\rm K}$ decays and 
constrained through other decays as reviewed in the following sections,
 we expect that 
the strategy $(\gamma,\beta)$ will take over the leading role.
Eventually the independent direct determinations of the five variables
in question will be crucial for the tests of the SM and its extensions.

\section{Radiative rare B decays}
{\it A. Ali and M. Misiak}

\begin{figure}[h]
\begin{center}
\includegraphics[width=75mm,angle=0]{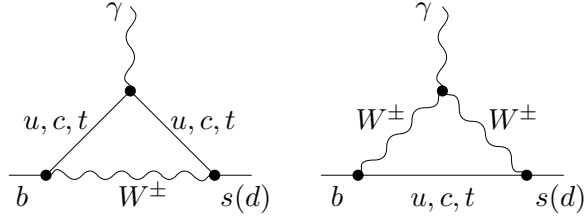}
\end{center}
\vspace*{-33mm}
\hspace*{55mm}  $\gamma$ \hspace{37mm} $\gamma$\\[9mm] 
\hspace*{44mm} $u,c,t$ \hspace{9mm}     $u,c,t$ \hspace{14mm} 
            $W^{\pm}$ \hspace{9mm}   $W^{\pm}$ \hspace{15mm}\\[4mm]
\hspace*{43mm} $b$ \hspace{1cm}  $W^{\pm}$  \hspace{5mm} $s(d)$ \hspace{6mm} 
               $b$ \hspace{7mm}   $u,c,t$   \hspace{6mm} $s(d)$\\[-1cm] 
\begin{center}
\caption{\it  Leading-order Feynman diagrams for ~$b \to s(d) \gamma$~ in the SM.}
\label{fig:SMdiag}
\end{center}
\vspace{-1cm}
\end{figure}

\vspace{2mm}
\noindent
The transitions $b \to s(d)\gamma$ and $b \to s(d) \ell^+ \ell^-$ receive
sizable contributions from loops involving the top quark
(Fig.~\ref{fig:SMdiag}).  Their dependence on $V_{ts}$ and $V_{td}$
may be used to test unitarity of the CKM matrix and to overconstrain
the Wolfenstein parameters $\bar{\rho}$ and $\bar{\eta}$. The
considered transitions manifest themselves in exclusive
$\overline{\rm B}$-meson decays like $\overline{\rm B} \to {\rm K}^\star \gamma$, $\overline{\rm B}
\to {\rm K}^* \ell^+ \ell^-$,~ $\overline{\rm B} \to \rho \gamma$~ and ~$\overline{\rm B}
\to \rho \ell^+ \ell^-$.~ The corresponding inclusive decays 
$\overline{\rm B} \to X_{s(d)} \gamma$ 
and $\overline{\rm B} \to X_{s(d)} \ell^+ \ell^-$ are experimentally
more challenging, but the theoretical predictions are significantly
more accurate,  thanks to the use of OPE and 
HQET. The exclusive processes remain interesting due to possible new
physics effects in observables other than just the total branching ratios
(photon polarization, isospin- and CP-asymmetries), as well as due to
information they provide on non-perturbative form-factors. This
information is particularly required in analyzing exclusive modes
generated by the $b \to d \gamma$ transition, in which case there is
little hope for an inclusive measurement.

In this section we discuss briefly the generic features of the CKM phenomenology in
the considered rare B-decays. The transitions $b \to s \gamma$ and
$b \to s \ell^+ \ell^-$ involve the CKM matrix elements from the
second and third column of this matrix, with the unitarity constraint
taking the form $\sum_{u,c,t} \lambda_i =0$, with
$\lambda_i=V_{ib}V^*_{is}$. This equation yields a unitarity triangle
which is highly squashed, as one of the sides of this triangle
$\lambda_u=V_{ub} V_{us}^* \simeq A\lambda^4 (\bar \rho -i \bar \eta)$
is doubly Cabibbo suppressed, compared to the other two sides
$\lambda_c
\simeq -\lambda_t =A\lambda^2 +...$. Hence, the transitions $b \to s \gamma$ and
$b \to s \ell^+ \ell^-$  are not expected to yield useful information on
the parameters $\bar \rho$ and $\bar \eta$, which define the apex of the
unitarity triangle of current interest (see Chapt.~1).  The test of unitarity for the $b 
\to s$ transitions in rare B-decays lies in checking the relation $\lambda_t
\simeq -\lambda_c$, which holds up to corrections of order $\lambda^2$. 

The impact of the decays $b \to d \gamma$ and $b \to d \ell^+ \ell^-$
on the CKM phenomenology is, however, quite different.  These
transitions involve the CKM matrix elements in the first and third
column, with the unitarity constraints taking the form $\sum_{u,c,t}
\xi_i =0$, with $\xi_i=V_{ib}V_{id}^*$. Now, all three matrix elements
are of order $\lambda^3$, with
$\xi_u \simeq A\lambda^3 (\bar \rho - i \bar \eta)$, 
$\;\xi_c \simeq -A\lambda^3$, 
and $\xi_t \simeq A\lambda^3(1-\bar \rho - i \bar \eta)$.  
This equation leads to the same unitarity triangle as studied through
the constraints $V_{ub}/V_{cb}$, $\Delta M_{B_d}$ (or $\Delta
M_{B_d}/\Delta M_{B_s}$). \linebreak Hence, the transitions $b \to d
\gamma$ and $b \to d \ell^+ \ell^-$ lead to complementary constraints on the CKM parameters
$\bar{\rho}$ and $\bar{\eta}$, as illustrated in the following.
Thus, the role of rare B-decays is that they provide complementary
constraints on the CKM matrix elements, hence test the CKM unitarity,
but they also constrain extensions of the Standard Model, and by that
token can act as harbinger of new physics.

A theoretical framework for analyzing the $b \to s \gamma$ transition
is set by the effective interaction Hamiltonian
\be \label{Heff}
{\cal H_{\rm eff}} =  -\f{4 G_F}{\sqrt{2}} V_{ts}^* V_{tb} \sum_{i=1}^8 C_i(\mu) Q_i.
\ee
%
The generic structure of the operators $Q_i$ is as follows:
\be \label{ops}
Q_i = \left\{ \begin{array}{ll}
(\bar{s} \Gamma_i c)(\bar{c} \Gamma'_{\underline{i}} b), & i=1,2, \\[3mm]
(\bar{s} \Gamma_i b) \sum_q (\bar{q} \Gamma'_{\underline{i}} q), 
& i=3,4,5,6,~~~~~~~~~ (q=u,d,s,c,b) \\[3mm]
\f{e m_b}{16 \pi^2} \bar{s}_L \sigma^{\mu \nu} b_R F_{\mu \nu}, & i=7, \\[3mm]
\f{g_s m_b}{16 \pi^2} \bar{s}_L \sigma^{\mu \nu} T^a b_R G^a_{\mu \nu},~~
& i=8. \end{array} \right.  
\ee
Here, $\Gamma_i$ and $\Gamma'_i$ denote various combinations of the
colour and Dirac matrices. Everything that is not important for $b \to
s \gamma$ at the leading order in ~$\alpha_{\rm em}$, ~$m_b/m_W$,
~$m_s/m_b$ ~and ~$V_{ub}/V_{cb}$ has been neglected in Eq.~(\ref{Heff}).

Perturbative calculations (see Ref.~[\ref{Buras:2002er}] and refs.\
therein) are used to find the Wilson coefficients in the
$\overline{\rm MS}$ scheme, at the renormalization scale $\mu_b \sim
m_b$
\be \label{cmb}
C_i(\mu_b) = C_i^{(0)}(\mu_b)+ \f{\alpha_s(\mu_b)}{4\pi} C_i^{(1)}(\mu_b)
+ \left( \f{\alpha_s(\mu_b)}{4\pi} \right)^2 C_i^{(2)}(\mu_b) + \ldots.
\ee
Here, $C_i^{(n)}(\mu_b)$ depend on $\alpha_s$ only via the ratio $\eta
\equiv \alpha_s(\mu_0)/\alpha_s(\mu_b)$, where $\mu_0 \sim m_W$. In
the Leading Order (LO) calculations, everything but $C_i^{(0)}(\mu_b)$
is neglected in Eq.~(\ref{cmb}). At the Next-to-Leading Order (NLO),
one takes $C_i^{(1)}(\mu_b)$ into account. The Wilson coefficients
contain information on the short-distance QCD effects due to hard
gluon exchanges between the quark lines of the leading one-loop
electroweak diagrams (Fig.~\ref{fig:SMdiag}).  Such effects enhance
the perturbative branching ratio ${\cal B}(b \to s \gamma)$ by roughly
a factor of three~[\ref{Bertolini:1986th}].

The same formalism applies to $b \to d \gamma$, too. The corresponding
operators $Q_i$ are obtained by replacing $\bar{s} \to \bar{d}$ in
Eq.~(\ref{ops}), and by including the $u$-quark analogues of
$Q_{1,2}$. The latter operators are no longer CKM-suppressed.  The
matching conditions $C_i(\mu_0)$ and the solutions of the RG equations,
yielding $C_i(\mu_b)$, coincide with those needed for the process $b
\to s \gamma$.

\subsection{Inclusive ~$\overline{\rm B} \to X_{s(d)} \gamma$~ decay}

The inclusive branching ratio ${\cal B}(\overline{\rm B} \to X_s \gamma)$ was
measured for the first time by CLEO in 1995~[\ref{Alam:1995aw}].
The present world averages
\bea
{\cal B}(\overline{\rm B} \to X_s \gamma~~ (E_\gamma > 1.6\;{\rm GeV})) &=&
\left( 3.28 \; {}^{+0.41}_{-0.36} \right) \times 10^{-4}, \label{av_1.6}
\\
{\cal B}(\overline{\rm B} \to X_s \gamma~~ (E_\gamma > ~~~{\textstyle\f{1}{20}}m_b
~~))  
 &=&
\left( 3.40 \; {}^{+0.42}_{-0.37} \right) \times 10^{-4}  \label{av_mb20}
\eea
are found from the following four measurements
\bea
{\cal B}(\overline{\rm B} \to X_s \gamma~ (E_\gamma > {\textstyle\f{1}{20}}m_b))
&=&
\left[ 3.88 \pm 0.36_{\rm stat} \pm 0.37_{\rm sys}~
\left({}^{+0.43}_{-0.23}\right)_{\rm theory} \right] \times 10^{-4},
\hspace{3mm} \mbox{(BABAR~[\ref{Aubert:2002pd}]),} \nonumber\\
{\cal B}(\overline{\rm B} \to X_s \gamma~ (E_\gamma > {\textstyle\f{1}{20}}m_b))
&=&
\left[ 3.21 \pm 0.43_{\rm stat} \pm 0.27_{\rm sys}~
\left({}^{+0.18}_{-0.10}\right)_{\rm theory} \right] \times 10^{-4},
\hspace{3mm} \mbox{(CLEO~[\ref{Chen:2001fj}]),} \nonumber\\
{\cal B}(\overline{\rm B} \to X_s \gamma~ (E_\gamma > {\textstyle\f{1}{20}}m_b))
&=&
\left[ 3.36 \pm 0.53_{\rm stat} \pm 0.42_{\rm sys}~
\left({}^{+0.50}_{-0.54}\right)_{\rm theory} \right] \times 10^{-4},
\hspace{3mm} \mbox{(BELLE~[\ref{Abe:2001hk}]),} \nonumber\\[2mm]
{\cal B}( b \to s \gamma) &=&
(3.11 \pm 0.80_{\rm stat} \pm 0.72_{\rm sys} ) \times 10^{-4},
\hspace{25mm} \mbox{(ALEPH~[\ref{Barate:1998vz}]),} \nonumber
\eea
in which full correlation of the ``theory'' errors has been assumed.
The averages (\ref{av_1.6}) and (\ref{av_mb20}) are perfectly 
consistent with the SM predictions~[\ref{Gambino:2001ew},\ref{Buras:2002tp}]
\bea \label{thSM_1.6}
{\cal B}(\overline{\rm B} \to X_s \gamma~~ (E_\gamma > 1.6\;{\rm GeV}))_{\rm SM}
&=& (3.57 \pm 0.30) \times 10^{-4},\\
{\cal B}(\overline{\rm B} \to X_s \gamma~~ (E_\gamma > ~~~{\textstyle\f{1}{20}}m_b~~))_{\rm SM}
&=& 3.70 \times 10^{-4} \label{thSM_mb20}.
\eea
By convention, contributions to ${\overline {\rm B}} \to X_s \gamma$ from the
intermediate real $\psi$ and $\psi^\prime$ are treated as background,
while all the continuum $c\bar{c}$ states are included assuming
quark-hadron duality. Non-continuum states other than $\psi$ and
$\psi^\prime$ have negligible effect.

When the theoretical result (\ref{thSM_1.6}) is reevaluated without
use of the CKM unitarity in the dominant contributions
(i.e. everywhere except for three small $(<2.5\%)$ corrections),
comparison with the experiment (\ref{av_1.6}) leads to the following
constraint on the CKM matrix elements
\be \label{bsgamma.constraint}
\left|\; 1.69 \; \lambda_u ~+~ 1.60 \; \lambda_c ~+~ 0.60 \; \lambda_t \; \right| 
~=~ (\; 0.94 ~\pm~ 0.07 \;) \; |V_{cb}|.
\ee
After using the numerical values of $\lambda_c \simeq |V_{cb}| = (41.0
\pm 2.1) \times 10^{-3}$ and $\lambda_u$ from the PDG~[\ref{Hagiwara:pw}],
this equation yields $\lambda_t \simeq -47 \times 10^{-3}$ with an
error of around 17\%. This is consistent with the unitarity relation
$\lambda_c \simeq -\lambda_t$. This relation, however, holds in the SM
with much better accuracy than what has just been derived from
Eq.~(\ref{bsgamma.constraint}).  On the other hand, if the SM with 3
generations is not valid, Eq.~(\ref{bsgamma.constraint}) is not valid
either.

Contrary to ${\cal B}(\overline{\rm B} \to X_s \gamma)$, the branching ratio
${\cal B}(\overline{\rm B} \to X_d \gamma)$, if measured, would provide us with
useful constraints on the Wolfenstein parameters 
$\bar \rho$ and $\bar \eta$. 
After using the CKM unitarity, it can be written as 
\begin{eqnarray}
\label{brxd}
{\cal B}({\overline{\rm  B}} \to X_d \gamma)
=\f{|\xi_t|^2}{|V_{cb}|^2}D_{t}+
\f{|\xi_u|^2}{|V_{cb}|^2}D_{u}+
\f{Re(\xi^{*}_t\xi_u)}{|V_{cb}|^2}D_{r}+
\f{Im(\xi^{*}_t\xi_u)}{|V_{cb}|^2}D_{i} \, .
\end{eqnarray}
The factors $\xi_i$ have been defined earlier.  The quantities $D_a$
$(a=t,u,r,i$), which depend on various input parameters such as $m_t$,
$m_b$, $m_c$, $\mu_b$ and $\as$, are given in Ref.~[\ref{Ali:1998rr}].
Typical values of these quantities (in units of $\lambda^4$) are: $
D_t=0.154, D_u=0.012, D_r=-0.028$, and $D_i=0.042$, corresponding to
the scale $\mu= 5$ GeV, and the pole quark mass ratio $m_c/m_b=0.29$.
The charge-conjugate averaged branching ratio
$\langle {\cal B}(B \to X_d \gamma) \rangle$ 
is obtained by discarding the last term on the right hand side of
Eq. (\ref{brxd}).

It is convenient to consider the ratio
\bea
\f{\langle {\cal B}(B \to X_d \gamma) \rangle}
  {\langle {\cal B}(B \to X_s \gamma) \rangle}
&=& 
\f{|\xi_t|^2}{|\lambda_t|^2}
+ \f{D_u}{D_t} \, \f{|\xi_u|^2}{|\lambda_t|^2}
+ \f{D_r}{D_t} \, \f{Re(\xi_t^* \xi_u)}{|\lambda_t|^2} \nonumber \\[2mm]
&=& \lambda^2  \,
\left[(1-\bar{\rho})^2 + \bar{\eta}^2 + \f{D_u}{D_t} (\bar{\rho}^2 +
\bar{\eta}^2) + \f{D_r}{D_t} (\bar{\rho}(1 - \bar{\rho})
 - \bar{\eta}^2)\right] + O(\lambda^4)\, \nonumber \\[2mm]
&\simeq & 0.036 ~~~[\mbox{for}~(\bar \rho ,\bar \eta)=(0.22,0.35)]~.
\label{bdgamckm}
\eea
The above result together with Eq.~(\ref{thSM_mb20}) implies $\langle
{\cal B}({\rm B} \to X_d \gamma) \rangle \simeq 1.3 \times 10^{-5}$ in the
SM. Thus, with $O(10^8)$ ${\rm B}\overline{\rm B}$ events already collected at the B
factories, $O(10^3)$ $b \to d \gamma$ decays are already
produced. However, extracting them from the background remains a
non-trivial issue.


Apart from the total branching ratios, the inclusive decays 
$\overline{\rm B} \to X_{s(d)} \gamma$ 
provide us with other observables that might be useful for the CKM
phenomenology. First, as discussed in  Chapt.~\ref{chap:III},
 the $\overline{\rm B} \to X_s \gamma$ photon spectrum is used to
extract the HQET parameters that are crucial for the determination of
$V_{ub}$ and $\vcb$.  Second, CP-asymmetries contain information on the CKM
phase. These asymmetries can be either direct (i.e. occur in the decay
amplitudes) or induced by the ${\rm B}\overline{\rm  B}$ mixing.

The mixing-induced CP-asymmetries in $\overline{\rm B} \to X_{s(d)} \gamma$ are
very small (${\cal O}(m_{s(d)}/m_b)$) in the SM, so long as the photon
polarizations are summed over.  It follows from the particular
structure of the dominant operator $Q_7$ in Eq.~(\ref{ops}), which
implies that photons produced in the decays of B and $\overline{\rm B}$ have
opposite {\em circular} polarizations. Thus, in the absence of new
physics, observation of the mixing-induced CP-violation would require
selecting particular {\em linear} photon polarization with the help of
matter-induced photon conversion into $e^+e^-$ pairs~[\ref{GP00}].  

The SM predictions for the direct CP-asymmetries read
\bea
\hspace*{-8mm} {\cal A}_{\rm CP} ({\rm B} \to X_s \gamma) 
\equiv \f{ \Gamma(\overline{\rm B} \to X_s \gamma) - \Gamma({\rm B} \to X_{\overline s}\, \gamma)}
         { \Gamma(\overline{\rm B} \to X_s \gamma) + \Gamma({\rm B} \to X_{\overline s}\, \gamma)}
\simeq \f{{\rm Im}(\lambda_t^*\lambda_u)D_i}{|\lambda_t|^2 \, D_t}  
\!\! &\simeq&  0.27 \, \lambda^2 \bar \eta ~\sim~ 0.5\%, \\[2mm]
\hspace*{-8mm} {\cal A}_{\rm CP} (\rm B \to X_d \gamma) 
\equiv \f{ \Gamma(\overline{\rm B} \to X_d \gamma) - \Gamma({\rm B} \to X_{\overline d}\, \gamma)}
         { \Gamma(\overline{\rm B} \to X_d \gamma) + \Gamma({\rm B} \to X_{\overline d}\, \gamma)}
\simeq \f{{\rm Im}(\xi_t^*\xi_u)D_i}{|\xi_t|^2 \, D_t}  
\!\! &\simeq& \!\! \f{-0.27 \, \bar\eta}{(1\!-\! \bar\rho)^2+\bar\eta^2}
\sim -13\%,
\eea
where $\bar \rho = 0.22$ and $\bar \eta = 0.35$ have been used in the
numerical estimates. As stressed in Ref.~[\ref{Ali:1998rr}], there is
considerable scale uncertainty in the above predictions, which would
require a NLO calculation of $D_i$ to be brought under theoretical
control. The smallness of ${\cal A}_{\rm CP} ({\rm B} \to X_s \gamma)$ is
caused by three suppression factors: $\lambda_u/\lambda_t$,
$\alpha_s/\pi$ and $m_c^2/m_b^2$. This SM prediction is consistent
with the CLEO bound \linebreak $-0.27 < {\cal A}_{\rm CP} ({\rm B} \to X_s \gamma) <
+0.10$~ at 95\% C.L.~[\ref{CLEO_CP}].

No experimental limit has been announced so far on either the
branching ratio ${\cal B}({\overline {\rm B}} \to X_d \gamma)$ or the CP
asymmetry ${\cal A}_{\rm CP} ({\rm B} \to X_d \gamma)$.  While
experimentally challenging, the measurement of these quantities might
ultimately be feasible at the B-factories
which would provide valuable and complementary constraints on the CKM
parameters.

\subsection{Exclusive radiative B decays}
The effective Hamiltonian sandwiched between
the~B-meson  and a single meson state (say, ${\rm K}^*$ or $\rho$
in the transitions ${\rm B} \to ({\rm K}^\star, \rho) \gamma$) can
be expressed in terms of
matrix elements of bilinear quark currents inducing heavy-light
transitions. These matrix elements are dominated by strong
interactions at small momentum transfer and cannot be calculated
perturbatively. They have to be obtained from a non-perturbative method,
such as the lattice-QCD and the QCD sum rule approach. As the inclusive 
branching ratio ${\cal B}({\rm B} \to X_s \gamma)$ in the SM is in striking 
agreement with  data, the role of the branching ratio ${\cal B}({\rm B} \to {\rm K}^* 
\gamma)$ is that it will determine the form factor governing the 
electromagnetic penguin transition, $T_1^{{\rm K}^*}(0)$. 

 To get a firmer theoretical prediction on the decay rate, one  
has to include the perturbative  QCD radiative corrections
arising from the vertex renormalization and the hard spectator
interactions. To incorporate  both types of QCD corrections, it is
helpful to use a factorization  Ansatz for the heavy-light transitions
at large recoil and at leading order in the inverse heavy meson mass,
introduced in Ref.~[\ref{Beneke:1999br_bis}]. Exemplified here by the ${\rm B} \to V
\gamma^*$ transition, a typical amplitude $f_k (q^2)$ can be written in
the form
\begin{equation}
f_k (q^2) = C_{\perp k} \xi_\perp(q^2) +  C_{\| k} \xi_\|(q^2) +
\Phi_B \otimes T_k (q^2) \otimes \Phi_V ,
\label{eq:fact-formula}
\end{equation}
where $\xi_\perp(q^2)$ and~$\xi_\|(q^2)$ are
the two independent form factors in these decays remaining in
the heavy quark and large energy limit; $T_k(q^2)$ is a hard-scattering kernel
calculated to
$O (\alpha_s)$; $\Phi_B$ and~$\Phi_V$ are the light-cone distribution
amplitudes of the~B- and~vector-meson, respectively, the symbol 
$\otimes$ 
denotes convolution with~$T_k$, and 
$C_k = 1 + O (\alpha_s)$ are the hard vertex renormalization
coefficients. In a number
of papers~[\ref{Beneke:2001at}--\ref{bdgAP}], 
the factorization
Ansatz of Eq.~(\ref{eq:fact-formula}) is shown to hold in
$O(\alpha_s)$, leading to the explicit $O(\alpha_s)$ 
corrections to the amplitudes ${\rm B} \to V \gamma$ and ${\rm B} \to V \ell^+ 
\ell^-$.

\begin{table}[h]
\begin{center}
\begin{tabular}{|l|l|l|}
\hline
Experiment & ${\cal B}_{\rm exp} ({\rm B}^0(\overline {{\rm B}}^0) \to {\rm K}^{*0} (\overline {\rm K}^{*0})
+ \gamma)$ &
${\cal B}_{\rm exp} ({\rm B}^\pm \to {\rm K}^{*\pm} + \gamma)$
\\ \hline
CLEO \protect[\ref{Coan:1999kh}] &
$(4.55^{+ 0.72}_{-0.68} \pm 0.34) \times 10^{-5}$ &
$(3.76^{+ 0.89}_{-0.83} \pm 0.28) \times 10^{-5}$ \\
BELLE \protect[\ref{NishidaS:2002}] &
$(3.91 \pm 0.23 \pm 0.25) \times 10^{-5}$ &
$(4.21 \pm 0.35 \pm 0.31) \times 10^{-5}$ \\
BABAR \protect[\ref{Aubert:2001}] &
$(4.23 \pm 0.40 \pm 0.22) \times 10^{-5}$ &
$(3.83 \pm 0.62 \pm 0.22) \times 10^{-5}$ \\
\hline 
\end{tabular}
\end{center}
\caption{\it Experimental branching ratios for the decays
 ${\rm B}^0(\overline{\rm B}^0) \to {\rm K}^{*0}(\overline {\rm K}^{*0}) \gamma$ and 
 ${\rm B}^\pm \to {\rm K}^{*\pm} \gamma$.}
\label{tab:Br-exp}
\end{table}   

We first discuss the exclusive decay ${\rm B} \to {\rm K}^* \gamma$, for which
data from the CLEO, BABAR, and BELLE measurements are available
and given in Table~\ref{tab:Br-exp} for the charge conjugated
averaged branching ratios. We note that the BELLE data alone has reached a 
statistical accuracy of better than  10\%.

Adding the statistical and systematic errors in quadrature, we get the
following world averages for the branching ratios:
\begin{eqnarray}
{\cal B}({\rm B}^0 \to {\rm K}^{*0} \gamma) &=& (4.08 \pm 0.26) \times 10^{-5} ~,
\nonumber\\
{\cal B}({\rm B}^\pm \to {\rm K}^{\pm} \gamma) &=& (4.05 \pm 0.35) \times 10^{-5} ~.
\label{bkstarbrs}
\end{eqnarray}
The two branching ratios are completely consistent with each other,
ruling out any significant isospin breaking in the respective decay 
widths, which is not expected in the SM~[\ref{Kagan:2001zk}] but
anticipated in some beyond-the-SM scenarios. Likewise, the
CP asymmetry in ${\rm B} \to {\rm K}^* \gamma$ decays, which in the SM is
expected to be of the same order of magnitude as for the inclusive decay,
namely ${\cal A}_{\rm CP} ({\rm B} \to {\rm K}^* \gamma) \leq 1\%$, is completely 
consistent with the present experimental bounds, the most stringent of 
which is posted by the BELLE collaboration~[\ref{NishidaS:2002}]:
${\cal A}_{\rm CP} ({\rm B} \to {\rm K}^* \gamma)=-0.022 \pm 0.048 \pm 0.017$.
In view of this, we shall concentrate in the following on the branching 
ratios in $\rm B \to {\rm K}^* \gamma$ decays to determine the form factors. 

Ignoring the isospin differences in the decay widths of
$\rm B \to {\rm K}^* \gamma$ decays, the branching ratios for ${\rm B}^\pm \to {\rm K}^{*\pm}
\gamma$ and $\rm B^0(\overline {\rm B}^{0}) \to {\rm K}^{*0}(\overline {\rm K}^{*0}) \gamma$ 
can be expressed as:
\begin{eqnarray}
{\cal B}_{\rm th} ({\rm B \to {\rm K}^{*}} \gamma) & = &
\tau_B \, \Gamma_{\rm th} ({\rm B \to {\rm K}^*} \gamma)
\label{eq:DW(B-Kgam)} \\
&=&
\tau_B \,\f{G_F^2 \alpha |V_{tb} V_{ts}^*|^2}{32 \pi^4} \,
m_{b, {\rm pole}}^2 \, M^3 \, \left [ \xi_\perp^{({\rm K}^*)} \right ]^2
\left ( 1 - \f{m_{{\rm K}^*}^2}{M^2} \right )^3
\left | C^{(0){\rm eff}}_7 +  A^{(1)}(\mu) \right |^2 , \nonumber
\end{eqnarray}
where~$G_F$ is the Fermi coupling constant,
$\alpha = \alpha(0)=1/137$ is the fine-structure constant,
$m_{b, {\rm pole}}$ is the pole $b$-quark mass,
$M$~and $M_{K^*}$ are the B- and ${\rm K}^*$-meson masses,
and~$\tau_B$ is the lifetime of the~${\rm B}^0$- or ${\rm B}^+$-meson.
The quantity $\xi_\perp^{K^*}$ is the soft part of the form factor
$T_1^{K^*}(q^2=0)$ in the ${\rm B} \to {\rm K}^* \gamma$ 
transition, to which the 
symmetries in the large energy limit apply. The two form factors 
$\xi_\perp^{K^*}$ and $T_1^{K^*}(q^2=0)$ are related by perturbative
$(O(\alpha_s))$ and power $(O(\Lambda_{\rm QCD}/m_b))$ 
corrections~[\ref{Beneke:2000wa}]. Thus, one could have equivalently 
expressed the 
$O(\alpha_s)$-corrected branching ratio for 
${\rm B} \to {\rm K}^* \gamma$ in terms of 
the QCD form factor  $T_1^{K^*}(q^2=0)$, and a commensurately modified 
expression for the explicit $O(\alpha_s)$ correction in the above 
equation~[\ref{Bosch:2001gv}]. In any case, the form factor
 $T_1^{K^*}(q^2=0)$ or
$\xi_\perp^{K^*}$ has to be determined by a non-perturbative method.

The function~$ A^{(1)}$ in Eq.~(\ref{eq:DW(B-Kgam)}) can be
decomposed into the following three components:
\begin{equation}
 A^{(1)} (\mu)  =   A_{C_7}^{(1)} (\mu) +
 A_{\rm ver}^{(1)} (\mu) +  A_{\rm sp}^{(1)K^*} (\mu_{\rm sp})~.
\label{eq:A1tb}
\end{equation}  
Here, $ A^{(1)}_{C_7}$ and $ A^{(1)}_{\rm ver}$ are the
$O (\alpha_s)$ (i.e. NLO) corrections due to the Wilson
coefficient~$C_7^{\rm eff}$
and in the $b \to s \gamma$ vertex, respectively, 
and $ A^{(1) K^*}_{\rm sp}$ is the ${\cal O} (\alpha_s)$ hard-spectator
correction to the ${\rm B} \to {\rm K}^* \gamma$ amplitude computed 
in~[\ref{Beneke:2001at}--\ref{bdgAP}].
This formalism leads to the following branching ratio 
for ${\rm B} \to {\rm K}^*
\gamma$ decays:
\begin{equation}
{\cal B}_{\rm th} ({\rm B} \to {\rm K}^* \gamma) \simeq
(7.2 \pm 1.1)\times 10^{-5} \,  
\left ( \f{\tau_B}{1.6~{\rm ps}} \right )
\left ( \f{m_{b,{\rm pole}}}{4.65~{\rm GeV}} \right )^2
\left ( \f{\xi_\perp^{(K^*)}}{0.35} \right )^2~,
\label{eq:Br-Ksgam}
\end{equation}
where the default values of the three input parameters are made explicit,
with the rest of the theoretical uncertainties indicated numerically;
the default value for the form factor $\xi_\perp^{(K^*)} (0)$ 
is based on the light-cone QCD sum rule estimates~[\ref{Ball:1998kk}]. 

The non-perturbative parameter~$\xi_\perp^{(K^*)} (0)$
can now be extracted from the  data on the branching ratios for ${\rm B} 
\to {\rm K}^* \gamma$ decays, given in Eq.~(\ref{bkstarbrs}), leading to the 
current world average $\langle {\cal B} ({\rm B} \to {\rm K}^* \gamma) \rangle =
(4.06 \pm 0.21) \times 10^{-5}$, which then yields  

\begin{equation}
\bar \xi_\perp^{(K^*)} (0) = 0.25 \pm 0.04 , \qquad
\left [ \bar T_1^{(K^*)} (0, \bar m_b) = 0.27 \pm 0.04 \right ] ~,
\label{eq:xi-Ks-average}
\end{equation}
where we have used the $O(\alpha_s)$ relation between the effective theory 
form factor $\xi_\perp^{(K^*)} (0)$ and the full QCD form factor 
$T_1^{(K^*)} (0, \bar m_b)$, worked out in~[\ref{Beneke:2000wa}].
This estimate is significantly smaller than
the corresponding predictions from the QCD sum rules analysis
$T_1^{(K^*)} (0) = 0.38 \pm 0.06$~[\ref{Ali:vd},\ref{Ball:1998kk}] and
from the lattice simulations 
$T_1^{(K^*)} (0) = 0.32^{+0.04}_{-0.02}$~[\ref{DelDebbio:1997kr}].
Clearly, more work is needed to calculate the 
${\rm B} \to {\rm K}^* \gamma$ decay 
form factors precisely. 

As already discussed, inclusive $b \to d \gamma$ transitions are not yet 
available experimentally. This lends great importance to the exclusive 
decays, such as ${\rm B} \to \rho \gamma, \omega \gamma$, to whose discussion we 
now turn. These decays differ from their ${\rm B} \to {\rm K}^* \gamma$ 
counterparts, in that the annihilation contributions are not 
Cabibbo-suppressed. In particular, the isospin-violating ratios
and CP-asymmetries in the decay rates involving the decays ${\rm B}^\pm 
\to \rho^\pm \gamma$ 
and ${\rm B}^0(\overline{{\rm B}^0}) \to \rho^0 \gamma$ are sensitive to
the penguin and annihilation interference in the amplitudes. 

We recall that ignoring the perturbative QCD corrections to the penguin
amplitudes the ratio of the branching ratios for the charged and neutral
B-meson decays in ${\rm B} \to \rho \gamma$ can be written 
as~[\ref{Ali:2000zu},\ref{Grinstein:2000pc}]
\begin{equation}
\f{{\cal B} ({\rm B}^- \to \rho^- \gamma)}{2 {\cal B} ({\rm B}^0 \to \rho^0 \gamma)}
\simeq \left |1 + \epsilon_A {\rm e}^{i \phi_A} \,
\f{V_{ub} V_{ud}^*}{V_{tb} V_{td}^*} \right |^2 ,
\label{eq:BR-ratio}
\end{equation}
where $\epsilon_A {\rm e}^{i \phi_A}$ includes the dominant
$W$-annihilation and possible sub-dominant long-distance contributions.
We shall use the value $\epsilon_A \simeq + 0.30 \pm 0.07$
for the decays ${\rm B}^\pm \to  \rho^\pm 
\gamma$~[\ref{Ali:1995uy},\ref{Khodjamirian:1995uc}],
obtained assuming 
factorization of the annihilation amplitude. The corresponding quantity 
for the decays ${\rm B}^0 \to 
\rho^0 \gamma$ is suppressed due to the electric charge of the spectator 
quark in ${\rm B}^0$ as well as by the unfavourable colour factors. Typical 
estimates for $\epsilon_A$ in ${\rm B}^0 \to \rho^0 \gamma$ put it at 
around 5\%~[\ref{Ali:1995uy},\ref{Khodjamirian:1995uc}]. 
 The strong interaction phase~$\phi_A$ vanishes in 
${\cal O} (\alpha_s)$ in the chiral limit and to 
leading twist~[\ref{Grinstein:2000pc}], giving theoretical credibility to
the factorization-based estimates. Thus, in the QCD factorization
approach the phase $\phi_A$ is expected to be small and one usually sets 
$\phi_A = 0$. Of course, $O(\alpha_s)$  vertex and hard 
spectator corrections generate non-zero strong phases, as discussed later.  
The isospin-violating correction depends on 
the unitarity triangle phase~$\alpha$ due to the relation:
\begin{equation}
\f{V_{ub} V_{ud}^*}{V_{tb} V_{td}^*} =
- \left | \f{V_{ub} V_{ud}^*}{V_{tb} V_{td}^*} \right |
{\rm e}^{i \alpha} \, .
\label{eq:UT-phase}
\end{equation}
The NLO corrections to the branching ratios of the exclusive decays
${\rm B}^\pm \to \rho^\pm \gamma$ and ${\rm B}^0 \to \rho^0 \gamma$ are derived
very much along the same lines as outlined for the decays ${\rm B} \to {\rm K}^*
\gamma$. Including the annihilation contribution, the ${\rm B} \to \rho
\gamma$ branching ratios, isospin- and CP-violating asymmetries are
given in~[\ref{Bosch:2001gv},\ref{bdgAP}].

Concentrating on the decays ${\rm B}^\pm \to \rho^\pm \gamma$, the expression 
for the ratio $R (\rho \gamma/{\rm K}^* \gamma) \equiv {\cal B}(B^\pm \to 
\rho^\pm \gamma)/{\cal B}(\rm B^\pm \to K^{*\pm} \gamma$) (where an
average over the charge-conjugated modes is implied) can be 
written as~[\ref{bdgAP}]
\beq
\hskip -0.3cm
R (\rho \gamma/{\rm K}^* \gamma) = S_\rho \left| V_{td} \over V_{ts} \right|^2
 {(M_B^2 - M_\rho^2)^3 \over (M_B^2 - M_{K^*}^2)^3 } \zeta^2
(1 + \Delta R)~,
\label{rapp}
\eeq
where $S_\rho = 1$ for the $\rho^\pm$  meson, and
$\zeta=\xi_{\perp}^{\rho}(0)/\xi_{\perp}^{K^*}(0)$, with
$\xi_{\perp}^{\rho}(0) (\xi_{\perp}^{K^*}(0))$ being the form factors
(at $q^2=0$) in the effective heavy quark theory for the decays ${\rm B} \to
\rho \gamma ({\rm B} \to {\rm K}^* \gamma)$. The quantity $(1 +\Delta R)$ entails
the explicit $O(\alpha_s)$ corrections, encoded through the functions
$A_{\rm R}^{(1)K^*}$, $A_{\rm R}^{(1) t}$ and $A_{\rm R}^{u}$, and the
long-distance contribution $L_{\rm R}^{u}$. For the decays ${\rm B}^\pm \to
\rho^\pm \gamma$ and ${\rm B}^\pm \to {\rm K}^{*\pm} \gamma$, this can be written
after charge conjugated averaging as
\bea
1 + \Delta R^\pm &=& \left| C_7^d + \lu L^u_R \over C_7^s \right|^2
           \left( 1 - 2 A^{(1)K^*}_R  {\Re C_7^s \over |C_7^s|^2} \right)
  \nonumber \\
 & & \hskip -0.5cm + {2\over |C_7^s|^2} \Re \left[ (C_7^d + \lu L^u_R)
 (A^{(1)t}_R + \lu^* A^u_R) \right]~.
\label{dr}
\eea
In the SM, $C_7^{d}=C_7$, as in the $b \to s \gamma$ decays; however, in 
beyond-the-SM scenarios, this may not hold making the decays ${\rm B} \to \rho 
\gamma$ interesting for beyond-the-SM searches~[\ref{Ali:2002kw}].
The definitions of the quantities $A^{(1)K^*}$, $A^{(1)t}$, $A^u$ and
$L^u_R = \epsilon_A \, C^{(0) {\rm eff}}_7$ can be seen in~[\ref{bdgAP}]. 
Their default values together
with that of $\zeta$ are summarized in Table~\ref{inputs}, where we have
also specified the theoretical errors in the more sensitive parameters
$\zeta$ and $L^u_R$.
\begin{table}[!t]
\begin{center}
\begin{tabular}{|l|l|}
\hline
 $\zeta = 0.76 \pm 0.10$ & $L^u_R = -0.095 \pm 0.022 $ \cr
 $A^{(1)K^*} = -0.113 - i 0.043$ & $A^{(1)t} = -0.114 - i 0.045$ \cr
 $A^u = -0.0181 + i 0.0211$ & \cr \hline
 $\eta_{tt} = 0.57$ & $\eta_{cc} = 1.38 \pm 0.53$ \cr
 $\eta_{tc} = 0.47 \pm 0.04$ & $\hat B_K = 0.86 \pm 0.15$ \cr
 $\eta_B = 0.55$ & $F_{B_d} \sqrt{\hat B_{B_d}} = 235 \pm 33 ^{+0}_{-24}\;
\mev$ \cr
 $\xi_s= 1.18 \pm 0.04^{+0.12}_{-0}$  & \cr \hline
 $\lambda = 0.221 \pm 0.002$ & $|V_{ub}/V_{cb}| = 0.097 \pm 0.010$ \cr   
 $\epsilon_K = (2.271 \pm 0.017) \; 10^{-3}$ & $\Delta M_{B_d} = 0.503 \pm
0.006 \; {\rm ps}^{-1}$ \cr
 $a_{\psi K_s} = 0.734 \pm 0.054$ & $\Delta M_{B_s} \geq 14.4 \; {\rm ps}^{-1}
\; (\cl{95})$   \cr
\hline
\end{tabular}
\end{center}
\caption{\it Theoretical parameters and measurements used in ${\rm B}\to  
\rho\gamma$ observables and in the CKM unitarity fits. For details and 
references, see~[\ref{Ali:2002kw},\ref{ALILOND}] }
\label{inputs}
\end{table}

 What concerns the quantity called $\zeta$, we note that there are 
several model-dependent estimates of the same in the literature. Some 
representative values are: $\zeta=0.76 \pm 0.06$ from the light-cone QCD 
sum rules~[\ref{Ali:1995uy}]; a theoretically improved estimate in the
same approach yields~[\ref{Ball:1998kk}]: $\zeta=0.75 \pm 0.07$;
$\zeta =0.88 \pm 0.02 (!)$  using hybrid QCD sum rules~[\ref{Narison:1994kr}],
and $\zeta=0.69 \pm 10\%$ in the quark model~[\ref{Melikhov:2000yu}]. Except 
for the hybrid QCD 
sum rules, all other approaches yield a significant SU(3)-breaking in the
magnetic moment form factors. In the light-cone QCD sum rule approach, 
this is anticipated due to the appreciable differences in the wave 
functions of the ${\rm K}^*$ and $\rho$-mesons. To reflect the current
dispersion in the theoretical estimates of $\zeta$, we take its value as
$\zeta=0.76 \pm 0.10$. A  lattice-QCD based estimate of the same is highly
desirable.   
 
The isospin breaking ratio 
\begin{equation}
\Delta (\rho\gamma)\equiv \frac{(\Delta^{+0} +\Delta^{-0})}{2},
\qquad \Delta^{\pm 0} = 
\frac{\Gamma (\rm B^\pm \to \rho^\pm \gamma)}
{ 2 \Gamma (\rm B^0 (\overline {\rm B}^0) \to \rho^0 \gamma)} -1
\end{equation}
 is given by
\bea
\Delta (\rho\gamma) &=& \left| C_7^d + \lu L^u_R \over C_7^d \right|^2
              \left( 1 - {2 \Re C_7^d ( A^{(1)t}_R + \lu^* A^u_R)\over
 |C_7^d|^2} \right) \nonumber \\
 & & \hskip -1cm +{2\over |C_7^d|^2} \Re \left[ (C_7^d + \lu L^u_R)
(A^{(1)t}_R
 + \lu^* A^u_R) \right]-1\; ,
\label{dnlo}
\eea
and the CP asymmetry $A^\pm_{CP} (\rho\gamma) = (\branch ({\rm B}^- \to \rho^- \gamma) -
\branch ({\rm B}^+ \to \rho^+ \gamma) )/ (\branch ({\rm B}^- \to \rho^- \gamma) + \branch ({\rm B}^+
\to \rho^+ \gamma) )$ is
\beq
A_{CP}^\pm (\rho\gamma) = - {2  \Im \left[ (C_7^d + \lu L^u_R) (A^{(1)t}_I +
 \lu^* A^u_I) \right]
                   \over |C_7^d + \lu L^u_R|^2} \; .
\label{acp}
\eeq
The observables $R^0 (\rho \gamma/{\rm K}^* \gamma)\equiv \bar \branch ({\rm B}^0
\to \rho^0 \gamma)/\branch (\overline {\rm B}^0 \to {\rm K}^{*0}\gamma)$ (where $\bar \branch$ is
the average of the ${\rm B}^0$ and $\overline {\rm B}^0$ modes) and $A^0_{CP} (\rho\gamma) =
(\branch ({\rm B}^0 \to \rho^0 \gamma) - \branch (\overline {\rm B}^0 \to \rho^0 \gamma) )/
(\branch ({\rm B}^0 \to \rho^- \gamma) + \branch (\overline {\rm B}^0 \to \rho^0 \gamma) )$ are
obtained from Eqs.~(\ref{rapp}, \ref{dr}, \ref{acp}) in the limit
$L^u_R = 0$ and $S_\rho = 1/2$.
\begin{figure}[!t]
\centerline{\mbox{\epsfxsize=15cm \hbox{\epsfbox{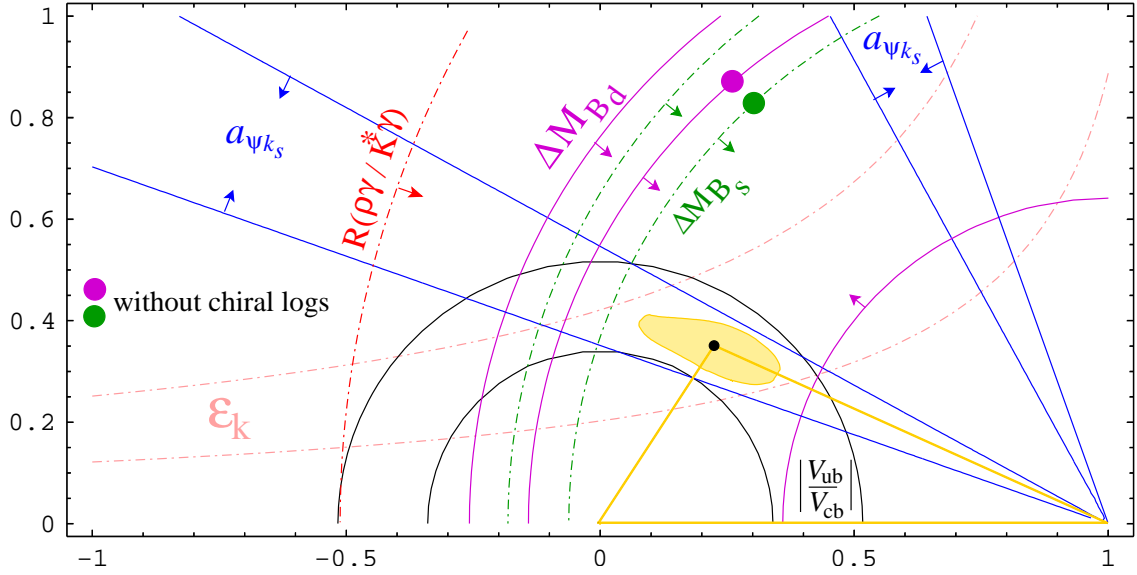}} }}
\vskip 0.2cm
%
\caption{Unitary triangle fit in the SM and the resulting 95\%
C.L. contour in the $\bar \rho$ - $\bar \eta$ plane. The impact of the
$R(\rho\gamma/{\rm K}^*\gamma)~<~0.047$ constraint is also shown (from 
Ref.~[\ref{Ali:2002kw}]).}
\label{fig:utsm}
\vskip -0.2cm
\end{figure}
The numerical estimates for the various observables depend, apart from
the hadronic parameters specific to the ${\rm B} \to V \gamma$
($V={\rm K}^*,\rho)$ decays, also on the CKM parameters, in particular $\bar
\rho$ and $\bar \eta$. A typical analysis of the constraints in the
$(\bar \rho, \bar \eta)$ plane from the unitarity of the 
CKM matrix~[\ref{Ali:2002kw}], 
including the measurements of the CP asymmetry
$a_{\psi K_s}$ in the decays 
${\rm B}^0/\overline{\rm B^0} \to J/\psi {\rm K}_s$ (and
related modes)~[\ref{Nir:2002gu}] is shown in \fig{fig:utsm}.
Note that for the hadronic parameters $F_{B_d} \sqrt{\hat{B}_{B_d}}$
 and $\xi_s$, the recent lattice estimates~[\ref{Lellouch:2002}] 
have been adopted that take into account
uncertainties induced by the so-called chiral 
logarithms~[\ref{Kronfeld:2002ab}].
These errors are extremely asymmetric and, once taken into account, reduce
sizeably the impact of the $\Delta M_{B_s}/ \Delta M_{B_d}$  lower bound
on the
unitarity triangle analysis, as shown in Fig.~\ref{fig:utsm}.
The $95\%$ CL contour is drawn taking into account chiral
logarithms uncertainties. The
fitted values for the Wolfenstein parameters are $\bar\rho = 0.22 \pm
0.07$ and $\bar\eta = 0.35 \pm 0.04$.  This yields $\Delta R^\pm
=0.055 \pm 0.130$ and $\Delta R^0=0.015 
\pm 0.110$~[\ref{bdgAP},\ref{Ali:2002kw}].  
The impact of the current upper limit
$R(\rho \gamma/{\rm K}^* \gamma) \leq 0.047$~[\ref{babar:jessop}] is also
shown. While not yet competitive to the existing constraints on the
unitarity triangle, this surely is bound to change with the
anticipated $O(1~({\rm ab})^{-1}))$ $\Upsilon(4S) \to {\rm B}\overline{\rm B}$ data
over the next three years at the B-factories.

Taking into account these errors and the uncertainties on the
theoretical parameters presented in Table~\ref{inputs}, leads to the
following SM expectations for the ${\rm B} \to ({\rm K}^*,\rho)
\gamma$ decays~[\ref{Ali:2002kw}]:
\bea
R^\pm (\rho \gamma /{\rm K}^* \gamma) &=& 0.023 \pm 0.012 \; ,\\
R^0 (\rho \gamma /{\rm K}^* \gamma) &=& 0.011 \pm 0.006 \; ,\\
\Delta (\rho \gamma ) &=& 0.04^{+0.14}_{-0.07} \; ,\\
A_{CP}^\pm (\rho\gamma) &=& 0.10^{+0.03}_{-0.02} \; , \\
A_{CP}^0 (\rho\gamma) &=& 0.06 \pm 0.02 \; .
\eea
The above estimates of $R^\pm (\rho \gamma /{\rm K}^* \gamma)$ and $R^0 (\rho \gamma /{\rm K}^* \gamma)$
can be combined with the measured branching ratios for ${\rm B} \to {\rm K}^* \gamma$ 
decays given earlier to yield:
\begin{equation}
{\cal B}({\rm B}^\pm \to \rho^\pm \gamma) = (0.93 \pm 0.49) \times 10^{-6}~,
\hspace{5mm} 
{\cal B}({\rm B}^0 \to \rho^0 \gamma) = (0.45 \pm 0.24) \times 10^{-6}~.
\label{eq:brhogamsm}
\end{equation}
The errors include the uncertainties on the hadronic parameters and the CKM
parameters $\bar \rho$, $\bar \eta$, as well as the current experimental
error on ${\cal B}({\rm B} \to {\rm K}^* \gamma)$.
While there is as yet no experimental bounds on the isospin- and 
CP-violating quantities, $\Delta (\rho \gamma )$, $A_{CP}^\pm (\rho\gamma)$ and $A_{CP}^0 
(\rho\gamma)$, the upper limits on the branching ratios $R^\pm (\rho \gamma /{\rm K}^* \gamma)$ 
and $R^0 (\rho \gamma /{\rm K}^* \gamma)$ have been significantly improved by the 
BABAR~[\ref{babar:jessop}] and BELLE~[\ref{NishidaS:2002}] collaborations 
recently. Averaged over the
charge conjugated modes, the current best upper limits 
are~[\ref{babar:jessop}]: 
${\cal B}({\rm B}^0 \to \rho^0 \gamma) < 1.4 \times
10^{-6}$, ${\cal B}({\rm B}^\pm \to \rho^\pm \gamma) < 2.3 \times
10^{-6}$ and ${\cal B}({\rm B}^0 \to \omega \gamma) < 1.2 \times
10^{-6}$ (at 90\% C.L.). They have been combined, using isospin   
weights for ${\rm B} \to \rho \gamma$ decays and assuming ${\cal B}({\rm B}^0 \to
\omega \gamma)={\cal B}({\rm B}^0 \to \rho^0 \gamma)$, to yield the
improved upper limit ${\cal B}({\rm B} \to \rho
\gamma) < 1.9 \times 10^{-6}$. The current measurements of the
branching ratios for ${\rm B} \to {\rm K}^* \gamma$ decays by 
BABAR~[\ref{Aubert:2001}], 
${\cal B}({\rm B}^0 \to {\rm K}^{*0} \gamma)=(4.23 \pm 0.40 
\pm 0.22) \times 10^{-5}$ and ${\cal B}({\rm B}^+ \to {\rm K}^{*+} \gamma)=(3.83 
\pm 0.62 \pm 0.22) \times 10^{-5}$, are then used to set an upper
limit on the ratio of the branching ratios $R(\rho \gamma/{\rm K}^*\gamma)
\equiv {\cal B}({\rm B} \to \rho \gamma)/{\cal  B}({\rm B} \to {\rm K}^* \gamma) < 0.047$
(at 90\% C.L.)~[\ref{babar:jessop}]. This bound is typically a factor
2 away from the SM estimates given above~[\ref{bdgAP},\ref{Ali:2002kw}]. 
However, in beyond-the-SM scenarios, this bound provides highly significant 
constraints on the relative strengths of the $b \to d \gamma$ and $b \to s 
\gamma$ transitions~[\ref{Ali:2002kw}].
\begin{figure}[!t]
 \centerline{\mbox{\epsfxsize=14,5cm \hbox{\epsfbox{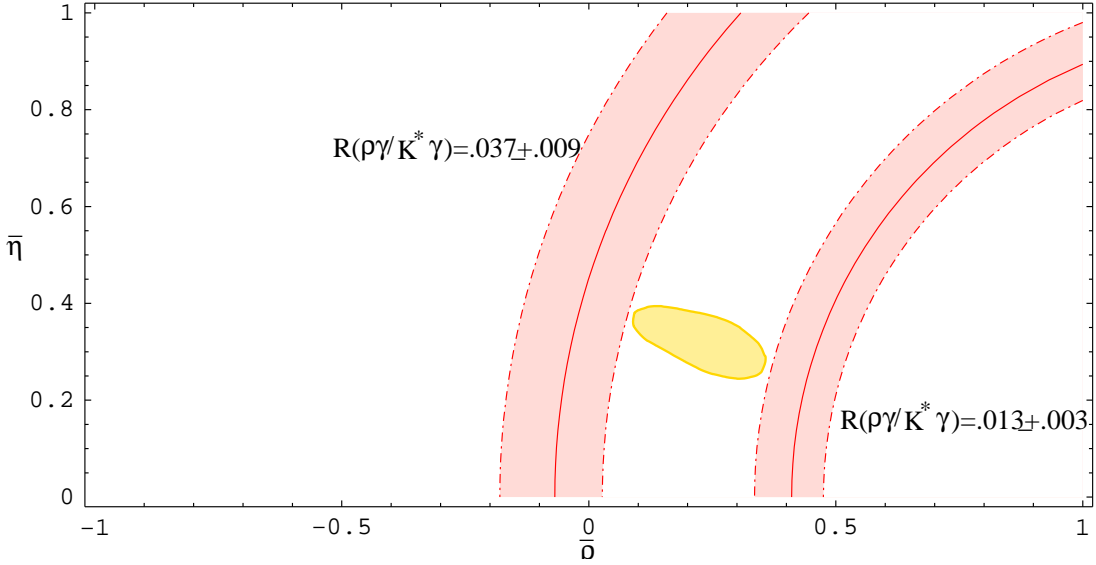}} }}
%
\caption{ Extremal values of $R(\rho\gamma/{\rm K}^*\gamma)$ t
hat are compatible with
the SM unitarity triangle analysis (from 
Ref.~[\ref{Ali:2002kw}]).}
\label{fig:rappsm}
\vskip -0.2cm
\end{figure}

The extremal values of $R(\rho \gamma/{\rm K}^* \gamma)$ compatible with the
SM UT-analysis are shown in \fig{fig:rappsm} where the bands
correspond to the values $0.037\pm 0.007$ and $0.013\pm 0.003$ (the
errors are essentially driven by the uncertainty on $\zeta$). The meaning
of this figure is as follows: any measurement of $R(\rho \gamma/{\rm K}^*
\gamma)$, whose central value lies in the range $(0.013,0.037)$ would
be compatible with the SM, irrespective of the size of the
experimental error. The error induced by the imprecise determination
of the isospin breaking parameter $\zeta$ limits currently the
possibility of having a very sharp impact from $R(\rho
\gamma/{\rm K}^*\gamma)$ on the UT analysis. This aspect needs further 
theoretical work.

\section{Weak phases from hadronic B decays}
{\it  M. Beneke, G. Buchalla (coordinator), M. Ciuchini, 
R. Fleischer, E. Franco, Y.-Y. Keum, G. Martinelli,
M. Pierini, J.L. Rosner and L. Silvestrini}

\vspace{2mm}

The next five contributions discuss the problem of extracting weak phases
from hadronic B decays. The emphasis is on determining
the CKM parameters $\gamma$ and $\alpha$, or equivalent constraints
on $\bar\rho$ and $\bar\eta$, from exclusive modes with two light
mesons in the final state, such as ${\rm B}\to\pi {\rm K}$ and ${\rm B}\to\pi\pi$.
This problem is difficult since the underlying weak
interaction processes are dressed by QCD dynamics, which is prominent
in purely hadronic decays.
Despite the general difficulty, there are several circumstances
that help us to control strong interaction effects and
to isolate the weak couplings:

\begin{itemize}
\item
{\bf Flavour symmetries:}
The impact of strong interactions may be reduced by eliminating
hadronic matrix elements through a combination of different channels,
exploiting approximate flavour symmetries of QCD. Important examples
are isospin, $U$-spin (doublet ($d$, $s$)) or, more generally,
$SU(3)_F$.
\item
{\bf Heavy-quark limit:} 
The fact that $m_b\gg\Lambda_{QCD}$
can be used to simplify the theoretical description of
QCD dynamics in B decays. Within this framework amplitudes
are expanded in $\Lambda_{QCD}/m_b$, long-distance and
short-distance contributions are factorized, and the latter can be
treated in perturbative QCD. As a result the impact of nonperturbative
hadronic physics is reduced.
\item 
{\bf Rich phenomenology:}
A large number of decay channels exists, which allows us
to explore different approaches, to apply various strategies based
on QCD flavour symmetries and to obtain cross-checks for dynamical
calculations based on factorization.
\end{itemize}

It has to be emphasized that this field is in a state of
ongoing development, both theoretically and experimentally.
On the theory side important questions still need further study
(general proof of factorization, light-cone dynamics of the B meson,
numerical accuracy of heavy-quark limit in various situations,
size of $SU(3)_F$-breaking corrections),
while many valuable new data continue to be collected
by the experiments.
It is worth noting that the approaches based on flavour symmetries and
those using dynamical calculations in the heavy-quark limit are complementary 
to each other. For instance, corrections from flavour symmetry breaking
can be estimated within factorization.
One may expect that the most important results might eventually
be obtained from the combined use of all the available options
mentioned above.

The following contributions summarize the status of
the subject as it was discussed at this workshop.
The contributions of J.L.~Rosner and R.~Fleischer highlight
strategies based on QCD flavour symmetries to extract $\alpha$ and
$\gamma$ from ${\rm B}\to\pi {\rm K}$, $\pi\pi$ decays. 
The status of factorization
is outlined by M.~Beneke. A critical point of view on extracting
$\gamma$ from global fits to hadronic modes
is presented by M.~Ciuchini et al.. Finally,
a phenomenological analysis based on the hypothesis of
hard-gluon dominance of ${\rm B}\to\pi$ form factors is described 
by Y.-Y.~Keum.

\subsection{Weak coupling phases}

{\it J.L. Rosner\footnote{
J.L.~Rosner would like to thank C.-W. Chiang, M. Gronau, Z. Luo, M. Neubert, and
D. Suprun for enjoyable collaborations on these subjects.}}


The phases of CKM matrix elements
describing charge-chan\-ging weak couplings of quarks are fundamental
quantities.  They are sometimes described in terms of angles $\alpha = \phi_2$,
$\beta = \phi_1$, and $\gamma = \phi_3$ in the unitarity triangle.
Now that BaBar and Belle are converging on a value of
$\sin(2 \beta)$, attention has turned to ways of
learning $\alpha$ and $\gamma = \pi - \beta - \alpha$. This summary
describes some recent work on the subject.

In Sec.\ \ref{sec:bpipi} we discuss ${\rm B}^0 \to \pi^+ \pi^-$ in the
light of recent measurements at BaBar~[\ref{Aubert:2002tv}] and 
Belle~[\ref{Bellealpha}] 
of time-dependent asymmetries.  This work was performed
in part in collaboration with M.~Gronau~[\ref{Gronau:2001cj},\ref{Gronau:2002cj},\ref{Gronau:2002gj}] and in part with Z.~Luo~[\ref{Luo:2001ek}].  
We then mention
how to learn $\gamma$ from various ${\rm B} \to {\rm K} \pi$ 
decays (Sec.\ \ref{sec:Bkpi},
collaboration with M.~Gronau~[\ref{Gronau:2001cj}] and with 
M.~Neubert~[\ref{Neubert:1998pt},\ref{Neubert:1998jq}]), 
$2 \beta + \gamma$ from
${\rm B} \to {\rm D}^{(*)} \pi$ 
(Sec.\ \ref{sec:BDpi}, collaboration with D.~Suprun and
C.W.~Chiang~[\ref{Suprun:2001ms}]), and $\alpha$ and $\gamma$ from
tree-penguin interference in ${\rm B} \to PP,~PV$ decays, where $P$ is a light
pseudoscalar and $V$ a light vector meson (Sec.\ \ref{sec:int}, collaboration
with C.W.~Chiang~[\ref{Chiang:2001ir}]).  Sec.\ \ref{sec:other} is a short
guide to other recent work, while we summarize in Sec.\ \ref{sec:sum}

\subsubsection{Determination of $\alpha$ from $\rm B^0 \to \pi^+ \pi^-$ 
decays \label{sec:bpipi}}

We regard $\alpha,\gamma$ as uncertain to about $\pi/4$: $126^\circ
\ge \alpha \ge 83^\circ$, $32^\circ \le \gamma \le 75^\circ$
[\ref{Gronau:2001cj}], in accord with $122^\circ \ge \alpha \ge 75^\circ$,
$37^\circ \le \gamma \le 80^\circ$~[\ref{Hocker:2001jb}].  If ${\rm B}^0 \to \pi^+
\pi^-$ were dominated by the ``tree'' amplitude $T$ with phase $\gamma
= {\rm Arg}(V_{ub}^* V_{ud})$, the parameter $\lpp \equiv e^{-2 i \beta}
A(\ob \to \pi^+ \pi^-)/A(\bo \to \pi^+ \pi^-)$ would be just $e^{2 i \alpha}$
and the indirect CP-violating asymmetry $\spp = 2 {\rm Im}\lpp/(1 + |\lpp|^2)$
would be $\sin 2 \alpha$.  Here
\beq
\frac{d \Gamma}{d t} \left\{ \begin{array}{c} \bo|_{t=0} \to f \\
\ob|_{t=0} \to f \end{array} \right\} \propto e^{- \Gamma t}[1 \mp \spp
\sin \Delta M_d t \pm \cpp \cos \Delta M_d t ]~~~,
\eeq
$\cpp = (1 - |\lpp|^2)/(1 + |\lpp|^2)$, and 
$\Delta \Gamma \simeq \Delta M_d/200$
has been neglected.  In the presence of non-zero
$\Delta \Gamma$ one can also measure $\appros = 2 {\rm Re} \lpp/(1+|\lpp|^2)$.
Since $|\spp|^2 + |\cpp|^2 + |\appros|^2 = 1$ one has $|\spp|^2 + |\cpp|^2 \le 1$.
However, one also has a penguin amplitude $P$ involving a $\bar b \to \bar d$
loop transition involving contributions $\sim$ $V_{ud}^* V_{ub}$,
$V_{cd}^* V_{cb}$, and $V_{td}^* V_{tb} = -V_{ud}^* V_{ub} - V_{cd}^* V_{cb}$.
The decay amplitudes are then
\beq
A(\bo \to \pi^+ \pi^-) = -(|T|e^{i \delta_T} e^{i \gamma} + |P|
e^{i \delta_P}),~
A(\ob \to \pi^+ \pi^-) = -(|T|e^{i \delta_T} e^{-i \gamma} + |P|
e^{i \delta_P}),
\eeq
where the strong phase difference $\delta \equiv \delta_P -\delta_T$.
It will be convenient to define $\rpp \equiv \ocb(\bo \to \pi^+
\pi^-)/\ocb(\bo \to \pi^+ \pi^-)_{\rm tree}$, where $\ocb$ refers to a branching
ratio averaged over $\bo$ and $\ob$.  One may use $\spp$ and $\cpp$ to learn
$\alpha,\delta$, resolving a discrete ambiguity with the help of 
$\rpp$~[\ref{Gronau:2002cj}].  Alternatively, one may directly 
use $\spp$, $\cpp$,
and $\rpp$ to learn $\alpha$, $\delta$, and $|P/T|$~[\ref{Gronau:2002gj},\ref{Charles:1998qx}].

Explicit expressions for $\rpp$, $\spp$ and $\cpp$ may be found 
in~[\ref{Gronau:2002cj},\ref{Gronau:2002gj}].  
In~[\ref{Gronau:2002cj}] we estimated
$|P/T| = 0.276 \pm 0.064$ (see also~[\ref{Beneke:2001ev}]), 
obtaining $|P|$ from
${\rm B}^+ \to {\rm K}^0 \pi^+$ via (broken) 
flavor SU(3) and $|T|$ from ${\rm B} \to \pi \ell
\nu$.  Plotting $\cpp$ against $\spp$ for various values of $\alpha$ in the
likely range, one obtains curves parametrized by $\delta$ which establish a
one-to-one correspondence between a pair $(\spp,\cpp)$ and a pair
$(\alpha,\delta)$ as long as $|\delta| \le 90^\circ$.  However, if $|\delta|$
is allowed to exceed about
$90^\circ$ these curves can intersect with one another, giving rise to a
discrete ambiguity corresponding to as much as $30^\circ$ uncertainty in
$\alpha$ when $\cpp = 0$.  In this case, when $\delta = 0$ or $\pi$, one
has $|\lpp|=1$ and $\spp = \sin2(\alpha + \Delta \alpha)$, where
$\tan(\Delta \alpha) = \pm (|P/T| \sin \gamma)/(1 \pm (|P/T| \cos \gamma)$
is typically $\pm 15^\circ$.  One can resolve the ambiguity
either by comparing the predicted $\rpp$ with experiment 
(see~[\ref{Gronau:2002cj}] for details) , or by comparing the allowed $(\rho,\eta)$
region with that determined by other observables~[\ref{Hocker:2001jb}].  An
example is shown in~[\ref{Gronau:2001cj}].

Once errors on $\rpp$ are reduced to $\pm 0.1$ (they are now about three times
as large~[\ref{Gronau:2002cj}]), a distinction between
$\delta = 0$ and $\delta = \pi$ will be possible when $\spp \simeq 0$,
as appears to be the case for BaBar~[\ref{Aubert:2002tv}].  For the Belle 
data~[\ref{Bellealpha}], which suggest $\spp <0$, the distinction becomes easier;
it becomes harder for $\spp > 0$.  With 100 fb$^{-1}$ at each of BaBar and
Belle, it will be possible to reduce $\Delta |T|^2/|T|^2$ from its present
error of 44\% and $\ocb(\bo \to \pi^+ \pi^-)$ from its present error of
21\% each to about 10\%~[\ref{Luo:2001ek}], which will go a long way toward
this goal.
In an analysis independent of $|P/T|$ performed since the workshop, the 
somewhat discrepant BaBar and Belle values of $\spp$ and $\cpp$, when averaged,
favor $\alpha$ between about $90^\circ$ and $120^\circ$ (see Fig.\ 1 
of~[\ref{Gronau:2002gj}]).

\subsubsection{Determination of $\gamma$ from ${\rm B} \to {\rm K} \pi$ 
decays\label{sec:Bkpi}}

\subsubsection*{$\gamma$ from $\bo \to {\rm K}^+ \pi^-$ and 
${\rm B}^+ \to {\rm K}^0 \pi^+$}
We mention some results of~[\ref{Gronau:2001cj}] on information
provided by ${\rm B}^0 \to {\rm K}^+ \pi^-$ decays, which involve both a penguin $P'$ and
a tree $T'$ amplitude.  One can use the flavor-averaged branching ratio $\ocb$
and the CP asymmetry in these decays, together with $P'$ information from the
${\rm B}^+ \to {\rm K}^0 \pi^+$ decay rate (assuming it is equal to the charge-conjugate
rate, which must be checked) and $T'$ information from ${\rm B} \to \pi \ell \nu$
and flavor SU(3), to obtain constraints on $\gamma$.  One considers the
ratio $R \equiv [\ocb(\bo \to {\rm K}^+ \pi^-)/\ocb(B^+ \to {\rm K}^0 \pi^+)][\tau_+/
\tau_0]$, where the $\rm B^+/B^0$ lifetime ratio $\tau_+/\tau_0$ is about 1.07. 
Once the error on this quantity is reduced to $\pm 0.05$ from its value of $\pm
0.14$ as of February 2002, which should be possible with 200 fb$^{-1}$ at
each of BaBar and Belle, one should begin to see useful constraints arising
from the value of $R$, especially if errors on the ratio $r \equiv |T'/P'|$
can be reduced with the help of better information on $|T'|$.

\subsubsection*{$\gamma$ from ${\rm B}^+ \to {\rm K}^+ \pi^0$ and 
${\rm B}^+ \to {\rm K}^0 \pi^+$}
One can use the ratio $R_c \equiv 2 \ocb(\rm B^+ \to {\rm K}^+ \pi^0)/\ocb(\rm 
B^+ \to {\rm K}^0
\pi^+)$ to determine $\gamma$~[\ref{Gronau:2001cj},\ref{Neubert:1998pt},\ref{Neubert:1998jq}].  Given the values as of February 2002, $R_c = 1.25 \pm 0.22$,
$A_c \equiv [{\cal B}(\rm B^- \to {\rm K}^- \pi^0) - {\cal B}(\rm B^+ \to {\rm K}^+ \pi^0)]/
\ocb(\rm B^+ \to {\rm K}^0 \pi^+) = -0.13 \pm 0.17$, and $r_c \equiv |T'+C'|/|p'| =
0.230 \pm 0.035$ (here $C'$ is a color-suppressed amplitude, while $p'$ is
a penguin amplitude including an electroweak contribution), and an 
estimate~[\ref{Neubert:1998pt},\ref{Neubert:1998jq}] of the electroweak penguin contribution,
one finds $\gamma \le 90^\circ$ or $\gamma \ge 140^\circ$ at the $1 \sigma$
level, updating an earlier bound~[\ref{Gronau:2001cj}] $\gamma \ge 50^\circ$.
A useful determination would involve $\Delta R_c = \pm 0.1$, achievable with
150 fb$^{-1}$ each at BaBar and Belle.

\subsubsection{Determination of $2 \beta + \gamma$ from 
${\rm B} \to {\rm D}^{(*)} \pi$ decays\label{sec:BDpi}}

The ``right-sign'' (RS) decay ${\rm B}^0 \to {\rm D}^{(*)-} \pi^+$, 
governed by the CKM
factor $V_{cb}^* V_{ud}$, and the ``wrong-sign'' (WS) decay $\ob \to
{\rm D}^{(*)-} \pi^+$, governed by $V_{cd}^* V_{ub}$, can interfere through
$\bo$--$\ob$ mixing, leading to information on the weak phase $2 \beta +
\gamma$.  One must separate out the dependence on a strong phase $\delta$
between the RS and WS amplitudes, measuring time-dependent observables
\beq
A_\pm(t) = (1+R^2) \pm (1 - R^2) \cos \Delta m t,~~
B_\pm(t) = - 2 R \sin(2 \beta + \gamma \pm \delta) \sin \Delta m t,
\eeq
where $R \equiv |{\rm WS/RS}| = r |V_{cd}^* V_{ub}/V_{cb}^* V_{ud}| \simeq 0.02
r$, with $r$ a parameter of order 1 which needs to be known better.  In
Ref.~[\ref{Suprun:2001ms}] we use the fact that $R$ can be measured in
the decay ${\rm B}^+ \to {\rm D}^{*+} \pi^0$ to conclude that with 250 million
${\rm B} \bar B$ pairs one can obtain an error of less than $\pm 0.05$
on $\sin(2 \beta + \gamma)$, which is expected to be greater than about
0.89 in the standard model.  Thus, such a measurement is not likely to
constrain CKM parameters, but has potential for an interesting
non-standard outcome.

\subsubsection{Determination of $\alpha$ and $\gamma$ from 
${\rm B} \to PP,~PV$ decays\label{sec:int}}

Some other processes which have a near-term potential for providing
information on tree-penguin interference (and hence on $\alpha$ and $\gamma$)
are the following~[\ref{Chiang:2001ir}: (1) the CP asymmetries in $\rm B^+ \to
\pi^+ \eta$ and $\pi^+ \eta'$; (2) rates in $\rm B^+ \to \eta' {\rm K}^+$ and
$\bo \to \eta' {\rm K}^0$; (3) rates in ${\rm B}^+ \to \eta {\rm K}^{*+}$ and 
$\bo \to \eta {\rm K}^{*0}$; and (4) rates in $\rm B^+ \to \omega {\rm K}^+$ 
and $\bo \to \omega {\rm K}^0$.
Other interesting branching ratios include those for $\bo \to \pi^-
{\rm K}^{*+}$, $\bo \to {\rm K}^+ \rho^-$, $\rm B^+ \to \pi^+ \rho^0$, 
$\rm B^+ \to \pi^+ \omega$, and $\rm B^{(+,0)} \to \eta' {\rm K}^{*(+,0)}$, 
with a story for each~[\ref{Chiang:2001ir}].  In order to see tree-penguin 
interference at the predicted level one needs to measure branching ratios 
at the level of $\Delta \ocb = (1-2) \times 10^{-6}$.

\subsubsection{References to other work \label{sec:other}}

For other recent suggestions on measuring $\alpha$ and $\gamma$, see
the review of~[\ref{Fleischer:2001zn}] and the contributions of
[\ref{Gronau:2001ff}] on the isospin triangle in 
$\rm B \to \pi \pi$ ($\alpha$),
[\ref{Gronau:1998vg},\ref{Atwood:2001ck}] on 
$\rm B^+ \to {\rm D} {\rm K}^+$ ($\gamma$),
[\ref{Kayser:1999bu}] on $\bo \to {\rm D} {\rm K}_S$ ($2 \beta + \gamma$),
[\ref{Buras:2000gc}]
on $\bo \to {\rm K} \pi$ ($\gamma$), [\ref{Fleischer:1999pa}] on $\bo \to \pi^+
\pi^-$ and $\rm B_s \to {\rm K}^+ {\rm K}_-$ ($\gamma$), 
and [\ref{Gronau:2000md}] on
$\bo \to {\rm K}^+ \pi^-$ and $\rm B_s \to {\rm K}^- \pi^+$ ($\gamma$).  
These contain references to earlier work.

\subsubsection{Summary \label{sec:sum}}

CKM phases will be learned in many ways.  While $\beta$ is well-known now and
will be better-known soon, present errors on $\alpha$ and
$\gamma$ are about $45^\circ$.  To reduce them to $10^\circ$ or less,
several methods will help.  (1) Time-dependent asymmetries in 
$\rm B^0 \to \pi^+ \pi^-$ already contain useful information.  
The next step will come when both
BaBar and Belle accumulate samples of at least 100 fb$^{-1}$.  (2) In
$\rm B^0 \to \pi^+ \pi^-$ an ambiguity between a strong phase $\delta$ 
near zero and one near $\pi$ (if the direct asymmetry parameter 
$C_{\pi \pi}$ is small) can be resolved experimentally, for example by 
better measurement of the $\rm B^0 \to \pi^+ \pi^-$ branching ratio and 
the $\rm B \to \pi \ell \nu$ spectrum.  (3) Several $\rm B \to {\rm K} \pi$ 
modes, when compared, can constrain
$\gamma$ through penguin-tree interference.  This has been recognized, for
example, in [\ref{Hocker:2001jb}].  (4) The rates in several $\rm B \to PP,~PV$
modes are sensitive to tree-penguin interference.  One needs to measure
branching ratios with errors less than $2 \times 10^{-6}$ to see such effects
reliably.


\boldmath
\subsection{Extracting  \boldmath{$\gamma$} through flavour-symmetry 
strategies}
\unboldmath

{\it R. Fleischer\footnote{
R. Fleischer would like to thank Andrzej Buras, Thomas Mannel and 
Joaquim Matias for
pleasant collaborations on the topics discussed below.}
}


An important element in the testing of the Kobayashi--Maskawa picture of
CP violation is the direct determination of the angle $\gamma$ of the 
unitarity triangle of the CKM matrix. Here the goal is to overconstrain 
this angle as much as possible. In the presence of new physics, discrepancies 
may arise between different strategies, as well as with the ``indirect'' 
results for $\gamma$ that are provided by the usual fits of the unitarity 
triangle, yielding at present $\gamma\sim 60^\circ$ 
[\ref{BUPAST},\ref{Lacker},\ref{ref:haricot}]. 

There are many approaches on the market to determine $\gamma$ (for
a detailed review, see Ref.~[\ref{RF-Phys-Rep}]). Here we shall focus on
$\rm B\to\pi {\rm K}$ modes [\ref{Gronau:2001cj},\ref{Buras:2000gc}], 
[\ref{GRL}--\ref{ital-corr}], which can be analysed 
through flavour-symmetry arguments and plausible dynamical assumptions, and 
the $U$-spin-related decays ${\rm B}_d\to \pi^+\pi^-$, 
${\rm B}_s\to {\rm K}^+{\rm K}^-$
[\ref{Fleischer:1999pa}]. 
The corresponding flavour-symmetry strategies allow the 
determination of $\gamma$ and valuable hadronic parameters with a ``minimal'' 
theoretical input. Alternative approaches, relying on a more extensive use of 
theory, are provided by the recently developed ``QCD factorization'' 
[\ref{Beneke:1999br_bis},\ref{Beneke:2001ev}] 
and ``PQCD'' [\ref{PQCD}] approaches, which allow furthermore a 
reduction of the theoretical uncertainties of the flavour-symmetry strategies 
discussed here. Let us note that these approaches are also particularly 
promising from a practical point of view: BaBar, Belle and CLEO-III may 
probe $\gamma$ through $\rm B\to\pi {\rm K}$ modes, whereas the 
$U$-spin strategy, requiring also a measurement of the ${\rm B}_s$-meson 
decay ${\rm B}_s\to {\rm K}^+{\rm K}^-$, is 
already interesting for run II of the Tevatron [\ref{FERMILAB}], and can be 
fully exploited in the LHC era [\ref{LHCB}]. A variant for the 
B-factories [\ref{U-variant}], 
where ${\rm B}_s\to {\rm K}^+{\rm K}^-$ is replaced by 
${\rm B}_d\to\pi^\mp {\rm K}^\pm$, points 
already to an exciting picture [\ref{FlMa2}]. 

\vspace{-1.5mm}

\subsubsection{Studies of $\rm B\to \pi {\rm K}$ decays}
Using the isospin flavour symmetry of strong interactions, relations between 
$\rm B\to\pi {\rm K}$ amplitudes can be derived, which suggest the following 
combinations to probe $\gamma$: the ``mixed'' $\rm B^\pm\to\pi^\pm {\rm K}$, 
${\rm B}_d\to\pi^\mp \rm {\rm K}^\pm$ 
system [\ref{PAPIII}--\ref{defan}], the 
``charged'' $\rm B^\pm\to\pi^\pm {\rm K}$, 
$\rm B^\pm\to\pi^0{\rm K}^\pm$ system 
[\ref{Neubert:1998pt},\ref{Neubert:1998jq},\ref{BF-neutral1}], and the 
``neutral'' ${\rm B}_d\to\pi^0 {\rm K}$, ${\rm B}_d\to\pi^\mp {\rm K}^\pm$ 
system [\ref{Buras:2000gc},\ref{BF-neutral1}].
Interestingly, already CP-averaged $\rm B\to\pi {\rm K}$ branching ratios
may lead to non-trivial constraints on $\gamma$ 
[\ref{FM},\ref{Neubert:1998pt},\ref{Neubert:1998jq}].
In order
to {\it determine} this angle, also CP-violating rate differences have
to be measured. To this end, we introduce the following observables
[\ref{BF-neutral1}]:
\begin{equation}\label{mixed-obs}
\left\{\begin{array}{c}R\\A_0\end{array}\right\}
\equiv\left[\frac{\mbox{BR}(\rm B^0_d\to\pi^-{\rm K}^+)\pm
\mbox{BR}(\overline{B^0_d}\to\pi^+{\rm K}^-)}
{\mbox{BR}({\rm B}^+\to\pi^+{\rm K}^0)+
\mbox{BR}(\rm B^-\to\pi^-\overline{{\rm K}^0})}\right]\frac{\tau_{B^+}}{\tau_{B^0_d}}
\end{equation}
\begin{equation}\label{charged-obs}
\left\{\begin{array}{c}R_{\rm c}\\A_0^{\rm c}\end{array}\right\}
\equiv2\left[\frac{\mbox{BR}({\rm B}^+\to\pi^0{\rm K}^+)\pm
\mbox{BR}({\rm B}^-\to\pi^0{\rm K}^-)}{\mbox{BR}({\rm B}^+\to\pi^+{\rm K}^0)+
\mbox{BR}({\rm B}^-\to\pi^-\overline{{\rm K}^0})}\right]
\end{equation}
\begin{equation}\label{neut-obs}
\left\{\begin{array}{c}R_{\rm n}\\A_0^{\rm n}\end{array}\right\}
\equiv\frac{1}{2}\left[\frac{\mbox{BR}({\rm B}^0_d\to\pi^-{\rm K}^+)\pm
\mbox{BR}(\overline{{\rm B}^0_d}\to\pi^+{\rm K}^-)}
{\mbox{BR}({\rm B}^0_d\to\pi^0{\rm K}^0)+
\mbox{BR}(\overline{{\rm B}^0_d}\to\pi^0\overline{{\rm K}^0})}\right].
\end{equation}

\newpage

If we employ the isospin flavour symmetry and make plausible dynamical 
assumptions, concerning mainly the smallness of certain rescattering 
processes, we obtain parametrizations of the following structure
[\ref{defan},\ref{BF-neutral1}] 
(for alternative ones, see Ref.~[\ref{neubert}]):
\begin{equation}\label{obs-par}
R_{({\rm c,n})},\, A_0^{({\rm c,n})}=
\mbox{functions}\left(q_{({\rm c,n})}, r_{({\rm c,n})},
\delta_{({\rm c,n})}, \gamma\right).
\end{equation}
Here $q_{({\rm c,n})}$ denotes the ratio of electroweak (EW) penguins to 
``trees'', $r_{({\rm c,n})}$ is the ratio of ``trees'' to QCD penguins, and
$\delta_{({\rm c,n})}$ the strong phase between ``trees'' and QCD penguins. 
The EW penguin parameters $q_{({\rm c,n})}$ can be fixed through theoretical 
arguments: in the mixed system [\ref{PAPIII}--\ref{GR}], we have $q\approx0$, 
as EW penguins contribute only in colour-suppressed form; in the charged and 
neutral $\rm B\to\pi {\rm K}$ systems, $q_{\rm c}$ and $q_{\rm n}$ can be fixed through 
the $SU(3)$ flavour symmetry without dynamical assumptions 
[\ref{Neubert:1998pt},\ref{Neubert:1998jq},\ref{Buras:2000gc},\ref{BF-neutral1}]. 
The $r_{({\rm c,n})}$ can be determined with 
the help of additional experimental information: in the mixed system, $r$ 
can be fixed through arguments based on 
factorization~[\ref{PAPIII},\ref{GR},\ref{Beneke:1999br_bis},\ref{Beneke:2001ev}] 
or $U$-spin~[\ref{Gronau:2000md}], whereas $r_{\rm c}$ and $r_{\rm n}$ can be 
determined from the CP-averaged $\rm B^\pm\to\pi^\pm\pi^0$ branching ratio by 
using only the $SU(3)$ flavour symmetry 
[\ref{GRL},\ref{Neubert:1998pt},\ref{Neubert:1998jq}]. 
The uncertainties arising in this programme 
from $SU(3)$-breaking effects can be reduced 
through the QCD factorization 
approach~[\ref{Beneke:1999br_bis},\ref{Beneke:2001ev}], 
which is moreover in 
favour of small rescattering processes. For simplicity, we shall neglect 
such FSI effects in the discussion given below. 

Since we are in a position to fix the parameters $q_{({\rm c,n})}$ 
and $r_{({\rm c,n})}$, we may determine $\delta_{({\rm c,n})}$ 
and $\gamma$ from the observables given in (\ref{obs-par}). This can 
be done separately for the mixed, charged and neutral $\rm B\to\pi {\rm K}$ 
systems. It should be emphasized that also CP-violating rate 
differences have to be measured to this end. Using 
just the CP-conserving observables $R_{({\rm c,n})}$, we may obtain 
interesting constraints on $\gamma$. In contrast to $q_{({\rm c,n})}$ 
and $r_{({\rm c,n})}$, the strong phase $\delta_{({\rm c,n})}$ suffers 
from large hadronic uncertainties. However, we can get rid of 
$\delta_{({\rm c,n})}$ by keeping it as a ``free'' variable, yielding 
minimal and maximal values for $R_{({\rm c,n})}$:
\begin{equation}\label{const1}
\left.R^{\rm ext}_{({\rm c,n})}\right|_{\delta_{({\rm c,n})}}=
\mbox{function}\left(q_{({\rm c,n})},r_{({\rm c,n})},\gamma\right).
\end{equation}
Keeping in addition $r_{({\rm c,n})}$ as a free variable, we obtain 
another -- less restrictive -- minimal value 
\begin{equation}\label{const2}
\left.R^{\rm min}_{({\rm c,n})}\right|_{r_{({\rm c,n})},\delta_{({\rm c,n})}}
=\mbox{function}\left(q_{({\rm c,n})},\gamma\right)\sin^2\gamma.
\end{equation}
These extremal values of $R_{({\rm c,n})}$ imply 
constraints on $\gamma$, since the cases corresponding to
$R^{\rm exp}_{({\rm c,n})}< R^{\rm min}_{({\rm c,n})}$
and $R^{\rm exp}_{({\rm c,n})}> R^{\rm max}_{({\rm c,n})}$
are excluded. Present experimental data seem to point towards values for
$\gamma$ that are {\it larger} than $90^\circ$, which would be in conflict
with the CKM fits, favouring $\gamma\sim60^\circ$ 
[\ref{BUPAST},\ref{Lacker},\ref{ref:haricot}]. 
Unfortunately, the present experimental uncertainties do not yet allow us 
to draw definite conclusions, but the picture should improve significantly 
in the future. 

An efficient way to represent the situation in the $\rm B\to\pi {\rm K}$ system 
is provided by allowed regions in the $R_{({\rm c,n})}$--$A_0^{({\rm c,n})}$
planes [\ref{FlMa1},\ref{FlMa2}], which can be derived 
within the Standard Model
and allow a direct comparison with the experimental data. A complementary
analysis in terms of $\gamma$ and $\delta_{\rm c,n}$ was performed in 
Ref.~[\ref{ital-corr}]. 
Another recent $\rm B\to\pi {\rm K}$ study can be found in 
Ref.~[\ref{Gronau:2001cj}], where the $R_{\rm (c)}$ were calculated for given 
values of $A_0^{\rm (c)}$ as functions of $\gamma$, and were compared with 
the B-factory data. In order to analyse $\rm B\to\pi K$ modes, also certain
sum rules may be useful [\ref{matias}].

\subsubsection{The ${\rm B}_d\to\pi^+\pi^-$ and the ${\rm B}_s\to {\rm K}^+{\rm K}^-$ decays}
As can be seen from the corresponding Feynman diagrams, ${\rm B}_s\to {\rm K}^+{\rm K}^-$ 
is related to ${\rm B}_d\to\pi^+\pi^-$ through an interchange of all down and 
strange quarks. The decay amplitudes read as follows [\ref{Fleischer:1999pa}]:
\begin{equation}\label{Bdpipi-ampl0}
A({\rm B}_d^0\to\pi^+\pi^-)\propto\left[e^{i\gamma}-d e^{i\theta}\right],\quad
A({\rm B}_s^0\to {\rm K}^+{\rm K}^-)\propto
\left[e^{i\gamma}+\left(\frac{1-\lambda^2}{\lambda^2}\right)
d'e^{i\theta'}\right],
\end{equation}
where the CP-conserving strong amplitudes $d e^{i\theta}$ and 
$d'e^{i\theta'}$ measure, sloppily speaking, ratios of penguin to tree 
amplitudes in ${\rm B}_d^0\to\pi^+\pi^-$ and ${\rm B}_s^0\to {\rm K}^+{\rm K}^-$, respectively. 
Using these general parametrizations, we obtain expressions for the 
direct and mixing-induced CP asymmetries of the following kind:
\begin{equation}\label{Bpipi-obs}
{\cal A}_{\rm CP}^{\rm dir}({\rm B}_d\to\pi^+\pi^-)=
\mbox{function}(d,\theta,\gamma),\,
{\cal A}_{\rm CP}^{\rm mix}({\rm B}_d\to\pi^+\pi^-)=
\mbox{function}(d,\theta,\gamma,\phi_d=2\beta)
\end{equation}
\begin{equation}\label{BsKK-obs}
{\cal A}_{\rm CP}^{\rm dir}({\rm B}_s\to {\rm K}^+{\rm K}^-)=
\mbox{function}(d',\theta',\gamma),\,
{\cal A}_{\rm CP}^{\rm mix}({\rm B}_s\to {\rm K}^+{\rm K}^-)=
\mbox{function}(d',\theta',\gamma,\phi_s\approx0).
\end{equation}

Consequently, we have four observables at our disposal, depending on six 
``unknowns''. However, since ${\rm B}_d\to\pi^+\pi^-$ and ${\rm B}_s\to {\rm K}^+{\rm K}^-$ are 
related to each other by interchanging all down and strange quarks, the 
$U$-spin flavour symmetry of strong interactions implies
\begin{equation}\label{U-spin-rel}
d'e^{i\theta'}=d\,e^{i\theta}.
\end{equation}
Using this relation, the four observables in (\ref{Bpipi-obs},\ref{BsKK-obs}) 
depend on the four quantities $d$, $\theta$, $\phi_d=2\beta$ and $\gamma$, 
which can hence be determined [\ref{Fleischer:1999pa}]. 
The theoretical accuracy is
only limited by the $U$-spin symmetry, as no dynamical assumptions about
rescattering processes have to be made. Theoretical considerations give us
confidence into (\ref{U-spin-rel}), as it does not receive $U$-spin-breaking 
corrections in factorization [\ref{Fleischer:1999pa}]. 
Moreover, we may also obtain 
experimental insights into $U$-spin breaking 
[\ref{Fleischer:1999pa},\ref{gronau-U-spin}]. 

The $U$-spin arguments can be minimized, if the 
${\rm B}^0_d$--$\overline{\rm B}^0_d$ 
mixing phase $\phi_d=2\beta$, which can be fixed through 
${\rm B}_d\to J/\psi {\rm K}_{\rm S}$, is used as an input. The observables 
${\cal A}_{\rm CP}^{\rm dir}({\rm B}_d\to\pi^+\pi^-)$ and
${\cal A}_{\rm CP}^{\rm mix}({\rm B}_d\to\pi^+\pi^-)$ allow us then to 
eliminate the strong phase $\theta$ and to determine $d$ as a function of
$\gamma$. Analogously, ${\cal A}_{\rm CP}^{\rm dir}({\rm B}_s\to {\rm K}^+{\rm K}^-)$ and 
${\cal A}_{\rm CP}^{\rm mix}({\rm B}_s\to {\rm K}^+{\rm K}^-)$ 
allow us to eliminate 
the strong phase $\theta'$ and to determine $d'$ as a function of
$\gamma$. The corresponding contours in the $\gamma$--$d$
and $\gamma$--$d'$ planes can be fixed in a {\it theoretically clean} way.
Using now the $U$-spin relation $d'=d$, these contours allow the 
determination both of the CKM angle $\gamma$ and of the hadronic quantities 
$d$, $\theta$, $\theta'$; for a detailed illustration, 
see Ref.~[\ref{Fleischer:1999pa}].
This approach is very promising for run II of the Tevatron and the experiments
of the LHC era, where experimental accuracies for $\gamma$ of 
${\cal O}(10^\circ)$ [\ref{FERMILAB}] and ${\cal O}(1^\circ)$ [\ref{LHCB}] 
may be achieved, respectively. It should be emphasized that not only 
$\gamma$, but also the hadronic parameters $d$, $\theta$, $\theta'$ are of 
particular interest, as they can be compared with theoretical predictions, 
thereby allowing valuable insights into hadron dynamics. For other recently 
developed $U$-spin strategies, the reader is referred 
to~[\ref{Gronau:2000md},\ref{U-spin-other}].

\subsubsection{The ${\rm B}_d\to\pi^+\pi^-$ and the 
${\rm B}_d\to \pi^\mp {\rm K}^\pm$ 
decays and implications for ${\rm B}_s\to {\rm K}^+{\rm K}^-$ decay}
A variant of the ${\rm B}_d\to \pi^+\pi^-$, ${\rm B}_s\to {\rm K}^+{\rm K}^-$ approach was developed
for the $e^+e^-$ B-factories [\ref{U-variant}], where ${\rm B}_s\to {\rm K}^+{\rm K}^-$ is not 
accessible: as ${\rm B}_s\to {\rm K}^+{\rm K}^-$ and ${\rm B}_d\to\pi^\mp {\rm K}^\pm$ are related to each 
other through an interchange of the $s$ and $d$ spectator quarks, we may 
replace the ${\rm B}_s$ mode approximately through its ${\rm B}_d$ counterpart, which has 
already been observed by BaBar, Belle and CLEO. Following these lines and 
using experimental information on the CP-averaged ${\rm B}_d\to\pi^\mp {\rm K}^\pm$ and 
${\rm B}_d\to\pi^+\pi^-$ branching ratios, the relevant hadronic penguin parameters 
can be constrained, implying certain allowed regions in observable space
[\ref{FlMa2}]. An interesting situation arises now in view of the recent 
B-factory measurements of CP violation in ${\rm B}_d\to\pi^+\pi^-$, 
allowing us to obtain new constraints on $\gamma$ as a function of the 
${\rm B}^0_d$--$\overline{\rm B}^0_d$ mixing phase $\phi_d$, which is 
fixed through 
${\cal A}_{\rm CP}^{\rm mix}({\rm B}_d\to J/\psi {\rm K}_{\rm S})$ 
up to a twofold ambiguity, $\phi_d\sim 51^\circ$ or $129^\circ$.
If we assume that ${\cal A}_{\rm CP}^{\rm mix}({\rm B}_d\to \pi^+\pi^-)$ is 
positive, as indicated by recent Belle data, and that $\phi_d$ is in 
agreement with the ``indirect'' fits of the unitarity triangle, i.e.\
$\phi_d\sim 51^\circ$, also the corresponding values for $\gamma$ 
around $60^\circ$ can be accommodated. On the other hand, for the second 
solution $\phi_d\sim129^\circ$, we obtain a gap around $\gamma\sim60^\circ$,
and could easily accommodate values for $\gamma$ larger than~$90^\circ$.  
Because of the connection between the two solutions for $\phi_d$ and the 
resulting values for $\gamma$, it is very desirable to resolve the twofold 
ambiguity in the extraction of $\phi_d$ directly. 
As far as ${\rm B}_s\to {\rm K}^+{\rm K}^-$ is concerned, the data on the 
CP-averaged ${\rm B}_d\to \pi^+\pi^-$, ${\rm B}_d\to\pi^\mp {\rm K}^\pm$
branching ratios imply a very constrained allowed region in the space of 
${\cal A}_{\rm CP}^{\rm mix}({\rm B}_s\to {\rm K}^+{\rm K}^-)$ and 
${\cal A}_{\rm CP}^{\rm dir}({\rm B}_s\to {\rm K}^+{\rm K}^-)$ within the 
Standard Model, thereby providing a narrow target range for run II of the 
Tevatron and 
the experiments of the LHC era [\ref{FlMa2}]. Other recent studies related to
${\rm B}_d\to\pi^+\pi^-$ can be found in 
Refs.~[\ref{Gronau:2001cj},\ref{Bpipi-recent}].


\boldmath
\subsection{Determining  \boldmath{$\,\gamma$} with QCD factorization}
\label{sec:benekesec}
\unboldmath

{\it M. Beneke \footnote{
M.~Beneke would like to thank Gerhard Buchalla, Matthias Neubert and Chris
Sachrajda for collaborations on the topics discussed in this article.}}

\subsubsection{Outline of the method}
The QCD factorization approach [\ref{Beneke:1999br_bis},\ref{BBNS2}] 
puts the well-known factorization
ansatz [\ref{Wirbel:1985ji}] 
for hadronic two-body B decay matrix elements on a firm
theoretical basis. It replaces the factorization ansatz by a
factorization formula that includes radiative corrections and
spectator scattering effects. Where it can be justified, 
the factorization ansatz emerges in the
simultaneous limit, when $m_b$ becomes large and when radiative
corrections are neglected.

The QCD factorization approach uses heavy 
quark expansion methods ($m_b\gg\Lambda_{\rm QCD}$) and 
soft-collinear factorization (particle energies $\gg\Lambda_{\rm QCD}$) 
to compute the matrix elements 
$\langle f|O_i|\bar {\rm B}\rangle$ relevant to hadronic B decays 
in an expansion in $1/m_b$ and 
$\alpha_s$. Only the leading term in $1/m_b$ assumes a simple 
form. The basic formula is 
\begin{eqnarray}
\label{fact}
\langle M_1 M_2|O_i|\overline{\rm B}\rangle 
&\hspace*{-0.2cm}=&\hspace*{-0.2cm} 
F^{B\to M_1}(0)\int_0^1 \!\!du\,T^I(u)
\Phi_{M_2}(u) \nonumber\\[0.0cm]
&&\hspace*{-2.8cm}
+\!\int \!d\xi du dv \,T^{II}(\xi,u,v)\,\Phi_B(\xi)\Phi_{M_1}(v) 
\Phi_{M_2}(u),
\end{eqnarray}
where $F^{B\to M_1}$ is a (non-perturbative) form factor, 
$\Phi_{M_i}$ and $\Phi_B$ are light-cone distribution 
amplitudes and $T^{I,II}$ are perturbatively calculable 
hard scattering kernels. Although not strictly proven to all orders 
in perturbation theory, the formula is presumed to be 
valid when both final state mesons are light. ($M_1$ is 
the meson that picks up the spectator quark from the B meson.) The formula 
shows that there is no long-distance interaction between the
constituents of the meson $M_2$ and the $(B M_1)$ system at leading 
order in $1/m_b$. This is the precise meaning of factorization. 
For a detailed discussion of (\ref{fact}) I refer 
to~[\ref{Beneke:2001ev},\ref{BBNS2}]. 
A summary of results that have been obtained in the QCD factorization 
approach is given in~[\ref{KEK}]. 

Factorization as embodied by (\ref{fact}) is not 
expected to hold at subleading order in $1/m_b$. 
Some power corrections related to scalar currents 
are enhanced by factors such as 
$m_\pi^2/((m_u+m_d)\Lambda_{\rm QCD})$. Some corrections 
of this type, in particular those related to scalar penguin 
amplitudes nevertheless appear to be calculable and turn out to be
important numerically. On the other hand, 
attempts to compute subleading 
power corrections to hard spectator-scattering in perturbation theory 
usually result in infrared divergences, which signal the breakdown 
of factorization. 
At least these effects should be estimated and included into the 
error budget. All weak 
annihilation contributions belong to this class of effects and often
constitute the dominant source of theoretical error.

\subsubsection{Uses of QCD factorization}


If the corrections to (\ref{fact}) were negligible and if all the
quantities in (\ref{fact}) were known or computed with sufficient
accuracy, the QCD factorization approach would allow one to determine
directly weak CP-violating phases from branching fraction or CP
asymmetry measurements, if the corresponding decay has two
interfering amplitudes with different phases. In practice, depending
on the particular decay mode, one is 
often far from this ideal situation. Depending on the theoretical
uncertainty or one's confidence in theoretical error estimates, I can
imagine the following uses of the QCD factorization approach, where in
ascending order one makes stronger use of theoretical rather than
experimental input.

\begin{itemize}
\item[1)] Many strategies to determine $\gamma$ are based on relating
  the strong interaction dynamics of different decay channels such
  that a sufficient set of measurements yields the weak phase 
  together with the strong amplitudes (see the contributions by
  Fleischer and Rosner in this Chapter). QCD factorization can
  complement this approach where it has to rely on assumptions. For
  instance, it may be used to estimate SU(3) flavour symmetry 
  breaking or to provide an estimate of small contributions to the decay
  amplitude that one would otherwise have to neglect to make use of
  the amplitude relations.
\item[2)] The QCD factorization approach generically predicts small
  strong rescattering phases, because rescattering is either
  perturbatively loop-suppressed, or power-suppressed in the
  heavy-quark limit. (Exceptions to the rule of small phases occur
  when the leading term in the $\alpha_s$- or $1/m_b$-expansion is
  suppressed, for instance by small Wilson coefficients.) Even if the
  quantitative prediction of the strong phases turns out to be
  difficult, the qualitative result of small phases can be used to
  sharpen the bounds on $\gamma$ that follow from some amplitude
  relations, or to turn the bounds into determinations of $\gamma$. An
  example of this in the context of a method suggested 
  in~[\ref{Neubert:1998jq}] 
  will be discussed below.
\item[3)] For predicting CP violation the ratio of two strong
  interaction amplitudes, $P/T$, (often a ratio of a pure 
  penguin and a dominantly tree amplitude, which are multiplied by
  different weak phases) is a particularly important quantity. While 
  $P/T$ is computed in the QCD factorization approach, one may decide
  to use only the calculation of the absolute value $|P/T|$ and 
  to dismiss the quantitative phase information. The rationale for
  this procedure could be that the calculation of the imaginary part is usually
  less accurate than the real part, because a one-loop calculation
  determines the phase only to leading order. For the same reason the
  value of the phase is more sensitive to uncalculable power
  corrections. In this procedure the phase information must be
  provided by an additional measurement, for instance a direct CP
  asymmetry. 
\item[4)] The full information of the QCD factorization approach is
  employed to compute two-body branching fractions as functions of 
  the parameters of the CKM matrix. Since the $b$ quark mass is not 
  very large it will be important to estimate the theoretical error
  from power corrections.
\end{itemize}
The development of QCD factorization has not yet reached the stage
where one can decide which of these strategies will turn out to be
most useful. (The strategy of choice obviously also depends on the
amount of experimental information available that would allow one to
drop one or the other piece of theoretical input.) 
Calculations of $\pi\pi$ and $\pi K$ final states showed
[\ref{Beneke:2001ev}] that one obtains naturally the right magnitude of penguin
and tree amplitudes. The accuracy of the calculation of strong phases
is less clear at present, but forthcoming measurements of direct CP 
asymmetries should shed light on this question. The current limits
favour small strong phases, but a quantitative comparison may require
a complete next-to-leading order calculation of the absorptive parts
of the amplitudes. It will also be important to clarify the relevance
of weak annihilation effects in the decay amplitudes. While the
current data do not favour the assumption of large annihilation
contributions, they can also not yet be excluded. Bounds on rare
annihilation-dominated decays will limit the corresponding amplitudes.

\subsubsection{Results related to the determination of $\,\gamma$}

The possibility to determine the CP-violating angle $\gamma$ 
by comparing the calculation of branching fractions 
into $\pi\pi$ and $\pi {\rm K}$ final states with the corresponding 
data has been investigated in~[\ref{Beneke:2001ev}] 
(see also [\ref{Du:2001hr}]). In the following I summarize the main
results, referring to~[\ref{Beneke:2001ev}] for details and to 
[\ref{Beneke:2002nj},\ref{Neubert:2002tf}] for partial 
updates of the analysis of~[\ref{Beneke:2001ev}].

\subsubsection*{$\gamma$ from CP-averaged charged $\rm B\to \pi {\rm K}$ decay}

The ratio 
\begin{equation}\label{Rst}
   R_* = 
   \frac{\mbox{Br}(\rm B^+\to\pi^+ {\rm K}^0)+\mbox{Br}(\rm B^-\to\pi^-\bar {\rm K}^0)}
        {2[\mbox{Br}(\rm B^+\to\pi^0 {\rm K}^+)+\mbox{Br}(\rm B^-\to\pi^0 {\rm K}^-)]}
   \,,
\end{equation}
currently measured as $0.71\pm 0.10$, provides a useful bound on 
$\gamma$~[\ref{Neubert:1998jq},\ref{neubert}]. 
The theoretical expression is
\begin{eqnarray}\label{expr}
   R_*^{-1} &=& 1 + 2\bar\varepsilon_{3/2}\cos\phi\,(q-\cos\gamma)
    + \bar\varepsilon_{3/2}^2 (1 - 2q\cos\gamma + q^2) \nonumber\\
&<& \left( 1 + \bar\varepsilon_{3/2}\,|q-\cos\gamma| \right)^2
   + \bar\varepsilon_{3/2}^2\sin^2\!\gamma,
\end{eqnarray}
where $\bar \epsilon_{3/2} e^{i\phi}$ is a tree-to-penguin ratio, 
whose magnitude can be fixed with SU(3) symmetry, 
and $q$ an electroweak penguin 
contribution, which can be determined theoretically. (In this
expression, a rescattering contribution $\varepsilon_a$, 
which QCD factorization predicts to be small, is neglected.)
The inequality is obtained by
allowing the relative strong phase $\phi$ to take any value. If $R_*$ is
smaller than one, the bound implies an exclusion region for 
$\cos\gamma$. The bound can be considerably sharpened, and the 
requirement $R_*<1$ relaxed, if the phase is
known to be small. QCD factorization as well as bounds on direct CP
asymmetries suggest that $\cos\phi>0.9$. In [\ref{Beneke:2001ev}] it was shown
that assuming the more conservative range  
$\cos\phi>0.8$, the measurement of $R_*$ combined with
$|V_{ub}/V_{cb}|$ provides a determination of $\gamma$ with a
theoretical error of about $10^\circ$, if $R_*$ is close to 1.

\subsubsection*{$\gamma$ from ${\rm B}_d(t)\to\pi^+\pi^-$ decay}

The QCD factorization approach allows us to interpret directly the 
mixing-induced and direct CP asymmetry in ${\rm B}_d\to\pi^+\pi^-$ decay without 
resort to other decay modes, 
since the tree and penguin amplitudes are both computed. 
The time-dependent asymmetry is defined by
\begin{eqnarray}
   A_{\rm CP}^{\pi\pi}(t)
   &=& \frac{\mbox{Br}({\rm B}^0(t)\to\pi^+\pi^-)
             - \mbox{Br}(\overline{\rm B}^0(t)\to\pi^+\pi^-)}
            {\mbox{Br}({\rm B}^0(t)\to\pi^+\pi^-)
             + \mbox{Br}(\overline{\rm  B}^0(t)\to\pi^+\pi^-)} \nonumber\\[0.2cm]
   &=& - S_{\pi\pi} \sin(\Delta M_B\,t)
    + C_{\pi\pi} \cos(\Delta M_B\,t). 
\end{eqnarray}
Assuming that the $\rm B\bar B$ mixing phase is determined experimentally 
via the mixing-induced CP asymmetry in ${\rm B}_d(t)\to J/\psi {\rm K}$ decay,
both $S_{\pi\pi}$ and $C_{\pi\pi}$ are measures of CP violation in the
decay amplitude and determine $\gamma$. In~[\ref{Beneke:2001ev}] it was 
shown that even a moderately accurate measurement of $S_{\pi\pi}$ 
translates into a stringent 
constraint in the $(\bar\rho,\bar\eta)$ plane. The predicted
correlation between $S_{\pi\pi}$ and $C_{\pi\pi}$ 
is shown in~[\ref{Beneke:2002nj}].

\subsubsection{$\gamma$ from CP-averaged $\rm B\to \pi {\rm K},\,\pi\pi$ decay}

\begin{figure}[t]
\begin{center}
\epsfig{file=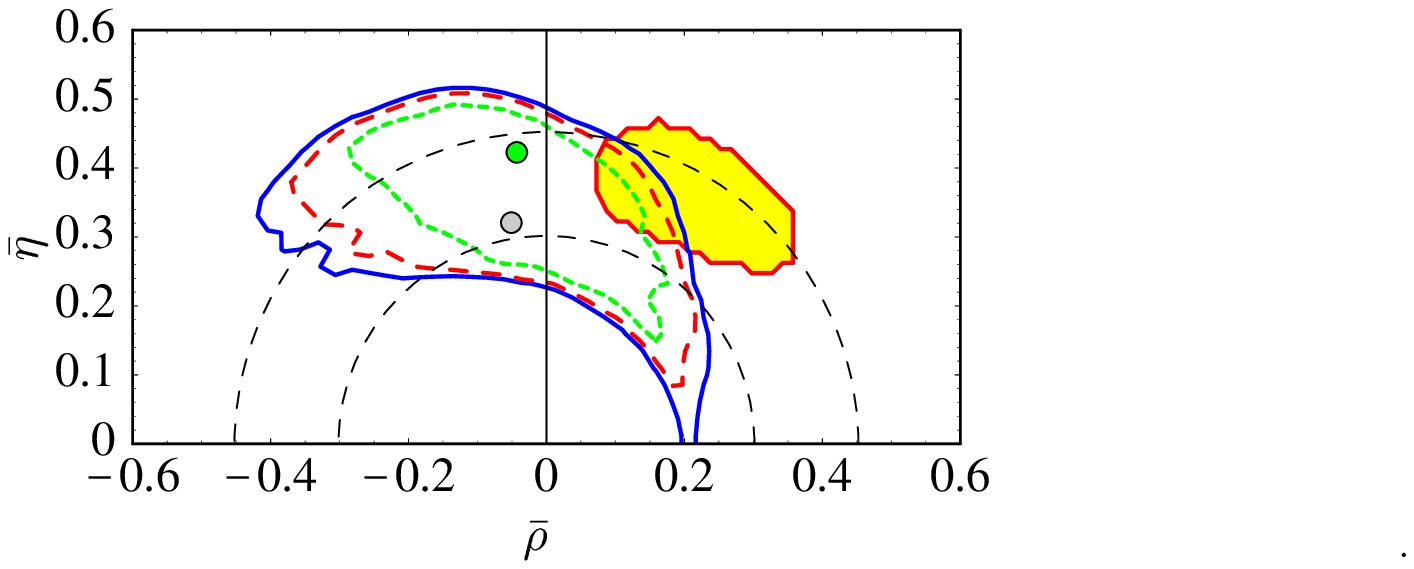,height=1.6in}
\caption{\it 95\% (solid), 90\% (dashed) and 68\% (short-dashed) confidence level 
contours in the $(\bar\rho,\bar\eta)$ plane obtained from a global 
fit to the CP averaged $\rm B\to\pi {\rm K},\pi\pi$ branching fractions, using 
the scanning method as described in~[\ref{Lacker}]. The darker dot shows the
overall best fit, whereas the lighter dot indicates the best fit for the
default hadronic parameter set. 
The light-shaded region indicates the region 
preferred by the standard global fit~[\ref{Lacker}], 
including  the direct measurement of $\sin(2\beta)$.}
\label{fig:pik}
\end{center}
\end{figure}

Since the branching fractions are computed as functions of 
the Wolfenstein parameters 
$(\bar\rho,\bar\eta)$, one can perform a fit of 
$(\bar\rho,\bar\eta)$ to the six measured CP-averaged 
$\rm B\to\pi\pi, \pi {\rm K}$ branching fractions. 
The result of this fit is 
shown in Fig.~\ref{fig:pik} based on 
the data as of May 2002. (The details of the
fit procedure can be found in~[\ref{Beneke:2001ev}], the input data in 
[\ref{Beneke:2002nj}]). The result of the fit is 
consistent with the standard fit 
based on meson mixing and $|V_{ub}|$. 
However, the $\pi\pi$, $\pi {\rm K}$ data persistently exhibit 
a preference for 
$\gamma$ near $90^\circ$, or, for smaller $\gamma$, smaller 
$|V_{ub}|$. The significance and interpretation of this observation
remains to be clarified.

\subsubsection{Weak annihilation}

Weak annihilation contributions are power-suppressed and not
calculable in the QCD factorization approach. (This is one of the
important differences between the QCD factorization approach and the
pQCD approach described by Y.~Keum in this Chapter.) The results discussed
above include an estimate of annihilation effects together with an 
uncertainty derived from a $\pm100\%$ variation of this estimate, 
encoded in the constraint $|\rho_A|<1$ for a certain weak
annihilation parameter~[\ref{Beneke:2001ev}]. 
Since this constraint is often a key factor in the
overall theoretical uncertainty estimate, it will be important to
obtain experimental information on weak annihilation. The current 
data on $\pi\pi$ and $\pi {\rm K}$ final states do not favour large
annihilation contributions, but also do not exclude this
possibility. The upper limits on annihilation-dominated charmless
decays are not yet tight enough to provide interesting constraints. 
However, we can adapt the estimate of annihilation contributions 
to $\bar {\rm B}_d\to \rm D^+ \pi^-$ performed in [\ref{BBNS2}] to the 
annihilation-dominated decay $\bar {\rm B}_d\to {\rm D}_s^+ {\rm K}^-$, recently
observed with a branching fraction $(3.8\pm 1.1)\cdot 10^{-5}$ [\ref{Bellebab}]. 
This results in a branching fraction estimate 
of $1.2\cdot 10^{-5}$ for central
parameters, or an upper limit 
$5\cdot 10^{-5}$ upon assigning a $100\%$ error to the
annihilation amplitude. While the annihilation mechanism in 
$\bar {\rm B}_d\to {\rm D}_s^+ {\rm K}^-$ is not identical to the 
dominant penguin annihilation term in $\rm B\to \pi {\rm K}$ decay, 
the comparison nevertheless supports the phenomenological treatment of 
annihilation suggested in~[\ref{Beneke:2001ev},\ref{BBNS2}].

\boldmath
\subsection{ $\rm B\to {\rm K}\pi$, charming penguins and the 
extraction of $\gamma$}
\unboldmath

{\it M. Ciuchini, E. Franco, G. Martinelli, M. Pierini and L. Silvestrini}

\subsubsection{Main formulae}
In this section we collect the main formulae for the amplitudes of
$\rm B\to {\rm K}\pi,\pi\pi$, introducing the parametrization used in the
analysis. We refer the reader to the literature for any detail
on the origin and the properties of these
parameters~[\ref{Ciuchini:1997hb},\ref{Ciuchini:1997rj},\ref{Buras:1998ra},\ref{Ciuchini:2001gv}]. From Ref.~[\ref{Buras:1998ra}], one
reads
\begin{eqnarray}
A({\rm B}_d\to {\rm K}^+\pi^-)&=& \frac{G_F}{\sqrt{2}}\Big(\lambda_t^s P_1-\lambda_u^s (E_1-P_1^{GIM})\Big)\nn\\
A(\rm B^+\to {\rm K}^+\pi^0)&=&\frac{G_F}{2}\Big(\lambda_t^s P_1-\lambda_u^s (E_1+E_2-P_1^{GIM}+A_1)\Big)+\Delta A\nn\\
A(\rm B^+\to {\rm K}^0\pi^+)&=& \frac{G_F}{\sqrt{2}}\Big(-\lambda_t^s P_1+\lambda_u^s (A_1-P_1^{GIM})\Big)+\Delta A\nn\\
A({\rm B}_d\to {\rm K}^0\pi^0)&=& \frac{G_F}{2}\Big(-\lambda_t^s P_1-\lambda_u^s (E_2+P_1^{GIM})\Big)+\Delta A\\
A({\rm B}_d\to \pi^+\pi^-)&=&\frac{G_F}{\sqrt{2}}\Big(\lambda_t^d (P_1+P_3)-\lambda_u^d (E_1+A_2-P_1^{GIM})-P_3^{GIM}\Big)\nn\\
A({\rm B}_d\to \pi^+\pi^0)&=& \frac{G_F}{2}\Big(-\lambda_u^d(E_1+E_2)\Big)+\Delta A\nn\\
A({\rm B}_d\to \pi^0\pi^0)&=&\frac{G_F}{2}\Big(-\lambda_t^s (P_1+P_3)-\lambda_u^s (E_2+P_1^{GIM}+P_3^{GIM}-A_2)\Big)+\Delta A\nn\,,
\end{eqnarray}

\vspace{2mm}

\noindent
where $\lambda_{q^\prime}^q=V_{q^\prime q}V_{q^\prime b}^*$. 
Neglecting the $A_i$, these parameters can be rewritten as
\begin{eqnarray}
&&E_1=a_1^c A_{\pi {\rm K}}\,,\quad E_2=a_2^c A_{K\pi}\,,\quad A_1=A_2=0\,,
\nn\\
&&P_1=a_4^c  A_{\pi K}+\tilde P_1\,,\quad P_1^{GIM}=(a_4^c-a_4^u)  A_{\pi K}+\tilde P_1^{GIM}\,.
\label{eq:amps}
\end{eqnarray}
The terms proportional to $a_i^q$ give the parameters computed in the limit $m_b\to\infty$ using QCD
factorization. Their definition, together with those of $A_{\pi K}$, $A_{K\pi}$, etc., can be found for instance in
Refs.~[\ref{Beneke:1999br_bis},\ref{BBNS2},\ref{Beneke:2001ev}], 
although power-suppressed terms included there, proportional to 
the chiral factors $r^\chi_{K,\pi}$,
should be discarded in eqs.~(\ref{eq:amps}). In our case, in fact, terms of $O(\Lambda_{QCD}/m_b)$
are accounted for by two phenomenological parameters: the charming-penguin parameter
$\tilde P_1$ and the GIM-penguin parameter $\tilde P_1^{GIM}$ . In $\rm B\to {\rm K}\pi$  there are no other contributions,
once flavour $SU(2)$ symmetry is used and few other doubly Cabibbo-suppressed terms, including corrections to
emission parametes $E_1$ and $E_2$, some annihilations ($A_1$) and the Zweig-suppressed contractions ($\Delta A$), are
neglected~[\ref{Buras:1998ra}]. On the contrary, 
further power-suppressed terms ($A_2$, $P_3$, $P_3^{GIM}$) enter the
$\rm B\to \pi\pi$ amplitudes, all with the same power of the Cabibbo angle. Therefore, these modes are subject to a larger uncertainty
than the $\rm B\to {\rm K}\pi$ ones.
\begin{table}
\begin{center}
\begin{tabular}{ccccc}
$\vert V_{cb}\vert\!\times\! 10^{3}$ & $\vert V_{ub}\vert\!\times\! 10^{3}$ &
$\hat B_K$ & $F_{B_d}\sqrt{B_d}$ (MeV) & $\xi$\\
\hline
$40.9\!\pm\! 1.0$&$3.70\!\pm\! 0.42$ & $0.86\!\pm\! 0.06\!\pm\!  0.14$& $230\!\pm\! 30\!\pm\! 15$ &
$1.16\!\pm\!0.03\!\pm\!0.04$\\[4pt]
$f_K(M_K^2)$ & ${\cal B}(\rm K^+\pi^-)\!\times\! 10^6  $ &  ${\cal B}({\rm K}^+\pi^0)\!\times\! 10^6  $ & ${\cal B}({\rm K}^0\pi^+)\!\times\! 10^6  $&
 ${\cal B}(K^0\pi^0)\!\times\! 10^6  $\\
\hline
$0.32\pm0.12$ & $18.6 \pm 1.1$  & $11.5 \pm 1.3$  & $17.9 \pm 1.7$ & $8.9 \pm 2.3$\\[4pt]
$f_\pi(M_\pi^2)$ & ${\cal B}(\pi^+\pi^-)\!\times\! 10^6 $ &  ${\cal B}(\pi^+\pi^0)\!\times\! 10^6  $ & ${\cal B}(\pi^0\pi^0)\!\times\!
 10^6 $ & \\
\hline
$0.27\pm0.08$&$5.2 \pm 0.6$ &$4.9\pm 1.1$ & $<\! 3.4\,${\small\it BaBar} &\\[4pt]
\end{tabular}

\vspace{-1mm}

\caption{\it Values of the input parameters used in our analysis. 
The CP-averaged branching ratios ${\cal B}$ are taken
from Ref.~[\ref{Patterson:fpcp}].}
\label{tab:expbr}
\end{center}
\end{table}

\vspace{-1mm}

Using the inputs collected in Table~\ref{tab:expbr},
we fit the value of the complex parameter $\tilde P_1=(0.13\pm 0.02)\, e^{\pm i (114\pm 35)^\circ}$. Notice that 
the sign of the phase is practically not constrained by the data. This result is
almost independent of the inputs used for the CKM parameters $\rho$ and $\eta$, namely whether
these parameters are taken from the usual unitarity triangle analysis 
(UTA)~[\ref{ref:haricot},\ref{Ciuchini:2001zf}] or only
the constraint from $\vert V_{ub}/V_{cb}\vert$ is used.

\begin{table}[htbp]
\begin{center}
\begin{tabular}{|c|cc|cc|} 
\hline
  Mode  & \multicolumn{2}{c}{UTA} & 
          \multicolumn{2}{c|}{$\vert V_{ub}/V_{cb}\vert$} \\ 
        & $\cal B$ ($10^{-6}$) & $\vert{\cal A}_{CP}\vert$ & 
        $\cal B$ ($10^{-6}$) & $\vert{\cal A}_{CP}\vert$ \\
\hline
 $\pi^+\pi^-$ & $8.9 \pm 3.3$ & $0.37\pm0.17$    & $8.7 \pm 3.6$ & $0.39\pm0.20$    \\
 $\pi^+\pi^0$ & $5.4 \pm 2.1$ & -- & $5.5 \pm 2.2$ & --    \\
 $\pi^0\pi^0$ & $0.44\pm 0.13$ & $0.61\pm0.26$   & $0.69 \pm 0.27$ & $0.45\pm0.27$  \\
\hline
 ${\rm K}^+\pi^-$   & $18.4 \pm 1.0$ & $0.21\pm0.10$   & $18.8 \pm 1.0$ & $0.21\pm0.12$   \\
 ${\rm K}^+\pi^0$   & $10.3 \pm 0.9$ & $0.22\pm0.11$   & $10.7 \pm 1.0$ & $0.22\pm0.13$   \\
 ${\rm K}^0\pi^+$   & $19.3 \pm 1.2$ & $0.00\pm0.00$   & $18.1 \pm 1.5$ & $0.00\pm0.00$   \\
 ${\rm K}^0\pi^0$   &  $8.7 \pm 0.8$ & $0.04\pm0.02$   &  $8.2 \pm 1.2$ & $0.04\pm0.03$   \\
\hline
\end{tabular}
\caption{\it Predictions for CP-averaged branching ratios 
$\cal B$ and absolute value of the
CP asymmetries $\vert {\cal A}_{CP}\vert$. The left (right) columns show
results obtained using constraints on the CKM parameters 
$\rho$ and $\eta$ obtained from the UTA
(the measurement of $\vert V_{ub}/V_{cb}\vert$). 
The last four channels are those used for fitting the charming penguin
parameter $\tilde P_1$.}
\label{tab:results}
\end{center}
\end{table}

\vspace{-2mm}

For the sake of simplicity, we also neglect here the contribution of $\tilde P_1^{GIM}$. The
$\rm B\to {\rm K}\pi$ data do not constrain this parameter very effectively, since its contribution
is doubly Cabibbo suppressed with respect to $\tilde P_1$. The remaining $\pi^+\pi^-$ mode
alone is not sufficient to fully determine the complex parameter $\tilde P_1^{GIM}$. It
is interesting, however, to notice that the GIM-penguin contribution is potentially
able to enhance the ${\cal B}(\rm B\to\pi^0\pi^0)$ up to few 
$\times 10^{-6}$~[\ref{Ciuchini:2001gv}].

Table~\ref{tab:results} shows the predicted values of the CP-averaged
branching ratios ${\cal B}$ and the absolute value of the CP-asymmetries $\vert{\cal A}_{CP}\vert$ for
the $\rm B\to K\pi$ and $\rm B\to\pi\pi$ modes, since the data are not able to fix the sign of asymmetries. 
Charming penguins are able to reproduce the ${\rm K}\pi$ data and are also consistent with the only $\pi\pi$
mode measured so far. It is interesting to notice that the latest measurements improve
the consistency, for a comparison see 
refs.~[\ref{Ciuchini:1997rj},\ref{Ciuchini:2001gv}].

\subsubsection{Remarks on different approaches}
Since the different approaches aiming at evaluating power-suppressed terms contain phenomenological parameters, it is
natural to ask whether, after all, they are equivalent or not, even if the physical mechanism invoked to introduce the
parameters is not the same. To answer this question, it is useful to compute the parameters $\tilde P_1$ and $\tilde
P_1^{GIM}$ within {\it improved} QCD factorization. They read
\begin{equation}
\tilde P_1 = r^\chi_K a_6^c A_{\pi K}+b_3 B_{\pi K}\,,~\tilde P_1^{GIM} = r^\chi_K (a_6^c-a_6^u) A_{\pi K}\,,
\end{equation}
where the functions $a_i^q$ ($b_i$) contain the complex parameter $\rho_H$ ($\rho_A$), see Ref.~[\ref{Beneke:2001ev}] for the
definitions. These two parameters account for chirally-enhanced terms, originating from hard-spectator interactions and annihilations
respectively, which are not computable within the {\it improved} QCD factorization.

The functional dependence of the amplitudes on the phenomenological parameters in the two approaches is different. For instance,
the GIM-penguin parameter is a pure short-distance correction in the {\it improved} QCD factorization, since
the $\rho_H$ dependence cancels out in the difference $a_6^c-a_6^u$. In practice, however, the main contribution of the
phenomenological parameters to the $\rm B\to K\pi$ amplitudes comes from the annihilation term $b_3$, i.e. from $\rho_A$ .
This term behaves effectively as the charming-penguin parameter, enhancing the Cabibbo-favored amplitude.

Notice that a vanishing $\rho_A$ (and $\rho_H$), which turns out to be compatible with the data, does not mean that the
phenomenological contribution is negligible. In fact, the parameters are defined so that the phenomenological terms are
functions of $X_{A(H)}=(1+\rho_{A(H)})\log(m_B/\mu_h)$, where the scale $\mu_h$ is 
assumed to be 0.5~GeV~[\ref{Beneke:2001ev}].

\boldmath
\subsubsection{On the possibility of extracting \boldmath$\gamma$}
\unboldmath

\begin{figure}[htbp] 
\begin{center}
\begin{tabular}{cc}
\includegraphics[width=7.3cm]{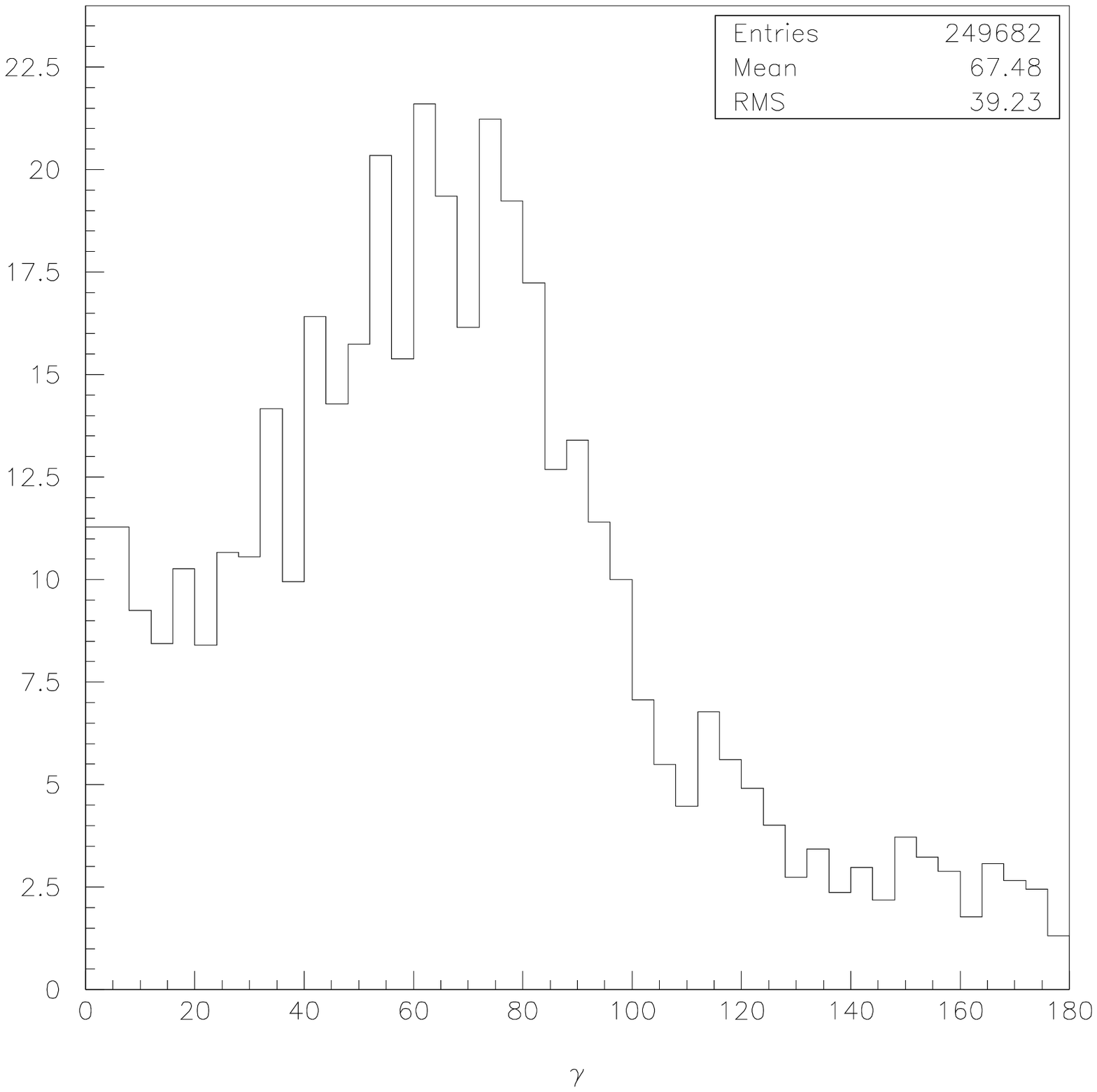} & 
\includegraphics[width=7.3cm]{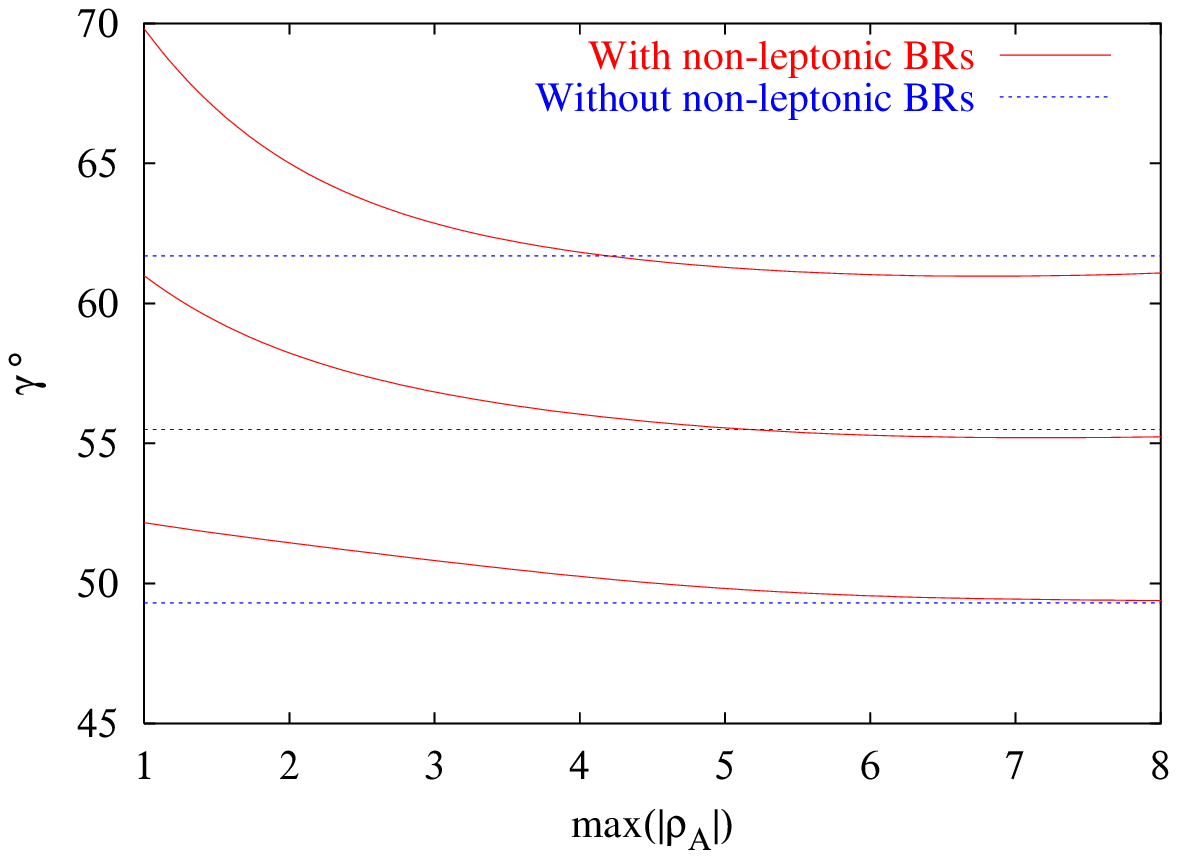}
\end{tabular}

\vspace{-2mm}

\caption{\it Fits of $\gamma$ from $\rm B\to K\pi$ using charming penguins (left) and UTA + {\it improved} QCD factorization as
a function of $\max\vert\rho_A\vert$ (right).}
\label{fig:fig}
\end{center}
\end{figure}

The presence of complex phenomenological parameters in the amplitudes makes the extraction of $\gamma$ very
problematic. Using the $\vert V_{ub}/V_{cb}\vert $-constrained fit, almost any value of $\gamma$ is
allowed, given the uncertainty on $\tilde P_1$, see Fig.~\ref{fig:fig} (left).
This seems a general problem which make us doubt recent claims
proposing non-leptonic B decays as an effective tool for the CKM matrix determination. Even more, we think that the
combination of the constraint from $\rm B\to K\pi$ decays on $\gamma$ with the others can even be misleading. The
reason is very simple: $\gamma$ is looked for through the effect of interefence terms in the branching ratios. The
presence of a competing amplitude with a new phase, i.e. the one containing the phenomenological parameter, makes
the extraction of $\gamma$ much more complicated. Although weak and strong phases can be disentangled in
principle, in practice we checked that not only the task is very difficult now, but the situation improves slowly as
data become more accurate, even when the CP asymmetries will be measured.

Concerning various analyses based on the {\it improved} QCD factorization claiming to
find a ``large'' value of $\gamma\sim 90^\circ$, we just notice that, as far as we know, they all assume the
bound $\vert\rho_A\vert <1$, suggested in 
Ref.~[\ref{Beneke:2001ev}] as a theoretical prejudice and
supported by the observation that even $\vert\rho_A\vert = 0$ produces a good fit to ${\cal B}(\rm B\to K\pi)$.
A better fit, however, can be obtained letting $\vert\rho_A\vert$ take values up to about $3$. As shown in the right plot
of Fig.~\ref{fig:fig}, by doing so, the contribution of the constraint from non-leptonic B decays to a global fit
of $\gamma$ becomes totally negligible. In other words, for $\vert\rho_A\vert\sim 3$, the annihilation amplitude
containing $\rho_A$ becomes competitive with the others, improving the fit to the ${\cal B}$s on the one hand and
weakening the predictivity on $\gamma$ on the other.

\vspace{-2mm}

\boldmath
\subsection{Determination of the weak phases  $\phi_2$ and $\phi_3$ 
from  $\rm B\to \pi\pi,K\pi$ in the  pQCD method}
\unboldmath

{\it Y.-Y. Keum\footnote{ 
Y.-Y. Keum would like to thank  G.~Buchalla
and  members of PQCD working group for
fruitful collaboration and joyful discussions.}.
}


In this section, we focus on the $\rm B \to \pi^{+}\pi^{-}$ and
${\rm K}\pi$ processes, providing promising strategies to determine
the weak phases of $\phi_2$ and $\phi_3$, 
by using the perturbative QCD method.
The perturbative QCD method (pQCD) has 
a predictive power demonstrated sucessfully 
in exclusive two body B-meson decays,
specially in charmless B-meson decay processes[\ref{pQCD}]. 
By introducing parton transverse momenta $k_{\bot}$, we can generate
naturally the Sudakov suppression effect due to resummation of large double 
logarithms $Exp[-{\alpha_s C_F \over 4 \pi} \ln^2({Q^2\over k_{\bot}^2})]$,
which suppress the long-distance contributions in the small $k_{\bot}$ region
and give a sizable average $<k_{\bot}^2> \sim \bar{\Lambda} M_B$. 
This can resolve the end point singularity problem and 
allow the applicability of pQCD to exclusive decays. We found that
almost all of the contribution to the exclusive matrix elements
come from the integration region where $\alpha_s/\pi < 0.3$ and 
the pertubative treatment can be justified.

In the pQCD approach, we can predict the contribution of non-factorizable
term and annihilation diagram on the same basis as the factorizable one.
A folklore for annihilation contributions is that they are negligible
compared to W-emission diagrams due to helicity suppression. 
However the operators $O_{5,6}$ with helicity structure $(S-P)(S+P)$
are not suppressed and give dominant imaginary values, 
which is the main source of strong phase in the pQCD approach.
So we have a large direct CP violation in 
$\rm B \to \pi^{\pm}\pi^{\mp}, {\rm K}^{\pm}\pi^{\mp}$,
since large strong phase comes from 
the factorized annihilation diagram, which can distinguish pQCD from
other models (see the previous two subsections). 

\subsubsection{Extraction of $\phi_2(=\alpha)$ from $\rm B \to \pi^{+}\pi^{-}$ decay}

Even though isospin analysis of $\rm B \to \pi\pi$ can provide a clean way
to determine $\phi_2$, it might be difficult in practice because of
the small branching ratio of $\rm B^0 \to \pi^0\pi^0$.
In reality to determine $\phi_2$, we can use the time-dependent rate
of $\rm B^0(t) \to \pi^{+}\pi^{-}$ including sizable penguin
contributions. In our analysis we use the c-convention.
The amplitude can be written as:
\beq
A(\rm B^0\to \pi^{+}\pi^{-})= 
-(|T_c|\,\,e^{i\delta_T} \, e^{i\phi_3} + |P_c|\, e^{i\delta_P})
\eeq
Penguin term carries a different weak phase than the dominant tree amplitude,
which leads to generalized form of the time-dependent asymmetry.

When we define $\rpp=\overline{Br}({\rm B}^0 \to \pi^{+}\pi^{-})/
\overline{Br}({\rm B}^0\to \pi^{+}\pi^{-})|_{tree}$, 
where $\overline{Br}$ stands for 
a branching ratio averaged over $\rm B^0$ and $\bar{\rm B}^0$, the explicit
expression for $S_{\pi\pi}$ and $C_{\pi\pi}$ are given by:
\beqa
R_{\pi\pi} &=& 1-2\,R_c\, \cos\delta \, \cos(\phi_1 +\phi_2) + R_c^2,  \\
R_{\pi\pi}S_{\pi\pi} &=& \sin2\phi_2 + 2\, R_c \,\sin(\phi_1-\phi_2) \,
\cos\delta - R_c^2 sin2\phi_1, \\
R_{\pi\pi}C_{\pi\pi} &=& 2\, R_c\, \sin(\phi_1+\phi_2)\, \sin\delta.
\eeqa
with $R_c=|P_c/T_c|$ and the strong phase difference
between penguin and tree amplitudes $\delta=\delta_P-\delta_T$.
The time-dependent asymmetry measurement provides two equations for
$C_{\pi\pi}$ and $S_{\pi\pi}$ in terms of $R_c,\delta$ and $\phi_2$.

If we know $R_c$ and $\delta$, then we can determine $\phi_2$ from the
experimental data on $C_{\pi\pi}$ versus $S_{\pi\pi}$. 

\begin{figure}[htbp] 
\begin{center}
\includegraphics[angle=-90,width=7cm]{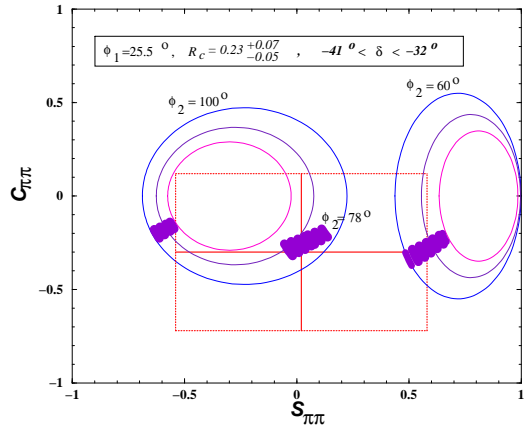} 

\vspace{-2mm}

\caption{\it Plot of $C_{\pi\pi}$ versus $S_{\pi\pi}$  for various values
of $\phi_2$ with $\phi_1=25.5^\circ$, $0.18 < R_c < 0.30$ and $-41^\circ <
\delta < -32^\circ$ in the pQCD method. Here we consider the allowed 
experimental ranges of BaBar measurment whinin $90\%$ C.L. 
Dark areas is allowed
regions in the pQCD method for different $\phi_2$ values.}
\label{fig:cpipi}
\end{center}
\end{figure}

Since the pQCD method provides $R_c=0.23^{+0.07}_{-0.05}$ and $-41^\circ
<\delta<-32^\circ$, the allowed range of $\phi_2$ at present stage is
determined as $55^\circ <\phi_2< 100^\circ$ as shown in Fig. \ref{fig:cpipi}. 
Since we have a relatively large
strong phase in pQCD, 
in contrast to the QCD-factorization ($\delta\sim 0^\circ$), 
we predict large direct CP violation effect of 
$A_{cp}(B^0 \to \pi^{+}\pi^{-}) = (23\pm7) \%$ which will be tested
by more precise experimental measurement in future. 
Since the data by Belle Collaboration [\ref{Bellealpha}] 
is located outside allowed physical regions, 
we only considered in the numerical analysis 
the recent BaBar measurement[\ref{babar}] with $90\%$ C.L. interval
taking into account the systematic errors:
\begin{itemize}
\item[$\bullet$]
$S_{\pi\pi}= \,\,\,\,\, 0.02\pm0.34\pm0.05$ 
\hspace{10mm} [-0.54,\hspace{5mm} +0.58]
\item[$\bullet$]
$C_{\pi\pi}=-0.30\pm0.25\pm0.04$ 
\hspace{10mm} [-0.72,\hspace{5mm} +0.12].
\end{itemize}
The central point of BaBar data corresponds to 
$\phi_2 = 78^\circ$ in the pQCD method. 

\subsubsection{Extraction of $\phi_3(=\gamma)$ 
from $\rm B^0 \to {\rm K}^{+}\pi^{-}$ and $\rm B^{+}\to {\rm K}^0\pi^{+}$ decays}

By using tree-penguin interference in $\rm B^0\to {\rm K}^{+}\pi^{-}(\sim
T^{'}+P^{'})$ versus $\rm B^{+}\to {\rm K}^0\pi^{+}(\sim P^{'})$, CP-averaged
$\rm B\to K\pi$ branching fraction may lead to non-trivial constraints
on the $\phi_3$ angle~[\ref{FM},\ref{Neubert:1998pt},\ref{Neubert:1998jq}]. 
In order to determine $\phi_3$, we need one more useful information 
on CP-violating rate differences[\ref{Gronau:2001cj}].
Let's introduce the following observables :
\beqa
R_K &=&{\overline{Br}(\rm B^0\to {\rm K}^{+}\pi^{-}) \,\, \tau_{+} \over
\overline{Br}(\rm B^+\to {\rm K}^{0}\pi^{+}) \,\, \tau_{0} }
= 1 -2\,\, r_K \, \cos\delta \, \, \cos\phi_3 + r_K^2 \nonumber \\
\cr
A_0 &=&{\Gamma(\rm \bar{B}^0 \to K^{-}\pi^{+}) - \Gamma(B^0 \to
K^{+}\pi^{-}) \over \Gamma(\rm B^{-}\to \bar{K}^0\pi^{-}) +
 \Gamma(B^{+}\to \bar{K}^0\pi^{+}) } \nonumber \\
&=& A_{cp}(\rm B^0 \to K^{+}\pi^{-}) \,\, R_K = -2 r_K \, \sin\phi_3 \,\sin\delta.
\eeqa
where $r_K = |T^{'}/P^{'}|$ is the ratio of tree to penguin amplitudes
in $\rm B\to K\pi$
and $\delta = \delta_{T'} -\delta_{P'}$ is the strong phase difference
between tree and penguin amplitudes.
After elimination of $\sin\delta$ in Eqs.~(8)--(9), we have
\beq
R_K = 1 + r_K^2 \pm \sqrt{4 r_K^2 \cos^2\phi_3 -A_0^2 \cot^2\phi_3}.
\eeq
Here we obtain $r_K = 0.201\pm 0.037$ 
from the pQCD analysis[\ref{pQCD}] 
and $A_0=-0.11\pm 0.065$ by combining recent BaBar
measurement on CP asymmetry of $\rm B^0\to K^+\pi^-$: 
$A_{cp}(\rm B^0\to K^+\pi^-)=-10.2\pm5.0\pm1.6 \%$ [\ref{babar}]
with present world averaged value of  $R_K=1.10\pm 0.15$~[\ref{rk}].

\begin{figure}[htbp]
\begin{center}
\includegraphics[angle=-90,width=9cm]{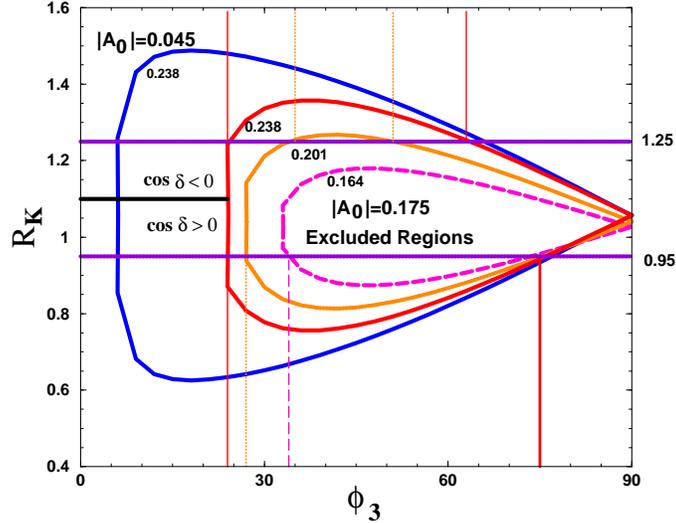} 
\caption{\it Plot of $R_K$ versus $\phi_3$ with $r_K=0.164,0.201$ and $0.238$.}
\end{center}
\label{fig:RK-lyu}
\end{figure}


From Table 2 of Ref.~[\ref{keum}],
we obtain $\delta_{P'} = 157^{\circ}$, $\delta_{T'} = 1.4^{\circ}$
and the negative $\cos\delta$: $\cos\delta= -0.91$.
As shown in Fig.~6.15, 
we can constrain the allowed range of 
$\phi_3$ within $1\,\sigma$ range of World Averaged $R_K$ as follows:
\begin{itemize}
\item[$\bullet$]For $\cos\delta < 0$, $r_K=0.164$: we can exclude
$0^{\circ} \leq \phi_3 \leq 6^\circ$. 
\item[$\bullet$]For $\cos\delta < 0$, $r_K=0.201$: we can exclude
$0^{\circ} \leq \phi_3 \leq 6^\circ$ and $ 35^\circ \leq \phi_3 \leq 51^\circ$. 
\item[$\bullet$]For $\cos\delta < 0$, $r_K=0.238$: we can exclude
$0^{\circ} \leq \phi_3 \leq 6^\circ$ and $ 24^\circ \leq \phi_3 \leq 62^\circ$.
\end{itemize}
When we take the central value of $r_K=0.201$,
$\phi_3$ is allowed within the ranges of $51^{\circ} \leq \phi_3 \leq
129^{\circ}$, which is consistent with the results of the model-independent
CKM-fit in the $(\bar\rho,\bar\eta)$ plane.

\subsubsection{Conclusion}
We discussed two methods to determine the weak phases 
$\phi_2$ and $\phi_3$ within the pQCD
approach through 1) Time-dependent asymmetries in $\rm B^0\to
\pi^{+}\pi^{-}$, 2) $\rm B\to K\pi$ processes via penguin-tree
interference. We can already obtain interesting bounds on $\phi_2$
and $\phi_3$ from present experimental measurements.
Our predictions within pQCD method is in good agreement with present
experimental measurements in charmless B-decays.
Specially our pQCD method predicted a large direct CP asymmetry
in $\rm B^0 \to \pi^{+}\pi^{-}$ decay, which will be a crucial touch stone
in order to distinguish our approach from others 
in future precise measurements.
More detail works on other methods in $\rm B\to K\pi$ and ${\rm D}^{(*)}\pi$ processes
will appear elsewhere [\ref{keum}].

\boldmath
\section{$\rm K\to\pi \nu\bar\nu$ decays}
{\it G.~Isidori and D.E. Jaffe}
\unboldmath

\vspace{-2mm}

\subsection{Theoretical description}

\begin{figure}[htbp] 
$$
\includegraphics[width=8cm]{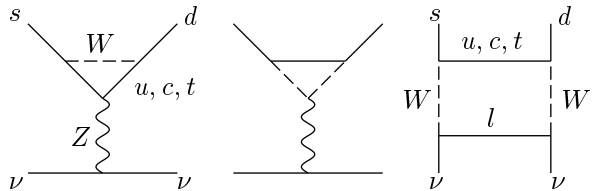}
$$

\vspace{-2mm}

\caption{\it One-loop diagrams contributing to the 
$s \to d \nu \bar{\nu}$ transition.}
\label{fig:Kpnn}
\end{figure}

The $s \to d \nu \bar{\nu}$ transition is one 
of the rare examples of weak processes whose 
leading contribution starts at ${\cal O}(G^2_F)$. At the one-loop 
level it receives contributions only from $Z$-penguin and 
$W$-box diagrams, as shown in Fig.~\ref{fig:Kpnn}, 
or from pure quantum electroweak effects.
Separating the contributions to the one-loop amplitude according to the 
intermediate up-type quark running inside the loop, we can write
%
%
\bea 
&& {\cal A}(s \to d \nu \bar{\nu}) = \sum_{q=u,c,t} V_{qs}^*V_{qd} {\cal A}_q  
 \quad \sim \quad \left\{ \begin{array}{ll} {\cal O}(\lambda^5
m_t^2)+i {\cal O}(\lambda^5 m_t^2)\    & _{(q=t)}   \\
{\cal O}(\lambda m_c^2 )\ + i{\cal O}(\lambda^5 m_c^2)     & _{(q=c)} \\
{\cal O}(\lambda \Lambda^2_{\rm QCD})     & _{(q=u)}
\end{array} \right. \quad
\label{uno}
\eea
%
where $V_{ij}$ denote the elements of the CKM matrix. 
The hierarchy of these elements 
would favour  up- and charm-quark contributions;
however, the {\em hard} GIM mechanism of the perturbative calculation
implies ${\cal A}_q \sim m^2_q/M_W^2$, leading to a completely 
different scenario. As shown on the r.h.s.~of (\ref{uno}), 
where we have employed the standard CKM phase convention 
(${\rm Im} V_{us}= {\rm Im} V_{ud}=0$) and 
expanded the $V_{ij}$ in powers of the Cabibbo angle, 
the top-quark contribution dominates both real and imaginary parts.
This structure implies several interesting consequences for
${\cal A}(s \to d \nu \bar{\nu})$: it is dominated by short-distance
dynamics, therefore its QCD corrections are small and calculable in perturbation theory; 
it is very sensitive to $V_{td}$, which is one of the less constrained CKM matrix elements;
it is likely to have a large CP-violating phase; it is very suppressed within the SM and thus 
very sensitive to possible new sources of quark-flavour mixing.

Short-distance contributions to ${\cal A}(s \to d \nu \bar{\nu})$,
are efficiently described, within the SM, by the 
following effective Hamiltonian [\ref{BB2}]
\be
{\cal H}_{eff} = \frac{G_F}{\sqrt 2} \frac{\alpha}{2\pi \sin^2\Theta_W}
 \sum_{l=e,\mu,\tau} \left[ \lambda_c X^l_{NL} + \lambda_t X(x_t) \right]
 (\bar sd)_{V-A}(\bar\nu_l\nu_l)_{V-A}~,
\label{eq:Heff} 
\ee
where $x_t=m_t^2/M_W^2$ and, as usual, $\lambda_q = V^*_{qs}V_{qd}$. 
The coefficients $X^l_{NL}$ and 
$X(x_t)$, encoding charm- and top-quark loop contributions, 
are known at the NLO accuracy in QCD [\ref{BB},\ref{MU}] and can be found 
explicitly in [\ref{BB2}]. The theoretical uncertainty in the dominant 
top contribution is very small and it is essentially determined by the 
experimental error on $m_t$. Fixing the $\overline{\rm MS}$ top-quark mass 
to ${\overline m}_t(m_t) = (166 \pm 5)$~GeV we can write
\be
X(x_t) = 1.51 \left[ \frac{ {\overline m}_t(m_t)}{166~\rm GeV} \right]^{1.15} = 
1.51 \pm 0.05~.
\ee
The simple structure of ${\cal H}_{eff}$ leads to two important  
properties of the physical ${\rm K}\to \pi \nu\bar \nu$ transitions:
\begin{itemize}
\item{} The relation between partonic and hadronic amplitudes 
is exceptionally accurate, since hadronic matrix elements
of the $\bar{s} \gamma^\mu d$ current between a kaon and a pion
can be derived by isospin symmetry from the measured ${\rm K}_{l3}$ rates. 
\item{} The lepton pair is produced in a state of definite CP 
and angular momentum, implying that the leading SM contribution 
to ${\rm K}_L \to \pi^0  \nu \bar{\nu}$ is CP-violating.
\end{itemize}

\medskip\noindent
The largest theoretical uncertainty 
in estimating ${\cal B}({\rm K}^+\to\pi^+ \nu\bar{\nu})$ 
originates from the charm sector.  
Following the analysis of Ref.~[\ref{BB2}],
the perturbative charm contribution is 
conveniently described in terms of the parameter 
\be
 P_0(X) = \frac{1}{\lambda^4}
\left[\frac{2}{3}X^e_{NL}+\frac{1}{3}X^\tau_{NL}\right] = 0.42 \pm 0.06~.
\label{eq:P0}
\ee 
The numerical error in the r.h.s. of Eq.~(\ref{eq:P0})  is obtained from a 
conservative estimate of NNLO corrections [\ref{BB2}]. 
Recently also non-perturbative effects introduced 
by the integration over charmed degrees of freedom
have been discussed [\ref{Falk_Kp}]. Despite a precise 
estimate of these contributions is not possible at present
(due to unknown hadronic matrix-elements), these can be 
considered as included in the uncertainty quoted in 
Eq.~(\ref{eq:P0}).\footnote{~The natural order of magnitude 
of these non-perturbative corrections, 
relative to the perturbative charm contribution 
is $m_K^2/(m_c^2 \ln(m^2_c/M^2_W)) \sim 2 \%$.}
Finally, we recall that genuine long-distance effects 
associated to light-quark loops are well below  
the uncertainties from the charm sector [\ref{LW}].

With these definitions the branching fraction of ${\rm K}^+\to\pi^+\nu\bar\nu$ 
can be written as   
\be
{\cal B}({\rm K}^+\to\pi^+\nu\bar\nu) = \frac{ \bar \kappa_+ }{ \lambda^2 } ~
\left[ ({\rm Im} \lambda_t)^2 X^2(x_t) +
\left( \lambda^4 {\rm Re}\lambda_c  P_0(X)+
       {\rm Re}\lambda_t X(x_t)\right)^2 \right]~, 
\label{eq:BRSM} 
\ee
where [\ref{BB2}]
\be 
\bar \kappa_+ = r_{K^+} \frac{3\alpha^2 {\cal B}({\rm K}^+\to\pi^0 e^+\nu)}{
2\pi^2\sin^4\Theta_W} = 7.50 \times 10^{-6} 
\ee
and $r_{K^+}=0.901$ takes into account the isospin breaking corrections 
necessary to extract the matrix element of the $(\bar s d)_{V}$ current 
from ${\cal B}({\rm K}^+\to\pi^0 e^+\nu)$ [\ref{MP}]. 

The case of ${\rm K}_L\to\pi^0 \nu\bar{\nu}$ is even cleaner from the
theoretical point of view [\ref{Litt}].  
Because of the  CP structure, only the imaginary parts in (\ref{eq:Heff}) 
--where the charm contribution is absolutely negligible--
contribute to ${\cal A}({\rm K}_2 \to\pi^0 \nu\bar{\nu})$. Thus 
the dominant direct-CP-violating component 
of ${\cal A}({\rm K}_L \to\pi^0 \nu\bar{\nu})$ is completely 
saturated by the top contribution, 
where  QCD corrections are suppressed and rapidly convergent. 
Intermediate and long-distance effects in this process
are confined only to the indirect-CP-violating 
contribution [\ref{BBSIN}] and to the CP-conserving one [\ref{CPC}],
which are both extremely small.
Taking into account the isospin-breaking corrections to the 
hadronic matrix element [\ref{MP}], we can write an
expression for the ${\rm K}_L\to\pi^0 \nu\bar{\nu}$ rate in terms of 
short-distance parameters, namely
\be
{\cal B}({\rm K}_L\to\pi^0 \nu\bar{\nu})_{\rm SM} =
 \frac{ \bar \kappa_L }{ \lambda^2 } ({\rm Im} \lambda_t)^2 X^2(x_t)
= 4.16 \times 10^{-10} \times \left[
\frac{\overline{m}_t(m_t) }{ 167~{\rm GeV}} \right]^{2.30} \left[ 
\frac{{\rm Im} \lambda_t }{ \lambda^5 } \right]^2~, 
\ee
which has a theoretical error below $3\%$.

At present the SM predictions of the two  ${\rm K}\to \pi\nu\bar\nu$ rates
are not extremely precise owing to the limited knowledge of both 
real and imaginary parts of $\lambda_t$. 
Taking into account all the indirect constraints in a global Gaussian fit, 
the allowed ranges read 
[\ref{rising},\ref{kettel}]\footnote{~As pointed out in Ref.~[\ref{kettel}], 
the errors in Eqs.~(\ref{BRK+nnt})--(\ref{BRKLnnt})
can be reduced if ${\rm Re}\lambda_t$ and  ${\rm Im}\lambda_t$
are directly extracted from  $a_{\rm CP} (B\to J/\Psi {\rm K}_S)$ and $\epsilon_K$;
however, this procedure introduces a stronger sensitivity to
the probability distribution of the (theoretical) estimate of $ B_K$.} 
\bea
{\cal B}({\rm K}^+ \to\pi^+ \nu\bar{\nu})^{ }_{\rm SM} &=& (0.72 \pm 0.21) 
\times 10^{-10}~, \qquad  \label{BRK+nnt}
\\
{\cal B}({\rm K}_L \to\pi^0 \nu\bar{\nu})^{ }_{\rm SM} &=& (0.28 \pm 0.10) 
\times 10^{-10}~. \qquad  \label{BRKLnnt}
\eea

The high accuracy of the theoretical predictions of ${\cal B}({\rm K}^+ \to\pi^+
\nu\bar{\nu})$ and ${\cal B}({\rm K}_L \to\pi^0 \nu\bar{\nu})$ in terms of modulus
and phase of $\lambda_t=V^*_{ts} V_{td}$ clearly offers
the possibility of very interesting tests of flavour dynamics.
Within the SM, a measurement of both channels would provide 
two independent pieces of information on the unitary triangle, 
or a complete determination of $\bar\rho$ and $\bar\eta$ from $\Delta S=1$
transitions. In particular, ${\cal B}({\rm K}^+\to\pi^+ \nu\bar\nu)$
defines an ellipse in the $\bar\rho$--$\bar\eta$ plane
and ${\cal B}({\rm K}^0_{\rm L}\to\pi^0 \nu\bar\nu)$ an horizontal
line (the height of the unitarity triangle). Note, in addition, 
that the determination of $\sin 2\beta$ which can be 
obtained by combining ${\cal B}({\rm K}^0_{\rm L}\to\pi^0 \nu\bar\nu)$
and ${\cal B}({\rm K}^+\to\pi^+ \nu\bar\nu)$ is extremely clean, being 
independent from uncertainties due to $m_t$ and $V_{cb}$ (contrary 
to the separate determinations of $\bar\rho$ and $\bar\eta$)~[\ref{BBSIN}].

In principle a very precise and highly non-trivial test of the 
CKM mechanism could
be obtained by the comparison of the following two sets of data [\ref{BBSIN}]:
the two ${\rm K}\to \pi\nu\bar\nu$ rates on one side,  the ratio 
$\Delta M_{B_d}/\Delta M_{B_s}$ and $a_{\rm CP}(B\to J/\Psi K_S)$ 
on the other side.
The two sets are determined by very different 
loop amplitudes ($\Delta S=1$ FCNCs and $\Delta B=2$ mixing)
and both suffer from very small theoretical uncertainties.
In particular, concerning the ${\rm K}^+\to\pi^+ \nu \bar\nu$ mode,
we can write~[\ref{rising}]
\be
{\cal B}({\rm K}^+\to\pi^+\nu\bar\nu) = \bar\kappa_+  |V_{cb}|^4 X^2(x_t) 
\left[ \sigma R_t^2 \sin^2\beta + \frac{1}{\sigma} 
\left(R_t \cos\beta + \frac{\lambda^4 P_0(X)}{|V_{cb}|^2 X(x_t)}
\right)^2 \right]~,
\ee
where $R_t$ is determined by $\Delta M_{B_d}/\Delta M_{B_s}$ 
[\ref{BB2}],\footnote{~As usual
we define $\xi=(F_{B_s}/F_{B_d})\sqrt{{B}_{B_s}/{B}_{B_d}}$
and $\sigma=1/( 1 - \frac{\lambda^2}{2})^2$.}
\be
R_t = \frac{ \xi  }{ \lambda } \sqrt{ \frac{ \Delta M_{B_d}}{ \Delta M_{B_s}} }
\ee
and $\sin\beta$ from $a_{\rm CP}(B\to J/\Psi {\rm K}_S)$. 
In the next few years, when the experimental determination of 
$a_{\rm CP}(B\to J/\Psi {\rm K}_S)$, $\Delta M_{B_d}/\Delta M_{B_s}$, and 
${\cal B}({\rm K}^+\to\pi^+\nu\bar\nu)$ will substantially improve, this 
relation could provide one of the most significant tests of the 
Standard Model in the sector of quark-flavour dynamics. 

Present experimental data on ${\rm K}\to \pi\nu\bar\nu$ rates
do not allow yet to fully explore the high-discovery potential 
of these CKM tests. Nonetheless, we stress that the evidence of the 
${\rm K}^+\to\pi^+ \nu\bar\nu$ decay obtained by BNL-E787 
already provides highly non-trivial constraints on realistic scenarios 
with large new sources of flavour mixing (see e.g. 
Ref.~[\ref{rising},\ref{BCIRS},\ref{GN}]).

\subsection{Experimental status and future prospects}

The Brookhaven experiment E787 [\ref{ref:kpnn.pnn1}]
searched for the decay ${\rm K}^+\to\pi^+ \nu\bar\nu$ by stopping
approximately 25\% of a 670, 710, 730 or 790 MeV$/c$\  ${\rm K}^+$ beam at $\sim 5$ MHz with 
$\sim 25\%$ $\pi^+$ contamination in a scintillating-fiber 
target along the axis of a 1-T solenoidal magnetic spectrometer.
The range ($R$), momentum ($P$) and energy ($E$) of charged
decay products are measured using the target, central drift chamber
and a cylindrical range stack composed of 21 layers of plastic
scintillator with two layers of tracking chambers.
Detection of the decay sequence $\pi^+\to\mu^+\to e^+$ in the
range stack provided a powerful tool against ${\rm K}^+\to\mu^+\nu(\gamma)$ decays.
A $4\pi$-sr calorimeter consisting of lead/scintillator layers in 
the barrel (14 radiation lengths) and undoped CsI crystals in the
endcap (13.5  radiation lengths) were used to veto photons
and suppress ${\rm K}^+\to\pi^+\pi^0$ background. Incident kaons were
detected and identified by \v Cerenkov, tracking and energy loss detectors
along the beam 
that aided in the suppression of backgrounds due
to scattered beam pions and the charge exchange process that resulted
in ${\rm K}^0_L\to\pi^+ \ell^-\nu$ decays ($\ell^- = e^-$,$\mu^-$) in the target.

\begin{figure}[t] 
\includegraphics[angle=0, height=.34\textheight]{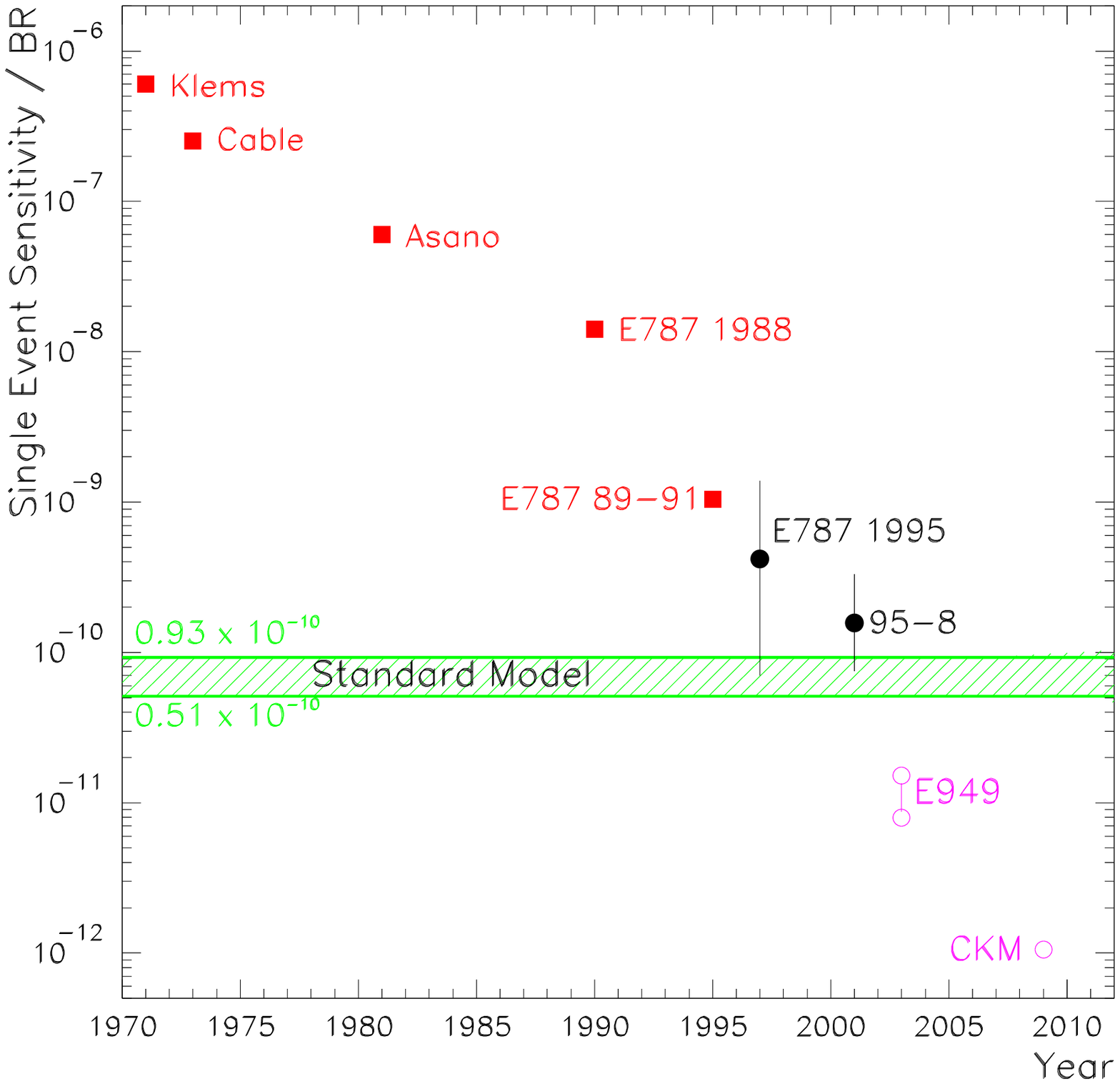}
\includegraphics[angle=0, height=.34\textheight]{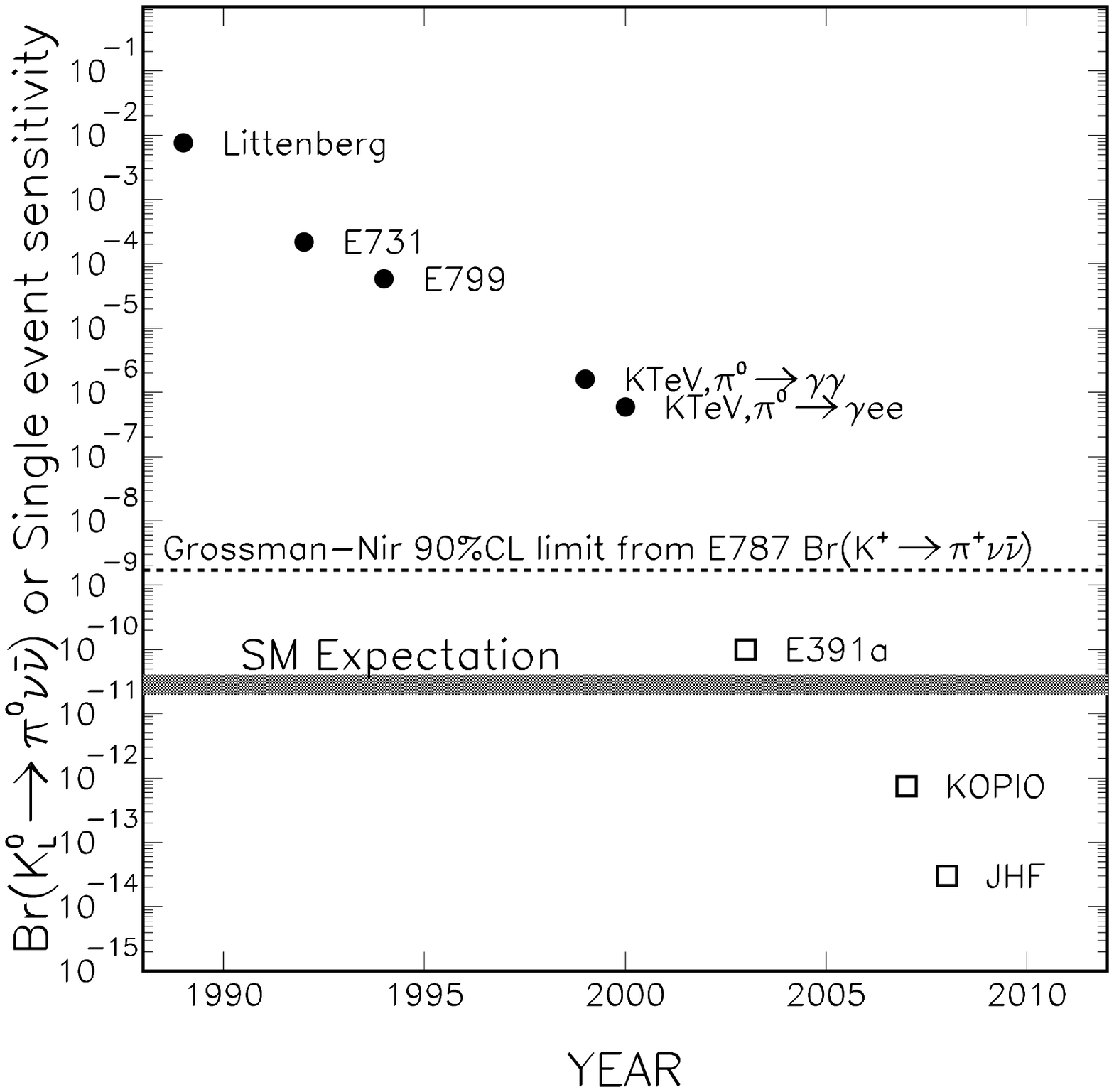}
  \caption{\it History and prospects for the study of  ${\cal B}({\rm K}^+\to\pi^+ \nu\bar\nu)$(left)
and {${\cal B}({\rm K}^0_{\rm L}\to\pi^0 \nu\bar\nu)$}(right).
The points with error bars are measured branching fractions, the solid points are
upper limits at 90\% CL and the open points or squares
are single event sensitivities.
The dashed line is a nearly model-independent limit
based on the E787's results for 
 ${\cal B}({\rm K}^+\to\pi^+ \nu\bar\nu)$~\protect[\ref{GN}].
The horizontal bands are the 68\% CL SM expectations.
\label{Fig:both.kpnn.history} }
\end{figure}

E787 has a long history, summarized in Fig.~\ref{Fig:both.kpnn.history},
that has lead to the development of a relatively robust analysis strategy.
The strategy begins with {\it a priori} identification of background sources
and development of experimental tools to suppress each background source
with at least two independent cuts. In the search for such rare processes,
background rejection cannot be reliably simulated, instead it is measured by 
alternatively inverting independent cuts and measuring the rejection of
each cut taking any correlations into account. To avoid bias, cuts are determined
using 1/3 of the data and then the backgrounds rates are measured with the remaining
2/3 sample. Background estimates are verified by loosening cuts and comparing
the observed and predicted rates, first in the 1/3 sample, then in the 2/3 sample.
Simulated signal events are used to measure the
geometrical acceptance for ${\rm K}^+\to\pi^+ \nu\bar\nu$ and the acceptance is
verified with a measurement of ${\cal B}({\rm K}^+\to\pi^+\pi^0)$.
The pre-defined signal region in $R$, $P$ and $E$ is not examined until 
all background estimates are verified. It is anticipated that similar strategies
will be employed in further investigations of ${\rm K}\to\pi \nu\bar\nu$ decays.

Brookhaven E787 was completed in 1998 and has observed two candidates for the
decay ${\rm K}^+\to\pi^+ \nu\bar\nu$ in the pion momentum region
211 to 229 MeV$/c$ with an estimated background of
$0.15\pm 0.05$ in a sample of $5.9\times 10^{12}$ stopped ${\rm K}^+$
that corresponds to [\ref{ref:kpnn.pnn1}]
\be
{\cal B}({\rm K}^+\to\pi^+ \nu\bar\nu) = (15.7{}^{+17.5}_{-8.2})\times 10^{-11}  \ .
\label{eq:E787}
\ee
The probability that the two candidates are entirely due to background is $0.02\%$ .
In addition  a search in the momentum interval 140 to 195  MeV$/c$ in 
a sample of $1.1\times 10^{12}$ stopped ${\rm K}^+$ yielded a single candidate
upon an estimated background of $0.73\pm0.18$
corresponding to a limit ${\cal B}({\rm K}^+\to\pi^+ \nu\bar\nu) < 420\times 10^{-11}$ at
90\% C.L.~[\ref{ref:kpnn.pnn2}]. Such a search below the peak of the two body ${\rm K}^+\to\pi^+\pi^0$
decays has the potential not only to augment the statistics of the higher momentum sample,
but also to investigate the shape of the $P(\pi^+)$ distribution predicted by the SM.
In addition, the search is somewhat complementary to that of the higher momentum interval
because the background is dominated by ${\rm K}^+\to\pi^+\pi^0$ decays in which the
charged pion undergoes a nuclear interaction in the target near the kaon decay vertex.

 E949 is an upgraded version of E787 with an expected net increase in sensitivity
of at least a factor of 5 based on 6000 hours of running time
or 5-10 SM events~[\ref{ref:kpnn.E949}]. The main detector
upgrades are an increased photon veto capability, both in the endcap and barrel regions,
as well as trigger and data acquisition improvements. 
E949 recently accumulated $1.9\times 10^{12}$
stopped kaons ($\sim 1/9$ of E949's goal)
and additional running is expected in 2003
assuming sufficient funding is forthcoming.

The CKM experiment at Fermilab expects to attain a single event
sensitivity of $1\times 10^{-12}$ that would correspond to 
$\sim 100$ ${\rm K}^+\to\pi^+ \nu\bar\nu$ events assuming the SM value of the 
branching fraction~[\ref{ref:kpnn.CKMexpt}].
Such a measurement would achieve a statistical precision comparable to
the current theoretical uncertainty in the branching fraction.
CKM departs from the E787/E949 technique by 
using kaon decays in flight in a 22 GeV$/c$, 50 MHz debunched beam with 60\% kaon purity.
The experiment will use photon veto technology similar to E787 and KTeV 
with the addition of ring-imaging \v Cerenkov detectors to aid in  kinematic suppression of 
backgrounds. The use of in-flight kaon decays means that the 
dominant ${\rm K}^+\to\pi^+\pi^0$ background in 
E787's search in the lower momentum region should not be present 
at CKM~[\ref{ref:kpnn.Diwan}].
CKM should be taking data in the second half of this decade.

\medskip

The progress concerning the neutral mode is much slower.
No dedicated experiment has started yet 
and the best direct limit is more than four orders of magnitude above the 
SM expectation [\ref{KTeV_nn}]. An indirect model-independent upper bound  on
$\Gamma({\rm K}_L\to\pi^0\nu\bar{\nu})$ can be 
obtained by the isospin relation~[\ref{GN}] 
\be
\Gamma({\rm K}^+\to\pi^+\nu\bar{\nu})~=  \Gamma({\rm K}_L\to\pi^0\nu\bar{\nu}) +
\Gamma({\rm K}_S\to\pi^0\nu\bar{\nu}) 
\label{Tri}
\ee
which is valid for any $s\to d \nu\bar\nu$ local operator of dimension 
$\leq 8$ (up to small isospin-breaking corrections).
Using the BNL-E787 result (\ref{eq:E787}), this implies 
${\cal B}({\rm K}_L\to\pi^0\nu\bar{\nu}) <  1.7 \times 10^{-9}~(90\%~{\rm CL})$.
Any experimental information below this figure can be translated into 
a non-trivial constraint on possible new-physics contributions to 
the $s\to d\nu\bar{\nu}$ amplitude.

\begin{figure}[t] 
\includegraphics[angle=0, height=.34\textheight]{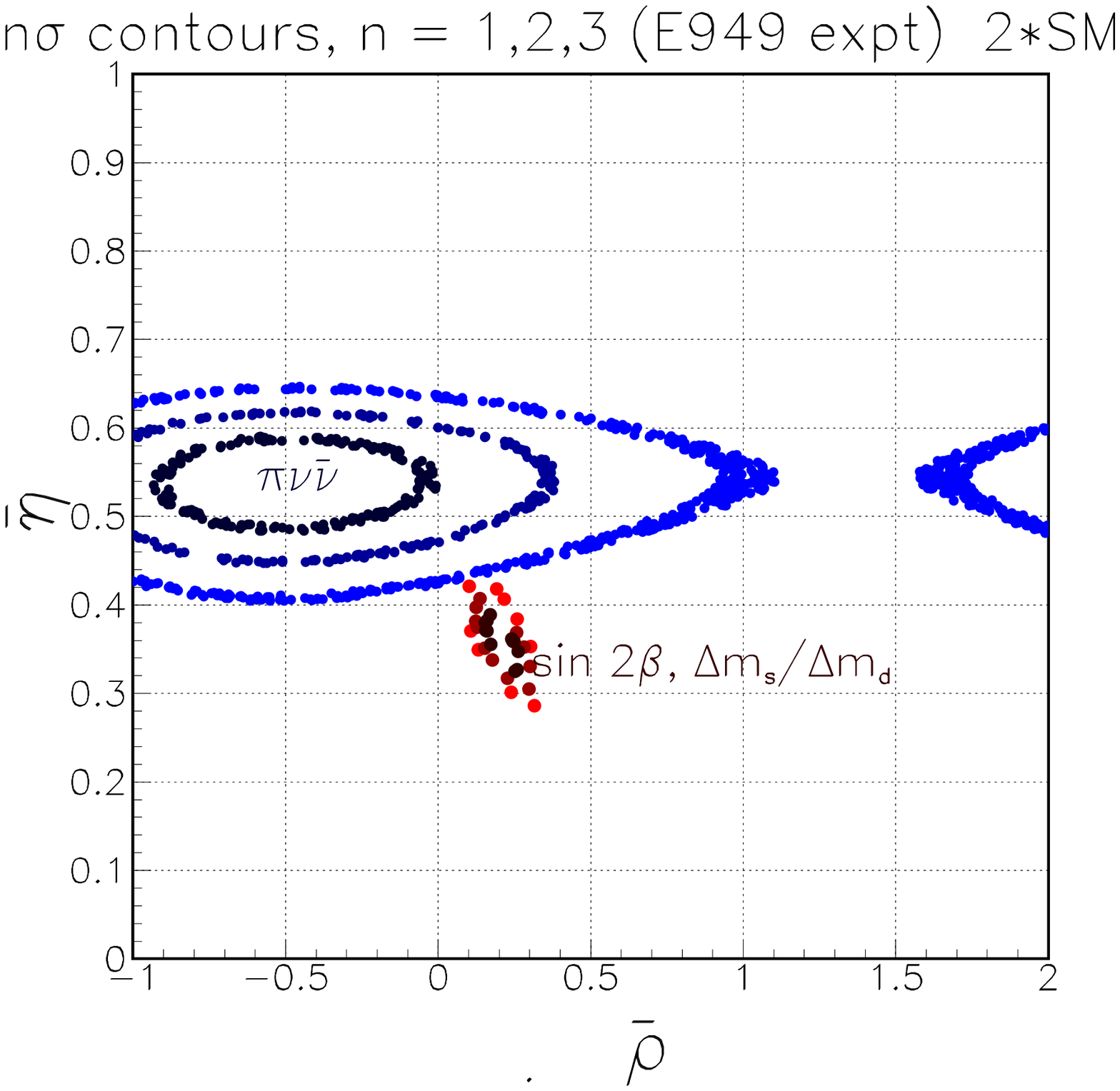}
\includegraphics[angle=0, height=.34\textheight]{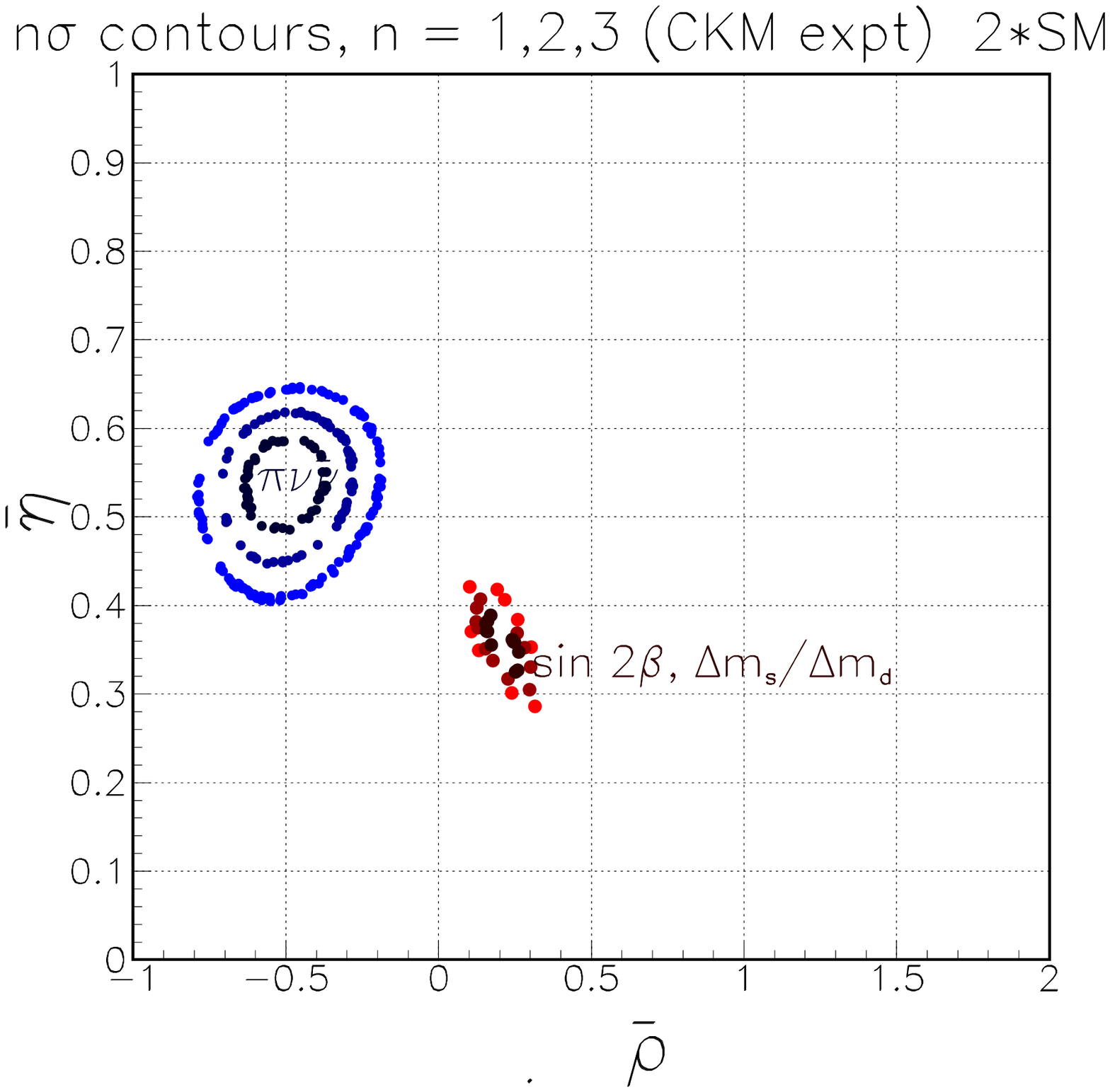}
  \caption{\it Comparison of the impact of hypothetical measurements of the 
           ${\rm K}\to\pi\nu\bar\nu$ branching fractions by E949 and KOPIO(left) or  CKM and KOPIO(right)
           in the $\bar\rho,\bar\eta$ plane
        with hypothetical measurements of  $\sin 2\beta$ and $\Delta M_s/\Delta M_d$.
        Contours at 68.3, 95.45 and 99.7\% CL are indicated for the K measurements.
        For the B measurements, the points indicating the three contours overlap.
        See text for details.
\label{Fig:kpnn.future} }
\end{figure}

The first ${\rm K}_L\to\pi^0\nu\bar{\nu}$ dedicated experiments are
E391a at KEK~[\ref{ref:kpnn.E391a}] and  KOPIO 
at Brook\-ha\-ven~[\ref{ref:kpnn.KOPIO}]. 
E391a is envisioned as a two-stage experiment and
will attempt to use a highly collimated ${\rm K}_{\rm L}^0$
 beam and a hermetic veto to observe high transverse momentum $\pi^0$
near the endpoint of the ${\rm K}^0_{\rm L}\to\pi^0 \nu\bar\nu$ spectrum 
with a technique similar to previous searches~[\ref{KTeV_nn}].
The first stage of E391a is regarded as a pilot experiment and
will use the KEK 12 GeV$/c$ proton beam and should begin data
taking in 2003. If successful, it could push
the limit on ${\cal B}({\rm K}^0_{\rm L}\to\pi^0 \nu\bar\nu)$ to within an
order of magnitude of the SM expectation (Fig.~\ref{Fig:kpnn.future}).
An aggressive second stage envisions  use of the high intensity
 50 GeV$/c$ proton beam from the
Japanese Hadron Facility(JHF) to reach a single event sensitivity of 
$3\times 10^{-14}$ or, equivalently, $\sim 1000$ SM events.

 The KOPIO experiment will attempt a new approach, using a microbunched,
low momentum beam, time-of-flight and a high precision
electromagnetic preradiator and calorimeter to fully reconstruct the
kinematics of the  ${\rm K}^0_{\rm L}\to\pi^0 \nu\bar\nu$ decay.
Coupled with highly efficient charged particle and photon vetoes,
KOPIO will be able to exploit the E787 strategy of independent
kinematic and veto cuts to measure all backgrounds with the data.
The goal of KOPIO is a single event sensitivity of $7.5\times 10^{-13}$
or the capability to obtain 40 SM events with a signal to background
of 2 corresponding to a precision on ${\cal J}$ or $\bar\eta$ of $\sim 10\%$.

As anticipated, one of the most interesting test of the  CKM mechanism could
be obtained by the comparison of the two ${\rm K}\to \pi\nu\bar\nu$ rates on one side
vs. the ratio $\Delta M_{B_d}/\Delta M_{B_s}$ and $a_{\rm CP}(\rm B\to J/\Psi K_S)$ on the other side.
As an illustration, in Figure~\ref{Fig:kpnn.future}
we consider the comparison of the two B-physics measurements, assumed to be 
$a_{\rm CP}(B\to J/\Psi {\rm K}_S) = 0.75\pm 0.02$ and $\Delta M_{B_d}/\Delta M_{B_s}= 17.0\pm 1.7\ {\rm ps}^{-1}$,
with the two ${\rm K}\to\pi\nu\bar\nu$ rates, both assumed to be twice the corresponding SM prediction.
The uncertainties on  ${\cal B}({\rm K}\to\pi\nu\bar\nu)$ measurements are those expected by 
E949, CKM and KOPIO experiments attaining their expected sensitivities.
The corresponding constraints in the $\bar\rho$--$\bar\eta$ plane have been derived  
assuming Gaussian uncertainties for all quantities, using the Bayesian statistics option of
the CKM fitter program~[\ref{ref:kpnn.CKMfitter}]. Negligible uncertainty in $|V_{cb}|$ 
is assumed in placing the ${\rm K}$ measurements in this {\em B-centric} rendering of the UT.
Note that the alternative, equally fundamental, parametrization  
of the UT using the $\lambda_t$ plane would 
remove the need for this assumption~[\ref{ref:kpnn.LSL.HF9}].
   
\newpage

\section*{References}
\addcontentsline{toc}{section}{References}

\vspace{7mm}

\renewcommand{\labelenumi}{[\theenumi]}
\begin{enumerate}

\item \label{BABAR}
The BaBar Physics Book, eds. P. Harrison and H. Quinn, 1998, SLAC report 504.

\vspace{3mm}

\item \label{LHCB}
B Decays at the LHC, eds. P. Ball, R. Fleischer, G.F. Tartarelli, 
P. Vikas and G. Wilkinson, \break
hep-ph/0003238.

\vspace{3mm}

\item \label{FERMILAB}
B Physics at the Tevatron, Run II and Beyond, K. Anikeev et al., 
hep-ph/0201071.

\vspace{3mm}

\item \label{BF97} 
{A.J.~Buras~and~R.~Fleischer,} Adv~Ser.~Direct.~High.~Energy~Phys.~{\bf 15}~(1998)~65;  \break
[hep-ph/9704376].

\vspace{3mm}

\item \label{Erice}
A.J. Buras, hep-ph/0101336, lectures at the International Erice School, 
August, 2000; \\
Y. Nir, hep-ph/0109090, lectures at 55th Scottish Univ. 
Summer School, 2001.

\vspace{3mm}

\item \label{BUPAST} A.J. Buras, F. Parodi and A. Stocchi, 
JHEP {\bf 0301} (2003) 029 [hep-ph/0207101].

\vspace{3mm}

\item \label{betaalpha}
J.M. Soares and L. Wolfenstein,  { Phys. Rev.} D~{\bf 47} (1993) 1021; 
Y. Nir and U. Sarid,  { Phys. Rev.} D~{\bf 47} (1993) 2818; 
Y. Grossman and Y. Nir,  { Phys. Lett.} B~{\bf 313} (1993) 126;
R. Barbieri, L.J. Hall and A. Romanino, { Phys. Lett.} B~{\bf 401} (1997) 47;
M. Ciuchini, E. Franco, L. Giusti, V. Lubicz and G. Martinelli, {Nucl. Phys. B~{\bf 573}} (2000) 201-222;
P. Paganini, F. Parodi, P. Roudeau and A. Stocchi, {Physica Scripta \bf { 58}} (1998) 556-569;
F. Parodi, P. Roudeau and A. Stocchi {Nuovo Cimento \bf{112A} (1999) 833.}

\vspace{3mm}

\item \label{Lacker}
A. H\"ocker, H. Lacker, S. Laplace and F. Le Diberder, 
Eur. Phys. J. C~{\bf 21} (2001) 225.

\vspace{3mm}

\item \label{BBSIN}
{ G. Buchalla and A.J. Buras}, 
 { Phys. Rev.} D~{\bf 54} (1996) 6782.

\vspace{3mm}

\item \label{Beneke:2001ev}
M. Beneke, G. Buchalla, M. Neubert, and C. Sachrajda,
Nucl.\ Phys.\ B~{\bf 606} (2001) 245.

\vspace{3mm}

\item \label{Gronau:2001cj}
M. Gronau and J. L. Rosner, 
Phys.\ Rev.\ D {\bf 65} (2002) 013004 [E: D~{\bf 65} (2002) 079901].

\vspace{3mm}

\item \label{Luo:2001ek}
Z. Luo and J. L. Rosner, Phys.\ Rev.\ D {\bf 65} (2002) 054027.

\vspace{3mm}

\item \label{BLO}
{ A.J. Buras, M.E. Lautenbacher and G. Ostermaier,}
{ Phys. Rev.} D~{\bf  50} (1994) 3433.

\vspace{3mm}

\item \label{B95}
{ A.J. Buras,} 
 { Phys. Lett.} B~{\bf 333} (1994) 476; 
{ Nucl. Instr. Meth.} {\bf A368} (1995) 1.

\vspace{3mm}

\item \label{Kayser}
R. Aleksan, B. Kayser and D. London, Phys. Rev. Lett. {\bf 73} (1994) 18; \\
J.P. Silva and L. Wolfenstein, { Phys. Rev.} D~{\bf  55} (1997) 5331; \\
I.I. Bigi and A.I. Sanda, hep-ph/9909479.

\vspace{3mm}

\item \label{UUT}
A.J. Buras, P. Gambino, M. Gorbahn, S. J\"ager and L. Silvestrini,
{ Phys. Lett.} B~{\bf 500} (2001) 161.

\vspace{3mm}

\item \label{ALILOND}
A. Ali and D. London, Eur. Phys. J. C~{\bf 9} (1999) 687 [hep-ph/9903535];
Phys. Rep. {\bf 320} (1999), 79 [hep-ph/9907243]; hep-ph/0002167; 
Eur. Phys. J. C~{\bf 18} (2001) 665.

\vspace{3mm}

\item \label{BF01}
A.J. Buras and R. Fleischer, {Phys.\ Rev.} D~{\bf 64} (2001) 115010.

\vspace{3mm}

\item \label{BCRS1}
A.J. Buras, P.H. Chankowski, J. Rosiek and {L}. Slawianowska, 
{ Nucl. Phys.} B~{\bf  619} (2001) 434.

\vspace{3mm}

\item \label{AI01}
G. D'Ambrosio and G. Isidori, { Phys. Lett.} B~{\bf 530} (2002) 108.
\vspace{3mm}

\newpage

\item \label{Branco}
G.C. Branco, L. Lavoura and J. Silva, (1999), CP Violation,
Oxford Science Publications, Press Clarendon, Oxford.
\vspace{3mm}

\item \label{BOBRNERE}
F.J. Botella, G.C. Branco, M. Nebot and M.N. Rebelo, 
Nucl. Phys. B~{\bf  651} (2003) 174  \break
[hep-ph/0206133].

\vspace{3mm}

\item \label{WO}
{ L. Wolfenstein}, { Phys. Rev. Lett.} {\bf 51} (1983) 1945.

\vspace{3mm}

\item \label{Bellealpha}
Belle Collaboration, K. Abe et al., 
{ Phys. Rev. Lett.} {\bf 89} (2002) 071801, 
hep-ex/0204002.

\vspace{3mm}

\item \label{BaBaralpha}
Talk by A. Farbin (BaBar Collaboration), XXXVIIth Recontres de Moriond,
Electroweak Interactions and Unified Theories, Les Arcs, France, 9-16 
March 2002, http://moriond.in2p3.fr/EW/2002/.

\vspace{3mm}

\item \label{ref:sin2b} 
                 Average from Y. Nir  hep-ph/0208080 based on:
                 R. Barate et al. (ALEPH Collaboration) {\it Phys. Lett.} 
                B~{\bf 492} (2000), 259; 
                 K. Ackerstaff et al. (OPAL Collaboration) 
           {\it Eur. Phys.} C~{\bf 5} (1998) 379;
                 T. Affolder at al. Phys. ReV. D61 (2000) 072005;
                 B. Aubert et al.   (Babar Collaboration) hep-ex/0207042;
                 K. Abe at al. (Belle Collaboration) hep-ex/0207098.

\vspace{3mm}

\item \label{ref:haricot}
M. Ciuchini, G. D'Agostini, E. Franco, V. Lubicz, G. Martinelli, 
F. Parodi, P. Roudeau, A. Stocchi,
JHEP {\bf 0107} (2001) 013 [hep-ph/0012308].

%
%
\vspace{3mm}

\item \label{Buras:2002er}
T.~Hurth, Rev.~Mod.~Phys. (to appear) [hep-ph/0212304]; 
A.~J.~Buras and M.~Misiak,
 Acta Phys. Pol. B 33 (2002) 2597 [hep-ph/0207131].

\vspace{3mm}

\item \label{Bertolini:1986th}
S.~Bertolini, F.~Borzumati and A.~Masiero,
Phys.\ Rev.\ Lett.\  {\bf 59} (1987) 180; \\
%
N.G.~Deshpande {\it et al.}, 
Phys.\ Rev.\ Lett.\  {\bf 59} (1987) 183.

\vspace{3mm}

\item \label{Alam:1995aw}
M.~S.~Alam {\it et al.}  (CLEO Collaboration),
Phys.\ Rev.\ Lett.\  {\bf 74} (1995) 2885.

\vspace{3mm}

\item \label{Aubert:2002pd}
B.~Aubert {\it et al.}  (BABAR Collaboration),
hep-ex/0207076.

\vspace{3mm}

\item \label{Chen:2001fj}
S.~Chen {\it et al.}  (CLEO Collaboration),
Phys.\ Rev.\ Lett.\  {\bf 87} (2001) 251807
[hep-ex/0108032].

\vspace{3mm}

\item \label{Abe:2001hk}
K.~Abe {\it et al.}  (BELLE Collaboration),
Phys.\ Lett.\ B {\bf 511} (2001) 151
[hep-ex/0103042].

\vspace{3mm}

\item \label{Barate:1998vz}
R.~Barate {\it et al.}  (ALEPH Collaboration),
Phys.\ Lett.\ B {\bf 429} (1998) 169.

\vspace{3mm}

\item \label{Gambino:2001ew}
P.~Gambino and M.~Misiak,
Nucl.\ Phys.\ B {\bf 611} (2001) 338
[hep-ph/0104034].

\vspace{3mm}

\item \label{Buras:2002tp}
A.J.~Buras, A.~Czarnecki, M.~Misiak and J.~Urban,
Nucl.\ Phys.\ B {\bf 631} (2002) 219
[hep-ph/0203135].
%

\vspace{3mm}

\item \label{Hagiwara:pw}
K.~Hagiwara {\it et al.}  (Particle Data Group Collaboration),
Phys.\ Rev.\ D {\bf 66} (2002) 010001.
%

\vspace{3mm}

\item \label{Ali:1998rr}
A.~Ali, H.~Asatrian and C.~Greub,
Phys.\ Lett.\ B {\bf 429} (1998) 87 
[hep-ph/9803314]; \\
A.~L.~Kagan and M.~Neubert,
Phys.\ Rev.\ D {\bf 58} (1998) 094012
[hep-ph/9803368].
%

\vspace{3mm}

\item \label{GP00} Y.~Grossman and D.~Pirjol, 
JHEP {\bf 0006} (2000) 02 
[hep-ph/0005069].
%

\vspace{3mm}

\item \label{CLEO_CP} T.E.~Coan {\it et al.} (CLEO Collaboration),
Phys.\ Rev.\ Lett.\ {\bf 86} (2001) 5661 
[hep-ex/0010075].
%
%

\vspace{3mm}

\item \label{Beneke:1999br_bis}
M.~Beneke, G.~Buchalla, M.~Neubert and C.~T.~Sachrajda,
Phys.\ Rev.\ Lett.\  {\bf 83} (1999) 1914 \break
[hep-ph/9905312].
%

\vspace{3mm}

\item \label{Beneke:2001at}
M.~Beneke, T.~Feldmann and D.~Seidel, Nucl.\ Phys.\ B {\bf 612} (2001) 25
[hep-ph/0106067].
%

\vspace{3mm}

\item \label{Bosch:2001gv}
S.~W.~Bosch and G.~Buchalla, Nucl.\ Phys.\ B {\bf 621} (2002) 459 
[hep-ph/01060].
%

\vspace{3mm}

\item \label{bdgAP}
A.~Ali and A.~Y.~Parkhomenko, Eur.\ Phys.\ J.\ C {\bf 23} (2002) 89 
[hep-ph/0105302].
%

\vspace{3mm}

\item \label{Coan:1999kh}
T.~E.~Coan {\it et al.}  (CLEO Collaboration),
Phys.\ Rev.\ Lett.\  {\bf 84} (2000) 5283 
[hep-ex/9912057].
%

\vspace{3mm}

\item \label{NishidaS:2002}
S.~Nishida (BELLE Collaboration),
Talk presented at the 31st.~International Conference on High Energy
Physics, July 24 - 31, 2002, Amsterdam, The Netherlands.

\vspace{3mm}

\item \label{Aubert:2001}
B. Aubert {\it et al.} (BABAR Collaboration)
Phys.\ Rev.\ Lett.\  {\bf 88} (2002) 101805 
[hep-ex/0110065].
%

\vspace{3mm}

\item \label{Kagan:2001zk}
A.~L.~Kagan and M.~Neubert,
Phys.\ Lett.\ B {\bf 539}, 227 (2002)
[hep-ph/0110078].
%

\vspace{3mm}

\item \label{Ball:1998kk}
P.~Ball and V.~M.~Braun,
Phys.\ Rev.\ D {\bf 58} (1998) 094016 
[hep-ph/9805422].
%

\vspace{3mm}

\item \label{Beneke:2000wa}
M.~Beneke and T.~Feldmann,
Nucl.\ Phys.\ B {\bf 592} (2001) 3 
[hep-ph/0008255].
%

\vspace{3mm}

\item \label{Ali:vd}
A.~Ali, V.~M.~Braun and H.~Simma, Z.\ Phys.\ C {\bf 63} (1994) 437 
[hep-ph/9401277].
%

\vspace{3mm}

\item \label{DelDebbio:1997kr}
L.~Del Debbio, J.~M.~Flynn, L.~Lellouch and J.~Nieves  (UKQCD
Collaboration),
Phys.\ Lett.\ B {\bf 416} (1998) 392 
[hep-lat/9708008].
%

\vspace{3mm}

\item \label{Ali:2000zu}
A.~Ali, L.~T.~Handoko and D.~London,
Phys.\ Rev.\ D {\bf 63} (2000) 014014 
[hep-ph/0006175].
%

\vspace{3mm}

\item \label{Grinstein:2000pc}
B.~Grinstein and D.~Pirjol,
Phys.\ Rev.\ D {\bf 62} (2000) 093002 
[hep-ph/0002216].
%

\vspace{3mm}

\item \label{Ali:1995uy}
A.~Ali and V.~M.~Braun, Phys.\ Lett.\ B {\bf 359} (1995) 223 
[hep-ph/9506248].
%

\vspace{3mm}

\item \label{Khodjamirian:1995uc}
A.~Khodjamirian, G.~Stoll and D.~Wyler, Phys.\ Lett.\ B {\bf 358} (1995) 129
[hep-ph/9506242].
%

\vspace{3mm}

\item \label{Ali:2002kw}
A.~Ali and E.~Lunghi,
DESY-02-089, hep-ph/0206242.
%

%
\vspace{3mm}

\item \label{Narison:1994kr}
S.~Narison,
Phys.\ Lett.\ B {\bf 327} (1994) 354 
[hep-ph/9403370].
%

\vspace{3mm}

\item \label{Melikhov:2000yu}
D.~Melikhov and B.~Stech,
Phys.\ Rev.\ D {\bf 62} (2000) 014006 
[hep-ph/0001113].
%

\vspace{3mm}

\item \label{Nir:2002gu}
Y.~Nir,
hep-ph/0208080.
%

\vspace{3mm}

\item \label{Lellouch:2002}
See, for example, L. Lellouch, plenary talk at the 31st.~International
Conference on High Energy Physics, July 24 - 31, 2002, Amsterdam, The
Netherlands.
%

\vspace{3mm}

\item \label{Kronfeld:2002ab}
A.~S.~Kronfeld and S.~M.~Ryan,
Phys.\ Lett.\ B {\bf 543} [hep-ph/0206058].
%

\vspace{3mm}

\item \label{babar:jessop} 
C. Jessop (BABAR Collaboration),
Talk presented at the 31st.~International Conference on High Energy
Physics, July 24 - 31, 2002, Amsterdam, The Netherlands.
%


%
\vspace{3mm}

\item \label{Aubert:2002tv}
BaBar Collaboration, B. Aubert et al., SLAC preprint SLAC-PUB-9229,
hep-ex/0205082, presented at 37th Rencontres de Moriond on Electroweak
Interactions and Unified Theories, Les Arcs, France, 9--16 Mar 2002.

\vspace{3mm}

\item \label{Gronau:2002cj}
M. Gronau and J. L. Rosner, Phys.\ Rev.\ D {\bf 65} (2002) 093012.

\vspace{3mm}

\item \label{Gronau:2002gj}
M. Gronau and J. L. Rosner, Phys.\ Rev.\ D {\bf 66} (2002) 119901. 

\vspace{3mm}

\item \label{Neubert:1998pt}
M. Neubert and J. L. Rosner, Phys.\ Lett.\ B {\bf 441} (1998) 403.

\vspace{3mm}

\item \label{Neubert:1998jq}
M. Neubert and J. L. Rosner, Phys.\ Rev.\ Lett.\ {\bf 81} (1998) 5076.

\vspace{3mm}

\item \label{Suprun:2001ms}
D. A. Suprun, C.-W. Chiang, and J. L. Rosner, Phys.\ Rev.\ D {\bf 65}
(2002) 054025.

\vspace{3mm}

\item \label{Chiang:2001ir}
C.-W. Chiang and J. L. Rosner, Phys.\ Rev.\ D {\bf 65} (2002) 074035.

\vspace{3mm}

\item \label{Hocker:2001jb}
A. H\"ocker, H. Lacker, S. Laplace, and F. Le Diberder, in {\it Proceedings of
the 9th International Symposium on Heavy Flavor Physics}, Pasadena, California,
10--13 Sep 2001, AIP Conf.\ Proc.\ {\bf 618} (2002) 27.

\vspace{3mm}

\item \label{Charles:1998qx}
J. Charles, Phys.\ Rev.\ D {\bf 59} (1999) 054007.

\vspace{3mm}

\item \label{Fleischer:2001zn}
R. Fleischer, in {\it Proceedings of the 9th International Symposium on
Heavy Flavor Physics}, Pasadena, California, 10--13 Sep 2001, AIP Conf.\ Proc.\
{\bf 618} (2002) 266.

\vspace{3mm}

\item \label{Gronau:2001ff}
M. Gronau, D. London, N. Sinha, and R. Sinha, Phys.\ Lett.\ B {\bf 514}
(2001) 315.

\vspace{3mm}

\item \label{Gronau:1998vg}
M. Gronau, Phys.\ Rev.\ D {\bf 58} (1998) 037301.

\vspace{3mm}

\item \label{Atwood:2001ck}
D. Atwood, I. Dunietz, and A. Soni, Phys.\ Rev.\ D {\bf 63} (2001) 036005.

\vspace{3mm}

\item \label{Kayser:1999bu}
B. Kayser and D. London, Phys.\ Rev.\ D {\bf 63} (2000) 116013.

\vspace{3mm}

\item \label{Buras:2000gc}
A. J. Buras and R. Fleischer, Eur.\ Phys.\ J. C {\bf 16} (2000) 97.

\vspace{3mm}

\item \label{Fleischer:1999pa}
R. Fleischer, Phys.\ Lett.\ B {\bf 459} (1999) 306.

\vspace{3mm}

\item \label{Gronau:2000md}
M. Gronau and J. L. Rosner, Phys.\ Lett.\ B {\bf 482} (2000) 71.

%
%

\vspace{3mm}

\item \label{RF-Phys-Rep}R.~Fleischer, 
Phys.\ Rep.\ {\bf 370} (2202) 537  [hep-ph/0207108].

\vspace{3mm}

\item \label{GRL}M.~Gronau, J.~L.~Rosner and D.~London,
{Phys.\ Rev.\ Lett.}\  {\bf 73} (1994) 21.

\vspace{3mm}

\item \label{PAPIII}R.~Fleischer,
{ Phys.\ Lett.}\ B {\bf 365} (1996) 399.

\vspace{3mm}

\item \label{FM}R.~Fleischer and T.~Mannel,
{Phys.\ Rev.}\ D {\bf 57} (1998) 2752.

\vspace{3mm}

\item \label{GR}M.~Gronau and J.~L.~Rosner,
{Phys.\ Rev.}\ D {\bf 57} (1998) 6843.

\vspace{3mm}

\item \label{defan}R.~Fleischer,
{Eur.\ Phys.\ J.}\ C {\bf 6} (1999) 451.


\vspace{3mm}

\item \label{neubert}M.~Neubert,
{JHEP} {\bf 9902} (1999) 014.

\vspace{3mm}

\item \label{BF-neutral1}A.~J.~Buras and R.~Fleischer,
{Eur.\ Phys.\ J.}\ C {\bf 11} (1999) 93.


\vspace{3mm}

\item \label{FlMa1}R.~Fleischer and J.~Matias,
Phys.\ Rev.\ D {\bf 61} (2000) 074004.

\vspace{3mm}

\item \label{ital-corr}M.~Bargiotti {\it et al.}, 
Eur.\ Phys.\ J.\ C {\bf 24} (2002) 361.



\vspace{3mm}

\item \label{PQCD}H.-n.~Li and H.~L.~Yu,
{Phys.\ Rev.}\ D {\bf 53} (1996) 2480; \\
Y.~Y.~Keum, H.-n.~Li and A.~I.~Sanda,
{Phys.\ Lett.}\ B {\bf 504} (2001) 6; \\
Y.~Y.~Keum and H.-n.~Li,
{Phys.\ Rev.}\ D {\bf 63} (2001) 074006.

\vspace{3mm}

\item \label{U-variant}R.~Fleischer,
{Eur.\ Phys.\ J.}\ C {\bf 16} (2000) 87.

\vspace{3mm}

\item \label{FlMa2}R.~Fleischer and J.~Matias,
Phys.\ Rev.\ D. {\bf 66} (2002) 054009, [hep-ph/0204101].

\vspace{3mm}

\item \label{matias}J.~Matias,
Phys.\ Lett.\ B {\bf 520} (2001) 131.

\vspace{3mm}

\item \label{gronau-U-spin}M.~Gronau,
{Phys.\ Lett.}\ B {\bf 492} (2000) 297.

\vspace{3mm}

\item \label{U-spin-other}R.~Fleischer,
{Eur.\ Phys.\ J.}\ C {\bf 10} (1999) 299,
Phys.\ Rev.\ D {\bf 60} (1999) 073008; \\
P.~Z.~Skands,
{JHEP} {\bf 0101} (2001) 008.

\vspace{3mm}

\item \label{Bpipi-recent}
M.~Gronau and J.~L.~Rosner,
Phys.\ Rev.\ D {\bf 65} (2002) 093012, 
113008 
and 
hep-ph/0205323; \\
C.-D.~L\"u and Z.-j.~Xiao,
Phys.\ Rev.\ D {\bf 66} (2002) 074011 [hep-ph/0205134].

\vspace{3mm}

\item \label{BBNS2}
M.~Beneke, G.~Buchalla, M.~Neubert and C.~T.~Sachrajda,
Nucl.\ Phys.\ B {\bf 591} (2000) 313.

\vspace{3mm}

\item \label{Wirbel:1985ji}
M.~Wirbel, B.~Stech and M.~Bauer,
Z.\ Phys.\ C {\bf 29} (1985) 637; \\
M.~Bauer, B.~Stech and M.~Wirbel,
Z.\ Phys.\ C {\bf 34} (1987) 103.
%


\vspace{3mm}

\item \label{KEK}
M.~Beneke, in: Proceedings of the 5th KEK Topical Conference:
Frontiers in Flavor Physics (KEKTC5), Tsukuba, Ibaraki, Japan, 20-22 Nov
2001 
[hep-ph/0202056].


\vspace{3mm}

\item \label{Du:2001hr}
D.~Du, H.~Gong, J.~Sun, D.~Yang and G.~Zhu,
Phys.\ Rev.\ D {\bf 65} (2002) 074001.

\vspace{3mm}

\item \label{Beneke:2002nj}
M.~Beneke, in: Proceedings of: 
Flavor Physics and CP Violation (FPCP), Philadelphia, 
Pennsylvania, 16-18 May 2002
[hep-ph/0207228].
\vspace{3mm}

\item \label{Neubert:2002tf}
M.~Neubert, in: Proceedings of the International Workshop on Heavy
Quarks and Leptons, Salerno, Italy, 27
May - 1 Jun 2002
[hep-ph/0207327].



\vspace{3mm}

\item \label{Bellebab}
BaBar Collaboration, hep-ex/0207053; P.~Krokovny [Belle Collaboration], 
talk at ICHEP2002, Amsterdam, July 2002.

%
%

\vspace{3mm}

\item \label{Ciuchini:1997hb}
M.~Ciuchini, E.~Franco, G.~Martinelli and L.~Silvestrini,
Nucl.\ Phys.\ B {\bf 501} (1997) 271 \break
[hep-ph/9703353].

\vspace{3mm}

\item \label{Ciuchini:1997rj}
M.~Ciuchini, R.~Contino, E.~Franco, G.~Martinelli and L.~Silvestrini,
Nucl.\ Phys.\ B {\bf 512} (1998) 3
[Erratum-ibid.\ B {\bf 531} (1998) 656]
[hep-ph/9708222].

\vspace{3mm}

\item \label{Buras:1998ra}
A.~J.~Buras and L.~Silvestrini,
Nucl.\ Phys.\ B {\bf 569} (2000) 3
[hep-ph/9812392].

\vspace{3mm}

\item \label{Ciuchini:2001gv}
M.~Ciuchini, E.~Franco, G.~Martinelli, M.~Pierini and L.~Silvestrini,
Phys.\ Lett.\ B {\bf 515} (2001) 33
[hep-ph/0104126].



\vspace{3mm}

\item \label{Patterson:fpcp} R.~Patterson, summary talk at FPCP02,
http://www.hep.upenn.edu/FPCP/talks/1803/180303Patterson.pdf.


\vspace{3mm}

\item \label{Ciuchini:2001zf}
M.~Ciuchini,
Nucl.\ Phys.\ Proc.\ Suppl.\  {\bf 109} (2002) 307
[hep-ph/0112133].
%
%

\vspace{3mm}

\item \label{pQCD}
Y.Y.~Keum, H.n.~Li and A.I.~Sanda, Phys. Lett. B~{\bf 504} (2001) 6;
Phys. Rev. D~{\bf 63} (2001) 054008; hep-ph/0201103; 
Y.Y.~Keum and H.n.~Li, Phys. Rev. D~{\bf 63} (2001) 074006; 
C.D.~Lu, K.~Ukai and M.Z.~Yang, Phys. Rev. D~{\bf 63} 
(2001) 074009; 
C.-H.~Chen, Y.Y.~Keum and H.-n. Li, 
Phys.Rev.D~{\bf 64} (2001) 112002; hep-ph/0204166;  
S.~Mishima, Phys.Lett. B~{\bf 521} (2001) 252. 

\vspace{3mm}

\item \label{babar}
BaBar Collaboration (B. Aubert et al.), 
hep-ex/0207055.


\vspace{3mm}

\item \label{rk}
R.~Bartoldus, talk on Review of rare two-body B decays at FPCP
workshop, May 17, 2002.

\newpage

\item \label{keum}
Y.Y.~Keum, Proceeding at the 3rd workshop on Higher Luminosity B Factory,
Aug. 6-7, 2002; hep-ph/0209002; hep-ph/0209014; 
Y.Y.~Keum et al., in preparation.

%
%

\vspace{3mm}

\item \label{BB2} 
 G. Buchalla and A.~J. Buras, Nucl. Phys. B {\bf 548} (1999) 309 [hep-ph/9901288].

\vspace{3mm}

\item \label{BB} 
 G. Buchalla and A.~J. Buras, Nucl. Phys. B {\bf 398} (1993) 285; 
 {\bf 400} (1993) 225;  
 {\bf 412} (1994) 106 [hep-ph/9308272];

\vspace{3mm}

\item \label{MU} 
M. Misiak and J. Urban, Phys. Lett. B. {\bf 451} (1999) 161
[hep-ph/9901278].

\vspace{3mm}

\item \label{Falk_Kp}
A.~F.~Falk, A.~Lewandowski and A.~A.~Petrov,
Phys. Lett. B. {\bf 505} (2001) 107 [hep-ph/0012099].

\vspace{3mm}

\item \label{LW} 
M. Lu and M. Wise, Phys. Lett. B. {\bf 324} (1994) 461
[hep-ph/9401204].

\vspace{3mm}

\item \label{MP} 
W.J. Marciano and Z. Parsa, Phys. Rev. D {\bf 53} (1996) R1.

\vspace{3mm}

\item \label{Litt} 
L. Littenberg, Phys. Rev. D{39} (1989) 3322.

\vspace{3mm}

\item \label{CPC} 
G. Buchalla and G. Isidori, Phys. Lett. B. {\bf 440} (1998) 170 [hep-ph/9806501]; \\
D. Rein and L.M. Sehgal, Phys. Rev. D{39} (1989) 3325.

\vspace{3mm}

\item \label{rising}
G.~D'Ambrosio and G.~Isidori,
Phys. Lett. B. {\bf 530} (2002) 108 [hep-ph/0112135].

\vspace{3mm}

\item \label{kettel}
S.~Kettell, L.~Landsberg and H.~Nguyen, hep-ph/0212321.


\vspace{3mm}

\item \label{BCIRS} 
A.~J. Buras {\it et al.}, Nucl. Phys. B {\bf 566} (2000) 3
[hep-ph/9908371].

\vspace{3mm}

\item \label{GN} 
Y. Grossman and Y. Nir, Phys. Lett. B {\bf 398} (1997) 163
[hep-ph/9701313].

\vspace{3mm}

\item \label{ref:kpnn.pnn1}{S. Adler {\it et al.} [E787 Collaboration], 
 Phys. Rev. Lett. {\bf 88}  (2002) 041803 and references therein.}

\vspace{3mm}

\item \label{ref:kpnn.pnn2}{S. Adler {\it et al.} [E787 Collaboration], 
 Phys. Lett. B~{\bf 537} (2002) 211 [hep-ex/0201037].}

\vspace{3mm}

\item \label{ref:kpnn.E949}{http://www.phy.bnl.gov/e949/ }

\vspace{3mm}

\item \label{ref:kpnn.CKMexpt}
{P.S. Cooper, Nucl. Phys. Proc. Suppl. {\bf 99B} (2001) 121; \\ 
http://www.fnal.gov/projects/ckm/Welcome.html }

\vspace{3mm}

\item \label{ref:kpnn.Diwan}{M.V.Diwan, hep-ex/0205089, 
La Thuile, Italy, 2002.}

\vspace{3mm}

\item \label{KTeV_nn}
A.~Alavi-Harati {\it et al.} [KTeV Collaboration],
Phys.\ Rev.\ D {\bf 61} (2000) 072006 [hep-ex/9907014].

\vspace{3mm}

\item \label{ref:kpnn.E391a}{http://psux1.kek.jp/$\sim$e391/ }

\vspace{3mm}

\item \label{ref:kpnn.KOPIO}{A. Konaka, 
hep-ex/9903016;  http://pubweb.bnl.gov/people/e926/ }

\vspace{3mm}

\item \label{ref:kpnn.CKMfitter}  
H. H\" ocker, H. Lacker, S. Laplace and F. Le Diberder, 
Eur. Phys. J. C~{\bf 21} (2001) 225, \break
hep-ph/0104062, 
http://ckmfitter.in2p3.fr/ 

\vspace{3mm}

\item \label{ref:kpnn.LSL.HF9}{L.S. Littenberg, hep-ex/0201026, 
to appear in the Proc. 9th Intl. Symp. on Heavy Flavor Physics. }
                                
\end{enumerate}

\newpage

\thispagestyle{empty}
~

\newpage

\chapter{SUMMARY AND HIGHLIGHTS}
\label{chap:VII}

With over two hundred participants, more than eighty presentations in 
plenary and parallel sessions, and twelve discussion sessions,
the first Workshop on the CKM Unitarity Triangle provided an
opportunity for an 
exploration of the status, open issues, and future directions in the
understanding of the quark mixing matrix. Thirteen more months for the 
preparation of this report have allowed to update results, refine 
most of the studies presented in February~2002, and add coherence to their 
presentation. It has also been an opportunity for the participants to 
discuss further, and to reach a consensus on several issues which had been 
debated at the Workshop. As a result, these proceedings could be written in 
the  form of a coherent document with a common signatory list
for each chapter, corresponding to the original working groups. On some 
issues we agreed to disagree: the continuation of this Workshop series 
will have to address these subjects.

The main goal of this first Workshop was to review  the status of the CKM 
Unitarity Triangle at the end 
of the B physics 
studies at LEP, SLD and CESR and during 
the hand-over of the responsibility for their 
continuation, with even higher accuracy and sensitivity, to the B factories
and the Tevatron. 
Chapter~1 introduces the CKM matrix and the Unitarity Triangle (UT), briefly 
describes the theoretical framework, and recalls the development of the 
B physics studies throughout a decade characterized by the operation of 
the LEP and SLC colliders, the Tevatron Run~I, and 
their role 
 in complementing the data obtained at CESR.

The cleanest way to measure the individual elements of the CKM mixing matrix 
is the determination of the yield of tree-level { semileptonic} 
processes which can be reliably computed.
These measurements have reached their full maturity thanks to both the increasingly 
large data sample available, and the advancements in the theoretical understanding.  
Chapter~2 is devoted to the determination of $|V_{us}|$ from $K_{\ell3}$ 
decays and that of $|V_{ud}|$ from super-allowed Fermi transitions and 
neutron beta decay. The relative accuracy on $|V_{us}|$ has now reached the 
1\% level while that on $|V_{ud}|$ is approaching a factor of twenty better. 
This requires  special attention in evaluating the theoretical uncertainties
that affect these determinations.
In fact, they represent a large fraction of the total uncertainty in both 
cases. A likely explanation of the present 2.2~$\sigma$ 
discrepancy in the unitarity relation $|V_{ud}|^2+|V_{us}|^2+|V_{ub}|^2 =1$
is an underestimate of these effects. 
In the absence of a clear indication of which 
uncertainty has been underestimated, 
it has been proposed to take the value 
$|V_{us}| = 0.2240 \pm 0.0036$ as a conservative estimate of $|V_{us}|$, which 
assumes  CKM unitarity. Possible improvements, which can be
expected for both $|V_{us}|$ and  $|V_{ud}|$ in the near future, are discussed. 
Particularly promising is the extraction of $|V_{us}|$ from $K_{\ell3}$ 
decays. Here we soon expect 
new, precise data with a consistent treatment of radiative corrections and, 
at the same time, new theoretical evaluations of the $SU(3)$ breaking effects. 
Concerning $|V_{ud}|$, a challenging opportunity for experiment is offered 
by the theoretically clean $\pi_{e 3}$ decay.

The other class of tree-level decays which is central to the CKM UT
 studies is represented by the $b \to c \ell \bar{\nu}$ and 
$b \to u \ell \bar{\nu}$ processes, which give access to $|V_{cb}|$ and 
$|V_{ub}|$, respectively. The discussion of the status of the extraction of 
these two CKM elements has attracted a large part of the participants and of 
the presentations at the Workshop. The third chapter of this 
report tries to summarize the status, 
with important updates 
from the Summer~2002 conferences.

With the statistical accuracy approaching the few percent level for 
some of these measurements, an important issue here is to carefully test the 
underlying theoretical assumptions. Both inclusive and exclusive
decays are routinely studied.
The measurement of $\vcb$ 
from exclusive decays is limited by the theoretical uncertainty on the
value of the form factor at maximum $q^2$ (see Table~\ref{tab_fin}). 
A detailed analysis of lattice QCD results and their uncertainty is
now available. Progress in  these studies, as well as more precise data from the B
factories, are expected. 
Important progress has been made during and after the Workshop on the 
$\vcb$ extraction from inclusive semileptonic decays. The current accuracy of 
experimental measurements is at the percent level. 
While  perturbative QCD corrections have been studied in  different
frameworks and the related uncertainty seems  under control,
the non-perturbative parameters appearing in the Operator Product
Expansion (OPE) have to be constrained using experimental data.
Several new analyses of the moments of distributions in semileptonic
and radiative decays have appeared in the last year. 
Their results demonstrate that a significant
 fraction of the uncertainty on $\vcb$, 
arising from these parameters, 
can be absorbed in the experimental uncertainty. As shown in Table~\ref{tab_fin}, the remaining 
theoretical uncertainty is now at the level of a percent. 
The consistent picture emerging from these preliminary studies represents a
remarkable success for the OPE and bolsters confidence in the 
inclusive $\vcb$ determination, as no violation of parton-hadron
duality has been  detected at the present level of accuracy. 

The CKM-suppressed counterpart of  the $b \to c \ell \bar{\nu}$ process is the 
$b \to u \ell \bar{\nu}$ decay which measures $\vub$, the smallest element in
the CKM mixing matrix. Since its rate is only about 1/60 of that for $b \to c$ 
transition, measuring this charmless decay accurately is a formidable experimental 
challenge. We have known for more than a decade that this process is present, 
thanks to  pioneering measurements  by ARGUS and CLEO.
The non-vanishing of $\vub$ is a pre-requisite for 
explaining CP violation within the Standard Model (SM). Knowing its magnitude 
accurately is a top priority for testing the CKM UT. In this respect,
significant progress has been made.
Theorists have devised a number of strategies to pin down the value of $\vub$
with good accuracy. Large data sets, complementary kinematical
characteristics  of the signal 
events at $\Upsilon(4S)$ and $Z^0$ energies, and experimental ingenuity have 
provided a significant set of results. None of them is approaching the accuracy 
obtained on $\vcb$ and the debate in the community on these fairly recent 
developments is still lively. But the results, 
obtained with very different methods, are all consistent and the overall accuracy 
amounts to better than 15\% (see Table~\ref{tab_fin}).

Another important area of  studies of tree-level B decays, discussed in Chapter~3, 
is represented by the determination of the exclusive and inclusive $b$-hadron lifetimes. 
Those for the lighter mesons are presently known to an accuracy of one 
percent, or better. For ${\rm B}^0_s$ and $b$-baryons important data has 
already been gathered and the Tevatron is expected to reach soon a similar 
accuracy. Lifetime differences between $b$-hadrons have been compared with 
expectations from OPE to test our understanding of 
the non-spectator contributions to beauty hadron decays. While this 
comparison has shown a consistent agreement between predictions and 
measurements in the meson sector, 
the measured ratio of baryon to meson lifetimes deviates from 
its initial expectation. 
This has been interpreted as a possible signal of problems in the underlying 
theory assumptions. However, 
much better  agreement with data is re-established once the NLO corrections are 
included.

There remain two CKM elements, $\vts$ and $\vtd$,
which have been so far accessed only  through 
box diagrams. They can be probed by  $\rm K^0-\overline K^0$ and 
${\rm B}^0_{d,s}-\overline {\rm B}^0_{d,s}$ mixing. The experimental status 
of these studies and that of the non-perturbative calculations of $\hat B_K$, 
$\sqrt{\hat B_{B_d}} F_{B_d}$, $\sqrt{\hat B_{B_s}} F_{B_s}$, and $\xi$ is 
reviewed  in Chapter~4. 

The theoretical discussion has centred on the determination of the
non-perturbative parameters for neutral meson mixing. For $F_{B_q}$ and
$\xi$, lattice calculations with two dynamical flavours of quarks are
becoming common and the first $2{+}1$ dynamical calculations have
appeared. Much attention has focused on the chiral extrapolations
needed to obtain the physical results, particularly for $\xi$, with a
final lattice value for UT fits given as
$\xi = 1.24(4)(6)$.
QCD sum rules give very consistent results for the B meson decay
constants, slightly less so for the $B$-parameters.
For neutral kaon mixing the benchmark lattice calculations are
quenched and lead to the  final result given in Table~\ref{tab_fin}.
The systematic uncertainty includes the estimate for quenching effects
and is considered to be very conservative (a more aggressive error estimate 
is given in Sec.~\ref{sec:KKbar-mixing} of Chapter 4). The best-developed alternative
technique to evaluate $\hat B_K$ is the large-$N_c$ expansion and
gives a consistent result, although the chiral corrections are more
than $100\%$ in this case.

The time structure of $\rm B^0$-$\overline{\rm B}^0$ oscillations has been precisely measured 
in the ${\rm B}_d^0$ sector. 
The LEP, Tevatron and SLC results are in excellent 
agreement with those obtained  at the B factories. After the inclusion of 
the latter, the oscillation frequency $\Delta M_d$ is known with a 
precision of about 1$\%$ (see Table~\ref{tab_fin}).
Further improvements are expected, 
which should bring the accuracy on $\Delta M_d$ to a few per mille.

On the other hand, 
 ${\rm B}_s^0$-$\overline{\rm B}_s^0$ oscillations have not been observed yet,
even though the experimental effort has allowed to largely exceed the anticipated 
sensitivity. Today we know that ${\rm B}_s^0$ mesons oscillate at least thirty times 
faster than ${\rm B}_d^0$ mesons. The final result of the searches at LEP and SLC is 
$\dms > 14.4 ~ \mbox{ps}^{-1}~\mbox{at 95\% C.L.}$,  with a sensitivity of 
$\dms = 19.2$~{ps}$^{-1}$. While 
this much sought-after phenomenon has so far eluded searches, 
the consequent lower limit on 
$\Delta M_s$ has already a significant impact on the determination of the 
UT parameters.
\begin{table}[t]
\begin{center}
\begin{tabular}{|c|c|c|c|c|}
\hline
                          Parameter                & Value  &   Experimental   &  Theory       \\
                                                   &        &   uncertainty   & uncertainty    \\ \hline
                          $\lambda$                & 0.2240 &    0.0036     &     -        \\
 \hline
$\left | V_{cb} \right | (\times 10^{-3})$ (excl.) &  42.1  &      1.1     &     1.9         \\
$\left | V_{cb} \right | (\times 10^{-3})$ (incl.) &  41.4  &      0.7     &     0.6       \\ \hline 
$\left | V_{ub} \right | (\times 10^{-4})$ (excl.) &  33.0  &     2.4     &     4.6       \\
$\left | V_{ub} \right | (\times 10^{-4})$ (incl.\, LEP) &  40.9  &      5.8     &     3.3       \\ 
$\left | V_{ub} \right | (\times 10^{-4})$ (incl.\,CLEO) &  40.8  &      5.2     &     3.9       \\ \hline
                
  $\Delta M_d~(\mbox{ps}^{-1})$      &  0.503 &      0.006  &      -        \\
                  $\Delta M_s~(\mbox{ps}^{-1})$      & $>$ 14.4 at 95\% C.L. & \multicolumn{2}{|c|}
                                                                           {sensitivity 19.2}  \\
            $F_{B_d} \sqrt{\hat B_{B_d}}$(MeV)       &  223   &      33      & 12  \\
$\xi=\frac{ F_{B_s}\sqrt{\hat B_{B_s}}}
                    { F_{B_d}\sqrt{\hat B_{B_d}}}$ & 1.24   &     0.04     & 0.06 \\ \hline
                         $\hat B_K$                &  0.86  &     0.06     &      0.14      \\ \hline
                        sin 2$\beta$               & 0.734   &     0.054 &        -       \\ \hline
\end{tabular} 
\caption[]{ \it {Main experimental and theoretical results entering the UT determination.
Other parameters of interest can be found in Chapters 2-5.
 \label{tab_fin}}}
\end{center}
\end{table}

The extraction of the UT parameters from all these inputs, 
within the Standard Model,  is discussed 
in Chapter~5, which represents a central part of this Workshop. 
Five quantities have been considered to constrain the upper apex of 
the UT in the 
$\bar{\rho}$-$\bar{\eta}$ plane. These are $\epsilon_K$, $|V_{ub}|/|V_{cb}|$, 
$\Delta M_d$, the limit on $\Delta M_s$ and sin 2$\beta$ from the measurement
of the CP asymmetry in the $J/\psi \rm K^0$ decays. 
Comparing the measured values of these observables with 
their theoretical predictions (in the SM or in a different model) yields
 a set of constraints,
which however depend on several parameters,
like quark masses, decay constants of B mesons and non-perturbative 
parameters, such as $\hat B_K$. Their values are constrained by both
measurements and theoretical calculations which are reviewed in Chapter~4.

Different methods have been proposed to combine this information and extract the 
UT parameters. They differ in the treatment of the 
theoretical uncertainties for which 
they adopt either a frequentist or  a Bayesian approach. 
Despite much interest in these studies, no systematic comparison 
of these methods had been performed before this Workshop. Moreover, 
different assumptions on the input parameters made any comparison of published 
results difficult. At the Workshop, different groups agreed to share a common 
set of input values (see Table~\ref{tab:inputs}), provided by the relevant working groups. 
In spite of using the same central values and errors, 
the likelihood functions associated with the input parameters
are different in the two approaches. 
As a consequence, the region defining the 95$\%$ (99$\%$) confidence level for the 
UT parameters is wider by 30$\%$ (20$\%$) in the frequentist as compared
to the Bayesian approach.  
Further tests have shown that, if the same likelihoods are used 
for input quantities, the output results become almost identical. The main origin of the 
difference between  the results in the Bayesian and the frequentist method is therefore the 
likelihood associated to the input quantities. But these differences will decrease 
progressively as the theoretical uncertainties will be 
reduced or related to experimental ones. 
An example of the latter is the extraction of $|V_{cb}|$ from inclusive 
decays, where --- as already mentioned --- 
experimental constraints from the moments have replaced
theoretical estimates in the aftermath of the Workshop.
It is also expected that additional inputs will be determined using unquenched 
Lattice simulations.

Independently of the statistical method adopted, a crucial outcome of these 
investigations is the remarkable agreement of the UT parameters, 
as determined by means of CP conserving quantities sensitive to the UT sides, with 
the CP violation measurements in the kaon sector ($\epsilon_K$) and in the B sector  
(sin2$\beta$). This agreement tells us that, at the present level of accuracy,
 the SM mechanism of flavour and CP violation describes  the data well.
 At the same time,  it is also an important test of the OPE, HQET and Lattice QCD, 
 on which the extraction of the CKM parameters rests. The present accuracy 
is at the 
10\% level; the B~factories and a next generation 
of facilities will improve the sensitivity of these tests by an order of magnitude.
The study of the impact of the uncertainties in the theoretical parameters on the UT 
fits has shown that the uncertainties in $\sqrt{\hat B_{B_d}} F_{B_d}$ have to be 
decreased by at least a factor of two in order to have a significant impact on the UT 
fits --- unless future calculations result in $\sqrt{\hat B_{B_d}} F_{B_d}$ values 
which differ significantly from present results. In the case of $\hat B_K$ and in 
particular $\xi$, even a modest reduction of the theoretical uncertainty could 
already have an important impact on the UT  fits. 

The output for various quantities of interest can be found in Table~\ref{tab:ckm_fit_final};
a pictorial representation of the fit is shown in Fig.~\ref{fig:bande}.
UT fits can also be used to obtain predictions for quantities that 
will only be measured in the future, such as the $\dms$ oscillation frequency, predicted to 
be $<22.2 ~\mbox{ps}^{-1}$, and the angle $\gamma$, predicted to be between 
49.0$^\circ$ and 77.0$^\circ$. These 95\% confidence levels ranges may be considered as 
a reference to which the direct measurements will need to be compared for identifying 
possible signals of New Physics.

While the determination of the triangle sides and the definition of the procedures 
for the UT fits had a central role at the Workshop, a number of topics, which will 
become of increasing importance at future meetings, started to be addressed. They are 
presented as individual contributions in Chapter~6. At this stage, general strategies 
for the determination of the UT need to be formulated. Preliminary 
studies show that the pairs of measurements  $(\gamma,\beta)$, $(\gamma,R_b)$ and 
$(\gamma,\overline{\eta})$ offer  the most  efficient sets of observables to determine 
$(\rhobar,\etabar)$. On the other hand the pair $(R_t,\beta)$ will play the leading role
in the UT fits in the coming years and for this reason it has been suggested to plot the
available constraints on the CKM matrix in the $(R_t,\beta)$ plane.
The present $(R_t,\beta)$ plot corresponding to the usual $(\rhobar,\etabar)$
plot can be found in Fig.~\ref{fig:rtbeta_real}.

There are more measurements of B~decays which are relevant to the CKM studies.
Radiative rare B-decays, very sensitive to  
New Physics in loops, have been reviewed for their potential relevance in probing the CKM UT. 
The combined studies of radiative decays into non-strange hadrons are sensitive to 
$(\rhobar,\etabar)$ and could in principle provide interesting constraints on the UT, 
provided that the theoretical errors can be reduced and that the branching fractions are 
accurately measured. 

The decays $\rm B \rightarrow \pi K$, $\rm B \rightarrow \pi \pi $ and $\rm B \rightarrow 
K \overline{K}$ are emphasised as useful tools for the determination of the 
angles $\alpha$ and $\gamma$. 
Here flavour symmetries and recent dynamical approaches like QCD factorization
and pQCD play the dominant role in the phenomenology. The status of 
QCD factorization has been outlined and also the critical points in view of a possible 
extraction of $\gamma$ have been described. 

Finally, the potential of the rare decays  
$\rm K^+ \rightarrow \pi^+ \nu \overline{\nu}$ and 
$\rm K_L \rightarrow \pi^0 \nu \overline{\nu}$ with respect to the determination of the 
UT has 
been discussed. These decays are essentially free of theoretical uncertainties but are 
subject to parametric uncertainties such as $m_c$ for 
$\rm K^+ \rightarrow \pi^+ \nu \overline{\nu}$ and $V_{cb}$ for both decays. These 
uncertainties should be reduced in the coming years so that the future measurements of 
these decays will provide very important independent measurements of $V_{td}$ and 
$\sin 2 \beta$ and more generally of the UT parameters. The comparison of these 
measurements with those obtained by B decays offers a very powerful tool for testing 
the SM and its extensions.

The CKM Workshop contributed to demonstrate the richness and variety of the 
physics landscape related to the CKM matrix and the Unitarity Triangle. 
Results from the LEP, SLC, CESR and Tevatron experiments have already 
provided us with an impressionistic 
outline of this landscape, which the B~factories and the Run~II at the 
Tevatron are now revealing in greater detail. While the CKM matrix appears 
likely to be the dominant source of flavour and CP violation, New Physics 
contributions may still  modify the shape 
of the UT 
and be revealed by  forthcoming studies. In this 
context the measurements of the angle $\gamma$ in non-leptonic B decays and 
those of $\Delta M_s$ will mark important new steps in the search for 
New Physics in the $\bar{\rho}$-$\bar{\eta}$ plane.
The present deviation from the SM expectation of 
the CP~asymmetry in ${\rm B}_d^0\to \phi {\rm K}_S$ also awaits a 
clarification, and the improved data on several rare decays will be very 
important in this programme. All this 
will be the subject of  future CKM Workshops.

\end{document}